# Roadmap on Neuromorphic Photonics

**Daniel Brunner[1,♦], Bhavin J. Shastri[2,♦], Mohammed A. Al-Qadasi[3], H. Ballani[4], Sylvain Barbay[5], Stefano Biasi[6], Peter Bienstman[7], Simon Bilodeau[8], Wim Bogaerts[7], Fabian Böhm[9], G. Brennan[4], Sonia Buckley[10], Xinlun Cai[11], Marcello Calvanese Strinati[12], B. Canakci[4], Benoit Charbonnier[13], Mario Chemnitz[14,15], Yitong Chen[16], Stanley Cheung[17], Jeff Chiles[10], Suyeon Choi[18], Demetrios N. Christodoulides[19], Lukas Chrostowski[3], J. Chu[4], J. H. Clegg[4], D. Cletheroe[4], Claudio Conti[12,20], Qionghai Dai[16], Luigi Di Lauro[22], Nikolaos-Panteleimon Diamantopoulos[23], Niyazi Ulas Dinc[24], Jacob Ewaniuk[2], Shanhui Fan[25], Lu Fang[26], Riccardo Franchi[6,✧], Pedro Freire[28], Silvia Gentilini[21], Sylvain Gigan[29], Gian Luca Giorgi[31], C. Gkantsidis[4], J. Gladrow[4], Elena Goi[32], M. Goldmann[1], A. Grabulosa[1], Min Gu[32], Xianxin Guo[33], Matěj Hejda[34], F. Horst[35], Jih-Liang Hsieh[24], Jianqi Hu[29], Juejun Hu[36], Chaoran Huang[37], Antonio Hurtado[38], Lina Jaurigue[39], K. P. Kalinin[4], Morteza Kamalian-Kopae[28], D. J. Kelly[4], Mercedeh Khajavikhan[19], H. Kremer[4], Jeremie Laydevant[40,41], Joshua C. Lederman[8], Jongheon Lee[19], Daan Lenstra[43], Gordon H.Y. Li[44], Mo Li[46,47], Yuhang Li[48], Xing Lin[26,27], Zhongjin Lin[11], Mieszko Lis[3], Kathy Lüdge[39], Alessio Lugnan[6], Alessandro Lupo[49], A. I. Lvovsky[50], Egor Manuylovich[28], Alireza Marandi[44,45], Federico Marchesin[7], Serge Massar[49], Adam N. McCaughan[10], Peter L. McMahon[40,42], Miltiadis Moralis-Pegios[51], Roberto Morandotti[22], Christophe Moser[24], David J. Moss[52], Avilash Mukherjee[3], Mahdi Nikdast[53], B.J. Offrein[35], Ilker Oguz[24], Bakhrom Oripov[10], G. O'Shea[4], Aydogan Ozcan[48], F. Parmigiani[4], Sudeep Pasricha[53], Fabio Pavanello[54], Lorenzo Pavesi[6], Nicola Peserico[55,69], L. Pickup[4], Davide Pierangeli[20,21], Nikos Pleros[51], Xavier Porte[38], Bryce A. Primavera[10], Paul Prucnal[8], Demetri Psaltis[24], Lukas Puts[43], Fei Qiao[26], B. Rahmani[4], Fabrice Raineri[5,68], Carlos A. Ríos Ocampo[56], Joshua Robertson[38], Bruno Romeira[57], Charles Roques-Carmes[25], Nir Rotenberg[2], A. Rowstron[4], Steffen Schoenhardt[32], Russell L. T. Schwartz[55,69], Jeffrey M. Shainline[10], Sudip Shekhar[3], A. Skalli[1], Mandar M. Sohoni[40], Volker J. Sorger[55,69], Miguel C. Soriano[31], James Spall[33], Ripalta Stabile[43], Birgit Stiller[58,59], Satoshi Sunada[60], Anastasios Tefas[51], Bassem Tossoun[61], Apostolos Tsakyridis[51], Sergei K. Turitsyn[28], Guy Van der Sande[62], Thomas Van Vaerenbergh[34], Daniele Veraldi[20], Guy Verschaffelt[62], E.A. Vlieg[35], Hao Wang[29,30], Tianyu Wang[63], Gordon Wetzstein[18], Logan G. Wright[64], Changming Wu[46], Chu Wu[26], Jiamin Wu[16], Fei Xia[29], Xingyuan Xu[65], Hangbo Yang[55,69], Weiming Yao[43], Mustafa Yildirim[24], S. J. Ben Yoo[66], Nathan Youngblood[67], Roberta Zambrini[31], Haiou Zhang[26] and Weipeng Zhang[8]**

1. Université Marie et Louis Pasteur, CNRS UMR 6174, institut FEMTO-ST, 25000 Besançon, France
2. Centre for Nanophotonics, Department of Physics, Engineering Physics & Astronomy, Queen's University, Canada
3. Department of Electrical and Computer Engineering, The University of British Columbia, Vancouver, Canada
4. Microsoft Research, Cambridge, UK
5. Université Paris-Saclay, CNRS, Centre de Nanosciences et de Nanotechnologies, France
6. Nanoscience Laboratory, Department of Physics, University of Trento, Italy
7. Photonics Research Group, Department of Information Technology, Ghent University/imec, Belgium
8. Princeton University, NJ, USA
9. Hewlett Packard Labs, Hewlett Packard Enterprise, Böblingen, Germany
10. National Institute of Standards and Technology, Boulder, CO, USA

11. State Key Laboratory of Optoelectronic Materials and Technologies, School of Electronics and Information Technology, Sun Yat-sen University, China
12. Enrico Fermi Research Center, Rome, Italy
13. Université Grenoble-Alpes, CEA, Leti, Grenoble, France
14. Leibniz-Institute of Photonic Technology, Jena, Germany
15. Institute of Applied Optics and Biophysics, Jena, Germany
16. Department of Automation, Tsinghua University, Beijing, China
17. Department of Electrical and Computer Engineering, North Carolina State University, NC, USA
18. Department of Electrical Engineering, Stanford University, CA, USA
19. University of Southern California, Los Angeles, CA, USA
20. Department of Physics, Sapienza University, Rome, Italy
21. Institute for Complex Systems, National Research Council, Rome, Italy
22. Institut National de la Recherche Scientifique-Énergie Matériaux Télécommunications (IN), Varennes, Canada
23. NTT Device Technology Labs, NTT Corporation, Atsugi, Kanagawa, Japan
24. Institute of Electrical and Microengineering, School of Engineering, École Polytechnique Fédérale de Lausanne, Switzerland
25. Edward L. Ginzton Laboratory, Stanford University, Stanford, CA, USA
26. Department of Electronic Engineering, Tsinghua University, China
27. Beijing National Research Center for Information Science and Technology, Tsinghua University, China
28. Aston Institute of Photonic Technologies (AiPT), Aston University, Birmingham, UK
29. Laboratoire Kastler Brossel, ENS-Universite PSL, CNRS, Sorbonne Université, France
30. Department of Precision Instruments, Tsinghua University, China
31. Instituto de Física Interdisciplinar y Sistemas Complejos (IFISC, UIB-CSIC), Campus Universitat de les Illes Balears, Spain
32. University of Shanghai for Science and Technology, China
33. Lumai Ltd., Wood Centre for Innovation, Oxford, UK
34. Hewlett Packard Labs, Hewlett Packard Enterprise, Diegem, Belgium
35. IBM Research Europe, Zürich, Switzerland
36. Department of Materials Science and Engineering, Massachusetts Institute of Technology, Cambridge, MA, USA
37. Department of Electronic Engineering, The Chinese University of Hong Kong, SAR China
38. Institute of Photonics, SUPA Dept of Physics, University of Strathclyde, Glasgow, UK
39. Technische Universität Ilmenau, Institut für Physik, Germany
40. School of Applied and Engineering Physics, Cornell University, Ithaca, NY, USA
41. USRA Research Institute for Advanced Computer Science, Mountain View, CA, USA
42. Kavli Institute at Cornell for Nanoscale Science, Cornell University, Ithaca, NY, USA
43. Eindhoven Hendrik Casimir Institute (EHCI), Eindhoven University of Technology, The Netherlands
44. Department of Applied Physics, California Institute of Technology, Pasadena, USA
45. Department of Electrical Engineering, California Institute of Technology, Pasadena, USA
46. Department of Electrical and Computer Engineering, University of Washington, WA, USA
47. Department of Physics, University of Washington, Seattle, WA, USA
48. University of California, Los Angeles, CA, USA
49. Université libre de Bruxelles, Belgium
50. Clarendon Laboratory, University of Oxford, UK
51. Department of Informatics, Center for Interdisciplinary Research and Innovation, Aristotle University of Thessaloniki, Greece
52. Optical Sciences Centre, Swinburne University of Technology, Australia
53. Colorado State University, Fort Collins, CO, USA
54. Université Savoie Mont Blanc, CNRS, Grenoble INP, CROMA, Grenoble, France

55. Florida Semiconductor Institute, University of Florida, FL, USA
56. Department of Materials Science and the Institute for Research in Electronics and Applied Physics, University of Maryland, College Park, MD, USA
57. International Iberian Nanotechnology Laboratory, Braga, Portugal
58. Max-Planck-Institute for the Science of Light, Erlangen, Germany
59. Institute of Photonics, Gottfried Wilhelm Leibniz University, Hannover, Germany
60. Institute of Science and Engineering, Kanazawa University, Japan
61. Hewlett Packard Labs, Hewlett Packard Enterprise, Santa Barbara, CA, USA
62. Applied Physics Research Group, Vrije Universiteit Brussel, Belgium
63. Department of Electrical and Computer Engineering, Boston University, Boston, MA, USA
    - ❖ Currently with Nanomaterials & Nanostructure Optics, Department of Electrical and Computer Engineering, Boston University, Boston, MA, USA
64. Department of Applied Physics, Yale University, New Haven, CT, USA
65. State Key Laboratory of Information Photonics and Optical Communications, Beijing University of Posts and Telecommunications, China
66. Department of Electrical and Computer Engineering, University of California, Davis, USA
67. Department of Electrical and Computer Engineering, University of Pittsburgh, PA, USA
68. Inphyni, Université Côte d'Azur, France
69. Electrical & Computer Engineering Department, University of Florida, FL, USA

♦ Guest editors of the Roadmap.

E-mails: daniel.brunner@femto-st.fr bhavin.shastri@queensu.ca

## Abstract

Neuromorphic photonics are processors inspired by the human brain and enabled by light (photons) instead of traditional electronics. Neuromorphic photonics and its associated concepts are experiencing a significant resurgence, building on foundational research from the 1980s and 1990s. This renewed momentum is driven by breakthroughs in photonic integration, nonlinear optics, and advanced materials, alongside the growing necessity of neuro-inspired computing in numerous applications of economic and societal relevance. The increasing demand for energy-efficient artificial intelligence (AI) solutions underscores the need for innovation and a cohesive vision to address key challenges, including scalability, energy efficiency, precision, and standardized performance benchmarks. Together, these efforts present an opportunity to establish a unique photonic advantage with practical, real-world applications.

This roadmap consolidates recent advances while exploring emerging applications, reflecting the remarkable diversity of hardware platforms, neuromorphic concepts, and implementation philosophies reported in the field. It emphasizes the critical role of cross-disciplinary collaboration in this rapidly evolving field.

The roadmap introduces various approaches to embedding the high-complexity transformations central to neuromorphic computing, focusing on frequency, delay, and spectral embeddings. This is followed by a discussion of architectures of photonic neural networks (PNNs) and an in-depth analysis of methods for implementing these architectures in photonic hardware. Dedicated sections delve into integrated photonic hardware, the realization of photonic weights and memories, and the optimization of training processes for photonic neuromorphic architectures. The roadmap concludes by exploring numerous potential applications, highlighting the challenges and advances necessary to transition neuromorphic photonic computing from a primarily academic pursuit to a technology with economic and societal impact. By synthesizing contributions from over 40 research teams, this roadmap aims to provide the photonics community with a comprehensive framework for unlocking the transformative potential of PNNs in advancing AI and beyond.

## Contents

# Frequency Multiplexed Photonic Neuromorphic Computing

**Alessandro Lupo[1] and Serge Massar[1]**
[1]Université libre de Bruxelles, Belgium (Laboratoire d'Information Quantique CP224, Av. F. D. Roosevelt 50, 1050 Bruxelles, Belgium)
[alessandro.lupo@ulb.be, serge.massar@ulb.be]

## Status

Frequency is among the most interesting and accessible degrees of freedom of light. Wavelength-division multiplexing (WDM) has been employed since the late 1980s as a natural way to multiply the capacity of optical communication channels [Oakley88, Faulkner89, Cusworth90]. Exploiting the frequency degree of freedom of light for information processing is less developed, in part because it is rather counterintuitive. In the past, the frequency domain has been used in the context of quantum optics, including for quantum key distribution [Merolla99] and for the generation of two-photon frequency entanglement [Olislager14]. These experiments were based on optical frequency combs whose lines were made to interfere in the spectral domain by electro-optical phase modulation. A similar scheme has been exploited to implement frequency multiplexed neuromorphic computing, where information is encoded in the comb line amplitudes and mixed by phase modulation, replicating the effect of a set of untrained connections in a neural network. This principle has been employed to realize both Reservoir Computers (RC) [Butschek22] and Extreme Learning Machines (ELM) [Lupo21A] which are, respectively, recurrent neural networks and feed-forward neural networks with one hidden layer. These schemes are economical in terms of hardware, as multiple "neurons" are processed in parallel using the same optical circuit, and multiple networks can be operated in parallel in different frequency bands as in [Lupo21B]. Furthermore, using a programmable spectral filter, one can readily apply output weights directly in the optical domain. These advantages have been demonstrated recently in a frequency multiplexed implementation of Deep Reservoir Computing [Lupo23].

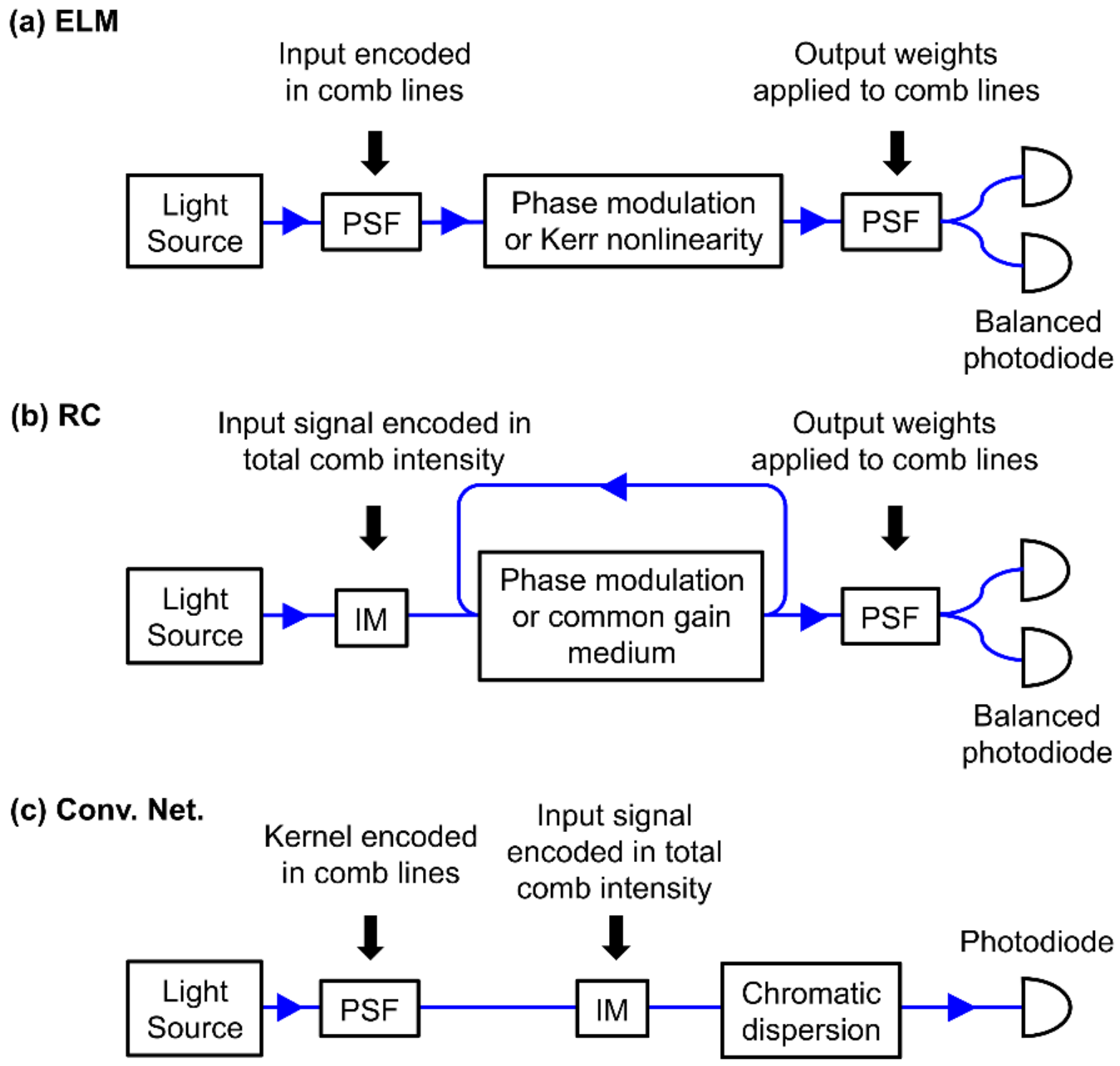

**Figure 1.** Schematic depiction of some frequency multiplexed neuromorphic computing systems. PSF: Programmable Spectral Filter, IM: Intensity Modulator. Depending on the specific application, the light source may be an optical comb (for instance an electro-optically modulated CW laser) or a high peak power pulsed laser. (a) ELM scheme based either on phase modulation [Lupo21A] or Kerr nonlinearity [Zhou22, Fischer23,Lee24, Zajnulina23]. (b) RC scheme based either on phase modulation [Butschek22] or on the mixing provided by a common gain medium [Li23]. (c) convolutional processor based on chromatic dispersion [Xu21].

Other systems based on frequency multiplexing but using alternative sources of mixing have been realized in the past years,- see, for example, a RC with neurons encoded in different modes of a Fabry-Perot laser [Li23], and a RC using nonlinear frequency mixing in a Lithium Niobate waveguide [Yildirim22]. The Kerr effect has also been studied as a frequency-mixing mechanism, both in a high-power regime (i.e. using short high peak power optical pulses) [Zhou22, Fischer23, Lee24] and a relatively low-power regime (i.e. using a CW laser) [Zajnulina23]. The frequency and time degrees of freedom can be combined, as in the high-speed convolution accelerator based on chromatic dispersion reported in [Xu21]. A summary of these schemes is reported in Fig. 1.

**Current and Future Challenges**

The field of frequency multiplexed neuromorphic information processing is in its early stage of development. The main interest in this approach is the possibility of high parallelism, making the systems economic in terms of hardware use. The main challenges, faced by all photonic neuromorphic computing systems, are a) to increase the complexity of the tasks that can be solved by using better hardware or new algorithms, and b) to improve scalability by reducing energy consumption and footprint while increasing speed of operation.

Concerning power consumption, mixing wavelengths in frequency multiplexed systems usually implies a high power cost. For instance, electro-optical phase modulation requires strong RF signals to achieve good mixing (up to almost 1W in Refs. [Lupo21A, Butschek22, Lupo23]), while exploiting optical nonlinearities such as the Kerr effect typically requires high optical powers (for instance the Kerr-induced nonlinear phase exploited to mix information in Ref. [Zhou22] is approximately 160 rad). Such a high nonlinearity requires long or highly nonlinear fibers or waveguides and high peak power pulsed lasers. Ref. [Zajnulina23] showed that useful information mixing already occurs with a Kerr-induced nonlinear phase of 0.3 rad (or lower), suggesting that it may be possible to avoid the use of high optical power. However, this concept has only been demonstrated on a 20-neuron network and is still to be validated on larger neuromorphic models.

Another challenge is the need for high-resolution and high-speed programmable spectral filters. These filters are required for manipulating light in the wavelength domain and are employed for different purposes by the implementations reported in the previous section. The most relevant filtering specifications are the minimum bandwidth and the settling time. The minimum bandwidth defines the minimum resolution at which the light spectrum can be manipulated, thus lower values enable a denser information encoding. The settling time defines the rate at which the filter can be updated (this quantity is not necessarily related to the overall information processing speed, for instance when the filter is used in the output layer). The most common commercial technology for programmable spectral filtering is based on liquid crystals (Waveshaper from Coherent). This commercial solution operates in the C and L bands and offers a minimum bandwidth of 10 GHz (equivalent to 0.08 nm) and a settling time of 500 ms. Assuming a 10 GHz channel spacing, the C-band could be divided into approximately 430 channels, representing, for example, neurons of a neuromorphic computing implementation.

Frequency multiplexed implementations often employ optical frequency combs, with comb lines made to interfere through frequency mixing. One of the challenges is the generation of broad and stable frequency combs (if the comb spacing is not stable, the mixing would generate new sidebands instead of producing interference among existing comb lines).

Other challenges, common to every photonic neuromorphic computing platform, are scalability and resistance to noise. Scalability is difficult to achieve in bulk systems, hence we foresee an increasing need for reliable photonic integration. The noise mitigation strategies strongly depend on the implementation, but in general, better performance comes at the cost of ADC/DAC converters with higher resolution, which implies a trade-off with power consumption.

### Advances in Science and Technology to Meet Challenges

Processing information in the frequency domain is not very intuitive. For this reason, there is probably significant scope for conceptual advances, such as the development of new neuromorphic algorithms and training methods, as well as alternative approaches to using the frequency degree of freedom.

Photonic integration is probably the key to increasing scalability. Progress in photonic integration will lead to less variability in component specifications and less optical losses, and of course reduced footprint. These improvements can be obtained both by increasing the reliability of fabrication processes and by implementing on-chip tuning mechanisms. For a preliminary proposal for an integrated RC based on frequency multiplexing, see [Kassa18]. Integrated optics can provide high-performance frequency combs, for example via soliton microcombs or mode-locked lasers, see [Chang22] for a review. In addition, programmable spectral filters with a density higher than the DWDM standard can be realized in integrated photonics, achieving sub-ms settling times, see, for example, [Jonuzi23].

As discussed above, frequency multiplexing systems can have a high power cost. It may be possible to circumvent this by using on-chip amplification and highly nonlinear waveguides, as well as approaches that require small nonlinear phases [Zajnulina23]. Alternative approaches to information mixing in the frequency domain, for instance exploiting chromatic dispersion [Xu21], or via the interaction with a common gain medium [Li23], may be useful.

### Concluding Remarks

Frequency multiplexing provides a powerful and natural way to enhance the bandwidth of optical information processing systems. This method has been applied intensively in telecommunication but its potential is only starting to be explored in the field of neuromorphic computing. Many commercially available solutions for manipulating the wavelength of light in WDM systems already exist, derived from the telecommunication industry. Further advances in integrated optics, for instance for programmable spectral filtering, could enable a wider implementation of frequency multiplexed systems for neuromorphic computing.

### Acknowledgments

We acknowledge funding from the European Union (grant 860830 “POST DIGITAL”) and from the FRS-FNRS (grants EOS O.0019.22F “PINCH” and J.01243.24 “FreqNeuroPhot”).

### References

[Butschek22] L. Butschek, A. Akrout, E. Dimitriadou, A. Lupo, M. Haelterman, and S. Massar, “Photonic reservoir computer based on frequency multiplexing,” Optics Letters, vol. 47, no. 4. Optica Publishing Group, p. 782, Feb. 03, 2022. doi: 10.1364/ol.451087.

[Chang22] L. Chang, S. Liu, and J. E. Bowers, “Integrated optical frequency comb technologies,” Nature Photonics, vol. 16, no. 2. Springer Science and Business Media LLC, pp. 95–108, Feb. 2022. doi: 10.1038/s41566-021-00945-1.

[Cusworth90] S. D. Cusworth, J. M. Senior, and A. Ryley, “Wavelength division multiple access on a high-speed optical fibre LAN,” Computer Networks and ISDN Systems, vol. 18, no. 5. Elsevier BV, pp. 323–333, Jun. 1990. doi: 10.1016/0169-7552(90)90120-h.

[Faulkner89] D. W. Faulkner, D. B. Payne, J. R. Stern, and J. W. Ballance, “Optical networks for local loop applications,” Journal of Lightwave Technology, vol. 7, no. 11. Institute of Electrical and Electronics Engineers (IEEE), pp. 1741–1751, 1989. doi: 10.1109/50.45897.

[Fischer23] B. Fischer et al., “Neuromorphic Computing via Fission-based Broadband Frequency Generation,” Advanced Science, vol. 10, no. 35. Wiley, Oct. 02, 2023. doi: 10.1002/advs.202303835.

[Jonuzi23] T. Jonuzi, A. Lupo, M. C. Soriano, S. Massar, and J. D. Domenéch, “Integrated programmable spectral filter for frequency-multiplexed neuromorphic computers,” Optics Express, vol. 31, no. 12. Optica Publishing Group, p. 19255, May 24, 2023. doi: 10.1364/oe.489246.

[Kassa18] W. Kassa, E. Dimitriadou, M. Haelterman, S. Massar, and E. Bente, “Towards integrated parallel photonic reservoir computing based on frequency multiplexing,” Neuro-inspired Photonic Computing. SPIE, May 21, 2018. doi: 10.1117/12.2306176.

[Lee24] K. F. Lee and M. E. Fermann, “Supercontinuum neural network and analog computing evaluation,” Physical Review A, vol. 109, no. 3. American Physical Society (APS), Mar. 20, 2024. doi: 10.1103/physreva.109.033521.

[Li23] R.-Q. Li, Y.-W. Shen, B.-D. Lin, J. Yu, X. He, and C. Wang, “Scalable wavelength-multiplexing photonic reservoir computing,” APL Machine Learning, vol. 1, no. 3. AIP Publishing, Jul. 18, 2023. doi: 10.1063/5.0158939.

[Lupo21A] A. Lupo, L. Butschek, and S. Massar, “Photonic extreme learning machine based on frequency multiplexing,” Optics Express, vol. 29, no. 18. Optica Publishing Group, p. 28257, Aug. 17, 2021. doi: 10.1364/oe.433535.

[Lupo21B] A. Lupo and S. Massar, “Parallel Extreme Learning Machines Based on Frequency Multiplexing,” Applied Sciences, vol. 12, no. 1. MDPI AG, p. 214, Dec. 27, 2021. doi: 10.3390/app12010214.

[Lupo23] A. Lupo, E. Picco, M. Zajnulina, and S. Massar, “Deep photonic reservoir computer based on frequency multiplexing with fully analog connection between layers,” Optica, vol. 10, no. 11. Optica Publishing Group, p. 1478, Nov. 06, 2023. doi: 10.1364/optica.489501.

[Merolla99] J.-M. Mérolla, Y. Mazurenko, J.-P. Goedgebuer, H. Porte, and W. T. Rhodes, “Phase-modulation transmission system for quantum cryptography,” Optics Letters, vol. 24, no. 2. Optica Publishing Group, p. 104, Jan. 15, 1999. doi: 10.1364/ol.24.000104.

[Oakley88] K. A. Oakley, “An economic way to see in the broadband dawn (passive optical network),” IEEE Global Telecommunications Conference and Exhibition. Communications for the Information Age. IEEE, 1988. doi: 10.1109/glocom.1988.26087.

[Olislager14] L. Olislager, E. Woodhead, K. Phan Huy, J.-M. Merolla, P. Emplit, and S. Massar, “Creating and manipulating entangled optical qubits in the frequency domain,” Phys. Rev. A, vol. 89, no. 5. American Physical Society (APS), May 22, 2014. doi: 10.1103/physreva.89.052323.

[Xu21] X. Xu et al., “11 TOPS photonic convolutional accelerator for optical neural networks,” Nature, vol. 589, no. 7840. Springer Science and Business Media LLC, pp. 44–51, Jan. 06, 2021. doi: 10.1038/s41586-020-03063-0.

[Yildirim22] M. Yildirim et al., “Nonlinear Optical Data Transformer for Machine Learning.” arXiv, 2022. doi: 10.48550/ARXIV.2208.09398.

[Zajnulina23] M. Zajnulina, A. Lupo, and S. Massar, “Weak Kerr Nonlinearity Boosts the Performance of Frequency-Multiplexed Photonic Extreme Learning Machines: A Multifaceted Approach.” arXiv, 2023. doi: 10.48550/ARXIV.2312.12296.

[Zhou22] T. Zhou, F. Scalzo, and B. Jalali, “Nonlinear Schrödinger Kernel for Hardware Acceleration of Machine Learning,” Journal of Lightwave Technology, vol. 40, no. 5. Institute of Electrical and Electronics Engineers (IEEE), pp. 1308–1319, Mar. 01, 2022. doi: 10.1109/jlt.2022.3146131.

# Time-multiplexed reservoir computing in the optical domain

**Kathy Lüdge, Lina Jaurigue**
Technische Universität Ilmenau, Institut für Physik (Weimarer Str. 25, 98693 Ilmenau)

[kathy.luedge@tu-ilmenau.de, Lina.jaurigue@ tu-ilmenau.de]

### Status

The concept of reservoir computing (RC) was introduced by Herbert Jaeger [1] to avoid the tedious training process in deep or recurrent neural networks. Since then it has been widely used with its biggest advantage being the possibility to use various physical systems as reservoirs [2]. With the introduction of time-multiplexing [3], a new road to using optical setups for realizations of RC was opened, mainly because single node setups with delayed feedback could now be used. The delayed feedback leads to a high dimensional phase space of the dynamical system and thus to a very complex transient system response in time. This response can be sampled in time and allows to realize single node reservoirs with a high number of virtual network nodes. In this scheme, it is not the spatial scale which limits the network dimension but only the details of output sampling and system timescale [4, 5]. A sketch of the time-multiplexed scheme is shown in Fig.1 which is, however, not limited to the depicted realization of a laser with optical self-feedback, but open to every dynamical system with complex transient response. Specifically in photonics, where other multiplexing techniques in frequency and space are possible [6], time-multiplexing can also be used as an add-on to other implementations by sampling the response multiple times during one input.

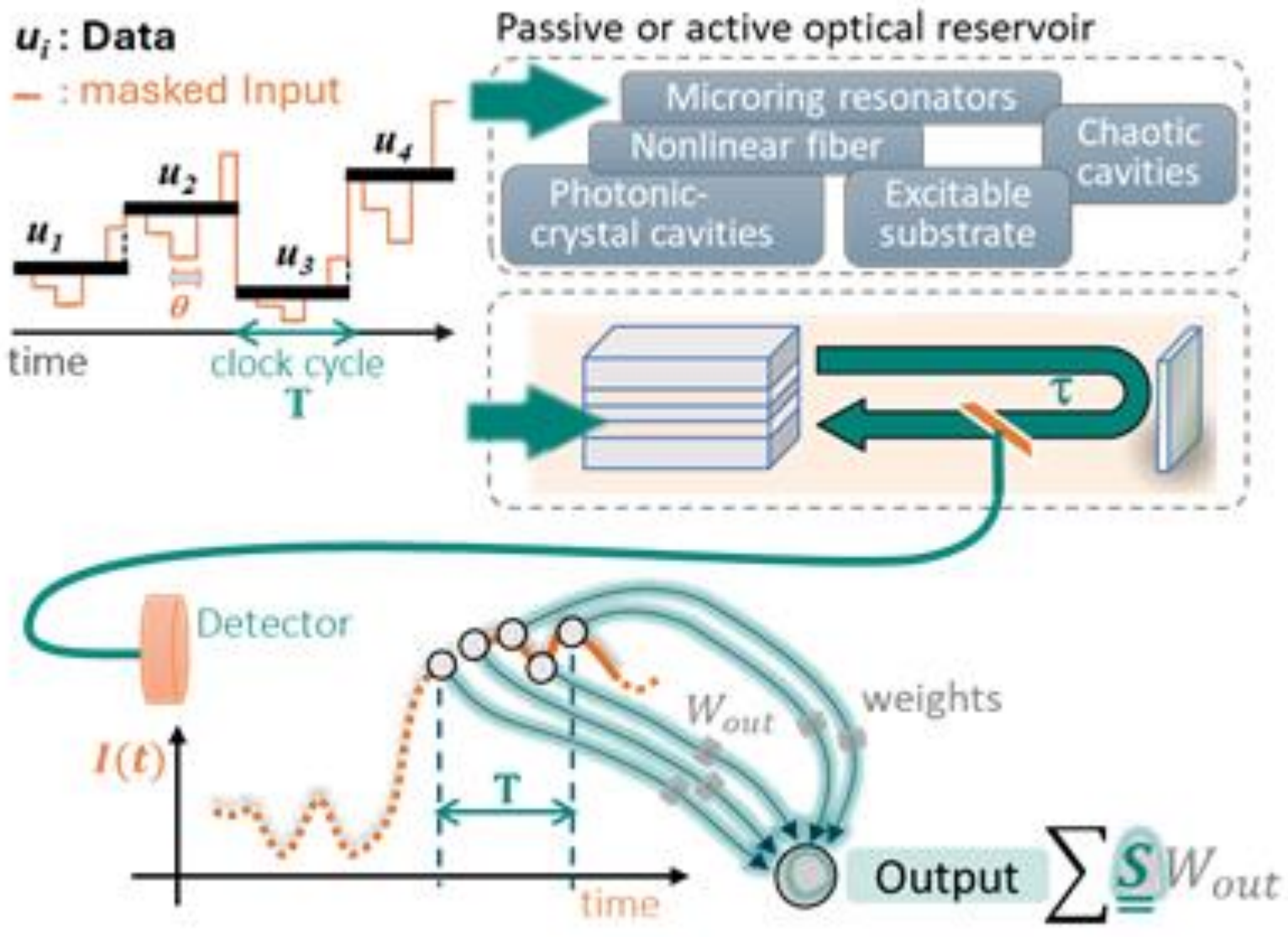

**Figure 1.** Sketch of the time multiplexing scheme which can be realized with passive or active cavities as well as with spiking systems [7, 8]. Inputs are injected in time after being multiplied with a mask. The time dependent system response to one input data $u_i$ is sampled multiple times during one clock cycle **T** (sketched exemplarily for a laser with self-feedback) which yields one row of the state matrix . The number of sampling points within **T** defines the virtual network size.

Efficient hyper-parameter optimization in physical RC is an issue with ongoing research. While those optimizations can easily be done in numerical simulations, the tuning of all the necessary parameters is not always possible in an optical setup. Using pre-processed data has shown to be very helpful to mitigate the hyper-parameter dependence[9]. Specifically in delay-based RCs, the hyper-parameters delay-time **τ** and injection rate 1/**T** determine the internal coupling topology. Historically, only two setting were chosen which were either resonant t = T[3, 10] or desynchronized ($\tau$ = T + $\theta$) [11]. As reported in [12], a resonant delay degrades the performance which was explained in [13, 14] by the dimensionality reduction of the accessible phase space. Further advances in finding optimal parameters for specific edge applications are envisioned if the possibility of changing the topology via the delay are utilized, i.e. the system response time determines the next neighbour coupling (due to the system's inertia) and the ratio $\tau$/T determines how the information of different inputs are mixed [9, 15].

## Current and Future Challenges

The motivation for developing optical computing approaches can generally be summarised in two points; high data processing speed and low energy consumption. For the utility of time-multiplexed reservoir computing in the optical domain, a key challenge lies in developing methods that yield an advantage in terms of these two points. To do this, it is critical to identify applications for which high speed optical data processing would be suitable and to then tailor the reservoir design, both in terms of topology and timescales, to this application. Therefore, a vital research question is the matching of timescales between task and reservoir. This is especially important for the design of energy efficient edge devices which need to measure and process data in real time. A delay-based laser setup, for example, reacts on nanosecond timescales which should match the input timescale at which the input data are generated.

For exploiting the full benefit of time-multiplexing, an extensive pre-processing of data as well as domain conversions (electrical optical) should be avoided. Envisioned real world applications in the optical domain include light based ranging (LIDAR), surveillance applications or channel equalization in optical data communications. While for the latter the timescales of task and reservoir response are both in the GHz range, the matching is a challenge for the others. Due to the sequential nature of data injection a higher dimensional reservoir (higher number of sampling points in time) is synonymous with slower injection rates. This can be circumvented by parallelization of multiple reservoirs, which in turn leads to a trade-off between data processing rates and the reservoir footprint. Consequently, a deeper understanding of the interplay between internal system dynamics and time-multiplexing is still needed in order to find new ways for optimizing the performance.

Another challenge of physical implementations of RC lies in the training process. In principle, training RC setups is very fast if realized on a conventional computer, because it only requires one linear regression step instead of the otherwise typical and time consuming training over multiple epochs. Nevertheless, so far this linear regression step cannot be realized fully optically in optical RC schemes and strong efforts in this direction are still needed.

Transferring the time-multiplexing approach to other neuromorphic computing schemes apart from RC is also an active research direction. In general, time-multiplexing can be used to extend the readout dimension in a large class of analog computing systems ranging from extreme learning machines [16] to convolutional neural networks [17].

## Advances in Science and Technology to Meet Challenges

Technologically, the dominating challenge is to realize analog computing edge devices with a small energy footprint and with sustainable materials. For this, new biologically inspired substrates need to be explored, as well as encoding schemes that allow for higher information density as for example by utilizing both phase and amplitude information of the light.

Especially in time-multiplexed systems, the input needs to be masked in order to guarantee a diverse response to the input (similar to the random input connections in Echo State Networks [1]). Up to now, this usually requires fast pattern generation and thus limits applicability for energy efficient hardware implementations. Recent developments target on circumventing this problem by using analog masks [18].

A third focus regarding technological implementations is related to the memory vs. non-linearity trade-off [19]. Promising approaches to overcome this limitation have to focus on external memory augmentation which is possible via input pre-processing (e.g. delayed input [9]) or output post-processing (delayed output [20]). Those schemes externally add memory to the RC without changing the reservoir. A sketch of these methods is shown in Fig.2. While optical systems have been pioneering the development of delay-based RCs[7], time-multiplexing can be applied to every dynamical system that shows a transient nonlinear response to an input. One complementary example are analog electronic implementations which are easy to realize in hardware, e.g. via thyristors or memristors, but respond on a much slower timescale [4].

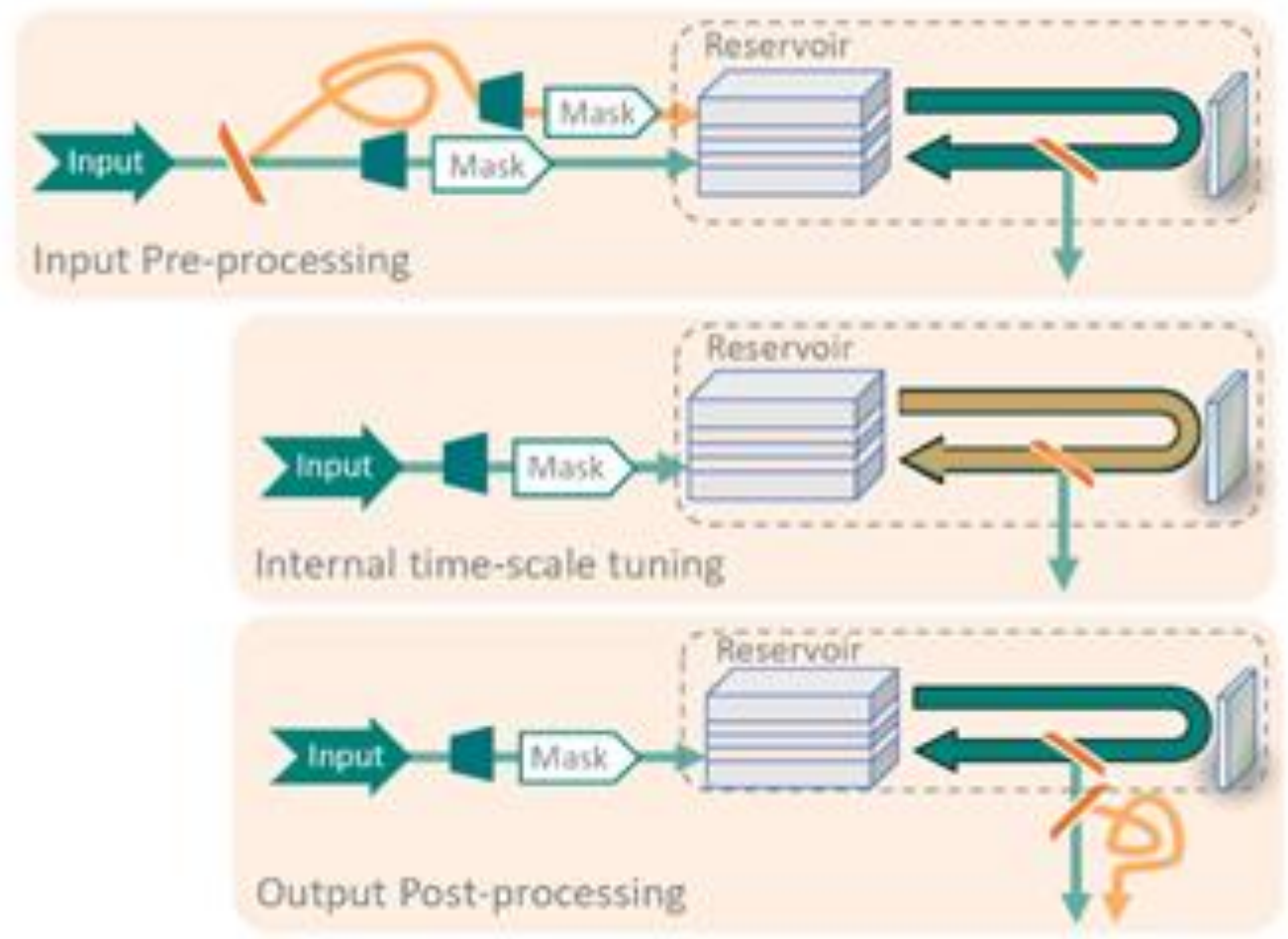

Figure 2. Sketch of 3 different schemes to augment the memory of a time-multiplexed RC setup. From top to bottom input delay, internal delay tuning and output delay are shown [9].

**Concluding Remarks**

Time-multiplexing in photonic neuromorphic computing schemes is a powerful concept that enlarges the underlying network dimension without increasing the spatial scale and at the same time allows to exploit the high data processing speed of optical implementations. It can easily be implemented with active delay-based setup as well as with passive nonlinear cavities. The best and most energy efficient realization of data injection and sampling is still an important research question. Currently innovative approaches for optimization using memory augmentation, efficient masking and new reservoir realizations are being explored. In order to pave the road to real world all optical RC applications, the matching between task and reservoir timescales is still an important issue including the identification of applications for which high speed optical data processing is mandatory.

**Acknowledgements**

*This work was supported by the EU horizon 2020* ***SpikePro*** *and the Carl Zeiß* ***Nexus*** *Program.*

## References

[1] H. Jaeger. *The "echo state" approach to analysing and training recurrent neural networks.* GMD Report **148**, GMD - German National Research Institute for Computer Science (2001). doi:10.24406/publica-fhg-291111.

[2] G. Tanaka, T. Yamane, J. B. Heroux, R. Nakane, N. Kanazawa, S. Takeda, H. Numata, D. Nakano, and A. Hirose. *Recent advances in physical reservoir computing: A review.* Neural Netw. **115**, 100 (2019). doi:10.1016/j.neunet.2019.03.005.

[3] L. Appeltant, M. C. Soriano, G. Van der Sande, J. Danckaert, S. Massar, J. Dambre, B. Schrauwen, C. R. Mirasso, and I. Fischer. I*nformation processing using a single dynamical node as complex system.* Nat. Commun. **2**, 468 (2011). doi:10.1038/ncomms1476.

[4] X. Liang, J. Tang, Y. Zhong, B. Gao, H. Qian, and H. Wu. *Physical reservoir computing with emerging electronics*. Nat. Electron. **7**, 193 (2024). doi:10.1038/s41928-024-01133-z.

[5] L. Mühlnickel, J. A. Jaurigue, L. C. Jaurigue, and K. Lüdge. *Reservoir computing using spin-VCSELs - the influence of timescales and data injection schemes*. Commun. Phys. **7**, 370 (2024).doi:10.1038/s42005-024-01858-5.

[6] Y. Bai, X. Xu, M. Tan, Y. Sun, Y. Li, J. Wu, R. Morandotti, A. Mitchell, K. Xu, and D. J. Moss. *Photonic multiplexing techniques for neuromorphic computing*. Nanophotonics **12**, 795 (2023). doi: 10.1515/nanoph-2022-0485.

[7] S. Abreu, I. Boikov, M. Goldmann, T. Jonuzi, A. Lupo, S. Masaad, L. Nguyen, E. Picco, G. Pourcel, A. Skalli, L. Talandier, B. Vettelschoss, E. A. Vlieg, A. Argyris, P. Bienstman, D. Brunner, J. Dambre, L. Daudet, J. D. Domenech, I. Fischer, F. Horst, S. Massar, C. R. Mirasso, B. J. Offrein, A. Rossi, M. C. Soriano, S. Sygletos, and S. K. Turitsyn. *A photonics perspective on computing with physical substrates.* Rev. Phys. **12**, 100093 (2024).doi:10.1016/j.revip.2024.100093.

[8] D. Brunner, M. C. Soriano, and G. Van der Sande. *Photonic Reservoir Computing,* Optical Recurrent Neural Networks. De Gruyter, Berlin, Boston, (2019). doi:10.1515/9783110583496.

[9] L. C. Jaurigue and K. Lüdge. *Reducing reservoir computer hyperparameter dependence by external timescale tailoring.* Neuromorph. Comput. Eng. **4**, 014001 (2024).doi:10.1088/2634-4386/ad1d32.

[10] K. Takano, C. Sugano, M. Inubushi, K. Yoshimura, S. Sunada, K. Kanno, and A. Uchida. *Compact reservoir computing with a photonic integrated circuit.* Opt. Express **26**, 29424 (2018).doi:10.1364/oe.26.029424.

[11] Y. Paquot, F. Duport, A. Smerieri, J. Dambre, B. Schrauwen, M. Haelterman, and S. Massar. *Optoelectronic reservoir computing.* Sci. Rep. **2**, 287 (2012). doi:10.1038/srep00287.

[12] A. Röhm, L. C. Jaurigue, and K. Lüdge. *Reservoir computing using laser networks.* IEEE J. Sel. Top. Quantum Electron. **26**, 7700108 (2019). doi:10.1109/jstqe.2019.2927578.

[13] F. Stelzer, A. Röhm, K. Lüdge, and S. Yanchuk. *Performance boost of time-delay reservoir computing by non-resonant clock cycle.* Neural Netw. **124**, 158 (2020). doi:10.1016/j.neunet.2020.01.010.

[14] F. Köster, S. Yanchuk, and K. Lüdge. *Insight into delay based reservoir computing via eigenvalue analysis.* J. Phys. Photonics **3**, 024011 (2021). doi:10.1088/2515-7647/abf237.

[15] T. Hülser, F. Köster, L. C. Jaurigue, and K. Lüdge. *Role of delay-times in delay-based photonic reservoir computing.* Opt. Mater. Express **12**, 1214 (2022). doi:10.1364/ome.451016.

[16] A. Lupo, L. Butschek, and S. Massar. *Photonic extreme learning machine based on frequency multiplexing.* Opt. Express **29**, 28257 (2021). doi:10.1364/oe.433535.

[17] X. Xu, M. Tan, B. Corcoran, J. Zou, A. Boes, T. G. Nguyen, S. T. Chu, B. E. Little, D. G. Hicks, R. Morandotti, A. Mitchell, and D. J. Moss. 1*1 tops photonic convolutional accelerator for optical neural networks.* Nature **589**, 44 (2021). doi:10.1038/s41586-020-03063-0.

[18] I. Bauwens, P. Bienstman, G. Verschaffelt, and G. Van der Sande. *Use of a simple passive hardware mask to replace the digital masking procedure in photonic delay-based reservoir computing.* IEEE J. Sel. Top. Quantum Electron. (2024). doi:10.1109/jstqe.2024.3451113.

[19] M. Inubushi and K. Yoshimura. *Reservoir computing beyond memory-nonlinearity trade-off.* Sci. Rep. **7**, 10199 (2017).

[20] J. A. Jaurigue, J. Robertson, A. Hurtado, L. C. Jaurigue, and K. Lüdge. *Postprocessing methods for delay-embedding and feature scaling of reservoir computers.* Commun. Eng. (2024). doi:10.21203/rs.3.rs-4741218/v1.

# Optical neural networks based on wavelength-division multiplexing

**Xingyuan Xu[1], David J. Moss[2]**

[1]State Key Laboratory of Information Photonics and Optical Communications, Beijing University of Posts and Telecommunications, Beijing 100876, China.

[2]Optical Sciences Centre, Swinburne University of Technology, Hawthorn, VIC 3122, Australia.

[ dmoss@swin.edu.au, xingyuanxu@bupt.edu.cn]

## Status

The neural network's ability to tackle complex tasks depends on its scale, defined by the number of neurons, synapses, and layers. Therefore, accelerating neural networks with optical hardware relies on achieving high levels of parallelism and throughput to effectively map input nodes and weighted synapses onto the physical optical systems. Wavelength-division multiplexing (WDM) exemplifies the advantages of light over electronics. The broad optical bands accommodate numerous wavelength channels, enabling parallel input nodes and weighted synapses, with the potential for clock rates reaching tens of gigahertz. Optical computing operations based on WDM utilize multi-wavelength sources combined with weight bands or wavelength-sensitive elements, including micro-ring resonators (MRR) [1-10], semiconductor optical amplifiers (SOA) [11], and phase change memory (PCM) [12, 13].

In 2011, Q. Xu et al. employed circuit employing a tunable Microring Resonator (MRR) for incoherent summation [1]. Following this, L. Yang et al. developed an optical matrix vector multiplier comprising cascaded lasers, modulators, wavelength multiplexers/demultiplexers, an MRR matrix, and photodetectors [2]. Input data vectors were encoded onto different wavelength powers, with weight vector elements determining MRR transmittance at specific positions. In 2014, A. Tait et al. proposed the broadcast-and-weight protocol [3], demonstrated later with an MRR weight bank in 2017 [4]. This protocol distributed input data across all wavelength channels, simultaneously adjusting their weights through power control. Recently, C. Huang et al. implemented a WDM-based Optical Neural Network (ONN) with a broadcast-and-weight architecture on a silicon photonic platform [10]. Input data from different neurons was multiplexed onto optical waveguides, with interconnections established via power splitters and weights controlled by MRR banks.

In 2020, Indium Phosphide (InP) platforms enabled the realization of a photonic feed-forward neural network [11], using Semiconductor Optical Amplifiers (SOAs) for loss compensation and weight adjustment. Additionally, Phase Change Memory (PCM) cells were explored for weighted interconnects, demonstrated by J. Feldmann et al. [12, 13]. An optical convolution accelerator demonstrated vector computing speeds of 11.3 Tera Operations Per Second (TOPS) [14], employing a time-wavelength interleaving technique for feature extraction from large-scale data. Input data modulation via digital-to-analog converters and convolutional kernel mapping onto microcomb lines facilitated convolution operations, particularly advantageous for convolutional neural networks due to reduced parametric complexity.

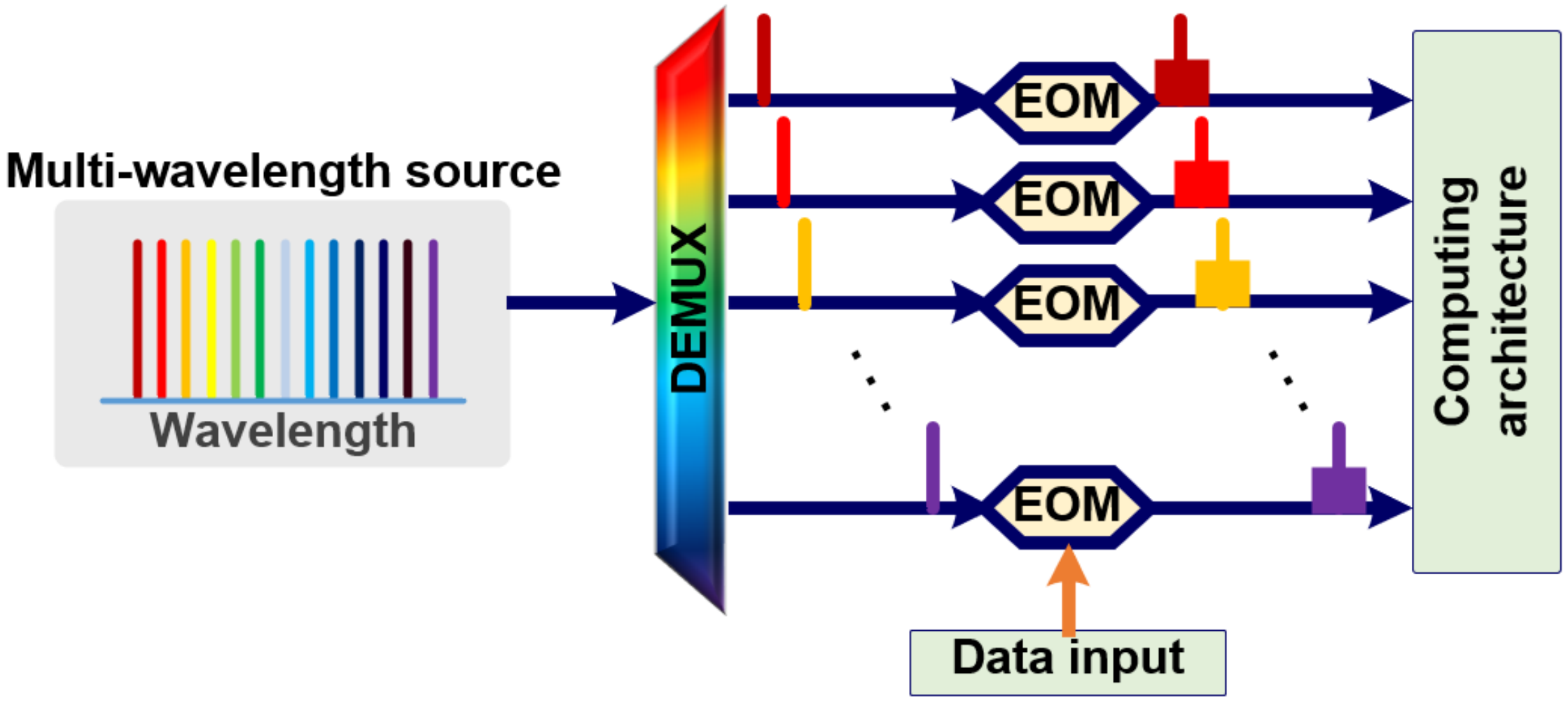

**Figure 1.** Architecture of WDM based computing systems.

## Current and Future Challenges

While Optical Neural Networks (ONNs) hold promise for achieving high computing power and energy efficiency, their analog nature suggests that hybrid opto-electronic neuromorphic hardware is likely the optimal solution, leveraging both the broadband capabilities of optics and the versatility of digital electronics. In such a setup, optics primarily handles computing operations while electronics manage data flow and storage. Despite notable progress, several challenges remain to be addressed for future ONN applications.

Firstly, dense integration of the entire photonics system is crucial to achieve competitive computing densities compared to electrical counterparts. Hybrid integration techniques are necessary to effectively utilize optics' broad bandwidths with Wavelength Division Multiplexing (WDM) techniques, integrating sources and multiply-and-accumulate units. Secondly, expanding the range of computing operations, including nonlinear neuron functions and Fourier transforms, on chip is essential to enhance ONN universality for diverse machine learning tasks. This requires advancements in computing architectures tailored to specific operations and high-nonlinearity component integration enabling nonlinear functions with low optical power. Moreover, as ONNs consist of massive programmable photonic units for high spatial-division parallelism, tailored algorithms are needed to address fabrication imperfections and on-chip cross-talk, ensuring fast-converging control of on-chip elements and network training.

Once these challenges are overcome, ONNs can seamlessly integrate with existing electronic hardware, significantly boosting overall system computing performance. This enhancement can dramatically accelerate the training speed of computationally intensive neural networks, paving the way for more sophisticated applications such as fully automated vehicles and real-time image/video processing.

## Advances in Science and Technology to Meet Challenges

Typical WDM systems comprise a multi-wavelength source generating parallel wavelength channels for data transmission or processing. These channels are then manipulated separately using wavelength multiplexers and demultiplexers. Integrated Optical Frequency Combs (OFCs) are crucial for creating the multi-wavelength source due to their compact design, unlike discrete laser arrays. OFCs ensure equal frequency intervals between wavelength channels, facilitating easy manipulation in the frequency domain [15-20]. Over the past two decades, advancements in photonic nanofabrication have led to integrated OFCs in various forms, offering benefits in system size, weight, power consumption, and cost. Integrated OFCs fall into several categories based on their physical origins: a) Kerr frequency combs, or microcombs, originate from parametric oscillation in integrated micro-ring resonators (MRRs) [17-20]. b) Mode-locked lasers utilize gain media like Erbium-doped fiber amplifiers and mode-locking mechanisms such as saturable absorbers to produce pulsed outputs. c) Electro-optic modulators use second-order nonlinearity to introduce sidebands around an optical carrier, requiring external RF sources.

Microcombs emerge as formidable contenders in the realm of integrated Optical Frequency Comb (OFC) sources, owing to their compact form factor and exceptionally wide bandwidths capable of spanning octaves, underpinned by robust broadband nonlinear gain. Originating from parametric oscillation within high-Q micro-resonators, these microcombs manifest in various structural formats, including integrated variants like micro-ring resonators and three-dimensional counterparts such as spheres or rods [17-20]. A critical determinant in the generation of microcombs lies in the attainment of substantial parametric gain, directly contingent upon the third-order nonlinearity of the material substrate and the Q factor of the resonator, indicative of diminished linear and nonlinear losses [17-20]. Within micro-resonators characterized by elevated Q factors, the optical intra-cavity field experiences pronounced resonance amplification, thereby facilitating the induction of nonlinear phenomena typically necessitating considerable optical power, such as modulation instability gain and subsequent parametric oscillation. Recent advancements in this domain have propelled microcombs towards increasing practical viability, rendering them as wideband, energy-efficient, compact, and readily scalable solutions amenable for integration into wavelength-division multiplexing systems [17-20].

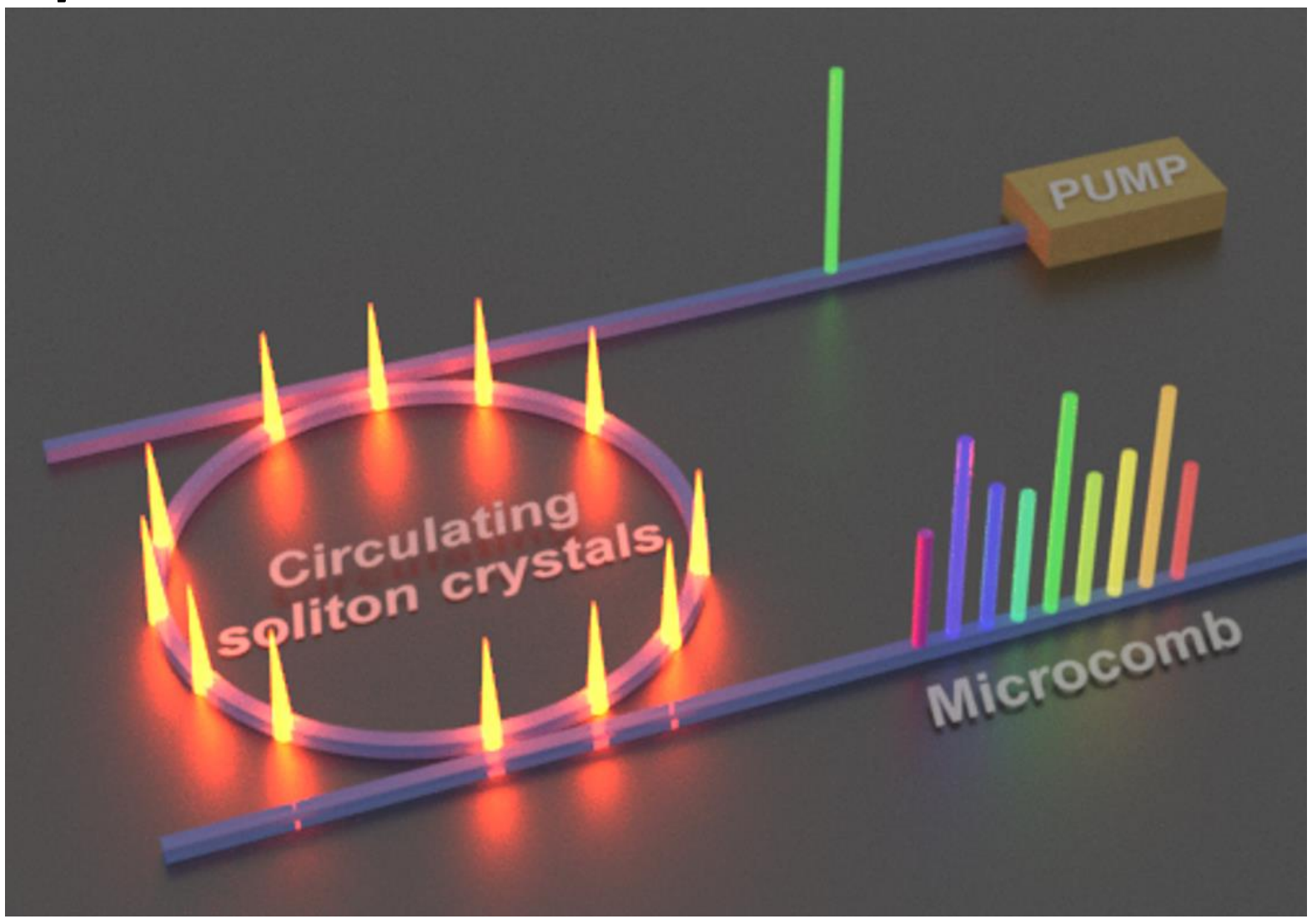

**Figure 2**. Illustration of microcomb generation.

## Concluding Remarks

Photonic multiplexing techniques have remarkable capacities for implementing the optoelectronic hardware that is isomorphic to neural networks, which can offer competitive performance in connectivity and linear operation of neural network. WDM is a unique technique enabled by optics, in contrast to electronics. Supported by the ultra-wide optical bandwidths up to 10s' of THz, 100's of wavelength channels can be established for parallel data processing of neural networks, thus leading to significantly enhanced computing speed—similar to the significantly enhanced data transmission capacity for WDM-based communications systems. With the recent advances of the chip-scale frequency combs, wideband and low-noise integrated optical sources are readily available, greatly promoting the potentials of WDM-based ONNs.

## References

[1] Q. Xu and R. Soref, "Reconfigurable optical directed-logic circuits using microresonator-based optical switches," *Opt. Express*, vol. 19, no. 6, pp. 5244–5259, 2011.

[2] L. Yang, R. Ji, et al., "On-chip CMOS-compatible optical signal processor," *Opt. Express*, vol. 20, no. 12, pp. 13560–13565, 2012.

[3] A. N. Tait, et al., "Broadcast and weight: an integrated network for scalable photonic spike processing," *J. Lightwave Technol.*, vol. 32, no. 21, pp. 4029–4041, 2014.

[4] A. N. Tait, et al., "Neuromorphic photonic networks using silicon photonic weight banks." *Sci. Rep*, vol. 7, no. 1, p. 7430, 2017.

[5] A. N. Tait, et al., "Multi-channel control for microring weight banks," *Opt. Express*, vol. 24, no. 8, pp. 8895–8906, 2016.

[6] A. N. Tait, et al., "Continuous calibration of microring weights for analog optical networks," *IEEE Photon. Technol. Lett.*, vol. 28, no. 8, pp. 887–890, 2016.

[7] A. N. Tait, et al., "Microring weight banks," *IEEE J. Sel. Top. Quantum Electron.*, vol. 22, no.6, pp. 312–325, 2016.

[8] C. Huang, S. Bilodeau, et al., "Demonstration of scalable microring weight bank control for large-scale photonic integrated circuits," *APL Photonics*, vol. 5, no. 4, p. 040803, 2020.

[9] S. Xu, J. Wang, and W. Zou, "Optical convolutional neural network with WDM-based optical patching and microring weighting banks," *IEEE Photonics Technol. Lett.*, vol. 33, no. 2, pp. 89–92, 2021.

[10] C. Huang, et al., "A silicon photonic-electronic neural network for fibre nonlinearity compensation," *Nat. Electron.*, vol. 4, no. 11, pp. 837-844, 2021.

[11] B. Shi, N. Calabretta, and R. Stabile, "InP photonic integrated multi-layer neural networks: Architecture and performance analysis," *APL Photonics*, vol. 7, no. 1, p. 010801, 2022.

[12] J. Feldmann, N. Youngblood, C. D. Wright, H. Bhaskaran, and W. H. P. Pernice, "All-optical spiking neurosynaptic networks with self-learning capabilities," *Nature*, vol. 569, no. 7755, pp. 208-214, 2019.

[13] J. Feldmann, N. Youngblood, M. Karpov, et al., "Parallel convolutional processing using an integrated photonic tensor core," *Nature*, vol. 589, no. 7840, pp. 52-58, 2021.

[14] X. Xu, M. Tan, B. Corcoran, et al., "11 TOPS photonic convolutional accelerator for optical neural networks," *Nature*, vol. 589, no. 7840, pp. 44-51, 2021.

[15] T. Udem, R. Holzwarth, and T. W. Hansch, "Optical frequency metrology," Nature, vol. 416, no. 6877, pp. 233-237, Mar 14, 2002.

[16] L. Chang, S. T. Liu, and J. E. Bowers, "Integrated optical frequency comb technologies," Nature Photonics, vol. 16, no. 2, pp. 95-108, Feb, 2022.

[17] A. Pasquazi, M. Peccianti, L. Razzari, D. J. Moss, S. Coen, M. Erkintalo, Y. K. Chembo, T. Hansson, S. Wabnitz, and P. Del'Haye, "Micro-combs: A novel generation of optical sources," Physics Reports 729, 1-81 (2018).

[18] T. J. Kippenberg, A. L. Gaeta, M. Lipson, and M. L. Gorodetsky, "Dissipative Kerr solitons in optical microresonators," Science 361 (2018).

[19] A. L. Gaeta, M. Lipson, and T. J. Kippenberg, "Photonic-chip-based frequency combs," Nat. Photonics 13, 158-169 (2019).

[20] D. J. Moss, R. Morandotti, A. L. Gaeta, and M. Lipson, "New CMOS-compatible platforms based on silicon nitride and Hydex for nonlinear optics," Nat. Photonics 7, 597-607 (2013).

# ARCHITECTURES

## Diffractive photonic computing units for advanced neural network architectures and systems

**Haiou Zhang[1], Chu Wu[1] and Xing Lin[1,2]**
[1]Department of Electronic Engineering, Tsinghua University, Beijing, 100084, China
[2]Beijing National Research Center for Information Science and Technology, Tsinghua University, Beijing, 100084, China
[ azhho@tsinghua.edu.cn, wuc23@mails.tsinghua.edu.cn, lin-x@tsinghua.edu.cn]

**Status**

Recent research on diffractive photonic computing and neural networks [1] has made significant progress in developing computing architectures and systems for versatile applications. This progress has been achieved through the use of advanced design methods and fabrication techniques with large-scale complex computation and high optical integrability. Based on the 3D and 2D modality of diffractive photonic computing units, the neural network architectures, including the convolutional neural networks (CNN), recurrent neural networks (RNN), networks of networks (NIN) [2], residual neural networks (Res-NN) [3], graphical neural networks (GNN) [4], and variational auto-encoders (VAE) [5] can be constructed. The deployment of diffractive neural networks allows for the creation of diffractive intelligent optoelectronic computing systems that can be used in various applications, such as high-speed visual information processing, high-throughput fiber optic communication, and microwave signal processing and sensing.

Fig. 1 shows the development of diffractive photonic computing units (DPU) from spatial light computing to on-chip integration. Diffractive photonic computing achieves the effective manipulation of large-scale photons during their complex optical field propagation by using diffractive elements. Weighted interconnections are established through interference superposition of modulated optical field between successive layers. This layered diffractive coefficients can be optimized with large-scale deep learning, allowing for efficient modulation of the optical field propagation to achieve the desired high-dimensional system mapping function. The 3D-printed diffractive photonic neural networks were originally designed with millions of neurons that had very low latency and power consumption. However, the weights were fixed once printed, which greatly limited their application areas. To overcome this limitation, the researchers used spatial light modulators (SLM) to create a large-scale neuromorphic photonics computational architecture. This architecture employed reconfigurable DPU with spatio-temporal multiplexing. The neural network architectures based on DPU has excellent wavelength expandability, allowing researchers to utilize the reconfigurable intelligence surface (RIS) for beam shaping and signal processing in the microwave band [6]. This technology is expected to be widely used in the fields of smart security and urban communication in the 6G era.

To improve device integration, researchers have started using diffractive optical elements (DOEs) [7] and subwavelength meta-structures [8] for the 3D integration of DPU. Meta-surfaces can enhance the capability of optical field modulation and enable joint modulation of multiple dimensions, including amplitude, phase, polarization, and angular momentum. For instance, the principle of Pancharatnam-Berry phasing can be used to simultaneously rotate the orientation of the nanorods of the micro-nanostructures and adjust their cross-sectional dimensions, length, and width to achieve joint modulation of amplitude and phase on a single neuron [9]. To create a more universal optoelectronic fusion computational processor, we need to integrate the DPU onto a chip in two dimensions. These diffractive meta-lines architecture [10,11] using a silicon photonic process has been shown to have tremendous potential for AI acceleration. However, the computational accuracy and throughput of the diffractive photonic chip needs further improvement due to the reduction in the number of neurons and the decrease in degree-of-freedom caused by the reduced dimensionality.

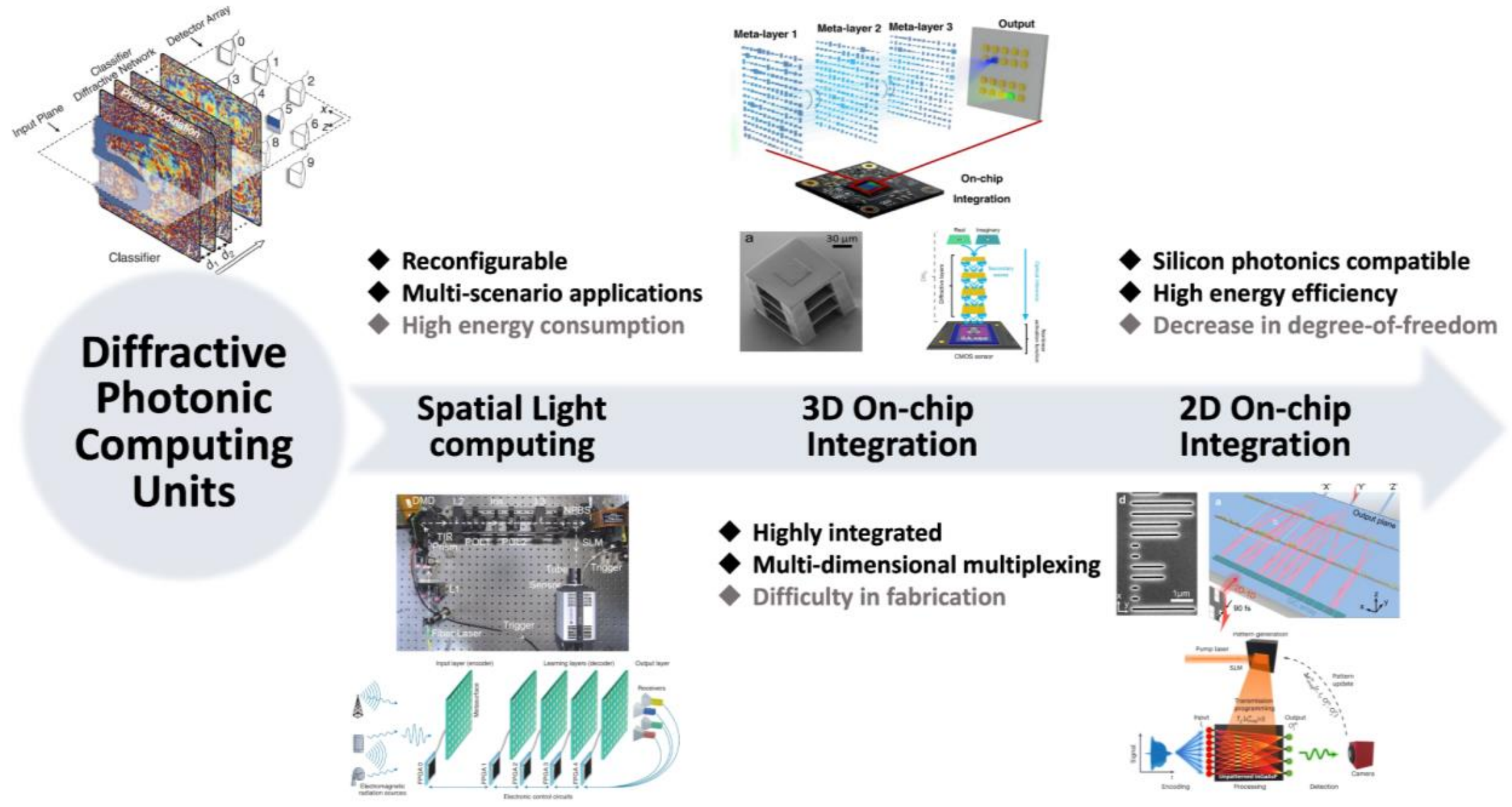

**Figure1.** Development of diffractive photonic computing units: From spatial light to on-chip integration. Diffractive neural networks have been extensively researched and discussed in various frequency bands, including terahertz, radio frequency, communication, and visible light bands. Efforts have been made to construct more robust and highly integrated diffractive photonic computing units.

### Current and Future Challenges

The ideal photonic neural network architecture and system should process information with sufficient computational accuracy, reconfigurability, high integration, low power consumption, and high energy efficiency. To construct large-scale advanced neural network architectures using DPU, the critical task is to create a matrix multiplier with sufficient computational accuracy. This multiplier should be capable of instantly generating matrix-matrix multiplication. However, diffractive photonic computing is an analog computing process that suffers from accuracy issues and layer-by-layer accumulation of errors. Therefore, the development of error correction algorithms is urgently needed. Additionally, for the architecture to be applicable to various scenarios, the device must be reconfigurable for adaptive tuning and training. In practice, optoelectronic fusion computing schemes are commonly used, where modulation is achieved through electrically controlled devices. For example, electrodes can be used to modulate the carrier concentration in Mach–Zehnder interferometer (MZI) or micro-ring resonators (MRR). Currently, light still experiences significant energy loss when propagating through a DPU. This issue should be addressed through material and architectural innovations. Without such innovations, the network will not be able to achieve sufficient depth. Gain control on a substrate of active materials is a promising approach [12]. Additionally, it is essential to investigate the feedback and storage mechanisms in constructing the photonic computing system, not only for neural networks with built-in memory but also for constructing Turing-complete general-purpose computational processors.

Generative AI often deals with tensor operations in thousands of dimensions, so it is necessary for our photonic computing to be scalable and capable of handling large data throughputs. Diffractive photonic computing currently can achieve high-throughput and large-scale integration with different multiplexing approaches. It maximizes the use of photons' properties and allows for the addition of parallel computing through schemes like spectral multiplexing [13], polarization multiplexing [14], and orbital angular momentum multiplexing [15]. The lack of efficient optical nonlinearities restricts neural networks to being linear, which limits both their accuracy and ability to handle complex tasks. Additionally, this lack of nonlinearity makes it challenging to develop optical logic gates.

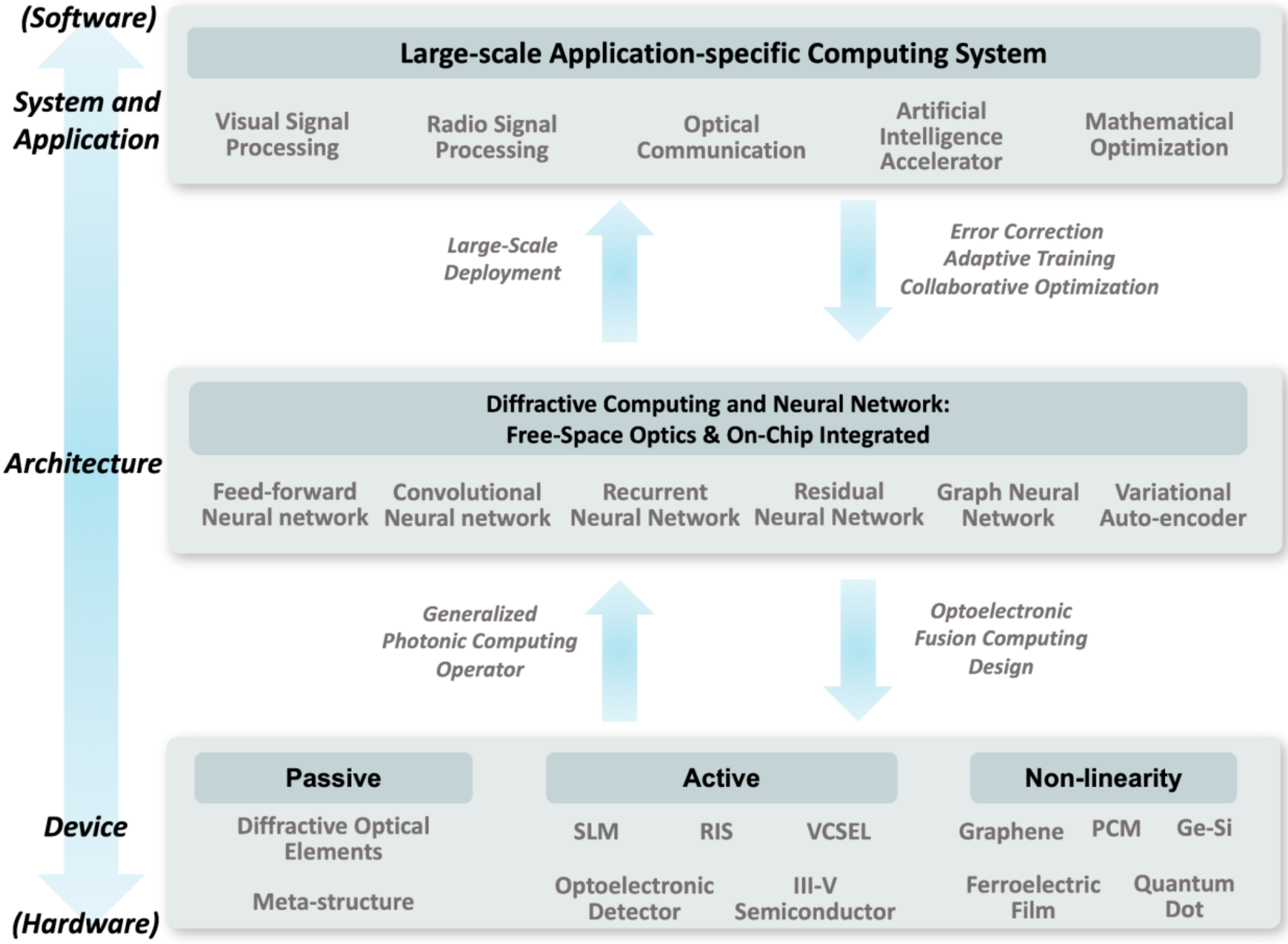

**Figure2.** Diffractive photonic computing systems and applications: Hardware-software co-design. This figure illustrates the components of diffractive photonic computing operators, architectures, and systems from bottom to top. It also provides an outlook on the potential for improvement at different levels.

### Advances in Science and Technology to Meet Challenges

To address the aforementioned challenges, researchers should prioritize establishing synergy between hardware and software [16]. Through this approach, they can coordinate the design of bottom-layer devices and top-layer architectures to improve computational accuracy and integration, among other features. Figure 2 illustrates a generalized operator library composed of various photonic computing devices. Passive modules with low energy consumption and low latency are suitable for fixed tasks. Active modules, on the other hand, can be utilized to program to switch between tasks. Additionally, there have been attempts at optoelectronic or all-optical nonlinearities, such as utilizing the nonlinearity of ferroelectric films [17] or the saturable absorption effect of graphene. The technology of structured phase change materials (PCM) [18] is promising due to its ability to change between crystalline and amorphous states in response to an energy impulse, resulting in a change in refractive index and a nonlinear impulse response during a phase change. This change is non-volatile, making it suitable for optical storage and reconfigurable optical computing. There has been a trend towards photonic computing involving electronic modulation of devices from a hardware perspective. Analog circuits are compatible with photonic computing and have become a research hotspot for the development of optoelectronic fusion computing chips, as they do not require AD/DA conversion.

At a higher level, a variety of photonic computing systems and applications can be achieved through the large-scale deployment of advanced neural network architectures. To increase the network scale, joint training and error correction of the system must be urgently addressed from a software perspective. In-situ training of photonic chips has been developed by researchers as a solution, where updated gradient values are computed directly by backlighting on the same hardware [19]. End-to-end large-scale network parameter updates are realized by dual adaptive training (DAT) in a dynamic error environment [20]. However, these methods, whether based on traditional neural networks or on full-field simulation of electromagnetic fields, are very energy- and time-consuming. Therefore, there is an urgent need to develop new architectures and mechanisms to design better photonic simulators and gradient solvers inspired by mathematical-physical methods. This involves integrating the entire process of photonic chip design, inference, and optimization into the hardware itself. Furthermore, the system can achieve higher computational efficiency by enabling clock alignment through hardware innovations at the architecture and device set level.

**Concluding Remarks**

In summary, diffractive photonic computing is a powerful tool for building non-Von Neumann computer architectures and generating disruptive innovations for artificial intelligence due to its high throughput and high parallelism superiority. To enhance the versatility and value of diffractive photonic computing, researchers should adopt a hardware-software co-design approach, integrating device optimization with system design. This involves constructing reconfigurable, integrable, and highly efficient error-correcting diffractive photonics computing chips within the context of optoelectronic fusion computing. We believe that in the future, diffractive photonic computing and neural networks will not only be able to process information at the edge and terminal equipment, but also hold the promise of realizing large-scale cloud computing, facilitating the application of visual and radio signal processing, optical communication, artificial intelligence accelerators, and mathematical optimizations.

**Acknowledgements**
*This work is supported by the National Natural Science Foundation of China (No. 62275139).*

## References

[1] X., Lin, Y., Rivenson, N. T., Yardimci, M., Veli, Y., Luo, M., Jarrahi, and A., Ozcan, “All-optical machine learning using diffractive deep neural networks,” Science, 361(6406), 1004-1008, 2018.

[2] T., Zhou, X., Lin, J., Wu, Y., Chen, H., Xie, Y., Li, J., Fan, H., Wu, L., Fang, and Q., Dai, “Large-scale neuromorphic optoelectronic computing with a reconfigurable diffractive processing unit,” Nature Photonics, 15(5), 367-373, 2021.

[3] H., Dou, Y., Deng, T., Yan, H., Wu, X., Lin, and Q., Dai, “Residual $D^2NN$: training diffractive deep neural networks via learnable light shortcuts,” Optics Letters, 45(10), 2688-2691, 2020.

[4] T., Yan, R., Yang, Z., Zheng, X., Lin, H., Xiong, and Q., Dai, “All-optical graph representation learning using integrated diffractive photonic computing units,” Science Advances, 8(24), eabn7630, 2022.

[5] Y., Chen, T., Zhou, J., Wu, H., Qiao, X., Lin, L., Fang, and Q., Dai, “Photonic unsupervised learning variational autoencoder for high-throughput and low-latency image transmission,” Science Advances, 9(7), eadf8437, 2023.

[6] C., Liu, Q., Ma, Z., Luo, Q., Hong, Q., Xiao, H., Zhang, L., Miao, W., Yu, Q., Cheng, L., Li, and T., Cui, “A programmable diffractive deep neural network based on a digital-coding metasurface array,” Nature Electronics, 5(2), 113-122, 2022.

[7] E., Goi, S., Schoenhardt, and M., Gu, “Direct retrieval of Zernike-based pupil functions using integrated diffractive deep neural networks,” Nature Communications, 13(1), 7531, 2022.

[8] X., Luo, Y., Hu, X., Ou, X., Li, J., Lai, N., Liu, X., Cheng, A., Pan and H., Duan, “Metasurface-enabled on-chip multiplexed diffractive neural networks in the visible,” Light: Science & Applications, 11(1), 158, 2022.

[9] A. C., Overvig, S., Shrestha, S. C., Malek, M., Lu, A., Stein, C., Zheng, and N., Yu, “Dielectric metasurfaces for complete and independent control of the optical amplitude and phase,” Light: Science & Applications, 8(1), 92, 2019.

[10] Z., Wang, L., Chang, F., Wang, T., Li, and T., Gu, "Integrated photonic metasystem for image classifications at telecommunication wavelength," Nature communications, 13(1), 2131, 2022.
[11] T., Fu, Y., Zang, Y., Huang, Z., Du, H., Huang, C., Hu, M., Chen, S., Yang, and H., Chen, "Photonic machine learning with on-chip diffractive optics," Nature Communications, 14(1), 70, 2023.
[12] T., Wu, M., Menarini, Z., Gao, and L., Feng, "Lithography-free reconfigurable integrated photonic processor," Nature Photonics, 17(8), 710-716, 2023.
[13] Z., Duan, H., Chen, and X., Lin, "Optical multi-task learning using multi-wavelength diffractive deep neural networks," Nanophotonics, 12(5), 893-903, 2023.
[14] J., Li, Y., Hung, O., Kulce, D., Mengu, and A., Ozcan, "Polarization multiplexed diffractive computing: all-optical implementation of a group of linear transformations through a polarization-encoded diffractive network," Light: Science & Applications, 11(1), 153, 2022.
[15] H., Wang, Z., Zhan, F., Hu, Y., Meng, Z., Liu, X., Fu, and Q., Liu, "Intelligent optoelectronic processor for orbital angular momentum spectrum measurement," PhotoniX, 4(1), 9, 2023.
[16] J., Laydevant, L., Wright, T., Wang, and P. L., McMahon, "The hardware is the software," Neuron, 112(2), 180-183, 2024.
[17] T., Yan, J., Wu, T., Zhou, H., Xie, F., Xu, J., Fan, L., Fang, X., Lin, and Q., Dai, "Fourier-space diffractive deep neural network," Physical review letters, 123(2), 023901, 2019.
[18] J., Feldmann, N., Youngblood, C. D., Wright, H., Bhaskaran, and W. H., Pernice, "All-optical spiking neurosynaptic networks with self-learning capabilities," Nature, 569(7755), 208-214, 2019.
[19] S., Pai, Z., Sun, T. W., Hughes, T., Park, B., Bartlett, I. A., Williamson, M., Minkov, M., Milanizadeh, N., Abebe, F., Morichetti, A., Melloni, S., Fan, O., Solgaard, and D. A., Miller, "Experimentally realized in situ backpropagation for deep learning in photonic neural networks," Science, 380(6643), 398-404, 2023.
[20] Z., Zheng, Z., Duan, H., Chen, R., Yang, S., Gao, H., Zhang, H., Xiong, and X., Lin, "Dual adaptive training of photonic neural networks," Nature Machine Intelligence, 5(10), 1119-1129, 2023.

# Analog Optical Computing

**H. Ballani, G. Brennan, B. Canakci, J. Chu, J. H. Clegg, D. Cletheroe, C. Gkantsidis, J. Gladrow, K. P. Kalinin, D. J. Kelly, H. Kremer, G. O'Shea, F. Parmigiani, L. Pickup, B. Rahmani, A. Rowstron**
Microsoft Research, 198 Science Park, Milton Road, Cambridge CB4 0AB, UK
[Hitesh.Ballani@microsoft.com, grace.brennan@microsoft.com, burcucanakci@microsoft.com, jiaqchu@microsoft.com, james.clegg@microsoft.com, daclethe@microsoft.com, christos.gkantsidis@microsoft.com, jannes.gladrow@microsoft.com, kkalinin@microsoft.com, doug.kelly@microsoft.com, t-hkremer@microsoft.com, gregos@microsoft.com, Francesca.Parmigiani@microsoft.com, lucinda.pickup@microsoft.com, t-brahmani@microsoft.com, antr@microsoft.com ]

**Status**

The rise of generative Artificial intelligence (AI) has both been driven by and been fuelling the exponential growth in compute capabilities of digital chips. Given the slowdown of Moore's law, these gains have been achieved by specialization of the hardware to the workloads, e. g. using sparsity and low bit precision arithmetic [1]. However, as we approach the fundamental limits of specialization, digital computing faces serious challenges in terms of scalability, performance, and sustainability [2]. There is, thus, an urgent need for non-traditional computing paradigms to keep up with the continuously increasing demand of machine learning (ML) and other computationally intensive workloads [3-12].

Here we comment on our efforts to build an Analog Optical Computer (AOC) to speed-up specialized computations. We estimate that AOC can offer more than 100x improvement in overall system efficiency in terms of Tera Operations per Second per Watt (TOPS/Watt at Int8 precision), as compared to state-of-the-art digital hardware, at scale. AOC uses optical and analog electronic technologies, respectively, to accelerate linear and non-linear compute primitives. Furthermore, these technologies are already (or soon to be) commodity, with an existing manufacturing ecosystem and operating at room temperature. While AOC is not a general-purpose computer, it is unique in that the exact same hardware can accelerate two computationally intensive verticals: machine learning (ML) inference and hard combinatorial optimization problems, building upon parallels in these almost independent communities. A critical innovation area in AOC is the codesign of hardware with these application verticals – a feature that has also been key to the synergistic success of digital chips like GPUs and deep learning. On the ML front, we have focussed on emerging analog-amenable machine learning that has the potential to translate the TOPS/Watt hardware gains into gains in terms of inferences per second per Watt. On the optimization front, we have developed novel optimization abstractions that take a big step in closing the gap between the expressiveness of the non-traditional hardware like AOC and the requirements of real-world optimization problems.

**Current and Future Challenges**

Non-traditional computing based on optics or analog electronics [3-12] has made a lot of progress over the past decade, but significant challenges remain. At the hardware level, planar optical technologies [6] offer the key advantage of component miniaturization, which is critical to scaling. However, they suffer from the fundamental challenge that precious on-chip real estate is used both for computing on and routing of data. 3-dimensional (3D) optics using surface-emitting source and modulators and detectors sidesteps this challenge, thus opening the path to a step change in performance gains as compared to digital chips. 3D demonstrations however, so far, are based on free-space implementations which are bulky. Furthermore, optical technologies are great for accelerating linear operations but, despite decades of impressive research, the potential for energy-efficient acceleration of non-linear and complicated operations is less promising. AOC tackles these hardware-level challenges by combining integrated 3D optics to accelerate linear operations and analog electronics for non-linear operations. This combination means the entire computation for ML and optimization problems, often very iterative in nature, can be done in the analog domain without any on-path digital conversions and without an explicit clock. The use of 3D optics allows us to implement a massive vector-matrix multiplier with compute-in-memory operation, thus alleviating the traditional IO bottleneck in digital chips. This all-analog, asynchronous and in-memory operation of our computer is key to achieving the two orders of magnitude gains in system efficiency.

There are challenges further up the stack too. Specifically, ultra-high TOPS/Watt is necessary but not sufficient to accelerate the target applications. In ML, prevalent models like auto-regressive transformers are IO-bound, so computers with impressive TOPS/Watt, by themselves, do not help towards radical end-to-end system performance improvements – particularly when the computations are noisy in nature. In optimization, there has been a lot of work on Ising machines [8-12], but their ability to efficiently accommodate real-world problems is still an unresolved challenge [13]. We believe that hardware-application codesign is critical to really take advantage of the speed-ups of non-traditional computers and compensate for their shortcomings. On the ML front, there are emerging models, e.g., energy-based [14] and diffusion models [15], that achieve excellent performance and functionality at the cost of increased operational intensity which is very conducive to the strengths and weaknesses of AOC. On the optimization front, we have proposed a more expressive abstraction for hard optimization problems – the quadratic unconstrained mixed optimisation (QUMO) abstraction [3] – that naturally maps to our hardware and is able to capture real world optimisation problems, for example, in finance, manufacturing, and healthcare sectors. Most importantly, the same AOC hardware caters to both these application segments.

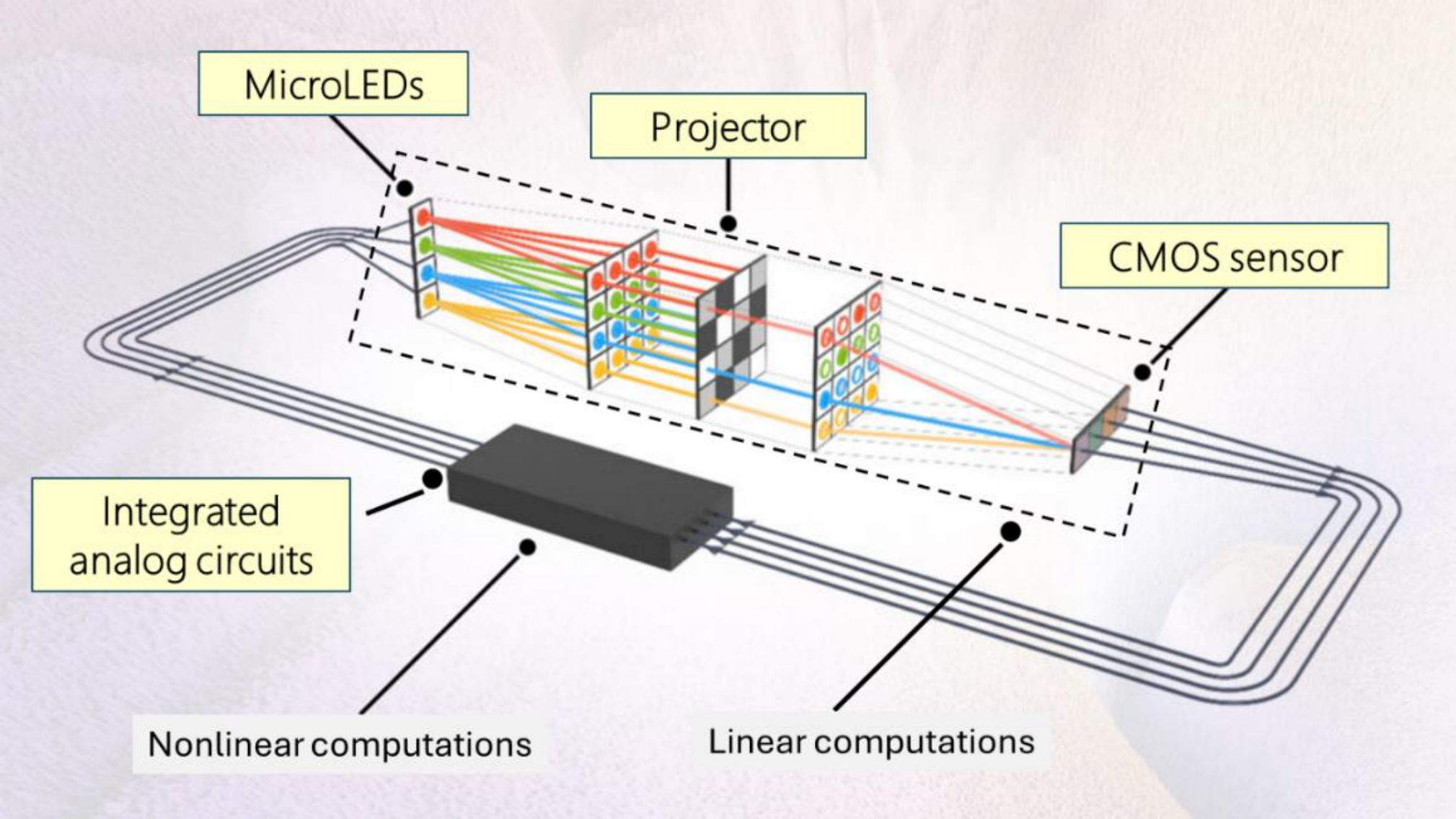

**Figure 1.** Schematic representation of the proposed AOC machine, highlighting that the linear operations are carried out by commodity optical technology and the nonlinear operations are carried out by commodity analog electrical technology. A free space architecture is highlighted for clarification purposes only.

### Advances in Science and Technology to Meet Challenges

The technologies underlying AOC hardware have been intentionally chosen such that they have an existing manufacturing ecosystem, allowing it to benefit from their low-cost and mass production. However, important technological progress is still needed to meet the performance required for computations which can be very different from the original performance specifications for these technologies in the consumer space, without sacrificing their potential scalability and cost. For instance, while micro-led-based light sources used in AOC can operate at GHz-bandwidth [16], their efficiency needs to be improved for the computer to achieve the target TOPS/Watts. Moreover, while optical fan-in and fan-out architectures have the potential of scaling, see e. g. [4], they still lack demonstrations at significant scales and need further technological enhancements. Additionally, to compete with the juggernaut of the GPU and digital ecosystem, existing silos between the algorithmic and ML experts and non-traditional hardware designers need to be broken. This would drive algorithmic and model innovations while also expanding the library of compute primitives that can be effectively built in such hardware. While we and a few other groups are fostering such joint communities, much more buy-in and scientific progress is still needed to create a sustaining flywheel that results in non-traditional computers powering the future of compute.

**Concluding Remarks**

By codesigning new machine learning models and optimisation abstractions and algorithms with hardware based on integrated 3D optical and analog technologies that are commodity, AOC has the potential to provide a 100x TOPS/Watt as compared to state-of-the-art digital computers and translate it into significant efficiency gains for promising ML models and real-world optimization problems.

**Acknowledgements**
*We acknowledge helpful discussions and support from our colleagues at Future AI Infrastructure at Microsoft Research Cambridge, UK, and at M365 Research, Michael Schapira and Natalya Berloff.*

**References**

[1] https://blogs.nvidia.com/blog/hot-chips-dally-research/

[2] A. S. Luccioni, S. Viguier, A. L. Ligozat, "Estimating the carbon footprint of bloom, a 176b parameter language model," Journal of Machine Learning Research, Volume 24, Issue 1, Article No.: 253, pp 11990–12004, 2023.

[3] K. Kalinin, G. Mourgias-Alexandris, H. Ballani, N. Berloff, J. Clegg, D. Cletheroe, C. Gkantsidis, I. Haller, V. Lyutsarev, F. Parmigiani, L. Pickup, A. Rowstron, "Analog Iterative Machine (AIM): using light to solve quadratic optimization problems with mixed variables", April 2023, arXiv:2304.12594.

[4] McMahon, P. L. The physics of optical computing. Nat Rev Phys 5, 717–734 (2023). https://doi.org/10.1038/s42254-023-00645-5.

[5] G. Wetzstein, A. Ozcan, S. Gigan, S. Fan, D. Englund, M. Soljacic, C. Denz, D. A. Miller, and D. Psaltis, Inference in artificial intelligence with deep optics and photonics. Nature 588, 39–47 (2020).

[6] B. J. Shastri, A. N. Tait, T. Ferreira de Lima, W. H. Pernice, H. Bhaskaran, C. D. Wright, and P. R. Prucnal, Photonics for artificial intelligence and neuromorphic computing. Nature Photonics 15, 102–114 (2021).

[7] S. Greengard, Photonic processors light the way. Communications of the ACM 64, 16–18 (2021).

[8] N. Mohseni, P. L. McMahon, and T. Byrnes, Ising machines as hardware solvers of combinatorial optimization problems. Nature Reviews Physics 4, 363–379 (2022).

[9] Inagaki T, Haribara Y, Igarashi K, Sonobe T, Tamate S, Honjo T, Marandi A, McMahon PL, Umeki T, Enbutsu K, Tadanaga O, Takenouchi H, Aihara K, Kawarabayashi KI, Inoue K, Utsunomiya S, Takesue H., A coherent Ising machine for 2000-node optimization problems. Science. 2016 Nov 4;354(6312):603-606. doi: 10.1126/science.aah4243. Epub 2016 Oct 20. PMID: 27811271. "A coherent Ising machine for 2000-node optimization problems". Science 354.6312 (2016).

[10] Kalinin, Kirill P., Amo, Alberto, Bloch, Jacqueline and Berloff, Natalia G.. "Polaritonic XY-Ising machine" Nanophotonics, vol. 9, no. 13, 2020, pp. 4127-4138. https://doi.org/10.1515/nanoph-2020-0162.

[11] D. Pierangeli, G. Marcucci, and C. Conti. "Large-scale photonic Ising machine by spatial light modulation". Physical Review Letters 122.21 (2019).

[12] Moy, W., Ahmed, I., Chiu, Pw. et al. A 1,968-node coupled ring oscillator circuit for combinatorial optimization problem solving. Nat Electron 5, 310–317 (2022). https://doi.org/10.1038/s41928-022-00749-3.

[13] DARPA QUICC program, https://www.darpa.mil/news-events/2021-10-04
[14] S. Bai, J. Z. Kolter, and V. Koltun, "Deep equilibrium models." Advances in neural information processing systems 32 (2019).
[15] Y. Song, J. Sohl-Dickstein, D. P. Kingma, A. Kumar, S. Ermon, B. Poole "Score-based generative modeling through stochastic differential equations", arXiv preprint arXiv:2011.13456 (2020).
[16] B. Pezeshki, A. Tselikov, R. Kalman and C. Danesh, "Wide and parallel LED-based optical links using multi-core fiber for chip-to-chip communications," 2021 Optical Fiber Communications Conference and Exhibition (OFC), San Francisco, CA, USA, 2021, pp. 1-3.

# Spatial photonic Ising machines for combinatorial optimization and spin physics

**Davide Pierangeli[1,2], Daniele Veraldi[1], Silvia Gentilini[2], Marcello Calvanese Strinati[3], and Claudio Conti[1,3]**

[1] Department of Physics, Sapienza University, 00185 Rome, Italy

[2] Institute for Complex Systems, National Research Council, 00185 Rome, Italy

[3] Enrico Fermi Research Center, 00184 Rome, Italy

[davide.pierangeli@roma1.infn.it daniele.veraldi@uniroma1.it silvia.gentilini@cnr.it marcello.calvanesestrinati@cref.it claudio.conti@uniroma1.it]

**Status**

Advances in critical areas such as artificial intelligence and communications require fast and scalable hardware for combinatorial optimization problems that are challenging to solve by conventional computers within a limited time. Ising Machines (IMs) [1] are gaining considerable attention as specialized hardware that could outperform traditional digital processors by implementing these NP-hard problems as an Ising model of interacting spins whose ground state represents the problem solution. IMs may drastically impact a myriad of applications. Photonic IMs [2] are especially promising, taking advantage of the unique features offered by the underlying optical system, such as its ultrafast dynamics, coherence, and optical parallelism, for accelerating the Ising computation.

Spatial Photonic Ising Machines (SPIMs) are a paradigm that encompasses IMs operating by spatial light modulation [3]. They are based on the finding that coherent optical propagation in free space maps the Ising model when spins are encoded through binary modes on the spatial profile of a laser beam. Specifically, spins are represented by binary phases multiplexed in space and the intensity on the detector gives the absolute value of the Ising Hamiltonian. Figure 1 illustrates the operating principle of a SPIM. The scheme leverages the spatial parallelism of free-space optics and the high pixel density of spatial light modulators (SLMs) to compute in parallel the Ising energy of thousands of interacting spins by the mere propagation of light. Computing the Ising energy is the building block of most heuristic algorithms for searching the ground state and for simulating spin systems at a finite temperature. This computation for $N$ spins requires $O(N^2)$ MAC operations on a digital processor, while on a SPIM the result is obtained by a single intensity measurement with a computational cost $O(1)$ almost independent of the system's size and low power consumption (*mW* laser light). Therefore, SPIMs showcase a remarkable advantage in scalability and energy efficiency, highlighting an alternative pathway to tackle extensive Ising problems. Large scales are readily achieved as SLMs allow the encoding of millions of switchable spins, a size challenging to realize with other photonic platforms. The first SPIM embedded more than 40,000 spins [3], breaking the size record of existing IMs. The device operates as a photonic annealer: the Ising energy is measured and its value is used to update iteratively the spin configuration via digital feedback by using a Metropolis-Hasting algorithm. This operating mechanism is general and can accelerate many other minimization algorithms. Adiabatic computation and simulated annealing have been implemented on a SPIM achieving remarkable accuracy [4]. In addition, the ground-state search on SPIMs can benefit from experimental factors. Their success probability has been enhanced by using physical noise within the experimental setup instead of random numbers generated digitally [5]. These attractive features make SPIMs a promising route for large-scale combinatorial optimization and simulation of spin-physics phenomena.

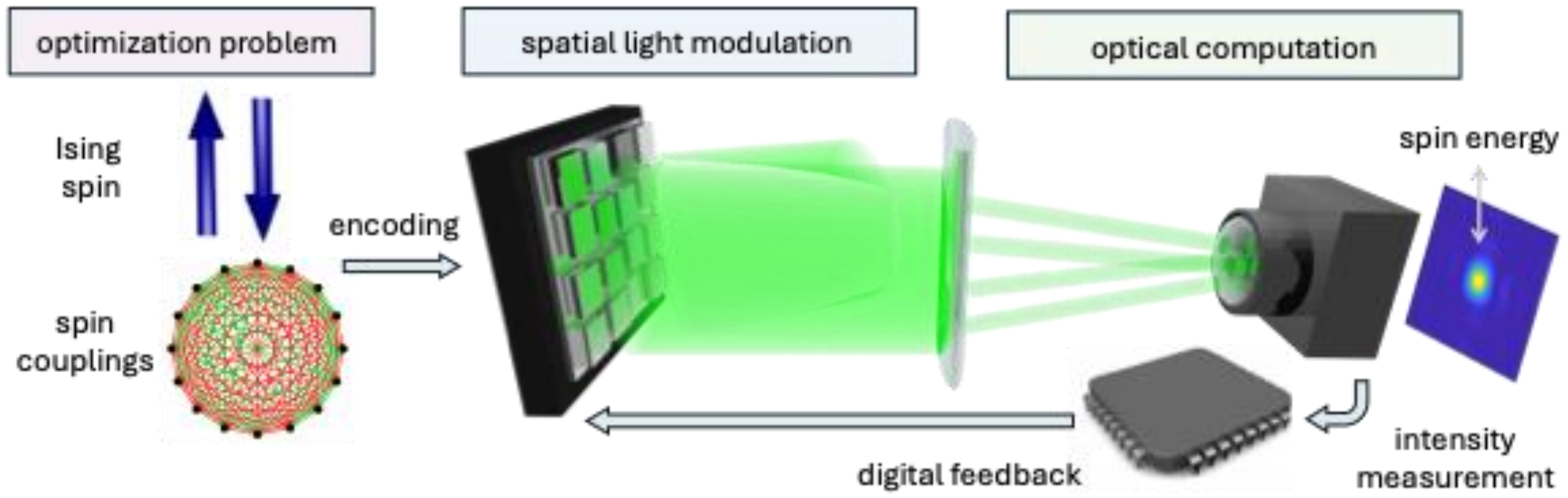

**Figure 1.** Concept and scheme of a SPIM. The combinatorial optimization problem is recast as a system of Ising spins interacting by a coupling matrix (graph). Both the binary spins and the couplings are encoded on the optical field by spatial modulation of the phase and/or amplitude. The spin energy is computed optically and obtained by measuring the optical intensity in the Fourier plane. Digital feedback uses the measurement to update the spin configuration through an iterative minimization algorithm until convergence to the ground state (problem solution).

## Current and Future Challenges

***Programming the spin couplings.*** The capability to program the spin couplings is crucial to implement on IMs the optimization problem of interest. In SPIMs, couplings are also realized optically and controlled by spatial light modulation with 8-bit resolution. Initially, the class of Mattis-type couplings is realized by amplitude modulation of the input beam [4, 5]. Various research groups have extended the implementable couplings by engineering the optical propagation [6, 7, 8, 9]. The Gauge transformation method [6] allows Mattis-type couplings by using a single phase-only SLM, greatly simplifying the setup and enabling accurate photonic simulation of various magnetic phase transitions. Combined with the intensity correlation method [7], the Gauge approach enables the implementation of quadratic unconstrained binary optimization (QUBO) problems expressed by low-rank and circulant matrices [10], including combinatorial tasks [11] and statistical learning [12]. Very recently, intense efforts have led to various fully programmable SPIMs based on matrix decomposition and multiplexing schemes [13, 14, 15]. SPIMs that exploit, respectively, the division of the focal plane [13], optical vector-matrix multipliers [14], and wavelength-division multiplexing [15], have demonstrated remarkable accuracy in solving arbitrary Max-Cut graphs with tens of spins. The challenge is now scaling up these more complex schemes to benchmark their performance and test their application to real-world tasks. Remarkably, large-scale full-rank coupling matrices have been realized in the case of gaussian random interactions by exploiting light transmission through a multiple scattering medium, which turns the SPIM into an optical spin-glass simulator [16, 17, 18]. Low-energy states [16] and spin glass dynamics of tunable complexity [17] have been demonstrated, offering new possibilities for photonic information storage [19].

***Speed up SPIM operation***. To compete with digital hardware, the challenge is reducing the SPIM iteration time by orders of magnitude. The SPIM run time is currently limited by the slow frame rate of liquid crystal SLMs (<1 kHz), which results in a time-to-solution of $10^1$ *s* for $10^3$ spins [13]. However, since the SPIM iteration time does not depend on the problem size and density, competitive time performance is expected at a scale of $10^5$ spins. This will require the design of dedicated algorithms that exploit further the SPIM spatial parallelism and minimize the number of iterations to reach the ground state such as genetic algorithms. Research aimed at overcoming the bottlenecks associated with the use of digital feedback, either through analog electronics [20] or optical cavities [21], is crucial to push SPIMs into the realm of ultrafast computation.

***Beyond the Ising Hamiltonian*.** The SPIM concept can be extended to realize both models with non-binary spins and higher-order Hamiltonians. SPIMs that realize four-body interactions by second-harmonic generation in nonlinear media [22] are promising for non-quadratic optimization, machine learning, and simulations of critical phenomena in magnetic materials [23]. The multi-level phase modulation of SLMs allows the encoding of clock and circular spins, enabling simulations of the XY Hamiltonian with programmable couplings [24] and its topological dynamics [25]. Potts and Heisenberg models have been implemented [26] by using a novel superpixel approach with a digital micromirror device (DMD). Furthermore, the SPIM scheme can be further developed to implement a spatial hyperspin machine to realize vector spin models in arbitrary dimensions and novel annealing methods [27], thus largely extending the scope of SPIM setups for spin physics. These findings are opening the new field of spatial photonic spin machines (SPSMs). Given the increasing complexity of high-dimensional models, SPSMs can offer even more significant computational advantages, standing out as a new powerful tool for statistical mechanics and an appealing alternative to high-performance computing. Mapping intractable computational tasks that arise in practical applications to these spin models is essential to unleash the full potential of SPSMs that will emerge.

## Advances in Science and Technology to Meet Challenges

A first advance towards an ultrafast SPIM is offered by upgrades in DMDs and MEMS-based SLMs. As the frame rate of spatial modulators approaches $10^5$ Hz, the SPIM computation time becomes comparable to conventional hardware at large scales. In this context, the breakthrough will be the development of electro-optic SLMs [27], which promise GHz frame rates. Electro-optic SPIMs are expected to deliver near-optimal solutions to large-scale optimization problems in *ms*, impacting ultrafast applications such as machine vision and communications. They can be readily integrated into compact and energy-efficient devices for widespread use in science and industry.

## Concluding Remarks

SPIMs are emerging as near-term hardware for combinatorial optimization and simulation of spin models, showcasing possible advantages at a large scale over digital processors. The field proliferates interesting developments. Further advantages and new possibilities are envisioned by using few-photon sources and quantum light to drive the setup. Advances in SLM technology may establish SPIMs as a leading non-von Neumann paradigm for ultrafast neuromorphic computing in the post-Moore's law era.

## Acknowledgements

We acknowledge funding from HORIZON EIC-2022-PATHFINDERCHALLENGES-01 HEISINGBERG Project No. 101114978 and the National Quantum Science and Technology Institute (NQSTI) under the National Recovery and Resilience Plan (NRRP) funded by the EU-NextGenerationEU, MUR Project PE0000023-NQSTI.

## References

[1] N. Mohseni, P. L. McMahon, and T. Byrnes, "Ising machines as hardware solvers of combinatorial optimization problems," Nature Reviews Physics, vol. 4, no. 6, p. 363-379, 2022.

[2] Y. Gao, G. Chen, L. Qi, W. Fu, Z. Yuan, and A.J. Danner, "Photonic Ising machines for combinatorial optimization problems," Applied Physics Reviews, vol. 11, no. 4, p. 041307, 2024.

[3] D. Pierangeli, G. Marcucci, and C. Conti, "Large-scale photonic Ising machine by spatial light modulation," Physical Review Letters, vol. 122, no. 21, p. 213902, 2019.

[4] D. Pierangeli, G. Marcucci, and C. Conti, "Adiabatic evolution on a spatial-photonic Ising machine," Optica, vol. 7, no. 11, p. 1535-1543, 2020.

[5] D. Pierangeli, G. Marcucci, D. Brunner, and C. Conti, "Noise-enhanced spatial-photonic Ising machine," Nanophotonics, vol. 9, no. 13, p. 4109-4116, 2020.

[6] Y. Fang, J. Huang, and Z. Ruan, "Experimental observation of phase transitions in spatial photonic Ising machine," Physical Review Letters, vol. 127, no. 7, p. 043902, 2021.

[7] J. Huang, Y. Fang, and Z. Ruan, "Antiferromagnetic spatial photonic Ising machine through optoelectronic correlation computing," Communications Physics, vol. 4, no. 1, p. 242, 2021.

[8] W. Sun, W. Zhang, Y. Liu, Q. Liu, and Z. He, "Quadrature photonic spatial Ising machine," Optics Letters, vol. 47, no. 6, p. 1498-1501, 2022.

[9] W. Fan, Y. Sun, X. Xu, D. W. Wang, S. Y. Zhu, H. Q. Lin, "Programmable Photonic Simulator for Spin Glass Models," 2023, arXiv:2310.14781.

[10] R. Z. Wang, J. S. Cummins, M. Syed, N. Stroev, G. Pastras, J. Sakellariou, S. Tsintzos, A. Askitopoulos, D. Veraldi, M. C. Strinati, S. Gentilini, D. Pierangeli, C. Conti, N. G. Berloff, Efficient computation using spatial-photonic Ising machines: Utilizing low-rank and circulant matrix constraints, 2024, arXiv:2406.01400.

[11] T. Sakabe, S. Shimomura, Y. Ogura, K. I. Okubo, H. Yamashita, H. Suzuki, and J. Tanida, "Spatial-photonic Ising machine by space-division multiplexing with physically tunable coefficients of a multi-component model," Optics Express, vol. 31, no. 26, p. 44127-44138, 2023.

[12] H. Yamashita, K. I. Okubo, S. Shimomura, Y. Ogura, J. Tanida, and H. Suzuki, "Low-rank combinatorial optimization and statistical learning by spatial photonic Ising machine," Physical Review Letters, vol. 131, n. 6, p. 063801, 2023.

[13] D. Veraldi, D. Pierangeli, S. Gentilini, M. C. Strinati, J. Sakellariou, J. S. Cummins, A. Kamaletdinov, M. Syed, R. Z. Wang, N. G. Berloff, D. Karanikolopoulos, P. G. Savvidis, C. Conti, "Fully Programmable Spatial Photonic Ising Machine by Focal Plane Division," 2024, arXiv:2410.10689.

[14] J. Ouyang, Y. Liao, Z. Ma, D. Kong, X. Feng, X. Zhang, X. Dong, K. Cui, F. Liu, W. Zhang, and Y. Huang, "On-demand photonic Ising machine with simplified Hamiltonian calculation by phase encoding and intensity detection," Communications Physics, vol. 7, no. 1, p. 168, 2024

[15] L. Luo, Z. Mi, J. Huang, and Z. Ruan, "Wavelength-division multiplexing optical Ising simulator enabling fully programmable spin couplings and external magnetic fields," Science Advances, vol. 9, no. 48, p. eadg6238, 2023.

[16] D. Pierangeli, M. Rafayelyan, C. Conti, and S. Gigan, "Scalable spin-glass optical simulator," Physical Review Applied, vol. 15, no. 3, p. 034087, 2021.

[17] M. Leonetti, E. Hörmann, L. Leuzzi, G. Parisi, and G. Ruocco, "Optical computation of a spin glass dynamics with tunable complexity," Proceedings of the National Academy of Sciences of the United States of America, vol. 118, n. 21, p. e2015207118, 2021.

[18] G. Jacucci, L. Delloye, D. Pierangeli, M. Rafayelyan, C. Conti, and S. Gigan, "Tunable spin-glass optical simulator based on multiple light scattering," Physical Review A, vol. 105, no. 3, p. 033502, 2022.

[19] M. Leonetti, G. Gosti, and G. Ruocco, "Photonic Stochastic Emergent Storage for deep classification by scattering-intrinsic patterns," Nature Communications, vol. 15, no. 1, p. 505, 2024.

[20] K. P. Kalinin, G. Mourgias-Alexandris, H. Ballani, N. G. Berloff, J. H. Clegg, D. Cletheroe, C. Gkantsidis, I. Haller, V. Lyutsarev, F. Parmigiani, L. Pickup, A. Rowstron, Analog Iterative Machine (AIM): using light to solve quadratic optimization problems with mixed variables, 2023, arXiv:2304.12594v2.

[21] M. Calvanese Strinati, D. Pierangeli, and C. Conti, “All-optical scalable spatial coherent Ising machine,” Physical Review Applied, vol. 15, no. 5, p. 054022, 2021.

[22] S. Kumar, H. Zhang, and Y. P. Huang, “Large-scale Ising emulation with four body interaction and all-to-all connections,” Communications Physics, vol. 3, no. 1, p. 108, 2020.

[23] S. Kumar, Z. Li, T. Bu, C. Qu, and Y. Huang, “Observation of distinct phase transitions in a nonlinear optical Ising machine,” Communications Physics, vol. 6, no. 1, p. 31, 2023.

[24] J. Ouyang, Y. Liao, X. Feng, Y. Li, K. Cui, F. Liu, H. Sun, W. Zhang, and Y. Huang, “Programmable and reconfigurable photonic simulator for classical $XY$ models,” Physical Review Applied, vol. 22, no. 2, p. L021001, 2024.

[25] J. Feng, Z. Li, L. Yuan, E. Hasman, B. Wang, X. Chen, “Spin Hamiltonians in the Modulated Momenta of Light,” 2024, arXiv:2405.00484v2.

[26] S. T. Yu, M. G. He, S. Fang, Y. Deng, Z. S. Yuan, “Spatial Optical Simulator for Classical Statistical Models,” Physical Review Letters, vol. 133, no. 23, p. 237101, 2024.

[27] M. Calvanese Strinati and C. Conti, “Hyperscaling in the Coherent Hyperspin Machine,” Physical Review Letters, vol. 132, no. 1, p. 017301, 2024.

[28] S. Trajtenberg-Mills, M. El Kabbash, C. J. Brabec, C. L. Panuski, I. Christen, and D. Englund, “LNoS: Lithium niobate on silicon spatial light modulator,” 2024, arXiv:2402.14608.

# Photonic extreme learning machine

**Stefano Biasi, Riccardo Franchi* and Lorenzo Pavesi**
Nanoscience Laboratory, Department of Physics, University of Trento, 38123 Trento, Italy
*Currently with Nanomaterials & Nanostructure Optics, Department of Electrical and Computer Engineering, Boston University, Boston, MA, USA

S.B. : stefano.biasi@unitn.it, R.F.: riccardo.franchi@unitn.it and L.P.: lorenzo.pavesi@unitn.it

## Status

The Extreme Learning Machine (ELM) is a computational paradigm that offers an efficient alternative to traditional neural networks and Support Vector Machine (SVM) models **[1]**. As shown in figure 1, ELM is based on a feed-forward neural network consisting of a single hidden layer where information is processed and sent to an output layer formed by at least a single output node **[2]**. The hidden layer nonlinearly maps input signals into a higher-dimensional computational space using random weights and an infinitely differentiable nonlinear function **[2]**. Training occurs exclusively in the output layer through a standard linearization process, such as the linear regression. In Von Neumann-based electronic hardware, ELMs typically involve matrix multiplication using random weights and nonlinear functions. Despite their optimal performances, electronic implementations become computationally expensive as data volume grows and face the memory-processor communication bottleneck **[3]**. This has led to alternative approaches to computing **[4,5]**. Since ELM does not require internal-interconnection tuning, it is suitable for photonic implementations, hence the name Photonic ELM (PELM) or Optical ELM (OELM).

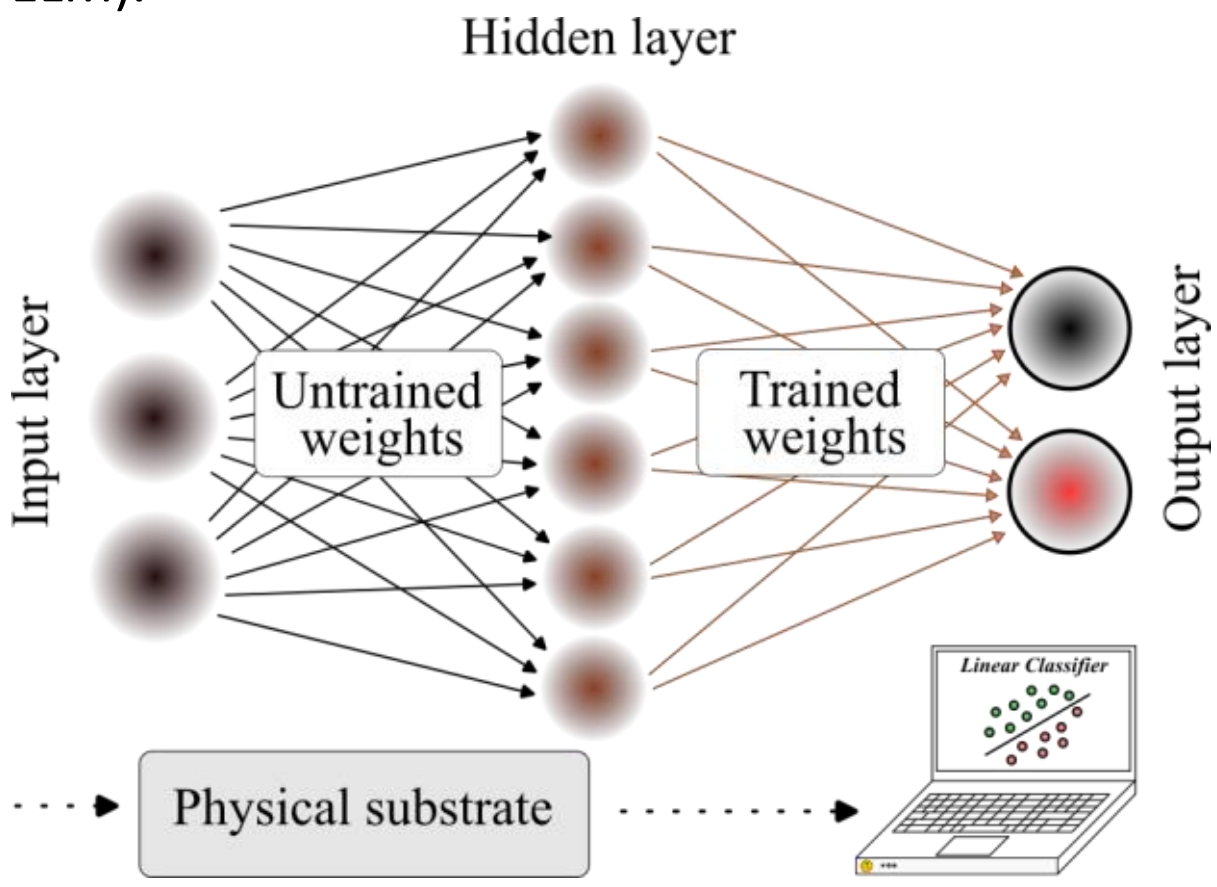

**Figure 1**: Sketch of an Extreme Learning Machine: input data undergoes a random dimensional expansion using untrained weights, followed by a nonlinear transformation in the hidden layer. Training involves adjusting the output weights during the readout phase. The untrained connection and/or encoding of the input information can be performed using a physical platform, with training typically executed offline by application of a linear transformation to the hidden layer output.

PELM has been demonstrated by using different platforms and input-data encoding methods. When free-space optics is used **[6]**, the encoding is performed at the input optical signal wavefront by a phase-only spatial light modulator (SLM) and the phase information is linearly self-mixed during light propagation (see figure 2 (a)). Fiber-optic based ELMs use different architectures: virtual neurons randomly projected into the input layer and nonlinearly processed by a lithium-niobate Mach-Zehnder modulator **[7]**; information encoded with programmable filters in the spectrum of a frequency comb and processed with phase modulator **[8]**; speckle dynamics in multimode fibers **[9,10]** (see figure 2 (b)); longitudinal modes of a Fabry-Perot laser **[11]**. In photonic integrated circuits (PICs), a PELM is demonstrated by using an array of microresonators for random input space expansion and integrated microheaters to encode the input **[12]** (see figure 2 (c)).

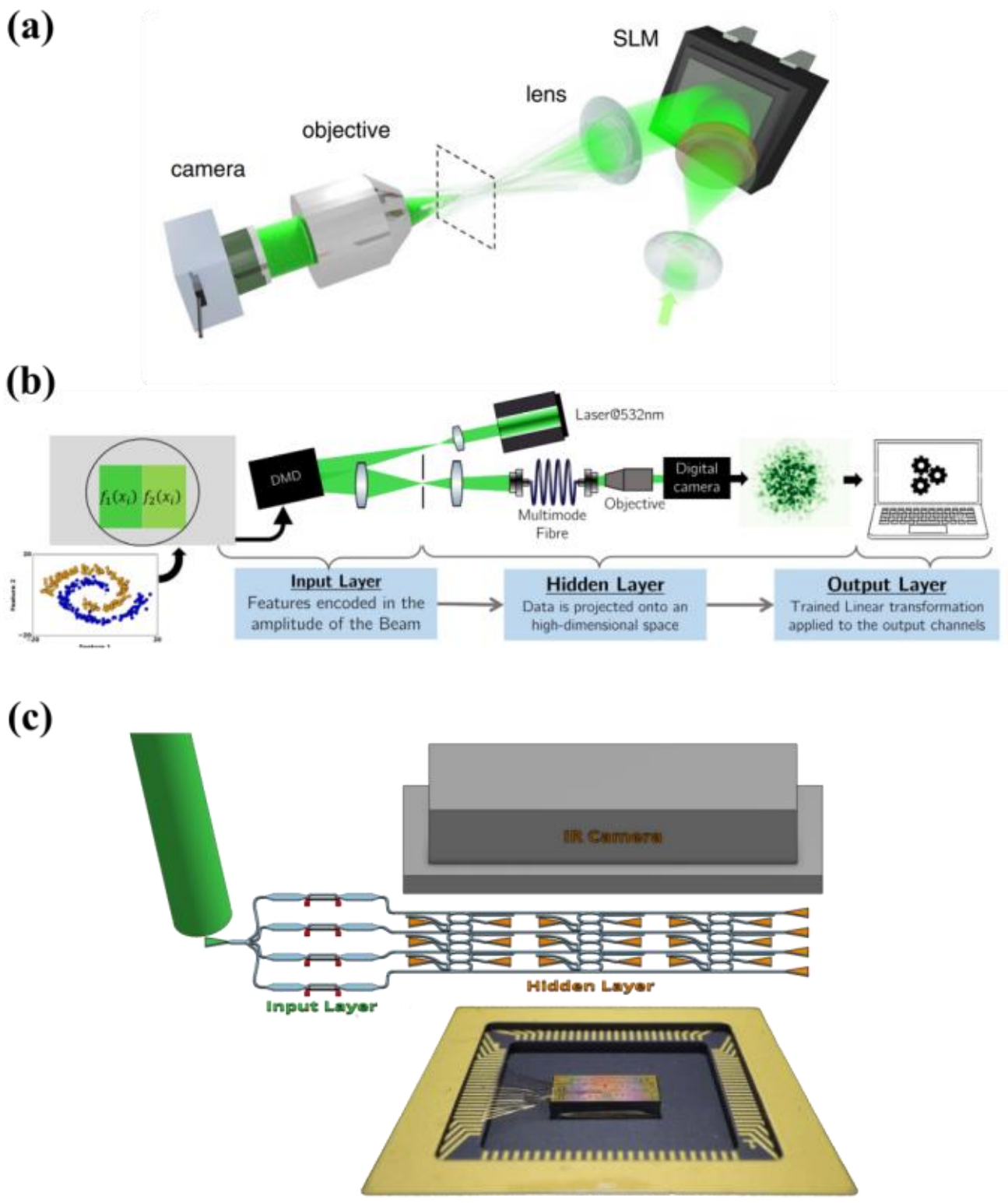

**Figure 2**: Three experimental implementations of a Photonic Extreme Learning Machine. (a) Bulk system: input data is encoded in the wavefront of a continuous-wave laser using a phase-only spatial light modulator. The phase information undergoes self-mixing during light propagation. The resulting optical signal in the far-field is detected by a camera which results in a nonlinear transformation of the output. Reprint by permission from **[6]**. (b) Fiber system: the input information is encoded by a digital micromirror. The modulated optical signal propagates in a multimodal fiber, generating a speckle pattern which is detected by a camera (nonlinear transformation). Reprint by permission from **[10]** (Creative Commons license CC BY 4.0). (c) Photonic integrated circuit: input data is encoded in the field amplitude of the optical signal with four Mach-Zehnder interferometers. The hidden layer consists of an array of 18 microresonators coupled to diffraction gratings. The input domain space is expanded by propagation in the microresonators, and a nonlinear function is applied by the camera which records the image of scattered light by the gratings **[12]**. In all these implementations, the trained weights are computed digitally and applied to the hidden layer intensities to obtain the prediction.

Most implementations use the square-law of photodetection as the nonlinear function. Few leverage the intrinsic nonlinearities of the materials **[13].** Examples include the Kerr nonlinearity of optical fiber **[14]** or of atomic vapor **[15]**. Finally, a quantum PELM model using bulk optics is presented in **[16]**. Here, the classical nonlinear function is replaced by quantum channel evolution followed by a measurement **[17]**.

**Current and Future Challenges**

Two fundamental aspects of ELM require further studies: the random expansion of spatial dimensions of the input space and the application of the nonlinear function. In bulk systems, mapping input signals into a higher-dimensional space is relatively straightforward. For instance, when a coherent laser beam scatters on a surface, the interference of scattered light creates large linear mappings that produce random patterns, known as speckles. A more limited speckle pattern can also be generated in multimodal fibers **[10]**. However, expanding input size in PICs is more complex, resulting in few hidden layer nodes.

Expansion strategies for fibers and PICs typically involve four approaches: spatially by increasing the physical size, virtually by using virtual nodes, in frequency through wavelength multiplexing techniques, in the optical field distribution by using different mode orders. These methods can be combined to create complex hyperspaces. In PICs, spatial size expansion is constrained by propagation losses. A hybrid approach, like space-wavelength multiplexing, seems to be the optimal **[18]**. However, the actual topology of the PIC strongly impacts the PELM performance **[12]**. Thus, a priori analysis of the network structure could reduce the number of required nodes and speed up the readout phase.

This task is challenging as it requires the actual modeling of the physical system. Moreover, existing PIC implementations only use the detector nonlinearities **[12]**, and there is a lack of studies based on the inherent nonlinearity of the waveguides and on the number of nodes to ensure optimal performance.
The role of the inherent nonlinearity of the material platform is another topic of research. Unlike in reservoir computing, where the nonlinearity relates to short-term memory, the nonlinearity in a PELM contributes to the high-dimensional input space expansion **[2]**. Indeed, some studies indicate limited learning capability of PELMs when light propagates in linear or weakly nonlinear systems **[13, 19]**. Controversially, recent experimental work claims that a weak nonlinear Kerr effect can be used efficiently for data processing **[14]**. Remarkably, the usual square-law nonlinearity of photodetectors is always present in the demonstrated PELMs, due to the offline data processing phase. However, the measurement of the optical field intensity yields a loss of the phase information, which in turn compresses the input space expansion.
Finally, a big open challenge is the all-optical approach. Currently, PELMs involve digitizing the hidden layer response for offline training and testing using software algorithms, like ridge regression, which ensures computational efficiency. Optical readout would accelerate data processing and reduce the amount of data to be stored, especially with large nonlinear input mappings. In PICs, this issue is also related to the input data encoding. Using conventional Mach-Zehnder modulators controlled by microheaters for readout and encoding hampers performance due to long thermalization times and thermal cross-talk **[12, 20].**

**Advances in Science and Technology to Meet Challenges**

In PELM based on bulk or fiber optics, input data encoding is often done using SLMs. Liquid crystal light modulators are versatile tools for generating arbitrary optical fields, enabling precise control over phase and amplitude. This allows the customization of degrees of freedom and a significant increase in encoding speed, which in turn is reflected in the possibility for PELM to process large datasets with numerous attributes at higher data rates while using simple interference and/or speckle pattern generation.
For PICs, improvements in fabrication methods, mostly hybrid approaches, could facilitate the realization of all-optical PELM. This requires the use of efficient and compact p-n junction modulators, which could mitigate issues associated with thermos-optic actuations. In addition, their large bandwidth (up to 50 GHz) could enable integration of the optical readout layer. Ideally, the training and testing could be performed optically via electronic pre-training, leveraging the amplitude and phase control of the optical fields exiting the hidden layer. Electronic pre-training involves initial photodiode detection of both amplitude and phase, followed by software linearization. As a result, optical tuning can be performed without reliance on complex backpropagation methods. Note that each node of the hidden layer may have a nonlinear function that is potentially different from the square-law of photodetectors. In both fiber-based and PIC-based systems, the significant progress of hybrid approaches in creating complex hyperspaces paves the way for high-performance implementations of PELMs with a reduced number of physical nodes. However, the scalability of any PELM requires concurrent consideration of the system topology, an area that has not been extensively explored theoretically. Finally, hybrid approaches based on phase change materials for non-volatile weights or two dimensional materials for large optical nonlinearities are unexplored research fields in PELM.

**Concluding Remarks**

Despite extensive efforts to implement the extreme learning machine algorithm in photonic hardware, the field is still in its infancy. In fact, a leading platform to implement PELM has not yet emerged. Ongoing advances in microelectronics and optoelectronics integration technology offer promising opportunities for developing novel and high-performance PELMs in bulk, fiber, and integrated optics.

However, a comparative study between the typical square-law nonlinearity of detectors and the intrinsic nonlinearity of the material combined with the number of active nodes required is still missing in the literature. Similarly, a comprehensive study of the role of topology in the input set expansion is lacking. Future developments and research may lead to the identification of an optimal platform and could enable PELMs for more complex classification tasks beyond the common neural network benchmarks.

**Acknowledgements**

This research was supported by the European Research Council under the Horizon 2020 research and innovation program (grant numbers 788793-BACKUP and 963463-ALPI) and by the Italian Ministry of Education, University, and Research under the project PRIN PELM (grant number 20177 PSCKT). S.B. acknowledges the co-financing of the European Union (FSE-REACT-EU) and the National Operational Programme on Research and Innovation 2014-2020 (DM1062/2021).

**References**

[1] G.-B. Huang, H. Zhou, X. Ding, and R. Zhang, "Extreme Learning Machine for Regression and Multiclass Classification," *IEEE Transactions on Systems, Man, and Cybernetics, Part B (Cybernetics)*, vol. 42, no. 2, pp. 513–529, Apr. 2012, doi: 10.1109/TSMCB.2011.2168604.

[2] G.-B. Huang, Q.-Y. Zhu, and C.-K. Siew, "Extreme learning machine: Theory and applications," *Neurocomputing*, vol. 70, no. 1, pp. 489–501, Dec. 2006, doi: 10.1016/j.neucom.2005.12.126.

[3] "Beyond von Neumann," *Nat. Nanotechnol.*, vol. 15, no. 7, pp. 507–507, Jul. 2020, doi: 10.1038/s41565-020-0738-x.

[4] A. Reuther, P. Michaleas, M. Jones, V. Gadepally, S. Samsi, and J. Kepner, "Survey and Benchmarking of Machine Learning Accelerators," in *2019 IEEE High Performance Extreme Computing Conference (HPEC)*, Sep. 2019, pp. 1–9. doi: 10.1109/HPEC.2019.8916327.

[5] D. Marković, A. Mizrahi, D. Querlioz, and J. Grollier, "Physics for neuromorphic computing," *Nat Rev Phys*, vol. 2, no. 9, pp. 499–510, Sep. 2020, doi: 10.1038/s42254-020-0208-2.

[6] D. Pierangeli, G. Marcucci, and C. Conti, "Photonic extreme learning machine by free-space optical propagation," *Photon. Res., PRJ*, vol. 9, no. 8, pp. 1446–1454, Aug. 2021, doi: 10.1364/PRJ.423531.

[7] S. Ortín *et al.*, "A Unified Framework for Reservoir Computing and Extreme Learning Machines based on a Single Time-delayed Neuron," *Sci Rep*, vol. 5, no. 1, p. 14945, Oct. 2015, doi: 10.1038/srep14945.

[8] A. Lupo, L. Butschek, and S. Massar, "Photonic extreme learning machine based on frequency multiplexing," *Opt. Express, OE*, vol. 29, no. 18, pp. 28257–28276, Aug. 2021, doi: 10.1364/OE.433535.

[9] S. Sunada, K. Kanno, and A. Uchida, "Using multidimensional speckle dynamics for high-speed, large-scale, parallel photonic computing," *Opt. Express, OE*, vol. 28, no. 21, pp. 30349–30361, Oct. 2020, doi: 10.1364/OE.399495.

[10] D. Silva, T. Ferreira, F. C. Moreira, C. C. Rosa, A. Guerreiro, and N. A. Silva, "Exploring the hidden dimensions of an optical extreme learning machine," *J. Eur. Opt. Society-Rapid Publ.*, vol. 19, no. 1, Art. no. 1, 2023, doi: 10.1051/jeos/2023001.

[11] M. Skontranis, G. Sarantoglou, K. Sozos, T. Kamalakis, C. Mesaritakis, and A. Bogris, "Multimode Fabry-Perot laser as a reservoir computing and extreme learning machine photonic accelerator," *Neuromorph. Comput. Eng.*, vol. 3, no. 4, p. 044003, Oct. 2023, doi: 10.1088/2634-4386/ad025b.

[12] S. Biasi, R. Franchi, L. Cerini, and L. Pavesi, "An array of microresonators as a photonic extreme learning machine," *APL Photonics*, vol. 8, no. 9, p. 096105, Sep. 2023, doi: 10.1063/5.0156189.

[13] G. Marcucci, D. Pierangeli, and C. Conti, "Theory of Neuromorphic Computing by Waves: Machine Learning by Rogue Waves, Dispersive Shocks, and Solitons," *Phys. Rev. Lett.*, vol. 125, no. 9, p. 093901, Aug. 2020, doi: 10.1103/PhysRevLett.125.093901.

[14] M. Zajnulina, A. Lupo, and S. Massar, “Weak Kerr Nonlinearity Boosts the Performance of Frequency-Multiplexed Photonic Extreme Learning Machines: A Multifaceted Approach.” arXiv, Dec. 19, 2023. doi: 10.48550/arXiv.2312.12296.

[15] N. A. Silva, V. Rocha, and T. D. Ferreira, “Optical Extreme Learning Machines with Atomic Vapors,” *Atoms*, vol. 12, no. 2, Art. no. 2, Feb. 2024, doi: 10.3390/atoms12020010.

[16] A. Suprano et al., “Experimental Property Reconstruction in a Photonic Quantum Extreme Learning Machine,” *Phys. Rev. Lett.*, vol. 132, no. 16, p. 160802, Apr. 2024, doi: 10.1103/PhysRevLett.132.160802.

[17] L. Innocenti, S. Lorenzo, I. Palmisano, A. Ferraro, M. Paternostro, and G. M. Palma, “Potential and limitations of quantum extreme learning machines,” *Commun Phys*, vol. 6, no. 1, pp. 1–9, May 2023, doi: 10.1038/s42005-023-01233-w.

[18] S. Biasi, G. Donati, A. Lugnan, M. Mancinelli, E. Staffoli, and L. Pavesi, “Photonic Neural Networks Based on Integrated Silicon Microresonators,” *Intelligent Computing*, vol. 3, p. 0067, Jan. 2024, doi: 10.34133/icomputing.0067.

[19] T. Zhou, F. Scalzo, and B. Jalali, “Nonlinear Schrödinger Kernel for Hardware Acceleration of Machine Learning,” *Journal of Lightwave Technology*, vol. 40, no. 5, pp. 1308–1319, Mar. 2022, doi: 10.1109/JLT.2022.3146131.

[20] S. Biasi, R. Franchi, D. Bazzanella, and L. Pavesi, “On the effect of the thermal cross-talk in a photonic feed-forward neural network based on silicon microresonators,” *Front. Phys.*, vol. 10, Dec. 2022, doi: 10.3389/fphy.2022.1093191.

# Optical to Digital Interfaces in Optical Neural Networks

**Ilker Oguz, Niyazi Ulas Dinc, Mustafa Yildirim, Jih-Liang Hsieh, Christophe Moser, Demetri Psaltis**
Institute of Electrical and Microengineering, School of Engineering, École Polytechnique Fédérale de Lausanne
{ilker.oguz, niyazi.dinc, mustafa.yildirim, jih-liang.hsieh, christophe.moser, demetri.psaltis}@epfl.ch

## Status

The recent resurgence in interest in the optical implementation of neural networks is motivated by the power efficiency and speed with which linear operations can be implemented. The advantage of optics for communicating data compared to an electronic implementation was presented as early as in 1980s[1]. In the optical implementation of neural networks, optics can be used as an interconnect technology, for instance, to transfer data from memory to processors. Alternatively, weighted optical interconnections can realize the linear part of system. A very interesting recent demonstration of the power efficiency of the optical implementation of linear operations showed that one photon of optical energy per weight is sufficient for a single neuron to operate accurately[2].

## Current and Future Challenges

Neural networks are nonlinear systems. The activation functions are generally a form of threshold that decides whether a neuron fires or not. The training of neural networks is also fundamentally a nonlinear process since the presentation of inputs and the evaluation of the corresponding outputs induces changes in the parameters of the system. The nonlinear portion of the computation can in principle be carried out by nonlinear optics. Indeed, early proposals suggested nonlinear optical devices for the implementation of the activation units and the learning function[3]. The nonlinear optics approach is very interesting but difficult to implement for a practical optical system that can compete with the very mature digital technology largely based on GPU's. Therefore, the solutions that are considered presently and for the foreseeable future are hybrid optical-digital systems, whereby the optical system performs linear operation and optical detection performs the non-linear operation. A schematic diagram of the hybrid approach is shown in Figure 1.

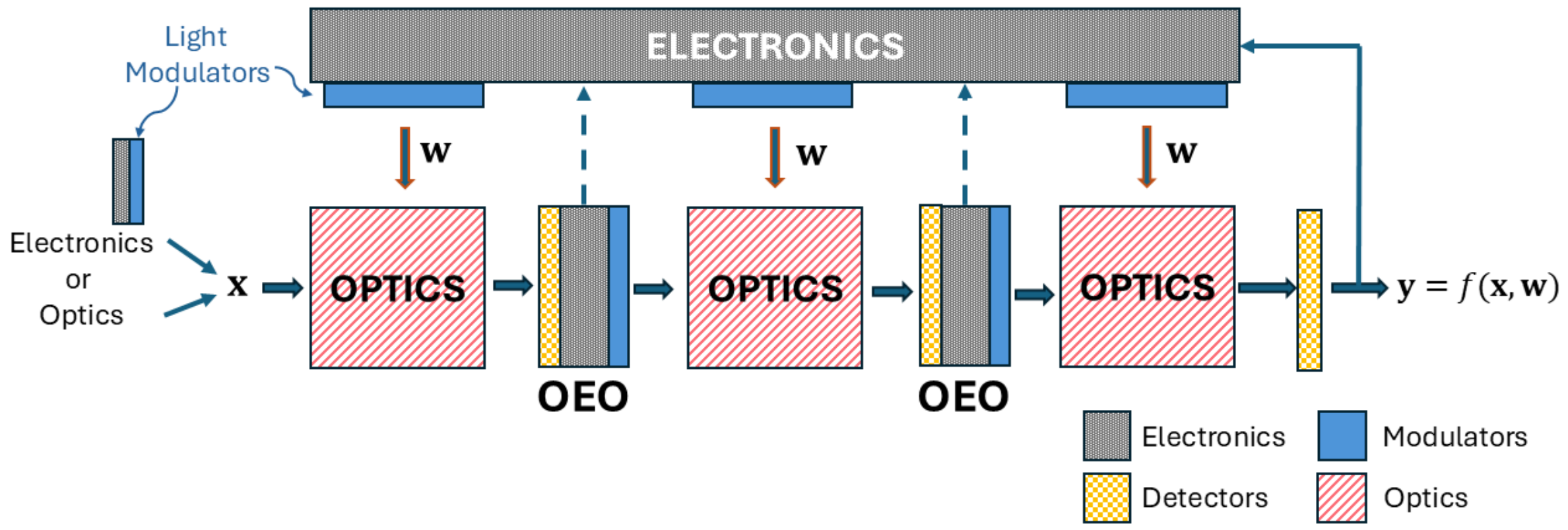

**Figure 1**. Hybrid optical-digital processor combining optical stages for the implementation of linear operations with digital processing, interfaced with light modulators and detectors.

## Advances in Science and Technology to Meet Challenges

The potential advantage of the inclusion of optics depends on the performance of the optical-electronic-optical (OEO) conversion. If the OEO system is not sufficiently power efficient, accurate and fast, then whatever advantage we might have from the inclusion of optics can be washed away. In fact, in order for the optics to become an integral part of GPU and TPU systems it must offer a strong advantage since digital technology is rapidly advancing and the software infrastructure has been developed for digital only systems. We outline in Figure 1 a possible architecture that provides a compelling advantage using optics along with GPUs and TPUs.

The diagram in Figure 1 is applicable to both 3D implementations and also 2D integrated optics approaches. Nevertheless, the devices used in most laboratory demonstrations are quite different. In the 2D integrated optics approach, devices developed for fiber optics telecommunications often include high speed modulators and detectors (> 1 GHz) while 3D free space implementation spatial light modulators (SLMs) and cameras developed for imaging and display applications are used. SLMs and cameras can support several million pixels (neurons) in an area approximately equal to 1 $cm^2$. However, the bandwidth of each channel (pixel) is limited to the kHz range since these devices were developed with optical displays for which 30Hz is often sufficient. Recent developments in integrated light modulators [4] can operate at 100 GHz bandwidth or more at low power consumption giving this approach for implementing OEO's an enormous advantage (6 orders of magnitude) over SLMs. This high speed and low power modulation advantage would be further boosted by accessing the third dimension. The approach of combining the high-speed integrated modulators with the parallelism of 2D SLMs was described in several papers recently [5], [6], [7], [8] and shows promise that optical systems based on the latter can outperform all digital systems in the coming years.

## Concluding Remarks

Competitiveness of optics for implementing NN's relies heavily on the interface devices between digital computer and the optical system. Most importantly, the power consumed for this interface should be minimized. Another approach to reach the same goal is to minimize the role of the digital system and the accompanying OEO's in the hybrid system of Figure 1. For example, random nonlinear transforms in physics can also be utilized for performing machine learning [9], [10], [11], combining rich optical connections with a very simple digital system. In addition, the recent demonstration of nonlinear operations using linear optics[12], [13] is a promising development, which performs some non-linear computations while in the optical domain, thus decreasing the number of OEO conversions. However, the computations required to implement the training of neural networks remain in the digital domain.

## Acknowledgements

I.O. is supported by Google Research, (Grant Number: 901381), M.Y. and N.U.D. are supported by the Swiss National Foundation, Sinergia grant, LION: Large Intelligent Optical Networks (CRSII5_216600).

## References

[1] M. R. Feldman, S. C. Esener, C. C. Guest, and S. H. Lee, 'Comparison between optical and electrical interconnects based on power and speed considerations', *Appl. Opt.*, vol. 27, no. 9, pp. 1742–1751, May 1988, doi: 10.1364/AO.27.001742.

[2] T. Wang, S.-Y. Ma, L. G. Wright, T. Onodera, B. C. Richard, and P. L. McMahon, 'An optical neural network using less than 1 photon per multiplication', *Nat. Commun.*, vol. 13, no. 1, Art. no. 1, Jan. 2022, doi: 10.1038/s41467-021-27774-8.

[3] K. Wagner and D. Psaltis, ‘Multilayer optical learning networks’, *Appl. Opt.*, vol. 26, no. 23, pp. 5061–5076, Dec. 1987, doi: 10.1364/AO.26.005061.

[4] C. Wang *et al.*, ‘Integrated lithium niobate electro-optic modulators operating at CMOS-compatible voltages’, *Nature*, vol. 562, no. 7725, pp. 101–104, Oct. 2018, doi: 10.1038/s41586-018-0551-y.

[5] C. L. Panuski *et al.*, ‘A full degree-of-freedom spatiotemporal light modulator’, *Nat. Photonics*, vol. 16, no. 12, pp. 834–842, Dec. 2022, doi: 10.1038/s41566-022-01086-9.

[6] F. Ashtiani, A. J. Geers, and F. Aflatouni, ‘An on-chip photonic deep neural network for image classification’, *Nature*, vol. 606, no. 7914, Art. no. 7914, Jun. 2022, doi: 10.1038/s41586-022-04714-0.

[7] Z. Chen *et al.*, ‘Deep learning with coherent VCSEL neural networks’, *Nat. Photonics*, vol. 17, no. 8, pp. 723–730, Aug. 2023, doi: 10.1038/s41566-023-01233-w.

[8] Z. Xu, T. Zhou, M. Ma, C. Deng, Q. Dai, and L. Fang, ‘Large-scale photonic chiplet Taichi empowers 160-TOPS/W artificial general intelligence’, *Science*, vol. 384, no. 6692, pp. 202–209, Apr. 2024, doi: 10.1126/science.adl1203.

[9] I. Oguz *et al.*, ‘Programming nonlinear propagation for efficient optical learning machines’, *Adv. Photonics*, vol. 6, no. 1, p. 016002, Jan. 2024, doi: 10.1117/1.AP.6.1.016002.

[10] A. Skalli *et al.*, ‘Photonic neuromorphic computing using vertical cavity semiconductor lasers’, *Opt. Mater. Express*, vol. 12, no. 6, pp. 2395–2414, Jun. 2022, doi: 10.1364/OME.450926.

[11] M. Rafayelyan, J. Dong, Y. Tan, F. Krzakala, and S. Gigan, ‘Large-Scale Optical Reservoir Computing for Spatiotemporal Chaotic Systems Prediction’, *Phys. Rev. X*, vol. 10, no. 4, p. 041037, Nov. 2020, doi: 10.1103/PhysRevX.10.041037.

[12] M. Yildirim, N. U. Dinc, I. Oguz, D. Psaltis, and C. Moser, ‘Nonlinear Processing with Linear Optics’, *ArXiv Prepr. ArXiv230708533*, 2023.

[13] F. Xia, K. Kim, Y. Eliezer, L. Shaughnessy, S. Gigan, and H. Cao, ‘Deep Learning with Passive Optical Nonlinear Mapping’. arXiv, Jul. 18, 2023. doi: 10.48550/arXiv.2307.08558.

# Towards scalable integration and learning for unconventional computing: the example of model free laser neural networks

**A. Grabulosa, A. Skalli, M. Goldmann, D. Brunner**
Institut FEMTO-ST, Université Franche Comte CNRS, UMR 6174, 25000 Besançon, France.
adria.grabulosa@femto-st.fr, anas.skalli@femto-st.fr, mirko@goldmann@gmail.com, daniel.brunner@femt-st.fr

## Status

With their joined memory topology and distributed large-scale transformations, Artificial intelligence (AI) and neural network (NN) concepts fundamentally differ from the Turing machine and the von Neumann architecture. Neural networks leverage, extensively, parallel and weighted communication between nonlinear elements, also called neurons. The major challenge for NN and NN-inspired computing hardware lies in achieving efficient parallel connectivity on a large scale, and to *program* this connectivity such that the system solves a particular task. As explained below, this is currently, and potentially fundamentally, out of reach considering classical 2D electronic integration approaches as well as current training concepts [1].

Implementing ANNs using photonic technology shows great promise in terms of scalability, speed, energy efficiency and parallel information processing [2]. This is true for, both, classical methods based on photonic integration via discrete components, as well as concepts that leverage more unconventional mappings between high-dimensional nonlinear hardware substrates and NN concepts. The underlying reason is mostly related to that, unlike electrons, photons are not electrically charged and hence do not suffer from inductive or capacitive energy dissipation. Additionally, the inherent parallelism of photonics allows modulating information (encoded in amplitude, phase, spectra and others) in both, the temporal and spatial domains without crosstalk in the linear parts of a photonic architecture, opening new avenues towards enhanced connectivity.

By now, unconventional implementations of photonic NNs [3] have reached sizes that are mostly out of reach for electronic and photonic non-digital hardware that follow a classical NN architecture. However, experimentally these are usually realized in free-space [3,4] or optical fibre [5], which is bulky in both cases. So far, such systems have been trained either by model-based approaches leveraging a digital-twin approximation of the unconventional hardware [6], or using model-free learning [4].

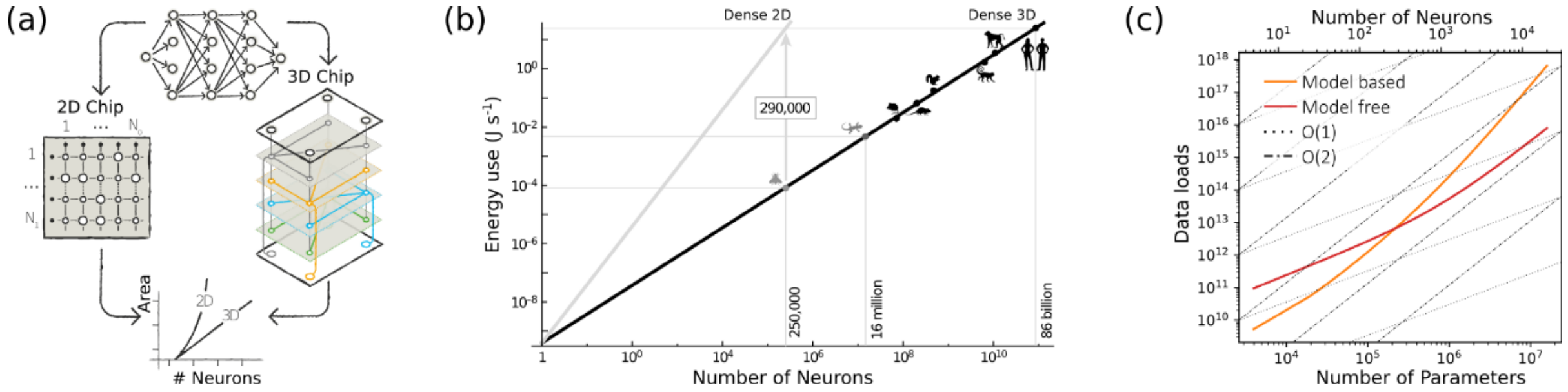

**Figure 1.** Scaling challenges for neural network computing. (a) The footprint of in-memory hardware for fully connected neural networ scales quadratic versus linear with the number of neurons in 2D and 3D integration, respectively. (b) The energy use in biological neura networks scales linear and in 2D electronics quadratic with the number of neurons, respectively. Adapted from [6]. (c) Training neura networks using model based or model free approaches has substantial impact on the scaling of von Neumann-like data loads.

## Current and Future Challenges

The scaling of NN hardware and computing concepts in terms of hardware footprint, energy consumption as well as training time is an essential aspect that determines the field's future. Continued miniaturization in electronic integrated circuits (ICs) appears to have reached its fundamental limit, and for various micro-processor performance metrics, exponential performance scaling has slowed down or come to a halt since around one decade [7]. As schematically illustrated in Fig. 1(a), a key challenge for NN hardware integration is achieving parallel interconnections scalably, which currently is limited to networks comprising around 1000 neurons. At the core of this challenge lies a fundamental quarrel between 2D geometry and dense network topologies. Schematically speaking, for routing networks in 2D the input and output neurons are arranged in columns and rows, and are connected by wires (electronics) or waveguides (photonics) within the space allocated in-between, forming so called cross-bar arrays. The area for such circuitry scales as the product between the input and output neuron numbers, hence footprint scaling is quadratic with network size. Yet, expanding the implementation into a third dimension fundamentally eases this scalability conflict; for example, input and output neurons then respectively occupy the 3D circuit's top and bottom 2D plane, while the circuit's volume can be utilized for out-of-plane interconnections [8]. Such or similar 3D routing may indeed be a fundamental prerequisite for achieving scalability and parallelism in photonic as well as electronic ICs. Yet, these circuits are hardly compatible with the 2D lithography concept developed for today's CMOS substrates.

Comparable scaling problems intricately linked to identical arguments of geometry are encountered with regards to the energy consumption of NN computing. Implementing dense network connections in the 2D in-memory computing setting of an electronic cross-bar array leads to a quadratic relationship between energy consumption and the number of neurons [9]. Importantly, this is in stark contrast to biological brains, where linear energy scaling has been proven for different species with brains varying by five orders of magnitude in size, see Fig. 2(b). Would one project the current quadratic scaling of current hardware onto biology, then the power budget of a human brain would only allow for operating the biological brain of a common housefly.

Finally, the story repeats for training costs when optimizing analogue in-memory computing systems, see Fig. 1(c). Currently, optimization using error back-propagation relies on a digital-twin approach, which uses the efficient unconventional computing substrate only in the forward direction for inference. The problem with error back-propagation is that in the backward direction one requires the derivatives of neuron activation functions to propagate, and such a symmetry breaking of $f(x)$ and $\partial f(x)/\partial x$ nonlinearity when going forward or respectively backwards through the substrates is physically forbidden in most settings. For the backward pass one therefore employs a differentiable digital model approximating the physical system, which can either be a model based on the set of corresponding physical equations or using a data-driven approach during which a multi-layer perceptron approximates the physical system [6]. Unfortunately, this potentially results in a vast computing overhead expended upon the auxiliary digital hardware used for approximating the unconventional system's gradients. These overheads can be of such scale that they challenge the very motivation of leveraging unconventional computing substrates.

## Advances in Science and Technology to Meet Challenges

While not yet at the same level of complexity and functionality, photonic ICs have significantly advanced and increasingly become a viable enabler to extend performance scalability of ICs in general. These ICs predominantly rely on complementary metal-oxide-semiconductor (CMOS) compatible technology.

Additive manufacturing via 3D printing stands out as an innovative tool for creating intricate 3D photonic components. Direct-laser writing (DLW) combined with two-photon polymerization (TPP), in particular polymer-based 3D printing due to inherently good compatibility with CMOS, enables fabricating 3D optical structures down to sub-µm scales. The importance of CMOS compatible processes and materials for photonic NN is fundamental for the maturation of next generation photonic NNs ICs. Figure 2(a) shows an electron beam microscopy micrograph of a 3D integrated circuit interfaced with semiconductor (GaAs) micro-lasers [10], demonstrating the compatibility of these processes to create a single, monolithic hybrid platform. Conceptually, the 3D waveguides interfaced on top of the micropillar arrays result in the optical coupling of the laser emission of the micro-cavities into the 3D waveguides.

Using spatially multiplexed modes of an injection locked large area vertical cavity surface emitting laser (LA-VCSEL), and by leveraging the inherent physics of the device as well as free-space propagation, we experimentally built a photonic neural network (PNN) where all components are realized in hardware using off-the-shelf, commercially available, low energy consumption components [3,11, 12]. The input and output weight matrices have been trained via iterative optimization based on evolutionary search algorithms or gradient descent using gradient estimation methods from reinforcement learning [13]. The system reached >98% accuracy in 6-bit header recognition tasks and promising initial results for the MNIST hand written digit recognition dataset. Crucially, our system performs classification at a high inference bandwidth of 15 kHz, which is not limited by the LA-VCSEL (GHz bandwidth) and could potentially increase towards the GHz range. Importantly, using this single-device approach 1000+ nodes are implemented fully in parallel. Since weights and connections are potentially passive in nature, the concept has the potential of linear scaling of its power consumption with the number of neurons in the network.

Finally, the concept also demonstrated the model free optimization of photonic NNs, with the potential benefits in terms of energy efficiency and reduced scaling of the optimization energy overhead. However, it has to be said that at the current stage the detailed analysis of scaling effects as well as the different contributions to the system's power consumption have not yet been evaluated in sufficient detail.

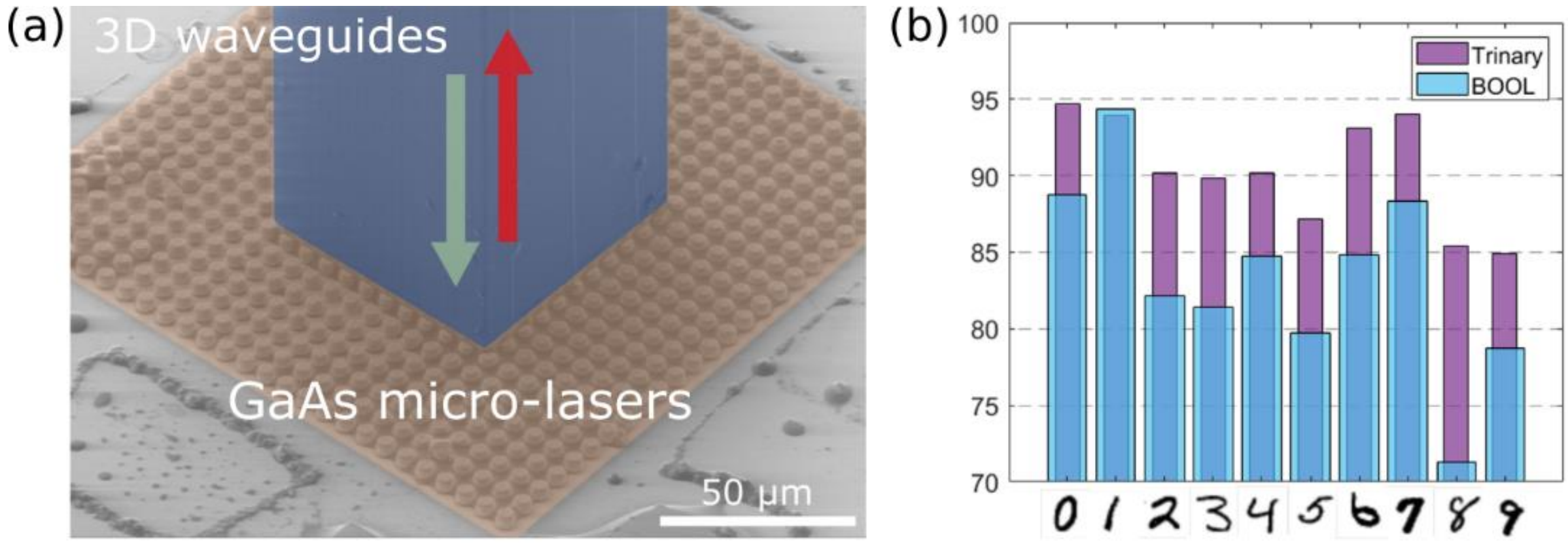

**Figure 2.** (a) CMOS compatible 3D photonic integration via additive two photon polymerization techniques. Reproduced with permission from [10]. (b) Unconventional and fully autonomous photonic computing substrates, here a LA-VCSEL, now approach the point where they provide relevant computational performance enhancement and can surpass the linear limit in standard benchmark tests.

**Concluding Remarks**

Combining high-dimensional unconventional hardware that leverages the inherent strengths of photonics with 3D interconnects will pave the way for future realizations of scalable, in-memory computing with NNs. Furthermore, to truly benefit from the advantages offered by these future systems in terms of energy efficiency, hardware-compatible and efficient training algorithms need to be developed. During this algorithmic development, a significant emphasis should be put on minimizing reliance on the traditional digital von Neumann architectures that currently constitute the primary bottleneck our innovations are trying to overcome.

An important realization is that today's optimization schemes leverage mathematical models for high performance on specific architectures and place a focus on compatibility with any conventional computer. Due to inherent fabrication imperfections, unconventional and analog physical neural networks will most likely require some kind of individual training for each new device, at least to a certain level. This increases training costs and power consumption, making energy cost scaling of the training much more crucial for such approaches. It is a potentially equally fundamental and philosophically interesting question how much time such unconventional hardware needs to sit inside a hardware school class will offset the ultimate energy balance between unconventional and conventional NN-inspired computing.

**Acknowledgements**
*ERC INSPIRE, Volkswagen Foundation NeuroQNet I&II.*

**References**
[1] W. M. M., "The chips are down for Moore's law," Nature **530**, 144–147 (2016).
[2] B. Shastri, et al., "Photonics for artificial intelligence and neuromorphic computing, Nature Photonics **15**, 102-114 (2021).
[3] A. Skalli, et al., to be submitted.
[4] J. Bueno, et al., "Reinforcement learning in a large-scale photonic recurrent neural network," Optica **5**, 756-760 (2018).
[5] L. Larger, et al., "Photonic information processing beyond Turing: an optoelectronic implementation of reservoir computing," Optics Express **20**, 3241-3249 (2013).
[6] L. G. Wright, et al., "Deep physical neural networks trained with backpropagation," Nature **601**, 549-555 (2022).
[7] https://www.karlrupp.net/2018/02/42-years-of-microprocessor-trend-data/
[8] S. Borkar, "Design challenges of technology scaling," IEEE Micro, vol. 19, no. 4, pp. 23–29, 1999.
[9] Boahen
[10] A. Grabulosa, J. Moughames, X. Porte, and D. Brunner, "Additive 3D photonic integration that is CMOS compatible", *Nanotechnology* **34**, 322002 (2023).
[11] X. Porte, A. Skalli, N. Haghighi, S. Reitzenstein, J. Lott, D. Brunner, "A complete, parallel, and autonomous photonic neural network in a semiconductor multimode laser," J. Phys. Photonics **3**, 024017 (2021).
[12] A. Skalli, X. Porte, N. Haghighi, S. Reitzenstein, J. Lott, D. Brunner, "Computational metrics and parameters of an injection-locked large area semiconductor laser for neural network computing." Optical Materials Express **12**, 2793-2804 (2022).
[13] Salimans, T., Ho, J., Chen, X., Sidor, S., & Sutskever, I. (2017). Evolution strategies as a scalable alternative to reinforcement learning. arXiv preprint arXiv:1703.03864.

# Hyperdimensional Photonic Neural Networks and Optics-informed Deep Learning

**Apostolos Tsakyridis, Miltiadis Moralis-Pegios, Anastasios Tefas and Nikos Pleros**
Department of Informatics, Center for Interdisciplinary Research and Innovation, Aristotle University of Thessaloniki, Thessaloniki, Greece
[atsakyrid@csd.auth.gr ; mmoralis@csd.auth.gr ; tefas@csd.auth.gr ; npleros@csd.auth.gr ]

**Status**

Following the pioneering attempts back in the '80s [1] and the deafening silence that followed for the next 25 years, the research efforts for transferring neural network (NN) concepts in the optical domain came to the fore stark again in mid-2010's [2]. This resurgence triggered a whole new scientific and technological area that is now typically referred to as neuromorphic photonics or photonic neural networks (PNNs). The main motivation for this impressive comeback is tightly bound to the booming of the Artificial Intelligence (AI) and Deep Learning (DL) compute models that designated the end of Moore's law: within the last 10 years, AI led computational power requirements to double every 3.4 months and computational energy consumption to 100's of MWh, declaring a clear need for a paradigm shift in computing hardware in order to avoid an energy collapse. The transition from digital to analog processing has been shown to comprise a highly promising pathway for significant energy savings, opening a new pathway for optics and setting the scene for analog photonic computational substrates. This requires, however, the convergence between DL architectural concepts and photonic circuitry, so that all necessary computational functions can be implemented via light. This perspective was strengthened by the growing maturity of integrated photonics, bearing bold promises for orders of magnitude improvement across almost all significant AI chipset performance metrics [2], as illustrated also in the spider diagram of Fig. 1(a).

It is true, however, that both digital and analog electronic AI chipsets made also significant progress in the meantime, while neuromorphic photonics faced important challenges that often turned their case into question [3]. After the bold headlines made by several start-up companies in the field a few years ago, PNNs got surrounded by a growing scepticism around i) the scalability perspectives of photonic Matrix-Vector-Multiply (MVM) circuitry towards supporting the massive amount of trainable parameters required by AI models, ii) their credentials to fulfil their promise for low-energy computing due to the need for high-speed Digital-to-Analog and Analog-to-Digital Conversion (DAC and ADC) stages at the interfaces of every photonic neural layer, and finally iii) the performance limitations stemming from the analog noise and the lower bit-resolution capabilities that can hardly exceed 4-5 bits for operational rates higher than 10 Gbaud [4].

**Current and Future Challenges**

Inspecting, however, the inner anatomy of the main PNN architectural and technological research efforts reveals a number of shortcomings that have ultimately led to the above criticism, with the main factors being the use of sub-optimal MVM layouts and the uncoupled trajectories of DL training models and analog photonic hardware [4]. Non-coherent perceptrons and MVM circuits were among the first that successfully demonstrated light-enabled AI tasks [2],[5],[6] exploiting a different wavelength per axon together with power addition between the different wavelengths at each neuron output, as shown in Fig. 1(b). The use of wavelength as function-enabling dimension negates, however, the traditional advantage of optics to utilize Wavelength Division Multiplexing (WDM) for parallelization purposes. On the other hand, coherent interferometric architectures require just a single wavelength and exploit the spatial dimension for performing multiple operations in parallel, as shown in Fig. 1(c), implying that they could in principle extend to WDM settings. These have, however, mainly focused on elementary matrix factorization theorems that lead to complex linear circuits of cascaded stages of 2x2 Mach-Zehnder interferometers (MZIs) [7],[8]. Despite the remarkable PNN deployments and applications presented with these configurations, there are certain inherent drawbacks associated with this scheme [4], including i) limited scalability, ii) incompatibility with WDM, iii) slow weight update rates, iv) non-restorable circuit fidelity that degrades with circuit dimensions, and v) a complex matrix programming process. Alternative schemes aiming at supporting a large amount of trainable parameters and increased NN dimensions invested in Time Domain Multiplexing (TDM); these layouts unfold each matrix into a vector that gets modulated into a respective optical time series via a fast modulator stage, producing at the output a time series of the weighted inputs that are then summed via integrator circuitry to yield the weighted input sum, as shown in Fig.1(d) [9]-[11]. On top of its scalability credentials, this approach avoids the use of high-speed and power-hungry ADCs but certainly yields increased latency values compared to WDM and coherent NN circuits.

In addition to the architectural challenges, PNNs find it rather hard to compete both with the versatile range of applications and the accuracy standards offered by digital AI chipsets. DL training models have obviously progressed so far within the realm of digital AI chipsets and are completely decoupled from the idiosyncrasy and limitations of analog photonic hardware [4], which brings the DL algorithmic framework confronted with a number of new unknown factors when migrating to photonic AI substrates, like noise, bandwidth, limited extinction ratios and photonic activations that do not always match typical NN activations [4].

**Advances in Science and Technology to Meet Challenges**

Overcoming these challenges has probably to proceed along a roadmap that leverages the classical multi-dimensional multiplexing advantages of optics into the PNN architectural framework [12], together with a hardware/software co-design approach for bridging training models with photons. The synergy of multiple dimensions can boost the scalability perspectives of photonic MVMs, as has been already indicated by initial efforts on space-wavelength PNNs [2] and time-wavelength PNNs [13], which allowed for the first time to break the 10 TOPs compute power limit. Time-space NNs were first demonstrated via coherent layouts that departed from typical MZI-based meshes [9],[11],[14],[15]; within this frame, the photonic Crossbar architecture demonstrated record-high 50Gbaud compute rates per axon and tiled matrix multiplication capabilities on a silicon chip [16] with and record-high fidelity close to 99.997% (Fig. 2(a)) [15], taking advantage of its profound loss, fidelity and programming benefits against MZI mesh architectures. The Crossbar architecture has been also shown theoretically to hold the credentials for encompassing the wavelength dimension in addition to its time-space multiplexing portfolio (Fig. 2(b)), transforming in this way photonic MVM circuitry into high-speed and scalable tensor multiplication modules [17], as shown in Fig. 2(c). The promise of time-space-wavelength division multiplexed (TSWDM) approaches for scalable PNNs has been pronounced more recently by a novel Arrayed Waveguide Grating Router (AWGR)-based PNN architecture [18] that showed record-high compute powers of 160TOPS, supporting $O(N^3)$ computations with $O(N^2)$ computational nodes.

This architectural line comes together with significant energy efficiency gains, since higher scalability implies that the signal remains longer in the optical domain and the power consumption of the electrical interfaces gets then shared among a higher number of optical Multiply-Accumulate (MAC) operations. TSWDM layouts negate also the need for high-speed ADCs and require just slow integrator circuitry with low power consumption for the accumulation function. An alternative energy saving perspective can be also shaped by using analog opto-electronic or all-optical activation functions [4] to avoid the need for high-speed ADCs at every single neural layer output. This approach is certainly associated with increased noise levels since the signal remains in the analog domain for more than one NN layer, implying that the accuracy performance of the complete NN might degrade significantly when following typical DL training models.

Performance degradation comes mainly from the missing association between typical DL training models and the analog photonic computational hardware, which can be only overcome through a systematic hardware-training model co-design approach. This has been clearly highlighted via innovative hardware-aware DL training algorithms that incorporate noise, bandwidth, and quantization limitations of the photonic hardware in the training process [19],[20]. These efforts led finally to the introduction of a new DL training framework termed as Optics-informed DL [4], which can also successfully adapt to non-typical activation functions that are realized via analog photonic elements. Combined with initial observations that only a small fraction of the NN layers within well-known benchmark models needs high bit-resolution capabilities [16], the introduction of Optics-informed DL models can allow for hybrid digital-analog photonic implementations that can support acceleration together with high-accuracy performance. This scheme employs a limited amount of NN layers realized via digital electronic hardware, with the remaining NN layers being implemented via photonics and adapting their computational line-rate to ensure the required bit-resolution on a per layer basis [16].

**Concluding Remarks**

Neuromorphic photonics emerged a few years ago with a great promise for changing the AI hardware landscape, but the expectations started quickly to blur when scalability, energy efficiency and performance of PNNs started to face significant hurdles. This stimulated, however, new research lines along TSWDM PNN architectures and Optics-informed DL training models, with their initial achievements providing solid indications that the field is entering the slope of enlightenment and holds significant credentials to finally shed light in the future of AI.

**Acknowledgements**
The work was in part funded by the EU Horizon projects PlasmoniAC (871391), SIPHO-G (101017194) and GATEPOST (101120938).

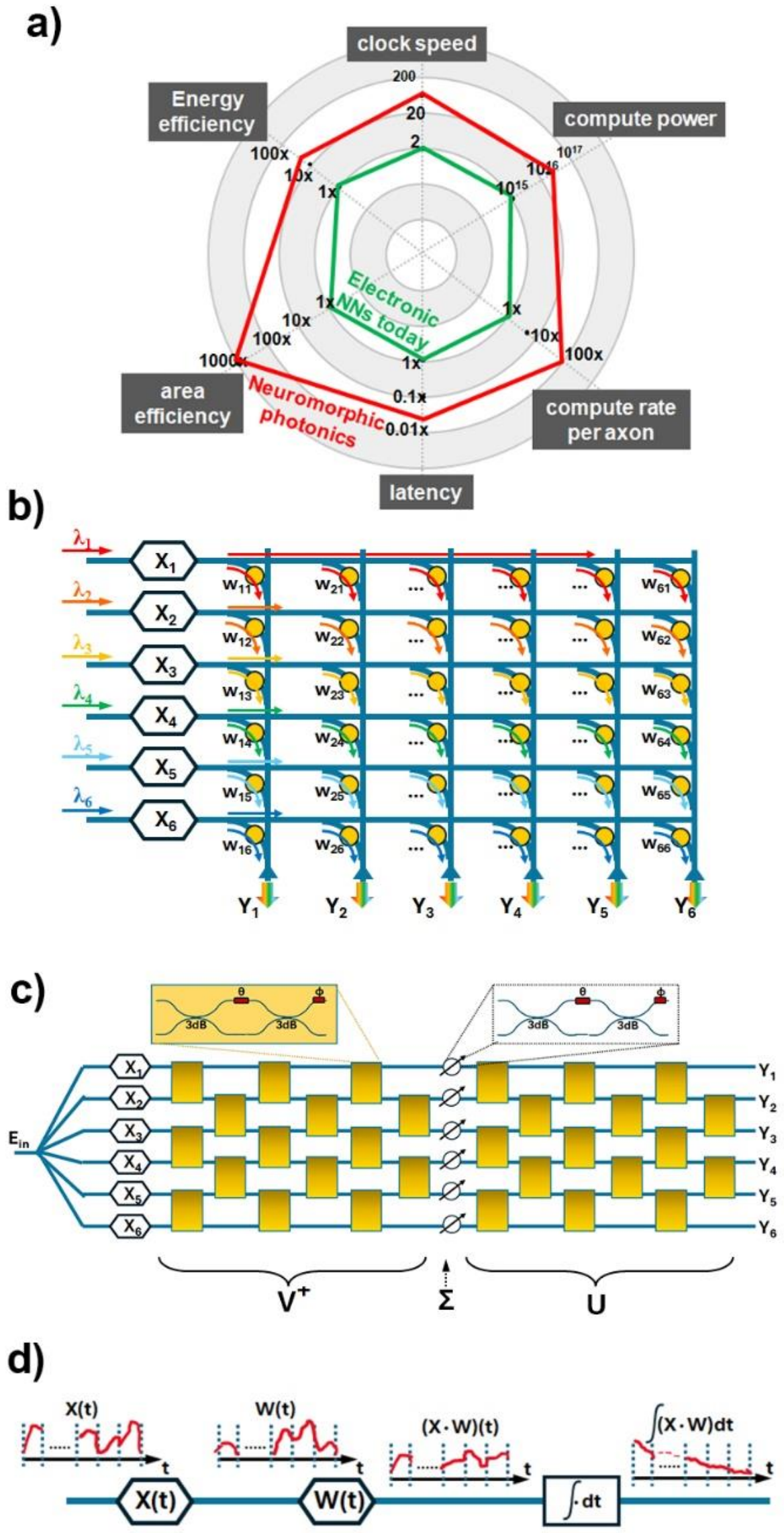

**Figure 1.** Optical MVM architectures a) non-coherent 6x6 MVM requiring 6 different wavelengths. Weighting nodes can be any type of components providing variable attenuation, like microring resonators or Phase Change Material (PCM) waveguides, b) coherent 6x6 MVM relying on the Singular-Value-Decomposition scheme and the Clements unitary MZI-based layout, c) TDM MVM architecture where X(t) is the input signal time vector synchronized with the respective flattened weight matrix that is encoded as a W(t) time vector.

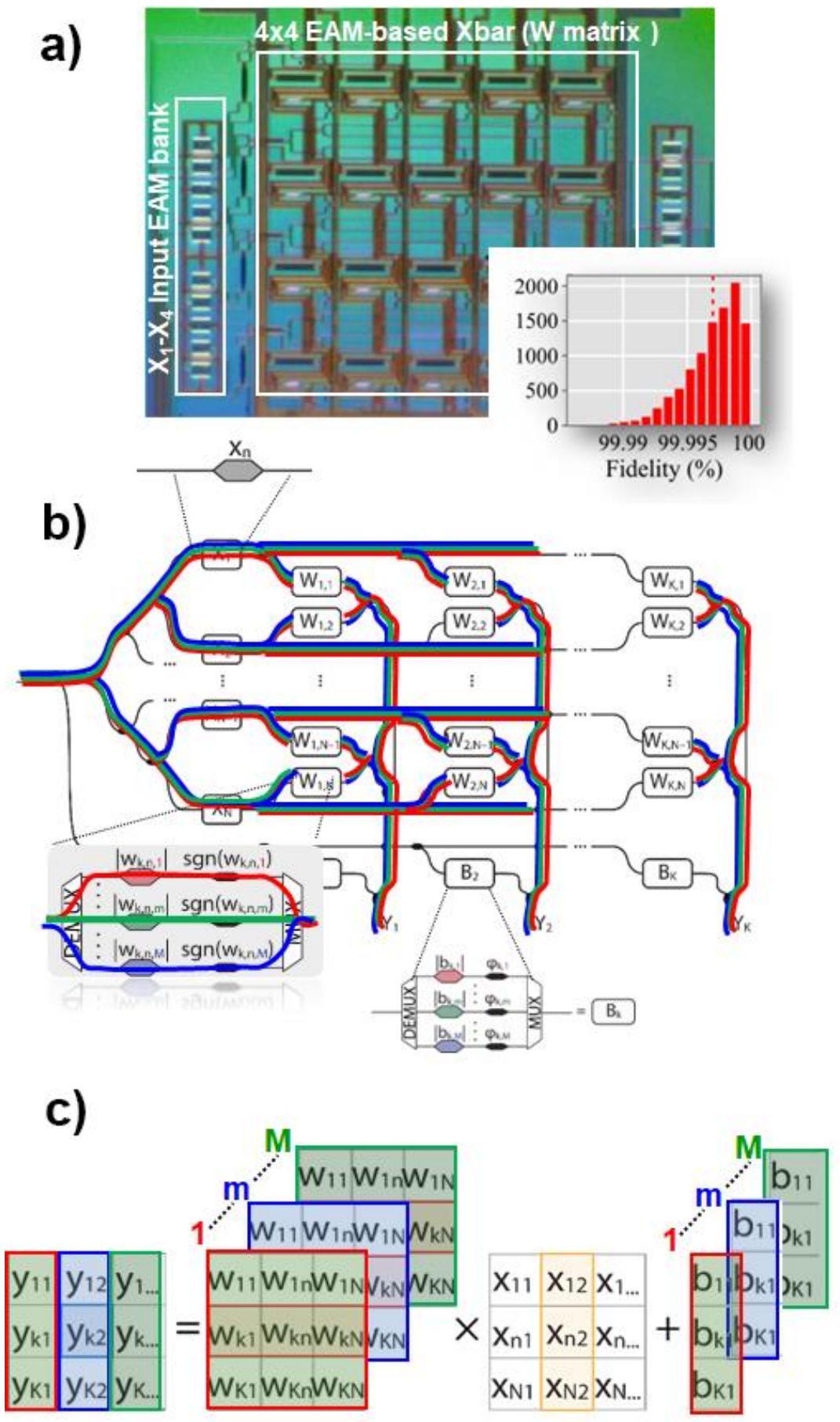

**Figure 2**. a) The first experimental prototype of a 4x4 silicon photonic Crossbar architecture exploiting SiGe Electro-Absorption Modulators (EAMs) for input and weight signal encoding, with the inset revealing the record-high fidelity of 99.997% obtained when fidelity restoration is applied **Erreur ! Source du renvoi introuvable.**, b) The modified photonic Crossbar architecture that can support TSWDM operation **Erreur ! Source du renvoi introuvable.**. It exploits multiple wavelengths by employing wavelength-specific EAM-based weighting nodes between a demux/mux silicon stage and provides fast weight update through the high-bandwidth EAM weights, c) The matrix-tensor-multiply operation enabled through TSWDM architectures, where every wavelength carries a different weight matrix.

## References

[1] Abu-Mostafa, Yaser S., and Demetri Psaltis. “Optical Neural Computers.” Scientific American 256, no. 3, pp. 88–95, 1987

[2] Thomas Ferreira de Lima, Bhavin J. Shastri, Alexander N. Tait, Mitchell A. Nahmias, and Paul R. Prucnal, "Progress in neuromorphic photonics", Nanophotonics, vol. 6, no. 3, pp. 577-599, 2017

[3] Kun Liao, Tianxiang Dai, Qiuchen Yan, Xiaoyong Hu, and Qihuang Gong, “Integrated Photonic Neural Networks: Opportunities and Challenges“, ACS Photonics, vol. 10, no. 7, pp. 2001–2010, 2023

[4] A. Tsakyridis, M. Moralis-Pegios, G. Giamougiannis, M. Kirtas, N. Passalis, A. Tefas, and N. Pleros, “Photonic neural networks and optics-informed deep learning fundamentals”, APL Photonics, vol. 9 no. 1, 011102, 2024

[5] Mario Miscuglio, Volker J. Sorger, “Photonic tensor cores for machine learning”, Appl. Phys. Rev. 1 7 (3): 031404, Sept. 2020

[6] Bin Shi, Nicola Calabretta, Ripalta Stabile, “InP photonic integrated multi-layer neural networks: Architecture and performance analysis”, APL Photonics 1 January 2022; 7 (1): 010801

[7] Y. Shen et al, “Deep learning with coherent nanophotonic circuits,” Nat. Photonics 11, 441–446 (2017)

[8] F. Shokraneh, S. Geoffroy-Gagnon, and O. Liboiron-Ladouceur, "The diamond mesh, a phase-error- and loss-tolerant field-programmable MZI-based optical processor for optical neural networks," Opt. Express 28, 23495-23508 (2020)

[9] R. Hamerly, L. Bernstein, A. Sludds, M. Soljačić, and D. Englund, “Large-Scale Optical Neural Networks Based on Photoelectric Multiplication”, Phys. Rev. X, vol. 9, no.2, 021032, 2019

[10] L. De Marinis et al., "A Codesigned Integrated Photonic Electronic Neuron," in IEEE J. of Quantum Electron., vol. 58, no. 5, pp. 1-10, Oct. 2022

[11] N. Youngblood, "Coherent Photonic Crossbar Arrays for Large-Scale Matrix-Matrix Multiplication," in IEEE J. of Sel. Topics in Quantum Electron., vol. 29, no. 2, pp. 1-11, March-April 2023

[12] B. Dong et al, “Higher-dimensional processing using a photonic tensor core with continuous-time data”, Nat. Photon. 17, 1080–1088 (2023)

[13] X. Xu et al., “11 TOPS photonic convolutional accelerator for optical neural networks”, Nature 589, 44–51 (2021)

[14] S. Xu, et al, “Optical coherent dot-product chip for sophisticated deep learning regression”, Light Sci Appl 10, 221 (2021)

[15] M Moralis-Pegios, G Giamougiannis, A Tsakyridis, D Lazovsky and N Pleros, “Perfect linear optics using silicon photonics”, arXiv preprint arXiv:2306.17728, 2023

[16] G. Giamougiannis, et. al., "Analog nanophotonic computing going practical: silicon photonic deep learning engines for tiled optical matrix multiplication with dynamic precision" Nanophotonics, vol. 12, no. 5, 2023

[17] A. Totovic et. al., “Programmable photonic neural networks combining WDM with coherent linear optics”, Sci Rep 12, 5605 (2022)

[18] C. Pappas, T. Moschos, A. Prapas, A. Tsakyridis, M. Moralis-Pegios, K. Vyrsokinos, and N. Pleros, "A 160 TOPS Multi-dimensional AWGR-based accelerator for Deep Learning," in Optical Fiber Communication Conference (OFC), San Diego, CA, USA, Mar. 2024

[19] M. Moralis-Pegios et. al., "Neuromorphic Silicon Photonics and Hardware-Aware Deep Learning for High-Speed Inference," in J. of Lightwave Technology, vol. 40, no. 10, pp. 3243-3254, 15 May15, 2022

[20] G. Mourgias-Alexandris, G. et. al., “Noise resilient and high-speed deep learning with coherent silicon photonics”, Nat Commun 13, 5572 (2022)

# Photonic Tensor Cores: Integration, Packaging, and Challenges

**Nicola Peserico[1,2], Russell L. T. Schwartz[1,2], Hangbo Yang[1,2], Volker J. Sorger[1,2]**
[1]University of Florida (Florida Semiconductor Institute, 1889 Museum Rd., Gainesville, FL, US)
[2]Electrical & Computer Engineering Department, University of Florida, FL, USA
nicolapeserico@ufl.edu, rschwartz2@ufl.edu, hangbo.yang@ufl.edu, volker.sorger@ufl.edu

## Status

Since the early works on optical neural processors, it has been clear to the scientific community that, scaling and integration would play a key role in the development of a Photonic Tensor Core (PTC) as hardware accelerators for Machine Learning [1]. Both digital and analog electronic circuits for Neural Networks (NN) benefit from the decades of integration and packaging solutions available for CMOS, including I/O interfaces, high-bandwidth memories (HBM), advanced packaging, etc. [2]. On the other hand, PTC processors, while promising high TOPS/W metrics, face module challenges; for instance to realize stand-alone PTC chiplets, beyond co-packaged optics (CPO) solutions, to laser-to-die integration such as enabled by photonic wire bonding (PWB) needs more yield test. While initial electronic-photonic heterogeneous die integration offers Tbps throughputs, yet current PIC integration concepts are mostly centered on relying on optical transceivers technology that was developed for datacenters hence is insufficiently covers AI architectures such as for large-scale language model ML accelerators [3]. This challenging path has brought about some highlighted successes (**Fig. 1**): from off-the-shelf experiments [4], recent works have shown the potential of integration and packaging. Different works presented integration of silicon photonic chip into PCB boards [5,6], capable of controlling, calibrating, and interact with the external electronic world. Furthermore, different approaches showed a path for the integration of the laser source such as heterogeneous laser integration to PWB [7,8]. Innovation in fabrication automation, advanced by the request for exponentially faster transceivers, made monolithic integration (electronics + photonics) a reality [9]. This unlocked emerging circuits design options and scaling mechanism for spiking neural networks [10], to the idea of a full 'black-box' PTC with just digital electronics I/O interfaces [11].

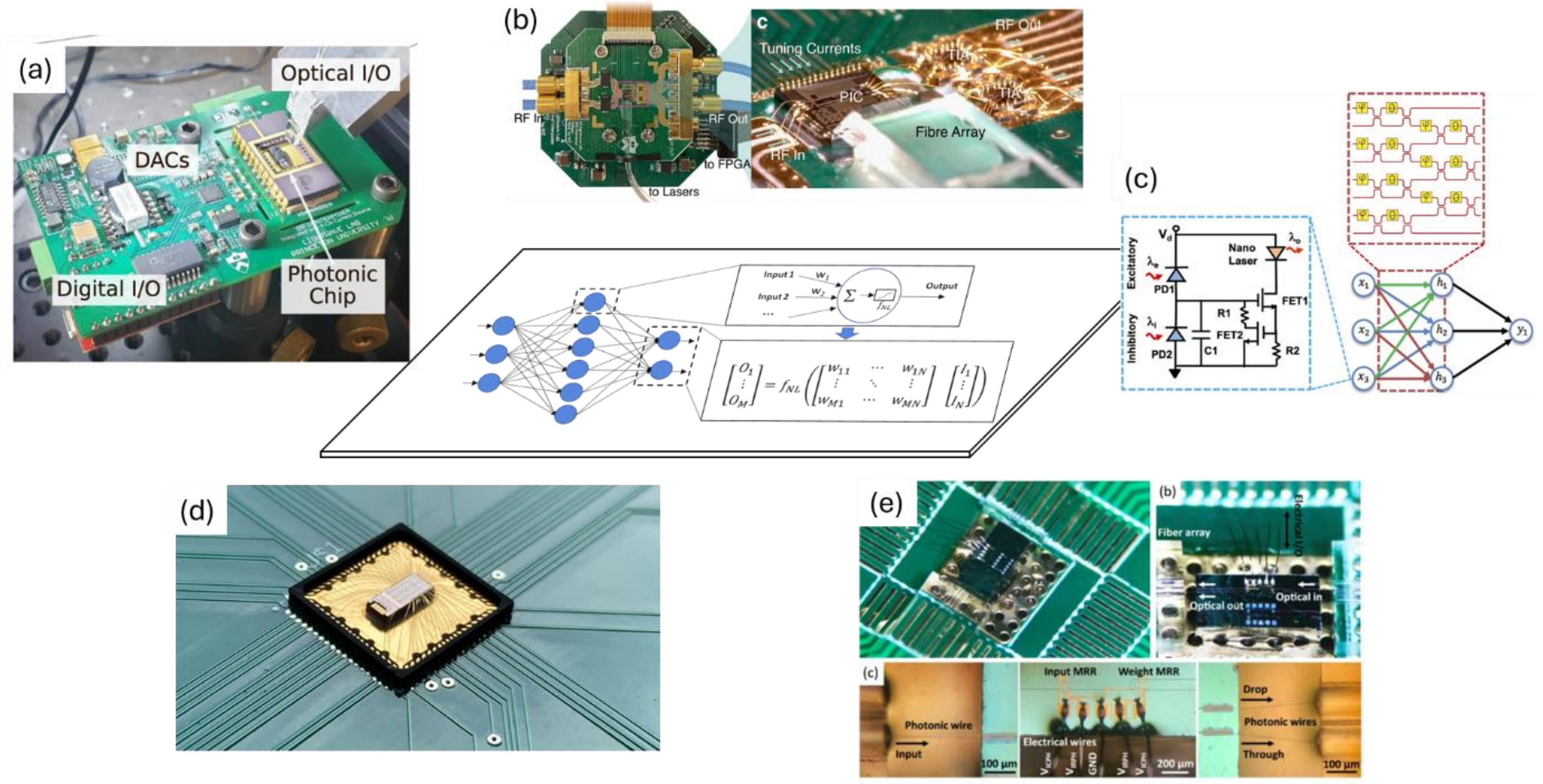

**Figure 1** (a) Photonic chip package and mounted on a controlling PCB [5]. (b) PIC mounted directly on a high-speed PCB board [6]. (c) Spiking neural network made possible by monolithic integration [10]. (d) Full optical chip, from laser to detector, in a single QFN carrier[7]. (e) PIC on PCB with Photonic Wire Bonding to optical fibers [8]. All rights belong to original publisher.

## Current and Future Challenges

| Maximized Inference Data Rates | | | | |
|---|---|---|---|---|
| **Component** | **Number** | **Speed** | **Bits** | **Throughput /channel** |
| Input Vector MZM | 3 | 25GHz | 6 | **150 Gbps** |
| Output PD | Compress to 3 | 25GHz Cap | 7 | **175 Gbps** |
| **Estimated Throughput** | | **0.975 Tbps** | | |

**Table 1**. Computation of a PTC throughput requirements for a simple 3x3 MVM, with 3 high speed input encoding modulators and corresponding 3 output photodetectors, working at the maximum available speed and bits.

With ventures starting to deploy photonic-based solutions as ML hardware accelerators, major challenges still remain and limit the full potential of this technology; for example, domain crossings (E/O and digital-to-analog) are energy demanding and intrinsically noisy, generating errors that will propagate throughout the entire NN. Improved modulation quality (lower insertion loss, higher extinction ratio and thermal stability), smaller footprint and higher energy efficiency, together with advanced CMOS nodes for ICs will improve the figures of merit for this aspect, reducing the cost for each domain transition. CPO solutions [12], a technology required nowadays by data center switches for interconnect, still must reach the PTC world, but supplying different benefits, from a tighter (e.g. 2.5D) integration of electronics and photonics to a larger bandwidth and speed in terms of (tera)-operations-per-second. Another challenge is related to memory: with modulators and detectors that can reach 20+GHz and DAC/ADC with 6+ bits, transferring the data to and from the PTC requires HBM modules [13] (see **Table 1**). Integration of such memories will require 2.5D or 3D heterogeneous integration, with complementary TSV connections to the interposer or board [14]. Finally, thermal and stress management must come into play as more electronics will be placed on the same package, together with other thermal sources such as lasers or heaters.

## Advances in Science and Technology to Meet Challenges

Over the last half a decade, research on PTCs has progressed and continues to advance at a steady pace. CPO can play a key role to ease the memory bottleneck, hence promoting major CMOS fabs to introduce silicon photonic lines into their available processes. Large scale integration has seen important steps forward, with thousands of optical components in a single chip coupled with electronics circuits, finally supported by photonic-enhanced EDA tools and PDKs. However, we are still far away from million components on-PIC, mainly because of large footprints from electrooptic modulators, especially those in silicon and lithium niobate (i.e. microring resonator-based modulators do offer footprint savings, but require cumbersome and power hungry tuning circuitry). Moreover, Verilog-A has started to appear for photonics [15], allowing for large scale PIC-EIC simulations. We can expect a more of such tools, combined with LVS and 3DHI options as well. Finally, innovation on novel materials have enabled new devices: phase-change materials, for example, allow low-consumption optical memory device on chip [16,17], reducing the need for memory refresh and so data transfer from and to the digital memory; ITO and TFLN have shown new possibilities for compact and fast modulators, respectively [18,19]. The integration of all new materials, co-packaging with electronic ICs, the connectivity via fiber optics or digital I/O will define the next generation of advancements.

## Concluding Remarks

Photonic hardware accelerators are a reality that have proven their potential in an exploding market for Machine Learning algorithms and applications. Novel hardware is being innovated at sustained pace, including fab process offerings to advanced packaging concepts, novel materials, and interconnections. However, standards of photonic packaging still need to be defined and rolled out.

## Acknowledgements

*The authors acknowledge the SRC/DARPA JUMP 2.0 funding support and CHIMES leadership.*

## References

[1] Shastri, B. J., Tait, A. N., Ferreira de Lima, T., Pernice, W. H., Bhaskaran, H., Wright, C. D., & Prucnal, P. R. (2021). Photonics for artificial intelligence and neuromorphic computing. *Nature Photonics*, *15*(2), 102-114.

[2] Seo, J. S., Saikia, J., Meng, J., He, W., Suh, H. S., Liao, Y., Hasssan, A., & Yeo, I. (2022). Digital versus analog artificial intelligence accelerators: Advances, trends, and emerging designs. IEEE Solid-State Circuits Magazine, 14(3), 65-79.

[3] Peserico, N., Shastri, B. J., & Sorger, V. J. (2023). Integrated photonic tensor processing unit for a matrix multiply: a review. Journal of Lightwave Technology, 41(12), 3704-3716.

[4] Xu, X., Tan, M., Corcoran, B., Wu, J., Boes, A., Nguyen, T. G., Chu, S.T., Little, B.E., Hicks, D.G., Morandotti, R., Mitchell, A., & Moss, D. J. (2021). 11 TOPS photonic convolutional accelerator for optical neural networks. Nature, 589(7840), 44-51.

[5] Lederman, J. C., Zhang, W., de Lima, T. F., Blow, E. C., Bilodeau, S., Shastri, B. J., & Prucnal, P. R. (2023). Real-time photonic blind interference cancellation. Nature communications, 14(1), 8197.

[6] Zhang, W., Lederman, J. C., Ferreira de Lima, T., Zhang, J., Bilodeau, S., Hudson, L., Tait, A., Shastri, B.J. & Prucnal, P. R. (2024). A system-on-chip microwave photonic processor solves dynamic RF interference in real time with picosecond latency. Light: Science & Applications, 13(1), 14.

[7] Luan, E., Yu, S., Salmani, M., Nezami, M. S., Shastri, B. J., Chrostowski, L., & Eshaghi, A. (2023). Towards a high-density photonic tensor core enabled by intensity-modulated microrings and photonic wire bonding. Scientific Reports, 13(1), 1260.

[8] Ma, X., Schwartz, R. L. T., Jahannia, B., Nouri, B. M., Dalir, H., Shastri, B. J., Peserico, N., & Sorger, V. J. (2023, November). Fully Integrated Photonic Tensor Core for Neural Network Applications. In 2023 IEEE Photonics Conference (IPC) (pp. 1-2). IEEE.

[9] Giewont, K., Nummy, K., Anderson, F.A., Ayala, J., Barwicz, T., Bian, Y., Dezfulian, K.K., Gill, D.M., Houghton, T., Hu, S. and Peng, B., 2019. 300-mm monolithic silicon photonics foundry technology. IEEE Journal of Selected Topics in Quantum Electronics, 25(5), pp.1-11.

[10] El Srouji, L., Krishnan, A., Ravichandran, R., Lee, Y., On, M., Xiao, X., & Ben Yoo, S. J. (2022). Photonic and optoelectronic neuromorphic computing. APL Photonics, 7(5).

[11] Peserico, N., de Lima, T. F., Prucnal, P., & Sorger, V. J. (2022). Emerging devices and packaging strategies for electronic-photonic AI accelerators: opinion. Optical Materials Express, 12(4), 1347-1351.

[12] Minkenberg, C., Krishnaswamy, R., Zilkie, A., & Nelson, D. (2021). Co-packaged datacenter optics: Opportunities and challenges. IET optoelectronics, 15(2), 77-91.

[13] Jun, H., Cho, J., Lee, K., Son, H. Y., Kim, K., Jin, H., & Kim, K. (2017, May). Hbm (high bandwidth memory) dram technology and architecture. In 2017 IEEE International Memory Workshop (IMW) (pp. 1-4). IEEE.

[14] Razdan, S., De Dobbelaere, P., Xue, J., Prasad, A., & Patel, V. (2022, May). Advanced 2.5 D and 3D packaging technologies for next generation Silicon Photonics in high performance networking applications. In 2022 IEEE 72nd Electronic Components and Technology Conference (ECTC) (pp. 428-435). IEEE.

[15] Singh, J., Morison, H., Guo, Z., Marquez, B.A., Esmaeeli, O., Prucnal, P.R., Chrostowski, L., Shekhar, S. and Shastri, B.J., 2022. Neuromorphic photonic circuit modeling in Verilog-A. APL Photonics, 7(4).

[16] Kari, S.R., Ocampo, C.A.R., Jiang, L., Meng, J., Peserico, N., Sorger, V.J., Hu, J. and Youngblood, N., 2023. Optical and electrical memories for analog optical computing. IEEE Journal of Selected Topics in Quantum Electronics, 29(2: Optical Computing), pp.1-12.

[17] Meng, J., Gui, Y., Nouri, B.M., Ma, X., Zhang, Y., Popescu, C.C., Kang, M., Miscuglio, M., Peserico, N., Richardson, K. and Hu, J., 2023. Electrical programmable multilevel nonvolatile photonic random-access memory. Light: Science & Applications, 12(1), p.189.

[18] Amin, R., Maiti, R., Carfano, C., Ma, Z., Tahersima, M.H., Lilach, Y., Ratnayake, D., Dalir, H. and Sorger, V.J., 2018. 0.52 V mm ITO-based Mach-Zehnder modulator in silicon photonics. Apl Photonics, 3(12).

[19] Lin, Z., Shastri, B.J., Yu, S., Song, J., Zhu, Y., Safarnejadian, A., Cai, W., Lin, Y., Ke, W., Hammood, M. and Wang, T., 2023. 65 GOPS/neuron Photonic Tensor Core with Thin-film Lithium Niobate Photonics. arXiv preprint arXiv:2311.16896.

# Photonic probabilistic computing engines for accelerating neuromorphic and physical system simulations

**Fabian Böhm[1], Guy Verschaffelt[2], Guy Van der Sande[2]**
[1]Hewlett Packard Labs, Hewlett Packard Enterprise, 71034 Böblingen, Germany [fabian.bohm@hpe.com]
[2]Applied Physics Research Group, Vrije Universiteit Brussel, Pleinlaan 2, 1050 Brussels, Belgium
[ fabian.bohm@hpe.com, guy.van.der.sande@vub.be, guy.verschaffelt@vub.be]

## Status

In simulations of physical and neuromorphic systems, there is often an apparent disconnect, as stochastic processes are emulated by deterministic digital computers using pseudorandom number generators (PRNGs). While PRNGs can mimic probabilistic behaviour, their use can easily become problematic, e.g., when long sequences of high-quality random numbers are needed at fast rates. For example, low-quality PRNGs can have adverse effects on the security of encryption methods [1]. Moreover, PRNGs can also become prohibitively resource intensive when simulating large-scale stochastic systems like the human brain, which contains billions of stochastic synapsis [2,3]. These shortcomings have motivated the development of probabilistic computing engines. Such computing hardware incorporates physical stochastic devices, such as noise sources, whose non-deterministic behaviour is harnessed for modelling probabilistic processes with high accuracy, speed, and efficiency. Optical systems have long been considered as stochastic devices, as they can generate random number sequences from a variety of stochastic processes, such as chaotic oscillations, multi-mode interferences, spontaneous emission, and parametric down-conversion [4-9]. Crucially, these processes can occur at high bandwidths, which allows for random number generation at rates that vastly outperforms PRNGs [4,5]. As illustrated in Fig. 1, optical stochastic devices can be broadly categorized by whether they generate continuous or binary random states. In the former, the output is continuously fluctuating and can be transformed into random bit sequences with appropriate sampling and post-processing. In the latter, also referred to as coinflip devices, the random process only yields binary outcomes.

While research had initially been focused on using individual optical systems for accelerating cryptographic applications [4], the recent surge in large-scale AI models is motivating the massively parallel use of stochastic devices for building neuromorphic systems [2]. Here, stochastic devices are employed as part of optical neural networks or Ising machines (see Fig.1b), e.g., to accelerate the sampling of neuron activation probabilities. By harnessing the efficiency and speed of photonic stochastic devices, probabilistic computing engines have shown the potential to accelerate inference and training of stochastic neural networks, while also reducing energy consumption [6,9-12]. Beyond neuromorphic computing, there are also promising applications in physical system simulations and combinatorial optimization. Here, probabilistic computing engines are used to accelerate sampling of complex probability distributions in place of computationally expensive Monte-Carlo simulations, e.g., for quantum many-body and chemical structure simulations [13,14]. However, a demonstration of large-scale probabilistic computing engines remains a challenge, and it is still an open question, whether probabilistic computing engines will eventually outperform digital computers.

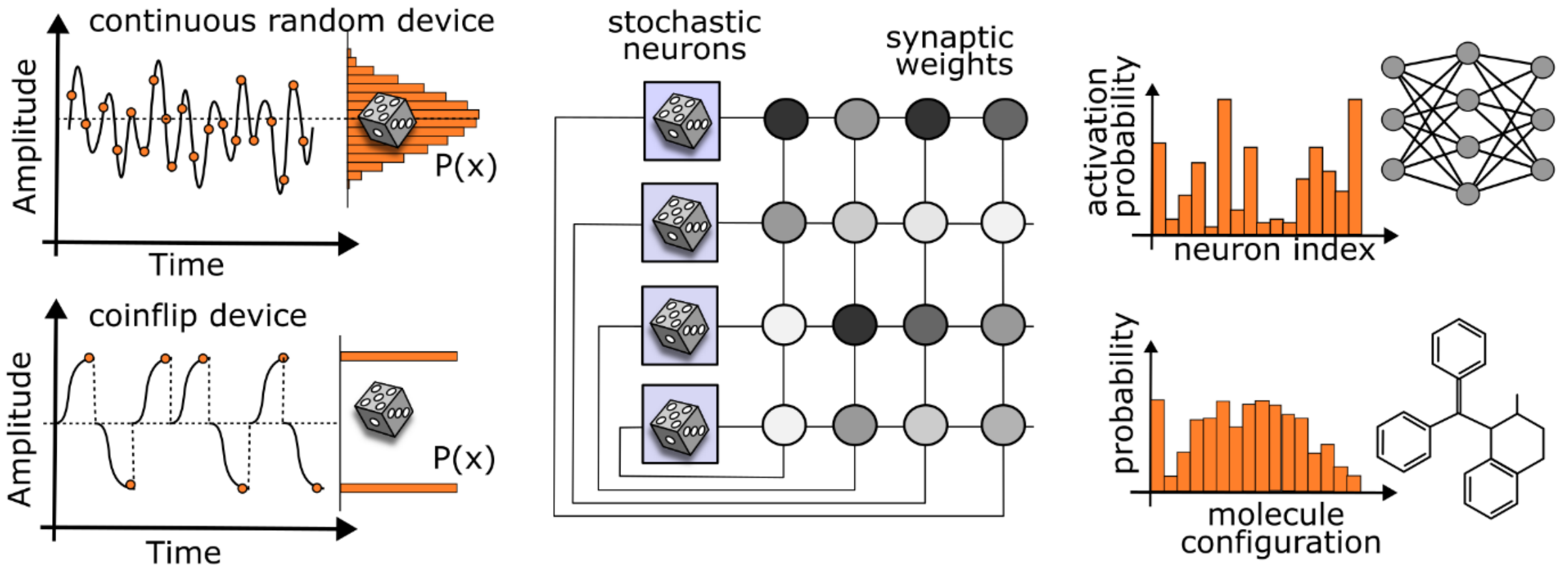

**Figure 1.** (a) Generation of random number sequences by sampling stochastic optical devices for continuous distributions (top) or binary distributions (bottom). (b) Exemplary schematic of a probabilistic computing engine for neuromorphic computing. Here, stochastic devices are used to emulate neurons, whose synaptic weights are embedded in a crossbar. (c) Exemplary applications of probabilistic computing engines for inferring probability distributions in stochastic neural networks (top) and molecular structure-based screening (bottom).

## Current and Future Challenges

As dedicated hardware accelerators, probabilistic computing engines must demonstrate the ability to perform statistical sampling at significantly higher rates than digital computers using PRNGs, while being similar in terms of footprint, energy consumption and cost as their digital counterparts. Crucially, they should be able to scale to large problem sizes, as probability distributions are often sampled from thousands of stochastic variables. While a few large-scale systems have been demonstrated at this point, they are still limited in speed or are emulated with digital computers [6,12]. This is in part because of well-known issues in building large-scale optical computing engines, such as optical neural networks and Ising machines. These issues are widely studied within the broader context of optical computing [15] and are briefly summarized in Fig.2. More specific to probabilistic computing engines are challenges related to the implementation of stochastic devices, their large-scale integration, and the embedding of statistical sampling applications.

### Fast and scalable stochastic devices

Currently, there is an apparent trade-off between the speed and the scalability of photonic stochastic devices. Devices based on amplified spontaneous emission and chaotic oscillations can generate samples at several hundred gigabits per second but require considerable power and device footprint for signal generation and for post-processing, thereby diminishing scalability. Coinflip devices based on bistable optical systems on the other hand can be realized as compact photonic integrated circuits but are limited in speed due to the need to be reset for the next coinflip operation.

### Problem embedding and tuneable probability distribution

For sampling applications, stochastic devices are often required to emulate different probability distributions, such as sigmoid or Gaussian distributions. Typically, a device's inherent stochastic distribution cannot support multiple target distributions and must be transformed with digital post-processing [4], which creates an overhead that can diminish the advantage of the optical stochastic device. At the same time, sampling applications must be embedded in the physical hardware, which can create an additional overhead. For stochastic Ising machines, for example, this embedding can considerably increase the required hardware resources and decrease computing performance [16].

### Device variability and error mitigation

Statistical sampling often requires that the random number sequence precisely follows a target distribution. Stochastic devices, on the other hand, can often exhibit device variability or systematic mismatches, that cause deviation from the target distribution. In error-sensitive tasks, such as the training of stochastic neural networks, such deviations can deteriorate performance and require error mitigation or calibration strategies [17].

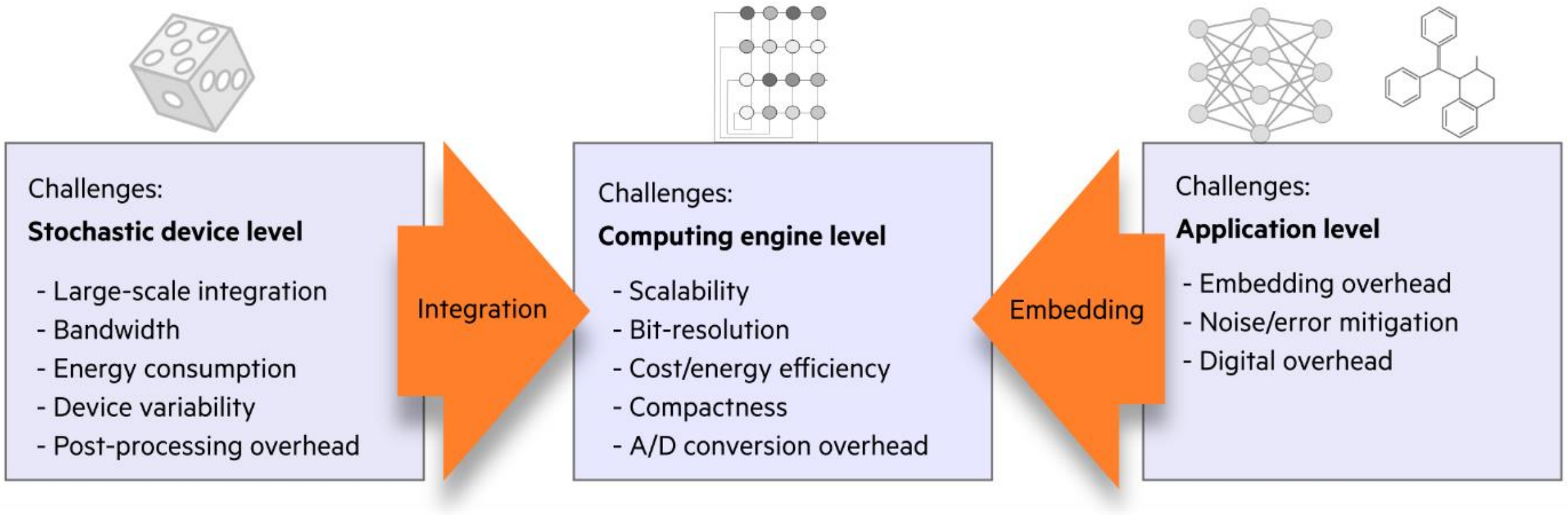

**Figure 2.** Summary of challenges related to building large-scale photonic probabilistic computing engines on the stochastic device level, the computing engine level, and the application level.

## Advances in Science and Technology to Meet Challenges

Moving forward, an important area of research will be the improvement in speed and energy-efficiency of photonic stochastic devices, together with their integration in large-scale computing systems. Coinflip devices based on optical bistable systems are an apparent fit for neuromorphic applications, as they naturally mimic the binary nature of activation in neurons. Devices based on bistable optical systems allow for tuneable sigmoid-type distributions used in many stochastic neural networks [6,8]. Integration in photonic devices appears feasible with existing fabrication methods, which will likely aid in reducing power consumption and improving scalability [9,18,19]. The emergence of fast coinflip devices based on VCSELs and bistable opto-electronic oscillators also promises improved random number generation rates [6,7]. Such systems operate at high bandwidths, which can yield coinflip operations at rates of a few gigabits per second. Another promising approach is the use of bistable devices in conjunction with Gaussian noise, such as opto-electronic feedback systems [6]. Here, a simple noise signal is used to drive a bistable system into continuous fluctuations that follow a sigmoid-type probability function. This type of operation can generate random numbers at high rates, as it foregoes the necessity to reset the coinflip operation. From an application perspective, different approaches have been brought forward to improve the embedding in probabilistic computing engines. This involves the inclusion of higher order variable interactions and the design of coinflip devices with more than two states [16,20]. Both approaches have shown to reduce embedding overhead and could thereby improve the sampling performance.

Crucially, regardless of the specific stochastic devices used in the engine, the success of probabilistic computing engines will be linked to overcoming the general challenges present in realizing large-scale photonic computing engines. For example, to embed practical stochastic neural networks, large-scale synaptic coupling elements must become available. For these reasons, hybrid systems are being considered as an alternative approach [3]. Contrary to the fully analog probabilistic computing sketched in Fig.1, hybrid systems primarily utilize CMOS technology for computation. Here, probabilistic processes are modelled with small-scale PRNGs. To ensure high-quality random numbers, these small-scale PRNGs are seeded with stochastic devices. This enables scalability without the issues of device variability, while still potentially reducing the energy and circuit footprint by orders of magnitude compared to a pure PRNG-based implementation. Here, photonic stochastic devices (e.g., devices based on spatio-temporal interference [5]) could be utilized for seeding a large set of PRNGs in parallel at high speeds.

## Concluding Remarks

Our universe is inherently probabilistic, from the quantum to the macroscopic level. Today, our desire to better understand these processes crucially depends on our ability to simulate large-scale stochastic systems with high efficiency. Faster and more efficient photonic probabilistic computing engines are therefore an important development to help sustain this drive for more accurate simulations. Neuromorphic computing in particular could greatly benefit from probabilistic computing engines. The current rise of large-scale generative AI models exemplifies the immense resources requirements of stochastic neural networks. Here, photonic probabilistic computing engines could considerably accelerate the simulation of stochastic neuromorphic systems and thereby reduce resource consumption. Because of the ubiquity of statistical sampling in various disciplines, the impact of photonic probabilistic computing engines could reach well beyond neuromorphic computing, e.g., in drug discovery, particle physics and finance. However, probabilistic computing is currently still in its early stages. As with most emerging non-von-Neumann computing schemes, it is still partially unclear which benefits over digital computers can be achieved. Crucially, this will involve the identification of real-world use cases that will see a large benefit from probabilistic computing engines.

## Acknowledgements

We acknowledge funding by the Research Foundation Flanders (FWO) under grants G028618N, G029519N and G006020N. Additional funding was provided by the EOS project "Photonic Ising Machines". This project (EOS number 40007536) has received funding from the FWO and F.R.S.-FNRS under the Excellence of Science (EOS) programme. We also acknowledge funding by the Defense Advanced Research Projects Agency (DARPA) under the Air Force Research Laboratory (AFRL) Agreement No. FA8650-23-3-7313. The views, opinions, and/or findings expressed are those of the authors and should not be interpreted as representing the official views or policies of the Department of Defense or the U.S. Government.

## References

[1] A. B., Orúe, H. L. Encinas, V. Fernández and F. Montoya, “A Review of Cryptographically Secure PRNGs in Constrained Devices for the IoT,” Proceedings of the Internal Joint Conference SOCO’17-ICEUTE’17, pp. 672-682, 2017.

[2] S. Misra et al., "Probabilistic Neural Computing with Stochastic Devices,” Adv. Mater., vol. 35, 2204569, 2022.

[3] N. S. Singh et al., "CMOS plus stochastic nanomagnets enabling heterogenous computers for probabilistic inference and learning,” Nat. Commun., vol. 15, 2685, 2024.

[4] J. D. Hart, Y. Terashima, A. Uchida, G. Baumgartner, T. E Murphy and R. Roy, „Recommendations and illustrations for evaluation of photonic random number generators,” APL Phot., vol. 2 090901, 2017.

[5] K. Kim et al., "Massively parallel ultrafast random bit generation with a chip-scale laser,” Science, vol. 371, pp. 948-952, 2021.

[6] F. Böhm, D. Alonso-Urquijo, G. Verschaffelt, G. Van der Sande, "Noise-injected analog Ising machines enable ultrafast statistical sampling and machine learning,” Nat. Commun., vol. 13, 5847, 2022.

[7] J. Zhao et al., "Fast all-optical random number generator,” 2022, arXiv2201.07616.

[8] C. Roques-Carmes et al., “Biasing the quantum vacuum to control macroscopic probability distributions,” Science, vol. 381, pp. 205-209, 2023.

[9] F. Brückerhoff-Plückelmann et al., “Probabilistic Photonic Computing with Chaotic Light,” 2024, arXiv2401.17915.

[10] B. Sutton, R. Faria, L. A. Ghantasala, R. Jaiswal, K. Camsari and S. Datta, “Autonomous Probabilistic Coprocessing With Petaflips per Second,” IEEE Access, vol. 8, pp. 157238-157252, 2020.

[11] N. A. Aadit et al., “Massively parallel probabilistic computing with sparse Ising machines,” Nat. Electron., vol. 5, pp. 460-468, 2022.

[12] S. Choi et al., “Photonic probabilistic machine learning using quantum vacuum noise,” 2024, arXiv2403.04731.

[13] H. Sakaguchi, K. Ogata, T. Isomura, S. Utsunomiya, Y. Yamamoto and K. Aihara, “Boltzmann Sampling by Degenerate Optical Parametric Oscillator Network for Structure-Based Virtual Screening,” Entropy, vol. 18(10), 365, 2016.

[14] S. Chowdhury, K. Y. Camsari and S. Datta, “Accelerated quantum Monte-Carlo with probabilistic computers,” Commun. Phys., vol. 6, 85, 2023.

[15] P. L. McMahon, “The physics of optical computing,” Nat. Rev. Phys., vol. 5, pp. 717-734, 2023.

[16] T. Bhattacharya et al., “Computing High-Degree Polynomial Gradients in Memory,” 2024, arXiv2401.16204.

[17] M. Benedetti, J. Realpe-Goméz, R. Biswas and A. Perdomo-Ortiz, “Estimation of effective temperatures in quantum annealers for sampling applications: A case study with possible applications in deep learning,” Phys. Rev. A, vol. 94, 022308, 2016.

[18] Y. Okawachi et al., „Demonstration of chip-based coupled degenerate optical parametric oscillators for realizing a nanophotonic spin-glass,” Nat. Commun., vol. 11, 4119, 2020.

[19] Z. Li et al., “Scalable On-Chip Optoelectronic Ising Machine Utilizing Thin-Film Lithium Niobate Photonics,” ACS Phot. Vol. 11(4), pp. 1703-1714, 2024.

[20] M. K. Bashar, A. Hasan and N. Shukla, “Designing a K-state P-bit Engine,” 2024, arXiv2403.06436.

# SUBSTRATES AND PHYSICS

## Linear and non-linear optical random projections for large-scale machine learning

**Fei Xia[1], Jianqi Hu[1], Hao Wang[1,2], Sylvain Gigan[1]**
1. Laboratoire Kastler Brossel, ENS-Universite PSL, CNRS, Sorbonne Université, Collège de France, Paris, 75005, France.
2. Department of Precision Instrument, Tsinghua University, Beijing 100084, China

Email address for each author:
F.X. fei.xia@lkb.ens.fr
J.H. jianqi.hu@lkb.ens.fr
H.W.h-wang20@mails.tsinghua.edu.cn
S.G. sylvain.gigan@lkb.ens.fr

### Status

Multiple light scattering is a pervasive phenomenon in optics, for instance when coherent light propagates through biological tissues, fog, or a sheet of paper, it generates a random speckle pattern at the output. Albeit inherently complex, this process is linear and coherent, and thus can be understood as a linear mapping that connects input and output wave fields. Intuitively, the linear nature is well described by a transmission matrix (TM), or likewise by a reflection matrix, which represents the way the medium deterministically mixes the input light into output speckles. In practical systems, experiments and theoretical studies reveal that the real and imaginary components of the entries in the TMs approximate random matrices with Gaussian independent and identical distributions [1]. This randomness and statistical properties, guaranteed by the laws of physics, have inspired the exploitation of scattering media for optical information processing [2], thanks to the connection with the signal processing concept of 'random projection'. This analogy has enabled various designs in optical computing that leverage optical scattering for optical random projection and beyond.

The concept has proven effective in analog compressed sensing, where each output speckle after passing through the complex medium acts as a random encoding of the input data, allowing a high-dimensional sparse image to be accurately reconstructed from a few measurements [3]. Beyond sensing, there has been a growing interest in utilizing optical random projection for machine learning applications, conceptually depicted in Fig. 1(a). Indeed, we have shown that linear random scattering is computationally equivalent to a single-layer neural network with all-to-all connectivities, which generates random speckle features to facilitate computing tasks, such as image classifications as shown in Fig. 1(b) [4]. The optical information processing by such a system approximates an elliptic kernel and can be generalized to a polynomial kernel [5], thereby favorably simplifying digital post-processing by implementing a kernel-based extreme learning machine. This notion continues to motivate new application scenarios, such as optical recurrent neural networks for time series processing, where the random medium constructs an optical reservoir for both conventional and next-generation reservoir computing [6, 7]. Furthermore, the exploitation of wave disorder also extends to Ising machines for accelerating combinatorial optimization [8] and optical direct feedback alignment for neural network training [9], among others. All these approaches embrace the large dimensions and fast speed of linear optical random projection [10][1], while demanding very low energy and memory costs, and minimal efforts for training or design.

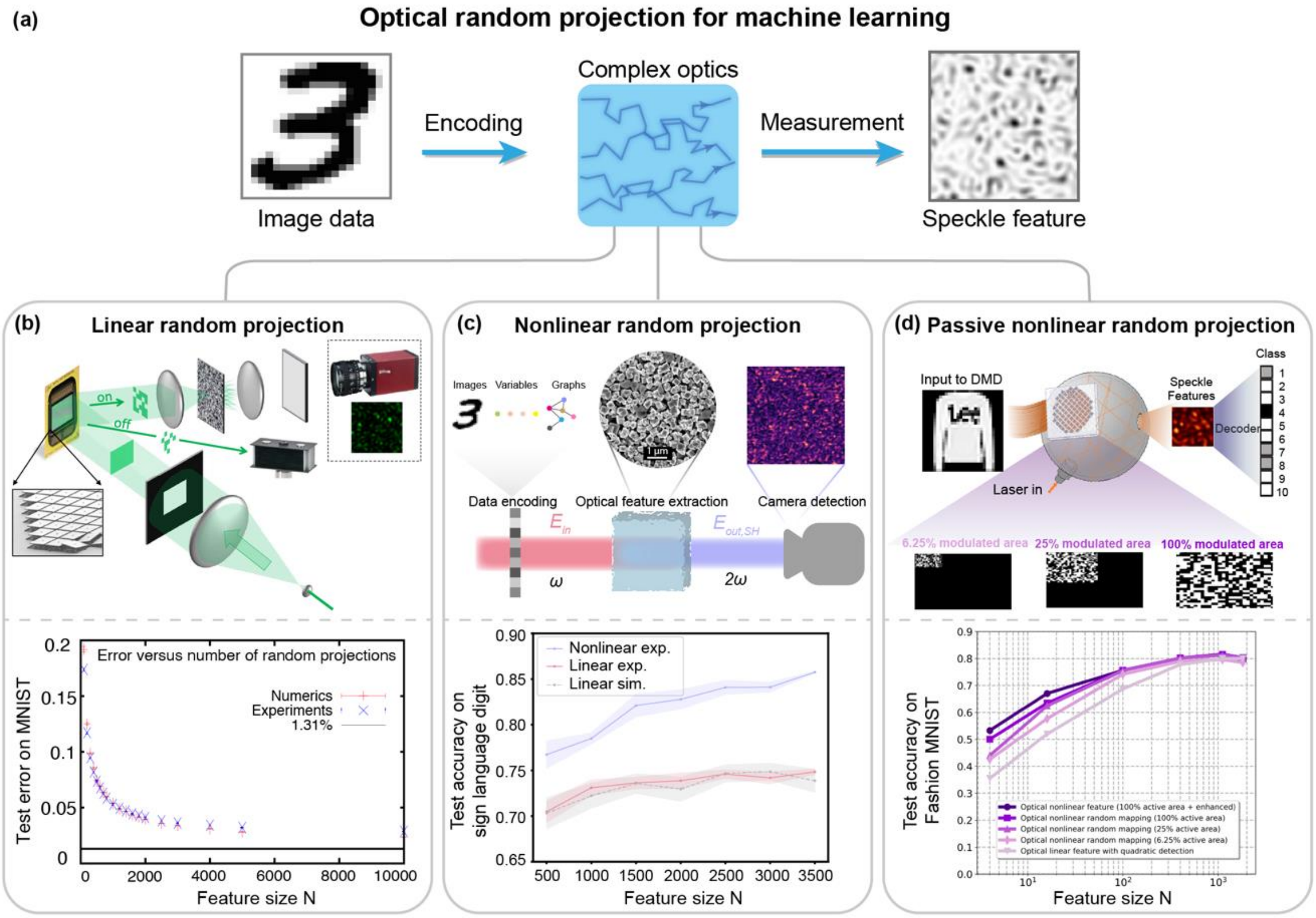

**Figure 1.** (a) After encoding data (e.g., an image) into an optical field (e.g., in its amplitude or phase), optical random projection can generate useful random speckle features at the output for various machine learning applications. (b) Linear multiple light scattering provides a random projection baseline [4]. (c) Nonlinear random projection based on a nonlinear complex medium (e.g., disordered lithium niobate nanocrystals), further enhances the computing expressivity [11]. (d) Passive nonlinear random projection based on a multiple-scattering cavity greatly compresses the data dimensions while retaining essential information that can be exploited by a decoder for diverse tasks [12].

## Current and Future Challenges

Although effective, linear multiple light scattering is limited within the scope of random projection. To approximate a wider range of mathematical operations, there is a pressing need to introduce internal nonlinearity, which, similarly to deeper layers in artificial neural networks, is necessary to enhance learning capability. To introduce nonlinearity, it should first be clarified that due to the nature of the optical system, nonlinearity can already be present during data loading and detection. During data loading into input devices, encoding nonlinearity can occur through procedures such as adding digital nonlinear transforms of the input data on a spatial light modulator (SLM) or through the process of phase encoding. During detection, detection nonlinearity can occur, such as the quadratic response due to intensity detection or other nonlinear responses by the detector. Besides these, to introduce nonlinearity between input and the detection, a case close to all-optical information processing, one can distinguish two main routes. A first straightforward approach is to leverage nonlinear optics to create nonlinear transformations of the input field, based on nonlinear optical materials to excite new optical frequencies. This has been demonstrated in the complex media domain, such as aggregates of lithium niobate nanocrystals [11] and silica multimode fibers [13]. The other direction is to encode information in light on terms other than the incident field. For example, we have recently demonstrated that by leveraging the encoding of data on the scattering potential and multiple scattering of light on between the SLM and a multiple-scattering cavity (Fig. 1d), we can create a passive nonlinearity that doesn't rely on material nonlinear responses, to create a nonlinear transformation between input data and the output speckle [12]. The second solution will be promising and inspiring for designing more power-efficient computing with enhanced performance due to the introduced nonlinearity. These show promise to expand various potential programmable dimensions (such as manipulating material properties and structural arrangement of light propagation) to increase the diversity of nonlinear transformations.

[1] Although optical random projection is linear, intensity detection adds a final quadratic non-linearity; however, pure linear random projections can be retrieved either via a holographic technique or through multiple measurements [10].

Besides nonlinearity, the trainability of optical weights is also desirable for approximating arbitrary functions optically. To achieve this, proposals such as sparing a port of the light modulators not only to be used solely as a data input device but also as an effective weight tuning device have been demonstrated in Ising machines [14], physical neural networks leveraging spectral shaping [15], and optical neural networks using complex media [16].

**Advances in Science and Technology to Meet Challenges**

While large-scale linear and nonlinear random projections can be computed by light scattering through disordered media, which is in principle very fast, the processes of data loading onto an SLM and detection by a camera eventually limit the speed of the entire optical computing system. Besides the bottleneck of digital-to-analog and analog-to-digital conversions, the most accessible apparatus used in this approach are fundamentally limited by their physical working principles. For instance, the speed of liquid-crystal SLMs is constrained by the response time of the molecules in switching orientations (typically below kHz), while the speed of digital micro-mirror devices is bounded by the mechanical flipping speed of mirrors (typically tens of kHz). On the detection side, both charge-coupled cameras and CMOS cameras have limited frame rates, typically up to kHz. Certainly, photonic computing systems with higher clock rates are demanded. To this end, one can for instance mention megapixel SLMs based on lithium niobate, which have recently been developed to reach GHz bandwidth, enabled by the electro-optic effect [17]. Likewise, silicon photonic detection arrays with GHz bandwidth [18] and single-photon avalanche diode cameras with sub-hundred-ps timing resolution [19] have also been demonstrated. The technological advancement of optical hardware is undeniably the driving force to boost the speed and energy efficiency of optical computing.

Beyond optical random projections, directly customizing the transformations between input and output will be desirable in many applications. This leads to the design and engineering of complex media as the optical surrogate for digital neural networks. In free space, multi-layer phase masks have been used to approximate target linear transformations [20] and even nonlinear processing [21]. In integrated photonics, inverse design has been an effective means to engineer the TM by patterning meta-structures in waveguides via nanofabrication [22]. Moreover, lithography-free, programmable control of the TM has also been demonstrated by modulating the real and/or imaginary parts of the refractive index locally in a slab waveguide [23, 24]. Besides its simplicity, this approach offers reconfigurability for training in situ or fine-tuning the model trained in silico. Nevertheless, compared to optical random projections in free space, the TM realizations in integrated photonics, though programmable, are still at relatively small scales. Therefore, merging the two pathways can be a viable solution toward large-scale programmable photonic computing [25, 26].

**Concluding Remarks**

In summary, we have discussed the current status and outlook of optical random projections for machine learning applications. In particular, we have seen how light scattering through complex media can be leveraged for large-scale computations and explored the possibilities of incorporating nonlinearity into the process. The various designs of nonlinearity, in general, pave the way toward enhanced expressivity and performance of optical neural networks. On a higher level, we may think of computing with complex media as part of the picture where the many degrees of freedom within the physical model can be leveraged into its computational model [27], and exploit the many parameters to approximate computing tasks needed in machine learning. Looking ahead, we envision the tailoring of more efficient optical processes to execute deep learning, with programmable controls, training, and inference at the scale matching the state of the art in artificial intelligence. This calls for the co-optimization of photonic systems and physics-aware algorithms.

**Acknowledgements**

F.X. acknowledges the funding from the Optica Foundation Challenge Award 2023. J.H. acknowledges the funding from Swiss National Science Foundation fellowship (P2ELP2_199825). H.W. acknowledges the funding from National Natural Science Foundation of China (623B2064). S.G. is a member of the institut Universitaire de France.

**References**

[1] S. M. Popoff, G. Lerosey, R. Carminati, M. Fink, A. C. Boccara, and S. Gigan, "Measuring the Transmission Matrix in Optics: An Approach to the Study and Control of Light Propagation in Disordered Media," *Physical Review Letters,* vol. 104, no. 10, p. 100601, 03/08/ 2010, doi: 10.1103/PhysRevLett.104.100601.

[2] S. Gigan, "Imaging and computing with disorder," *Nature Physics,* vol. 18, no. 9, pp. 980-985, 2022/09/01 2022, doi: 10.1038/s41567-022-01681-1.

[3] A. Liutkus, D. Martina, S. Popoff, G. Chardon, O. Katz, G. Lerosey, S. Gigan, L. Daudet, and I. Carron, "Imaging With Nature: Compressive Imaging Using a Multiply Scattering Medium," *Scientific Reports,* vol. 4, no. 1, p. 5552, 2014/07/09 2014, doi: 10.1038/srep05552.

[4] A. Saade, F. Caltagirone, I. Carron, L. Daudet, A. Drémeau, S. Gigan, and F. Krzakala, "Random projections through multiple optical scattering: Approximating Kernels at the speed of light," in *2016 IEEE International Conference on Acoustics, Speech and Signal Processing (ICASSP)*, 20-25 March 2016 2016, pp. 6215-6219, doi: 10.1109/ICASSP.2016.7472872.

[5] R. Ohana, J. Wacker, J. Dong, S. Marmin, F. Krzakala, M. Filippone, and L. Daudet, "Kernel Computations from Large-Scale Random Features Obtained by Optical Processing Units," in *ICASSP 2020 - 2020 IEEE International Conference on Acoustics, Speech and Signal Processing (ICASSP)*, 4-8 May 2020 2020, pp. 9294-9298, doi: 10.1109/ICASSP40776.2020.9053272.

[6] M. Rafayelyan, J. Dong, Y. Tan, F. Krzakala, and S. Gigan, "Large-Scale Optical Reservoir Computing for Spatiotemporal Chaotic Systems Prediction," *Physical Review X,* vol. 10, no. 4, p. 041037, 11/20/ 2020, doi: 10.1103/PhysRevX.10.041037.

[7] H. Wang, J. Hu, Y. Baek, K. Tsuchiyama, M. Joly, Q. Liu, and S. Gigan, "Optical next generation reservoir computing," *arXiv preprint arXiv:2404.07857,* 2024.

[8] D. Pierangeli, M. Rafayelyan, C. Conti, and S. Gigan, "Scalable Spin-Glass Optical Simulator," *Physical Review Applied,* vol. 15, no. 3, p. 034087, 03/30/ 2021, doi: 10.1103/PhysRevApplied.15.034087.

[9] J. Launay, I. Poli, K. Müller, G. Pariente, I. Carron, L. Daudet, F. Krzakala, and S. Gigan, "Hardware beyond backpropagation: a photonic co-processor for direct feedback alignment," *arXiv preprint arXiv:2012.06373,* 2020.

[10] R. Ohana, D. Hesslow, D. Brunner, S. Gigan, and K. Müller, "Linear optical random projections without holography," *Optics Express,* vol. 31, no. 16, pp. 25881-25888, 2023/07/31 2023, doi: 10.1364/OE.496224.

[11] H. Wang, J. Hu, A. Morandi, A. Nardi, F. Xia, X. Li, R. Savo, Q. Liu, R. Grange, and S. Gigan, "Large-scale photonic computing with nonlinear disordered media," *Nature Computational Science*, vol. 4, no. 6, pp. 429-439, 2024/06/01 2024, doi: 10.1038/s43588-024-00644-1.

[12] F. Xia, K. Kim, Y. Eliezer, S. Han, L. Shaughnessy, S. Gigan, and H. Cao, "Nonlinear optical encoding enabled by recurrent linear scattering," *Nature Photonics*, vol. 18, no. 10, pp. 1067-1075, 2024/10/01 2024, doi: 10.1038/s41566-024-01493-0.

[13]U. Teğin, M. Yıldırım, İ. Oğuz, C. Moser, and D. Psaltis, "Scalable optical learning operator," *Nature Computational Science,* vol. 1, no. 8, pp. 542-549, 2021/08/01 2021, doi: 10.1038/s43588-021-00112-0.

[14]G. Jacucci, L. Delloye, D. Pierangeli, M. Rafayelyan, C. Conti, and S. Gigan, "Tunable spin-glass optical simulator based on multiple light scattering," *Physical Review A,* vol. 105, no. 3, p. 033502, 03/03/ 2022, doi: 10.1103/PhysRevA.105.033502.

[15]L. G. Wright, T. Onodera, M. M. Stein, T. Wang, D. T. Schachter, Z. Hu, and P. L. McMahon, "Deep physical neural networks trained with backpropagation," *Nature,* vol. 601, no. 7894, pp. 549-555, 2022/01/01 2022, doi: 10.1038/s41586-021-04223-6.

[16]F. Xia, Z. Wang, L. Wright, T. Onodera, M. Stein, J. Hu, P. McMahon, and S. Gigan, *Hardware-efficient large-scale reconfigurable optical neural network (ONN) using complex media* (SPIE Nanoscience + Engineering). SPIE, 2023.

[17]S. Trajtenberg-Mills, M. ElKabbash, C. J. Brabec, C. L. Panuski, I. Christen, and D. Englund, "LNoS: Lithium Niobate on Silicon Spatial Light Modulator," *arXiv preprint arXiv:2402.14608,* 2024.

[18]C. Rogers, A. Y. Piggott, D. J. Thomson, R. F. Wiser, I. E. Opris, S. A. Fortune, A. J. Compston, A. Gondarenko, F. Meng, X. Chen, G. T. Reed, and R. Nicolaescu, "A universal 3D imaging sensor on a silicon photonics platform," *Nature,* vol. 590, no. 7845, pp. 256-261, 2021/02/01 2021, doi: 10.1038/s41586-021-03259-y.

[19]K. Morimoto, A. Ardelean, M.-L. Wu, A. C. Ulku, I. M. Antolovic, C. Bruschini, and E. Charbon, "Megapixel time-gated SPAD image sensor for 2D and 3D imaging applications," *Optica,* vol. 7, no. 4, pp. 346-354, 2020/04/20 2020, doi: 10.1364/OPTICA.386574.

[20]X. Lin, Y. Rivenson, N. T. Yardimci, M. Veli, Y. Luo, M. Jarrahi, and A. Ozcan, "All-optical machine learning using diffractive deep neural networks," *Science,* vol. 361, no. 6406, pp. 1004-1008, 2018, doi: doi:10.1126/science.aat8084.

[21]M. Yildirim, N. U. Dinc, I. Oguz, D. Psaltis, and C. Moser, "Nonlinear processing with linear optics," *Nature Photonics*, vol. 18, no. 10, pp. 1076-1082, 2024/10/01 2024, doi: 10.1038/s41566-024-01494-z.

[22]V. Nikkhah, A. Pirmoradi, F. Ashtiani, B. Edwards, F. Aflatouni, and N. Engheta, "Inverse-designed low-index-contrast structures on a silicon photonics platform for vector–matrix multiplication," *Nature Photonics,* 2024/02/16 2024, doi: 10.1038/s41566-024-01394-2.

[23]T. Onodera, M. M. Stein, B. A. Ash, M. M. Sohoni, M. Bosch, R. Yanagimoto, M. Jankowski, T. P. McKenna, T. Wang, and G. Shvets, "Scaling on-chip photonic neural processors using arbitrarily programmable wave propagation," *arXiv preprint arXiv:2402.17750,* 2024.

[24]T. Wu, M. Menarini, Z. Gao, and L. Feng, "Lithography-free reconfigurable integrated photonic processor," *Nature Photonics,* vol. 17, no. 8, pp. 710-716, 2023/08/01 2023, doi: 10.1038/s41566-023-01205-0.

[25]J. Moughames, X. Porte, M. Thiel, G. Ulliac, L. Larger, M. Jacquot, M. Kadic, and D. Brunner, "Three-dimensional waveguide interconnects for scalable integration of photonic neural networks," *Optica,* vol. 7, no. 6, pp. 640-646, 2020/06/20 2020, doi: 10.1364/OPTICA.388205.

[26]Z. Xu, T. Zhou, M. Ma, C. Deng, Q. Dai, and L. Fang, "Large-scale photonic chiplet Taichi empowers 160-TOPS/W artificial general intelligence," *Science,* vol. 384, no. 6692, pp. 202-209, 2024, doi: doi:10.1126/science.adl1203.

[27]H. Jaeger, B. Noheda, and W. G. van der Wiel, "Toward a formal theory for computing machines made out of whatever physics offers," *Nature Communications,* vol. 14, no. 1, p. 4911, 2023/08/16 2023, doi: 10.1038/s41467-023-40533-1.

# Diffractive Optical Networks for Visual Computing and Computational Imaging

**Yuhang Li and Aydogan Ozcan**
Department of Electrical and Computer Engineering, University of California, Los Angeles, CA 90095, USA.
Department of Bioengineering, University of California, Los Angeles, CA 90095, USA.
California NanoSystems Institute (CNSI), University of California, Los Angeles, CA 90095, USA.
[ yuhangli@g.ucla.edu, ozcan@ucla.edu ]

## Status

Computers are integral to nearly every aspect of modern life and are now omnipresent globally. In the digital age, applications that process large volumes of data impose severe demand on computing hardware, necessitating not only reduced latency and higher storage capacity but also greater bandwidth, cost-effectiveness, and the ability to perform complex artificial intelligence tasks with energy efficiency[1]. To address some of these challenges, significant research efforts are underway to develop new approaches using optics and photonics-based computing technologies, prompted by the need for faster and more energy-efficient computing solutions. Exploration of optical computing approaches is motivated by several critical limitations faced by electronics, notably the requirement for a significant amount of multicasting, resulting in a distributed communication burden[2]. Besides, with the inherent advantages of light, including its parallelism, lower energy consumption, and higher bandwidth, neuromorphic photonics emerged as a promising solution, potentially surpassing traditional computing limitations through the deployment of light-based elements[3]. With advances in artificial intelligence, neuromorphic computing is gaining significant attention, representing a shift from software-based machine learning and deep learning algorithms to optics/photonics-based hardware implementations. By leveraging their immense bandwidth and parallelism, optical neural networks aim to enable a new spectrum of applications that are currently unattainable with conventional computing technologies. Additionally, they could be applied in areas beyond the typical realm of deep learning, enabling computation with raw data in the analog domain (i.e., before the actual measurement, digitization, and storage take place), providing extreme speed advantages. These analog optical networks also help with the capture of the most useful features of interest while compressing information through visual processors that harness different degrees of freedom carried by light. Some of these visual computing and computational imaging applications require rapid response times, extensive bandwidth, and minimal energy consumption[4], [5], [6], [7]; see, e.g., **Figure 1**.

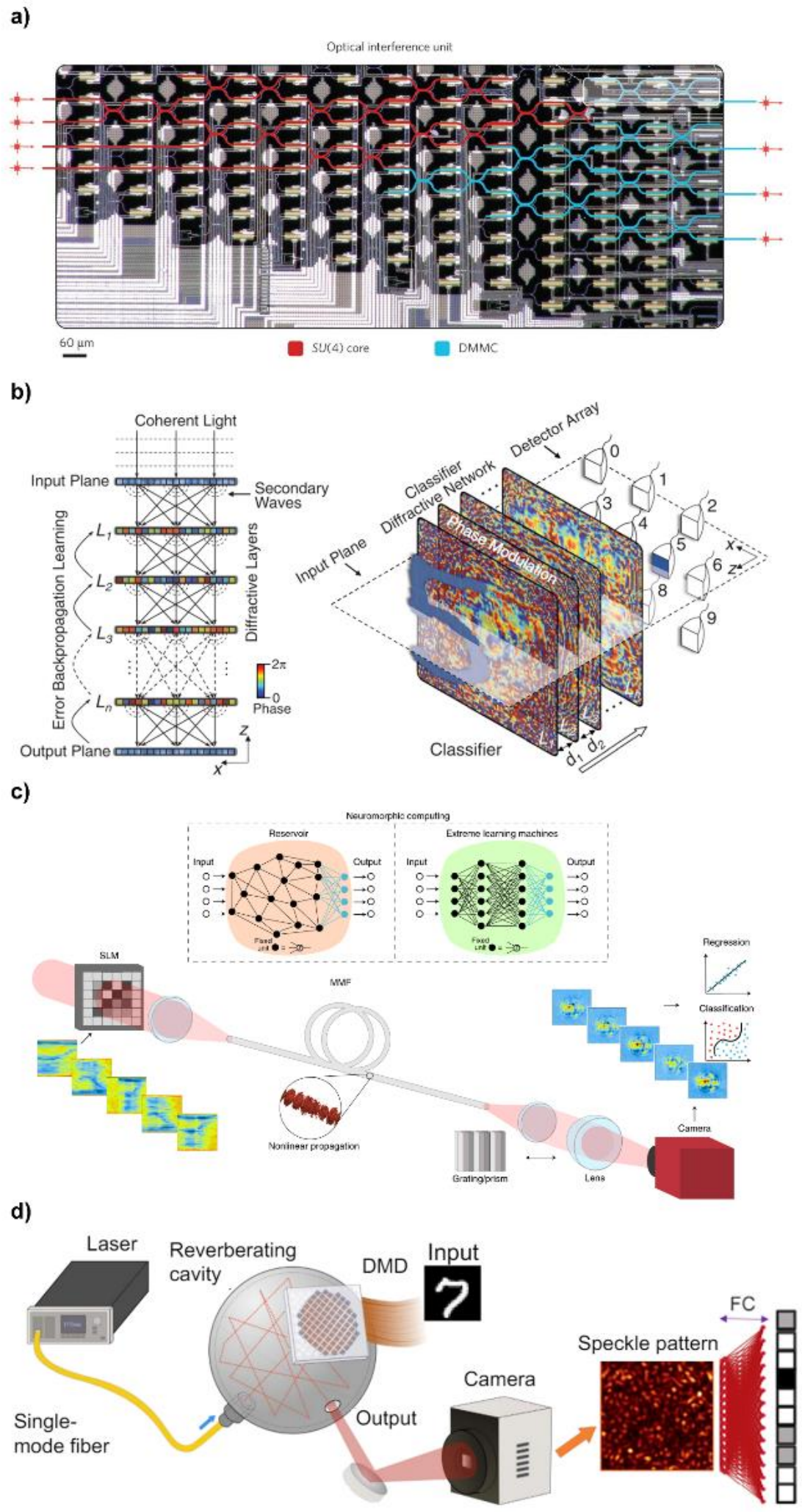

**Figure 1. Various configurations of neuromorphic optical computing.** (a) Optical neural network based on a nanophotonic circuit. (b) Diffractive deep neural network ($D^2NN$) architecture. (c) Optical computing framework based on spatiotemporal effects in multimode fibres. (d) Image classification using a tuneable cavity.

## Current and Future Challenges

In neuromorphic optical/photonic computing and its applications in visual computing and computational imaging fields, a particularly promising yet challenging area of research is the ability to see and sense through scattering media. This capability could be transformative for various applications in e.g., medical imaging, autonomous vehicle navigation, and security systems, by enabling enhanced image reconstruction methods in environments obscured by random and unknown diffusers[8]. Nevertheless, several significant challenges must be addressed to fully leverage this potential.

There is a wide array of neuromorphic optical and photonic devices that aim to address some of these challenges, including free-space-based and on-chip integrated photonics implementations. Free-space methods naturally align well with the 2D and 3D nature of a scene, but can face 3D alignment and fabrication challenges, especially at shorter wavelengths of operation. For sensing through random diffusers, the development of innovative optical structures that can precisely manipulate light is crucial for recovering the spatial and/or spectral details that are scrambled due to random scattering. Additionally, neuromorphic optical devices that rely on the complex behavior of light, such as diffraction, multi-path interference and scattering, necessitate precise algorithms capable of accurately simulating these physical phenomena with computational models representing the light interactions within the optical system. For instance, accurately modeling diffusive media within these systems is crucial for developing capabilities to image and sense through random diffusers. These algorithms must adeptly optimize the parameters of an optical network to effectively perform a given computational imaging and sensing task. Furthermore, developing optical technologies that can dynamically recalibrate in real time and operate reliably under varying scattering conditions is crucial for practical applications that involve time-varying aberrations and distortions.

Finally, transitioning from laboratory prototypes to scalable, cost-effective deployments/solutions remains a formidable hurdle. Neuromorphic optical and photonics technologies must be manufacturable at scale without prohibitive costs. Moreover, these systems must be compact and energy-efficient to be feasible for massive deployment in consumer products and industrial applications. Developing scalable fabrication techniques that maintain the precision required for optical components is critical for the widespread adoption and success of neuromorphic optical computing systems – which is especially challenging for 3D optical structures that operate at shorter wavelengths, covering, e.g., the visible band.

## Advances in Science and Technology to Meet Challenges

To overcome some of the challenges discussed earlier associated with imaging and sensing through random diffusers in the neuromorphic visual computing field, substantial advancements in multiple scientific and technological domains, e.g., materials science, optical engineering, computational algorithms, and system integration, are necessary. Recent advances in this field introduced diffractive visual processors for imaging through random, unknown diffusers using diffractive deep neural networks ($D^2NN$)[8], [9], eliminating the need for digital computation. This approach reconstructs the input visual information represented in the phase and/or amplitude of the optical field, by processing it in the analog domain through a series of diffractive surfaces. Each one of these passive surfaces consists of thousands of field modulation units at the sub-wavelength scale (~λ/2), meticulously optimized through deep learning and iterative error back-propagation algorithms to adjust the amplitude and/or phase profile of each diffractive layer. As light travels in 3D between these engineered surfaces, the diffractive layers collectively transform the input optical field into an output profile dictated by the specific task. These diffractive visual processors not only allow for imaging of unknown objects through random and unknown diffusers but also extend their applications to areas such as Quantitative Phase Imaging (QPI), eliminating the need for the laborious iterative reconstruction algorithms typically required for optical phase retrieval[10]. Additionally, diffractive visual processors can utilize a single-pixel detector and broadband illumination to enable direct classification of unknown objects, even through random/unknown phase diffusers[11]. A similar approach with a single-pixel detector was demonstrated to rapidly identify hidden defects or objects within a 3D sample, bypassing the need for extensive sample scanning or image processing[12]; defects/artifacts or hidden structures within such 3D objects were identified, in a snap-shot, with a single-pixel diffractive visual processor operating at the terahertz part of the spectrum, which can see through objects that are normally opaque in the visible or IR bands. These diffractive visual processors can also be effectively combined with electronic neural networks to create hybrid electronic-optical computing systems. This integration significantly improves the system's capabilities to image and sense through random diffusers and occlusions[13], [14] (see Fig. 2).

In general, neuromorphic optical devices face bit depth limitations due to limited resolution and potential fabrication errors. Recent studies have started to incorporate these constraints during the design process of visual processors and evaluate the impact of limited bit depth on the system performance[13]. Transitioning these optical computing technologies from laboratory prototypes to market-ready products requires advancements in scalable and cost-effective fabrication[15], [16] and system design[17], [18], along with improvements in optical sources[19] and materials[20] to minimize costs while boosting performance.

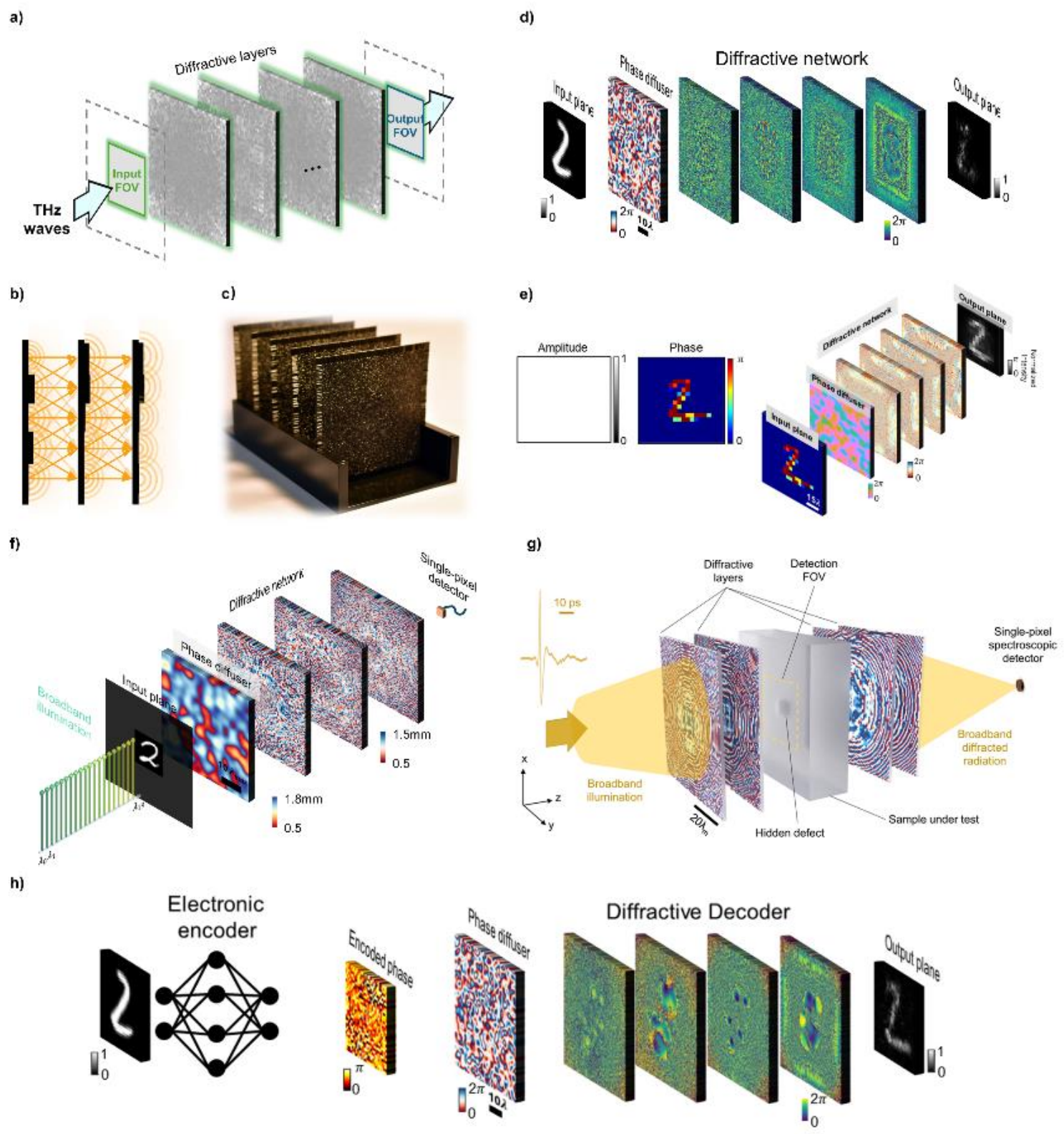

**Figure 2. Overview of Diffractive Deep Neural Networks (D²NNs) for Visual Computing and Computational Imaging.** (a) The schematic of a D²NN. (b) Physical principle of light-matter interactions within a D²NN. (c) A photograph of a 3D-printed D²NN prototype. (d) Seeing through random diffusers without a digital computer (e) Quantitative phase imaging (QPI) through random diffusers using diffractive networks. (f) All-optical image classification through unknown random diffusers using a single-pixel diffractive network. (g) Rapid sensing of hidden objects and defects using a single-pixel diffractive terahertz processor. (h) Optical information transfer through random unknown diffusers using electronic encoding and diffractive decoding.

## Concluding Remarks

Neuromorphic optical computing emerges as a cornerstone of innovation, with potentially transformative applications in visual computing and computational imaging fields. Neuromorphic optical devices not only leverage existing algorithmic methodologies for their programming and training through the rapid advancements in deep learning but also have direct access to and process raw optical data in the analog domain, extending beyond the traditional scope of machine learning algorithms applied after the information is measured, digitized, and stored/transmitted. Despite some impending challenges, its inherent features of immense bandwidth and parallelism—enhancing speed, efficiency, and the ability to manage large amounts of 2D/3D optical data— mark it as a potentially transformative approach in visual computing and computational imaging fields. A collaborative effort across multiple disciplines—from e.g., photonics and materials science to computer science and mechanical engineering could further enhance neuromorphic optical systems, opening new avenues for innovations across diverse domains and applications, from healthcare to autonomous systems, among many others.

**Acknowledgements**

*The authors acknowledge the support of the U.S. Department of Energy (DOE), Office of Basic Energy Sciences, Division of Materials Sciences and Engineering under award no. DE-SC0023088.*

## References

[1] D. V. Christensen *et al.*, “2022 roadmap on neuromorphic computing and engineering,” *Neuromorph. Comput. Eng.*, vol. 2, no. 2, p. 022501, Jun. 2022, doi: 10.1088/2634-4386/ac4a83.

[2] B. J. Shastri, C. Huang, A. N. Tait, T. F. de Lima, and P. R. Prucnal, “Silicon Photonics for Neuromorphic Computing and Artificial Intelligence: Applications and Roadmap,” in *2022 Photonics & Electromagnetics Research Symposium (PIERS)*, Apr. 2022, pp. 18–26. doi: 10.1109/PIERS55526.2022.9792850.

[3] B. J. Shastri *et al.*, “Photonics for artificial intelligence and neuromorphic computing,” *Nat. Photonics*, vol. 15, no. 2, pp. 102–114, Feb. 2021, doi: 10.1038/s41566-020-00754-y.

[4] Y. Shen *et al.*, “Deep learning with coherent nanophotonic circuits,” *Nature Photon*, vol. 11, no. 7, pp. 441–446, Jul. 2017, doi: 10.1038/nphoton.2017.93.

[5] X. Lin *et al.*, “All-optical machine learning using diffractive deep neural networks,” *Science*, vol. 361, no. 6406, pp. 1004–1008, Sep. 2018, doi: 10.1126/science.aat8084.

[6] U. Teğin, M. Yıldırım, İ. Oğuz, C. Moser, and D. Psaltis, “Scalable optical learning operator,” *Nat Comput Sci*, vol. 1, no. 8, pp. 542–549, Aug. 2021, doi: 10.1038/s43588-021-00112-0.

[7] F. Xia, K. Kim, Y. Eliezer, L. Shaughnessy, S. Gigan, and H. Cao, “Deep Learning with Passive Optical Nonlinear Mapping.” arXiv, Jul. 18, 2023. Accessed: Jul. 26, 2023. [Online]. Available: http://arxiv.org/abs/2307.08558

[8] Y. Luo *et al.*, “Computational imaging without a computer: seeing through random diffusers at the speed of light,” *eLight*, vol. 2, no. 1, p. 4, Dec. 2022, doi: 10.1186/s43593-022-00012-4.

[9] Y. Li, Y. Luo, B. Bai, and A. Ozcan, “Analysis of Diffractive Neural Networks for Seeing Through Random Diffusers,” *IEEE J. Select. Topics Quantum Electron.*, vol. 29, no. 2, pp. 1–20, 2022, doi: 10.1109/JSTQE.2022.3194574.

[10] Y. Li, Y. Luo, D. Mengu, B. Bai, and A. Ozcan, “Quantitative phase imaging (QPI) through random diffusers using a diffractive optical network.” arXiv, Jan. 19, 2023. doi: 10.48550/arXiv.2301.07908.

[11] B. Bai *et al.*, “All-optical image classification through unknown random diffusers using a single-pixel diffractive network,” *Light Sci Appl*, vol. 12, no. 1, Art. no. 1, Mar. 2023, doi: 10.1038/s41377-023-01116-3.

[12] J. Li *et al.*, “Rapid sensing of hidden objects and defects using a single-pixel diffractive terahertz sensor,” *Nat Commun*, vol. 14, no. 1, p. 6791, Oct. 2023, doi: 10.1038/s41467-023-42554-2.

[13] Y. Li, T. Gan, B. Bai, Ç. Işıl, M. Jarrahi, and A. Ozcan, “Optical information transfer through random unknown diffusers using electronic encoding and diffractive decoding,” *AP*, vol. 5, no. 4, p. 046009, Aug. 2023, doi: 10.1117/1.AP.5.4.046009.

[14] M. S. S. Rahman, T. Gan, E. A. Deger, Ç. Işıl, M. Jarrahi, and A. Ozcan, “Learning diffractive optical communication around arbitrary opaque occlusions,” *Nat Commun*, vol. 14, no. 1, p. 6830, Oct. 2023, doi: 10.1038/s41467-023-42556-0.

[15] F. Han, S. Gu, A. Klimas, N. Zhao, Y. Zhao, and S.-C. Chen, “Three-dimensional nanofabrication via ultrafast laser patterning and kinetically regulated material assembly,” *Science*, vol. 378, no. 6626, pp. 1325–1331, Dec. 2022, doi: 10.1126/science.abm8420.

[16] A. Grabulosa, J. Moughames, X. Porte, M. Kadic, and D. Brunner, “Additive 3D photonic integration that is CMOS compatible,” *Nanotechnology*, vol. 34, no. 32, p. 322002, May 2023, doi: 10.1088/1361-6528/acd0b5.

[17] J. Feldmann *et al.*, “Parallel convolutional processing using an integrated photonic tensor core,” *Nature*, vol. 589, no. 7840, pp. 52–58, Jan. 2021, doi: 10.1038/s41586-020-03070-1.

[18] J. Moughames *et al.*, “Three-dimensional waveguide interconnects for scalable integration of photonic neural networks,” *Optica, OPTICA*, vol. 7, no. 6, pp. 640–646, Jun. 2020, doi: 10.1364/OPTICA.388205.

[19] Z. Zhou *et al.*, “Prospects and applications of on-chip lasers,” *eLight*, vol. 3, no. 1, p. 1, Jan. 2023, doi: 10.1186/s43593-022-00027-x.

[20] Y. Zuo *et al.*, “All-optical neural network with nonlinear activation functions,” *Optica*, vol. 6, no. 9, p. 1132, Sep. 2019, doi: 10.1364/OPTICA.6.001132.

# Analog photoelectronic computing processors

**Yitong Chen[1], Jiamin Wu[1], Fei Qiao[2], Lu Fang[2], Qionghai Dai[1]**
[1]Department of Automation, Tsinghua University, Beijing, China.
[2]Department of Electronic Engineering, Tsinghua University, Beijing, China.

[yitongchen@sjtu.edu.cn; wujiamin@tsinghua.edu.cn; qiaofei@tsinghua.edu.cn; fanglu@tsinghua.edu.cn; qhdai@tsinghua.edu.cn]

**Status**

Among all-analog computing devices, all-optical architectures have attracted enormous attention because of its ultra-high speed and energy efficiency[1], especially considering the bloom of large artificial intelligence (AI) models[2], [3] and high-throughput imaging processing[4]. However, it has long been faced with problems of limited computational complexity, due to optical storage, integration scale and demands for high instantaneous power to achieve optical nonlinearity[5]–[9]. Combining with electronic devices can overcome many of these restrictions, but traditional photoelectronic computing using optical and digital electronic computing leads to impairment in the end-to-end speed and energy efficiency, due to the frequent analog-to-digital conversions (ADCs). The latency and energy consumption of optical computing can be as low as 17 fJ/MAC[10], while one ADC consumes orders of magnitude higher[11].

All-analog photoelectronic computing provides an exceeding solution to this dilemma (Fig. 1). By fusing optical and analog electronic computing on a single chip, the costly ADCs are maximum avoided when retaining the advantages of both photonic and electronic computing. Reported all-analog chip combining electronics and light (ACCEL) uses a diffractive neural network for parallel large-scale passive computing and analog electronic circuits with a photodetector array for nonlinearity and sequential calculation. In this way, ACCEL implements multi-layer nonlinear reconfigurable neural networks for various tasks while achieving a systemic computing speed of $4.55\times10^3$ TOPS and a systemic energy efficiency of $7.48\times10^4$ TOPS/W[12] orders of magnitude faster and energy-efficient than cutting-edge photonic and electronic processors.

Meanwhile, all-analog photoelectronic computing provides an efficient way for optical computing to connect to existing digital systems. There is no doubt that modern digital computing systems have developed to quite high maturity in not only industry, but also daily life. It means that an all-optical computing system requires to show potential to establish a comparably thorough system before attracting adequate industrial research input and practical applications. It is fatal in the development of AI architectures, other than input from academy. Fortunately, all-analog photoelectronic computing provides an elegant solution to it by bridging optical computing and mature digital systems with analog electronic circuits. As much as possible processing are implemented before conversions so the cost of ADCs are reduced by more than 98% without impairment in accuracy [12]. With such ultra-efficient interface to digital systems, analog photoelectronic computing benefits from the mature existing digital systems for much easier implementation of complicated tasks.

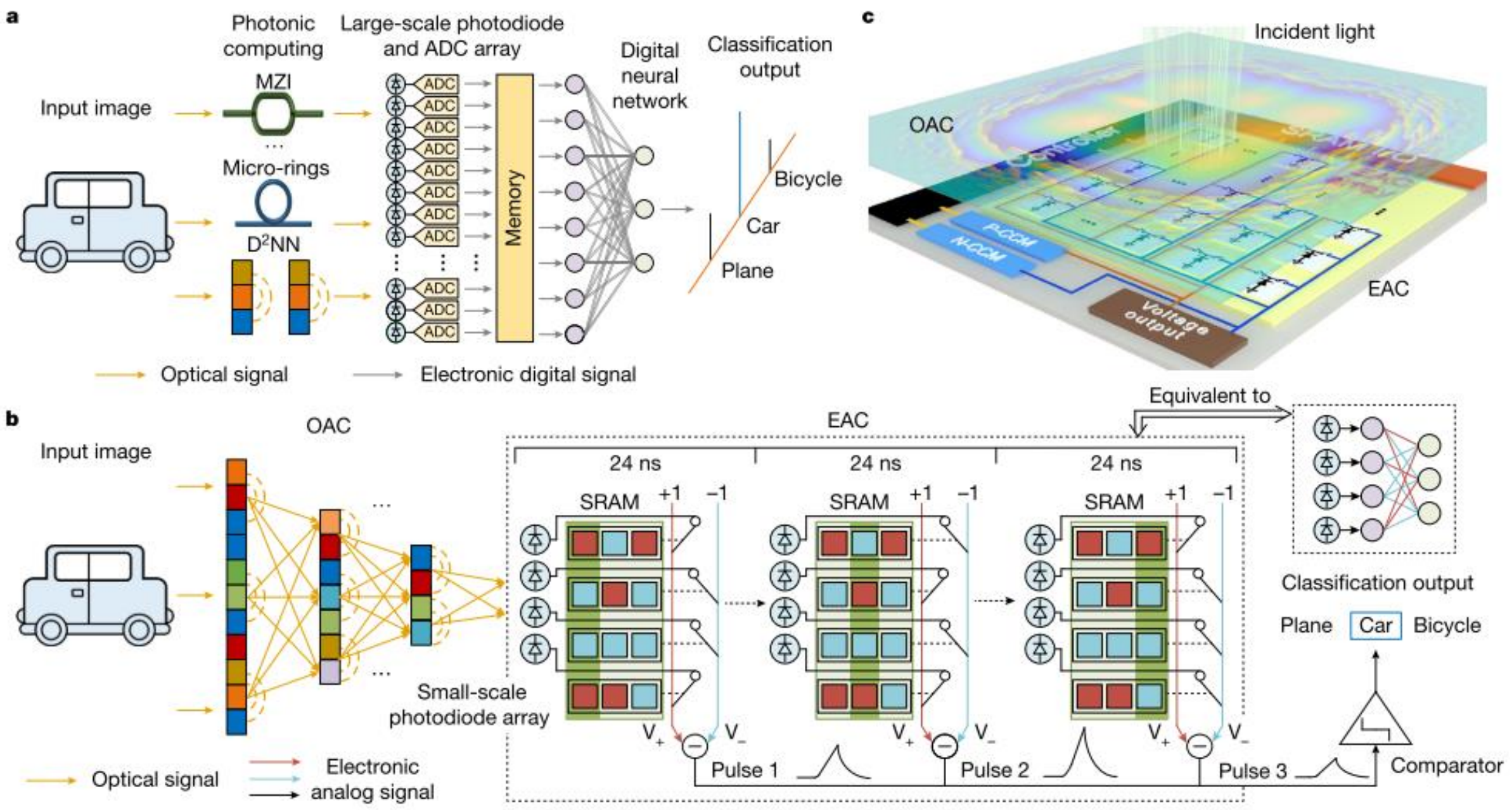

**Figure 1.** **a**, The workflow of traditional optoelectronic computing, including large-scale photodiode and ADC arrays. **b**, The workflow of ACCEL. **c**, Schematic of ACCEL with an OAC integrated directly in front of an EAC circuit for high-speed, low-energy processing of vision tasks. MZI, Mach–Zehnder interferometer; D2NN, diffractive deep neural network . Images reprinted with permission from Springer Nature.

## Current and Future Challenges

The principle challenges in analog photoelectronic computing development includes:

- Connection ways between the optical and analog electronic parts

It indicates the integration way for the photoelectronic parts on a single chip and determines the respective functions of each parts playing in the network. One of the reported ways is to convert the optical signals output from the photonic part into currents via photodetectors for the following electronic computing. In this way, both the optical and electronic analog parts implements several layers in the network respectively and therefore may achieve larger network scales. If 3D integration and bonding can be implemented, higher integration density can be obtained, e.g. <0.01 $mm^2$ per device[13]. The disadvantage of this method is that it at present requires digital control units for the analog electronic parts to recharge the computing lines or update the weights. Then the computing latency is constrained by the clock frequency. Using the electronic signals converted from the optical parts to modulate other pump light signals and back to complete the optical computing is one way to avoid the restriction. Such chips are reported chip to classify images at 570 ps/frame but restricted in relatively primary tasks[14]. Intrinsically, if the input signals are modulated with digital circuits instead of sensing from natural scenes, it remains to be restricted by the clock frequency.

- Photo-electronic-photo recurrent

In devices mentioned above, the optical part usually provides high-speed parallel linear multiply and add (MAC) operations and the analog electronic part provides nonlinear computing with optional storage. While modern large neural networks usually include multilayers, which usually require repeated switch between these functions. Space-efficient integration of light source, photodetectors, optical computing paths and analog electronic circuits on a single chip and the frequent switch between them remains a big challenge[15].

- Metrics for cross-model evaluation

When pushing the practical application of analog photoelectronic computing, we find it critical to establish general metrics to evaluate the chips based on different physical models. The traditional wide-accepted metrics for chips are the computing speed (MAC/s), energy efficiency (MAC/J), computing density (MAC/$mm^2$), etc. Although MAC has precise definition in digital systems, it is risky to directly transfer them to analog computing. It leads to paradox that an analog chip has much higher MAC numbers but achieves lower classification accuracy in the same task than a digital chip. Effective MAC considering different physical models and performance, e.g. calculation precision or classification accuracies, are required for credible comparison.

**Advances in Science and Technology to Meet Challenges**

Along with the above challenges, all-analog photoelectronic computing poses high demands for the following parts, including photo-electronic conversion, energy loss and stability of optical systems, and the overall package:

- Response speed

Considering the constant light speed, the response latency of the passive optical part usually relies on the path length. While the response latency of the analog electronic parts signally relies on the process precision. For example, in the reported work in 2023[12] , the reset time of the computing lines accounted for more than 50% of the overall photoelectronic computing time, which is approximately proportional to the capacitance of the photodiode, therefore proportional to the size of the photodiode. The size of the photodiode used in [12] is 14μm×8μm each, and the reported size in cutting-edge commercial image sensors can be as small as 0.56μm× 0.56μm [16], which indicates that the use of state-of-the-art image sensor technology can potentially reduce the reset time to 0.28%.

- The sensitivity of photoelectronic conversion

It determines the loss from light to electrical current. In order to achieve ultra-fast processing speed, the computing and exposure time of each frame are both reduced acutely. To maintain the sufficient SNR in ultra-short exposure time, sensitive photoelectronic convertors and low-loss bonding are imperative.

- Substrate compatibility

Optical paths and electronic circuits usually use different substrates, such as silicon dioxide, silicon-on-insulator (SOI), lithium niobate[17], etc. Some methods to integrate and overall package these different materials have been published in recent years (Fig. 2) [18], but it still requires to be transformed into stable procedures for mass production of analog photoelectronic devices.

- Correction of error accumulation

In digital computers, 0 plus 0 always equals 0 but in analog computing, almost 0 plus almost 0 could equal non-zero and cause errors. The error accumulation during the all-analog calculation becomes more remarkable as the analog photoelectronic computing architecture allows multilayer connections. Fortunately, the electronic parts provides easy reconfigurability for pre-calibration[19] and in-process training[12] to reduce the experimental error. But the method of online autonomous correction still requires end-to-end migration to nonlinear systems [20].

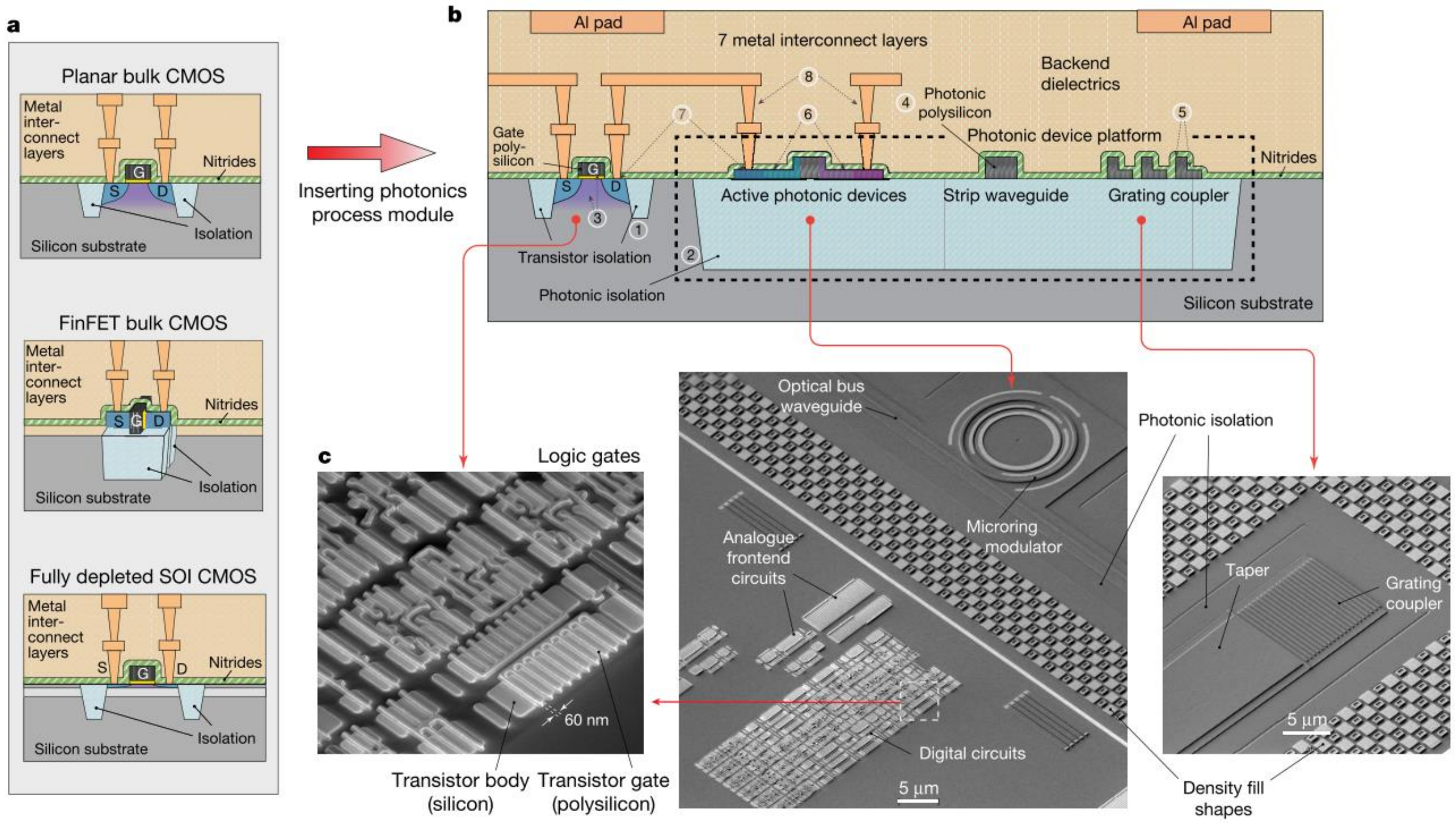

**Figure 2.** Photonic integration with nanoscale transistors. a, Illustration of three major deeply scaled CMOS processes: planar bulk CMOS, FinFET bulk CMOS, and fully depleted SOI CMOS. b, Integration of a photonics process module into planar bulk CMOS with photonic devices implemented in an optimized polysilicon film (220 nm) deposited on a photonic trench filled with silicon oxide (about 1.5 µm). c, Scanning electron micrographs of different photonic and electronic blocks. Images reprinted with permission from Springer Nature.

## Concluding Remarks

Analog photoelectronic computing processors, effectively fills many technical gaps of all-optical and all-analog electronic computing, providing a promising new route for ultra-high performance processors. Through the fusion of optical and analog electronic computing, the bottlenecks of speed and energy consumption of ADC are bypassed, meanwhile many of the advantages of digital computing, such as nonlinearity, storage, reconfigurablity are retained. Although analog photoelectronic computing has already achieved exceeding end-to-end computing speed and efficiency, it is only the beginning for us to reveal the enormous potential of this comprehensive architecture. The idea of all-analog photoelectronic computing hopefully opens the door of flexible access from optical computing to the existing extremely mature digital systems, and therefore accelerates such new processors to daily life and establish the joint development environment of both academy and industry much sooner.

## Acknowledgements
This work was supported by the National Natural Science Foundation of China (62088102, 62125106, 92164203, 62222508,62071272) and the Project of MOST (2021ZD0109901, 2020AA0105500).

## References

[1] Wetzstein G, Ozcan A, Gigan S, Fan S, Englund D, Soljačić M, Denz C, Miller D A B and Psaltis D 2020 Inference in artificial intelligence with deep optics and photonics *Nature* **588** 39-47
[2] Bueno J, Maktoobi S, Froehly L, Fischer I, Jacquot M, Larger L and Brunner D 2018 Reinforcement learning in a large-scale photonic recurrent neural network. *Optica* **5** 756-760.
[3] Xu Z, Zhou T, Ma M, Deng C, Dai Q and Fang L 2024 Large-scale photonic chiplet Taichi empowers 160-TOPS/W artificial general intelligence *Science* **384** 202-209
[4] Wu J, Guo Y, Deng C, Zhang A, Qiao H, Lu Z, Xie J, Fang L and Dai Q 2022 An integrated imaging sensor for aberration-corrected 3D photography *Nature* **612** 62–71

[5] Fu T, Zang Y, Huang Y, Du Z, Huang H, Hu C, Chen M, Yang S and Chen H 2023 Photonic machine learning with on-chip diffractive optics *Nat. Commun.* **14** 70
[6] Chen Y, Zhou T, Wu J, Qiao H, Lin X, Fang L and Dai Q 2023 Photonic unsupervised learning variational autoencoder for high-throughput and low-latency image transmission *Sci. Adv.* **9** eadf8437
[7] Ono M, Hata M, Tsunekawa M, Nozaki K, Sumikura H, Chiba H and Notomi M 2020 Ultrafast and energy-efficient all-optical switching with graphene-loaded deep-subwavelength plasmonic waveguides *Nat. Photonics* **14** 37–43
[8] Cheng Y, Zhang J, Zhou T, Wang Y, Xu Z, Yuan X and Fang L 2024 Photonic neuromorphic architecture for tens-of-task lifelong learning. *Light Sci. Appl.* **13** 56
[9] Lin X, Rivenson Y, Yardimci N T, Veli M, Luo Y, Jarrahi M, and Ozcan A 2018 All-optical machine learning using diffractive deep neural networks *Science* **361** 1004-8
[10] Feldmann J, Youngblood N, Karpov M, Gehring H, Li X, Stappers M, Gallo M L, Fu X, Lukashchuk A, Raja A S, Liu J, Wright C D, Sebastian A, Kippenberg T J, Pernice W H P and Bhaskaran H 2021 Parallel convolutional processing using an integrated photonic tensor core Nature 589 52–8
[11] Shen Y, Zhu Z, Liu S and Yang Y 2018 A Reconfigurable 10-to-12-b 80-to-20-MS bandwidth scalable SAR ADC *IEEE Transactions on Circuits and Systems I: Regular Papers* **65** 51-60
[12] Chen Y, Nazhamaiti M, Xu H, Meng Y, Zhou T, Li G, Fan J, Wei Q, Wu J, Qiao F, Fang L and Dai Q 2023 All-analog photoelectronic chip for high-speed vision tasks *Nature* **623** 48-57
[13] Chen Z, Sludds A, Davis R III, Christen I, Bernstein L, Ateshian L, Heuser T, Heermeier N, Lott J A, Reitzenstein S, Hamerly R and Englund D 2023 Deep learning with coherent VCSEL neural networks *Nat. Photonics* **17** 723–30
[14] Ashtiani F, Geers A J, and Aflatouni F 2022 An on-chip photonic deep neural network for image classification *Nature* **606** 501–6
[15] Zhou T, Lin X, Wu J, Chen Y, Xie H, Li Y, Fan J, Wu H, Fang L and Dai Q 2021 Large-scale neuromorphic optoelectronic computing with a reconfigurable diffractive processing unit *Nat. Photonics* **15** 367–73
[16] Clara S 2022 Commercializes World's Smallest Pixel in New 200MP Image Sensor with Superior Low-light Performance for High-end Smartphones (Calif.) (available at: https://www.ovt.com/press-releases/omnivision-commercializes-worlds-smallest-pixel-in-new-200mp-image-sensor-with-superior-low-light-performance-for-high-end-smartphones/) (Accessed 11 April 11 2024)
[17] Hu Y, Yu M, Buscaino B, Sinclair N, Zhu D, Cheng R, Shams-Ansari A, Shao L, Zhang M, Kahn J M and Lončar M 2022 High-efficiency and broadband on-chip electro-optic frequency comb generators *Nat. Photonics* **16** 679–85
[18] Atabaki A H, Moazeni S, Pavanello F, Gevorgyan H, Notaros J, Alloatti L, Wade M T, Sun C, Kruger S A, Meng H, Qubaisi K A, Wang I, Zhang B, Khilo A, Baiocco C V, Popović M A, Stojanović V M and Ram R J 2018 Integrating photonics with silicon nanoelectronics for the next generation of systems on a chip *Nature* **556** 349-54
[19] Yuan X, Wang Y, Xu Z, Zhou T and Fang L 2023 Training large-scale optoelectronic neural networks with dual-neuron optical-artificial learning *Nat. Commun.* **14** 7110
[20] Pai S, Sun Z, Hughes T W, Park T, Bartlett B, Williamson I A, Minkov M, Milanizadeh M, Abebe N, Morichetti F, Melloni A, Fan S, Solgaard O and Miller D A B 2023 Experimentally realized in situ backpropagation for deep learning in photonic neural networks. *Science* **380** 398-404

# Nanophotonic spike-based sensing and computing

**Bruno Romeira**
International Iberian Nanotechnology Laboratory, Av. Mestre José Veiga s/n, 4715-330 Braga, Portugal
bruno.romeira@inl.int

**Status**

Nanophotonic spike-based sensing and computing aims to push the frontiers of brain-inspired science by using miniscule excitable spiking emitter and sensory (detector) neurons in a hardware-oriented approach. The insect's tiny eye-brain biological system outperforms the most powerful sensors and computers in routine tasks such as real-time sensory data processing, perception, and motor control, while maintaining an extremely low energy budget (~50 mW) [1]. Even a simple living organism like the jellyfish without a centralized brain can learn its environment [2], despite having only a thousand spiking neurons and a few sensory eyes. These natural intelligent organisms feature biological spiking neurons, which are small (micrometre-sized), energy-efficient (estimated at ~1 fJ/spike), and exhibit unique biophysical diversity, namely when placed close to the sensory cells, promoting robust in-sensory computation. A paradigm shift in smart sensors will involve bringing intelligence closer to the source of data, enabling seamless integration of sensing, computing, and decision-making on a single miniaturized chip platform.

Over 13 years ago, the first reports of micrometer-sized spiking lasers [3],[4] and photodetectors [5], [6] were published (see Fig. 1), providing expectations that sensing and light conversion into spikes within a single hardware could be achieved. Despite the impressive advances in photonic integrated circuits over the last decade, such as the silicon photonics (SiPhot) and the indium phosphide (InP) platform, integrating these large optoelectronic components into energy-efficient and scalable neural network sensory systems remains a formidable challenge. As illustrated in Fig. 1, efforts are underway to reduce the size of such components while emulating sensory neuronal functions. This is achieved through the utilization of light-matter interactions at the nanoscale. Sub-micrometer semiconductor emitters (nanolasers [7], [8], [9], nanoLEDs [10], [11] and nanocavities [12], [13]), and photodetecting nanostructures [6], [14], [15], [16], [17] capable of operating in the few photon regime, are currently being developed using III-V, Si, SiGe, and 2D materials. Nonlinear phenomena in such nanophotonic structures, such as saturable absorption, two-photon absorption, bistability, negative differential resistance, are being considered to achieve the nonlinear properties needed to emulate the complex neurosynaptic functions. This promises components featuring high-speed, low-power, sensory and spike-based dynamic processing capabilities. Nonetheless, achieving advanced sensory, nonlinear spiking, and synaptic properties at the nanoscale (<<1 μm) remains challenging, awaiting demonstration in a single nanophotonic platform.

Here, it is presented a brief outline of the research challenges to develop energy-efficient semiconductor photonic emitter and sensory neurons down to 100 nanometers in size, while preserving their nonlinear spiking characteristics. Exploring new neuromorphic in-sensory nanophotonic circuits which integrate sensing and spike encoding, as in biological organisms, will enable neural circuits for spatio-temporal processing with low-latency (<ms) and low-power (<mW) in a compact chip (<$mm^3$), yet to be demonstrated in hardware. This will revolutionize in-sensor neuromorphic spike-based processing for edge AI computing [18], enabling fast real-time learning in dynamic and complex environments with resource-constrained endpoint hardware.

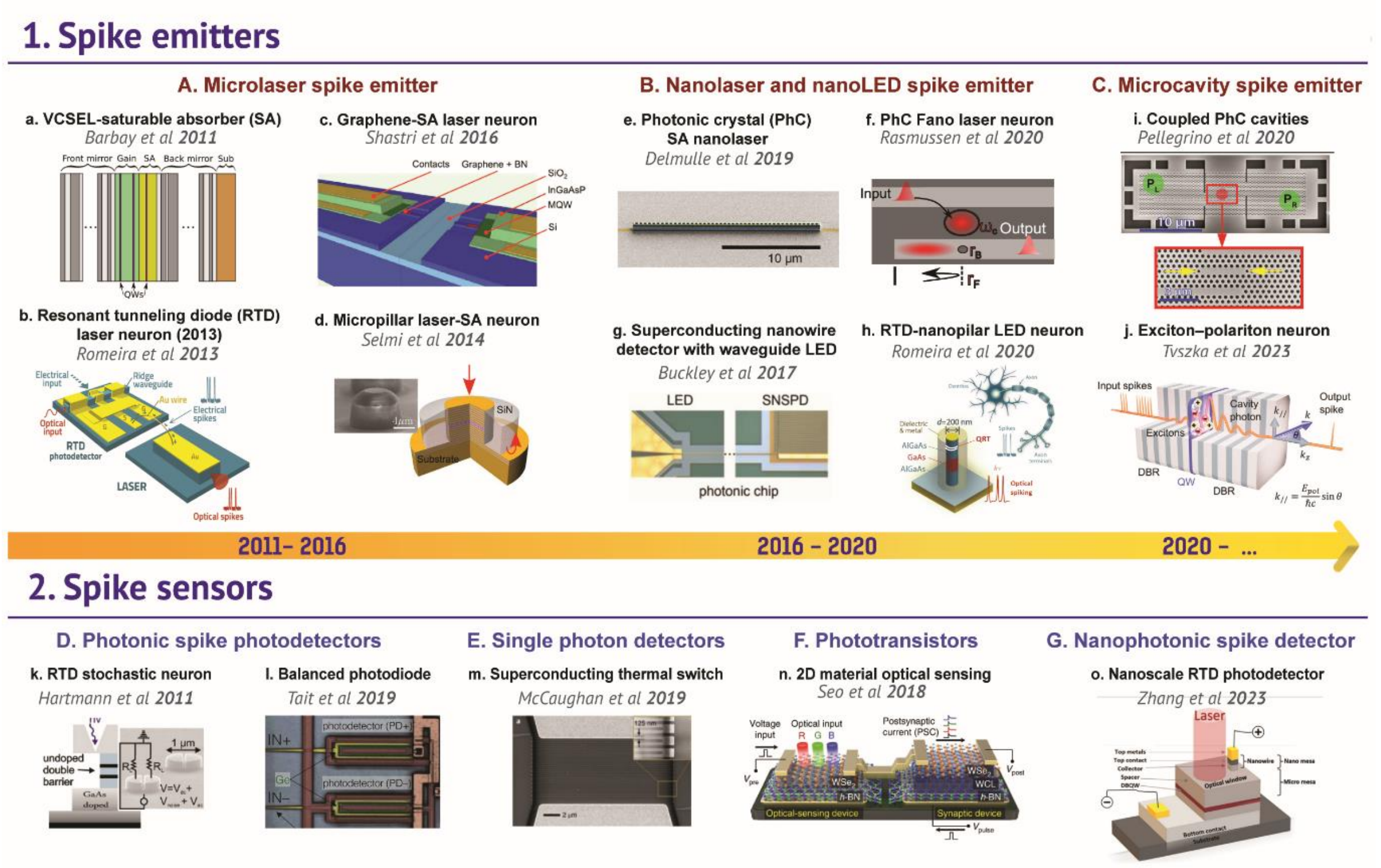

**Figure 1.** Nanophotonic spike-based sensing and computing key developments in recent years with a focus on micro- and nanoscale emitter and sensor spike-based artificial neurons. Panel (a) reprinted with permission from OPTICA [3]. Panel (b) reprinted with permission from OPTICA [5], Panel (c) reprinted with permission from Springer Nature [4], (d) reprinted with permission from American Physical Society [7], (e) reprinted with permission from OPTICA [8], (f) reprinted with permission from OPTICA [9], (g) reprinted with permission from AIP Publishing [10], (h) reprinted with permission from De Gruyter [11], (i) reprinted with permission from AIP Publishing [12], (j) reprinted with permission from Wiley-VCH GmbH, (k) reprinted with permission from AIP Publishing [6], (l) reprinted with permission from American Physical Society [14], (m) reprinted with permission from Springer Nature [15], (n) reprinted with permission from Springer Nature [16], (o) reprinted with permission from IOP Publishing [17].

## Current and Future Challenges

The key challenges in merging sensing with spike-based computing capabilities in a miniaturized chip are:

- Reducing the size of spiking neurons
- Achieving dynamic complex functions of spiking neurons
- Integrating synapses onto neurons
- Incorporating multimodal sensing
- Interconnecting nanoscale neurons using light
- Ensuring chip integration and scalability.

Each of these challenges will be briefly discussed.

*Reducing the size of spiking neurons*

The fundamental component of a neuromorphic spike-based system is the artificial nanophotonic spiking neuron. However, even a single edge-emitting laser, with a few hundred micrometers long, needs milliwatts of electrical power to reach threshold. This energy demand, corresponding to a few pJ/bit at 10 Gb/s data rates, far exceeds the optical energy needed for photodetection (~1000 photons or ~0.13 fJ/bit for a thermal-noise limited receiver). Efforts are ongoing, aiming at shrinking the laser size by 100-fold while maintaining spiking neuron functions. Addressing several challenges in these nanolasers is needed, including overcoming the diffraction limit of light, maintaining nonlinear properties at small scales, achieving efficient emission via improved quantum efficiency, and realizing efficient electrical pumping with low series resistance. For short-distance on-chip spike-based applications, incoherent nanoLEDs could replace nanolasers. NanoLEDs operate without a threshold, enabling improved efficiency at low injection, and potentially simpler fabrication, higher yield, less complex driving circuitry, and higher thermal stability.

*Achieving dynamic complex functions of spiking neurons*
A single biological neuron is a thousand times more complex than any state-of-the-art artificial neuron [19]. Implementing neuron models in nanophotonic hardware such as Hodgkin-Huxley, FitzHugh-Nagumo, or Izhikevich [20] (beyond the typical integrate-and-fire model) is needed to minimize computational cost and achieve biological realism in physical hardware. To achieve highly complex nonlinear neuron functions it is necessary to investigate nonlinear phenomena in nanophotonic structures such as saturable absorption, two-photon absorption, bistability, plasmonic, and negative differential resistance. The challenge is to realize the diverse neuronal functions, including chattering, phasic, triggered, integrator, tonic, mixed mode, and oscillatory, etc., spiking patterns, within the same nanoscale photonic platform in a controllable and programmable way.

*Integrating synapses onto neurons*
To emulate photonic neural synapses, memristive crossbar arrays are often used. Efforts are underway to incorporate synaptic memory functions directly into the spiking neurons. This could lead to synapses without the need of additional weight circuit connections, thus reducing the latency and power consumption in the neuron-synapse communication link, while providing local processing of information.

*Incorporating multimodal sensing*
Nowadays, smart sensors use sequential stages of sensor, memory, and machine learning (in the cloud). A large bottleneck is data transfer between sensor and computer. Exploring new in-sensory components integrating sensing and spike processing within a single unit will enable the encoding of analogue signals into spiking signals, resulting in computing at the edge with on-chip amplification and pre-processing of incoming sensor data. Incorporating multimodal sensing (like light, RF, magnetic, strain, tactile, and electrophysiology) directly onto artificial neurons is essential in future smart sensors. Specifically, in light-sensing related applications, the challenge includes the development of few-photon artificial visual sensory neurons using photosensitive nanostructures to replicate the unique properties of biological eye-brain visual organisms, including high sensitivity, hyperspectral colour vision, high dynamic range, and polarization sensitivity.

*Interconnecting nanoscale neurons using light*
Interconnecting nanoscale neurons using light is essential for mimicking brain circuitry, characterized by hierarchical structures of sub-circuits. Both near-field light broadcasting and far-field optical connectivity represent significant challenges due to the nanoscale dimensions of emitter and receiver neurosynaptic nodes and the low-photon operating regimes. Achieving spatial parallelism and fan-in/out of neural nodes involves a combination of free-space and 2D/3D waveguiding connectivity. Developing nanowaveguides (sub-μm) with low-loss is crucial for the connection of sub-circuits, enabling them to operate with overlapping light signals and transmitting efficiently their total input/output to other sub-circuits.

*Ensuring chip integration and scalability*
Achieving advanced sensory, nonlinear spiking, and synaptic functions within a single monolithic nanophotonic integrated chip presents a significant challenge. A promising approach involves modular construction of neuromorphic chips by assembling multiple smaller chiplets, similar to LEGO bricks. The challenge consists in achieving large-scale heterogeneous integration by combining chips of various semiconductor photonic materials (e.g., III-V, Si, SiGe, 2D) using advanced packaging techniques for the next generation of nanophotonic chips.

## Advances in Science and Technology to Meet Challenges

To overcome the challenges of scalable and low-power nanophotonic spike-based sensing and computing architectures, Table 1 summarizes a number of key features of integrated nanophotonic devices, the challenges to developing the neurosynaptic devices, their interconnectivity and advanced integration towards larges neural networks, and the scientific and technological advances that will be required to meet the challenges.

**Table 1.** Summary of some key integrated spike-based emitting and sensing features, challenges, and required scientific and technological advances for nanophotonic spike-based sensing and computing chip platforms.

| **Features** | **Challenges** | **Required scientific and technological advances** |
|---|---|---|
| Neuron emitters *power consumption* | • Idle state voltage and current required to drive the nanophotonic neurons limits minimum energy expenditure to 100 fJ | • Miniaturize nanolaser sources to consume sub-µA (towards sub-fJ operation)<br>• Use of incoherent light sources without a current threshold (e.g., nanoLEDs) for short-distance on-chip neural networks |
| Neuron emitters *optical output* | • Nanoscale LEDs/lasers emit low power (nW-µW) and exhibit low external quantum efficiency <<0.1 | • Develop methods for light extraction (nanostructuring, nanocavities) to improve external quantum efficiency >0.2<br>• Improve internal quantum efficiency using surface passivation methods<br>• Combine nanolight sources with external components (e.g., on-chip microlenses) |
| Neuron emitters *speed* | • Fast timescales (sub-ns) results in limited energy per emitted spike below 1 fJ/spike<br>• Nanodevices have small horizontal cross section and therefore suffer from high ohmic contact resistance | • Use of nanocavities to enhance light-matter interaction (Purcell effect)<br>• Incorporate tunnel junctions in nanoscale semiconductor emitters to reduce contact resistance |
| Neuron detectors *sensitivity* | • Achieve extremely high sensitivity<br>• Operation down to the few-photon or single-photon regime | • Use internal gain amplification effects to increase optical sensitivity<br>• Using novel combination of materials (e.g., 2D and III-V) to increase optical detector responsivity<br>• Use nanostructuring and metamaterials to enhance absorption |
| Neuron emitters/detectors *nonlinearity and dynamic spike function diversity* | • Achieve heterogeneous neurons with rich spiking nonlinearity (e.g., spiking, bursting, mixed mode, oscillatory, etc.)<br>• Achieve heterogeneous neurons with programmable and reconfigurable functions within a single architecture | • Introduce in the nanophotonic nanostructures nonlinear properties such as saturable absorption, two-photon absorption, bistability, plasmonic, negative differential resistance<br>• Combine optical, electrical, magnetic, optomechanical control on the same chip for programmable functions |
| Integrating synapses onto neurons | • Synapses and neurons made of different materials<br>• Exploring fast writing/erasing operations, along with short/long retention times, remains a challenge in synapses due to the speed/retention trade-off of memory technologies | • Integrate directly the synaptic function onto neurons avoiding the need of additional weight circuit connections<br>• Explore novel memories approaches (e.g., quantum effects), concepts (3D integration) and materials (e.g., 2D) |
| Incorporating light and multimodal sensing onto neurons | • Develop few-photon artificial sensory neurons with multi-properties, as found in eye-brain living organisms<br>• Integrate other sensing functions | • Incorporate high sensitivity, hyperspectral colour vision, high dynamic range, and polarization sensitivity<br>• Incorporate multimodal sensing onto spiking neurons: light, RF, magnetic, strain, tactile, and electrophysiology |
| Interconnectivity using light | • Difficulties in emitter-receiver communication due to small dimensions and low-photon operating regimes<br>• 2D connectivity does not provide scalable fan-in and fan-out | • Implement near-field light broadcasting approaches<br>• Implement 2D multilayer arrays connected in 3D with receptive light fields<br>• Use 3D waveguides (photonic wires)<br>• Use external components (e.g., microlenses) |
| Chip integration | • All previous features in a single monolithic integrated chip is extremely difficult | • Modular assembly (LEGO-like) of multiple smaller chiplets using heterogeneous integration |

## Concluding Remarks

The roadmap outlined here emphasizes the challenges of developing energy-efficient semiconductor nanophotonic emitter and sensory neurons with advanced nonlinear properties suitable for spike-based sensing and computing. Despite the remarkable advances in photonic integrated circuits, such as the silicon photonics and the indium phosphide platform, integrating optoelectronic components into energy-efficient and scalable light-based neural network nanosystems remains a challenge. The community will benefit from a collaborative and coordinated effort to develop small-sized spiking nanophotonic neurons, achieve dynamic complex functions, incorporate synaptic memory functions and multimodal sensing directly onto neurons, efficiently interconnect nanoscale neurons using near-field and far-field light in both 2D and 3D configurations, and ensuring advanced heterogeneous and modular chip integration of nanodevices (chiplets) with programmable and reconfigurable neurosynaptic functionalities. Overcoming these challenges will enable the realization of compact, low-power, low-latency, high-speed neuromorphic spike-based processing for edge AI computation, enabling applications where fast real-time learning of spatio-temporal information in dynamic and complex environments is required, emulating what living organisms already do.

**Acknowledgements**
Financial support by European Union, H2020-FET-OPEN framework programme, Project 828841 – ChipAI, Horizon Europe, project 101046790 – InsectNeuroNano, and Fundação para a Ciência e a Tecnologia (FCT), Portugal, project 2022.03392.PTDC – META-LED.

## References

[1] H. Haberkern and V. Jayaraman, “Studying small brains to understand the building blocks of cognition,” *Curr. Opin. Neurobiol.*, vol. 37, pp. 59–65, 2016.
[2] J. Bielecki, S. K. Dam Nielsen, G. Nachman, and A. Garm, “Associative learning in the box jellyfish Tripedalia cystophora,” *Curr. Biol.*, vol. 33, no. 19, pp. 4150-4159.e5, Oct. 2023.
[3] S. Barbay, R. Kuszelewicz, and A. M. Yacomotti, “Excitability in a semiconductor laser with saturable absorber,” *Opt. Lett.*, vol. 36, no. 23, pp. 4476–4478, Dec. 2011.
[4] B. J. Shastri, M. A. Nahmias, A. N. Tait, A. W. Rodriguez, B. Wu, and P. R. Prucnal, “Spike processing with a graphene excitable laser,” *Sci. Rep.*, vol. 6, no. 1, p. 19126, 2016.
[5] B. Romeira, J. Javaloyes, C. N. Ironside, J. M. L. Figueiredo, S. Balle, and O. Piro, “Excitability and optical pulse generation in semiconductor lasers driven by resonant tunneling diode photo-detectors,” *Opt. Express*, vol. 21, no. 18, 2013.
[6] F. Hartmann, L. Gammaitoni, S. Höfling, A. Forchel, and L. Worschech, “Light-induced stochastic resonance in a nanoscale resonant-tunneling diode,” *Appl. Phys. Lett.*, vol. 98, no. 24, p. 242109, Jun.

2011.
[7] F. Selmi, R. Braive, G. Beaudoin, I. Sagnes, R. Kuszelewicz, and S. Barbay, “Relative Refractory Period in an Excitable Semiconductor Laser,” *Phys. Rev. Lett.*, vol. 112, no. 18, p. 183902, May 2014.
[8] M. Delmulle, S. Combrie, F. Raineri, and A. De Rossi, “Design of a Nanolaser for Neuromorphic Computing,” in *Frontiers in Optics + Laser Science APS/DLS*, 2019, p. JW4A.68.
[9] T. S. Rasmussen, Y. Yu, and J. Mork, “All-optical non-linear activation function for neuromorphic photonic computing using semiconductor Fano lasers,” *Opt. Lett.*, vol. 45, no. 14, pp. 3844–3847, Jul. 2020.
[10] S. Buckley *et al.*, “All-silicon light-emitting diodes waveguide-integrated with superconducting single-photon detectors,” *Appl. Phys. Lett.*, vol. 111, no. 14, p. 141101, 2017.
[11] B. Romeira, J. M. L. Figueiredo, and J. Javaloyes, “NanoLEDs for energy-efficient and gigahertz-speed spike-based sub-λ neuromorphic nanophotonic computing,” *Nanophotonics*, vol. 9, no. 13, pp. 4149-4162., 2020.
[12] D. Pellegrino *et al.*, “Mode-field switching of nanolasers,” *APL Photonics*, vol. 5, no. 6, 2020.
[13] K. Tyszka *et al.*, “Leaky Integrate-and-Fire Mechanism in Exciton–Polariton Condensates for Photonic Spiking Neurons,” *Laser & Photonics Rev.*, vol. 17, no. 1, p. 2100660, 2023.
[14] A. N. Tait *et al.*, “Silicon Photonic Modulator Neuron,” *Phys. Rev. Appl.*, vol. 11, no. 6, p. 64043, Jun. 2019.
[15] A. N. McCaughan *et al.*, “A superconducting thermal switch with ultrahigh impedance for interfacing superconductors to semiconductors,” *Nat. Electron.*, vol. 2, no. 10, pp. 451–456, 2019.
[16] S. Seo *et al.*, “Artificial optic-neural synapse for colored and color-mixed pattern recognition,” *Nat. Commun.*, vol. 9, no. 1, p. 5106, 2018.
[17] Q. R. A. Al-Taai *et al.*, “Optically-triggered deterministic spiking regimes in nanostructure resonant tunnelling diode-photodetectors,” *Neuromorphic Comput. Eng.*, vol. 3, no. 3, p. 34012, Sep. 2023.
[18] C. Posch, T. Serrano-Gotarredona, B. Linares-Barranco, and T. Delbruck, “Retinomorphic Event-Based Vision Sensors: Bioinspired Cameras With Spiking Output,” *Proc. IEEE*, vol. 102, no. 10, pp. 1470–1484, 2014.
[19] D. Beniaguev, I. Segev, and M. London, “Single cortical neurons as deep artificial neural networks,” *Neuron*, vol. 109, no. 17, pp. 2727-2739.e3, 2021.
[20] E. M. Izhikevich, *Dynamical Systems in Neuroscience The Geometry of Excitability and Bursting*. MIT Press, 2010.

# Spiking lasers as artificial optical neurons for neuromorphic photonic systems

**Matěj Hejda[1], Lukas Puts[2], Nikolaos-Panteleimon Diamantopoulos[3], Weiming Yao[2], Sylvain Barbay[4], Benoit Charbonnier[5], Fabrice Raineri[4,6], Daan Lenstra[2], Antonio Hurtado[7]**

1. Hewlett Packard Labs, Hewlett Packard Enterprise, 1831 Diegem, Belgium
2. Photonic Integration Group, Department of Electrical Engineering, Eindhoven University of Technology, 5600 MB Eindhoven, The Netherlands
3. NTT Device Technology Labs, NTT Corporation, 3-1 Morinosato-Wakamiya, Atsugi, Kanagawa, 243-0198 Japan
4. Université Paris-Saclay, CNRS, Centre de Nanosciences et de Nanotechnologies, Palaiseau, France
5. Université Grenoble-Alpes, CEA, Leti, Grenoble, France
6. Inphyni, Université Côte d'Azur, France
7. Institute of Photonics, SUPA Dept of Physics, University of Strathclyde, Glasgow, United Kingdom

[matej.hejda@hpe.com, l.puts@tue.nl, np.diamantopoulos@ntt.com, w.yao@tue.nl, sylvain.barbay@c2n.upsaclay.fr, benoit.charbonnier@cea.fr, fabrice.raineri@univ-cotedazur.fr, dlenstra@tue.nl, antonio.hurtado@strath.ac.uk ]

## Status

Recent developments in the domain of ultra-large transformer-based machine learning models have demonstrated the remarkable capabilities and broad universality of these models for generative artificial intelligence (AI). However, the sheer scale and extensive computing resource requirements of these models render them less suitable for deployment in edge or energy-/latency-constrained settings. As an alternative approach to deep artificial neural network-powered models running on digital hardware, *neuromorphic engineering* aims to mimic some of the working principles of biological brains to devise digital or mixed-signal hardware and algorithms for highly energy-efficient AI computation. In a strict sense, neuromorphism implies a (dynamical) isomorphism with biological neurons. Therefore, a typical neuromorphic system implements a network of processing nodes (neurons) with high degree of processing parallelism, where the nodes compute dynamically and communicate using events (typically called spikes) encoded in time with a high degree of sparsity. In other words, the neuromorphic hardware provides a substrate for realising spiking neural networks (SNNs).

Furthermore, thanks to many advantageous properties of optical systems, such as low-loss signal propagation, extensive bandwidth, and suitability of multiplexing over additional degrees of freedom of lightwaves, and thanks to advances in photonic integration, photonics is now rapidly coming into the spotlight as a promising substrate for unconventional optical computing. Therefore, photonic devices capable of producing excitable spikes represent one of the key building blocks for light-enabled neuromorphic systems. The dynamical isomorphism to biological spiking is typically captured in phenomenological models such as leaky integrate-and-fire (LIF) neurons. LIF neurons exhibit, among others, two key properties: leaky integration of inputs over time, and excitable thresholding and spiking. Some of the first reports on optical LIF dynamics have been in a fibre-based platform [1]. Nowadays, the landscape of excitable photonic devices encompasses a broad variety of technologies. Typically, spiking optical devices are implemented by *(a)* achieving excitability optically in either active or passive photonic devices, or by *(b)* achieving excitability electrically, coupling a light-source to an excitable electronic circuit and performing E/O conversion. In this article, we will focus solely on excitable optical spiking obtained in lasers. Since photonic integrated circuits (PICs) represent arguably the most promising platform for realising photonic SNNs powered by excitable lasers thanks to the size-weight-power (SWaP) optimised character of PICs, we will primarily focus on integration-friendly solutions.

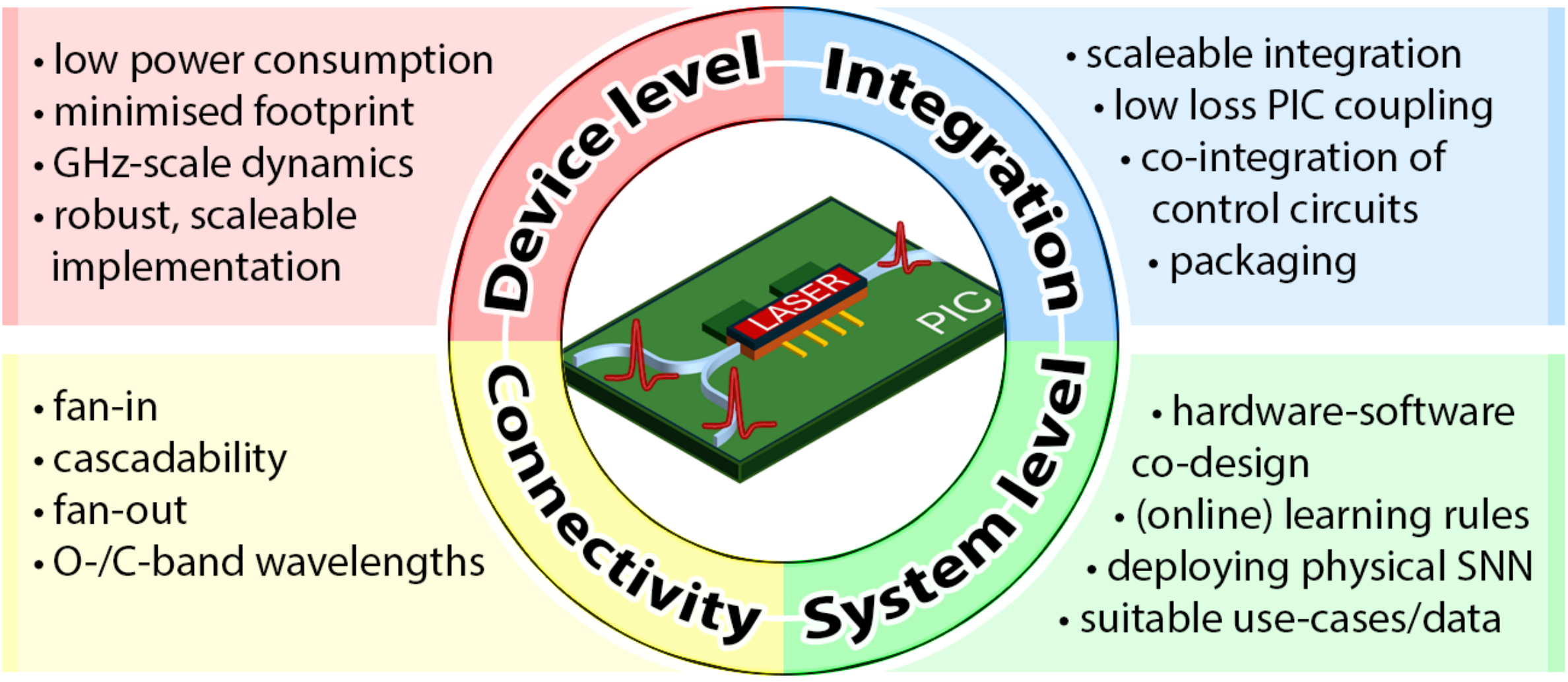

**Figure 1.** Overview of challenges of realising an ideal photonic spiking neuron using a laser.

**Current and Future Challenges**

We can consider the current challenges of excitable lasers for neuromorphic photonics using a bottom-up approach, from individual laser to a system-level perspective (Figure 1).
On the **device level**, an ideal excitable laser neuron should offer: *(a)* a minimised footprint alongside reduced power consumption. For example, nanoscale photonic crystal lasers confine light at wavelength-scale with low threshold current, yet the very low energy is not favourable to the strong nonlinear dynamics underpinning spiking and may be masked by noise. Moreover, introducing saturable absorption (SA) within their tiny volume makes fabrication more complex; *(b)* The device should have a robust implementation and operational scheme that allows for scalability. For example, optical injection allows to introduce excitable regimes in (integrated) lasers [2], but also introduces dependence on an additional external CW laser and high sensitivity to injection signal properties; *(c)* The spiking laser should exhibit high-speed (>10 GHz) neuron-like excitable dynamics to achieve low energy/spike, and well defined spiking threshold with LIF or resonate-and-fire character.
On the **integration** level, efficient solutions are required for scaling up the number of neurons (that is, fabrication/integration of dedicated spiking micro/nano lasers on a chip with efficient pumping schemes). Most photonic spiking hardware demonstrations so far have relied on a single photonic integration platform (e.g., Si or InP), while achievable processing speeds have mostly been limited to ~GHz rates or below. Future challenges shall include incorporating emerging heterogeneous photonic integration platforms for taking full advantage of state-of-the-art PICs, including low-loss co-integration with Si/SiN components as well as interfacing and packaging with required device control and stabilisation circuits. This would make photonic spiking processors a viable candidate for a wide range of photonic AI applications.

In terms of **connectivity and system-level**, spiking lasers should allow for fan-in, exhibit cascadability (re-eliciting of spikes from upstream signals) and allow for fan-out. Demonstrations of both integrated cascaded spiking [3] and fan-in/out remain less explored, particularly in experiments. For practicality, operation on standard telecom wavelengths (O/C-bands) is preferred to allow building upon mature optical telecommunications technologies. More complex PIC implementations that will functionally interconnect many spiking lasers represent a key challenge, followed by demonstrations of practical system-level implementations towards functional photonic SNNs. Finally, advantageous algorithms for spike-based photonic systems remain an open research question, particularly with respect to on-chip or local learning rules that would allow us to seize the full processing speed of photonics.

**Advances in Science and Technology to Meet Challenges**

Recent works have reported promising progress towards chip-scale excitable lasers. The first heterogeneously-integrated spiking laser has been demonstrated using a low power-consumption membrane III-V on Si technology, by integrating an optical feedback section for high-speed spiking dynamics [4], [5]. The laser was capable of producing optical spikes with demonstrated rates >10 GHz and <pJ/spike energies by using either electrical or optical inputs. Alternatively, a spiking laser was recently theoretically reported in a III-V/Si platform that allows for dense integration of lasers on SOI optical circuits [6]. This multi-section (gain-SA-quenching regions) InP spiking laser allows for waveguide-interfaced all-optical excitatory and inhibitory inputs [7]. Furthermore, a monolithically integrated two-section (gain-SA) spiking laser was demonstrated on a commercially available active-passive InP integration platform. Noise-triggered quasi-random optical excitability was observed theoretically and experimentally [8] with hundreds of MHz spike firing rates. Phase-space analysis for control and design parameters of these spiking lasers was also recently reported [9], allowing for device and operation optimization towards the excitable dynamic regime.

For nanoscale photonic crystal resonators, a drastic improvement in footprint and energy budget is expected in networks of such spiking nanolasers. Since the lasing current threshold scales with the field effective volume, in the order of $(\lambda/n)^3$, values in ranges as low as µA and fJ/bit were achieved [10]. Hybrid integration now allows to bring these essential features of photonic crystals in silicon PICs, providing an essential step toward networks of interconnected lasers [11]. Very recently, the challenge of implementing SA was solved by modification of the structure in [11], introducing SA-based spiking with excitability threshold in the fJ range, with ~100ps spikes and few GHz spike firing rates [12], [13].

Finally, for VCSEL-neurons, injection locking through an optically wire-bonded PIC was recently proposed [14]. Since current demonstrations of spiking VCSELs relied on 2.5Gbps-bandwidth telecom devices [15], using optimised VCSEL technologies offers prospects for enhanced bandwidth (>10 GHz) and efficient operation. Alternatively, for operation without injection, excitability was also reported in evanescently-coupled VCSELs [16], and micropillar-SA lasers [17] which deliver fast (hundreds of ps) single-mode optical spikes. Inputs can be provided through pump pulses (electrical or optical), i.e. incoherently, or in a coherent way (at microlaser cavity resonance) [18]. The excitable dynamics of micropillar-SA lasers can further be combined with a delayed feedback to form an all-optical spiking buffer memory, where the pulse patterns converge towards multistable asymptotic regular/irregular states [19].

**Table 1.** Comparison of current metrics in selected spiking lasers.

| Device | Technology | Spike FWHM | Spiking rate | Energy/spike | Type, ref |
|---|---|---|---|---|---|
| LD+feedback | heterog. III-V/Si | ~10 ps | >10 GHz | few fJ | Exper. [4], [5] |
| multi-section | heterog. III-V/Si | ~50 ps | <2.5 GHz | 10s of fJ | Theory [7] |
| two-section | monolithic InP | ~75ps | few GHz | ~ 1 pJ | Exper. [8] |
| nano-PhC | heterog. III-V/Si | 100 ps | few GHz | few fJ | Exper. [12] |
| µ-pillar | III/V | ~200 ps | ~3 GHz | 50 fJ | Exper. [18] |
| spiking VCSEL | III/V | ~100 ps | <3 GHz | ~100 fJ | Exper. [15] |

**Concluding Remarks**

Spiking light sources represent one of the fundamental building blocks for neuromorphic photonic systems. While studies on excitability in various classes of lasers have been ongoing for over two decades, we are nowadays witnessing their maturing into chip-scale, integrated devices. On individual device level, latest works on spiking lasers are now approaching desirable device-level performance metrics. Therefore, the next immediate challenges lie in further process developments to yield reliable spiking lasers that can be produced, efficiently integrated, and operated at higher volume scale. While heterogenous III-V/Si is currently the most studied platform in this scope, we envision other emerging platforms such as SiN to also garner a larger share of research interest. Beyond individual devices and heavily resource-constrained circuits (containing at maximum few neurons), larger chip-scale photonic SNNs and more complex functional spiking laser arrangements remain to be explored. Even with a constrained number of neurons, spike-coded photonic neuromorphic computing was shown to enable practical information processing [20]. Once the larger functional spiking laser arrangements are achieved, efficient full-scale system-level implementations will need to be explored. Finally, implementation of larger, complex, and practical computing problems and AI tasks on photonic neuromorphic hardware poses an open, promising direction of further research.

**Acknowledgements**

M.H., L.P., W.Y. and A.H. would like to acknowledge the EIC Pathfinder Open project “SpikePro” (101129904). A.H. acknowledges support from the UKRI Turing AI Acceleration Fellowship (EP/V025198/1).

**References**

[1] D. Rosenbluth, K. Kravtsov, M. P. Fok, and P. R. Prucnal, ‘A high performance photonic pulse processing device’, *Opt. Express*, vol. 17, no. 25, p. 22767, Dec. 2009, doi: 10.1364/OE.17.022767.

[2] K. Alexander, T. Van Vaerenbergh, M. Fiers, P. Mechet, J. Dambre, and P. Bienstman, ‘Excitability in optically injected microdisk lasers with phase controlled excitatory and inhibitory response’, *Opt. Express*, vol. 21, no. 22, p. 26182, Nov. 2013, doi: 10.1364/OE.21.026182.

[3] M. A. Nahmias *et al.*, ‘A Laser Spiking Neuron in a Photonic Integrated Circuit’. Dec. 15, 2020. Accessed: Dec. 17, 2021. [Online]. Available: http://arxiv.org/abs/2012.08516

[4] N.-P. Diamantopoulos, S. Yamaoka, T. Fujii, H. Nishi, T. Segawa, and S. Matsuo, ‘All-Optical Spiking Membrane III-V Laser on Si’, in *CLEO 2023*, San Jose, CA: Optica Publishing Group, 2023, p. STu4P.3. doi: 10.1364/CLEO_SI.2023.STu4P.3.

[5] N.-P. Diamantopoulos, T. Fujii, S. Yamaoka, H. Nishi, and S. Matsuo, ‘Ultrafast Electro-Optic Spiking Membrane III-V Lasers on Silicon Utilizing Integrated Optical Feedback’, *J. Light. Technol.*, vol. 42, no. 22, p. 7776–7784, Jul. 2024, doi: 10.1109/JLT.2024.3428532.

[6] K. Mekemeza-Ona, B. Charbonnier, and K. Hassan, ‘Design of an Integrated III–V on silicon semiconductor laser for spiking neural networks’, in *2021 IEEE International Interconnect Technology Conference (IITC)*, Kyoto, Japan: IEEE, Jul. 2021, pp. 1–2. doi: 10.1109/IITC51362.2021.9537482.

[7] K. Mekemeza-Ona, B. Routier, and B. Charbonnier, ‘All optical Q-switched laser based spiking neuron’, *Front. Phys.*, vol. 10, p. 1017714, Nov. 2022, doi: 10.3389/fphy.2022.1017714.

[8] L. Puts, D. Lenstra, K. Williams, and W. Yao, ‘Measurements and modeling of a monolithically integrated self-spiking two-section laser in InP’, *IEEE J. Quantum Electron.*, pp. 1–1, 2022, doi: 10.1109/JQE.2022.3224786.

[9] L. Puts, D. Lenstra, K. Williams, and W. Yao, ‘Phase-space analysis of a two-section InP laser as an all-optical spiking neuron: dependency on control and design parameters’. arXiv, 2024. [Online]. Available: https://arxiv.org/abs/2404.12771

[10] E. Dimopoulos *et al.*, ‘Experimental demonstration of a nanolaser with a sub-µA threshold current’, *Optica*,

vol. 10, no. 8, p. 973, Aug. 2023, doi: 10.1364/OPTICA.488604.
[11] G. Crosnier *et al.*, ‘Hybrid indium phosphide-on-silicon nanolaser diode’, *Nat. Photonics*, vol. 11, no. 5, pp. 297–300, May 2017, doi: 10.1038/nphoton.2017.56.
[12] M. Delmulle *et al.*, ‘Excitability in a PhC Nanolaser with an Integrated Saturable Absorber’, in *2023 Conference on Lasers and Electro-Optics Europe & European Quantum Electronics Conference (CLEO/Europe-EQEC)*, Munich, Germany: IEEE, Jun. 2023, pp. 1–1. doi: 10.1109/CLEO/Europe-EQEC57999.2023.10231939.
[13] B. Garbin *et al.*, ‘Reconfigurable photonic neuron’, in *Integrated Optics: Devices, Materials, and Technologies XXVIII*, SPIE, Mar. 2024, p. PC128890H. doi: 10.1117/12.3012122.
[14] M. Hejda *et al.*, ‘Optical spike amplitude weighting and neuromimetic rate coding using a joint VCSEL-MRR neuromorphic photonic system’, *Neuromorphic Comput. Eng.*, vol. 4, no. 2, p. 02411, 2024, doi: 10.1088/2634-4386/ad4b5b.
[15] M. Hejda, J. Robertson, J. Bueno, and A. Hurtado, ‘Spike-based information encoding in vertical cavity surface emitting lasers for neuromorphic photonic systems’, *J. Phys. Photonics*, vol. 2, no. 4, p. 044001, Aug. 2020, doi: 10.1088/2515-7647/aba670.
[16] M. Hejda, M. Vaughan, I. Henning, R. Al-Seyab, A. Hurtado, and M. Adams, ‘Spiking Behaviour in Laterally-Coupled Pairs of VCSELs With Applications in Neuromorphic Photonics’, *IEEE J. Sel. Top. Quantum Electron.*, pp. 1–10, 2022, doi: 10.1109/JSTQE.2022.3218950.
[17] K. Alfaro-Bittner, S. Barbay, and M. G. Clerc, ‘Pulse propagation in a 1D array of excitable semiconductor lasers’, *Chaos Interdiscip. J. Nonlinear Sci.*, vol. 30, no. 8, p. 083136, Aug. 2020, doi: 10.1063/5.0006195.
[18] V. A. Pammi, K. Alfaro-Bittner, M. G. Clerc, and S. Barbay, ‘Photonic Computing With Single and Coupled Spiking Micropillar Lasers’, *IEEE J. Sel. Top. Quantum Electron.*, vol. 26, no. 1, pp. 1–7, Jan. 2020, doi: 10.1109/JSTQE.2019.2929187.
[19] S. Terrien *et al.*, ‘Equalization of pulse timings in an excitable microlaser system with delay’, *Phys. Rev. Res.*, vol. 2, no. 2, p. 023012, Apr. 2020, doi: 10.1103/PhysRevResearch.2.023012.
[20] A. Masominia, L. E. Calvet, S. Thorpe, and S. Barbay, ‘Online spike-based recognition of digits with ultrafast microlaser neurons’, *Front. Comput. Neurosci.*, vol. 17, Jul. 2023, doi: 10.3389/fncom.2023.1164472.

# Neuromorphic Photonics with Vertical Cavity Surface Emitting Lasers (VCSELs)

**Joshua Robertson[1], Xavier Porte[1] and Antonio Hurtado[1]**
[1]Institute of Photonics, SUPA Department of Physics, University of Strathclyde, Technology and Innovation Centre, 99 George Street, Glasgow, G1 1RD, United Kingdom
joshua.robertson@strath.ac.uk, javier.porte-parera@strath.ac.uk,
antonio.hurtado@strath.ac.uk

## Status

Vertical Cavity Surface Emitting Lasers (VCSELs) are one of the most broadly applied, and commonly found semiconductor lasers due to their unique properties. VCSELs have vertically orientated cavities that make use of high-reflectivity Distributed Bragg Reflector (DBR) mirrors to sandwich gain material into an extremely compact structure with emission perpendicular to the substrate (Fig. 1(a)). This design benefits from high-yield, mature commercial fabrication, that allows for efficient testing of devices directly on wafer for low-cost manufacturing with a simple pathway to array structures and optical fibre coupling. Thanks to the maturity of the VCSEL platform, devices can have >30% wall plug efficiency with very low lasing threshold currents, multiple polarization states and close to ideal Gaussian emission profiles. The VCSEL platform also boasts high modulation capability (offering 10s GHz bandwidths), making it widely adopted in optical communication systems, e.g. optical interconnects in data centres and fibre-optic telecommunication networks, and sensing functionalities, e.g. Light Detection and Ranging (LIDAR), mobile phone and automotive sensors. There is therefore great potential to add intelligence and computing capabilities to the ubiquitous, key-enabling VCSEL technology platform.

As a result, in recent years, VCSELs have seen increasing investigation as neuromorphic processing platforms and artificial neuronal models, where non-linear interactions within the laser cavity can be exploited to provide neuron-like functionalities for computing. VCSELs, subject to the injection of optical inputs, not only possess the ability to act as high-speed, efficient non-linear transformers, but naturally exhibit a full set of dynamical properties (e.g. excitability, mode competition, chaos). This renders VCSELs appropriate for use as optical spiking neurons and their integration into photonic spiking neural networks (SNNs) able to perform complex computational tasks efficiently using ultrafast neural-like optical spikes, and also for information processing systems using laser-based photonic reservoir computing (RC) techniques, such as the exemplar systems depicted in Fig. 1. Such VCSEL systems have achieved operation with multi-GHz rate optical inputs, and high efficiency with ~10fJ estimated energy per nonlinear transformation (see [1] and references therein). These facts show there is great potential for systems based upon VCSELs to deliver neuromorphic information processing functionalities on an ultrafast, highly-efficient platform that already sees large adoption in sensing and communication applications. VCSELs are therefore an exciting technology that can push compact, fast and energy-efficient photon-enabled neuromorphic systems to strategic domains, such as edge-computing and data-centre technologies.

## Current and Future Challenges

Main challenges facing VCSEL-based neuromorphic photonic systems include the scalability, connectivity and miniaturization of the platform (see Fig. 2). Neuromorphic systems typically require large fan-in/fan-out and cascadable outputs to operate, especially when working in traditional neural network architectures. However, to date a majority of experimental neuromorphic VCSEL demonstrators, such as those aiming at developing light-enabled VCSEL-based photonic RC and photonic spike-processing systems and SNNs, are realised with optical fibre and free-space optical components. Whilst these allow hardware-friendly lab-based experimental systems, these can also be bulky and difficult to scale. Hence, the development of low footprint photonic integrated circuits (PICs) incorporating VCSELs and permitting the scaling of multiple devices into compact system architectures are key challenges for the future. In this context, the vertical-light emission of these semiconductor lasers might add extra difficulties that will need to be considered.

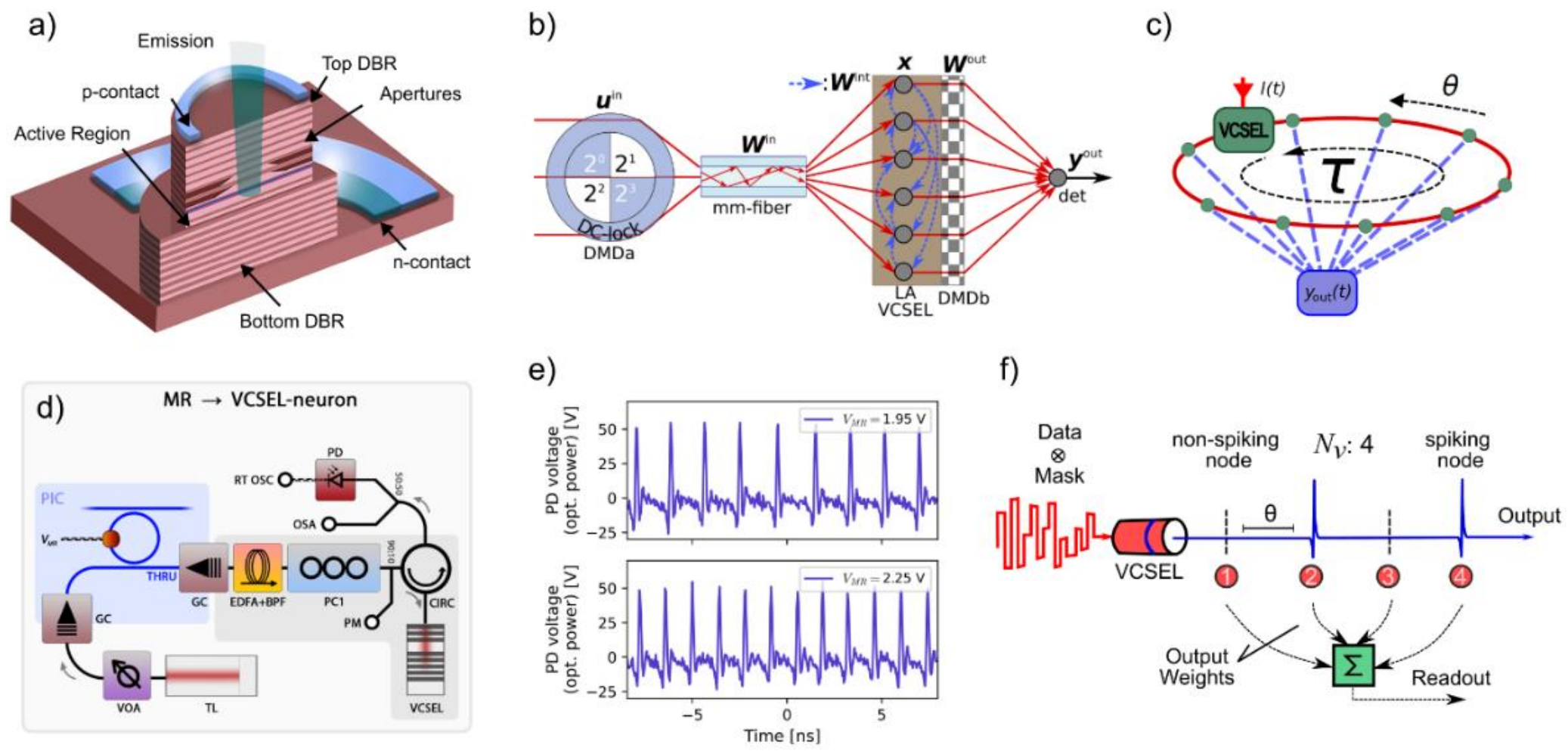

**Figure 1.** (a) Structure of a VCSEL and (b-f) exemplar VCSEL neuromorphic photonics systems; (b) Photonic RC system built with a large-area VCSEL, including a schematic of the network's architecture (adapted from [4]); (c) Schematic of a photonic time-delay RC system built using a single VCSEL (adapted from [1]); (d-e) Micro-ring resonator (MRR) coupled to a VCSEL for tuneable optical spiking rate-coding. (d) Experimental setup. (e) Time traces for two MRR bias cases yielding different spiking frequencies (adapted from [8]); (f) Schematic of a photonic spiking neural network based on a single spiking VCSEL and time-multiplexed optical inputs (adapted from [16]).

Further challenges for neuromorphic photonic systems with VCSELs might include system control and stability. For example, in current VCSEL-based photonic RC systems and spiking VCSEL optical neurons, the control of both external (e.g. optical injection strength, initial frequency detuning) and specific device parameters (e.g. applied bias current) is key. Interestingly, VCSELs enable the use of multiple magnitudes for computing, e.g. optical wavelength, light intensity, polarization, etc.; however, to access each of these, an element of programmability and/or stabilisation is required, especially as hardware-complexity and processing demands increase. There are therefore challenges in correctly designing and engineering controls to minimise adverse effects in future VCSEL-based neuromorphic computing platforms. Another potential issue to address is the development of new VCSEL structures that require optimisation for application in neuromorphic photonic functionalities. These could include for instance the development of systems with precise emission wavelength control, coupled-device architectures. Such novel optimised VCSEL systems could enable new functionalities beyond those currently directly possible with standard devices (typically optimised for communications or sensing functions).

Finally, the full potential of operating with high-speed spike-based VCSEL neuromorphic photonic platforms, while extremely promising, needs to be further determined and established. This will include the development of novel algorithms (e.g. data-encoding, spiking learning rules, etc.) that make beneficial the use of the (optical) spiking representation, and the subsequent implementation of such learning processes directly in hardware. These aspects remain a challenge for the community.

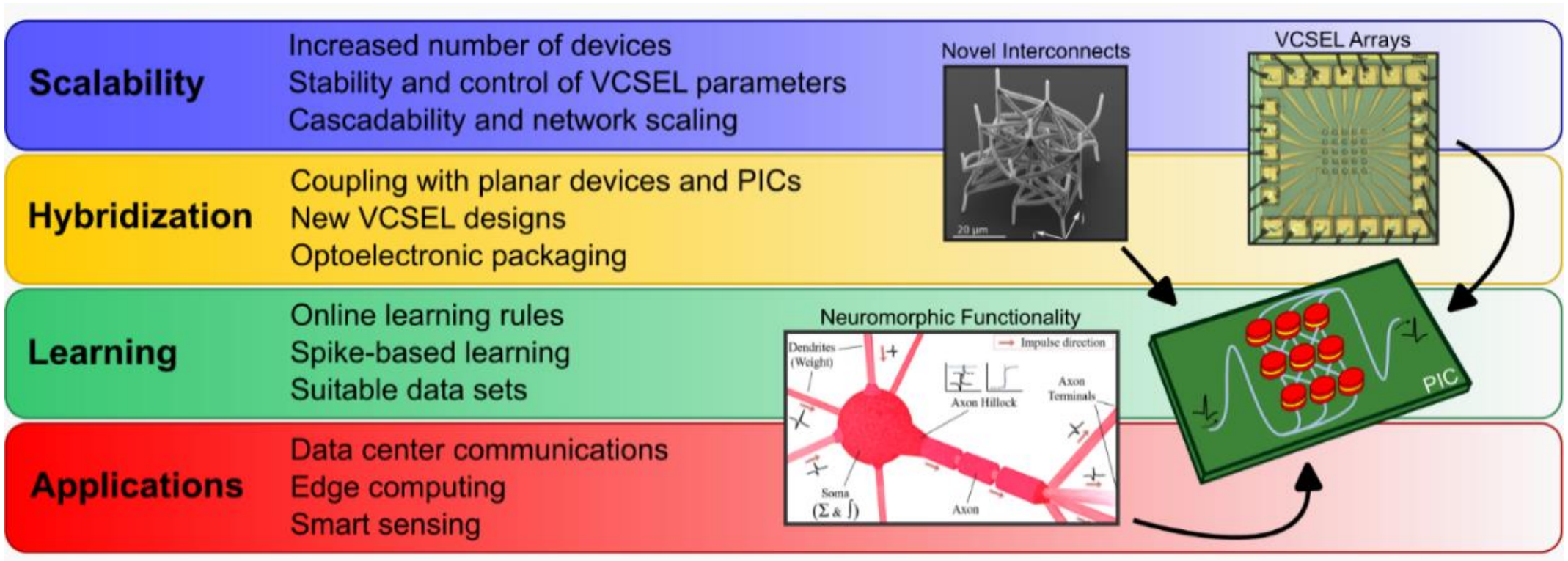

**Figure 2.** Challenges facing VCSEL-based neuromorphic photonic systems and foreseen application domains (adapted from [5,9,10]).

## Advances in Science and Technology to Meet Challenges

One approach that currently strives to increase scalability and network connectivity of VCSEL-based neuromorphic is research into photonic RC systems that use single devices (with time- or spatial-multiplexing protocols) to build full photonic neural networks. These systems enable hardware-friendly implementations with simplified controls and training requirements [2-4]. Recent works on time-multiplexed single-VCSEL RCs (Fig. 1(c)) have shown high performance at time-series prediction and classification tasks, while operating with GHz-rate inputs [2,3]. Similarly, spatially-multiplexed photonic RCs with a single multi-transverse mode large-area VCSEL (Fig. 1(b)) demonstrated successful operation in a 6-bit header recognition task whilst offering simplified in-hardware training [4].

VCSEL arrays are also being investigated for the enhancement of scalability/connectivity and system integration. Reports on 5x5 VCSEL arrays have achieved optical coupling to external lasers as well as intra-array coupling, enabling their use for spatio-temporal photonic RC [5,6]. A 5x5 VCSEL array was also used to build a homodyne-based optical neural network, with VCSELs operating as coherent transmitters and weighting elements with real-time programmability, offering increased processing performance [7]. Other efforts towards the integration of VCSELs into chip-scale neuromorphic photonic systems include the coupling of spiking VCSELs with on-chip silicon micro-ring resonators [8] (Fig. 1(d-e)), and the development of compact photonic neural networks with micropillar VCSELs interconnected by polymer waveguide architectures [9].

VCSELs are also investigated for use as optical spiking neurons and SNNs [10-20]. Experimental works demonstrated a wide range of neuronal behaviours, such as refractoriness, integrate-/resonate-and-fire, in VCSELs at high-speeds [10-12] (see also [1] and references therein), with their successful application to complex tasks, e.g. pattern recognition [10], XOR classification [13], image processing for target detection/tracking [14]. The use of VCSELs as synaptic elements for weighting and regeneration of optical spikes [15] was also reported. Furthermore, a GHz-rate photonic SNN built with a single spiking-VCSEL was recently demonstrated (Fig. 1(f)), revealing accurate performance in complex classification tasks, and operation with new spike-tailored learning procedures dramatically reducing training costs [16]. Multi-section VCSELs with saturable absorbing regions have also been investigated in theory [17][18] (but also experimentally using optically-pumped micropillar VCSELs [11]) for use as optical spiking neurons and SNNs. Recent theoretical works have also described the use of evanescently-coupled VCSELs [19] and coupled micropillar VCSEL systems [20] for communication of spiking signals that eliminate coherent optical injection requirements. Future work in VCSEL design/fabrication is needed to develop the aforementioned (and other) device structures to validate experimentally these promising theoretical results.

## Concluding Remarks

Given their unique properties (e.g. vertical emission, compactness, high-speed, efficiency) and low manufacturing costs, VCSELs have become a crucial and ubiquitously deployed photonic technology platform. VCSELs are widely used in strategic sectors ranging from sensing (e.g. LIDAR, automotive sensors), computing (e.g. optical interconnects in datacentres and computing systems) and communication technologies (e.g. optical telecom networks). There is therefore much to be gained by adding 'intelligence' and neuromorphic processing capabilities into all aforementioned VCSEL-enabled technologies. Remarkably, VCSELs can exhibit multiple nonlinear optical responses (e.g. oscillations, chaos, nonlinear switching, excitability) which have sparked increasing research interest into the development of nonlinear transformers and optical spiking neurons with VCSELs for use as nodes in photonic neural networks. Multiple VCSEL-based neuromorphic systems have been reported, including photonic spike-based processing modules and photonic RC systems, demonstrating impressive performance in complex tasks (e.g. dataset classification, time-series prediction, image/video processing). Further research is now required to tackle existing challenges regarding scalability and network connectivity, system integration and control, fabrication of optimised VCSEL designs and development of novel training protocols to ensure the future impact of the highly-promising neuromorphic photonic VCSEL platform across strategic sectors (e.g. datacentre technologies, edge-computing, sensing and communications, computing/AI hardware).

**Acknowledgements**

The authors acknowledge this work was supported by the UKRI Turing AI Acceleration Fellowships Programme (EP/V025198/1), and by the EU Pathfinder Open project 'SpikePro'.

**References**

1. *A. Skalli, J. Robertson, D. Owen-Newns, M. Hejda, X. Porte, S. Reitzenstein, A. Hurtado, and D. Brunner, "Photonic neuromorphic computing using vertical cavity semiconductor lasers", Opt. Mat. Exp., 12, 2395 (2022)*
2. *J. Vatin, D. Rontani, and M. Sciamanna, "Experimental reservoir computing using VCSEL polarization dynamics," Opt. Express, 27, 18579 (2019).*
3. *J. Bueno, J. Robertson, M. Hejda, and A. Hurtado, "Comprehensive performance analysis of a vcsel-based photonic reservoir computer," IEEE Phot. Tech. Lett., 33, 920 (2021).*
4. *X. Porte, A. Skalli, N. Haghighi, S. Reitzenstein, J.A. Lott, and D. Brunner, "A complete, parallel and autonomous photonic neural network in a semiconductor multimode laser," IOP JPhys Photonics 3, 024017 (2021).*
5. *T. Heuser, M. Pflüger, I. Fischer, J.A. Lott, D. Brunner, and S. Reitzenstein, "Developing a photonic hardware platform for brain-inspired computing based on 5×5 VCSEL arrays," IOP JPhys: Photonics 2, 044002 (2020).*
6. *M. Pflüger, D. Brunner, T. Heuser, J.A. Lott, S. Reitzenstein, and I. Fischer, "Experimental reservoir computing with diffractively coupled VCSELs", Opt. Letts., 49, 2285 (2024)*
7. *Z. Chen, A. Sludds, R. Davis III, I. Christen, L. Bernstein, L. Ateshian, T. Heuser, N. Heermeier, J.A. Lott, S. Reitzenstein, R. Hamerly, and D. Englund, "Deep learning with coherent VCSEL neural networks", Nat. Photon., 17, 723, (2023)*
8. *M. Hejda, E.A. Doris, S. Bilodeau, J. Robertson, D. Owen-Newns, B.J. Shastri, P.R. Prucnal, and A. Hurtado, "Optical spike amplitude weighting and neuromimetic rate coding using a joint VCSEL-MRR neuromorphic photonic system", IOP Neuromorph. Comp. Eng., in press (2024).*
9. *A. Grabulosa, J. Moughames, and X. Porte, M. Kadic, and D. Brunner "Additive 3D photonic integration that is CMOS compatible", Nanotechnology 34, 322002 (2023).*
10. *J. Robertson, M. Hejda, J. Bueno, and A. Hurtado, "Ultrafast optical integration and pattern classification for neuromorphic photonics based on spiking VCSEL neurons," Sci. Reps. 10, 6098 (2020).*

11. *F. Selmi, R. Braive, G. Beaudoin, I. Sagnes, R. Kuszelewicz, and S. Barbay, "Relative refractory period in an excitable semiconductor laser," Phys. Rev. Lett. 112, 183902 (2014).*

12. *A. Dolcemascolo, B. Garbin, B. Peyce, R. Veltz, and S. Barland, "Resonator neuron and triggering multipulse excitability in laser with injected signal," Phys. Rev. E 98, 062211 (2018).*

13. *Y. Zhang, S. Xiang, X. Cao, S. Zhao, X. Guo, A. Wen, and Y. Hao, "Experimental demonstration of pyramidal neuron-like dynamics dominated by dendritic action potentials based on a VCSEL for all-optical XOR classification task," Photon. Res. 9, 1055 (2021).*

14. *J. Robertson, P. Kirkland, G. Di Caterina and A. Hurtado, "VCSEL-based photonic spiking neural networks for ultrafast detection and tracking", IOP Neuromorph. Comp. Eng., 4, 014010 (2024).*

15. *J.A. Alanis, J. Robertson, M. Hejda, A. Hurtado, "Weight adjustable photonic synapse by nonlinear gain in a vertical cavity semiconductor optical amplifier", Appl. Phys. Lett., 119, 201104 (2021).*

16. *D. Owen-Newns, J. Robertson, M. Hejda, A. Hurtado, "Photonic spiking neural networks with highly efficient training protocols for ultrafast neuromorphic computing systems", Intelligent Computing 2, 0031 (2023)*

17. *M.A. Nahmias, B.J. Shastri, A.N. Tait, and P.R. Prucnal, "A Leaky Integrate-and-Fire Laser Neuron for Ultrafast Cognitive Computing," IEEE J. Sel. Top. Quantum Electron. 19(5), 1–12 (2013).*

18. *Z. Song, S. Xiang, Z. Ren, G. Han, and Y. Hao, "Spike sequence learning in a photonic spiking neural network consisting of VCSELs-SA with supervised training," IEEE J. Sel. Top. Quantum Electron. 26, 1–9 (2020).*

19. *M. Hejda, M. Vaughan, I. Henning, R. Al-Seyab, A. Hurtado and M. Adams, "Spiking Behaviour in Laterally-Coupled Pairs of VCSELs With Applications in Neuromorphic Photonics," in IEEE J. Sel. Top. Quantum Electron., 29, 1700210 (2023).*

20. *V.A. Pammi, K. Alfaro-Bittner, M.G. Clerc, and S. Barbay, "Photonic computing with single and coupled spiking micropillar lasers," IEEE J. Sel. Top. Quantum Electron. 26, 1500307 (2019).*

# Laser synchronization via programmable photonic circuits

**Jongheon Lee[1], Demetrios N. Christodoulides[1,2], Mercedeh Khajavikhan[1,2]**

[1]Ming Hsieh Department of Electrical and Computer Engineering, University of Southern California, Los Angeles, California 90089, USA

[2]Department of Physics and Astronomy, University of Southern California, Los Angeles, CA, 90089, USA

khajavik@usc.edu

## Status

Synchronization of coupled oscillator systems has a rich history, dating back to Christiaan Huygens' observation of the synchronization of pendulum clocks in 1665 [1]. Synchronization manifests across diverse scales of existence, from the coordinated firing of neurons in the brain to the orbital resonances of celestial bodies, revealing a profound underlying order that governs the complex interplay of systems in both nature and technology. In physics, a collection of independent nonlinear oscillators typically exhibits a chaotic behavior, yet, when these oscillators are interconnected, they can manifest a coherent state [2]. In the context of technological applications, one intriguing domain where synchronization of nonlinear oscillators can address critical needs is in the development of chip-scale high-power lasers. One area where synchronization of nonlinear oscillators can address critical needs is in the development of chip-scale high power laser arrays.

Recent years have witnessed a growing interest in realizing chip-scale lasers with higher radiance–defined as power per unit area per unit solid angle. To transform this goal into reality, two principally different approaches are actively being pursued. One strategy revolves around leveraging a cavity that can support a single spatially extended mode, as for example in Photonic Crystal Surface Emitting Lasers (PCSELs) [3]. An alternative approach involves establishing synchronization within an array of fully integrated coupled lasers [4]. While the former strategy has been extensively studied in the past, the latter has been gaining substantial momentum due to its potential for scaling up radiance through the introduction of additional lasing units towards the goal of 1kW single mode emission.

Typically, an array of uncoupled lasers operates incoherently, where adding more lasing elements does not enhance the radiance, but merely increases the total output power. Coupling these lasers together, for example by means of a common cavity, transforms the individual laser fields into components of supermodes, thus facilitating phase-locking and coherent beam combining. However, the simultaneous lasing of multiple supermodes can also lead to a chaotic emission due to their mutually incoherent nature. Consequently, for radiance scaling, the common cavity must be judiciously tailored to foster operation in a single supermode. While over years several strategies have been pursued for enabling single mode common cavities in free space settings [5,6], most current integrated laser arrays are based on coupling to nearest neighbors that leads to multimode lasing. Recently, inspired by the developments in topological physics, supersymmetric transformations, and non-Hermitian physics several innovative methodologies have been proposed to enable coherent beam combining of large laser arrays on chip [7-11].

## Current and Future Challenges

Neuromorphic computing and programmable photonic mesh lattices can provide a universal framework for implementing any given common cavity architecture on a chip. This approach is motivated by recent advancements in optical neural networks and photonic artificial intelligence accelerators, where programmable circuits are leveraged to implement optical deep neural networks [12,13]. Generally, synchronization tend to rely on coupling through an external linear circuit. This function can be mathematically represented by a matrix, where the diagonal elements indicate self-coupling, and the off-diagonal components signify the mutual coupling between pairs of lasers within the array. To realize this coupling matrix in its most generic form, one can employ a linear optical circuit that consists of a network of directional couplers, phase shifters, and linear gain/loss elements also known as photonic mesh lattices. The rationale behind the proposed approach is that the matrix representing the coupling between lasers can be implemented by a linear photonic circuit. Using singular value decomposition, an arbitrary matrix $M=U\Sigma V^*$ can be factorized as the product of two unitary matrices (U and V), which can be realized using an array of directional couplers and phase shifters, and a diagonal matrix ($\Sigma$) that includes output coupling [12,14]. While commercial products have been developed for photonic accelerators and neural networks, their application in laser phase locking remains unexplored.

As an example, all-to-all coupling, where each laser is uniformly coupled to every other, is a highly sought-after configuration for coherent beam combining. This arrangement leads to single-mode and in-phase operation, allowing all gain elements to contribute evenly to the total output power. Such a coupling can be realized by a unity matrix (an N-by-N matrix with identical elements). Directly implementing all-to-all coupling on a chip typically demands complex designs, including 3D patterning or intersecting waveguides [8,15]. However, a programmable photonic network offers a viable solution to achieve this coupling scheme in a two-dimensional layout [12]. Figure 1 illustrates a potential design where a series of DFB lasers are interconnected through a photonic mesh network. To facilitate power circulation in a specified direction within this network, a direction-dependent loss can be introduced into the photonic mesh network or the lasers, as shown in the upper waveguide in Fig. 1 [16].

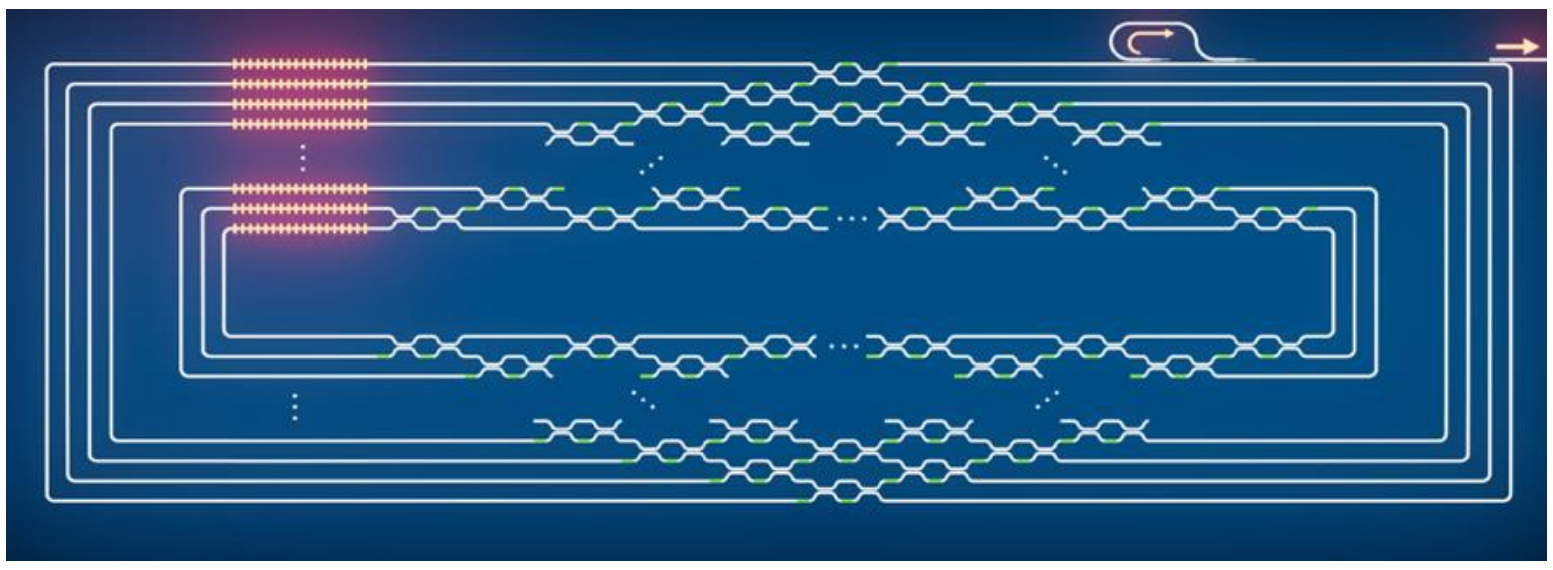

**Fig. 1**. An array of DFB lasers that are coupled through an external photonic mesh network. By adjusting the phases through micro heaters, the mesh network can be reconfigured to implement various network topologies.

The universal common cavity method holds promise for ushering a new era in laser array synchronization. It also prompts several important questions: (i) Do certain configurations of the coupling matrix better support specific synchronization behaviors in lasers? (ii) Given the coupling matrix's capacity for infinite decomposition, what criteria should guide the selection of the most effective representation? (iii) How does the optical path length within the external circuit influence the synchronization dynamics? (iv) Is it feasible to design photonic mesh networks that offer enhanced resistance to the random phase shifts experienced by lasers? (v) Are there network designs capable of enhancing error resilience? (vi) Can these mesh networks be adapted for use on active platforms? In particular, in the context of laser synchronization, the resilience of the array to random fluctuations of phases is an important property. In most cases these variations are caused by environmental factors like temperature fluctuations, but sometimes also the nonlinearities manifest themselves in the form of random phases across the array, too. Generally, under nominal phase conditions, a judiciously designed system will direct all or very large portion of the power of N lasers into a single waveguide/mode, hence increasing the radiance by a factor of N, whereas random phase fluctuations will lead to a deviation from the perfect radiance scaling condition. Previous studies have shown that it is possible to strategically add “recycling mirrors” to a common cavity in order to reduce the sensitivity to such random effects [17,18]. The recycling mirror, if properly positioned in the cavity, leads to a non-zero mode discrimination that can further enhances through the nonlinearity of the gain medium, thus leading to single mode operation. The net effect is an increase of the power of the desired mode even when the phase space is fully explored. In the integrated mesh network laser arrangement, power recycling can be done, for example, by manipulating the diagonal matrix in the SVD factorization.

## Advances in Science and Technology to Meet Challenges

To date, research on programmable photonic mesh networks has primarily focused on their application within fully integrated optical neural networks (ONNs) on silicon chips. In these setups, light is introduced into the circuit through a single, precisely tuned laser. Conversely, this current study examines DFB lasers which are influenced by feedback capable of modifying their lasing frequency through injection. It has been observed in previous research on fiber and solid-state lasers that the lasing gain elements within an external cavity navigate the frequency spectrum to lock onto a mode exhibiting minimal loss. Essentially, the photonic network plays a crucial role in setting the lasing frequency of the array upon achieving full synchronization. A key objective of our research is to explore the concept of frequency agility and assess its significance in the context of coherent beam combining. This necessitates treating the laser as a photonic entity that seeks out the mode with the lowest loss across the spectrum.

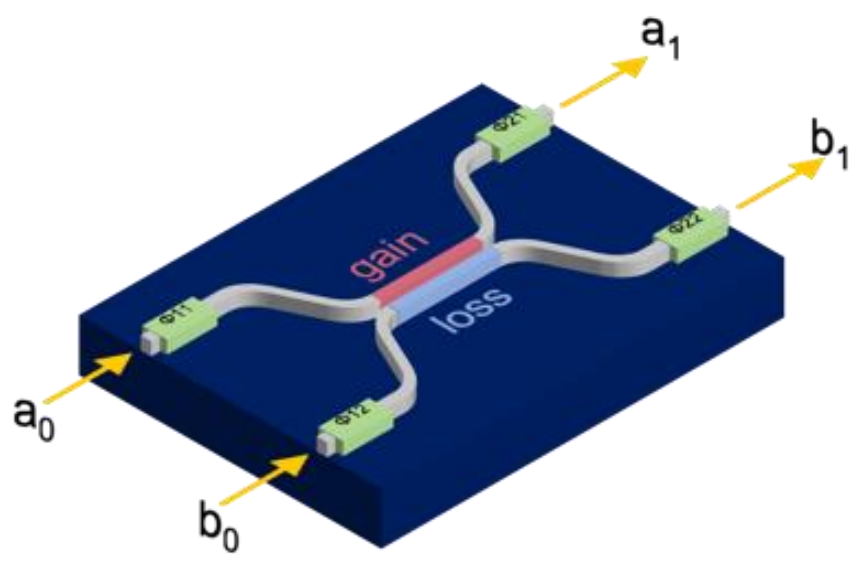

**Fig. 2** a PT symmetric directional coupler.

The prospect of replacing the intrinsically passive photonic mesh networks with active circuits. While several matrix factorization techniques exist for implementing an arbitrary matrix in terms of the product of non-Hermitian building blocks, they do not ideally map to the function of known photonic components. Nevertheless, recent numerical studies have shown that the functionality of a photonic mesh network can be delivered with an array of non-Hermitian parity-time (PT) symmetric directional couplers where the power splitting ratios are found through optimization techniques (see Fig. 2) [19]. It will be of interest to search for ways to algorithmically factorize an arbitrary matrix in terms of such non-Hermitian couplers. The non-Hermitian implementation of the photonic mesh network entails several advantages in terms of footprint, switching speed (if active control is intended), and power consumption.  To start with the state-of-the-art Joule heaters that are typically used for programming the passive mesh networks are reported to have a phase shift with a power requirement of the order of 20 mW and a switching time of a few microseconds, with the reported length of the heater to be a few hundreds of micrometers [20]. On the other hand, for the gain of 80 per cm, a PT coupler at a length of 25 µm requires ~220 µW of power to amplify a 1 mW signal. However, the average power required per PT coupler is merely ~50 µW. Even at a quantum efficiency of 10%, the required power is ~0.5 mW, which is still considerably lower than what is reported for phase shifters. Semiconductor amplifiers can also be modulated at a sub-nanosecond time scale [21]. One additional benefit of this approach is the possibility of implementing the entire laser array system using III-V semiconductor materials in a monolithic fashion. In this approach, waveguides can be realized using quantum well intermixing (QWI) methods [22], which alter the refractive index of III-V materials through inducing defects, or selective area regrowth. It should be noted that the entire network can be implemented using varying level of gain and without introducing loss. The use of non-Hermitian networks to implement the programmable photonic circuit brings about fundamental questions as what part is the laser and what is the external cavity while in turn opens avenues for controlling modal response of extremely large photonic lasers on chip.

## Concluding Remarks

Photonic mesh networks and neuromorphic computing present transformative opportunities for advancing on-chip laser brightness scaling, while simultaneously providing a deeper understanding of the fundamental mechanisms governing synchronization dynamics in networks with short to intermediate time-delay feedback. Achieving robust and efficient synchronization in such systems requires a comprehensive exploration and optimization of several critical parameters. Key challenges include mitigating the impact of design deviations on the resilience of synchronization and ensuring stable operation across various configurations.

Addressing these challenges necessitates innovative strategies, such as reintroducing a controlled portion of optical power back into the network to reinforce synchronization, incorporating synthetic gauge fields into the coupling matrix to enable precise control over the network's phase dynamics, and enhancing the frequency selectivity of the array to preferentially amplify the mode with the highest gain. These approaches not only improve the stability of laser arrays but also lay the groundwork for harnessing emergent nonlinear effects to fine-tune performance.
Such investigations are poised to significantly advance the stability, scalability, and efficiency of on-chip laser technologies, opening new avenues for their integration into high-performance photonic systems. By bridging fundamental research with practical design considerations, this work contributes to the broader effort of developing next-generation photonic platforms capable of addressing diverse challenges in science and technology.

**References**

[1] C. Huygens, *Œuvres complètes de Christiaan Huygens: Correspondance 1664-1665* (Martinus Nijhoff, 1893), Vol. 5.

[2] L. M. Pecora and T. L. Carroll, *Synchronization in chaotic systems,* Phys. Rev. Lett. 64, 8 (1990).

[3] R. Morita, T. Inoue, M. D. Zoysa, *et al., Photonic-crystal lasers with two-dimensionally arranged gain and loss sections for high-peak-power short-pulse operation,* Nat. Photonics 15.4, 311–318 (2021).

[4] A. Brignon, ed., *Coherent Laser Beam Combining* (Wiley, 2013).

[5] C. J. Corcoran and F. Durville, *Experimental demonstration of a phase-locked laser array using a self-Fourier cavity,* Appl. Phys. Lett. 86 (2005).

[6] M. Khajavikhan, J. R. Leger., *Modal Analysis of Path Length Sensitivity in Superposition Architectures for Coherent Laser Beam Combining*, Selected Topics in Quantum Electronics, IEEE 15 281(2009).

[7] M. P. Hokmabadi, N. S. Nye, R. El-Ganainy, et al., *Supersymmetric laser arrays*, Science 363, 623 (2019).

[8] Y. G. Liu, P. S. Jung, M. Parto, et al. *Gain-induced topological response via tailored long-range interactions*, Nat. Phys. 17, 704 (2021).

[9] Y. G. Liu, Y. Wei, O. Hemmatyar, et al. *Complex skin modes in non-Hermitian coupled laser arrays*, Light. Sci. & Appl. 11, 336 (2022).

[10] M. A. Bandres, S. Wittek, G. Harari, et al., *Topological insulator laser: Experiments*, Science 359, eaar4005 (2018).

[11] J. H. Choi, W. Hayenga, Y. Liu, et al., *Room temperature electrically pumped topological insulator lasers*, Nat. Commun. 12.1, 3434 (2021)

[12] Y. Shen, N. C. Harris, S. Skirlo, et al., *Deep learning with coherent nanophotonic circuits*, Nat. Photonics 11, 441 (2017).

[13] W. Bogaerts, D. Pérez, J. Capmany, et al., *Programmable photonic circuits*, Nature 586, 207 (2020).

[14] M. Reck, A. Zeilinger, H. J. Bernstein, et al., *Experimental realization of any discrete unitary operator,* Phys. Rev. Lett. 73, 58 (1994).

[15] J. Moughames, X. Porte, M. Thiel, et al., *Three-dimensional waveguide interconnects for scalable integration of photonic neural networks,* Optica 7, 640 (2020).

[16] J. Ren, Y. G. Liu, M. Parto, et al., *Unidirectional light emission in PT-symmetric microring lasers,* Opt. Express 26, 27153-27160 (2018).

[17] M. Khajavikhan and J. R. Leger, *Modal analysis of path length sensitivity in superposition architectures for coherent laser beam combining*, IEEE J. Sel. Top. Quantum Electron. 15, 281 (2009).

[18] M. Khajavikhan, K. John, and J. R. Leger, *Experimental measurements of supermodes in superposition architectures for coherent laser beam combining*, IEEE J. Quantum Electron. 46, 1221 (2010).

[19] H. Deng and M. Khajavikhan, *Parity–time symmetric optical neural networks*, Optica 8, 1328 (2021).
[20] M. Jacques, A. Samani, E. El-Fiky, et al., *Optimization of thermo-optic phase-shifter design and mitigation of thermal crosstalk on the SOI platform*, Opt. Express 27, 10456 (2019).
[21] C. Ironside, *Semiconductor integrated optics for switching light* (IOP Publishing, 2021).
[22] P. Aleahmad, M. Khajavikhan, D. N. Christodoulides, et al., *Integrated multi-port circulators for unidirectional optical information transport,* Sci. Rep. 7, 2129 (2017).

# Microcomb-based neuromorphic processing devices

**Luigi Di Lauro[1], David J. Moss[2], Roberto Morandotti[1]**

[1]Institut National de la Recherche Scientifique-Énergie Matériaux Télécommunications (IN), 1650 Boulevard Lionel Boulet, Varennes, QC J3X 1P7, Canada.

[2]Centre of Excellence in Optical Microcombs for Breakthrough Science (COMBS), Swinburne University of Technology, Hawthorn, Vic 3122, Australia.

Luigi.DiLauro@inrs.ca, Roberto.Morandotti@inrs.ca

## Status

Optical frequency combs are rapidly emerging as pivotal tools for developing new neuromorphic processing devices. They provide new means to emulate fundamental neuronal operations, such as integration and matrix multiplications, as well as to realize fully connected artificial neural networks (ANN) [1-3]. Combs are characterized by evenly spaced discrete optical frequency lines generated through mode-locking techniques in laser cavities [4,5] or by leveraging nonlinear parametric processes such as four-wave mixing, in integrated microring resonators (MRRs) [6-10. The latter provide platforms with compact footprints, minimal nonlinear absorption, and low linear loss for energy-efficient microcombs generation [11,12]

MRRs have been employed to develop optical convolutional accelerators (CAs) as preprocessing units for ANNs. Convolution is a fundamental operation in neuromorphic devices for various applications, such as image (e.g., edge detection, blurring) and signal processing (e.g., noise reduction, filtering) [1]. CAs are designed to extract meaningful information from large volumes of data, thereby reducing complexity and overall processing time, ultimately increasing processing speed. They achieve this by producing new datasets representing simplified features or patterns extracted from the original input (see Fig.1 (a) for an illustrative example of convolution operation).

For instance, CAs can be implemented using microcombs through time and wavelength interleaving techniques, using a similar method to demonstrate optical perceptrons (i.e., the artificial neurons) with MRRs [13]. The latter approach utilizes programmable optical filters to flatten the microcomb spectrum, delay individual lines, and apply the kernel multiplications to the data. At the same time, the extracted feature output is recombined via high-speed photodetection. Similarly, kernel multiplications can be implemented via MRR-based weight banks, simultaneously enabling spectrum slicing, kernel weight multiplication, and recombination.

CAs can be interfaced with multilayer perceptrons to enable the realization of fully connected microcomb-based ANNs capable of processing optical information at ultra-high speeds [1] (see Fig. 1 (b)). This results in an integrated neuromorphic device that can optically process large-scale data in parallel, achieving accuracy exceeding 90% at terabit/s speeds [14]. Their applications include ultrahigh bandwidth real-time video/image processing, LIDAR for self-driving cars, holography, automated in-vitro cell growth, and medical diagnostics.

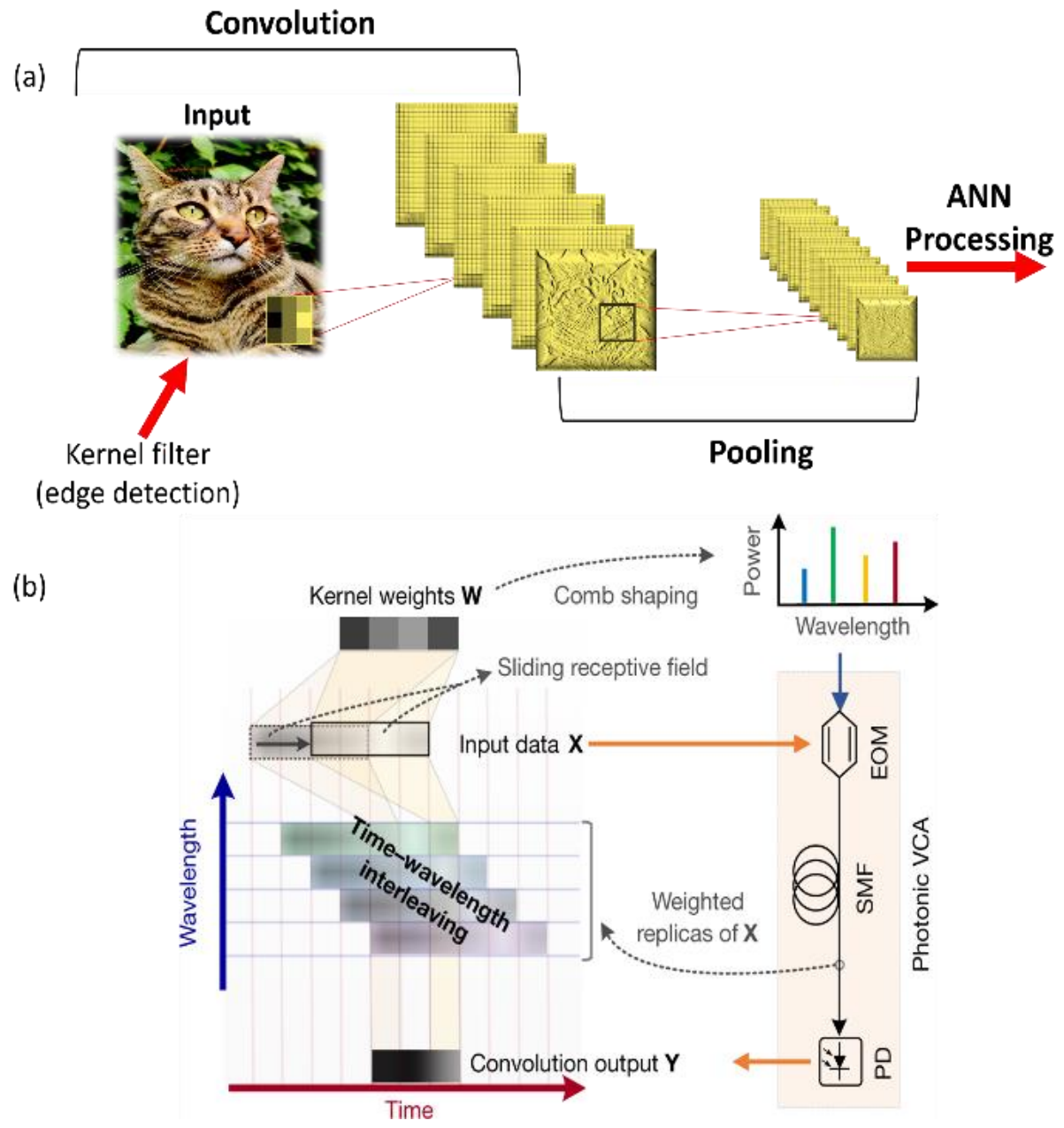

**Figure 1.** (**a**) The process begins with feature extraction (the edge of the picture in this case) using convolutional filtering on an input image of, e.g., a cat, represented as a matrix of RGB values. The kernel, utilized for feature extraction, is a lower-dimensional matrix displayed at the bottom right of the original picture, which multiplies the values of the original image matrix. Each subsequent subplot, shown in the center of (a), displays an individual feature map generated by convolving the input image with the feature detection kernel. Following the convolution operation, max pooling is applied to create a downscaled (down-sampling) version of the convolution output, effectively reducing dimension and complexity of the original image. The pooled output on the right provides a summarized version of the features extracted by the convolutional operation, which is then fed to an ANN for further processing. (**b**) The optical convolution operation is practically performed using microcombs. The convolutional kernel, represented by a weight vector W of length R, is encoded into the optical power of microcomb lines using spectral shaping. The input waveform X is simultaneously sent to different wavelengths using electro-optical modulation, creating replicas weighted by W. These replicas undergo a dispersive delay with a step equal to the symbol duration of X, achieving both time and wavelength interleaving. Finally, the delayed and weighted replicas are combined through high-speed photodetection, resulting in a convolution between X and W for a specific convolution window or receptive field in each time slot.". (b) is adapted from [1] with permission from Springer Nature.

## Current and Future Challenges

The field of neuromorphic photonics, harnessing microcombs generated within integrated platforms, is in a constant state of evolution. Despite their impressive performance, a pressing need exists to address the gap stemming from the increasing computational demands of emerging smart applications and services, such as the Internet of Things. While increasing the convolutional/processing layers and network components can potentially enhance the processing power of such neuromorphic devices, it necessarily introduces additional complexity that impacts scalability, a crucial aspect of systems integration. Moreover, further components (e.g., additional neurons) naturally impact the latency and processing speed of such devices, both of which are crucial to maintaining efficiency in processing technologies.

Considering the previously mentioned aspects, the majority of current implementations of microcomb-based CNs and ANNs rely on time and wavelength interleaving techniques. Although effective, these methods significantly increase processing latency and speed, which are further deteriorated by the several electro-optical and digital-analog conversions required for data encoding and readout.

These interrelated challenges restrict the processing capability that optics can offer for neuromorphic devices despite the potential for delivering extremely low throughput latency in the picosecond range and processing speeds exceeding terabit/s.

Researchers have explored and implemented parallelization strategies to enhance speed while ensuring low complexity and latency, including simultaneous data encoding via wavelength division multiplexing. Although this approach offers an efficient means to address these challenges, it necessitates additional encoding and readout components (such as intensity modulators, waveform generators, waveshapers, and additional detection schemes), which can be impractical. In addition, transitioning from lab and controlled experimental settings to real-world applications introduces limitations for deploying such neuromorphic photonic devices and integrating them with other systems and technologies, such as those already embedded in telecommunication infrastructures. Ensuring the resilience of microcomb sources against noise, external factors, and environmental conditions is particularly crucial for maintaining the accuracy and reliability of information processing in various applications. Moreover, controlling parameter fluctuations, such as temperature-induced changes in the refractive index of MRRs, which can affect weighting operations in CAs or modify the activation function of perceptrons in ANNs, requires real-time and precise readjustments for each component of the device to manage optical power fluctuations and maintain coherence.

**Advances in Science and Technology to Meet Challenges**

Among the various proposed approaches to addressing the challenges above, programmable photonic circuits (PICs) [15] present an efficient, reliable, and high-performing solution for developing new microcomb-based neuromorphic processing devices. This is possible by exploiting the advancements in CMOS (complementary metal-oxide-semiconductor)–silicon photonics technology, which enables extensive integration of multiple components on a single platform (chip), including optical spectral shapers, modulators, dispersive media, demultiplexers, and photodetectors.

Precise PIC parameter tuning and control can be achieved through optimization algorithms, such as particle swarm optimization or genetic algorithms (GAs) [16,17]. Drawing inspiration from evolution and natural selection, these algorithms can efficiently explore the extensive parameter space of such devices (see an example of the application of GAs in Fig. 2 (a)). Recent demonstrations have proven the effectiveness of GAs in generating and customizing stable and coherent microcombs in integrated MRRs [18] (see Fig. 2 (b)). This capability ensures adaptability and stability across diverse settings and environmental conditions, making them well-suited for next-generation neuromorphic photonic processing platforms, including CAs and (deep) ANNs [2].

Pruning techniques are increasingly being employed in the construction of new neuromorphic photonic integrated devices [19,20] to enhance the processing performance of integrated ANNs while maintaining a low-complexity network design with a sufficient number of layers. Pruning identifies and eliminates redundant or unnecessary connections between neurons without notably affecting performance. This procedure yields a more compact and computationally efficient network that enhances latency, speed, and scalability.

Moreover, the incorporation of phase-change materials (PCMs) within PICs has garnered significant interest [21,22]. These materials have demonstrated their potential as optimal candidates for implementing various functionalities, including all-optical encoding, memory storage, nonlinear activation functions, and network weighting operations. Their integration with PCIs can also mitigate bottlenecks resulting from multiple electro-optical conversions inherent in conventional encoding techniques, reducing throughput latency and increasing processing speeds up to terahertz.

PICs-PCMs with pruning and optimization algorithms offer a potential solution to the current limitations of microcomb-based neuromorphic processing technologies. Combining these methods and technologies not only improves the realization of integrated neuromorphic photonic systems but also significantly improves computational complexity, memory requirements, and inference time, enhancing suitability for deployment across resource-constrained platforms, large-scale data centers, and telecommunication infrastructures.

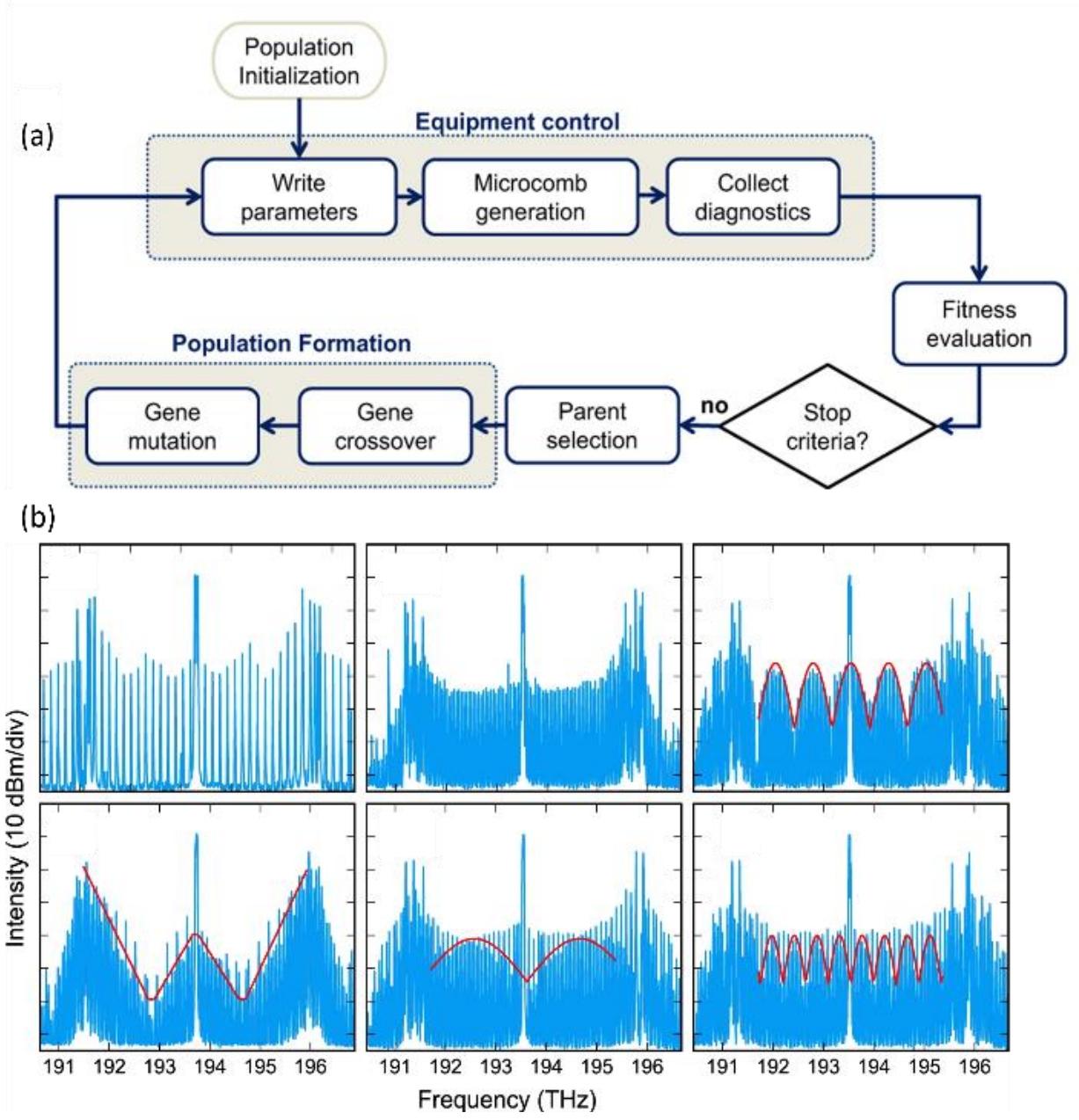

Figure 2. (**a**) Block diagram for implementing GAs aimed at microcomb generation and tailoring within MRRs. The optimization procedure starts with a population of individuals, denoted as P, where each individual is represented by a vector containing parameters to be optimized, commonly referred to as genes. The algorithm unfolds through two primary iterative steps: i) Evaluation of each individual's performance using a fitness function that integrates various objectives and constraints concerning to specific features of the microcomb. ii) Selection of individuals after crossover and mutation operations to form a new population with potentially enhanced genes, a process known as population formation. This iterative cycle continues with the fitness of the new population assessed at each step. The algorithm stops upon identifying the best population, achieving the maximum fitness function value after a predefined number of iterations. Ultimately, this iterative process yields the optimized set of setup parameters necessary to attain the desired state of the target comb. (**b**) Different microcombs achieved experimentally in a MRR through GAs by imposing specific requirements on line spacing and spectral envelope obtaining the desired output. (a) and (b) are adapted from [18], with permission from Springer Nature.

## Concluding Remarks

In the rapidly evolving landscape of neuromorphic photonic processing devices, microcomb-based technologies are contributing to revolutionizing data processing for smart applications. MRRs provide novel means to realize nonlinear activation functions and emulate fundamental neuronal operations by leveraging the unique properties of optical frequency combs. This, in turn, has paved the way for the development of highly efficient optical CAs and ANNs capable of handling large amounts of data generated by new services (e.g., the Internet of Things). However, the growing performance demands of such applications introduce new stringent requirements regarding speed, latency, and scalability.

To address these challenges, recent advancements in PCIs and PCMs, coupled with pruning techniques, offer alternatives to realizing novel microcomb-based neuromorphic devices. Furthermore, employing evolutionary algorithms enables autonomous optimization of device operation across various settings. These emerging methodologies provide precise control and tunability, which are crucial for adapting to dynamic environmental conditions and enhancing network performance while tackling scalability and advancing system integration. Ultimately, this can lead to innovative neuromorphic photonic technologies that meet increasing performance demands and contribute to environmental sustainability across various domains, including telecommunications, medical diagnostics, and computer vision, ushering in a new era of optical data processing.

## Acknowledgements

This work was supported by the Natural Sciences and Engineering Research Council of Canada (NSERC) through the Alliance and Discovery Grants Schemes, by the MESI PSR-SIIRI Initiative in Quebec, by the Canada Research Chair Program, and by the Mitacs Elevate Postdoctoral Fellowship Program, as well as the Australian Research Council Centre of Excellence in Optical Microcombs for Breakthrough Science (COMBS) CE230100006.

**References**

[1] X. Xu *et al.*, "11 TOPS photonic convolutional accelerator for optical neural networks," *Nature*, vol. 589, no. 7840, pp. 44–51, 2021.

[2] S. Xia *et al.*, "Deep-learning-empowered synthetic dimension dynamics: morphing of light into topological modes," *Adv. Photonics*, vol. 6, no. 2, p. 026005, 2024.

[3] B. Fischer *et al.*, "Neuromorphic Computing via Fission-based Broadband Frequency Generation," *Adv. Sci.*, vol. 10, no. 35, p. 2303835, 2023.

[4] A. Aadhi *et al.*, "Mode-locked laser with multiple timescales in a microresonator-based nested cavity," *APL Photonics*, vol. 9, no. 3, p. 031302, 2024.

[5] M. Peccianti *et al.*, "Demonstration of a stable ultrafast laser based on a nonlinear microcavity," *Nat. Commun.*, vol. 3, no. 1, p. 765, 2012.

[6] A. Pasquazi *et al.*, "Micro-combs: A novel generation of optical sources," *Phys. Rep.*, vol. 729, pp. 1–81, 2018.

[7] H. Bao *et al.*, "Laser cavity-soliton microcombs," *Nat. Photonics*, vol. 13, no. 6, pp. 384–389, 2019.

[8] M. Rowley *et al.*, "Self-emergence of robust solitons in a microcavity," *Nature*, vol. 608, no. 7922, pp. 303–309, 2022.

[9] M. Monika *et al.*, "Quantum state processing through controllable synthetic temporal photonic lattices," *Nat. Photonics* (2024). doi:10.1038/s41566-024-01546-4.

[10] H. Yu, *et al.*, "Quantum key distribution implemented with d-level time-bin entangled photons," *Nat. Commun.* vol. 16, no. 171 (2025).

[11] D. J. Moss, R. Morandotti, A. L. Gaeta, and M. Lipson, "New CMOS-compatible platforms based on silicon nitride and Hydex for nonlinear optics," *Nat. Photonics*, vol. 7, no. 8, pp. 597–607, 2013.

[12] L. Di Lauro *et al.*, "Parametric control of thermal self-pulsation in micro-cavities," *Opt. Lett.*, vol. 42, no. 17, pp. 3407–3410, 2017.

[13] X. Xu *et al.*, "Photonic perceptron based on a Kerr Microcomb for high-speed, scalable, optical neural networks," *Laser Photonics Rev.*, vol. 14, no. 10, p. 2000070, 2020.

[14] M. Tan *et al.*, "Photonic signal processor based on a Kerr microcomb for real-time video image processing," *Commun. Eng.*, vol. 2, no. 1, pp. 1–13, 2023.

[15] W. Bogaerts *et al.*, "Programmable photonic circuits," *Nature*, vol. 586, no. 7828, pp. 207–216, 2020.

[16] B. Wetzel *et al.*, "Customizing supercontinuum generation via on-chip adaptive temporal pulse-splitting," *Nat. Commun.*, vol. 9, no. 1, pp. 1–10, 2018.

[17] B. Fischer *et al.*, "Autonomous on-chip interferometry for reconfigurable optical waveform generation," *Optica*, vol. 8, no. 10, p. 1268, 2021.

[18] Celine Mazoukh *et al.*, "Genetic algorithm-enhanced microcomb state generation. " vol. 7, no. 81, pp. 1-11, 2024.

[19] X. Geng, J. Gao, Y. Zhang, and D. Xu, "Complex hybrid weighted pruning method for accelerating convolutional neural networks," *Sci. Rep.*, vol. 14, no. 1, p. 5570, 2024.

[20] Y. Shi *et al.*, "Neuroinspired unsupervised learning and pruning with subquantum CBRAM arrays," *Nat. Commun.*, vol. 9, no. 1, p. 5312,. 2018.

[21] P. Prabhathan *et al.*, "Roadmap for phase change materials in photonics and beyond," *iScience*, vol. 26, no. 10, p. 107946, 2023.

[22] W. Zhang, R. Mazzarello, M. Wuttig, and E. Ma, "Designing crystallization in phase-change materials for universal memory and neuro-inspired computing," *Nat. Rev. Mater.*, vol. 4, no. 3, pp. 150–168, 2019.

# Ultrafast Neural Networks using Lithium Niobate Nonlinear Photonics

**Gordon H.Y. Li**, Department of Applied Physics, California Institute of Technology, Pasadena 91125, CA, USA, E-mail: ghli@caltech.edu
**Alireza Marandi**, Department of Applied Physics, California Institute of Technology, Pasadena 91125, CA, USA; and Department of Electrical Engineering, California Institute of Technology, Pasadena 91125, CA, USA, E-mail: marandi@caltech.edu.

## Status

In recent years, the use of optics and photonics in special-purpose hardware accelerators for neural networks has experienced increased prevalence [1]. There have been promising efforts towards accelerating linear operations in neural networks, such as matrix multiplications [2] and convolutions [3], directly in the optical domain. However, current optical neural networks (ONNs) are largely bottlenecked by costly optoelectronic conversions or slow optical nonlinearities for performing neuron activations [4]. This prevents ONNs from harnessing the full optical bandwidth and potential ultrafast operation available to light-based computing. To overcome this bottleneck, recent advances in lithium niobate photonics [5] have enabled unprecedentedly strong and ultrafast optical nonlinearities, which presents an exciting opportunity for the next generation of truly end-to-end and all-optical neural networks with exceptionally high computational clock rates.

Lithium niobate has long been a workhorse material for telecommunications modulators [6] due to its large electro-optic coefficient and wide transparency window at optical wavelengths. More recently, it has found applications in computing due to its strong and near-instantaneous $\chi^{(2)}$ optical nonlinearity. This enables efficient parametric processes such as second harmonic generation (SHG) and degenerate optical parametric amplification (DOPA). In addition, lithium niobate is ferroelectric, which allows for more flexible quasi-phase matching of $\chi^{(2)}$ processes at a wide range of desired wavelengths via periodically poled lithium niobate (PPLN). Early work on coherent Ising machines (CIMs), which can be considered as special types of recurrent neural networks for combinatorial optimization, utilized DOPA in free-space PPLN crystals [7, 8]. The performance and scalability of CIMs was later improved by using an optical fiber platform with DOPA in weakly-guiding PPLN waveguides [9-11] albeit using digital electronic matrix multiplication. The system was also further extended to implement spiking neural networks [12, 13].

However, these free space and weakly guiding PPLN devices require relatively high pulse energies for nonlinear operations due to poor spatio-temporal light confinement and lack of dispersion engineering. A critical breakthrough was the introduction of low loss and strongly guiding PPLN nanophotonic waveguides in a thin-film lithium niobate platform [5]. The nanoscale dimensions greatly enhance the light intensity and allow for dispersion engineering of PPLN waveguides to achieve ultrabroadband and efficient $\chi^{(2)}$ processes by utilizing femtosecond laser pulses. Record SHG efficiency ($> 2600\%/W/cm^2$) [14] and DOPA gain ($> 50\,dB/cm$) [15] with large phase-matching bandwidth ($> 10\,THz$) were demonstrated, which enable the possibility of terahertz clock rates that were previously unattainable for ONNs with reasonable energies.

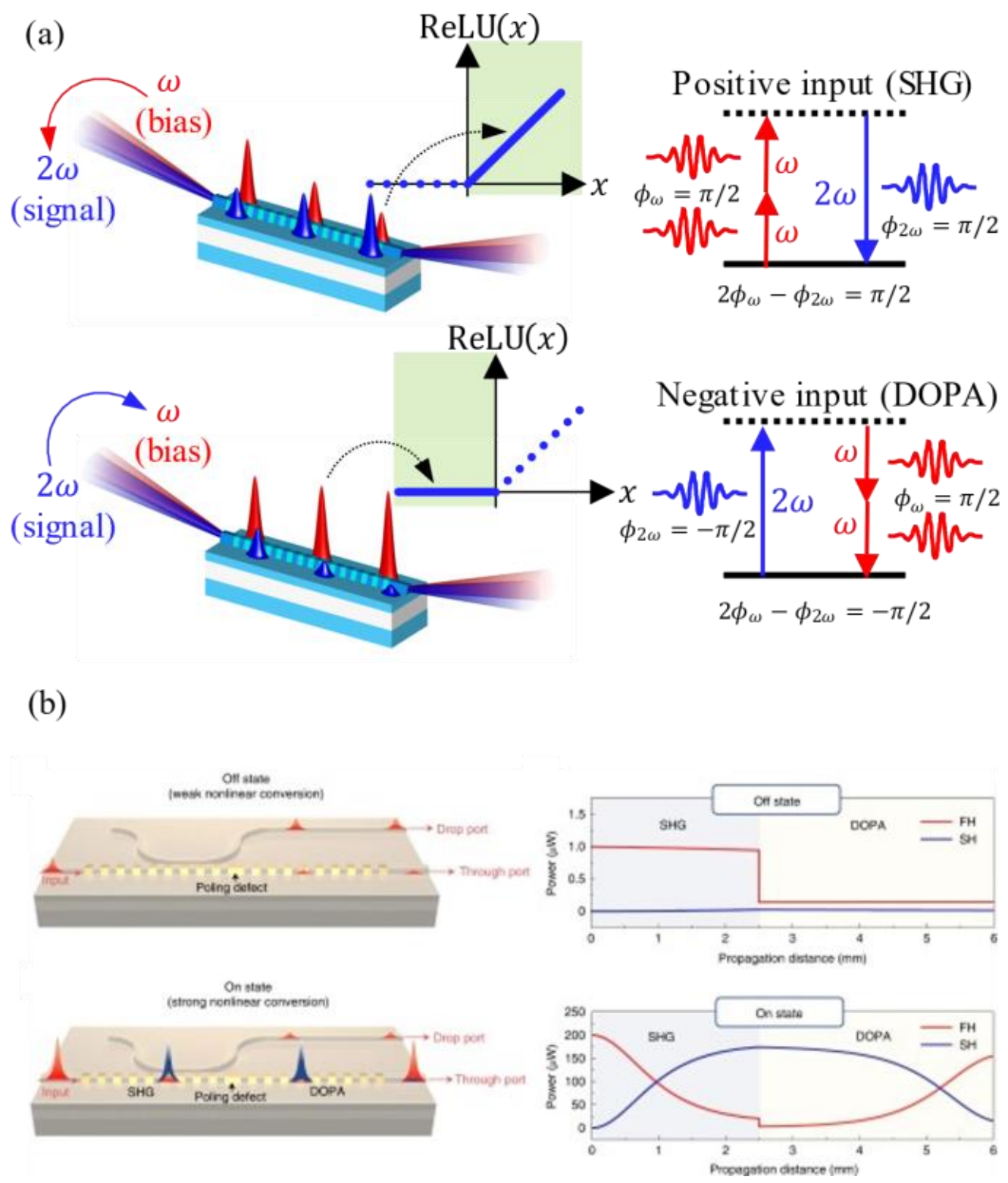

**Figure 1.** PPLN nanophotonic waveguides implementing important all-optical neural network functionalities based on nonlinear optical processes such as (a) ReLU nonlinear activation function for deep learning and (b) switching or modulation for signal routing in neural networks. Image (a) Adapted from [4] and image (b) adapted from [16]

## Current and Future Challenges

Current research focuses on how to harness the strong and ultrafast nonlinearity of PPLN nanophotonic waveguides to develop the fundamental devices and building-blocks needed for ONNs. For example, it was demonstrated that the rectified linear unit (ReLU) function $ReLU(x) = (0, x)$, which is the most widely used nonlinear activation function for deep neural networks, can be implemented by carefully tuning the interplay of SHG and DOPA between femtosecond laser pulses as shown in Fig. 1(a) [4]. This approach achieved an energy per activation of $16\ fJ$ and time per activation of $75\ fs$ (equivalent to $> 13\ THz$ maximum clock rate), hence achieving a record energy-time product per activation of $1.2 \times 10^{-27} J\ s$ among any other hardware implementations. Another important functionality demonstrated using PPLN nanophotonic waveguides was all-optical switching [16]. This is necessary for the ultrafast modulation and routing of optical signals between neurons for synaptic connections, and the multiplexing/demultiplexing of ultrafast input/output signals. The operating principle was based on introducing a poling defect to engineer the interplay between SHG and DOPA, as shown in Fig. 1(b), and achieved a state-of-the-art energy-time product of $3.7 \times 10^{-27} J\ s$.

The main future challenge will be to translate these record-breaking device-level metrics and potential terahertz clock rates to system-level ONN architectures. As such, it appears inefficient to simply use existing neural network architectures that were optimized for digital electronics due to the vastly different operating regime compared to analog optics. One key issue will be how to deal with noise and imperfections that are far lower in digital electronics. This will require innovations in noise-robust neural network architectures and training algorithms tailored for optics [17]. Another issue is the unfavourable size of the nonlinear photonic devices (micro- to milli-meter scale) compared to nanoscale transistors and the limited number of photonic components ($< 10^3$) that can currently be integrated on-chip in thin-film lithium niobate, which is far less than in electronics ($> 10^{10}$). One effective way to address this limitation is through time-multiplexing, which introduces virtual neurons via time-delayed feedback of a single nonlinear node. This can greatly reduce the overall hardware complexity, whilst preserving the advantages of ultrafast computational clock rates. An early proof-of-concept demonstration in an optical fiber platform shows that photonic neural cellular automata using SHG is a promising candidate for an ONN architecture that best harnesses the $\chi^{(2)}$ nonlinear optical processes for neural network computations such as image classification tasks [18].

## Advances in Science and Technology to Meet Challenges

Moving towards an all-optical neural network with ultrafast computational clock rates will require the monolithic integration of various components on chip in thin-film lithium niobate including PPLN sections, electro-optic modulators, optical cavities, couplers, and ultrashort pulse light sources. Each of these components have been demonstrated separately [5], however, the integration of many components to form large-scale circuits will require advances in current fabrication techniques to yield a functioning system-on-chip with competitive computing performances. Furthermore, solutions for co-packaging and control of both slow and high-speed RF electronics will need to be developed. Methods for active stabilization will need to be employed to ensure the correct long-term operation of the neural network.

An example of what we envision a feasible and near-term time-multiplexed all-optical neural network to look like is shown in Fig. 2. It consists of two synchronously-pumped pulsed optical parametric oscillators with different cavity lengths and high-speed programmable couplings between the cavities, which enables both linear and nonlinear operations to be performed by properly controlling the pulses. This architecture enables ultrafast neural network computations that can accommodate an arbitrary number of neurons $N$ whilst using only a constant number of hardware components and achieve fully programmable all-to-all connections requiring $O(N)$ cavity roundtrips.

Finally, more work is needed to better understand the computational power and complexity of neural networks based on $\chi^{(2)}$ optical nonlinearities, such as networks of coupled optical parametric oscillators. A clear picture of the algorithmic pitfalls and advantages of strong and ultrafast $\chi^{(2)}$ nonlinearities will help inform the optimal choice of neural network architecture. This will also aid in the identification of both new and existing applications that can most benefit from all-optical neural networks with ultrafast clock rates.

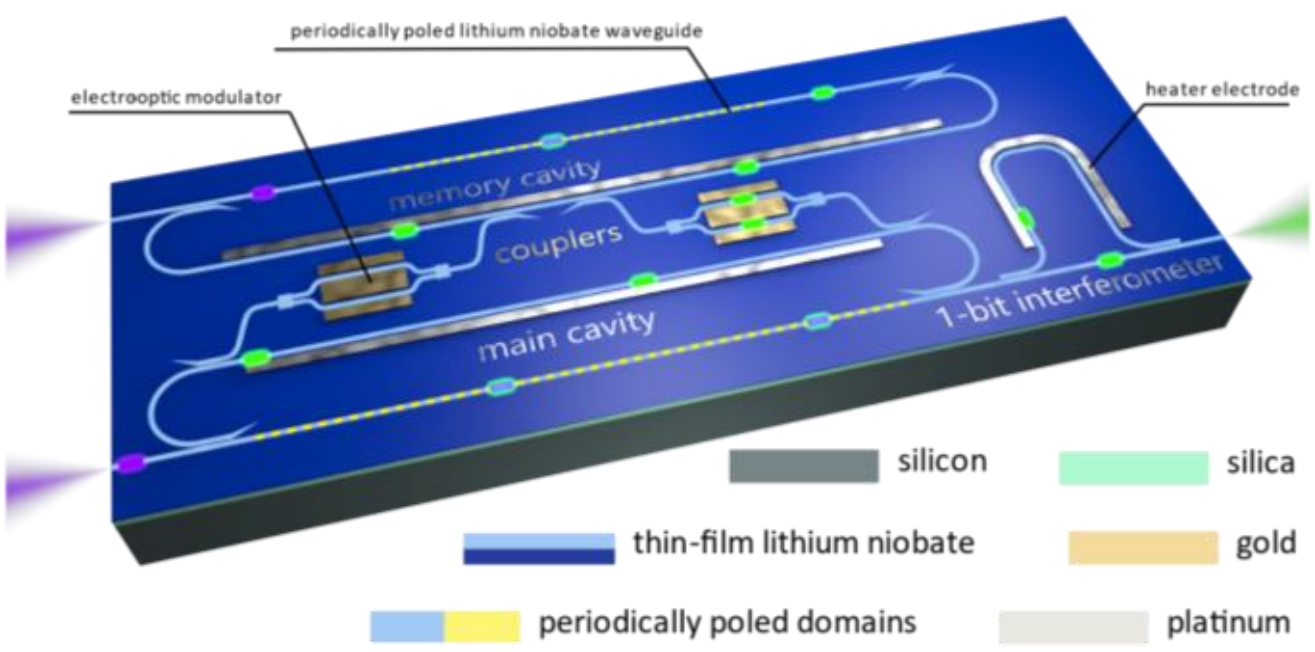

Figure 2. A schematic for a potential time-multiplexed all-optical neural network monolithically integrated on-chip in thin-film lithium niobate, combining various devices such as PPLNs, electro-optic modulators, optical cavities, and couplers.

## Concluding Remarks

In conclusion, we have summarized the history, current status, and future challenges of using $\chi^{(2)}$ optical nonlinearities in lithium niobate towards ultrafast neural network computations. This nascent field is rapidly developing owing to recent breakthroughs in lithium niobate nanophotonics at the device-level. Future work advancing system-level architectures will enable the next generation of end-to-end and all-optical neural networks with beyond terahertz computational clock rates.

### Acknowledgements

The authors acknowledge support from ARO grant no. W911NF-23-1-0048, NSF grant no. 1846273 and 1918549, and Center for Sensing to Intelligence at Caltech. G.H.Y.L acknowledges support from the Quad Fellowship.

## References

[1] X. Sui, Q. Wu, J. Liu, Q. Chen, and G. Gu, "A review of optical neural networks," IEEE Access, vol. 8, pp. 70 773–70 783, 2020.

[2] Y. Shen, N. C. Harris, S. Skirlo, M. Prabhu, T. Baehr-Jones, M. Hochberg, X. Sun, S. Zhao, H. Larochelle, D. Englund, and M. Soljačić, "Deep learning with coherent nanophotonic circuits," *Nature Photonics*, vol. 11, no. 7, pp. 441–446, 2017.

[3] J. Feldmann, N. Youngblood, M. Karpov, H. Gehring, X. Li, M. Stappers, M. Le Gallo, X. Fu, A. Lukashchuk, A. S. Raja, J. Liu, C. D. Wright, A. Sebastian, T. J. Kippenberg, W. H. P. Pernice, and H. Bhaskaran, "Parallel convolutional processing using an integrated photonic tensor core," *Nature*, vol. 589, no. 7840, pp. 52–58, 2021.

[4] G. H.Y. Li, R. Sekine, R. Nehra, R. M. Gray, L. Ledezma, Q. Guo, and A. Marandi, "All-optical ultrafast ReLU function for energy-efficient nanophotonic deep learning," *Nanophotonics*, vol. 12, no. 5, pp. 847–855, 2022.

[5] D. Zhu, L. Shao, M. Yu, R. Cheng, B. Desiatov, C. Xin, Y. Hu, J. Holzgrafe, S. Ghosh, A. Shams-Ansari, E. Puma, N. Sinclair, C. Reimer, M. Zhang, and M. Loncar, "Integrated photonics on thin-film lithium niobate," *Advances in Optics and Photonics*, vol. 13, no. 2, pp. 242–352, 2021.

[6] C. Wang, M. Zhang, X. Chen, M. Bertrand, A. Shams-Ansari, S. Chandrasekhar, P. Winzer, and M. Loncar, "Integrated lithium niobate electro-optic modulators operating at CMOS-compatible voltages," *Nature*, vol. 562, no. 7725, pp. 101–104, 2018.

[7] A. Marandi, Z. Wang, K. Takata, R. L. Byer, and Y. Yamamoto, "Network of time-multiplexed optical parametric oscillators as a coherent Ising machine," *Nature Photonics*, vol. 8, no. 12, pp. 937–942, 2014.

[8] K. Takata, A. Marandi, R. Hamerly, Y. Haribara, D. Maruo, S. Tamate, H. Sakaguchi, S. Utsunomiya, and Y. Yamamoto, "A 16-bit coherent Ising machine for one-dimensional ring and cubic graph problems," *Scientific Reports*, vol. 6, no. 1, p. 34089, 2016.

[9] P. L. McMahon, A. Marandi, Y. Haribara, R. Hamerly, C. Langrock, S. Tamate, T. Inagaki, H. Takesue, S. Utsunomiya, K. Aihara et al., "A fully programmable 100-spin coherent Ising machine with all-to-all connections," *Science*, vol. 354, no. 6312, pp. 614–617, 2016.

[10] T. Inagaki, Y. Haribara, K. Igarashi, T. Sonobe, S. Tamate, T. Honjo, A. Marandi, P. L. McMahon, T. Umeki, K. Enbutsu et al., "A coherent Ising machine for 2000-node optimization problems," *Science*, vol. 354, no. 6312, pp. 603–606, 2016.

[11] T. Honjo, T. Sonobe, K. Inaba, T. Inagaki, T. Ikuta, Y. Yamada, T. Kazama, K. Enbutsu, T. Umeki, R. Kasahara et al., "100,000-spin coherent Ising machine," Science Advances, vol. 7, no. 40, p. eabh0952, 2021.

[12] T. Inagaki, K. Inaba, T. Leleu, T. Honjo, T. Ikuta, K. Enbutsu, T. Umeki, R. Kasahara, K. Aihara, and H. Takesue, "Collective and synchronous dynamics of photonic spiking neurons," *Nature Communications*, vol. 12, no. 1, p. 2325, 2021.

[13] T. Makinwa, K. Inaba, T. Inagaki, Y. Yamada, T. Leleu, T. Honjo, T. Ikuta, K. Enbutsu, T. Umeki, R. Kasahara, K. Aihara, and H. Takesue, "Experimental observation of chimera states in spiking neural networks based on degenerate optical parametric oscillators," *Communications Physics*, vol. 6, no. 1, p. 121, 2023.

[14] C. Wang, C. Langrock, A. Marandi, M. Jankowski, M. Zhang, B. Desiatov, M. M. Fejer, and M. Loncar, “Ultrahigh-efficiency wavelength conversion in nanophotonic periodically poled lithium niobate waveguides,” *Optica*, vol. 5, no. 11, pp. 1438–1441, 2018.
[15] L. Ledezma, R. Sekine, Q. Guo, R. Nehra, S. Jahani, and A. Marandi, “Intense optical parametric amplification in dispersion-engineered nanophotonic lithium niobate waveguides,” *Optica*, vol. 9, no. 3, pp. 303–308, 2022.
[16] Q. Guo, R. Sekine, L. Ledezma, R. Nehra, D. J. Dean, A. Roy, R. M. Gray, S. Jahani, and A. Marandi, “Femtojoule femtosecond all-optical switching in lithium niobate nanophotonics,” *Nature Photonics*, vol. 16, no. 9, pp. 625–631, 2022.
[17] L. G. Wright, T. Onodera, M. M. Stein, T. Wang, D. T. Schachter, Z. Hu, and P. L. McMahon, “Deep physical neural networks trained with backpropagation,” *Nature*, vol. 601, no. 7894, pp. 549–555, 2022.
[18] G. H.Y. Li, C. R. Leefmans, J. Williams, R. M. Gray, M. Parto, and A. Marandi, “Deep learning with photonic neural cellular automata,” *arXiv preprint* arXiv:2309.13186, 2023.

# Nanoprinted Neural Networks

**Elena Goi[1,2], Steffen Schoenhardt[1,2], Min Gu[1,2]**
[1]School of Artificial Intelligence Science and Technology, University of Shanghai for Science and Technology, 516 Jungong Road, Shanghai 200093, PR China
[2]Institute of Photonic Chips, University of Shanghai for Science and Technology, 516 Jungong Road, Shanghai 200093, PR China
[elenagoi@usst.edu.cn, steffenschoenhardt@usst.edu.cn, gumin@usst.edu.cn]

## Status

Nanoprinting, the additive manufacturing of free-form structures with nanometre feature sizes using lithographic methods based on two-photon polymerisation[1], is now a mature fabrication technology, capable of producing structures of great complexity in a single fabrication step and with resolutions sufficient for metasurfaces in the NIR and visible wavelength regimes[2]. This allows for the design of 3-Dimensional optical elements not feasible with traditional subtractive or 2-Dimensional lithographic methods and is compatible with fabrication on near arbitrary substrates including commercial CMOS imaging sensors.
Two distinct implementations of Neural Networks fabricated with nanoprinting methods have emerged in the recent years. On one hand, we have nanoprinted holographic Neural Networks, which utilize the interaction of light fields with thin scatterers to realise optical convolutions and matrix multiplications in free space. By employing a custom nanoprinting system with galvo-dithered symmetry correction of the writing voxel, Goi et. al. achieved multi-layered holographic perceptrons[3,4], where the neurons have a lateral diameter of down to 400 nm and axial resolution as high as 10 nm. In this way, they realise compact optical Neural Networks with neuron densities of $6.25*10^8$ neurons/cm$^2$ directly printed on commercial CMOS chips (**Figure 1a**, **1b**). On the other hand, nanoprinting also enables the realisation of complex 3D-routed waveguide based photonic interconnects that define an optical signal path in a volume such that it directly resembles the connections of dendrites and axons in the brain. A fractal topology of connected waveguides with diameters of approximately 1 µm was shown by Moughames et. al.[5], implementing compact parallel couplers with a single input and up to 81 output channels as bifurcation layer, as well as HAAR filters arranged in large arrays for spatial filtering in connecting layers of convolutional Neural Networks (**Figure 1c**). Yu et al. created an 8-point Euclidean Steiner tree network of waveguides with smallest feature size of ~200 nm based on a branching axon and dendrite optimization approach[6] (**Figure 1d**). These 3D networks were demonstrated to host topological nontrivial Dirac-like conical dispersion in their photonic band structure, which the authors envision to employ as platform for optical convolutional Neural Networks.
This development shows that nanoprinting, while initially being employed to facilitate miniaturisation and integration, is evolving into a tool that enables research towards harvesting the complex physics of light in compact neuromorphic photonic systems.

## Current and Future Challenges

- *Error-prone performance*

The performance of nanoprinted holographic networks strongly depends on the precise fabrication of their diffractive elements[7]. The most critical parameter for the performance of the network is thereby the diameter of the diffractive neuron, which at the same time is the parameter most difficult to precisely control[8]. Similarly, for nanoprinted networks of waveguides it is difficult to precisely maintain the diameter of the waveguides over large lengths. Further, due to their high aspect ratios, nanoprinted waveguide networks are prone to deform during development.

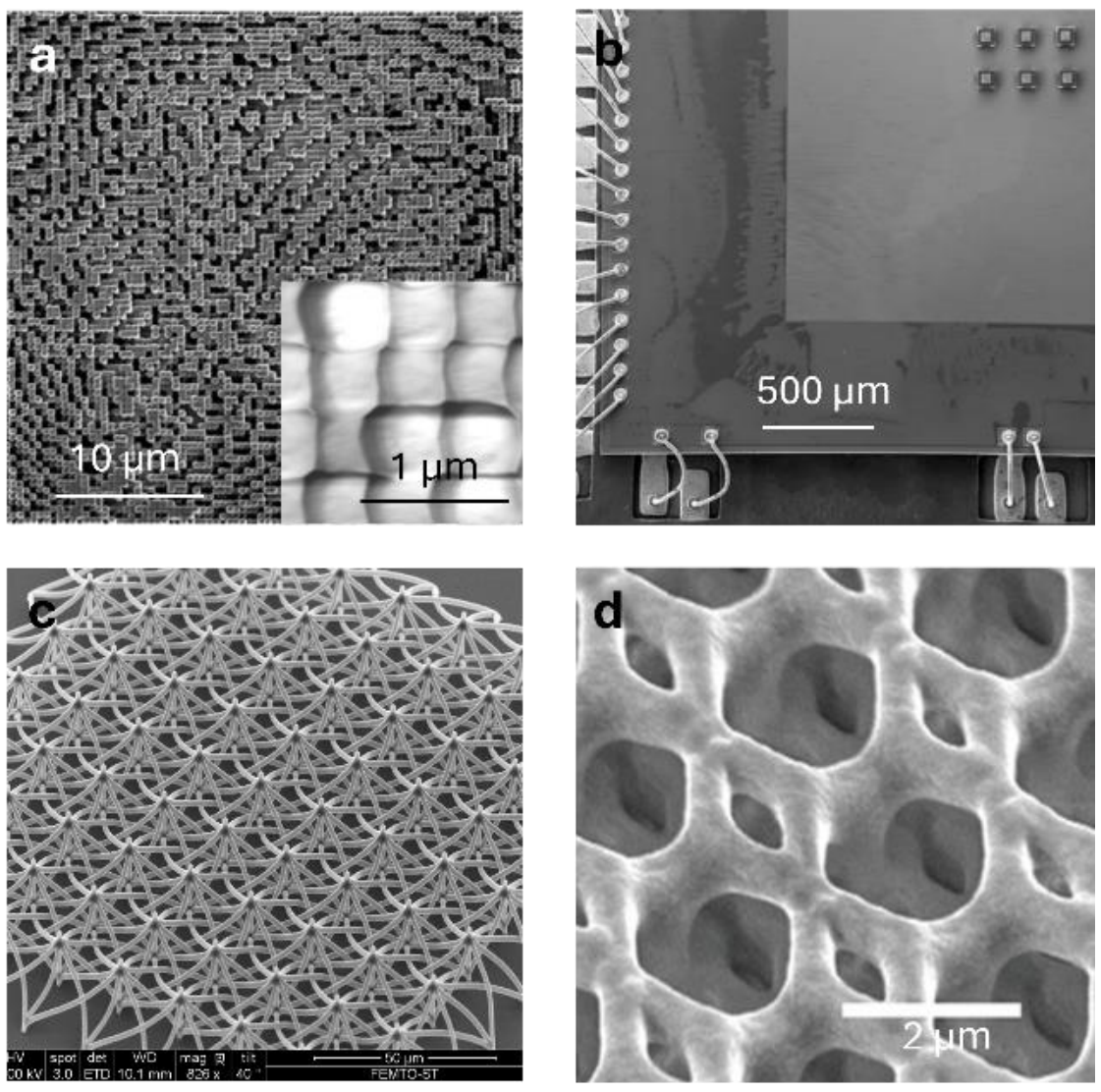

**Figure 1. a**) Scanning electron microscopy (SEM) image of a nanoprinted holographic perceptron. The insert shows an AFM topographical image of a section. **b**) SEM image of a commercial CMOS sensor with an array of holographic perceptrons nanoprinted on the surface. Adapted with permission from Ref.[3]. **c**) SEM micrograph of an array of 1×9 couplers hosting nine elements, including an inset with higher magnification Adapted with permission from Ref.[5]. **d**) SEM image of nanoprinted Steiner tree network. The scale bars are 5 µm. Adapted with permission from Ref.[6].

- *Static elements*

Imperfections during fabrication are common to impair the performance of optical networks of all sorts and are typically addressed by dynamical control of parameters determining their behaviour. For 3D nanoprinted Neural Networks there are currently no established means to tune the behaviour of the printed network after fabrication. Equivalently, dynamic reconfiguration of the nanoprinted Neural Networks is not possible with current methods.

- *Limited models for nanophotonics*

Nanoprinted Neural Networks are typically modelled in-silico, i.e. on a computer that calculates the propagation of light through the network with idealised and strongly simplified models[5,9]. These models are based on interactions of light fields with homogenous linear media. There are currently no means to efficiently model more complex problems, like the propagation of optical fields through an inhomogeneous nonlinear layer or near-field effects in compact nanoprinted systems, which are compatible with the iterative training approaches common to the design of Neural Networks.

- *Scalable optical nonlinear hidden layers*

For complex decision making in Neural Networks, the implementation of a nonlinear thresholding function is essential. While there is a range of optical nonlinear effects that can be observed in a wide range of settings, a notable nonlinear effect typically requires high signal intensities or long propagation lengths. This is not compatible with polymer-based nanoprinted Neural Networks, which typically do not exceed a volume of 300x300x300 $\mu m^3$. Further, the lack of efficient tools to physically accurately model the interaction of optical nonlinear materials with light in the near field, prohibits the implementation in a static nanoprinted network.

**Advances in Science and Technology to Meet Challenges**

- *In operando monitoring of nanoprinting process*

To minimize fabrication errors, beyond optimisation of the nanoprinting systems stability in terms common to nanofabrication methods, such as mechanical and thermal stability or stability of the laser source, nanoprinting offers the possibility to monitor the result of a fabrication *in operando*[10]. By continuous monitoring of the printed structures, characterising the results in the context of precise molecular models (**Figure 2a**) and under consideration of environmental and system conditions[11], it should therefore be possible to develop advanced nanoprinting protocols that correct for errors in the printed structures in an iterative approach.

- *New training approaches*

In order to advance nanoprinted Neural Networks beyond their current performance, it is important to equip them with some level of optical nonlinearity. For static nanoprinted networks it is key to derive new models for the interaction of light fields with distributed nonlinear media that are compatible with current in-silico training methods. More promising would be to train the nanoprinted networks in-situ[12,13], thus taking into account the complex physics involved, which would however require a means to reconfigure nanoprinted networks efficiently, locally and dynamically.

- *Reconfigurability*

There are several emerging directions with the potential to introduce reconfigurability into nanoprinted Neural Networks. 4D printing is a 3D printing technique for fabricating structures that can be reconfigured by applying different stimuli, such as for example heat, humidity, magnetic or electric fields[14]. In the context of nanoprinted Neural Networks contained in relatively small volumes, this method is interesting to facilitate global changes across the network (**Figure 2b**), for example to tune the network into a certain wavelength response[15]. More targeted tuning of a nanoprinted network can be achieved by deposition of reconfigurable materials such as phase change materials, which then can be locally reconfigured through laser irradiation to modify the behaviour of the network. A similar path is provided through recent advances in the field of optical data storage, where information can be densely stored in a polymer through targeted laser irradiation[16,17]. Further, recently described light-matter interactions on an atomistic level may pave a way towards dynamical reconfigurability of nanoprinted Neural Networks with topologically structured light[18,19].

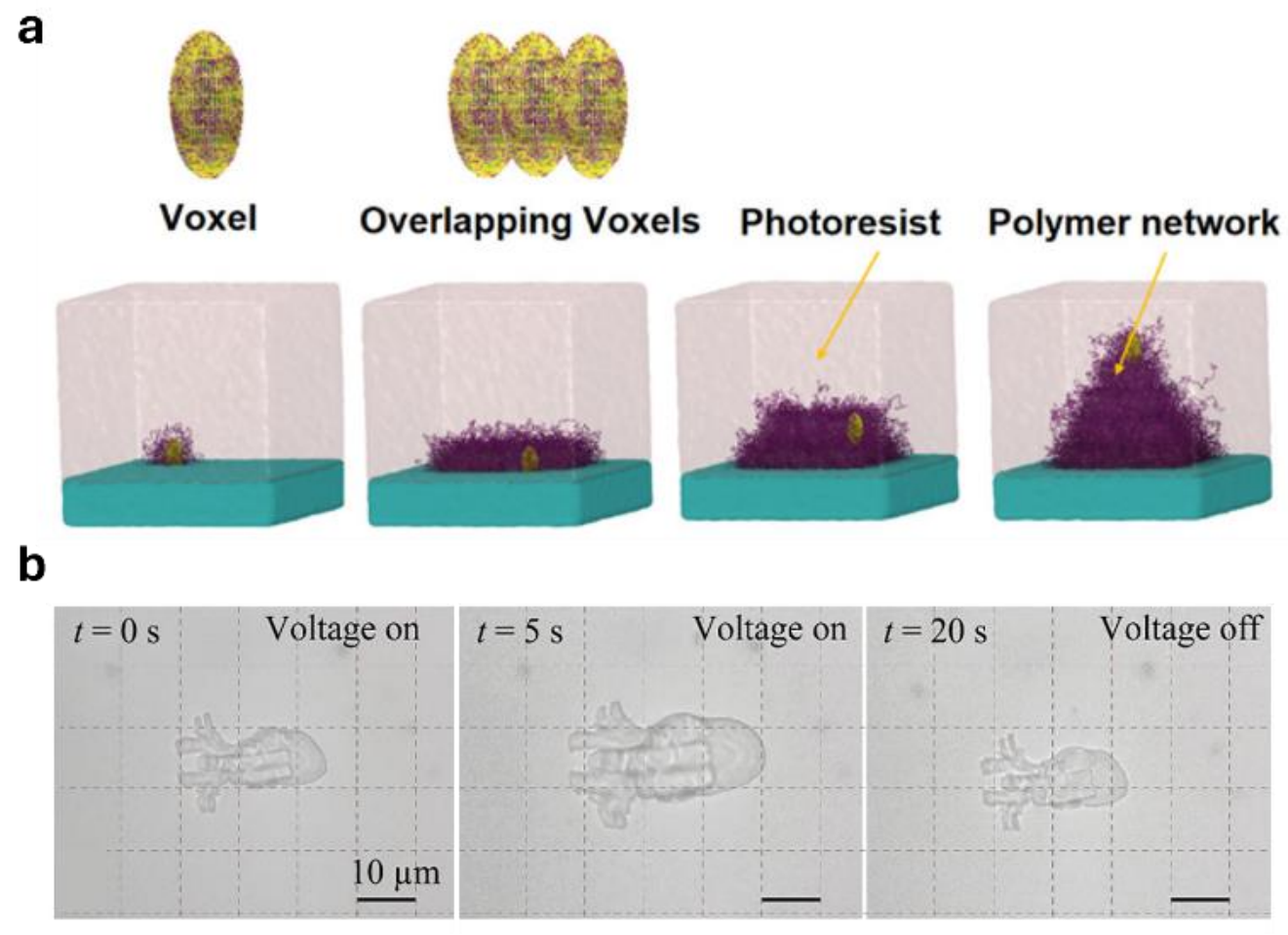

**Figure 2**. **a**) Model of the time evolution of the fabrication process during TPN. The yellow ellipsoidal volume represents the laser exposure volume (voxel) and overlapping voxels allow the fabrication of polymer networks with different geometries. Adapted with permission from Ref.[11]. **b**) Study of the electrical-mechanical response of a heart pumping model nanoprinted with optical force brush under electrical stimulation. Adapted with permission from Ref.[15].

## Concluding Remarks

Nanoprinted Neural Networks are compact neuromorphic network structures created by means of laser-based nanolithography. The sub-micron feature size achievable with this fabrication method not only opened additive manufactured Neural Networks up for information processing in the VIS/NIR wavelength regime, but also allowed for creation of networks with record high neuron densities[4]. While today nanoprinting methods are relatively mature, nanoprinted Neural Networks are prone to fabrication errors. Combined with the lack of reconfigurability available for nanoprinted networks, this represents a key issue for their performance and limits their large-scale implementation. Since models for interaction of light with nonlinear materials cannot efficiently be implemented with the in-silico training approach used to design nanoprinted networks, current implementations are linear and hence limited in their applicability to more complex problems. Despite these challenges, we currently witness that nanoprinted Neural Networks are evolving into a platform for leveraging the complex physics of light in compact neuromorphic photonic systems.

**Acknowledgements**
E.G. acknowledges the funding support from the Shanghai Natural Science Foundation (21ZR1443400), Shanghai Rising-Star Program (21QA1403600) and the Natural Science Foundation of Shanghai (21ZR1443400). M.G. acknowledges the support from the Science and Technology Commission of Shanghai Municipality (project no. 21DZ1100500), the Shanghai Municipal Science and Technology Major Project, and the Shanghai Frontiers Science Center Program (2021-2025 No. 20).

## References

[1] S. Maruo, O. Nakamura, and S. Kawata, "Three-dimensional microfabrication with two-photon-absorbed photopolymerization," Opt. Lett. **22**(2), 132 (1997).

[2] Z. Gan, Y. Cao, R. a Evans, and M. Gu, "Three-dimensional deep sub-diffraction optical beam lithography with 9 nm feature size.," Nat. Commun. **4**(May), 2061 (2013).

[3] E. Goi, X. Chen, Q. Zhang, B. P. Cumming, S. Schoenhardt, H. Luan, and M. Gu, "Nanoprinted high-neuron-density optical linear perceptrons performing near-infrared inference on a CMOS chip," Light Sci. Appl. **10**(10), 40 (2021).

[4] E. Goi, S. Schoenhardt, and M. Gu, "Direct retrieval of Zernike-based pupil functions using integrated diffractive deep neural networks," Nat. Commun. **13**(1), 7531 (2022).

[5] J. Moughames, X. Porte, M. Thiel, G. Ulliac, L. Larger, M. Jacquot, M. Kadic, and D. Brunner, "Three-dimensional waveguide interconnects for scalable integration of photonic neural networks," Optica **7**(6), 640–646 (2020).

[6] H. Yu, Q. Zhang, B. P. Cumming, E. Goi, J. H. Cole, H. Luan, X. Chen, and M. Gu, "Neuron-Inspired Steiner Tree Networks for 3D Low-Density Metastructures," Adv. Sci. **8**(19), 2100141 (2021).

[7] D. Mengu, Y. Luo, Y. Rivenson, and A. Ozcan, "Analysis of diffractive optical neural networks and their integration with electronic neural networks," IEEE J. Sel. Top. Quantum Electron. **26**(1), 3700114 (2019).

[8] E. Goi, M. Chen, S. Schoenhardt, and M. Gu, "Impact of common fabrication errors on the performance of diffractive neural networks," in *SPIE/COS Photonics Asia, 2022* (2022), p. 123180A.

[9] X. Lin, X. Lin, Y. Rivenson, N. T. Yardimci, M. Veli, Y. Luo, M. Jarrahi, and A. Ozcan, "All-optical machine learning using diffractive deep neural networks," Science **361**(6404), 1004–1008 (2018).

[10] M. Elmeranta, G. Vicidomini, M. Duocastella, A. Diaspro, and G. de Miguel, "Characterization of nanostructures fabricated with two-beam DLW lithography using STED microscopy," Opt Mater Express **6**(10), 3169–3179 (2016).

[11] E. Sedghamiz, M. Liu, and W. Wenzel, "Challenges and limits of mechanical stability in 3D direct laser writing," Nat. Commun. **13**(1), 2115 (2022).

[12] T. Zhou, X. Lin, J. Wu, Y. Chen, H. Xie, Y. Li, J. Fan, H. Wu, L. Fang, and Q. Dai, "Large-scale neuromorphic optoelectronic computing with a reconfigurable diffractive processing unit," Nat. Photonics **15**(5), 367–373 (2021).

[13] T. W. Hughes, M. Minkov, Y. Shi, and S. Fan, "Training of photonic neural networks through in situ backpropagation," Optica **5**(7), 864–871 (2018).

[14] D. Saritha and D. Boyina, "A concise review on 4D printing technology," Mater. Today Proc. **46**, 692–695 (2021).

[15] C. Yi, S. Qu, Y. Wang, H. Qi, Y. Zhang, and G. J. Cheng, "Optical force brush enabled free-space painting of 4D functional structures," Sci. Adv. **9**(38), eadg0300 (2023).

[16] S. Lamon, Q. Zhang, H. Yu, and M. Gu, "Neuromorphic Optical Data Storage Enabled by Nanophotonics: A Perspective," ACS Photonics **11**(3), 874–891 (2024).

[17] M. Zhao, J. Wen, Q. Hu, X. Wei, Y.-W. Zhong, H. Ruan, and M. Gu, "A 3D nanoscale optical disk memory with petabit capacity," Nature **626**(8000), 772–778 (2024).

[18] N. I. Zheludev and G. Yuan, "Optical superoscillation technologies beyond the diffraction limit," Nat. Rev. Phys. **4**(1), 16–32 (2022).

[19] N. I. Zheludev and K. F. MacDonald, "The Birth of Picophotonics," Opt Photon News **34**(9), 34–41 (2023).

# The Advent of Nonlinear Extreme Learning Machines

**Mario Chemnitz[1,2]**
[1]Leibniz-Institute of Photonic Technology, Albert-Einstein-Str. 9, 07745 Jena, Germany
[2] Institute of Applied Optics and Biophysics, Philosophenweg 7, 07743, Jena, Germany.

E-mails: mario.chemnitz@leibniz-ipht.de

**Abstract**

Nonlinear optical extreme learning machines (ELMs) leverage the optics-owned dynamics of nonlinear wave propagation in complex media for neuromorphic computing without the need for traditional node-based architectures. This article discusses the challenges in interpretability, performance scaling, and interconnectivity, with future advancements required in controlling nonlinear wave dynamics, improving system efficiency, and developing theoretical frameworks for understanding their computational capacity.

**Status**

Nonlinearity lies at the heart of the vast functionality of artificial neural networks (ANNs). Yet, while optics is rich in nonlinearities, integrating those in optical hardware is considered too power-hungry and sophisticated for top-down model-inspired designs [1]. Nonlinear optical extreme learning machines (ELM) may break with those assumptions by leveraging the computational power intrinsic to nonlinear wave propagation in complex media for scalable, energy-efficient computing. Extreme learning machines [2] were introduced to computer science in 2006 as randomly connected feed-forward neural networks conditioned by training only the output layer, similar to reservoir computers without recurrence. Marcucci et al. have found the ELM concept well suited to exploit optical wave propagation in nonlinear (Schrödinger) systems for neuromorphic computing without the need for active electronic or photonic nodes [3]. Instead, the computations are carried out passively through the natural dynamics of a complex system, leveraging the intrinsic strengths of light as an information medium, including multi-dimensionality, complex-number space, and foremost optical nonlinearity.

Recent experimental work illustrates this concept, e.g., by spatial information processing through nonlinear mode coupling in multimode fibers [4], second harmonic generation in scattering media [5], or frequency domain information processing through frequency mixing in single-mode optical fibers [6], [7], [8] or lithium niobate waveguides [9]. Information is stored in the phase and amplitude of femtosecond optical pulses, which offer a broad spectral bandwidth and high peak power, even at low pulse energies. As these pulses propagate through a medium, the encoded modes undergo complex nonlinear interactions, coupling spatial modes or generating new frequencies. These interactions map the stored information into an output spectrum or mode profile spanning the system's hyperspace. The resulting modes represent arbitrary but fixed nonlinear compositions of the input features. Through supervised training, these selected modes are superimposed, forming a nonlinear decision boundary that enables the system to perform various computational tasks using a single physical unit. One key advantage of nonlinear ELMs is their unique mechanism for performance scaling: As the nonlinear interactions in the system increase, so too does the system's computational power. Such enhancement has been theoretically predicted first by Marcucci et al. [3], and indicatively shown experimentally by Tegin et al. in nonlinear function regression [4], Fischer et al. in solving the n-bit XOR [6], and by Wang et al. [5] across various task-dependent benchmarks, which all improved with increasing the system nonlinearity. This scalability, combined with the ability to emulate entire neural networks within a single physical unit, makes nonlinear ELMs a special platform that largely capitalizes on the unique strengths of optical nonlinearity.

### Current and Future Challenges

In addition to their tremendous advantages in trainability, adaptability, and generalization, nonlinear ELMs face significant challenges in interpretability, performance scaling, and interconnectivity, rendering them uncompetitive with current top-down approaches.

**Interpretability** – Quantifying the computational capabilities of physical wave reservoirs has been a long-standing challenge, as their fully analog, continuous nature makes computational metrics such as FLOPs, TOPS, or MACs difficult or impossible to access. Instead, performance is often evaluated empirically using traditional task-specific benchmarks, such as chaotic time-series prediction, logistic regression, data (e.g., IRIS, WINE), or image classification (MNIST, ImageNet). The system that beats the most benchmarks to a decent degree wins, even though the reasons why it works or why it is limited remain unclear. Unlike delay-based photonic RCs, in which performance can be brought to light, e.g., by bifurcation or memory capacity analysis, photonic ELMs lack such a general performance analysis framework. The most pressing need seems to be to quantify the number of independent modes, or "nodes", that span the higher dimensional feature space. This information may be easier accessible in multimode fibers, where the number of higher-order modes can be estimated, but is hidden for systems that operate on a mode continuum, such as supercontinuum fiber processors or scattering media. In addition, the field longs for task-independent benchmarks that are scalable in a number of parameters and hardness as uniform measures across systems.

**Performance** – Related to the lack of interpretability, performance is often reported, if at all, in terms of task-specific energy per inference under various bold assumptions. Initial ELM experiments in fibers do indeed indicate a potential for energy-efficient, low-latency neuromorphic processing. For instance, emulating neural networks through nonlinear mixing of frequency modes has been demonstrated in a commercial nonlinear single-mode fiber at pulse energies of below 90 pJ [6] (same order of magnitude as demonstrated with second-harmonic generation in lithium-niobate waveguides [9]). This is 1-3 orders of magnitude lower than a single inference requires on a typical GPU depending on the dataset, considering the smallest ANN sizes required to solve the task. This, together with the potential to scale their computational power with system nonlinearity, paves a unique way toward energy-efficient computing. However, at the same time, latencies are bound to the propagation lengths (i.e., ~5 ns per meter in silica fibers) and processing rates to I/O speeds of the adaptive optics in use. Hence, unlocking ELMs to save energy requires efficient ways to encode and decode fresh information into and from each pulse individually at MHz to GHz pulse rates, which poses a significant technological challenge. In frequency-domain processing, current off-the-shelf spectral encoders, known as WaveShapers, feature only Hz update rates and limited spectral coverage. Spatial mode processing may gain here from the tens of kHz refresh rates of micromirror devices and similar read-out rates of super-CMOS cameras. Yet, nonlinear spatial mode processing suffers from higher energy requirements (100's nJ to µJ per pulse) and spatial mode collapse when scaled in nonlinear performance [10].

**Interconnectivity** – Nonlinear ELM approaches, similar to RCs, are driven by offline training on measured data, which dramatically limits their use for efficient, low-latency applications. Initial work has successfully implemented online training using optical weight banks on the readout side, representing a step towards all-optical operation[8], [11]. However, scaling such approaches to arbitrary M-to-N weight-and-sum operations for cascading reservoirs remains a critical challenge. Furthermore, nonlinear ELMs based on optical nonlinearities face the additional challenge of maintaining peak power for subsequent layers. This is particularly severe in frequency domain systems operating over a wide bandwidth, e.g., using supercontinuum [6] or second-harmonic generation [9], where power and information are distributed over remote channels and are difficult to collect for further processing layers. Finally, the advantages of waveguide integration in terms of stability, environmental robustness, and coupling efficiency are nullified when adaptive free-space optics are used on the input and output sides.

## Advances in Science and Technologies to Meet Challenges

The future of wave-based neuromorphic computing lies in advancements to harness and control complex nonlinear wave dynamics for enhancing performance in ELMs and beyond. In particular, such control may help to nudge a complex system to achieve higher accuracies and improve the interconnectivity by channeling energy into output modes that can be better relayed. Recent works have successfully tailored nonlinear system dynamics, e.g., by carefully conditioning the input wavefront for feed-forward training in MMFs[12] or by enriching the nonlinear interactions in supercontinuum generation through adaptive pulse splitting[13]. A novel emerging approach lies in optimizing the system-internal parameters locally. Promising techniques have just recently been explored and include dynamic tuning of the modal dispersion via (a) local temperature control on liquid-core fibers [14], [15], (b) locally distributed mechanical stresses in multimode glass fibers[16], or local photo-refractive index change on waveguide layers [17], [18]. Both strategies, i.e., optimizing the system's excitation or inner propagation properties, may be utilized for bias training and lead to unprecedented control over the system's "random" projections.

On another note, unlocking the potential energy benefit of nonlinear ELMs requires pulse-wise encoding and decoding techniques. While challenging to achieve in the spatial domain, frequency-domain processing with single-mode fibers or waveguides may leverage ultrafast RF electronics in combination with dispersive Fourier transform techniques[19] for GHz-rate spectral encoding[20] and shot-to-shot spectral measurements. In addition, exploring highly nonlinear materials, such as softglasses[21] or liquids[15] with hundreds of times higher nonlinearity than fused silica, is expected to significantly reduce power consumption per inference and provide more scope for scaling computational power through nonlinearity.

Finally, novel strategies to address interpretability must be advised through theoretical groundwork in computer sciences and physics. Existing pragmatic approaches, such as identifying ANN primitives that match the accuracy of the analog machine under test[6], should be replaced by more rigorous approaches. Inspiring physics-inspired approaches include the use of Fisher information[22], [23] or state entropy [24] to infer information preservation or generalization capabilities of the system.

## Closing remarks

Nonlinear ELMs represent a promising avenue for nonlinear photonics as a means of ultrafast and even energy-efficient computing, especially if their unique performance scaling with system nonlinearity can be exploited on a shot-to-shot basis. Frequency-domain processing offers this potential over spatial mode approaches, while spatial domain processing remains more scalable in terms of the number of parameters. Nevertheless, the challenges related to interpretability and online functionality continue to pose difficulties across ELM platforms, which present fresh opportunities for research and model advancement. It is particularly important to highlight the novel possibilities for unconventional computing that stem from the direct, trainable, and energy-efficient access to any non-linear mapping through these nonlinear ELMs. This advancement paves the way for innovative approaches to neuromorphic computing models, thus creating new pathways for research and model development.

## References

[1] R. S. Tucker and K. Hinton, "Energy Consumption and Energy Density in Optical and Electronic Signal Processing," *IEEE Photonics J.*, vol. 3, no. 5, pp. 821–833, Oct. 2011, doi: 10.1109/JPHOT.2011.2166254.

[2] G.-B. Huang, Q.-Y. Zhu, and C.-K. Siew, "Extreme learning machine: Theory and applications," *Neurocomputing*, vol. 70, no. 1–3, pp. 489–501, Dec. 2006, doi: 10.1016/j.neucom.2005.12.126.

[3] G. Marcucci, D. Pierangeli, and C. Conti, "Theory of Neuromorphic Computing by Waves: Machine Learning by Rogue Waves, Dispersive Shocks, and Solitons," *Physical Review Letters*, vol. 125, no. 9, p. 093901, Aug. 2020, doi: 10.1103/PhysRevLett.125.093901.

[4] U. Teğin, M. Yıldırım, İ. Oğuz, C. Moser, and D. Psaltis, "Scalable optical learning operator," *Nat Comput Sci*, vol. 1, no. 8, pp. 542–549, Aug. 2021, doi: 10.1038/s43588-021-00112-0.

[5] H. Wang *et al.*, "Large-scale photonic computing with nonlinear disordered media," *Nat Comput Sci*, vol. 4, no. 6, pp. 429–439, Jun. 2024, doi: 10.1038/s43588-024-00644-1.

[6] B. Fischer *et al.*, "Neuromorphic Computing via Fission-based Broadband Frequency Generation," *Advanced Science*, vol. 10, no. 35, p. 2303835, Dec. 2023, doi: 10.1002/advs.202303835.

[7] T. Zhou, F. Scalzo, and B. Jalali, "Nonlinear schrödinger kernel for hardware acceleration of machine learning," *Journal of Lightwave Technology*, vol. 40, no. 5, pp. 1308–1319, Mar. 2022, doi: 10.1109/JLT.2022.3146131.

[8] M. Zajnulina, A. Lupo, and S. Massar, "Weak Kerr Nonlinearity Boosts the Performance of Frequency-Multiplexed Photonic Extreme Learning Machines: A Multifaceted Approach," Dec. 19, 2023, *arXiv*: arXiv:2312.12296. doi: 10.48550/arXiv.2312.12296.

[9] M. Yildirim *et al.*, "Nonlinear optical feature generator for machine learning," *APL Photonics*, vol. 8, no. 10, p. 106104, Oct. 2023, doi: 10.1063/5.0158611.

[10] U. Tegin, M. Yildirim, I. Oguz, C. Moser, and D. Psaltis, "Scalable optical learning operator," *Nature Computational Science*, vol. 1, no. 8, pp. 542–549, Aug. 2021, doi: 10.1038/s43588-021-00112-0.

[11] X. Porte, A. Skalli, N. Haghighi, S. Reitzenstein, J. A. Lott, and D. Brunner, "A complete, parallel and autonomous photonic neural network in a semiconductor multimode laser," 2021.

[12] I. Oguz *et al.*, "Forward–forward training of an optical neural network," *Opt. Lett.*, vol. 48, no. 20, p. 5249, Oct. 2023, doi: 10.1364/OL.496884.

[13] B. Wetzel *et al.*, "Customizing supercontinuum generation via on-chip adaptive temporal pulse-splitting," *Nat Commun*, vol. 9, no. 1, p. 4884, Nov. 2018, doi: 10.1038/s41467-018-07141-w.

[14] R. Scheibinger, J. Hofmann, K. Schaarschmidt, M. Chemnitz, and M. A. Schmidt, "Temperature-sensitive dual dispersive wave generation of higher-order modes in liquid-core fibers," *Laser & Photonics Reviews*, vol. 17, no. 1, p. 2100598, 2023.

[15] M. Chemnitz, S. Junaid, and M. A. Schmidt, "Liquid-Core Optical Fibers—A Dynamic Platform for Nonlinear Photonics," *Laser & Photonics Reviews*, p. 2300126, Jul. 2023, doi: 10.1002/lpor.202300126.

[16] Z. Finkelstein, K. Sulimany, S. Resisi, and Y. Bromberg, "Spectral shaping in a multimode fiber by all-fiber modulation," *APL Photonics*, vol. 8, no. 3, p. 036110, Mar. 2023, doi: 10.1063/5.0121539.

[17] T. Onodera *et al.*, "Scaling on-chip photonic neural processors using arbitrarily programmable wave propagation," Feb. 27, 2024, *arXiv*: arXiv:2402.17750. Accessed: Sep. 22, 2024. [Online]. Available: http://arxiv.org/abs/2402.17750

[18] M. Nakajima, K. Tanaka, and T. Hashimoto, "Neural Schrödinger Equation: Physical Law as Deep Neural Network," *IEEE Trans. Neural Netw. Learning Syst.*, vol. 33, no. 6, pp. 2686–2700, Jun. 2022, doi: 10.1109/TNNLS.2021.3120472.

[19] K. Goda and B. Jalali, "Dispersive Fourier transformation for fast continuous single-shot measurements," *Nature Photon*, vol. 7, no. 2, pp. 102–112, Feb. 2013, doi: 10.1038/nphoton.2012.359.

[20] S. Thomas, A. Malacarne, F. Fresi, L. Potì, and J. Azaña, "Fiber-based programmable picosecond optical pulse shaper," *Journal of Lightwave Technology*, vol. 28, no. 12, pp. 1832–1843, 2010, doi: 10.1109/JLT.2010.2048700.

[21] T. Sylvestre *et al.*, "Recent advances in supercontinuum generation in specialty optical fibers [Invited]," *J. Opt. Soc. Am. B, JOSAB*, vol. 38, no. 12, pp. F90–F103, Dec. 2021, doi: 10.1364/JOSAB.439330.

[22] D. Bouchet, S. Rotter, and A. P. Mosk, "Maximum information states for coherent scattering measurements," *Nat. Phys.*, vol. 17, no. 5, pp. 564–568, May 2021, doi: 10.1038/s41567-020-01137-4.

[23] S. Rotter, "The concept of Fisher information in scattering problems and neural networks," in *Nonlinear Optics and its Applications 2024*, SPIE, Jun. 2024, p. PC1300401. doi: 10.1117/12.3022535.

[24] G. Zhang, "Correlation between entropy and generalizability in a neural network," Jul. 05, 2022, *arXiv*: arXiv:2207.01996. doi: 10.48550/arXiv.2207.01996.

# Optoacoustic neuromorphic photonics

**Birgit Stiller[1,2]**
[1] Max-Planck-Institute for the Science of Light, Staudtstr. 2, 91058 Erlangen, Germany
[2] Institute of Photonics, Gottfried Wilhelm Leibniz University, Welfengarten1A, 30176 Hannover, Germany
[ birgit.stiller@mpl.mpg.de ]

## Status

Optical fibers carry optical information over long lengths and are able to transmit information with low loss and at a high bandwidth. One option to process and manipulate optical information in photonic waveguides or optical fibers is to use optoacoustic interactions, specifically stimulated Brillouin-Mandelstam scattering (SBS). The latter is a nonlinear optical effect that couples an optical pump wave, a counterpropagating optical Stokes wave and a traveling acoustic wave together [1]. In practice, a strong optical pump can be backreflected by the acoustic waves and the backscattering may be so strong that it even limits the maximum power that can be sent through the fiber. The field of SBS and optomechanical interactions has recently been very active in exploring new microstructures and materials for photonic integrated chips as well as optical fibers [2-9].

Very recently, optoacoustic signal processing has been demonstrated for the use in photonic neuromorphic computation [10-16,18-21,24]. The fact that this interaction can be controlled by external optical pulses without transducers makes it particularly interesting for the combination with different photonic systems and neural network structures. As acoustic waves have very different time scales, they can offer a new dimension to the optical domain adding a latency component due to the slow acoustic velocity, for instance as recurrent functionality in the system.

To fully harness the potential of photonic computing, we need to address the ongoing challenge of implementing photonic memories, particularly high-speed and coherent random-access memory. This is a building block that optoacoustics can offer: an all-optically controlled coherent photonic memory [10-16,18-20]. Via the effect of SBS, the photonic-phononic memory coherently transfers optical information to traveling acoustic waves, that propagate at 5 orders of magnitude lower speed. The optoacoustic memory fulfils several key requirements for an optical high-performance random-access memory due to its coherence, on-chip compatibility, frequency selectivity and high bandwidth.

The concept is outlined in Fig. 1a. An optical data stream is sent into the optical fiber or photonic chip and converted into an acoustic wave via a counterpropagating optical control pulse. The information is then kept in the acoustic waves until a second control pulse “read” enters the medium and transforms the information back into the optical domain. The first experiments were shown in highly nonlinear fiber [10] and then in photonic waveguides on a chip [11]. The memory was proven to be coherent in amplitude and phase [11] and provides a high bandwidth up to GHz [12,13]. In fact, it is possible to store optical pulses down to 150 ps pulse width which is possible because of the waveguide nature of the system (Fig. 1d). The memory can be accessed at high speed [13] and can be operated at different frequency channels at the same time with negligible crosstalk, even at frequency channels as close as few GHz [14]. It was also experimentally demonstrated that one can access and read out the memory several times and all-optically controlled at several positions on a photonic chip [15]. Because the acoustic waves have a determined direction, this type of photonic memory is nonreciprocal and the scheme is not impacted by simultaneously counterpropagating optical data pulses [16]. This allows also for the nonreciprocal manipulation of orbital angular momentum modes [17].

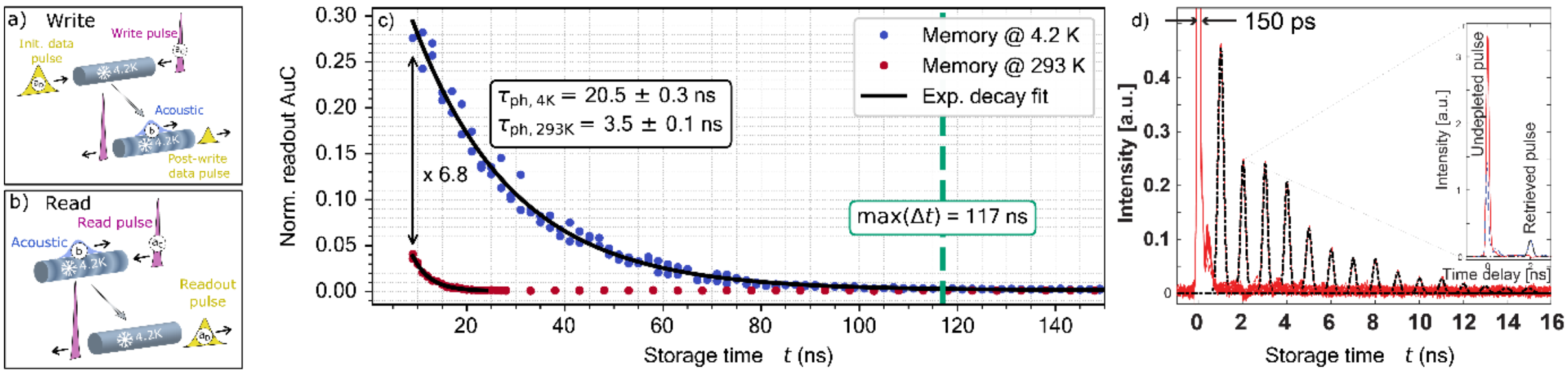

**Fig. 1** – a) and b) Photonic storage in acoustic waves [19]: a) Write process, b) Read-out process; c) Storage time up to over 140 ns in a cryogenic Brillouin-based memory [19]; d) Storage of 150ps-long optical pulses [12].

## Current and Future Challenges

There are some challenges to be addressed when using SBS. One of them is the narrow linewidth of the Brillouin resonance which is a feature for Brillouin lasers but can be a nuisance for signal processing. The intrinsic linewidth (~30 MHz in silica) limits the length of the optical pulses to several nanoseconds. However, it has been recently shown that compensating the narrow spectral linewidth with a large gain leads to operation with short optical pulses. In Fig. 1d, you can observe for instance the storage of 150ps-long optical pulses for over 15 ns, demonstrating a decent operation bandwidth of several GHz [12].

A limiting factor can be the acoustic lifetime of several nanoseconds. This inherent decay is due to the damping of the acoustic waves while propagating through the material. Therefore, the optical information that is encoded in the acoustic waves would be lost and a way of preserving the coherent acoustic vibration is required. This challenge has been tackled in several ways, an active and a passive method is briefly described here. In the active method, the acoustic waves that contain the optical information are reinforced optically [18]. These optical refresh pulses scatter from the existent acoustic waves with the encoded information and transfer energy to them. Note that the information is not written several times, the information is only written once and then amplified. Experimentally, storage times up to 40 ns for the amplitude and phase encoded information has been shown. The passive method involves bringing the optical system to moderate cryogenic temperatures of 4 K. The optical signal can then be retrieved up to 140 ns later than the original pulse which is more than an order of magnitude of the acoustic lifetime at room temperature [19] (Fig. 1c). This improvement in acoustic lifetime could be interesting for further signal processing schemes as it allows for a higher throughput of optical data that can be processed.

Another challenge is the upscaling of the information density, which can be addressed, for instance, with coherent information encoding. Very recently, the storage of the first quadrature-phase-shift-keying (QPSK) has been demonstrated, at room temperature as well as cryogenic temperatures [20]. The forthcoming challenge will be to upscale the density to higher-order QAM signals.

## Advances in Science and Technologies to Meet Challenges

Building signal processing schemes using acoustic waves can have great potential. Recently several building blocks towards this approach have been experimentally shown. An optoacoustic recurrent operator (OREO) was experimentally demonstrated in Ref. [21] to process contextual information such as time series signals. So far optical recurrent neural networks were mostly implemented with microring structures [22] or delay lines [23], which may put limits on bandwidth, multifrequency operations in the synthetic domain and sometimes requires slow tuning based on heaters. OREO however, can be operated in a single optical fiber or photonic waveguide. The concept of OREO links information of an optical pulse sequence via interaction acoustic waves. The information of one optical computation step is linked via acoustic waves to the subsequent one and can therefore manipulate the later computational steps. The concept is shown in Fig. Stiller1a. A control optical pulse processes a data pulse via an acoustic wave that persists in the medium. A second processing step with a new data pulse and a new control pulse creates a second acoustic wave that interacts with the previous acoustic wave. This interaction can be repeated and the processing steps are linked over several layers as long as the acoustic wave has not decayed.

The acoustic waves link the optical pulses in the time series, capturing their information and alter subsequent operations. OREO was implemented to show a recurrent dropout and to predict up to 27 patters in a time series (Fig. 2a). For 9 different patterns, an accuracy of over 99% could be achieved using SBS, whereas an accuracy of 45% (Fig. 2b) with potential improvement to 92% for 27 patterns is demonstrated (Fig. 2c). OREO can be introduced as a bi-directional perceptron for new classes of optical neural networks.

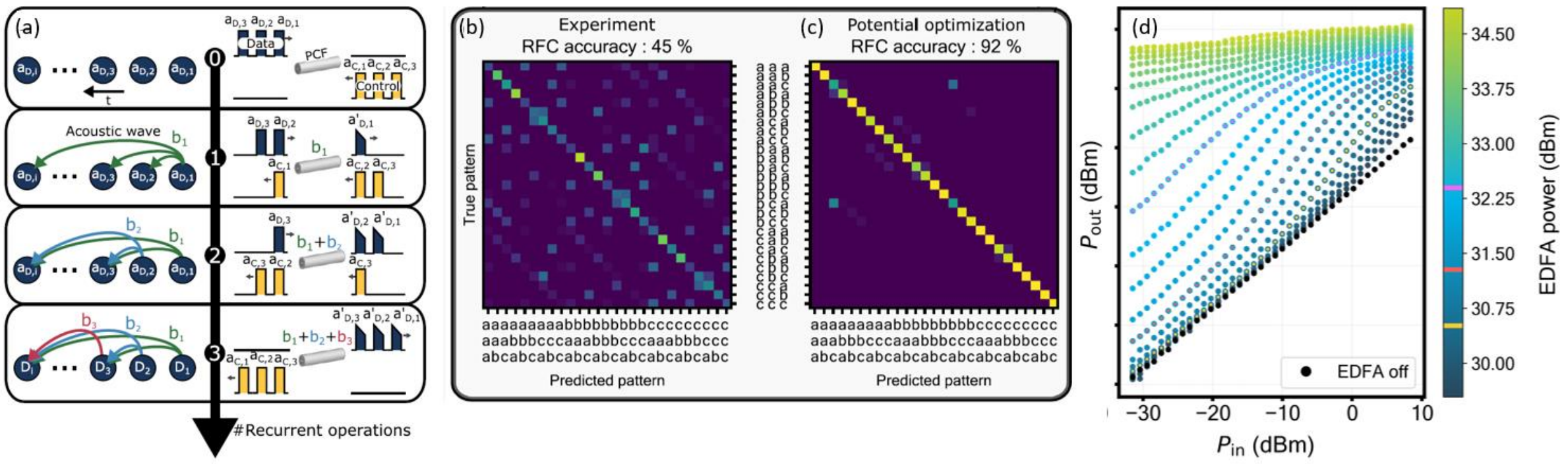

**Fig. 2** - (a) Concept of an optoacoustic recurrent operator [21]; (b) Pattern recognition task, RCF accuracy 45% for the experiment [21], (c) Pattern recognition task with optimized experimental parameters [21], (d) Different photonic activation functions using a double Brillouin amplifier setup [24].

The interaction of optical and acoustic waves can not only introduce a latency element to a photonic system but also serve with its nonlinearity because SBS is a nonlinear optical effect. In Ref. [24], a photonic activation function based on traveling sound waves is experimentally shown. The implemented all-optical nonlinear activation function has a programmable nonlinearity, low insertion loss, preserves coherence and is frequency selective and compatible with on-chip designs.

The scheme is based on a double Brillouin-amplifier setup with an added feature that makes the nonlinearity of the system depend on the optical input data. Different activation functions with a LEAKYRELU, SIGMOID, and QUADRATIC shape were demonstrated by only tuning the optical control pump power (Fig. 2d). The design not only allows for tailoring the photonic activation function but also compensates for the insertion loss automatically by providing net optical gain as high as 20 dB which may pave the way for deep optical neural networks. The inherent frequency-selectivity of SBS makes the optoacoustic activation function particularly suitable for frequency-basis information encoding, which was experimentally shown in a dual-frequency proof-of-concept [24].

Optoacoustic interaction in photonic computing schemes is a versatile add-on and provides features that can be all-optically configured. As this interaction is recently being investigated towards the quantum regime [25-29], there is also an avenue towards quantum signal processing.

**References**:

[1] C. Wolff, M. J. A. Smith, B. Stiller, and C. G. Poulton, “Brillouin scattering - theory and experiment: tutorial”, Journal of the Optical Society of America B, 38 (4), 1243-1269 (2021).

[2] B. J. Eggleton, et al., “Brillouin integrated photonics”, Nature Photonics 13, 664–677 (2019). https://doi.org/10.1038/s41566-019-0498-z

[3] N. T. Otterstrom et al., “A silicon Brillouin laser”, Science 360, 1113-1116 (2018) DOI: 10.1126/science.aar6113
[4] C. C. Rodrigues, et al., “On-Chip Backward Stimulated Brillouin Scattering in Lithium Niobate Waveguides”, arXiv:2311.18135
[5] K. Ye, et al., “Surface acoustic wave stimulated Brillouin scattering in thin-film lithium niobate waveguides”, arXiv:2311.14697
[6] R. Pant, et al., "On-chip stimulated Brillouin scattering," Opt. Express 19, 8285-8290 (2011).
[7] R. Botter et al., “Guided-acoustic stimulated Brillouin scattering in silicon nitride photonic circuits”, Sci. Adv.8, eabq2196 (2022).DOI:10.1126/sciadv.abq2196
[8] S. Gundavarapu, et al., “Sub-hertz fundamental linewidth photonic integrated Brillouin laser”, Nature Photon 13, 60–67 (2019). https://doi.org/10.1038/s41566-018-0313-2
[9] A. Geilen, A. Popp, D. Das, S. Junaid, C.G. Poulton, M. Chemnitz, C. Marquardt, M. A. Schmidt, and B. Stiller, “Exploring extreme thermodynamics in nanoliter volumes through stimulated Brillouin-Mandelstam scattering”, *Nature Physics*, https://doi.org/10.1038/s41567-023-02205-1 (2023).
[10] Z. Zhu et al., “Stored Light in an Optical Fiber via Stimulated Brillouin Scattering”, *Science* 318, 1748-1750 (2007).DOI:10.1126/science.1149066
[11] M. Merklein, B. Stiller, K. Vu, S. J. Madden, and B. J. Eggleton, “A chip-integrated coherent photonic-phononic memory“, *Nature Communications* 8, 574 (2017), doi:10.1038/s41467-017-00717-y.
[12] B. Stiller, K. Jaksch, J. Piotrowski, M. Merklein, K. Vu, P. Ma, S. J. Madden, C. G. Poulton and B. J. Eggleton, “Brillouin light storage for 100 pulse widths”, *npj Nanophotonics* 1, 5 (2024). https://doi.org/10.1038/s44310-024-00004-x (2024). arXiv:2308.01009.
[13] J. Piotrowski, M. K. Schmidt, B. Stiller, C. G. Poulton, M. J. Steel, “Picosecond acoustic dynamics in stimulated Brillouin scattering”, *Optics Letters 46, 2972-2975* (2021). arXiv:2103.05732.
[14] B. Stiller*, M. Merklein*, C. G. Poulton, K. Vu, P. Ma, S. J. Madden, and B. J. Eggleton, “Crosstalk-free multi-wavelength coherent light storage via Brillouin interaction”, *APL Photonics* Vol. 4, Issue 4, 040802 (2019). arXiv:1803.08626.
[15] B. Stiller, M. Merklein, C. Wolff, K. Vu, C. G. Poulton, S. J. Madden, and B. J. Eggleton, “On-chip multi-stage optical delay based on cascaded Brillouin light storage“, *Optics Letters*, Vol. 43, Issue 18, 4321-4324 (2018).
[16] M. Merklein, B. Stiller, K. Vu, P. Ma, S. J. Madden, and B. J. Eggleton, “On-chip broadband nonreciprocal light storage”, *Nanophotonics* (2020), doi.org/10.1515/nanoph-2020-0371. arXiv:1806.00146.
[17] X. Zeng, P. St.J. Russell, C. Wolff, M. H. Frosz, G. K. L. Wong, and B. Stiller, “Nonreciprocal vortex isolator by stimulated Brillouin scattering in chiral photonic crystal fibre”, *Science Advances* 8(42), eabq606 (2022). arXiv:2203.03680.
[18] B. Stiller, M. Merklein, C. Wolff, K. Vu, P. Ma, S. J. Madden, and B. J. Eggleton, „Coherently refreshed acoustic phonons for extended light storage”, *Optica* 7 (5), 492-497 (2020). arxiv:1904.13167.
[19] A. Geilen, S. Becker, and B. Stiller, “High-speed coherent photonic random-access memory in long-lasting sound waves “, accepted for *ACS Photonics*, arXiv:2311.06219 (2024).
[20] O. Saffer, J. H. Marines Cabello, S. Becker, A. Geilen, and B. Stiller, “Brillouin-based storage of QPSK signals with fully tunable phase retrieval”, arXiv:2410.05156 (2024).
[21] S. Becker, D. Englund, and B. Stiller, “An optoacoustic field-programmable perceptron for recurrent neural networks”, *Nature Communications* 15, 3020 (2024). https://doi.org/10.1038/s41467-024-47053-6.
[22] Tait, A. N. et al., “Neuromorphic photonic networks using silicon photonic weight banks”, *Sci. Rep.* 7, 7430 (2017).
[23] D. Brunner, et al., “Tutorial: Photonic neural networks in delay systems”, *J. Appl. Phys*. 124, 152004 (2018).
[24] G. Slinkov, S. Becker, D. Englund, and B. Stiller, “All-optical nonlinear activation function based on stimulated Brillouin scattering”, arXiv:2401.05135 (2024).
[25] C. Zhu, C. Genes and B. Stiller, “Optoacoustic entanglement in a continuous Brillouin-active solid state system”, accepted for *Physical Review Letters*, arXiv:2401.10665 (2024).
[26] L. Blázquez Martínez, P. Wiedemann, C. L. Zhu, A. Geilen, B. Stiller, “Optoacoustic cooling of traveling hypersound waves”, *Physical Review Letters,* 132, 023603 (2024). arXiv:2305.19823.

[27] E. A. Cryer-Jenkins, G. Enzian, L. Freisem, N. Moroney, J. J. Price, A. Ø. Svela, K. D. Major, and M. R. Vanner, "Second-order coherence across the Brillouin lasing threshold," *Optica* 10, 1432-1438 (2023).
[28] G. Enzian, L. Freisem, J. J. Price, A. Ø. Svela, J. Clarke, B. Shajilal, J. Janousek, B. C. Buchler, P. K. Lam, and M. R. Vanner, "Non-Gaussian Mechanical Motion via Single and Multiphonon Subtraction from a Thermal State", *Phys. Rev. Lett.* 127, 243601 (2021).
[29] J. Zhang, C. Zhu, C. Wolff and B. Stiller, "Quantum coherent control in pulsed waveguide optomechanics", *Physical Review Research* 5, 013010 (2023). arXiv:2203.16946.

# Superconducting Optoelectronic Neural Networks

**Jeffrey M. Shainline, Bryce A. Primavera and Jeff Chiles**
National Institute of Standards and Technology (325 Broadway, Boulder, CO, 80305, USA)
[ jeffrey.shainline@nist.gov , bryce.primavera@nist.gov , jeff.chiles@nist.gov ]

## Status

Semiconductors, superconductors, and photonic integrated circuits have complimentary features, yet these technologies have not been monolithically integrated to exploit all their capabilities simultaneously. Computational hardware combining these attributes would bring performance advantages for neuromorphic supercomputers [1]. Current hardware for AI is limited by communication, and this limitation arises from the digital computing paradigm itself [2, 3]. The operations performed by digital systems are poorly matched to the analog computations, distributed memory, and parallel communication leveraged by neural systems. As models continue to grow, the number of processors required increases, and communication between processors becomes a severe bottleneck. This constraint can be overcome with light-based communication. Optical interconnects do not exhibit wiring parasitics, so each processing element can make thousands of direct connections, eliminating the need for data transfer over a shared routing network. Furthermore, communication at the single-photon level is achievable with superconducting detectors. These detectors require cryogenic operation at 4K, but the cooling energy is more than offset by the reduction in optical power [4]. Cryogenic operation also enables superconducting circuit elements such as Josephson junctions, which readily implement the nonlinear operations required for neural networks [5]. Josephson junctions have the highest speed over energy quotient of any active circuit element, so there is no known method to compute faster with less energy. Systems combining single-photon communication with Josephson-junction computation are known as superconducting optoelectronic networks (SOENs).

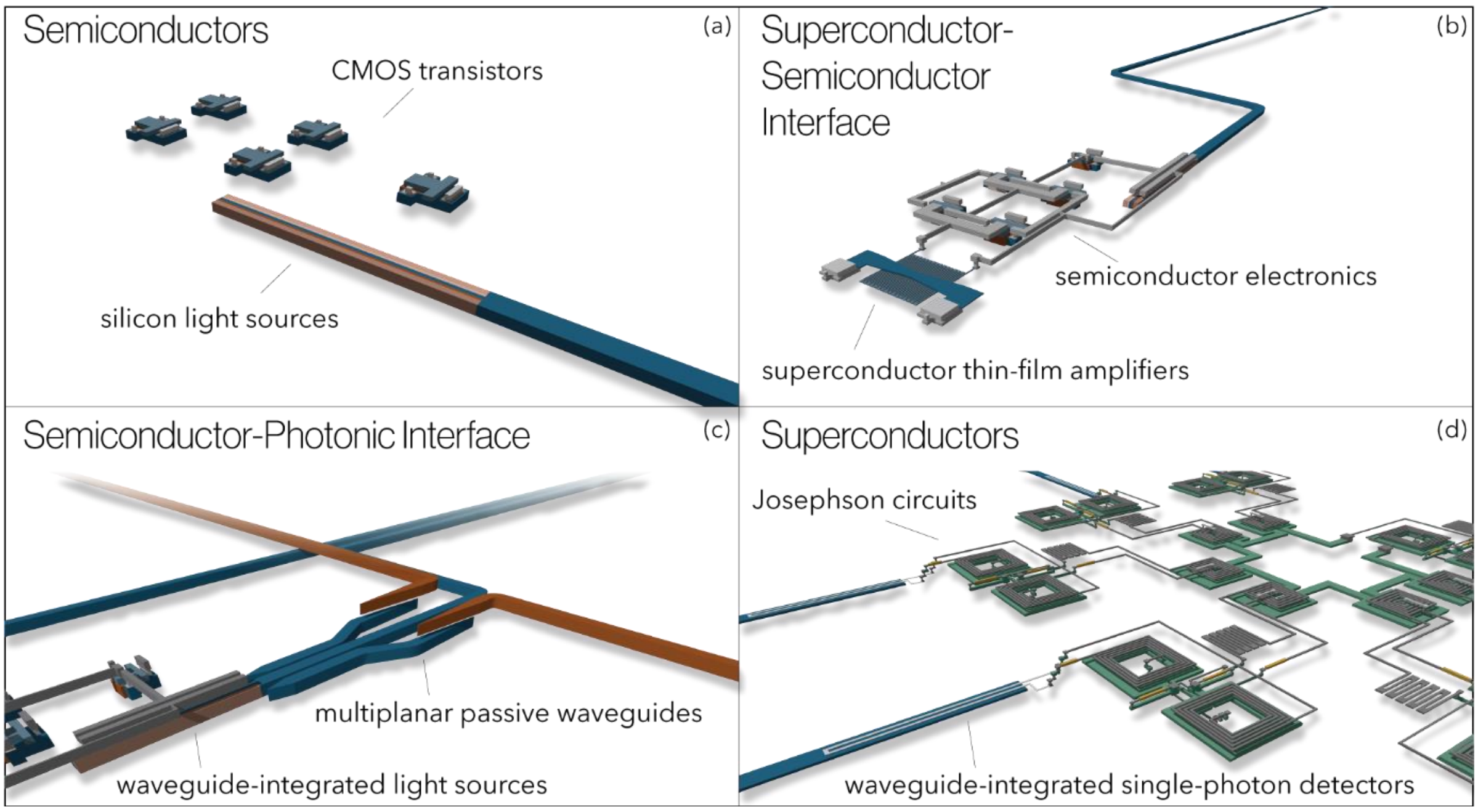

**Figure SEQ Figure * ARABIC 1**. The semiconductor-superconductor-photonic process. (a) The first process module includes CMOS transistors and silicon light sources. (b) The second process module forms the superconductor-semiconductor interfaces. (c) The semiconductor-photonic interface, where waveguide-coupled light sources meet multiplanar routing waveguides. (d) Superconducting circuitry, including single-photon detectors and Josephson circuits.

With SOENs, analog synaptic, dendritic, and neuronal computations are implemented with Josephson junctions, while communication is carried out with semiconductor transmitters and light sources. Optical spike events fan out across a network of nanophotonic and fiber-optic waveguides. Our team has demonstrated the computations of synapses and dendrites with superconducting optoelectronic hardware, including synaptic weighting, signal integration, and nonlinear transfer functions [6]. We have also demonstrated co-located synaptic memory cells [7]; all-silicon, waveguide-integrated optical links [8]; superconductor-semiconductor interfaces [9]; multi-planar waveguides for interconnects [10, 11]; and low-loss, high-bandwidth fiber-to-waveguide interfaces for coupling on-chip nanophotonic waveguides to fiber optics for long-range connections [12]. Here we outline a roadmap to bring this technology from a nascent state in the lab to an impactful presence in the world.

**Current and Future Challenges**

One motivation for SOENs comes from comparison to established hardware. When performing multiply-accumulate operations (MACs), as are used in deep learning, SOENs require $1x10^{-14}$ joules to perform an 8-bit MAC. A GPU requires $2x10^{-13}$ joules for an 8-bit instruction in the best case [16]. The benefits of SOENs for communication are even more significant. For SOENs to communicate an 8-bit number across the network requires $9x10^{-14}$ joules, while the analogous operation in a GPU system is DRAM memory access, which uses $5x10^{-12}$ joules per bit [17], or $4x10^{-11}$ for 8 bits. Thus, SOENs can perform similar operations to GPUs with four hundred times less energy per operation, and that includes a factor of 500 to keep the SOEN system cold.

While this simple comparison to GPUs indicates energy efficiency advantages, the complete benefits of SOENs will be realized with neuromorphic architectures and learning algorithms. To make further progress on these fronts, the complete semiconductor-superconductor-photonics fabrication process must be realized. The elements of this process are shown in Fig. 1. They include CMOS electronics and silicon light sources (Fig. 1a); superconducting thin-film amplifiers (Fig. 1b); multi-planar photonics (Fig. 1c); superconducting single-photon detectors (Fig. 1d); and superconducting Josephson-junction circuits (Fig. 1d).

We envision SOENs being fabricated through a series of five processing modules in a commercial foundry. CMOS electronics form the transmitter circuits that generate light when a neuron fires. Light sources for scalable communication may also be based on silicon, which is possible at low temperature [8]. The first module, semiconductor device fabrication, requires the highest temperatures. The transistors and their wiring layers can withstand the thermal demands of the remainder of the process. The signals generated by the millivolt superconducting circuits must be capable of switching silicon transistor gates with thresholds close to 1V. This functionality is possible with superconducting thin-film amplifiers [9]. These amplifiers are fabricated in the second process module. For each neuron to communicate to thousands of others, dense photonic connectivity is required. The third module consists of multiple planes of photonic waveguides that couple the semiconductor light sources to the communication routing infrastructure [10, 11]. Single-photon sensors are the receivers in our synapses, fabricated in the fourth module. The electrical signals from these detectors couple to Josephson circuits for synaptic and dendritic processing of photon detection events as well as synaptic plasticity circuits. The fifth module is for Josephson circuits.

**Advances in Science and Technology to Meet Challenges**

Looking to the future, we aspire to achieve very large computational systems. Scaling into the third dimension is necessary, both on the wafer and in multi-wafer systems. Fortunately, achieving multiple planes of waveguides, single-photon detectors, and Josephson junctions is straightforward, and all these feats have been accomplished by either our group [10, 11, 13] or others [14, 15]. In this process, only one plane of MOSFETs and light sources is required; only the superconductor and passive photonic modules must be repeated. Three-dimensional integration is much more straightforward when superconductors are employed than with semiconductors alone. Additionally, low power dissipation of superconducting electronics ensures heating in multiple planes of active components is not a problem.

The envisioned technology progression is shown in Fig. 2. Wafer-scale systems can be realized in a 45nm technology node. Although the process requires multiple fabrication modules, it could be inexpensive due to the simplicity of all steps. Each wafer-scale system would house a million neurons and a billion synapses. Massive supercomputing systems comprising many such wafers interconnected with optical fibers and free-space communication will be possible, enabling cognitive computing at an unprecedented scale, leading to exciting opportunities for scientific discovery, computation capable of addressing currently intractable problems, and immense opportunities for industry.

The full power of neuromorphic hardware will not be harnessed until more sophisticated algorithms are developed that leverage the time domain and photonic movement of information across space. Despite significant progress, there has not been a breakthrough that puts spiking neuromorphic hardware in contention to dethrone matrix-vector-based deep learning, even though deep learning ignores much of the elegance and sophistication of biological neural processors and omits the efficient use of space [18, 19] and time [20] that is central to cognition. Developing algorithms that leverage these attributes and can be efficiently implemented on hardware could initiate a sea change in artificial intelligence and cognitive computing.

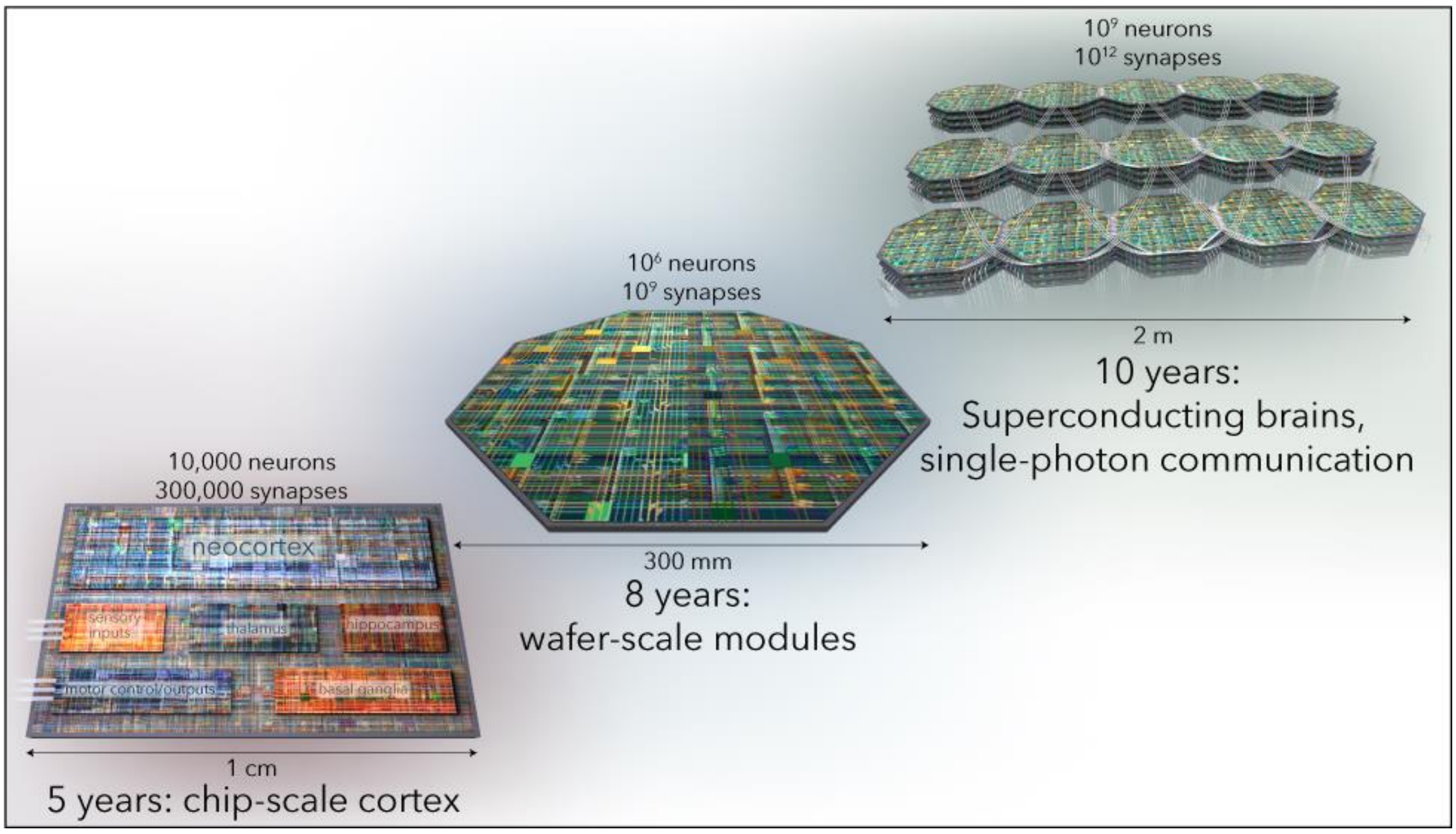

**Figure 2.** Supercomputing timeline. With this hardware, we will initially develop a small-scale cortex on a chip with 10,000 neurons and 300,000 synapses. This feasibility demonstration will enable further work with a foundry partner. Within 10 years, profound new computational capabilities will be enabled through multi-wafer systems interconnected with fiber optics, approaching the physical limits of cognition.

**Concluding Remarks**

This is a moonshot project with significant scientific challenges and potentially immense benefits to society. Existing neuromorphic hardware does not approach the device and circuit complexity nor does it encompass the system scale of biological neural systems. With SOENs, there is a path to a hardware process with the potential to match the sophistication of the human brain. Demonstration of a small-scale cortex on a chip would enable the community to test the hypothesis that artificial circuits with numerous plasticity mechanisms, sophisticated dendrites, single-photon communication, and high-speed superconducting computation will be useful for information processing. This cortex would provide a testbed for development of spiking algorithms unhindered by communication bottlenecks and accelerated by the speed of superconducting electronics. Realizing systems with 10 billion neurons (roughly the number in the human neocortex) will be possible in a cube 2m on a side. The entire system would consume 1MW or less, depending on the achievable light source efficiency. This is a factor of 30 less power than existing digital supercomputers. This superconducting optoelectronic brain would operate more than a hundred thousand times faster than the human brain and a billion times faster than a digital computer emulating neural function at this scale.

**Acknowledgements**
This is a contribution of NIST, an agency of the U.S. government, not subject to copyright.

**References**

[1] J.M. Shainline. "Optoelectronic intelligence," Applied Physics Letters, vol. 118, no. 16, pp. 160501, 2021. Accessed: May 12, 2024. doi: 10.1063/5.0040567. [Online]. Available: https://pubs.aip.org/aip/apl/article/118/16/160501/236342

[2] J V´egh. "How Amdahl's law limits performance of large artificial neural networks," Brain Informatics, vol. 6, pp. 1–11, 2019. Accessed: May 12, 2024. doi: 10.1186/s40708-019-0097-2. [Online]. Available: https://link.springer.com/article/10.1186/s40708-019-0097-2

[3] Lohn, Andrew, and Micah Musser. "AI and compute: How much longer can computing power drive artificial intelligence progress?" Center for Security and Emerging Technology (2022). Accessed: May 12, 2024. [Online]. Available: https://cset.georgetown.edu/wp-content/uploads/AI-and-Compute-How-Much-Longer-Can-Computing-Power-Drive-Artificial-Intelligence-Progress.pdf

[4] B.A. Primavera and J.M. Shainline. "Considerations for neuromorphic supercomputing in semiconducting and superconducting optoelectronic hardware," Frontiers in Neuroscience, vol. 15, pp. 732368, 2021. Accessed: May 12, 2024. doi: 10.3389/fnins.2021.732368. [Online]. Available: https://www.frontiersin.org/journals/neuroscience/articles/10.3389/fnins.2021.732368/full

[5] J.M. Shainline, S.M. Buckley, A.N. McCaughan, J. Chiles, A. Jafari-Salim, R.P. Mirin, and S.W. Nam, "Circuit designs for superconducting optoelectronic loop neurons," Journal of Applied Physics, vol. 124, pp. 15 (2018). Accessed: May 12, 2024. doi: 10.1063/1.5038031 [Online]. Available: https://pubs.aip.org/aip/jap/article/124/15/152130/348058

[6] S. Khan, B.A. Primavera, J. Chiles, A.N. McCaughan, S.M. Buckley, A.N. Tait, A. Lita, J. Biesecker, A. Fox, D. Olaya, R.P. Mirin, S.W. Nam, and J.M. Shainline, "Superconducting optoelectronic single-photon synapses," Nature Electronics, vol. 5, pp. 650-659, (2022). Accessed: May 12, 2024. doi: 10.1038/s41928-022-00840-9. [Online]. Available: https://www.nature.com/articles/s41928-022-00840-9

[7] B.A. Primavera, S. Khan, R.P. Mirin, S.W. Nam, and J.M. Shainline, "Programmable Superconducting Optoelectronic Single-Photon Synapses with Integrated Multi-State Memory," *arXiv preprint arXiv:2311.05881* (2023). Accessed: May 12, 2024. doi: 10.48550/arXiv.2311.05881. [Online]. Available: https://arxiv.org/abs/2311.05881

[8] S. Buckley, J. Chiles, A.N. McCaughan, G. Moody, K.L. Silverman, M.J. Stevens, R.P. Miring, S.W. Nam, and J.M. Shainline, "All-silicon light-emitting diodes waveguide-integrated with superconducting single-photon detectors," Applied Physics Letters, vol. 111, no. 14 (2017). Accessed: May 12, 2024. doi: 10.1063/1.4994692. [Online]. Available: https://pubs.aip.org/aip/apl/article/111/14/141101/34053

[9] A.N. McCaughan, V.B. Verma, S.M. Buckley, J.P. Allmaras, A.G. Kozorezov, A.N. Tait, S.W. Nam, and J.M. Shainline. "A superconducting thermal switch with ultrahigh impedance for interfacing superconductors to semiconductors," Nature electronics, vol. 2, pp. 451-456, (2019). Accessed: May 12, 2024. doi: 10.1038/s41928-019-0300-8. [Online]. Available: https://www.nature.com/articles/s41928-019-0300-8

[10] J. Chiles, S. Buckley, N. Nader, S.W. Nam, R.P. Mirin, and J.M. Shainline, "Multi-planar amorphous silicon photonics with compact interplanar couplers, cross talk mitigation, and low crossing loss," APL Photonics,

vol. 2, no. 11, (2017). Accessed: May 12, 2024. doi: 10.1063/1.5000384 [Online]. Available: https://pubs.aip.org/aip/app/article/2/11/116101/122779
[11] J. Chiles, S. Buckley, S.W. Nam, R.P. Mirin, and J.M. Shainline, "Design, fabrication, and metrology of 10× 100 multi-planar integrated photonic routing manifolds for neural networks," APL Photonics, vol. 3, no. 10, (2018). Accessed: May 12, 2024. doi: 10.1063/1.5039641 [Online]. Available: https://pubs.aip.org/aip/app/article/3/10/106101/123050
[12] S. Khan, S.M. Buckley, J. Chiles, R.P. Mirin, S.W. Nam, and J.M. Shainline, "Low-loss, high-bandwidth fiber-to-chip coupling using capped adiabatic tapered fibers," APL Photonics, vol. 5, no. 5, (2020). Accessed: May 12, 2024. doi: 10.1063/1.5145105. [Online]. Available: https://pubs.aip.org/aip/app/article/5/5/056101/123222
[13] V.B. Verma, F. Marsili, S. Harrington, A.E. Lita, R.P. Mirin, and S.W. Nam, "A three-dimensional, polarization-insensitive superconducting nanowire avalanche photodetector," Applied Physics Letters, vol. 101, no. 25, (2012). Accessed: May 12, 2024. doi: 10.1063/1.4768788. [Online]. Available: https://pubs.aip.org/aip/apl/article/101/25/251114/26354
[14] S.K. Tolpygo, V. Bolkhovsky, R. Rastogi, S. Zarr, A.L. Day, E. Golden, T.J. Weir, A. Wynn, and L.M. Johnson, "Planarized fabrication process with two layers of SIS Josephson junctions and integration of SIS and SFS π-junctions," IEEE Transactions on Applied Superconductivity, vol. 29, no. 5, pp. 1-8, (2019). Accessed: May 12, 2024. doi: 10.1109/TASC.2019.2901709. [Online]. Available: https://ieeexplore.ieee.org/abstract/document/8666798
[15] T. Ando, S. Nagasawa, N. Takeuchi, N. Tsuji, F. China, M. Hidaka, Y. Yamanashi, N. Yoshikawa, "Three-dimensional adiabatic quantum-flux-parametron fabricated using a double-active-layered niobium process," Superconductor Science and Technology, vol. 30, no. 7, pp. 075003, (2017). Accessed: May 12, 2024. doi: 10.1088/1361-6668/aa6ef4. [Online]. Available: https://iopscience.iop.org/article/10.1088/1361-6668/aa6ef4/meta
[16] S. Shankar and A. Reuther, "Trends in Energy Estimates for Computing in AI/Machine Learning Accelerators, Supercomputers, and Compute-Intensive Applications," *2022 IEEE High Performance Extreme Computing Conference (HPEC)*, Waltham, MA, USA, 2022, pp. 1-8. Accessed: May 12, 2024. doi: 10.1109/HPEC55821.2022.9926296. [Online]. Available: https://ieeexplore.ieee.org/document/9926296
[17] Chatterjee, Niladrish, Mike O'Connor, Donghyuk Lee, Daniel R. Johnson, Stephen W. Keckler, Minsoo Rhu, and William J. Dally, "Architecting an energy-efficient dram system for GPUs." In 2017 IEEE International Symposium on High Performance Computer Architecture (HPCA), pp. 73-84. IEEE, 2017. Accessed: May 12, 2024. doi: 10.1109/HPCA.2017.58. [Online]. Available: https://ieeexplore.ieee.org/abstract/document/7920815
[18] D.S. Bassett, D.L. Greenfield, A. Meyer-Lindenberg, D.R. Weinberger, S.W. Moore, and E.T. Bullmore, "Efficient physical embedding of topologically complex information processing networks in brains and computer circuits," PLoS computational biology, vol. 6, no. 4, pp. e1000748, (2010). Accessed: May 12, 2024. doi: 10.1371/journal.pcbi.1000748. [Online]. Available: https://journals.plos.org/ploscompbiol/article?id=10.1371/journal.pcbi.1000748
[19] Sporns, Olaf. *Networks of the Brain*. MIT press, 2016.
[20] Buzsáki, György. *Rhythms of the Brain*. Oxford university press, 2006.

# Quantum Photonic Reservoir Computing

**Miguel C. Soriano, Gian Luca Giorgi, and Roberta Zambrini**
Instituto de Física Interdisciplinar y Sistemas Complejos (IFISC, UIB-CSIC),
Campus Universitat de les Illes Balears E-07122, Palma de Mallorca, Spain
[miguel@ifisc.uib-csic.es, gianluca@ifisc.uib-csic.es, roberta@ifisc.uib-csic.es]

**Status**

Quantum photonic reservoir computing is a rapidly evolving interdisciplinary field that brings together concepts from quantum computing and simulations, photonics, and reservoir computing. By integrating these disciplines, the principles of quantum mechanics can be harnessed to process information in a highly parallel and efficient manner. This field was triggered by the proposal to extend the reservoir computing paradigm to the quantum regime [1] and quickly extended to the photonic domain [2, 3].

This innovative framework leverages the unique properties of quantum systems, as quantum coherence, superposition, entanglement, interference, and an exponentially large Hilbert space, to achieve unprecedented processing capabilities powered by reservoir computing (RC). RC is a supervised machine learning approach tailored for time series processing, rooted in recurrent neural networks (RNNs), and can be implemented in software or directly on a physical substrate. The advent of photonic implementations of RC allowed for ultra-fast demonstrators for spoken digits recognition or time series prediction [4]; this development also widened the ways to design reservoirs, including implementations using time or frequency multiplexing. The recent interest in quantum substrates is driven by the potential to further expand the possibilities and designs of RC and to boost the number of neurons exponentially with respect to the physical network units. In this context, *quantum photonic* reservoir computing, schematically illustrated in Fig. 1, promises several key advantages: high expressivity, scalability, high-speed processing, room-temperature operation, and low power consumption. The parallel processing capabilities inherent in quantum photonic systems have the potential to surpass the efficiency of classical computers. Furthermore, photonic systems can operate at extremely high rates, outpacing their electronic counterparts. Notably, the lower power consumption of these systems, especially when utilizing passive components, presents a significant advantage that has already been exploited in classical photonic approaches to RC.

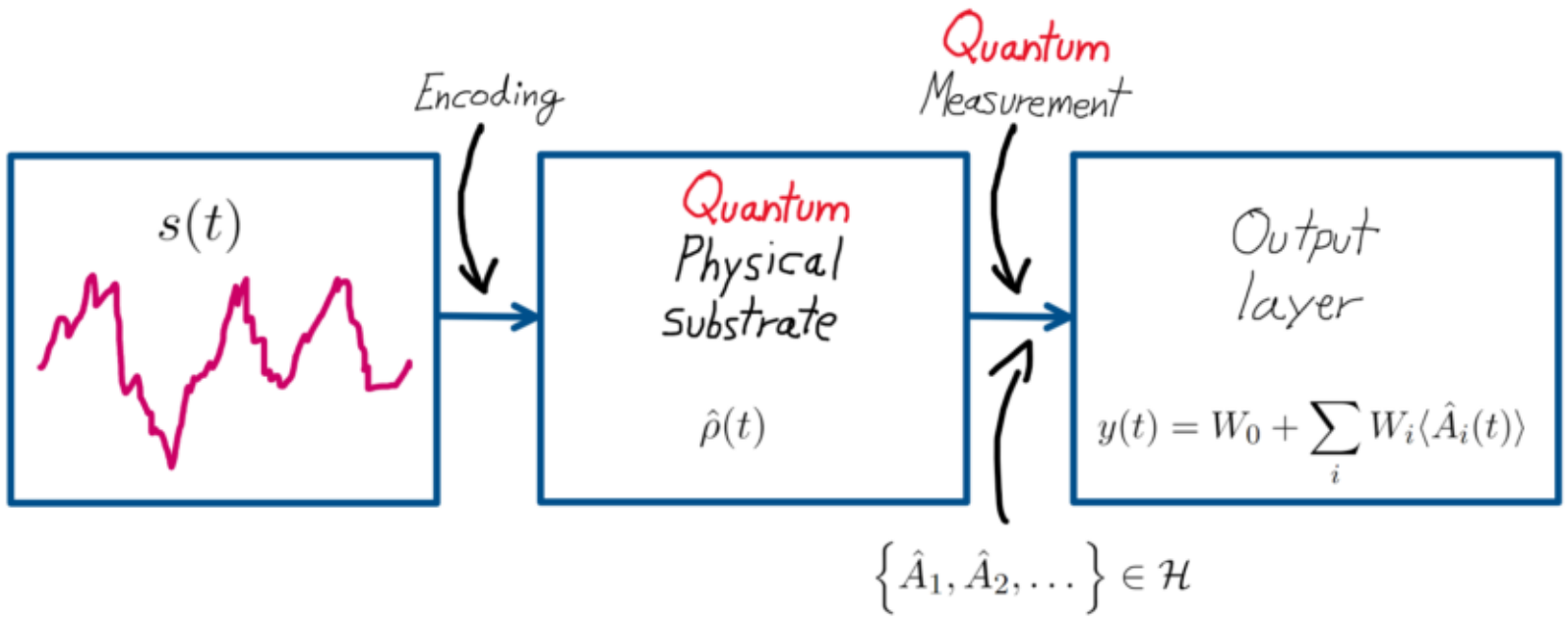

**Figure 1.** Schematic representation of quantum reservoir computing. In this example, a classical signal *s(t)* is encoded in a quantum physical substrate whose dynamics is described by the evolution of the density matrix $\rho(t)$. The output *y(t)* of the quantum reservoir computer is then constructed as a weighted linear combination of system observables $A_i(t)$, obtained via quantum measurement. The output weights $W_i$ are trained through supervised learning to approximate a target output, tailored to the specific computing task.

Moreover, the intrinsic robustness of the reservoir computing paradigm to noise and errors renders it suitable for noisy intermediate-scale quantum (NISQ) computing platforms [5, 6]. The applications of quantum photonic reservoir computing can span multiple domains, including real-time processing tasks such as time-series analysis in finance, weather forecasting, and health monitoring. Additionally, by exploiting the principles of quantum mechanics, this technology holds promise for developing innovative sensors with unique functionalities in metrology and sensing applications.
While quantum photonic reservoir computing is still an emerging field, it has the potential to revolutionize the way we process information and tackle complex problems, offering a novel perspective on the future of computing and problem-solving.

**Current and Future Challenges**

First proposals for quantum photonic reservoir computing have been of theoretical nature and validated with numerical simulations [2, 7]. The initial challenges from the theoretical point of view included finding ways to encode and decode information, such that the quantum reservoirs would operate in the non-linear input-output regime [8, 9], and the ability to incorporate non-destructive measurement protocols [7]. Recent proposals have shown successful ways to address these challenges in parametric processes. For example, some utilize squeezed light in a continuous-variable description based on homodyne detection [7], while others are based on single-photon detection and occupation probabilities [10, 11]. A boost in performance can also be offered by hybrid classical-quantum architectures.

The quest for experimental demonstrations has naturally followed after the first theoretical works, with significant efforts to identify suitable technological platforms. In this context, we would like to highlight recent work that has demonstrated essential parts of a photonic quantum reservoir computer based on a photonic quantum memristor, which possesses the desired properties of non-linearity and memory [10]. Other suggested technological platforms include systems based on e.g. exciton-polaritons in semiconductor quantum dots [12].

Several challenges lie ahead to validate the theoretical and numerical predictions in a practical system. One notable hurdle is scalability, which is essential for a quantum reservoir to have a large number of accessible degrees of freedom. Depending on the implementation, the number of degrees of freedom can scale exponentially or polynomially with the system size. For instance, a quadratic scaling with the number of modes was reported in [3] using Gaussian states of light, implying that computationally hard tasks require around 10 modes to be solved accurately.

In the quantum reservoir computing literature, applications related to either classical tasks [13] or quantum tasks [6, 14] have been suggested. Meanwhile, classical tasks have been demonstrated, for instance in superconductor-based quantum reservoir computing. In contrast, experimental demonstrations of quantum tasks have yet to be achieved.

**Advances in Science and Technology to Meet Challenges**

Quantum photonic reservoir computing has benefited from recent advancements in quantum technologies and will continue to do so as the field evolves. These improvements will enable the development of operative hardware implementations of this technique. One area of progress is in integrated systems for quantum photonics, which shares common ground with similar trends in classical photonic reservoir computing. Other platforms being perfected that have a great potential for quantum photonic reservoir computing include frequency comb-based quantum networks [15] and exciton-polariton systems in quantum dots [2]. A significant challenge that needs to be addressed in these hardware platforms is the efficient injection and extraction of information. Figure 2 illustrates a theoretical suggestion for a possible implementation of quantum photonic reservoir computing where the use of multimode light for the reservoir allows for a powerful solution to continuous monitoring. Moreover, current strategies employed in generic photonic quantum computing, such as the use of clusters and teleportation, could also be considered for quantum reservoir computing [16].

A relevant issue in quantum information processing concerns the fact that to find the expectation value of system observables, many copies (ensemble) of the system itself are required. This is particularly pertinent in the context of quantum machine learning in general and quantum RC in particular, as accurate knowledge of such observables is essential for the construction of the output layer. A strategy for determining them with high accuracy while minimizing the number of measurements that need to be performed was proposed in [17]. These results suggest that methods based on quantum trajectories deserve to be further explored.

Finally, while we have focused the discussion on systems that operate in the visible or infrared ranges, microwave photonics is an interesting complementary approach [18] to the ones discussed here.
By benefiting from these scientific and technological advancements, the development of quantum photonic reservoir computing can be accelerated.

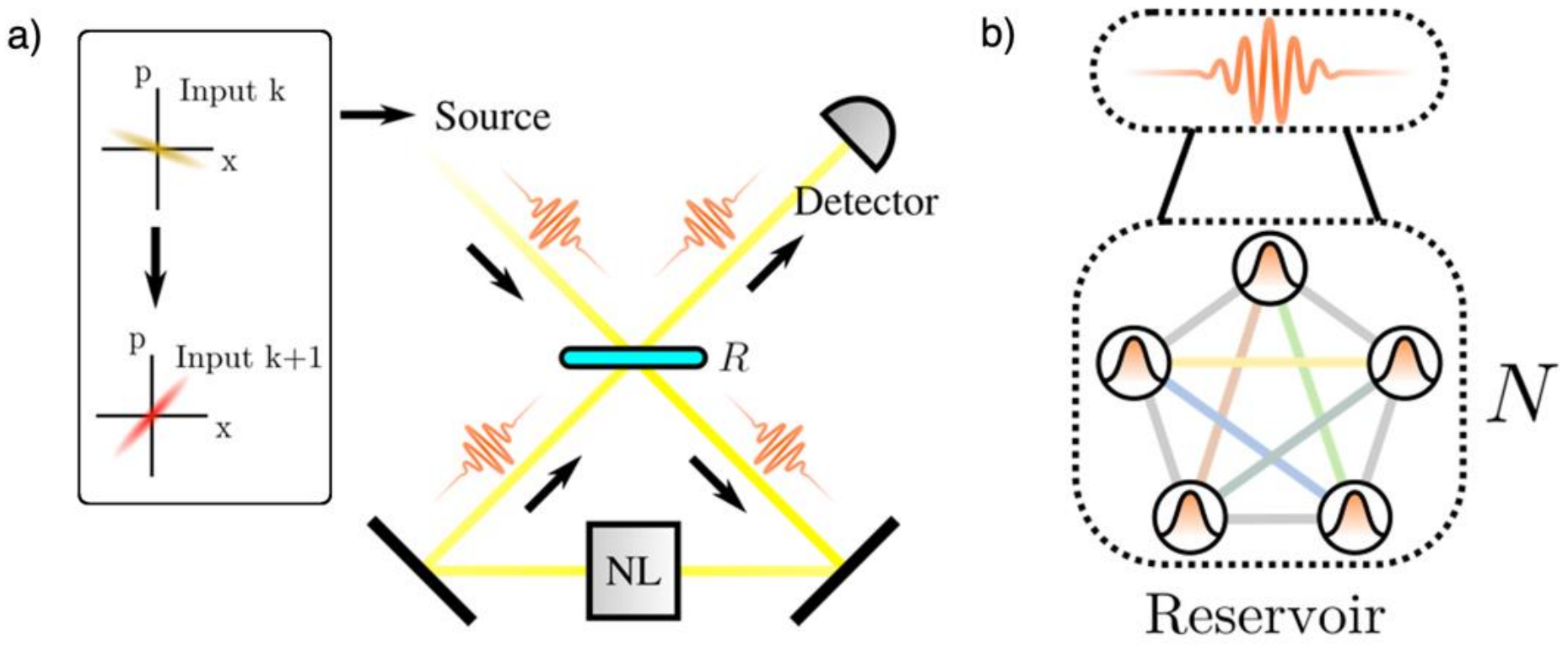

**Figure 2**. (a) Schematic representation of a photonic quantum reservoir computer based on a frequency comb generated by short light pulses. The external information input is encoded in the angle of squeezing of each optical pulse. The optical setup consists of a nonlinear element within a feedback loop and a homodyne detector. (b) The reservoir comprises multiple frequency modes that are coupled, forming a complex quantum network. Image adapted from [GarciaBeni24].

## Concluding Remarks

The field of quantum reservoir computing is still at an early stage of development, yet its maturity can be observed in the first proof-of-principle experiments already reported. Among the promising avenues for the efficient implementation and application to timely and meaningful tasks are photonic platforms, in conjunction with complementary approaches based on superconducting qubits [19]. Furthermore, quantum photonic reservoir computing has the potential to facilitate fruitful interactions with other photonic quantum RNNs, such as those described in [20], which are designed to address temporal tasks and face comparable challenges to those described here. The possibility to boost other quantum technologies in quantum sensing or communications, as well as the possible interplay with quantum simulations are key avenues to explore. In conclusion, the combination of the advantages of photonics (rapid processing speed, low energy consumption, low decoherence) with the unique characteristics of quantum mechanics has the potential to become one of the most powerful unconventional computing protocols to be implemented in the near future.

## Acknowledgements

We acknowledge the Spanish State Research Agency, through the María de Maeztu project CEX2021-001164-M funded by the MICIU/AEI/10.13039/501100011033, the COQUSY project PID2022-140506NB-C21 and -C22 funded by MICIU/AEI/10.13039/501100011033, and the INFOLANET project PID2022-139409NB-I00 funded by MICIU/AEI/10.13039/501100011033, MINECO through the QUANTUM SPAIN project, and EU through the RTRP - NextGenerationEU within the framework of the Digital Spain 2025 Agenda. The CSIC Interdisciplinary Thematic Platform (PTI) on Quantum Technologies in Spain is also acknowledged.

## References

[1] K. Fujii and K. Nakajima, “Harnessing Disordered-Ensemble Quantum Dynamics for Machine Learning”, Physical Review Applied, vol. 8, no. 2, p. 024030, Aug. 2017, doi:10.1103/PhysRevApplied.8.024030

[2] S. Ghosh, A. Opala, M. Matuszewski, T. Paterek, and T. C. Liew, “Quantum Reservoir Processing,” npj Quantum Information, vol. 5, p. 35, Apr. 2019, doi:10.1038/s41534-019-0149-8

[3] J. Nokkala et al., “Gaussian states of continuous-variable quantum systems provide universal and Versatile Reservoir Computing,” Communications Physics, vol. 4, p. 53, Mar. 2021, doi:10.1038/s42005-021-00556-w

[4] D. Brunner, M. C. Soriano, and G. V. der Sande, Eds., Photonic Reservoir Computing: Optical Recurrent Neural Networks. Berlin, Boston: De Gruyter, Jul. 2019, doi:10.1515/9783110583496

[5] P. Mujal et al., “Opportunities in Quantum Reservoir Computing and Extreme Learning Machines,” Advanced Quantum Technologies, vol. 4, no. 8, p. 2100027, Aug. 2021, doi:10.1002/qute.202100027

[6] S. Ghosh, K. Nakajima, T. Krisnanda, K. Fujii, and T. C. H. Liew, “Quantum Neuromorphic Computing with Reservoir Computing Networks,” Advanced Quantum Technologies, vol. 4, no. 9, p. 2100053, Sep. 2021, doi:10.1002/qute.202100053

[7] J. García-Beni, G. L. Giorgi, M. C. Soriano, and R. Zambrini, “Scalable photonic platform for real-time quantum reservoir computing,” Physical Review Applied, vol. 20, no. 1, p. 014051, Jul. 2023, doi:10.1103/physrevapplied.20.014051

[8] P. Mujal, J. Nokkala, R. Martínez-Peña, G. L. Giorgi, M. C. Soriano, and R. Zambrini, “Analytical evidence of nonlinearity in qubits and continuous-variable quantum reservoir computing,” Journal of Physics: Complexity, vol. 2, no. 4, p. 045008, Nov. 2021, doi:10.1088/2632-072X/ac340e

[9] H. Xu, T. Krisnanda, R. Bao, and T. C. H. Liew, “The roles of Kerr nonlinearity in a bosonic quantum neural network,” New Journal of Physics, vol. 25, no. 2, p. 023028, Feb. 2023, doi:10.1088/1367-2630/acbc43

[10] M. Spagnolo et al., “Experimental photonic quantum memristor,” Nature Photonics, vol. 16, no. 4, pp. 318–323, Mar. 2022, doi:10.1038/s41566-022-00973-5

[11] J. Dudas et al., “Quantum Reservoir Computing implementation on coherently coupled quantum oscillators,” npj Quantum Information, vol. 9, no. 1, p. 64, Jul. 2023, doi:10.1038/s41534-023-00734-4

[12] A. Opala and M. Matuszewski, “Harnessing Exciton-Polaritons for Digital Computing, neuromorphic computing, and optimization [invited],” Optical Materials Express, vol. 13, no. 9, p. 2674, Aug. 2023, doi:10.1364/ome.496985

[13] A. Labay-Mora, J. García-Beni, G. L. Giorgi, M. C. Soriano and R. Zambrini, “Neural networks with quantum states of light,” Philosophical Transactions Royal Society A, vol. 382, p. 20230346, Jan. 2025, doi:10.1098/rsta.2023.0346

[14] J. Nokkala, “Online quantum time series processing with random oscillator networks,” Scientific Reports, vol. 13, p. 7694, May 2023, doi:10.1038/s41598-023-34811-7

[15] J. Nokkala et al., “Reconfigurable optical implementation of Quantum Complex Networks,” New Journal of Physics, vol. 20, no. 5, p. 053024, May 2018, doi:10.1088/1367-2630/aabc77

[16] J. García-Beni, I. Paparelle, V. Parigi, G. L. Giorgi, M. C. Soriano, and R. Zambrini, “Quantum machine learning via continuous-variable cluster states and teleportation,” arXiv preprint arXiv:2411.06907

[17] P. Mujal, R. Martínez-Peña, G. L. Giorgi, M. C. Soriano, and R. Zambrini, “Time-series quantum reservoir computing with weak and projective measurements,” npj Quantum Information, vol. 9, p. 16, Feb. 2023, doi:10.1038/s41534-023-00682-z

[18] A. Senanian et al., “Microwave signal processing using an analog quantum reservoir computer,” Nature Communications, vol. 15, p. 7490, Aug. 2024, doi:10.1038/s41467-024-51161-8

[19] F. Hu et al., “Tackling Sampling Noise in Physical Systems for Machine Learning Applications: Fundamental Limits and Eigentasks,” Physical Review X, vol. 13, p. 041020, Oct. 2023, doi:10.1103/PhysRevX.13.041020

[20] R. De Prins, G. Van der Sande, and P. Bienstman, “A recurrent Gaussian quantum network for online processing of quantum time series,” Scientific Reports, vol. 14, p. 12322, May 2024, doi:10.1038/s41598-024-61004-7

# Quantum photonic neural networks

**Jacob Ewaniuk[1,2] and Nir Rotenberg[1,3]**
[1] Centre for Nanophotonics, Department of Physics, Engineering Physics & Astronomy, 64 Bader Lane, Queen's University, Kingston, Ontario, Canada K7L 3N6
[2] jacob.ewaniuk@queensu.ca
[3] nir.rotenberg@queensu.ca

## Status

Quantum photonic neural networks (QPNNs) are reconfigurable nonlinear photonic circuits designed in analogy with conventional neural networks. In its simplest form, the architecture, visualized in Fig. 1a, is comprised from a series of layers of reconfigurable linear interferometric meshes (that act like weights) interspaced with single-site optical nonlinearities (activation functions), through which Fock states flow [1]. In practice, the whole network should be integrated within a photonic chip, ideally together with high-quality single photon sources and, for some applications, detectors.

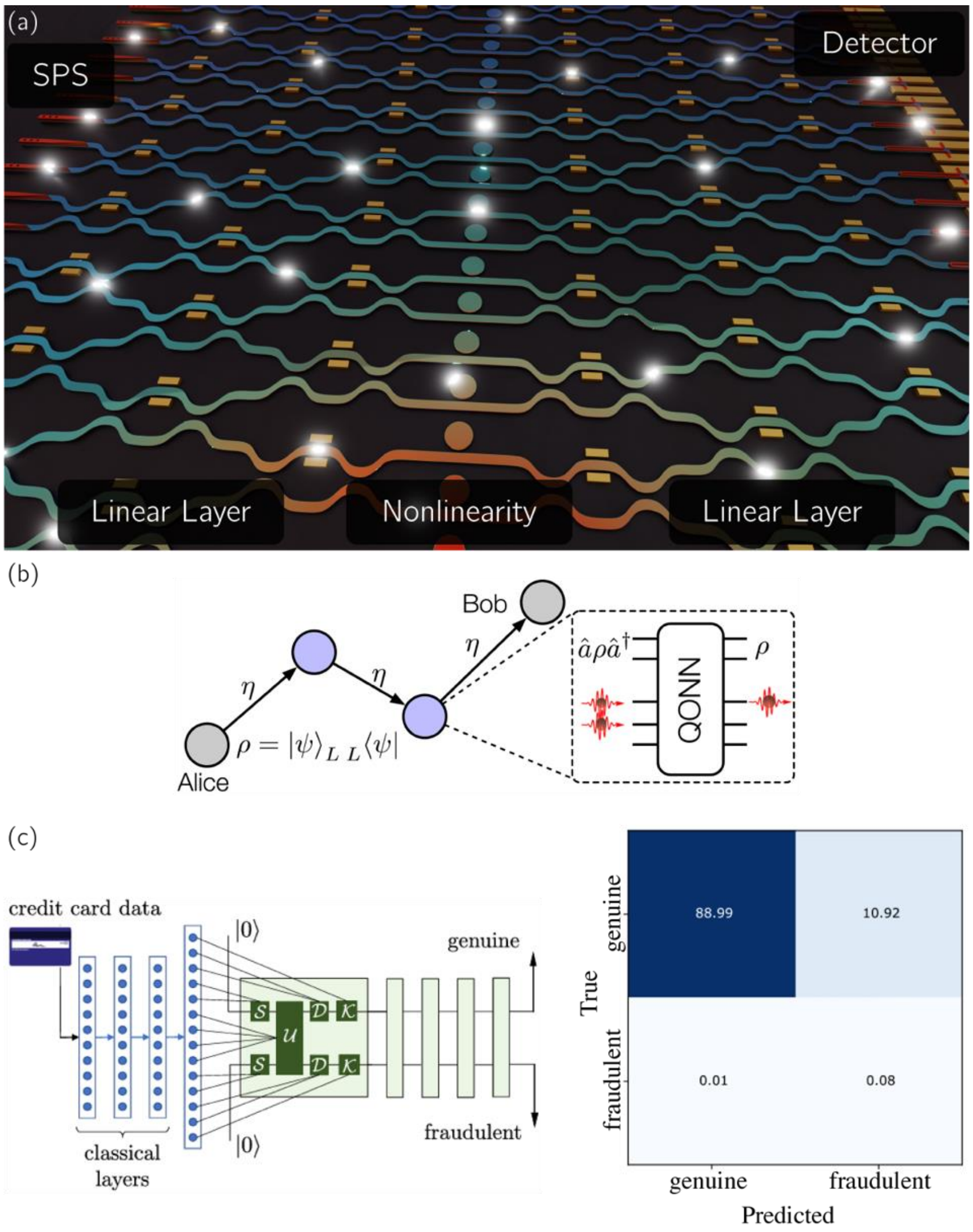

**Figure 1.** (a) Exemplary visualization of a QPNN with all its components integrated on a single photonic chip. (b) Application of a QPNN as a one-way quantum repeater node. Using ancillary photons and optical modes, the QPNN is trained to repair the logical state after it experiences loss while traversing through communication channels with transmissivity . Reprinted from [1]. (c) Using QPNNs for fraud detection. Classical layers are used to pre-process the data such that it can be encoded in quantum states that then traverse through this continuous-variable network, eventually classifying the data as genuine or fraudulent. Reprinted from [5].

The exploration of QPNNs has only just began, and at the scale of the full networks has been largely limited to simulations and theoretical studies. Although nascent, these studies already predict the power and potential of QPNNs to revolutionize a variety of quantum and machine learning (ML) technologies. Moreover, these simulations provide a roadmap highlighting the photonic building blocks and enabling technologies that must be developed for large-scale QPNNs to be viable.

The excitement surrounding QPNNs arises from their potential to disrupt current quantum technologies (Fig. 1). Already in the pioneering work that introduced QPNNs, Steinbrecher *et al*. demonstrated that an ideal network was capable of deterministically (i.e., perfectly, every time) performing quantum information processing (QIP) tasks ranging from the two-qubit gates needed for quantum computation to entanglement generation and detection for quantum communication and sensing [1], an example of which is shown in Fig. 1b. Subsequently, QPNNs were predicted to achieve optimal cloning of quantum states [2], perform quantum tomography [3], and enhance typical ML tasks such as image recognition [4] or fraud detection (c.f., Fig. 1c) [5]. By combining optical nonlinearities with quantum properties such as entanglement and quantum interference, QPNNs can learn to perform these tasks with near-perfect success rates and fidelities, well beyond what is possible with classical or even linear quantum photonic circuits.

Even better, QPNNs can learn to overcome circuit imperfections, including unavoidable losses, imperfect photon routing and sub-optimal nonlinearities [6]. Realistic components, however, impose limitations on the QPNN architecture. The introduction of losses, for example, unavoidably reduces the success rate of the network, though its (conditional) fidelity need not decrease (in the case where all photons successfully traverse the network). Since losses scale with circuit length, the maximum size of a realistic network depends on the amount of loss. For example, considering QPNNs designed for two-qubit operations, those constructed from state-of-the-art silicon (silicon nitride) circuits have 0.3 dB/cm [7] (0.01 dB/cm [8]) losses, such that 3% (0.1%) of the photons will be lost in a 2-layer network, which increases to 10% (0.3%) at 6 layers. Conversely, overcoming sub-optimal (i.e., weaker) nonlinearities requires additional layers, creating a delicate balance between the different imperfections that results in optimal network geometries [6]. As we discuss below, this suggests that hybrid approaches may be vital to the realization of QPNNs, with their scalability directly linked to the quality of each individual component and the ways in which they are interfaced. Finally, we note that although here we mainly discuss QPNNs based on discrete-variable QIP, much of what we conclude directly applies to their continuous-variable counterparts that use squeezed states [5].

**Current and Future Challenges**

At this early stage, many challenges to the development of both the theory and hardware for QPNNs remain. Although, as noted above, several powerful applications of QPNNs have already been identified, this list is by no means comprehensive. Even as more applications are identified, it also remains to be seen which network architecture is best for each (starting from discrete versus continuous networks), in much the same way that the strengths of the many different types of classical photonic neural networks discussed in this roadmap are suitable for different tasks. Additionally, efficient training algorithms are needed for QPNNs. These will likely require efficient quantum state tomography, possibly enhanced by machine learning [9], and full or partial *in situ* training approaches [10], as simulations of quantum networks on classical machines rapidly become computationally prohibitive as the networks scale.

Quantum photonic neural networks are inherently complex devices, comprised of multiple delicate photonic elements (c.f., Fig. 1a). Many components, such as low-loss photonic waveguides, reconfigurable phase shifters, directional couplers, single-photon sources and detectors already exist, yet often on disparate platforms. For example, the low-loss elements that make up a reconfigurable linear mesh, which acts as the weights for the network, are typically based on a silicon or silicon nitride platform with thermo-optical phase shifters [7]. In contrast, on-demand single-photon sources are typically based on III-V semiconductor quantum dots [11], while high-efficiency single-photon detectors are comprised of superconducting nanowires [12], both of which operate at cryogenic temperatures. To address this disparity, much of the current research focuses on creating cryogenically-compatible linear photonic elements, with a secondary focus on the hybrid integration of the various different platforms and components, as we outline below.

Conversely, the quantum optical nonlinearities necessary for QPNNs do not currently exist. Models of discrete-variable QPNNs are currently based on Kerr nonlinearities that impart a $\frac{n(n-1)\varphi}{2}$ phase shift on a passing $n$-photon state. An ideal nonlinearity requires that $\varphi = \pi$ and does not distort nor reshape the photon pulse, although near-ideal operational fidelities are possible for $\varphi = \frac{\pi}{10}$ [6]. However, since the best demonstrations of few-photon nonlinear phase shifts, to date, remain 4 orders-of-magnitude below the ideal [13], increasing this strength or finding alternative optical nonlinearities remains an active challenge. In a similar vein, continuous-variable QPNNs require deterministic non-Gaussian operations at the few-photon level that, currently, can only be applied probabilistically [14].

**Advances in Science and Technology to Meet Challenges**

As outlined above, three main areas of technological development are required for the realization of an operational QPNN: (1) cryogenically-compatible photonic elements, (2) hybrid integration and, (3) deterministic few-photon nonlinearities. While some progress has been made towards each of these areas, as shown by the examples in Fig. 2, much remains to be done, as we outline here.

Cryogenic photonic elements must be compact, both to reduce losses and allow the QPNNs to scale. Current technology based on electro-optical devices, for example on the lithium niobate platform, is cryogenically-compatible, low-loss and can be fabricated into circuits [15]. Yet, these devices have very large footprints, often with dimensions in the 100s of microns, limiting circuit sizes. Recently, nano-mechanical devices have been fabricated, where applied electric fields induce a motion that, in turn, modulates their optical properties, as shown in Fig. 2b [16]. While these are relatively small, may be relatively fast, and have been successfully interfaced with deterministic single-photon sources, the technology required to integrate many of these into a single device has yet to be developed.

Likewise, hybrid integration of quantum photonic circuits is also in its infancy. Several approaches have been explored, to date, including pick-and-place (c.f., Fig. 2a) [8] and nanomanipulator transfer processes [17]. These focus on interfacing linear photonic circuitry with single-photon sources or detectors and represent a promising first step towards large-scale integration. Yet, several aspects must be improved. First, the quantum emitters used must have lifetime-limited transitions if they are to emit indistinguishable photons, likely requiring electrical gating of the waveguides in which they are embedded. Hence, simultaneous electronic and photonic integration will be required. Second, the insertion losses the hybrid systems must be reduced, far surpassing the current best coupling efficiency of 24% [17].

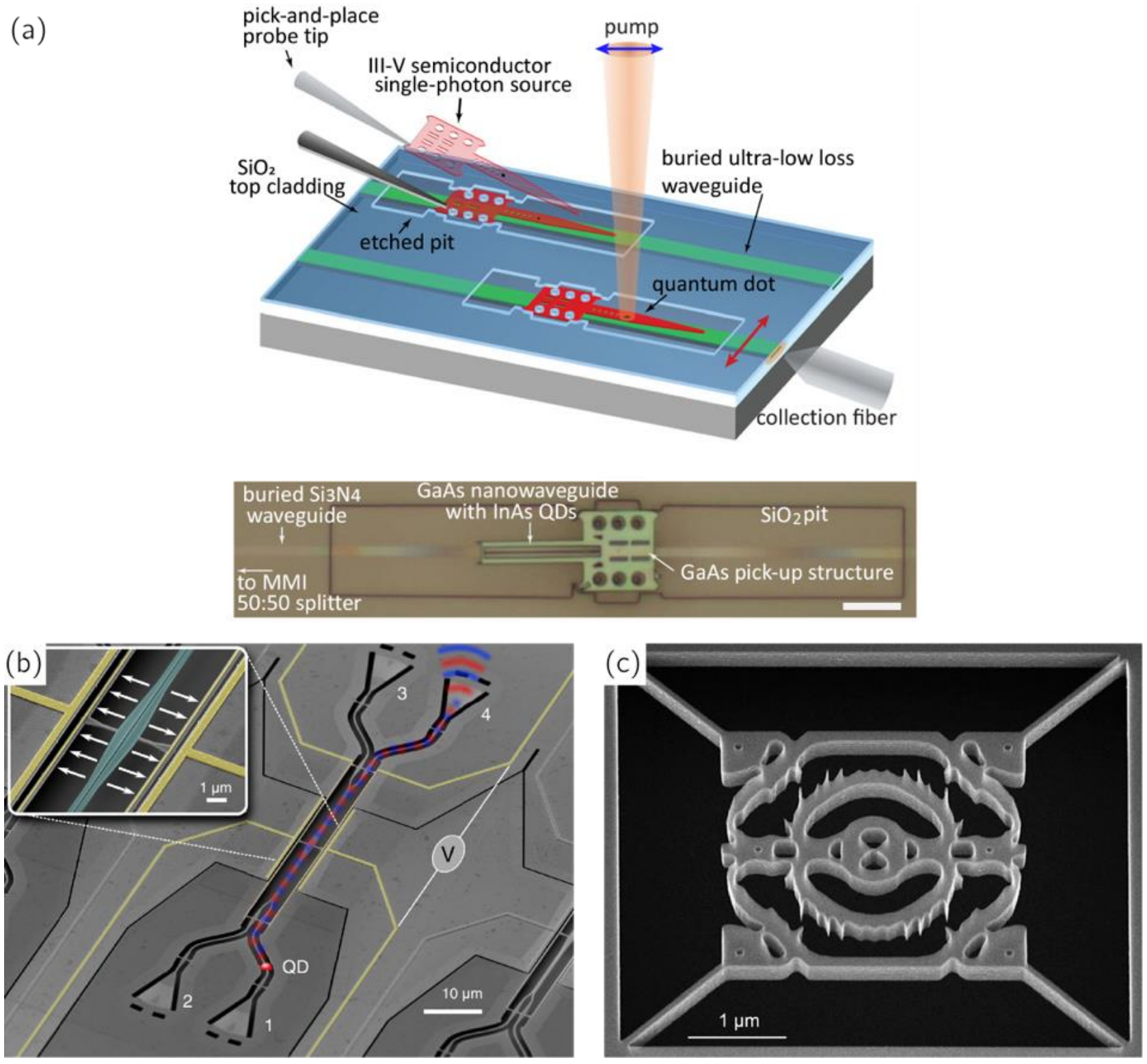

**Figure 2.** (a) Hybrid integration of a GaAs photonic element, including an InAs quantum dot single-photon source, on a silicon nitride platform. A pick-and-place process is applied to attach the single-photon source, which is terminated with a mode transformer designed to couple emitted photons to the silicon nitride waveguide. Reprinted from [8]. (b) Nano-opto-electromechanical single-photon router fabricated on a GaAs platform, integrated with an InAs quantum dot single-photon source. When a voltage is applied across the electrodes (yellow), mechanical motion is induced at the waveguides to modulate the photon routing. Reprinted with permission from [16] © Optical Society of America. (c) SEM image of a topology-optimized dielectric bowtie cavity fabricated in silicon. Reprinted from [19].

Finally, an efficient (and integrated) few-photon nonlinearity that can act as an activation function (or perform an on-demand non-Gaussian transformation) is still missing. To bring Kerr nonlinearities to the requisite strengths, high-quality photonic cavities with ultrasmall mode volumes (e.g., quality factor on the order of $10^9$ with mode volume of $2.5 \times 10^{-5}\lambda^3$in silicon) must be created [18]. Recent advances in inverse design, coupled with nanofabrication, bring this closer to reality, as shown in Fig. 2c [19], yet further enhancement and interfacing with waveguides are still required. Alternatively, a strong nonlinear scattering of two-photon pulses has recently been observed from well-coupled and coherent quantum dots [20]. A more systematic exploration of these effects, with extension to larger photon number states, may well identify nonlinearities suitable for the realization of QPNNs. This further motivates the development of hybrid integration approaches as noted above.

## Concluding Remarks

The first set of theoretical studies of QPNNs showed us that these devices may far outperform their linear or classical counterparts, providing ML or QIP functionalities beyond what is currently possible. Continuing on this path will likely uncover new areas and fields that these networks may impact, but will also likely teach us how they may best be constructed or even trained. At the same time, we are only now beginning to put together the building blocks for QPNNs, from their linear meshes, to sources and detectors, and finally the photonic nonlinearities that will drive them. Already, the first steps are impressive, with new hardware touching on these aspects under constant development. Yet, reaching the exacting standards of quantum technologies, and integrating the disparate components, remain as open challenges to be overcome if we are to unlock the full potential of these networks.

## Acknowledgements

The authors gratefully acknowledge financial support from the Natural Sciences and Engineering Council of Canada (NSERC), the National Research Council of Canada (NRC) and Queen's University.

## References

[1] G. R. Steinbrecher, J. P. Olson, D. Englund, and J. Carolan, "Quantum optical neural networks," *npj Quantum Information,* vol. 5, no. 1, p. 60, 2019, doi: 10.1038/s41534-019-0174-7.

[2] D. Stanev, N. Spagnolo, and F. Sciarrino, "Deterministic optimal quantum cloning via a quantum-optical neural network," *Phys. Rev. Res.,* vol. 5, no. 1, p. 013139, 2023, doi: 10.1103/PhysRevResearch.5.013139.

[3] Y. Zuo, C. Cao, N. Cao, X. Lai, B. Zeng, and S. Du, "Optical neural network quantum state tomography," *Advanced Photonics,* vol. 4, no. 2, p. 026004, 2022, doi: 10.1117/1.AP.4.2.026004.

[4] R. Parthasarathy and R. T. Bhowmik, "Quantum Optical Convolutional Neural Network: A Novel Image Recognition Framework for Quantum Computing," *IEEE Access,* vol. 9, pp. 103337-103346, 2021, doi: 10.1109/ACCESS.2021.3098775.

[5] N. Killoran, T. R. Bromley, J. M. Arrazola, M. Schuld, N. Quesada, and S. Lloyd, "Continuous-variable quantum neural networks," *Phys. Rev. Res.,* vol. 1, no. 3, p. 033063, 2019, doi: 10.1103/PhysRevResearch.1.033063.

[6] J. Ewaniuk, J. Carolan, B. J. Shastri, and N. Rotenberg, "Imperfect Quantum Photonic Neural Networks," *Advanced Quantum Technologies,* vol. 6, no. 3, p. 2200125, 2023, doi: 10.1002/qute.202200125.

[7] J. Wang, F. Sciarrino, A. Laing, and M. G. Thompson, "Integrated photonic quantum technologies," *Nature Photonics,* vol. 14, no. 5, pp. 273-284, 2020, doi: 10.1038/s41566-019-0532-1.

[8] A. Chanana *et al.*, "Ultra-low loss quantum photonic circuits integrated with single quantum emitters," *Nature Communications,* vol. 13, no. 1, p. 7693, 2022, doi: 10.1038/s41467-022-35332-z.

[9] G. Torlai, G. Mazzola, J. Carrasquilla, M. Troyer, R. Melko, and G. Carleo, "Neural-network quantum state tomography," *Nature Physics,* vol. 14, no. 5, pp. 447-450, 2018, doi: 10.1038/s41567-018-0048-5.

[10] L. G. Wright *et al.*, "Deep physical neural networks trained with backpropagation," *Nature,* vol. 601, no. 7894, pp. 549-555, 2022, doi: 10.1038/s41586-021-04223-6.

[11] P. Senellart, G. Solomon, and A. White, "High-performance semiconductor quantum-dot single-photon sources," *Nature Nanotechnology,* vol. 12, no. 11, pp. 1026-1039, 2017, doi: 10.1038/nnano.2017.218.

[12] L. You, "Superconducting nanowire single-photon detectors for quantum information," *Nanophotonics,* vol. 9, no. 9, pp. 2673-2692, 2020, doi: doi:10.1515/nanoph-2020-0186.

[13] V. Venkataraman, K. Saha, and A. L. Gaeta, "Phase modulation at the few-photon level for weak-nonlinearity-based quantum computing," *Nature Photonics,* vol. 7, no. 2, pp. 138-141, 2013, doi: 10.1038/nphoton.2012.283.

[14] N. Namekata, Y. Takahashi, G. Fujii, D. Fukuda, S. Kurimura, and S. Inoue, "Non-Gaussian operation based on photon subtraction using a photon-number-resolving detector at a telecommunications wavelength," *Nature Photonics,* vol. 4, no. 9, pp. 655-660, 2010, doi: 10.1038/nphoton.2010.158.

[15] E. Lomonte *et al.*, "Single-photon detection and cryogenic reconfigurability in lithium niobate nanophotonic circuits," *Nature Communications,* vol. 12, no. 1, p. 6847, 2021, doi: 10.1038/s41467-021-27205-8.

[16] C. Papon *et al.*, "Nanomechanical single-photon routing," *Optica,* vol. 6, no. 4, pp. 524-530, 2019, doi: 10.1364/OPTICA.6.000524.

[17] I. E. Zadeh *et al.*, "Deterministic Integration of Single Photon Sources in Silicon Based Photonic Circuits," *Nano Letters,* vol. 16, no. 4, pp. 2289-2294, 2016, doi: 10.1021/acs.nanolett.5b04709.

[18] M. Heuck, K. Jacobs, and D. R. Englund, "Controlled-Phase Gate Using Dynamically Coupled Cavities and Optical Nonlinearities," *Phys. Rev. Lett.,* vol. 124, no. 16, p. 160501, 2020, doi: 10.1103/PhysRevLett.124.160501.

[19] M. Albrechtsen *et al.*, "Nanometer-scale photon confinement in topology-optimized dielectric cavities," *Nature Communications,* vol. 13, no. 1, p. 6281, 2022, doi: 10.1038/s41467-022-33874-w.

[20] H. Le Jeannic *et al.*, "Dynamical photon-photon interaction mediated by a quantum emitter," *Nature Physics,* vol. 18, no. 10, pp. 1191-1195, 2022, doi: 10.1038/s41567-022-01720-x.

# Integrated Photonic Reservoir Computing

**Peter Bienstman**
Department of Information Technology, Ghent University/imec, Technologiepark-Zwijnaarde 126, B-9052 Gent, Belgium

Peter.Bienstman@UGent.be

## Status

Reservoir computing [1-3] (Figure 1) arose in the early 2000s as a method to simplify the training of recurrent neural networks for time series processing. Rather than optimising all the internal weights in the recurrent network (the so-called reservoir), these weights are instead randomly initialised and kept fixed. The only system parameters that are trained are the weights in the readout layer, which performs a simple linear combination of the time traces of all the internal nodes. This training is done by minimising the squared distance between the actual output time trace and the desired time trace, and this can be achieved using a closed-form solution involving matrix pseudo-inverses. Since no iterative method like backpropagation is needed, this simplifies training enormously.

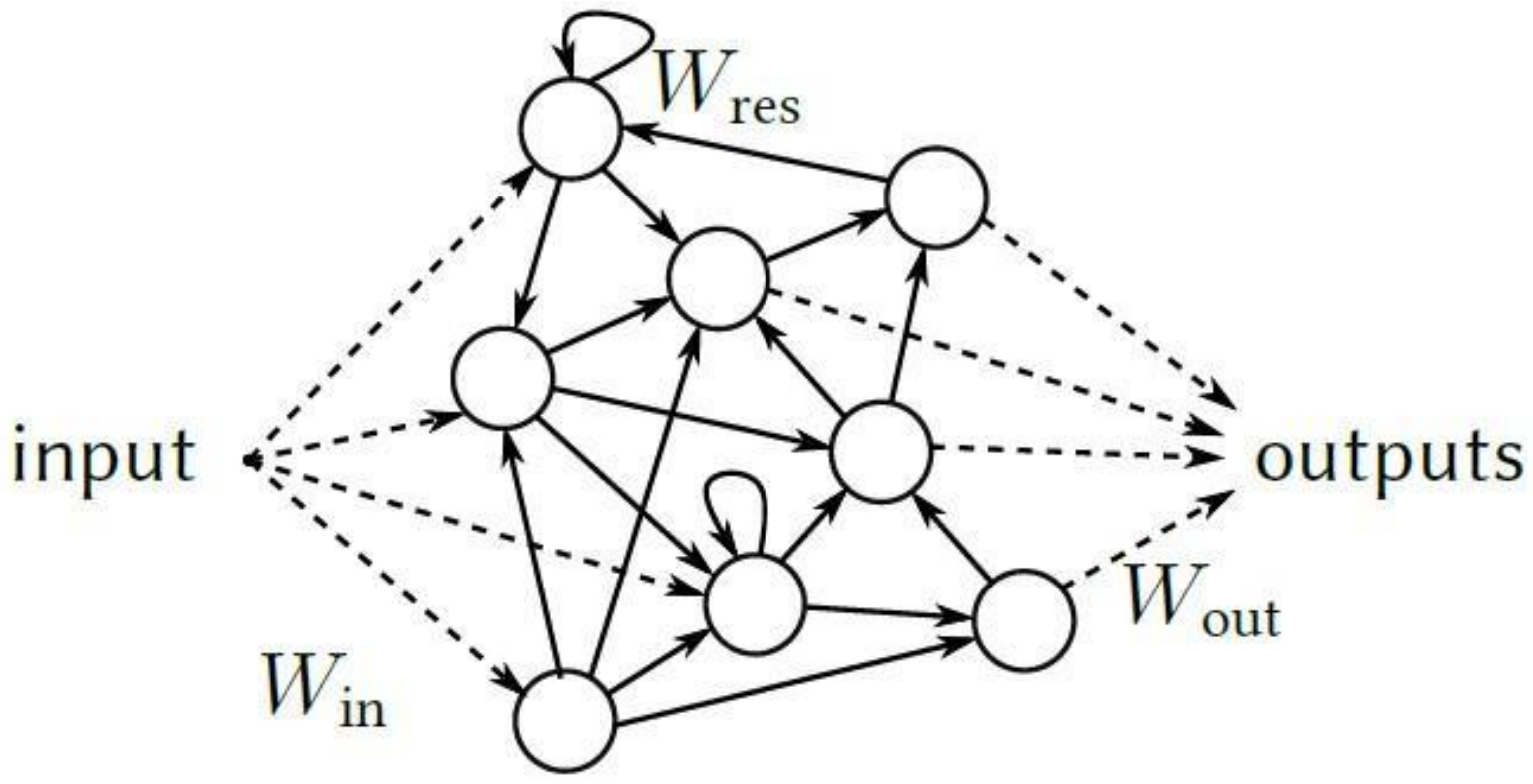

**Figure 1.** In reservoir computing, the internal reservoir weights Wres are randomly initialised and left untrained. Typically, only the output weights Wout are trained through a simple closed-form linear regression expression. The input weights Win are randomly generated as well, although in some cases they can be trained.

The reservoir paradigm lends itself naturally to hardware implementations, since a large class of nonlinear dynamical systems can function as a reservoir. Recently, photonic implementations of reservoir computing have been the subject of active research. Different flavours have been investigated [4,5], e.g. fibre-based systems and free-space systems. Here however, we will focus on integrated chip-based approaches, e.g. in silicon photonics, which can be compact, robust, high-speed and cheap to fabricate in volume.

Although recurrent neural networks and reservoir computing have recently been overshadowed by feedforward neural networks and deep learning, both in the context of software and hardware implementations, they nevertheless can be very relevant for a number of applications. First, because the recurrence in effect 'recycles' the neurons over time, the network does not need to be as large. Indeed, there are several non-trivial applications where reservoirs of a few dozen nodes can achieve respectable performance. Second, hardware implementations do not require stringent fabrication tolerances, since the network is random anyway. Any deviations will be taken into account during training, resulting in a set of bespoke weights for each chip. Finally, a hardware implementation in photonics makes a lot of sense for applications where the input is already in the optical domain (such that no electro-optical conversion penalty needs to be paid) and where traditional electronic implementations are limited in terms of bandwidth and/or latency.

**Current and Future Challenges**

The field has been transitioning from synthetic academic benchmarks (like chaotic time series prediction, or bit-level temporal tasks [6]) to more industrially relevant applications, especially those where the input is in the optical domain and that require high throughput. An obvious example is the processing of telecommunications signals, either in undoing nonlinear dispersion in fibre links [7-9], or in identifying anomalies [10]. Another class of promising problems lies in image recognition, e.g. in the analysis of flow cytometry data [11]. Still, reservoir systems are not as general-purpose as e.g. matrix multiplication accelerators, and the challenge is to keep expanding the number of relevant use cases for this technology.

A second issue is related to the efficient training of the reservoir weights, especially when they are implemented optically on-chip, in a so-called optical readout scheme. Although in theory there exists a closed-form solution for the weights, this relies of full observability of all the internal reservoir states, which is in practice not always the case. Additionally, when we are working with coherent signals but are only interested in the amplitude of the final output, linear regression in the complex domain needs to impose a certain phase to the target signal, which is too restrictive. Therefore, novel iterative training schemes need to be investigated, like bespoke ones for optical reservoir readouts [12], or techniques adapted from other fields like particle swarm optimisation [13] or (augmented) direct feedback alignment [14-15].

Finally, more work is needed in exploiting other degrees of freedom that optics offers, like in designing chips that can perform telecommunication tasks on several wavelength division multiplexing (WDM) channels in parallel.

**Advances in Science and Technology to Meet Challenges**

Apart from the issues mentioned above, optimising the on-chip losses is of paramount importance to improve the attractiveness of reservoir computing in commercial applications. This is where further technological evolution can play an important role, e.g. by optimising the fabrication process to reduce scattering losses due to roughness. In this context, the switch to lower-contrast material systems like SiN instead of SOI can play an important role. Finally, better designs of basic building blocks (splitters, combiners, waveguide crossings, …) can also help with the robustness and the losses. It is worth noting however that there is an intrinsic loss mechanism in the optical readout of reservoir computing, where a combiner tree coherently adds the weighted signals coming from different nodes, and each combiner stage inherently has an average 3 dB radiation loss. Switching to systems with reduced coherence or to multimodal systems could be an option here, at least for some applications.

Additionally, in most implementations, the only nonlinearity present in integrated photonic reservoir computing is that of the photodetector. However, for some applications it is worthwhile investigating additional nonlinearities inside the reservoir itself. Further development of nonlinear materials [16] and of heterogeneous integration will be very important here.

**Concluding Remarks**

Integrated photonic reservoir computing has gained interest as an alternative technique to process time-varying signals. This is especially true for those applications where the input is in the optical domain, and where high throughput and low latency is required. Examples of such use cases include nonlinear dispersion compensation in high-speed telecommunication links, and the processing of flow cytometry images. Work is underway to address further challenges like improved training schemes for optical readout, decreased losses and better scaling, and the integration of nonlinear materials and devices.

## Acknowledgements

This work was supported by the European Commission under the H2020 framework program (Nebula, Neoteric, Phoenics, Prometheus, Neuropuls, Neho, Respite), the Flemish FWO projects G006020N and 3S044419, and the Belgian EOS project G0H1422N.

## References

[1] M. Wolfgang, T. Natschläger, and H. Markram, "Real-time computing without stable states: a new framework for neural computation based on perturbations," Neural Comput. 14(11), 2531–2560 (2002).

[2] D. Verstraeten, B. Schrauwen, M. D'Haene, and D. Stroobandt, "An experimental unification of reservoir computing methods," Neural Netw. 20(3), 391–403 (2007).

[3] M. Lukoševičius and H. Jaeger, "Reservoir computing approaches to recurrent neural network training," Comput. Sci. Rev. 3(3), 127–149 (2009).

[4] G. V. der Sande, D. Brunner, and M. C. Soriano, "Advances in photonic reservoir computing," Nanophotonics 6(3), 561–576 (2017).

[5] A. Lugnan, A. Katumba, F. Laporte, M. Freiberger, S. Sackesyn, C. Ma, E.J.C. Gooskens, J. Dambre, P. Bienstman, "Photonic neuromorphic information processing and reservoir computing", APL Photonics, 5(020901) (2020).

[6] K. Vandoorne, P. Mechet, T. V. Vaerenbergh, M. Fiers, G. Morthier, D. Verstraeten, B. Schrauwen, J. Dambre, and P. Bienstman, "Experimental demonstration of reservoir computing on a silicon photonics chip," Nat. Commun. 5(1), 3541 (2014).

[7] S. Sackesyn, C. Ma, J. Dambre, and P. Bienstman, "Experimental realization of integrated photonic reservoir computing for nonlinear fiber distortion compensation," Opt. Express 29(20), 30991–30997 (2021).

[8] K. Sozos, A. Bogris, P. Bienstman, G. Sarantoglou, S. Deligiannidis, C. Mesaritakis, "High-speed photonic neuromorphic computing using recurrent optical spectrum slicing neural networks", Communications Engineering, 1, doi:10.1038/s44172-022-00024-5 (2022).

[9] A. Argyris, J. Bueno, and I. Fischer, "Photonic machine learning implementation for signal recovery in optical communications," Sci. Rep. 8(1), 8487 (2018).

[10] G. von Huenefeld, G. Ronninger, P. Safari, I. Sackey, R. Thomas, P. Cegielski, S. Suckow, E. Seker, D. Stahl, S. Masaad, E.J.C. Gooskens, P. Bienstman, C. Schubert, J. Fischer, R. Freund, "Enabling optical modulation format identification using an integrated photonic reservoir and a digital multiclass classifier", ECOC, Switzerland, p.Tu 5.19 (2021).

[11] M. Gouda, A. Lugnan, J. Dambre, G. V. Branden, C. Posch, P. Bienstman, "Improving the classification accuracy in label-free flow cytometry using event-based vision and simple logistic regression", IEEE Journal on Selected Topics in Quantum Electronics, doi:10.1109/JSTQE.2023.3244040 (2023).

[12] C. Ma, J. Van Kerrebroeck, H. Deng, S. Sackesyn, E.J.C. Gooskens, B. Bai, J. Dambre, P. Bienstman, "Integrated photonic reservoir computing with an all-optical readout", Optics Express, 31(21), p.34843-34854 (2023).

[13] X. Zuo, L. Pei, B. Bai, J. Wang, J. Zheng, T. Ning, F. Dong, Z. Zhao, "Integrated Silicon Photonic Reservoir Computing with PSO Training Algorithm for Fiber Communication Channel Equalization", Journal of Lightwave Technology, doi: 10.1109/JLT.2023.3270025 (2023).

[14] M. Nakajima, K. Inoue, K. Tanaka, Y. Kuniyoshi, T. Hashimoto, and K. Nakajima, "Physical deep learning with biologically inspired training method: gradient-free approach for physical hardware", Nat Commun, 13(1) 1 (2022).

[15] M. J. Filipovich, Z. Guo, M. Al-Qadasi, B.A. Marquez, H.D. Morison, V.J. Sorger, P.R. Prucnal, S. Shekhar, B.J. Shastri, "Silicon photonic architecture for training deep neural networks with direct feedback alignment", Optica, 9(12) 1323–1332 (2022).
[16] M. Wuttig, H. Bhaskaran, T. Taubner, "Phase-change materials for non-volatile photonic applications", Nature Photon, 11(8), 465–476 (2017).

# Programmable Photonic Integrated Circuits for Signal Processing

**Wim Bogaerts**, Ghent University – IMEC
Wim.bogaerts@ugent.be (ORCID: 0000-0003-1112-8950)

**Status**

Photonics presents an exceptional platform for analog signal processing, with robustness to noise and crosstalk and very high signal bandwidths. We can separate photonic signal processing into two classes: (1) processing of optical spectra (filtering, spectrometry, …) which span a wide wavelength band, but which change relatively slowly over time, and (2) processing of high-speed time signals, as microwave signals modulated on an optical carrier (often called 'microwave photonics' [1].) Processing functions include signal generation, conversion between the electrical and optical domain, (cross-)modulation, amplification, frequency conversion and filtering.

Photonic integrated circuits (PIC) implement these functions on the surface of a chip, which provides a compact and stable platform, with a steady growth in component count. Most photonic circuits today only perform the specific functions for which they have been designed, and even with an increasing use of electrical tuners to adjust the circuit parameters, new functionality requires the design and fabrication of a new chip.

Programmable photonics takes a significant step further, by also manipulating the optical connections on the chip to define new functionalities [2]. A typical programmable PIC defines its connections through a waveguide mesh of electrically controlled optical gates that control the coupling between waveguides and their relative phase delay. This implements passive connectivity, distribution and even wavelength filtering. A fully programmable optical signal processor would look like Figure 1, combining a fully reconfigurable waveguide mesh with active optical functions. Today, no such chips with full functionality have been demonstrated. On one hand, there are passive waveguide meshes that can be reconfigured into arbitrary circuits , acting as a switch or wavelength filter, but where the active functions remain off-chip [3], [4]. On the other hand, some chips combine multiple active functions with some degree of reconfigurability, but without full control over the optical path [5], [6].

To realize the full potential of programmable photonics, the scale of the circuits needs to increase, all passive and active functions should be combined on the chip, and the photonics should be closely integrated with its driver electronics and software layers to provide the users with all the knobs to control the flow of light in software and define their own functionality. This functionality can be very diverse, allowing the chip to function as a transceiver, a spectrometer, a microwave filter, a switch, or a neural network, effectively enabling both classes of signal processing on the same platform.

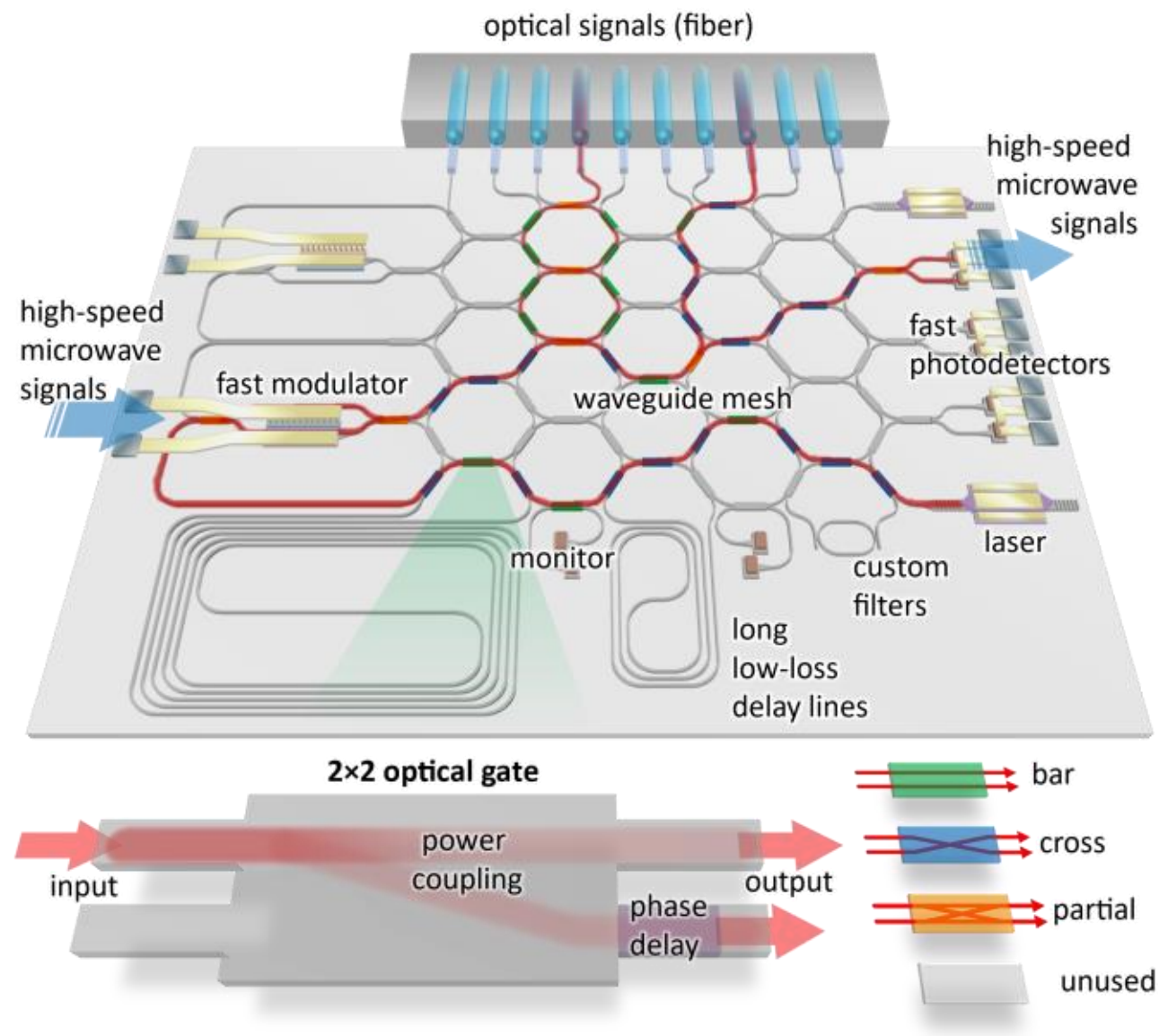

**Figure 1**. Programmable photonic chip with recirculating waveguide mesh. The optical gates control the coupling between the waveguides and their relative phase delay. The mesh is connected to input/output ports and functional building blocks.

## Current and Future Challenges

To realize programmable photonics, a complete technology stack is needed, from the PIC itself, over its driver electronics, to the algorithms and programming layers to support the users. This is schematically represented in Figure 2. On the hardware side, we see challenges within the photonic circuit, but also with the electronics and the packaging. On the PIC itself, the focus is on the key building block: the optical gate. This typically consists of electro-optic phase shifters, and to scale these to larger circuits, they must have low optical loss, low power consumption, fast response time, and a small footprint with short optical path length. They must also be compatible with the existing active building blocks on the platform, and preferably compatible with standard CMOS electronic drivers. Today, no electro-optic phase shifter technology meets all those requirements.

When arranging these gates into waveguide meshes, different connection topologies are possible. Current meshes are quite simple and uniform, based on either a unidirectional or a recirculating connectivity. But these are limited in the functionality they can implement. To go beyond these early examples requires theoretical frameworks for a vast unexplored design space.

The photonic chip is complemented by driver electronics: each gate needs a driver, and there will also be monitors. This requires a large number of electrical connections. As all the building blocks on the PIC are analog, the electronics need sufficient precision, and will have to cope with variability, crosstalk and aging. The photonic-electronic subcircuits will therefore need calibration routines.

This brings us to the many software challenges for programmable photonics. During operation, control routines should keep the circuit in a stable state. Some simpler (forward-only) mesh topologies can be controlled quite easily, but the more complex meshes are more difficult. Just the issue of where to best incorporate optical monitors for control loops is an open question.

Configuration routines should help the user to program the circuit to perform its desired function. Routing connectivity is a first class of problems, but more advanced functions such as delay lines or filters become more complex, especially we move away from a uniform waveguide mesh. Today, programming of a photonic chip is very much like writing machine code, setting the state of each gate, with some simple assembly-language routines to define basic functions. More formal descriptions are needed to make such functions truly useful and shield the users from the specific chip implementations.

**Advances in Science and Technology to Meet Challenges**

Recent technological progress has focused largely on the hardware, and especially the PIC platform. The increased use of tuners pushes the development of better electro-optic phase shifters. Heaters are the workhorse today, but even with recent advances they have too high power consumption and thermal crosstalk. Electro-optic effects used for high-speed modulators are also not suitable: carriers introduce too high loss, and the Pockels effect is generally too weak, making the phase shifter too long. Effects like strain (e.g. actuated by piezo actuators) suffer from the same problem. [7]. Specialty materials, such as barium titanate [8], phase-change materials [9], or liquid crystals [10], offer strong phase shifts, but are still immature. Micromechanical structures (MEMS) have been integrated onto silicon photonic, but still need to prove their worth in terms of robustness [11]. The choice of phase shifters for the optical gates impacts the functionality. In particular, its optical path length will limit the free-spectral range (FSR) of a filter circuit. Current demonstrations or recirculating meshes are limited to an FSR < 100GHz [4].

In parallel, the integration of electronics with photonics is seeing a rapid boost. Most larger circuit demonstrations are controlled through printed circuit boards (PCB) with off-the-shelf drivers, although we start seeing custom-designed drivers with large channel counts [12]. These represent a costly development, especially as each of the actuation mechanisms has its own peculiar driver requirements. These 100s or 1000s of driver channels need to be connected to the photonic chip, which is a challenge in its own right. While some platforms support monolithic photonic-electronic integration [13], the most common approach is through packaging, such as flip-chipping between two chips [12] or through an interposer [14], or even simple wirebonding.

On the algorithmic part, progress is relatively slow. Underlying mathematical underpinnings for programmable photonics have been mostly focused on forward-only meshes, especially as certain topologies allow for efficient ‘self-configuration’ (i.e. configuration using simple feedback loops, rather than calibration routines) [15], [16]. For recirculating meshes, the topology is more complicated, which makes it more difficult to define fundamental mathematical properties. This can be done for path routing [17], which can be approached through a graph-based representation [18], [19]. More complex functions, such as filters with a specific pass band, are mostly defined by hand. And this is just for regular mesh architectures. The exploration of non-regular meshes is only just starting…

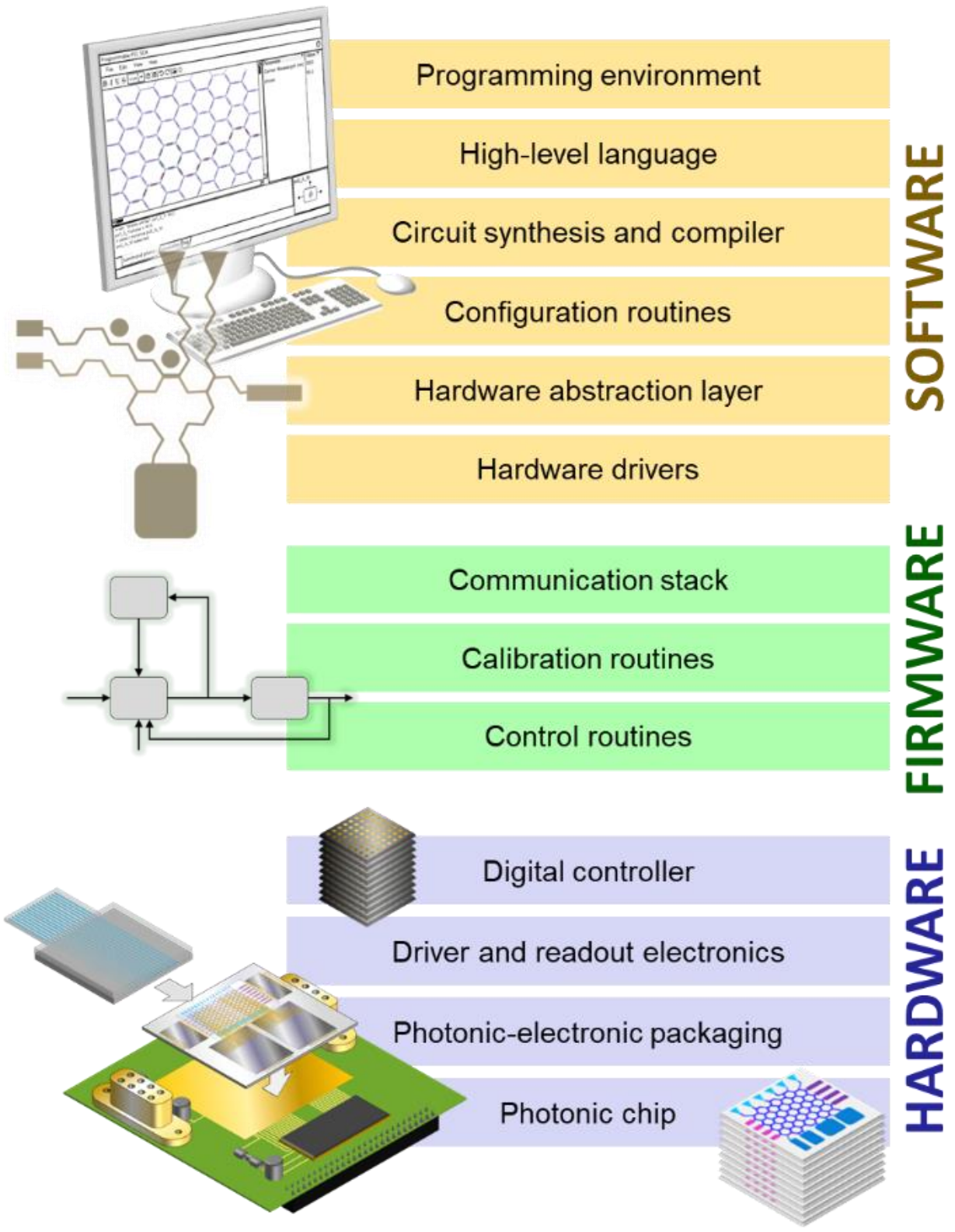

**Figure 2.** Technology stack for programmable photonics, combining photonics and electronics hardware, as well as software layers to provide different levels f abstraction and control to developers.

## Concluding Remarks

Programmable photonics as a practical technology in its infancy. Today, we see a gradient from widely tunable application-specific PICs (e.g. for microwave signal processing) over forward-only meshes (e.g. for matrix-vector multiplication or qubit processing) to the emerging general-purpose programmable photonic circuits. A true fully programmable photonic signal processor still has to be realized, but it is there that the most useful potential lies. Such a processor could open up the field of photonics to a more diverse range of applications, both for optical signal processing (spectrometers,…) as for microwave signal processing. Developments using a programmable chip take much less iteration time than spinning out a new chip design for each test, and these chips can be directly useful in low-volume applications [20]. And a software interface could expose the technology to a much wider engineering community. Just like off-the-shelf field-programmable gate arrays (FPGA), digital signal processors (DSP) and microcontrollers have kickstarted a wave of innovation based on electronics, programmable photonics could have the same effect for the world of optics and photonics.

## Acknowledgements

The author is supported by the European Union through the ERC PhotonicSWARM project (grant 725555), the H2020 projects MORPHIC (grant 780283) and PHORMIC (grant 101070322), the Flemish Research Foundation (FWO-Vlaanderen) through the project GRAPHSPAY (grant G020421N), and the US Air Force Office of Scientific Research (AFOSR) through award FA8655-21-1-7035.

## References

[1] D. Marpaung, J. Yao, and J. Capmany, "Integrated microwave photonics," *Nature Photonics*, vol. 13, no. 2, pp. 80–90, Feb. 2019, doi: 10.1038/s41566-018-0310-5.

[2] W. Bogaerts *et al.*, "Programmable photonic circuits," *Nature*, vol. 586, no. 7828, pp. 207–216, 2020, doi: 10.1038/s41586-020-2764-0.

[3] D. Pérez-López *et al.*, "General-purpose programmable photonic processor for advanced radiofrequency applications," *Nat Commun*, vol. 15, no. 1, p. 1563, Feb. 2024, doi: 10.1038/s41467-024-45888-7.

[4] Y. Zhang, X. Chen, L. Van Iseghem, I. Zand, H. Salmanian, and W. Bogaerts, “A compact programmable silicon photonic circuit,” in *IEEE Silicon Photonics Conference*, Tokyo, Japan, Apr. 2024, p. Postdeadline paper.

[5] X. Guo *et al.*, “Versatile silicon microwave photonic spectral shaper,” *APL Photonics*, vol. 6, no. 3, p. 036106, Mar. 2021, doi: 10.1063/5.0033516.

[6] H. Deng *et al.*, “Single-Chip Silicon Photonic Processor for Analog Optical and Microwave Signals.” arXiv, Nov. 14, 2023. Accessed: Mar. 23, 2024. [Online]. Available: http://arxiv.org/abs/2311.09258

[7] M. Dong *et al.*, “High-speed programmable photonic circuits in a cryogenically compatible, visible–near-infrared 200 mm CMOS architecture,” *Nat. Photon.*, vol. 16, no. 1, pp. 59–65, Jan. 2022, doi: 10.1038/s41566-021-00903-x.

[8] J. Geler-Kremer *et al.*, “A ferroelectric multilevel non-volatile photonic phase shifter,” *Nat. Photon.*, vol. 16, no. 7, pp. 491–497, Jul. 2022, doi: 10.1038/s41566-022-01003-0.

[9] A. Shafiee, S. Pasricha, and M. Nikdast, “A Survey on Optical Phase-Change Memory: The Promise and Challenges,” *IEEE Access*, vol. 11, pp. 11781–11803, 2023, doi: 10.1109/ACCESS.2023.3241146.

[10] L. Van Iseghem *et al.*, “Low power optical phase shifter using liquid crystal actuation on a silicon photonics platform,” *Opt. Mater. Express*, vol. 12, no. 6, p. 2181, Jun. 2022, doi: 10.1364/OME.457589.

[11] N. Quack *et al.*, “Integrated silicon photonic MEMS,” *Microsyst Nanoeng*, vol. 9, no. 1, p. 27, Mar. 2023, doi: 10.1038/s41378-023-00498-z.

[12] C. V. Poulton *et al.*, “Coherent LiDAR With an 8,192-Element Optical Phased Array and Driving Laser,” *IEEE J. Select. Topics Quantum Electron.*, pp. 1–8, 2022, doi: 10.1109/JSTQE.2022.3187707.

[13] K. Giewont *et al.*, “300-mm Monolithic Silicon Photonics Foundry Technology,” *IEEE J. Select. Topics Quantum Electron.*, vol. 25, no. 5, pp. 1–11, Sep. 2019, doi: 10.1109/JSTQE.2019.2908790.

[14] W. Bogaerts *et al.*, “Scaling programmable silicon photonics circuits,” in *Silicon Photonics XVIII*, G. T. Reed and A. P. Knights, Eds., San Francisco, United States: SPIE, Mar. 2023, p. 1. doi: 10.1117/12.2647293.

[15] D. A. B. Miller, “Self-aligning universal beam coupler.,” *Optics express*, vol. 21, no. 5, pp. 6360–70, 2013, doi: 10.1364/OE.21.006360.

[16] S. Bandyopadhyay, R. Hamerly, and D. Englund, “Hardware error correction for programmable photonics,” *Optica*, vol. 8, no. 10, p. 1247, Oct. 2021, doi: 10.1364/OPTICA.424052.

[17] Z. Gao *et al.*, “Provable Routing Analysis of Programmable Photonic Circuits,” *J. Lightwave Technol.*, pp. 1–12, 2024, doi: 10.1109/JLT.2024.3385338.

[18] X. Chen, P. Stroobant, M. Pickavet, and W. Bogaerts, “Graph Representations for Programmable Photonic Circuits,” *Journal of Lightwave Technology*, vol. 38, 2020.

[19] F. V. Kerchove, X. Chen, D. Colle, W. Tavernier, W. Bogaerts, and M. Pickavet, “An Automated Router With Optical Resource Adaptation,” *J. Lightwave Technol.*, vol. 41, no. 18, pp. 5807–5819, Sep. 2023, doi: 10.1109/JLT.2023.3275385.

[20] W. Bogaerts and A. Rahim, “Programmable Photonics: An Opportunity for an Accessible Large-Volume PIC Ecosystem,” *IEEE Journal of Selected Topics in Quantum Electronics*, vol. 26, no. 5, p. 1, 2020, doi: 10.1109/JSTQE.2020.2982980.

# Scaling Coherent Integrated Photonic Neural Networks for Artificial Intelligence Applications

**Mahdi Nikdast and Sudeep Pasricha**
Colorado State University, Fort Collins, CO 80524, United States of America
Mahdi.Nikdast@colostate.edu, Sudeep.Pasricha@colostate.edu.

**Status**

Deep neural networks (DNNs) have significantly advanced the field of artificial intelligence (AI) across several application domains such as image recognition, natural language processing (NLP), and autonomous systems, to name a few. As these applications continue to grow increasingly complex, fuelled in part by, for example, the rising demand for NLP-based services like ChatGPT, the size and complexity of DNNs are also on the rise. Accordingly, there is a need to optimize DNN software implementations and especially the underlying hardware for scalable and energy-efficient DNN training and inferencing. Recent efforts to enhance DNN efficiency have focused on domain-specific AI accelerators, which employ closely integrated data processing units arranged in a systolic array. However, as Moore's Law nears its limits [1], electronic accelerators encounter significant obstacles stemming from the slowdown in CMOS scaling and the limitations of low-bandwidth metallic interconnects. Consequently, the ongoing advancement of AI is hindered by the substantial energy overhead incurred during the training and inference of growing DNNs on electronic processors [2].

Limitations of electronic DNN accelerators have urged exploration into utilizing emerging technologies for DNN acceleration, opening new avenues of research. Among different candidates, silicon photonics has shown significant promise due to its CMOS compatibility and offering high-bandwidth chip-scale communication and the ability to realize optical-domain computation using photonic devices, to reduce computational complexity by taking advantage of the natural parallelism of optics [3]. By leveraging optical interconnects for communication and photonic devices for computation, silicon-photonic-based neural network accelerators (SPNNAs) can achieve orders of magnitude better energy efficiency for performing computationally expensive multiply-and-accumulate (MAC) operations [2] – [5], which are the most power-hungry and common operations in DNNs [2]. Among possible SPNNA implementations, coherent SPNNAs (C-SPNNAs), which operate on a single wavelength, have an advantage over noncoherent SPNNAs in terms of eliminating power hungry wavelength-conversion steps and multiple wavelength sources, and avoiding inter-channel crosstalk noise. Fig. 1 presents an overview of a multi-layer C-SPNNA with $N_1$ inputs, $N_2$ outputs, and $M$ layers. Each layer includes an optical-interference unit (OIU) implemented using an array of Mach–Zehnder interferometer (MZI) devices with a specific architecture, connected to a nonlinear activation unit (NAU) using an optical-gain (amplification) unit (OGU). Using this organization, any weight matrix corresponding to a linear multiplier in the fully connected layer of a multi-layer perceptron can be factorized into two unitary matrices and one diagonal matrix using singular value decomposition, as shown in Fig. 1.

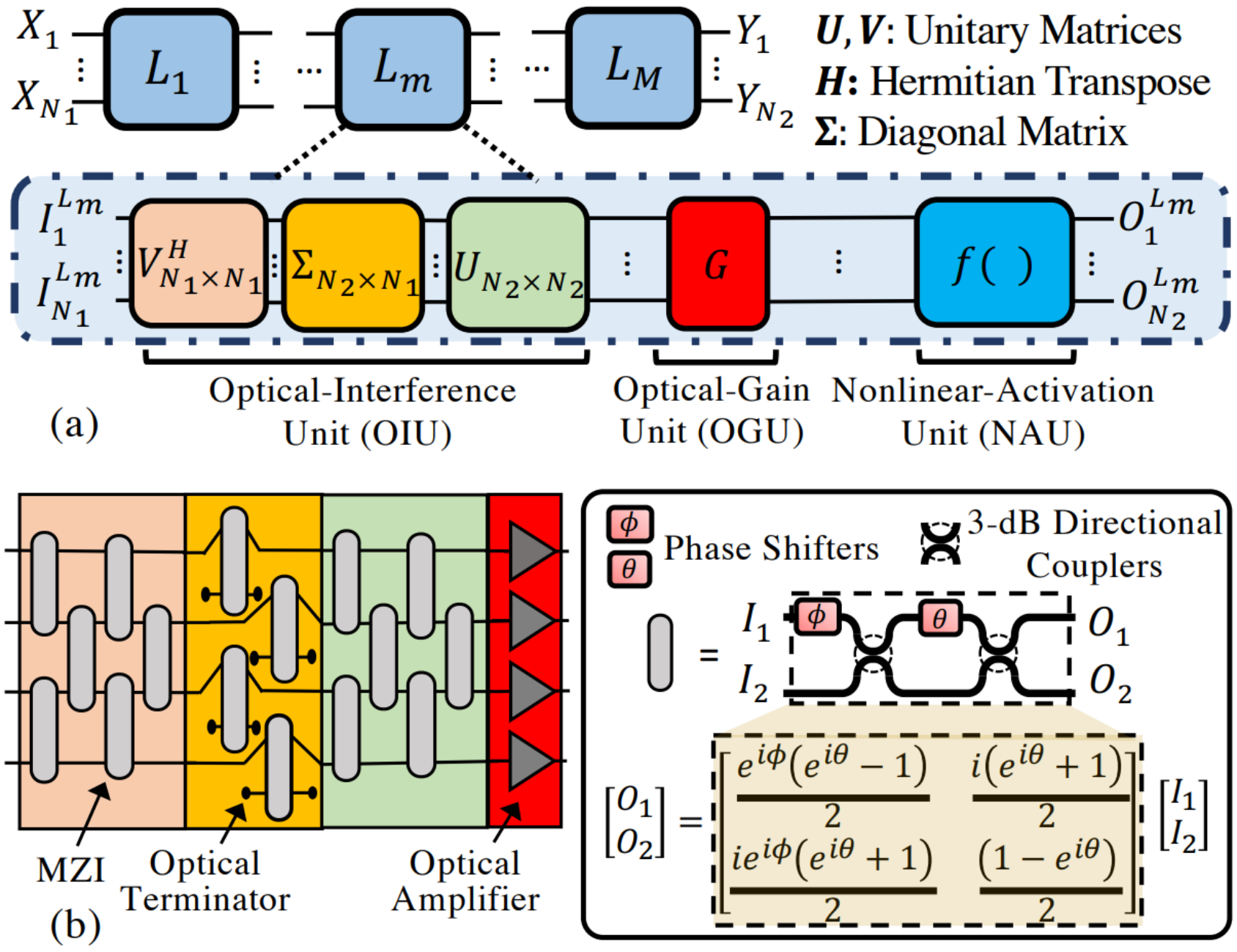

**Figure 1.** An overview of a coherent SPNNA with $N_1$ inputs, $N_2$ outputs, and $M$ layers. (b) An optical-interference unit architecture (left) based on [5] with $N_1 = N_2 = 4$, considered as an example, and the underlying 2×2 MZI multiplier with the corresponding MZI transfer function (right) [Figure is from https://arxiv.org/abs/2204.03835; no permission is required based on the CC BY-NC-SA 4.0 DEED licence].

## Current and Future Challenges

Despite promising C-SPNNA developments, both in academia [2] and industry [6] – [8], several challenges must be addressed to further scale up C-SPNNAs capable of achieving high efficiencies, comparable to their electronic counterparts. Photonic devices are intrinsically bulky and suffer from optical losses and crosstalk noise, the impact of which can accumulate as C-SPNNAs scale up, hence degrading the signal-to-noise ratio (SNR) and the inferencing accuracy in such networks, as demonstrated in [9]. Note that adding Semiconductor Optical Amplifiers (SOAs) to C-SPNNAs to compensate for optical losses is impractical because of the Spontaneous Emission Noise (ASE) in SOAs—especially at low optical inputs—while they can also amplify coherent crosstalk noise. In addition, the large footprint of MZI devices often used in C-SPNNAs (e.g., an MZI is a few hundred micrometer long) and the number of such devices required to implement a fully connected layer, which grows at the rate of $O(N_1^2 + N_2^2)$, not only limit the size of C-SPNNA implementable on a wafer as the network scales up but also further complicates device control, which is performed through an electronic controller in C-SPNNAs, and device-to-device matching due to the large number of devices [10], [11]. Furthermore, the underlying devices in C-SPNNAs are susceptible to uncertainties stemming from optical lithographic imperfections and thermal crosstalk and can experience imprecisions due to non-uniform insertion losses and quantization errors due to low-precision encoding in the tuned parameters (e.g., phase angles). For example, it was demonstrated in [12] that the inferencing accuracy in a C-SPNNA can degrade by 46%, even when the imperfection parameters are restricted within a small range. Due to optical losses and imperfections discussed, existing C-SPNNAs are often designed to be small (as compared to a flatten network) where MAC operations can be broken down to be performed on a smaller C-SPNNA unit in which different parameters (e.g., weight parameters) and their adjustments will be carried out using a digital memory interacting with the photonic network through multiple Digital-to-Analog Convertors (DACs) and Analog-to-Digital Convertors (ADCs) [13]. Nevertheless, the power consumption and latency associated with using DACs and ADCs further degrade the performance and efficiency of C-SPNNAs. Last, a C-SPNNA that performs based on combining a photonic (for computation) and an electronic (for control) subsystem is more susceptible to malicious hardware and software security attacks where attackers can profit from data exchanges between the two signal domains and act on the integration interfaces [14].

## Advances in Science and Technology to Meet Challenges

To achieve greater scalability in C-SPNNAs capable of handling complex and realistic DNNs with high energy efficiencies, it is essential to engage in co-design and co-optimization efforts spanning the entire system stack. This includes refining photonic device designs and implementations to developing photonic-friendly optimization mechanisms tailored for C-SPNNAs. A good initial step is to analyse the impact of optical losses, crosstalk noise, and imperfections (due to fabrication-process variations, thermal variations, etc.) comprehensively in C-SPNNAs, to fully understand the costs (e.g., scalability constraints, power overhead) imposed by such inefficiencies in C-SPNNAs [9], [12]. Such analysis can help designers better understand and explore the design space of C-SPNNAs, informing different design choices and strategies (from device to system level) when developing C-SPNNAs. The work in [15] indicated that by a better design of photonic devices, enabled by inverse design techniques, C-SPNNAs can achieve better scalability with much higher accuracies. It also showed that the footprint of a photonic device (e.g., MZIs in coherent C-SPNNAs) is mainly constrained by the required phase shifters, hence highlighting the need for more efficient and compact phase shifters. Therefore, C-SPNNAs can benefit from a more compact design of photonic devices with minimized losses and noise, which can be addressed (to some extent) by inverse design techniques. To address the impact of process variations, there is a need for comprehensive modelling of different nonuniformities along with associated statistics (e.g., distributions) in optical lithography processes. Such models can be then utilized during the design of C-SPNNAs to realize variation-aware photonic devices and robust networks. For example, the work in [16], [17] showcased the promise of fabrication-process-variation-aware layout and design of photonic devices in C-SPNNAs, and demonstrated an average increase of 72% in the inferencing accuracy of a C-SPNNA. The work in [18] also demonstrated the promise of design-time and run-time DNN model pruning to further reduce the size and improve the energy efficiency and robustness in C-SPNNAs. To address signal domain inconsistencies and avoid conversions, photonic memories can be developed to help increase energy efficiency in C-SPNNAs, as discussed in [19], [20]. Last, there is a need for low-cost and nondisruptive bias control and mitigation mechanisms (e.g., on-chip health control mechanisms) to not only preserve C-SPNNA performance under inevitable imperfections but also to retrieve performance and operation under potential hardware and software attacks (e.g., when an attacker induces thermal crosstalk for misclassification in a C-SPNNA), as described in [14].

## Concluding Remarks

Coherent silicon-photonic-based neural network accelerators (C-SPNNAs) offer significant performance gains to accelerate emerging DNNs, in particular for performing compute-intensive multiply-and-accumulate (MAC) operations with low latency and high energy efficiencies. In this article, we reviewed some of the critical challenges in further scaling C-SPNNAs and improving energy efficiency in such networks, including optical loss, crosstalk noise, device footprint, sensitivity to fabrication-process and thermal variations, phase noise and control, and hardware and software security concerns in C-SPNNAs. In addition, we proposed several existing solutions and those that will be necessary to tackle different challenges while emphasizing the essential role of cross-layer co-design and co-optimization. Additionally, we explored the limitations posed by utilizing digital memories in conjunction with photonic compute substrates, highlighting the necessity for fast and scalable photonic memory to facilitate energy-efficient photonic computation. The discussions in this article can help C-SPNNA designers to better understand different challenges currently facing further development of such systems, hence creating new research avenues to address these challenges. Finally, we recognize that other challenges exist, such as the development of a fully optical nonlinear unit, but for the sake of brevity, they are not addressed in this article.

**Acknowledgements**
Some of the work discussed in this article was supported in part by the United States National Science Foundation (NSF) under grant numbers CCF-1813370, CCF-2006788, and CNS-2046226.

## References

[1] S. Pasricha and M. Nikdast, “A survey of silicon photonics for energy efficient manycore computing,” IEEE Design and Test of Computers, vol. 37, no. 4, pp. 60–81, 2019.

[2] F. Sunny, E. Taheri, M. Nikdast, S. Pasricha, “A survey on silicon photonics for deep learning,” ACM Journal of Emerging Technologies in Computing Systems (JETC), vol. 17, no. 4, article no. 61, pp. 1–57, July 2021.

[3] Y. Shen, N. C. Harris, S. Skirlo, M. Prabhu, T. Baehr-Jones, M. Hochberg, X. Sun, S. Zhao, H. Larochelle, D. Englund, M. Soljačić, “Deep learning with coherent nanophotonic circuits,” Nature Photonics, vol. 11, pp. 441–446, 2017.

[4] F. Shokraneh, S. Geoffroy-Gagnon, and O. Liboiron-Ladouceur, “The diamond mesh, a phase-error-and loss-tolerant field-programmable MZI-based optical processor for optical neural networks,” Optics Express, vol. 28, no. 16, pp. 23 495–23 508, 2020.

[5] W. R. Clements, P. C. Humphreys, B. J. Metcalf, W. S. Kolthammer, and I. A. Walmsley, “Optimal design for universal multiport interferometers,” Optica, vol. 3, no. 12, pp. 1460–1465, 2016.

[6] Lightelligence, [Online]: https://www.lightelligence.ai/ Accessed on: April 22, 2024.

[7] Lightmatter, [Online]: https://lightmatter.co/, Accessed on: April 22, 2024.

[8] Cognifiber, [Online]: https://www.cognifiber.com/ Accessed on: April 22, 2024.

[9] A. Shafiee, S. Banerjee, K. Chakrabarty, S. Pasricha and M. Nikdast, “Analysis of Optical Loss and Crosstalk Noise in MZI-based Coherent Photonic Neural Networks,” in *IEEE/Optica Journal of Lightwave Technology*, 2024, doi: 10.1109/JLT.2024.3373250.

[10] A. Shafiee, S. Banerjee, K. Chakrabarty, S. Pasricha, and M. Nikdast, “LoCI: An analysis of the impact of optical loss and crosstalk noise in integrated silicon-photonic neural networks,” *ACM Great Lakes Symposium on VLSI (GVLSI)*, June 2022, pp. 351–355.

[11] S. Banerjee, M. Nikdast, and K. Chakrabarty, “Modeling silicon-photonic neural networks under uncertainties,” *IEEE/ACM Design, Automation and Test in Europe (DATE) Conference and Exhibition*, Grenoble, France, March 2021, pp. 98–101.

[12] S. Banerjee, M. Nikdast, and K. Chakrabarty, “Characterizing coherent integrated photonic neural networks under imperfections,” *IEEE/OSA Journal of Lightwave Technology (JLT)*, vol. 41, no. 5, pp. 1464–1479, 2023.

[13] Q. Cheng, J. Kwon, M. Glick, M. Bahadori, L. P. Carloni and K. Bergman, "Silicon Photonics Codesign for Deep Learning," *in Proceedings of the IEEE*, vol. 108, no. 8, pp. 1261–1282, Aug. 2020.

[14] F. G. De Magalhaes, M. Nikdast and G. Nicolescu, “Integrated Photonic AI Accelerators Under Hardware Security Attacks: Impacts and Countermeasures,” *IEEE International Midwest Symposium on Circuits and Systems (MWSCAS)*, Tempe, AZ, USA, 2023, pp. 806–810.

[15] A. Shafiee, S. Banerjee, B. Charbonnier, S. Pasricha, and M. Nikdast, “Compact and Low-Loss PCM-based Silicon Photonic MZIs for Photonic Neural Networks,” *IEEE Photonics Conference (IPC)*, Orlando, FL, November 2023.

[16] A. Mirza, A. Shafiee, S. Banerjee, K. Chakrabarty, S. Pasricha, and M. Nikdast, “Characterization and optimization of coherent MZI-based nanophotonic neural networks under fabrication non-uniformity,” *IEEE Transactions on Nanotechnology (TNANO)*, vol. 21, pp. 763–771, 2022.

[17] Z. Ghanaatian, A. Shafiee, and M. Nikdast, “Variation-Aware Layout and Design Optimization of Silicon Photonic Mach–Zehnder Interferometers,” *IEEE Photonics Conference (IPC)*, Orlando, FL, November 2023.

[18] S. Banerjee, M. Nikdast, S. Pasricha, and K. Chakrabarty, “Pruning Coherent Integrated Photonic Neural Networks,” *in IEEE Journal of Selected Topics in Quantum Electronics*, vol. 29, no. 2: Optical Computing, pp. 1-13, March-April 2023, Art no. 6101013.

[19] A. Shafiee, B. Charbonnier, S. Pasricha, and M. Nikdast, “Design Space Exploration for PCM-based Photonic Memory,” *ACM Great Lakes Symposium on VLSI (GLSVLSI)*, Knoxville, TN, June 2023.

[20] F. Sunny, A. Shaifee, B. Charbonnier, M. Nikdast, and S. Pasricha, "COMET: A Cross-Layer Optimized Optical Phase Change Main Memory Architecture," *IEEE/ACM Design, Automation and Test in Europe (DATE) Conference and Exhibition*, Valencia, Spain, March 2024.

# Non-coherent Integrated Photonics for Scalable AI and Deep Neural Network Acceleration

**Sudeep Pasricha and Mahdi Nikdast**
Department of Electrical and Computer Engineering, Colorado State University, USA
sudeep@colostate.edu, mahdi.nikdast@colostate.edu

## Status

In recent years, deep neural networks (DNNs) have enabled cutting-edge artificial intelligence (AI) applications, including ChatGPT and other large language models (LLMs) for natural language processing, perception systems in autonomous vehicles, online search and recommendation systems, protein structure prediction and genomic analysis, and network anomaly detection [1]-[6]. Today, the penetration and growth of AI are largely limited by hardware capabilities. The CMOS-based electronic backbones used in state-of-the-art GPUs and AI accelerators (such as Google's TPUs) are hitting fundamental performance and energy limits, due to the ending of Moore's Law and related trends, such as Denard's scaling (power density) and Koomey's law (instructions per Joule) [7]. Improving hardware scalability for AI has thus becomes an immediate and significant challenge.

Among emerging technologies, integrated photonics offers the ability to communicate, compute, and even access memory at light speeds. Several high-speed, energy-efficient, and low-cost integrated photonic devices and integrated circuits have been implemented in recent years with CMOS-compatible manufacturing techniques. The resulting integrated photonics solutions have surpassed latencies of electronic systems by an order of magnitude or more [8]. Further, power dissipation in such integrated photonics platforms scales almost linearly with clock frequency, whereas electrical circuits exhibit a quadratic relationship [9]. The ability to leverage multiplexing techniques such as wavelength-division multiplexing (WDM) and mode-division multiplexing (MDM) further enable much higher bandwidth densities than electronic systems. With several foundries developing silicon photonics process design kits, the integrated photonic industry is rapidly moving towards standardization, not unlike the fabless semiconductor industry [10].

Several recent efforts have demonstrated integrated photonics-based approaches for accelerating matrix-vector multiplications (MVMs), vector dot products, accumulations, and non-linear activations across AI applications, using either coherent or non-coherent optical signal characteristics. Compared to coherent alternatives, non-coherent approaches provide greater flexibility to employ multiplexing approaches, utilize more compact devices, and possess greater cost scalability when implementing larger AI applications. Many non-coherent photonic accelerators have been proposed for diverse DNN applications, including convolutional neural networks (CNNs) [11], recurrent neural networks (RNNs) [12], transformers [13], and graph neural networks (GNNs) [14]. These and other non-coherent accelerators often utilize microring resonators (MRRs) to encode neural network weights and activations via amplitude modulation, due to their wavelength-selective transmission characteristics. The narrow 3-dB bandwidth of MRRs allows their use in WDM, where multiple MRRs are arranged in banks to modulate/filter multiple different wavelengths, to improve computational throughout. Continued advances with non-coherent integrated photonics remain essential to supporting larger and more complex AI applications in the future.

## Current and Future Challenges

Several challenges with non-coherent integrated photonics have come to the forefront with the recent activities related to AI acceleration.

The response of MRRs is susceptible to environmental perturbations, such as those from semiconductor fabrication-process variations and runtime thermal variations. These variations change the performance and characteristics of MRRs, thereby impacting the correctness of wavelength-selective operations performed with them. Readjusting the device response can be done via either thermo-optic (TO) or electro-optic (EO) tuning approaches. TO approaches involve integrated filament microheaters that generate heat proportional to the square of the bias voltage. Such microheaters can induce large changes in effective refractive index to compensate for variations. However, TO tuning is relatively slow, with support for limited modulation rates of a few hundred KHz, and also contributes to thermal crosstalk. EO tuning involves doping silicon waveguides and applying an external electric field to manipulate carrier concentration to adjust effective refractive index. EO tuning can support GHz level modulation but has a lower tuning range compared to TO tuning.

Optical losses are another challenge that limits achievable precision and scalability with optical non-coherent accelerator platforms. Losses during coupling of optical signals from external sources, and within devices used in optical computations cause attenuations in optical signal intensities reaching the outputs. This creates challenges when scaling up hardware implementations to support larger DNNs. Specifically, the optical intensities required at the photodetectors to detect an optical signal with a desired resolution (in terms of bits) places a limit on the size of the matrix that can be computed optically. Optical losses have also been shown to contribute to a significant drop in DNN inference accuracy. Optical signal attenuation due to losses can be compensated by using semiconductor optical amplifiers (SOAs) that can amplify their input signals. However, current SOAs have high energy overheads and non-linear gain-current curves requiring careful calibration.

High energy consumption is another challenge. Tuning mechanisms can be very power hungry, e.g., TO tuning requires a high continuous external biasing and power supply (~mW level). As implementations scale up and the number of MRRs increases, tuning energy overhead becomes significant. Compensating for increased losses with scaling either requires more SOAs or increasing laser power, both of which negatively impact energy efficiency. Interfacing with digital electronic components also requires using costly conversion steps with DAC and ADC circuits operating at full data rate (tens of gigahertz), which is currently energy inefficient.

**Advances in Science and Technology to Meet Challenges**

New materials and devices are needed to improve the characteristics of optical components in non-coherent AI accelerator implementations. Emerging photonic DAC devices that combine digital to analog conversion and electro-optic modulation [15] have shown promise in reducing area and power consumption while supporting high sampling rates, high precision, and low distortion. Similarly, heterogeneous modulators leveraging III-V MOSCAP and 2D materials have recently achieved compact layouts with gigahertz-level cut-off frequency, high modulation efficiency, and sub-picojoule-per-bit power consumption. Improved designs are needed for on-chip light sources to eliminate complex fiber packaging and alignment processes, integrated frequency combs for WDM sources, programmable nonlinear units, modulators for high-bandwidth processing, and efficient ADC/DAC and photodetectors, specifically targeting non-coherent optical computing.

To reduce the reliance on electronic memories (and associated optoelectronic conversion overheads) during AI acceleration, realizing data storage in the optical domain is a promising direction. This idea has been recently explored by storing neuron weights in DNNs within phase change materials (PCMs) integrated into non-coherent DNN accelerators [12]. In these implementations, precomputed weights for a given DNN inference task can be stored in PCMs integrated within the optical computing components (such as MRRs, waveguides) to enable efficient real-time inference for any given input (activations) to the DNN. A recent work also showed how optical main memory based on PCMs can be realized [16]. By integrating such optically controlled memories with optical AI accelerators, it can be possible to significantly improve computation throughput and energy efficiency.

New cross-layer design techniques are needed that can co-optimize hardware devices/circuits/architectures with AI software. Recent efforts have illustrated the promise of co-optimizing MRR device designs and their tuning circuits with their physical layout and architecture design to reduce crosstalk, losses, variation susceptibility, and energy consumption [17]. Co-optimizing DNN hyperparameters and parameters with non-coherent photonics hardware design is also promising. Recent efforts that employ advanced DNN quantization techniques to reduce parameter bit-width requirements, DNN parameter sparsification techniques, efficient mapping approaches, and photonic hardware-aware training techniques can preserve DNN inference accuracy while reducing photonic component complexity [17]-[19].

Lastly, new methods are needed to efficiently integrate photonic components with electronic components (e.g., signal conversion circuits, control systems). Such approaches can employ 2.5D photonic interposer-based integration to realize high yield AI accelerators at scale [20]. Monolithic fabrication in advanced CMOS nodes and 3D stacking can also improve system throughput and performance densities. Better electronic-photonic design automation techniques will become essential for such advanced integration methods to improve designer productivity and efficiency, for instance by supporting automated circuit layout generation and fast photonic circuit simulations.

**Concluding Remarks**

Non-coherent integrated photonic platforms can provide exceptional advantages for AI acceleration, including high bandwidth, low latency, and high energy-efficiency. These innovative hardware platforms can overcome the biggest technological bottlenecks in state-of-the-art electronic AI platforms. But many open challenges remain, related to robustness to variations and losses, energy overheads, and scalability. While the implementation efficiency of operations required for AI inference continues to improve, implementing training directly in photonic hardware is still a challenge due to the difficulties of in-situ implementation of the backpropagation algorithm with its gradient calculations and bidirectional propagation requirements. Nonetheless, new ideas and developments remain poised to enhance the scalability and applicability of non-coherent integrated photonics to accelerate an ever-growing library of diverse AI applications. Integrated photonics will be one of the most compelling technologies in the long term to meet the increasing demands of emerging AI applications and high throughput computing.

**Acknowledgements**
This work is sponsored in part by the National Science Foundation under grant numbers CCF-1813370, CCF-2006788, and CNS-2046226.

## References

[1] T. Wu, S. He, J. Liu, S. Sun, K. Liu, Q. Han, Y. Tang, "A brief overview of ChatGPT: The history, status quo and potential future development", IEEE/CAA Journal of Automatica Sinica, 10(5), pp.1122-1136, 2023.

[2] J. Dey, S. Pasricha, "Robust Perception Architecture Design for Automotive Cyber-Physical Systems", IEEE Computer Society Annual Symposium on VLSI (ISVLSI), 2022.

[3] D. Gufran, S. Pasricha, "FedHIL: Heterogeneity Resilient Federated Learning for Robust Indoor Localization with Mobile Devices", ACM Transactions on Embedded Computing Systems (TECS), 2023.

[4] A. Balasubramaniam, F. Sunny, S. Pasricha, "R-TOSS: A Framework for Real-Time Object Detection using Semi-Structured Pruning", IEEE/ACM DAC, 2023.

[5] S. Huang, J. Yang, S. Fong, Q. Zhao, "Artificial intelligence in the diagnosis of COVID-19: challenges and perspectives", International journal of biological sciences, 17(6), p.1581, 2021.

[6] V. K. Kukkala, S. V. Thiruloga, and S. Pasricha, “LATTE: LSTM Self-Attention based Anomaly Detection in Embedded Automotive Platforms”, ACM Transactions on Embedded Computing Systems (TECS), Volume 20, Issue 5s, Oct 2021.

[7] S. Pasricha, M. Nikdast, “A Survey of Silicon Photonics for Energy Efficient Manycore Computing” IEEE Design and Test, vol. 37, no. 4, pp. 60-81, Aug 2020.

[8] F. Sunny, E. Taheri, M. Nikdast, S. Pasricha, “A Survey on Silicon Photonics for Deep Learning”, ACM Journal on Emerging Technologies in Computing Systems (JETC), Vol. 17, Iss. 4, Oct 2021.

[9] Z. Ying et al., “Electronic-photonic arithmetic logic unit for high-speed computing,” Nature Communications, vol. 11, no. 1, p. 2154, 2020.

[10] A. E.-J. Lim et al., “Review of silicon photonics foundry efforts,” IEEE Journal of Selected Topics in Quantum Electronics, vol. 20, no. 4, pp. 405–416, 2013.

[11] F. Sunny, A. Mirza, M. Nikdast, S. Pasricha, “CrossLight: A Cross-Layer Optimized Silicon Photonic Neural Network Accelerator”, IEEE/ACM Design Automation Conference (DAC), 2021.

[12] F. Sunny, M. Nikdast and S. Pasricha, “RecLight: A Recurrent Neural Network Accelerator With Integrated Silicon Photonics”, IEEE Computer Society Annual Symposium on VLSI (ISVLSI), 2022.

[13] S. Afifi, F. Sunny, M. Nikdast, S. Pasricha, “TRON: Transformer Neural Network Acceleration with Non-Coherent Silicon Photonics”, ACM GLSVLSI, 2023.

[14] S. Afifi, F. Sunny, A. Shafiee, M. Nikdast, S. Pasricha, “GHOST: A Graph Neural Network Accelerator using Silicon Photonics”, ACM Transactions on Embedded Computing Systems (TECS), 2023.

[15] J. Meng, M. Miscuglio, J. George, A. Babakhani, V. Sorger, “Electronic Bottleneck Suppression in Next-Generation Networks with Integrated Photonic Digital-to-Analog Converters”, Advanced photonics research, 2(2), p.2000033, 2021.

[16] F. Sunny, A. Shafiee, B. Charbonnier, M. Nikdast, S. Pasricha, “COMET: A Cross-Layer Optimized Optical Phase Change Main Memory Architecture”, IEEE/ACM DATE, Mar 2024.

[17] F. Sunny, M. Nikdast, S. Pasricha, “Cross-Layer Design for AI Acceleration with Non-Coherent Optical Computing”, ACM GLSVLSI, 2023.

[18] F. Sunny, M. Nikdast and S. Pasricha, “A Silicon Photonic Accelerator for Convolutional Neural Networks with Heterogeneous Quantization”, ACM GLSVLSI, 2022.

[19] F. Sunny, M. Nikdast, and S. Pasricha, “SONIC: A Sparse Neural Network Inference Accelerator with Silicon Photonics for Energy-Efficient Deep Learning”, IEEE/ACM Asia & South Pacific Design Automation Conference (ASPDAC), Jan 2022.

[20] F. Sunny, E. Taheri, M. Nikdast, S. Pasricha, “Silicon Photonic Network-on-Interposer Design for Energy Efficient Convolutional Neural Network Acceleration on 2.5D Chiplet Platforms”, IEEE/ACM DATE, Mar 2024.

# Hybrid integration for photonic computing

**Zhongjin Lin[1], Xinlun Cai[1], Lukas Chrostowski[2]**
[1]State Key Laboratory of Optoelectronic Materials and Technologies, School of Electronics and Information Technology, Sun Yat-sen University, Guangzhou 510275, China
[2]Department of Electrical and Computer Engineering, The University of British Columbia, Vancouver, V6T 1Z4, Canada
[linzhj56@mail.sysu.edu.cn, caixlun5@mail.sysu.edu.cn, lukasc@ece.ubc.ca]

## Status

Each breakthrough in computing technologies can significantly advance the development of our world. Currently, electronic processors dominate as the primary computing technologies in our daily lives. However, with the explosive growth of computing demand for artificial intelligence, meeting this demand becomes increasingly challenging for electronic processors due to limitations in speed and energy efficiency caused by Joule heating, electromagnetic crosstalk, parasitic capacitance, and so on. To address these issues, recently several new computing technologies have been proposed and demonstrated [1-3]. Among them, the photonic processor stands out as a promising technology for processing artificial intelligence tasks.

Photonic processors used for artificial neural networks offer several advantages, such as low energy consumption, low latency, and parallel computing. Many photonic integrated circuit (PIC) material platforms can be implemented to realize photonic processors, including indium phosphide (InP), silicon nitride (SiN), thin-film lithium niobate (TFLN), and silicon-on-insulator (SOI). Each material platform has its own advantages and disadvantages for realizing photonic computing [4]. For example, InP has the inherent ability to realize active components such as light sources, optical amplifiers, and photodetectors, but it entails a complex fabrication process, higher cost, and smaller wafer size. SiN can provide high-performance comb sources, which are important for wavelength-division multiplexing (WDM)-based photonic neural networks, but it cannot achieve fast modulation speed. TFLN, as a new PIC material platform, can achieve an ultra-fast modulation speed with a low insertion loss but faces challenges in high integration density. SOI can achieve high integration density and large-scale integration, but it has a high insertion loss at fast modulation speed.

Hybrid integration enables the leveraging of the strengths of multiple material platforms while circumventing their drawbacks through the integration of two or more PIC chips into a single package [5]. Introducing hybrid integration technologies, photonic processors [6] (see Fig. 1) can achieve a larger computing scale, faster weight update speed, higher computing density, and full analog computing which has not been using a single PIC material platform. In this roadmap, we will introduce the challenges and implemented technologies of hybrid integration.

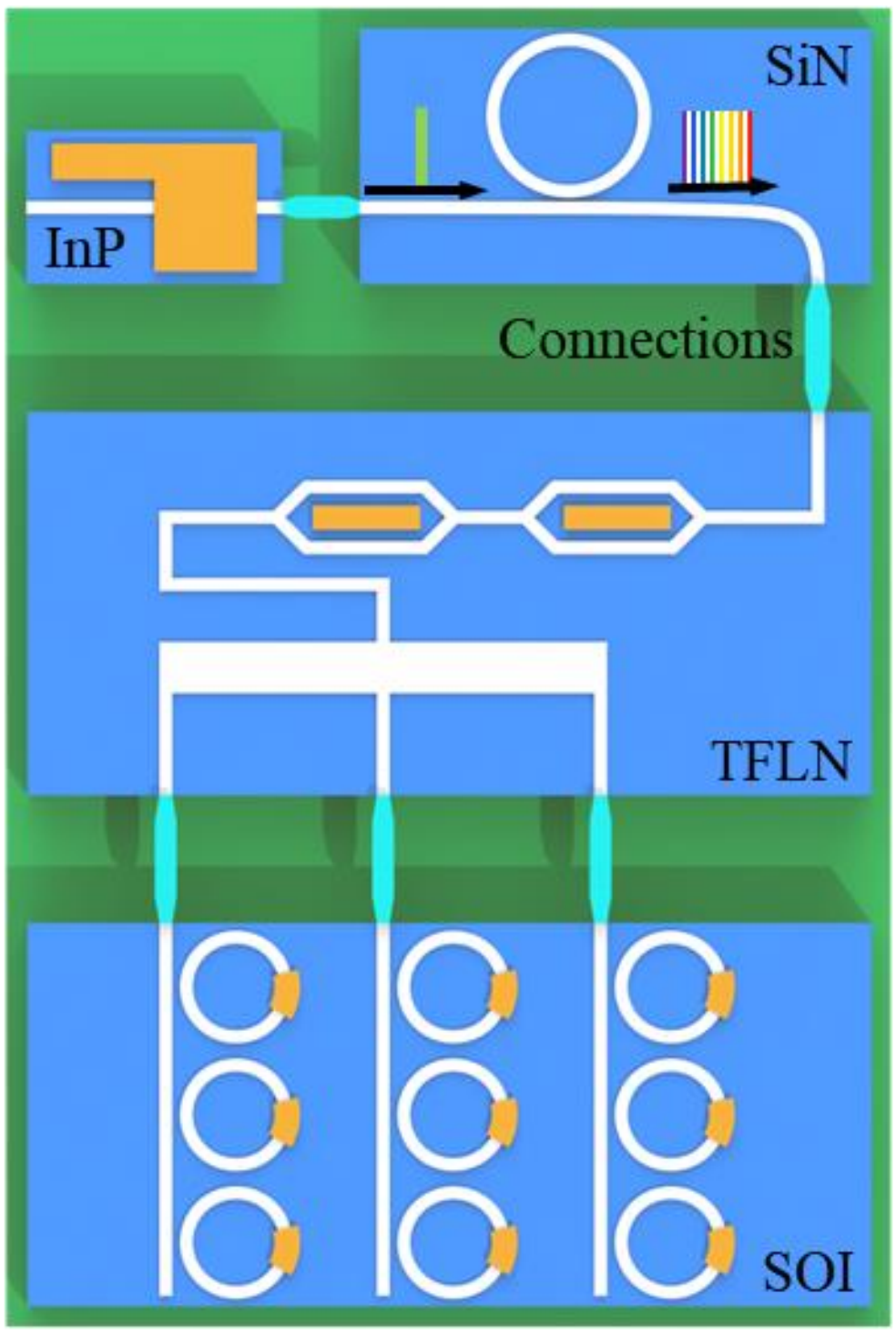

**Figure 1.** A schematic of the hybrid integration of indium phosphide (InP), silicon nitride (SiN), thin-film lithium niobate (TFLN), and silicon-on-insulator (SOI) chips to build a high-performance photonic processor for neural networks.

### Current and Future Challenges

The hybrid integration of InP, SiN, TFLN, and SOI chips (see Fig. 1) enables the photonic processor to incorporate lasers, comb sources, time-division multiplexing (TDM)-based computing units, and wavelength-division multiplexing (WDM)-based computing units, respectively [6]. With the advantage of large computing scale and fast weight update speed, TDM-based computing units using TFLN chips are suitable for building the neural network from the input layer to the hidden layer 1. With the advantage of high computing density, WDM-based computing units using SOI chips are suitable for building the neural network from hidden layer to hidden layer or output layer. Therefore, a powerful photonic processor can be achieved by hybrid integration technologies.

The challenges of the hybrid integration include:

- High-precision alignment. Typically, the alignments of optical input/output (I/O) ports between two photonic chips need to achieve a precision less than 100 nm, owing to small sizes of optical waveguide modes. Additionally, since two photonic chips often consist of numerous optical I/O ports that need to be interconnected to construct powerful photonic processors for artificial neural networks, this poses a higher challenge in alignment for hybrid integration.
- Stability. Unlike heterogeneous integration, hybrid integration may be sensitive to vibration and thermal expansion. For example, if two silicon photonic chips which consist of 50 optical I/O ports with a 127-µm period are connected, a misalignment of approximately 800 nm would happen when the temperature increases 50 °C due to the thermal expansion.

### Advances in Science and Technology to Meet Challenges

Hybrid integration can be realized by flip-chip bonding, micro-transfer printing, photonic wire bonding technologies:

- Flip-chip bonding. This hybrid integration technology connects two photonic chips through solder bumps [7, 8], boasting a history of more than six decades, making it a highly mature process. Although initially proposed for the hybrid integration of electronic chips, this technology has been widely implemented to connect InP chips (lasers and photodetectors) with TFLN or SOI chips [9]. However, due to its low alignment precision, it is constrained in efficiently coupling light from one chip to another chip through grating couplers, which typically necessitate an alignment precision better than 1 µm.
- Micro-transfer printing. This technology enables the hybrid integration by picking up a chip from its native substrate using an elastomer stamp, and then rapidly and precisely transferring it onto the target photonic chip [10, 11]. Using an adhesive bonding agent, the adhesion of two photonic chips can be strong enough to meet the requirement of stability.
- Photonic wire bonding. This technology builds optical interconnects by using three-dimensional (3D) polymer waveguides which are fabricated by two-photon polymerization [12-14]. Photonic wire bonding includes two steps: assembling photonic chips and fabricating 3D polymer waveguides. It allows the assembly of two photonic chips with an alignment precision lower than 10 µm. Despite this, low insertion losses can still be achieved by shaping the 3D polymer waveguides between the optical I/O ports of the two photonic chips. Additionally, this technology enables interconnections to be insensitive to vibration and thermal expansion.

**Concluding Remarks**

Hybrid integration takes the advantages of multiple material platforms, allowing powerful photonic processors to be built for large-scale artificial neural networks. It can be realized by flip-chip bonding, micro-transfer printing, and photonic wire bonding technologies. Although it has challenges in high-precision alignment requirements and stability, these challenges can be addressed by future technologies. Hybrid integration will bring a new breakthrough for photonic computing.

**Acknowledgements**

*National Natural Science Foundation of China (62405382); Natural Science Foundation of Guangdong Province,China (2514050001995).*

## References

[1] X. Liu and K. K. Parhi, "DNA Memristors and Their Application to Reservoir Computing," *ACS Synth Biol,* vol. 11, pp. 2202-2213, Jun 17 2022.

[2] S. Kan, K. Nakajima, T. Asai, and M. Akai-Kasaya, "Physical Implementation of Reservoir Computing through Electrochemical Reaction," *Adv Sci (Weinh),* vol. 9, p. e2104076, Feb 2022.

[3] B. J. Shastri, A. N. Tait, T. Ferreira de Lima, W. H. P. Pernice, H. Bhaskaran, C. D. Wright, and P. R. Prucnal, "Photonics for artificial intelligence and neuromorphic computing," *Nature Photonics,* vol. 15, pp. 102-114, 2021.

[4] J. E. Bowers and A. Y. Liu, "A comparison of four approaches to photonic integration," in *2017 Optical Fiber Communications Conference and Exhibition (OFC)*, 2017, pp. 1-3.

[5] P. Kaur, A. Boes, G. Ren, T. G. Nguyen, G. Roelkens, and A. Mitchell, "Hybrid and heterogeneous photonic integration," *APL Photonics,* vol. 6, p. 061102, 2021.

[6] Z. Lin, B. J. Shastri, S. Yu, J. Song, Y. Zhu, A. Safarnejadian, W. Cai, Y. Lin, W. Ke, M. Hammood, T. Wang, M. Xu, Z. Zheng, M. Al-Qadasi, O. Esmaeeli, M. Rahim, G. Pakulski, J. Schmid, P. Barrios, W. Jiang, H.Morison, M. Mitchell, X. Guan, N. AF. Jaeger, L. A. Rusch, S. Shekhar, W. Shi, S. Yu, X. Cai, L. Chrostowski "120 GOPS Photonic tensor core in thin-film lithium niobate for inference and in situ training," *Nature Communications, vol. 15, no. 9081, pp. 1-10, 2024*.

[7] M. J. Wale, "Self aligned, flip chip assembly of photonic devices with electrical and optical connections," in *40th Conference Proceedings on Electronic Components and Technology*, 1990, pp. 34-41.

[8] W. Ke, Y. Lin, M. He, M. Xu, J. Zhang, Z. Lin, S. Yu, and X. Cai, "Digitally tunable optical delay line based on thin-film lithium niobate featuring high switching speed and low optical loss," *Photonics Research,* vol. 10, pp. 2575-2583, 2022.

[9] Y. Wang, S. S. Djordjecvic, J. Yao, J. E. Cunningham, X. Zheng, A. V. Krishnamoorthy, M. Muller, M.-C. Amann, R. Bojko, and N. A. Jaeger, "Vertical-cavity surface-emitting laser flip-chip bonding to silicon photonics chip," in *2015 IEEE Optical Interconnects Conference (OI)*, 2015, pp. 122-123.

[10] A. Carlson, A. M. Bowen, Y. Huang, R. G. Nuzzo, and J. A. Rogers, "Transfer printing techniques for materials assembly and micro/nanodevice fabrication," *Advanced Materials,* vol. 24, pp. 5284-5318, 2012.

[11] J. Yoon, S. M. Lee, D. Kang, M. A. Meitl, C. A. Bower, and J. A. Rogers, "Heterogeneously integrated optoelectronic devices enabled by micro-transfer printing," *Advanced Optical Materials,* vol. 3, pp. 1313-1335, 2015.

[12] M. Mitchell, B. Lin, I. Taghavi, S. Yu, D. Witt, K. Awan, S. Gou, J. Young, and L. Chrostowski, "Photonic Wire Bonding for Silicon Photonics III-V Laser Integration," pp. 1-2, 2021.

[13] S. Juodkazis, "Laser polymerized photonic wire bonds approach 1 Tbit/s data rates," *Light Sci Appl,* vol. 9, p. 72, 2020.

[14] Z. Lin, S. Yu, Y. Chen, W. Cai, B. Lin, J. Song, M. Mitchell, M. Hammood, J. Jhoja, N. A. F. Jaeger, W. Shi, and L. Chrostowski, "High-performance, intelligent, on-chip speckle spectrometer using 2D silicon photonic disordered microring lattice," *Optica,* vol. 10, p. 497, 2023.

# Future trends for III/V photonic circuits for optical computing

**Ripalta Stabile**
Eindhoven Hendrik Casimir Institute (EHCI), Technische Universiteit Eindhoven, 5600 MB, Eindhoven, The Netherlands
r.stabile@tue.nl

**Status**

Neuromorphic photonics rises as a new research field [1], aiming to transfer the high-bandwidth and low-energy interconnect credentials in photonics to the field of neuromorphic computing. This is strongly supported by the level of maturity reached by photonic integration technologies [2][3], which makes available high-performance and sophisticated integrated circuits as never seen before.

With respect to the mature Silicon photonics and to promising technologies still in an early stage of development (such as Lithium Niobate on Insulator), the Indium Phosphide (InP) material platform offers a complete set of functionalities, enabled by the monolithic integration of active and passive components on-chip [3]. In particular, the integration of active elements provides both gain for on-chip loss compensation, opening to synaptic operation scalability, and triggered non-linearities, making it the ideal platform for investigating neuromorphic photonic architectures.

Notable examples of neuromorphic photonic implementations include coherent optical linear units based on in-phase and quadrature modulator schemes [4]: When combined with MZI-Semiconductor Optical Amplifier (SOA) based optical nonlinear functions, this architecture offers up to 97.24% accuracy for MNIST digit recognition [5]. The combination of coherent optics with wavelength division multiplexing (WDM) also opens to a multifunctional programmable neural network platform [6]. The exploitation of a cross-connect architecture, by using arrayed waveguide gratings (AWGs) and SOA technologies, enables the realization of high-data throughput matrix multiplication units [7], trading high dynamic range (up to 9 bit resolution) with energy efficiency (~tens of pJ/MAC). Operating the SOA both in its linear and non-linear regime makes possible the demonstration of fully monolithically integrated SOA-based neurons, exploiting SOA-based wavelength converters as non-linear functions [8]. The versatility offered by the cross-connect architecture is further exploited to perform convolution neural networks on-chip for a competitive computational speed of 10.24 TMACs/s and an end-to-end energy efficiency of 0.26 pJ/MAC [9]. Reservoir computing, a particular type of recurrent neural network, has also been demonstrated, based on the richer dynamics of a network of coupled SOAs [10]. Superior performance could be further achieved when using non-linear resonators on III-V platforms [11], suggesting how noteworthy is pursuing this technology.

**Current and Future Challenges**

InP wafers of 6 inches are now commercially available, opening a route to productivity and lower costs per chip. The generic III-V photonic integration is an excellent platform for developing cutting edge devices that generate, modulate and detect light on-chip, however it falls short of the compactness requirements for higher computational density and energy efficiency. In fact, the low refractive index contrast typical of the InP layer stack requires larger radius waveguide bends and bigger components overall, steadily increasing the final circuit size.

The accumulated noise and signal degradation intrinsic to the analog processing still limit the overall system resolution. In SOA-based deep neural networks, the SOAs themselves introduce noise, which builds up in cascades of these same components. However, we have recently demonstrated that noise compression happens in an SOA-based all-optical neuron, when exploiting multi-wavelength to single-wavelength conversion as non-linear function, which suggests that neural networks with arbitrary depth are possible [12]. The relative intensity noise of InP integrated lasers still limits the maximum effective number of bits, but this could be readily mitigated via hybrid integration. In addition, state-of-the-art InP modulators have been recently demonstrated to reach 100 GHz with a dynamic extinction ratio much higher than 10 dB [13], supporting analog processing.

Based on the highlighted challenges, the interfacing of InP technology to complementary material platforms is identified as the necessary step to enable next generation InP neuromorphic photonics.

**Advances in Science and Technology to Meet Challenges**

A new III-V technology platform is rising that is based on the exploitation of InP Membranes On Silicon, also called IMOS platform. IMOS is attractive as it allows the integration of ultra-compact components, due to its intrinsic high refractive index, and the availability of native active devices [14]. Moreover, the vertical integration of InP membranes on top of electronic ICs is expected to enable the development of large-scale, high-speed and complex optoelectronic circuitries [15]. Figure 1a provides the lateral view of an SOA in the IMOS platform, after bonding the InP layer on top of the Silicon layer via a BCB polymer layer and patterning. For the same circuit architectures, in IMOS a four-fold improvement in compactness is readily possible when including active devices, and a >10-fold improvement for passive circuitries only. Values of SOA bias currents 35% lower than in generic InP are already possible [16], accordingly improving the energy efficiency per operation. Coherent approaches for IMOS Photonic Neural Networks (PNNs) could also be pursued in an energy efficient and ultra-compact fashion, by exploiting the strong thermo-optic effect in InP and minimizing the heat capacity [17].

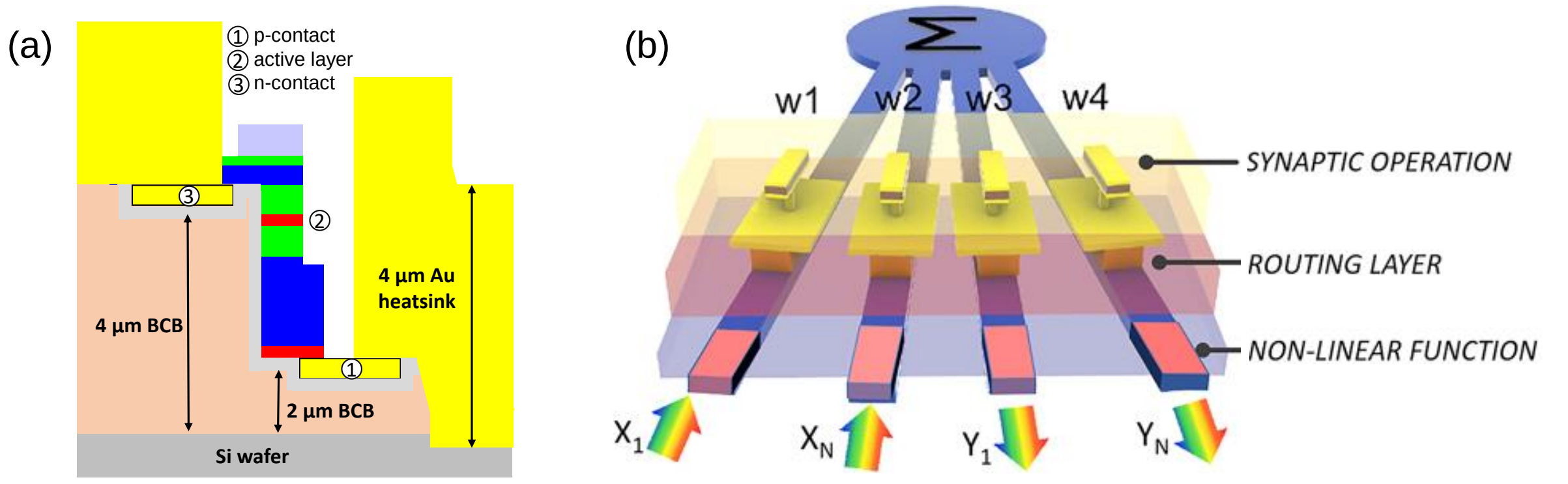

Figure 1. Schematic illustration of a) the final cross-section of an SOA with heat sink in IMOS platform and of b) the envisioned photonic neuron in a 3D fashion, showing three different layers with different functionalities.

Hybrid integration schemes are overall expected to enable a step-change in PNN performance. Multi-layer $HfO_2/Al_2O_3$ memristors embedded with III-V/Si photonics are expected to provide high-speed power-efficient phase tuners, on-chip non-volatile memory and co-integration of all necessary components onto a single chip [18]. We also envision a novel and advanced concept of 3D photonic neuron [19], made of three planes stacked in a 3D fashion, each with a different functionality: synaptic operation, routing and non-linearity (Figure 1b). In particular, the integration of electrically controlled non-volatile Sb-based layers onto low-loss SiN waveguides is expected to produce record-low-loss non-volatile photonic structures [20]. On the other side, InP nano-photonic crystals are envisioned to be used for ultra-low-energy ultra-fast photonic fan-in, gating and activation function units [11], after transferring this technology under (or onto) the SiN Sb-loaded waveguide platform. The envisioned 3D photonic neural network platform will deliver the most compact and complete functional set for neuromorphic computations that, interconnected to form programmable meshes [2], is foreseen to release sub-fJ/MAC energy efficiencies and peta-scale computational power.

**Concluding Remarks**

The need of on-chip gain for scalability has so far been leveraged to demonstrate both linear and non-linear units in InP PNNs.  A number of breakthroughs are though required to unlock the potential of neuromorphic photonics. We foresee that the development of advanced hybrid integrated schemes – involving InP as key player – will enable peta-scale computation and fJ/operation energy consumption, shaping the future of neuromorphic computing.

**Acknowledgements**

This research was supported by the Netherlands Organization of Scientific Research (NWO) under the Zwaartekracht programma, “Research Centre for Integrated Nanophotonics”, Grant No. 024.002.033. The author thanks B. Shi and Y. Jiao for fruitful discussion.

## References

[1]. B.J. Shastri, B.J., A.N. Tait, T. Ferreira de Lima, , W.H.P. Pernice, H. Bhaskaran, C.D. Wright and P.R. Prucnal, “Photonics for artificial intelligence and neuromorphic computing”, *Nat. Photonics*, 15, 102–114 (2021).

[2]. W. Bogaerts, D. Pérez, J. Capmany, D. A. B. Miller, J. Poon, D. Englund, F. Morichetti, and A. Melloni, “Programmable photonic circuits”, *Nature*, 586, 207–216 (2020).

[3]. M. Smit et al.,“An introduction to InP-based generic integrated technology”, *IOP Semicond. Sci. Technol.*, 29(8), 083001 (2014).

[4]. G. Mourgias-Alexandris, A. Totovic, A. Tsakyridis, N. Passalis, K. Vyrsokinos, A. Tefas, and N. Pleros, “Neuromorphic photonics with coherent linear neurons using dual-IQ modulation cells”, *J. Lightwave Technol.* 38(4), 811–819 (2020).

[5]. G. Mourgias-Alexandris, A. Tsakyridis, N. Passalis, A. Tefas, K. Vyrsokinos, and N. Pleros, “An all-optical neuron with sigmoid activation function”, *Opt. Express* 27, 9620–9630 (2019).

[6]. A. Totovic, G. Giamougiannis, A. Tsakyridis, D. Lazovsky, and N. Pleros , “Programmable photonic neural networks combining WDM with coherent linear optics”, *Sci. Rep.*, 12, 5605 (2022).

[7]. B. Shi, N. Calabretta, and R. Stabile, “Deep neural network through an InP SOA-based photonic integrated cross-connect,” *IEEE J. Sel.* Top. Quantum Electron., 26(1), 7701111 (2020).

[8]. B. Shi, N. Calabretta, and R. Stabile, “InP photonic integrated multi-layer neural networks: Architecture and performance analysis”, *APL Photonics*, vol. 7, no. 1, p. 10801 (2022).

[9]. B. Shi, N. Calabretta, R. Stabile, “Photonic Convolutional Neural Networks though SOA-based Cross-connect”, *IEEE J. Sel. Top. Quantum Electron. Special issue on Optical Computing*, vol. 29, no. 2 (2023).

[10]. K. Vandoorne, J. Dambre, D. Verstraeten, B. Schrauwen and P. Bienstman, "Parallel Reservoir Computing Using Optical Amplifiers," in *IEEE Transactions on Neural Networks*, vol. 22, no. 9, pp. 1469-1481 (2011).

[11]. L. Constans, S. Combrié, X. Checoury, G. Beaudoin, I. Sagnes, F. Raineri, A. De Rossi, “III-V/Silicon hybrid non-linear nanophotonics in the context of on-chip optical signal processing and analog computing”, *Frontiers in Physics*, 7, pg. 133 (2019).

[12]. B. Shi, N. Calabretta, and R. Stabile, “Emulation and Modelling of SOA-based All-Optical Photonic Integrated Deep Neural Network with Arbitrary Depth”, *IOP Neuromorphic Computing Engineering*, 2, 034010 (2022).

[13]. J. A. Hillier *et al*., "A 100 GBaud co-planar stripline Mach-Zehnder modulator on Indium Phosphide platform," *Opto-Electronics and Communications Conference (OECC)*, Shanghai, China, 2023, pp. 1-3 (2023).

[14]. Y. Jiao et al., “Indium Phosphide Membrane Nanophotonic Integrated Circuits on Silicon,” *Phys. Status Solidi Appl. Mater. Sci.*, vol. 217, no. 3, pp. 1–12 (2020).

[15]. S. Abdi *et al*., "Research Toward Wafer-Scale 3D Integration of InP Membrane Photonics With InP Electronics," in *IEEE Transactions on Semiconductor Manufacturing*, 37(3) (2024).

[16]. D. W. Feyisa, S. Abdi, R. van Veldhoven, N. Calabretta, Y. Jiao and R. Stabile, "Low Polarization Sensitive O-band SOA on InP Membrane for Advanced Photonic Integration," in *Journal of Lightwave Technology*, 3369232 (2024).

[17]. Y. Wang, V. Dolores-Calzadilla, K. A. Williams, M. K. Smit and Y. Jiao, "Ultra-Compact and Efficient Microheaters on a Submicron-Thick InP Membrane," in *Journal of Lightwave Technology*, vol. 41, no. 6, pp. 1790-1800 (2023).

[18]. Cheung, S., Tossoun, B., Yuan, Y. et al., "Energy efficient photonic memory based on electrically programmable embedded III-V/Si memristors: switches and filters", Commun Eng 3, 49 (2024).
[19]. R. Stabile, G. Dabos, C. Vagionas, B. Shi, N. Calabretta, and N. Pleros, "Neuromorphic photonics: 2D or not 2D?", *J. Appl. Phys.*, vol. 129, no. 20, p. 200901 (2021).
[20]. M. Delaney, I. Zeimpekis, D. Lawson, D. W. Hewak, and O. L. Muskens, "A new family of ultralow loss reversible phase-change materials for photonic integrated circuits: $Sb_2S_3$ and $Sb_2Se_3$", *Adv. Funct. Mater.*, 30, 2002447 (2020).

# Silicon photonics-electronics co-integration for photonic computing

**Avilash Mukherjee[1], Mohammed A. Al-Qadasi[1], Mieszko Lis[1], Lukas Chrostowski[1], Bhavin J. Shastri[2] and Sudip Shekhar[1]**
[1]University of British Columbia, [2]Queen's University

avilash@ece.ubc.ca, alqadasi@ece.ubc.ca, mieszko@ece.ubc.ca, lukasc@ece.ubc.ca, bhavin.shastri@queensu.ca, sudip@ece.ubc.ca

**Status**

Deep Neural Networks (DNN) for image recognition [1] and language translation [2] have sparked the development of DNN accelerators [3], [4] which significantly outperform general-purpose CPUs. However, as DNNs increase in size and scale [2], digital DNN accelerators face high data transfer costs [5]. Thus, several prior works have suggested using emerging memory technologies [6], analog computing [7], [8], and photonic computing [5], [9]-[11] for further DNN acceleration.

Silicon Photonics (SiP) [12] has been traditionally used for communication [13]-[15] across large distances, where optical fiber communication is more efficient than electrical wire communication. A surge in data bandwidth requirements due to the deployment of large language models [2], datacenter disaggregation [14], and limits in electrical interconnect scaling [16], [17], are causing deployment of SiP transceivers in data centers [13]-[15], and network-on-package/chip (NoP/NoC) [11], [15].

Alberio [5] leverages waveguide splitters to broadcast data and wavelength division multiplexing (WDM) to perform multiply and add (MAC) operations in Mach-Zehnder modulators (MZM). DEAP-CNN [9] performs multiplication via micro-ring resonators (MRRs) and addition on a photodiode (PD). Flumen [11] uses a Mach-Zehnder interferometer (MZI) array with an 8-λ WDM for MAC operations in the photonic domain.

The proposed SiP accelerators promise significant speedup for weight-stationary tasks over digital DNN accelerators [5], [9], [10]. However, higher energy consumption [5], [10] and low precision [10] remain as challenges.

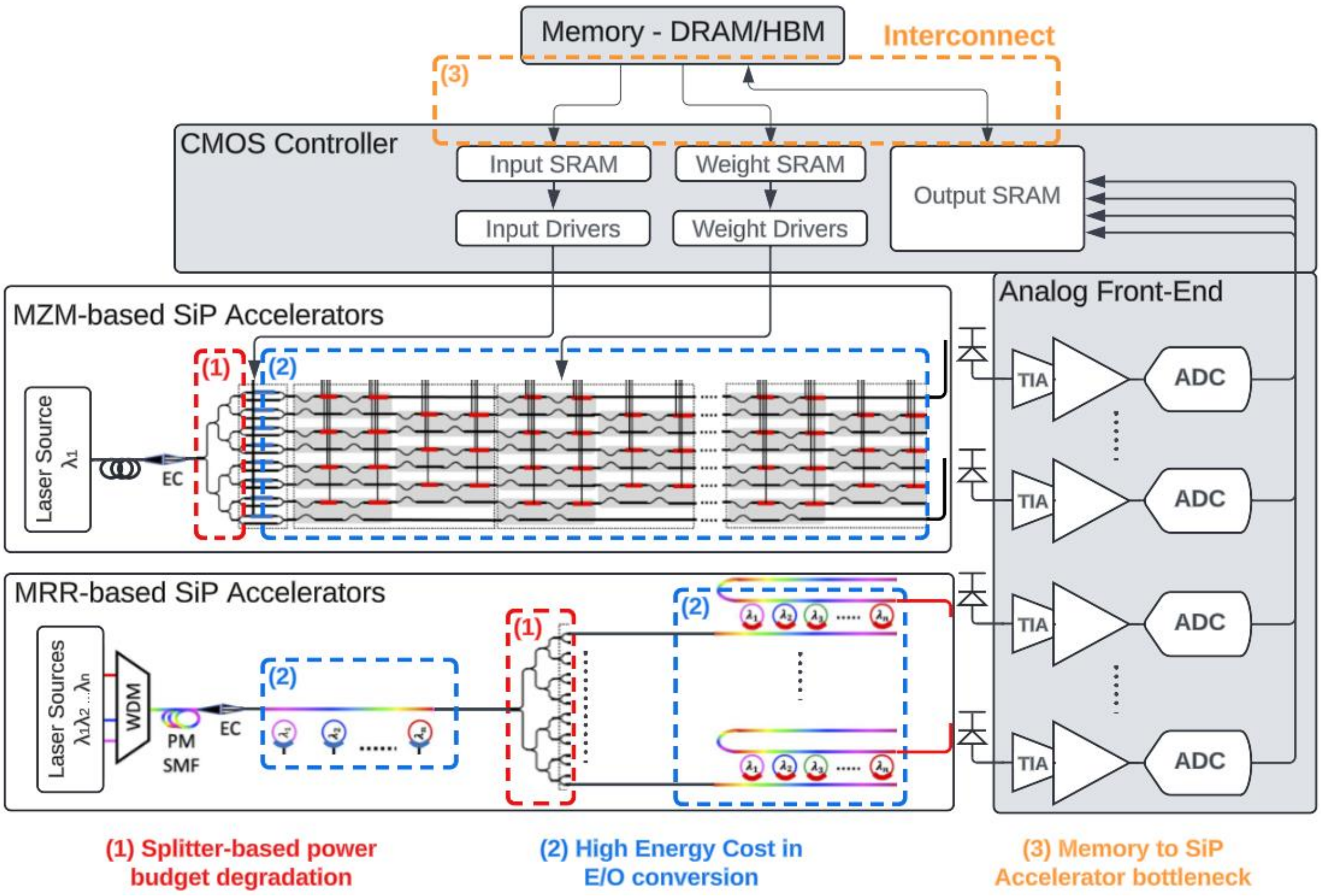

**Fig. 1.** General SiP Accelerator system, showing the MZM and MRR-based implementations of SiP accelerators, AFE, CMOS controller and the interconnect connecting to the memory system. Current challenges in SiP accelerators include (i) power budget degradation due to splitters, (ii) high energy cost in E/O conversion for both inputs and weights, and (iii) SiP to Memory bottleneck. Adapted from [10].

## Current and Future Challenges

We show a general SiP accelerator with MZM/MRR-based implementations, highlighting the current challenges in Fig.1. The broadcast operation through waveguide splitters severely degrades the photonic link budget [10]. Consequently, the scalability of SiP accelerators is constrained by the minimum optical power required by the analog-frontend (AFE) to provide digital bits reliably [10]. The achievable output bit precision for different array sizes of SiP accelerators is depicted in Fig.2(a).

Optical power loss degrades the bit precision significantly in the prior works. In Alberio [5], waveguide splitting causes 9.5dB optical power loss. Additional losses due to couplers, MZMs, and waveguide grating used for WDM further decrease the optical power at the PD to -11dBm from an initial laser power of 15.7dBm/λ [5], limiting output precision to 4 bits. Losses in the 8x8 MZI array in Flumen [11] allow only 5 bits of output precision [10]. Similarly, DEAP-CNN [9] incurs at least -10.8dB optical power loss due to waveguide splitters.

Furthermore, electrical-to-optical (E/O) conversion consumes significant energy in the SiP accelerators, consuming more than 80% of energy in an SiP MAC operation, as shown in Fig.2(b). Currently, weight tuning contributes to most of the energy consumption. However, input modulation energy can become significant as the bit precision increases.

While SiP accelerators focus on MAC operation, the data transfer bottleneck to the SiP accelerators impedes the speedup [18] and energy benefits. The data for MAC operation goes through the memory interconnect to reach the SiP accelerator, and then MAC outputs traverse the interconnect again to be stored in memory. Significantly higher energy is also consumed at a system level as the SiP interconnect energies are higher than the energy/MAC in SiP accelerators (Fig.2(c)).

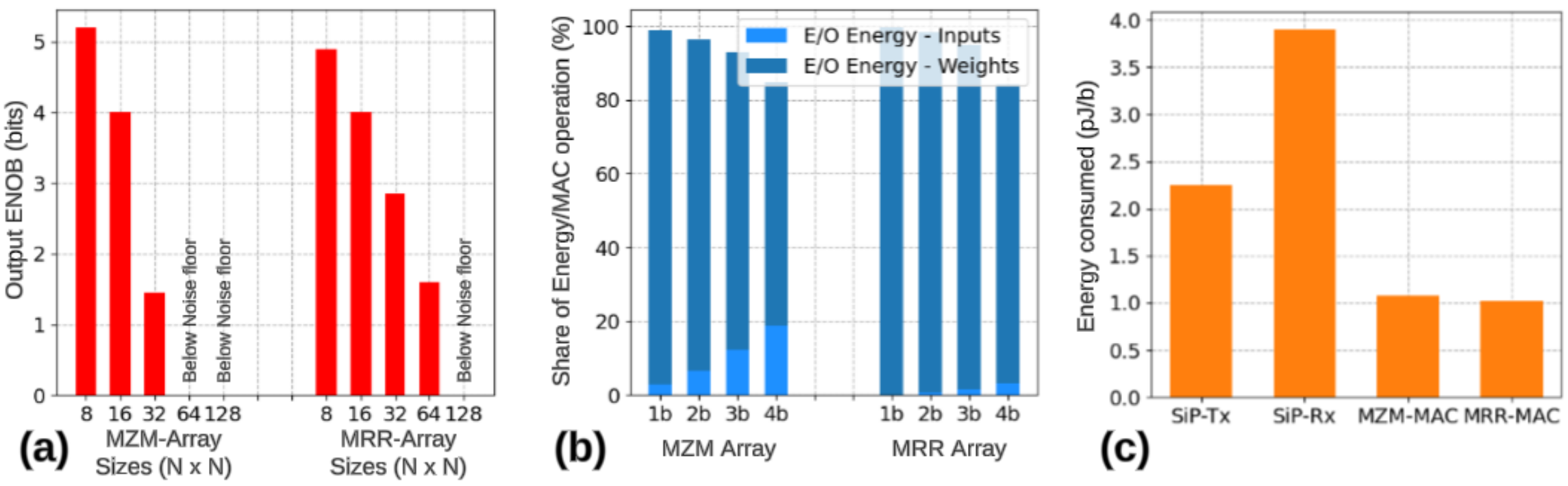

**Fig. 2.** (a) Effective number of bits (ENOB) for the output activations for different MZM/MRR-array sizes (N) in SiP accelerators. Number of neurons in the SiP accelerator arrays is N x N, and ranges from 64 to 16384 in our analysis. Calculations assume 10dBm laser power per λ with 10% WPE, a PD responsivity of 1.2A/W, data rate of 10GS/s, and AFE specifications in [10]. (b) Percentage of energy/MAC operation in E/O conversion for Input and Weights for MZM/MRR-based implementations with thermal phase shifters across MAC bit precisions 1b-4b. Adapted from [10]. (c) Energy/bit for SiP transmitters [15], SiP receivers [19], and MAC operations in MZM/MRR-based SiP accelerators using thermal phase shifters [10].

## Advances in Science and Technology to Meet Challenges

Advancements in SiPs to reduce optical loss are paramount for implementing SiP accelerators with higher precision. To that end, other avenues to scale SiP accelerators should be explored further. Scaling SiP accelerators via increased WDM might be attractive, as current SiP transmitters have demonstrated an 8-λ WDM [14], while optimized MRR designs can theoretically support up to 108-λ [9]. However, advances in multi-λ lasers with high efficiency and high power in each wavelength and CMOS controllers to support numerous MRRs for larger WDM while maintaining weight linearity and ensuring minimal ring crosstalk across all λ are required to increase WDM.

MZM and MRR tuning energy can be improved through insulation by introducing trenches, under-cuts, and back-side substrate removal, reducing energy/MAC by more than 60% for SiP accelerators [10]. However, the CMOS controller should include the effects of self-heating for robust operation.

Advances in packaging can further reduce E/O and system energy consumption. A 2.5D/3D integrated solution, such as the TSMC COUPE [20], where the PIC is implemented in an SOI process, and an advanced FinFET CMOS process is stacked with the PIC, is ideal for providing the short interconnects between the electronics and the photonic circuits.

## Concluding Remarks

In this brief, we summarized the challenges in SiP accelerators, focusing on link-budget degradation, energy costs in E/O conversion, and the memory to SiP accelerator bottleneck. These challenges affect the bit precision achieved in SiP accelerators and incur latency and energy costs at the system level. Currently, SiP accelerators with arrays sized 32x32 are feasible for 2-3 bits of output precision (Fig. 2(a)).

We outline potential strategies to enhance the scalability of SiP accelerators for large DNN sizes. At the device level, utilizing efficient optical phase shifters, modulators, and multi-wavelength lasers, and at the assembly level, advanced packaging techniques can significantly reduce E/O conversion and data transfer energy. At the system level, better co-design of electronics and photonics to manage noise, linearity, and voltage swing considerations can further enable scalability. However, increasing the bit precision of the SiP accelerators remains a difficult challenge.

## References

[1] K. He, X. Zhang, S. Ren and J. Sun, "Deep Residual Learning for Image Recognition," 2016 IEEE Conference on Computer Vision and Pattern Recognition (CVPR), Las Vegas, NV, USA, 2016, pp. 770-778.

[2] J. Achiam, S. Adler, S. Agarwal, L. Ahmad, I. Akkaya, F.L. Aleman, D. Almeida, J. Altenschmidt, S. Altman, S. Anadkat, R. Avila. GPT-4 technical report. arXiv preprint arXiv:2303.08774. 2023 Mar 15.

[3] N. Jouppi, G. Kurian, S. Li, P. Ma, R. Nagarajan, L. Nai, N. Patil, S. Subramanian, A. Swing, B. Towles, C. Young, X. Zhou, Z. Zhou, and D. A. Patterson. 2023. TPU v4: An Optically Reconfigurable Supercomputer for Machine Learning with Hardware Support for Embeddings. In Proceedings of the 50th Annual International Symposium on Computer Architecture (ISCA '23). Association for Computing Machinery, New York, NY, USA, Article 82, 1–14.

[4] B. Keller et al., "A 95.6-TOPS/W Deep Learning Inference Accelerator With Per-Vector Scaled 4-bit Quantization in 5 nm," in IEEE Journal of Solid-State Circuits, vol. 58, no. 4, pp. 1129-1141, April 2023.

[5] K. Shiflett, A. Karanth, R. Bunescu and A. Louri, "Albireo: Energy-Efficient Acceleration of Convolutional Neural Networks via Silicon Photonics," 2021 ACM/IEEE 48th Annual International Symposium on Computer Architecture (ISCA), Valencia, Spain, 2021, pp. 860-873.

[6] A. Mukherjee, K. Saurav, P. Nair, S. Shekhar and M. Lis, "A Case for Emerging Memories in DNN Accelerators," 2021 Design, Automation & Test in Europe Conference & Exhibition (DATE), Grenoble, France, 2021, pp. 938-941.

[7] Ambrogio, S., Narayanan, P., Tsai, H. et al. Equivalent-accuracy accelerated neural-network training using analogue memory. Nature 558, 60–67 (2018).

[8] H. Jia et al., "Scalable and Programmable Neural Network Inference Accelerator Based on In-Memory Computing," in IEEE Journal of Solid-State Circuits, vol. 57, no. 1, pp. 198-211, Jan. 2022.

[9] V. Bangari et al., "Digital Electronics and Analog Photonics for Convolutional Neural Networks (DEAP-CNNs)," in IEEE Journal of Selected Topics in Quantum Electronics, vol. 26, no. 1, pp. 1-13, Jan.-Feb. 2020, Art no. 7701213.

[10] M. A. Al-Qadasi, L. Chrostowski, B. J. Shastri, S. Shekhar; Scaling up silicon photonic-based accelerators: Challenges and opportunities. APL Photonics 1 February 2022; 7 (2): 020902.

[11] K. Shiflett, A. Karanth, R. Bunescu, and A. Louri. 2023. Flumen: Dynamic Processing in the Photonic Interconnect. In Proceedings of the 50th Annual International Symposium on Computer Architecture (ISCA '23). Association for Computing Machinery, New York, NY, USA, Article 75, 1–13.

[12] Shekhar, S., Bogaerts, W., Chrostowski, L. et al. Roadmapping the next generation of silicon photonics. Nat Commun 15, 751 (2024).

[13] H. Li et al., "A 3-D-Integrated Silicon Photonic Microring-Based 112-Gb/s PAM-4 Transmitter With Nonlinear Equalization and Thermal Control," in IEEE Journal of Solid-State Circuits, vol. 56, no. 1, pp. 19-29, Jan. 2021.

[14] C. S. Levy et al., "8-λ × 50 Gbps/λ Heterogeneously Integrated Si-Ph DWDM Transmitter," in IEEE Journal of Solid-State Circuits, vol. 59, no. 3, pp. 690-701, March 2024.

[15] D. F. Logan et al., "800 Gb/s Silicon Photonic Transmitter for CoPackaged Optics," 2020 IEEE Photonics Conference (IPC), Vancouver, BC, Canada, 2020, pp. 1-2.

[16] M. Rakowski et al., "45nm CMOS — Silicon Photonics Monolithic Technology (45CLO) for Next-Generation, Low Power and High Speed Optical Interconnects," 2020 Optical Fiber Communications Conference and Exhibition (OFC), San Diego, CA, USA, 2020, pp. 1-3.

[17] J. Jaussi et al., "26.2 A 205mW 32Gb/s 3-Tap FFE/6-tap DFE bidirectional serial link in 22nm CMOS," 2014 IEEE International Solid-State Circuits Conference Digest of Technical Papers (ISSCC), San Francisco, CA, USA, 2014, pp. 440-441.

[18] C. Demirkiran, F. Eris, G. Wang, J. Elmhurst, N. Moore, N. C. Harris, A. Basumallik, V. J. Reddi, A. Joshi, and D. Bunandar. 2023. An Electro-Photonic System for Accelerating Deep Neural Networks. J. Emerg. Technol. Comput. Syst. 19, 4, Article 30 (October 2023), 31 pages.

[19] H. Li, C. -M. Hsu, J. Sharma, J. Jaussi and G. Balamurugan, "A 100-Gb/s PAM-4 Optical Receiver With 2-Tap FFE and 2-Tap Direct-Feedback DFE in 28-nm CMOS," in IEEE Journal of Solid-State Circuits, vol. 57, no. 1, pp. 44-53, Jan. 2022.

[20] H. Hsia et al., "Integrated Optical Interconnect Systems (iOIS) for Silicon Photonics Applications in HPC," 2023 IEEE 73rd Electronic Components and Technology Conference (ECTC), Orlando, FL, USA, 2023, pp. 612-616.

# Neuromorphic Computing in 3D Electronic-Photonic-Ionic-Integrated Circuits: Roadmap for Brain-Derived Neuromorphic Computing

**S. J. Ben Yoo**
sbyoo@ucdavis.edu
Department of Electrical and Computer Engineering, University of California, Davis, CA 95616, United States of America

**Status**

Neuromorphic [1] computing adopts its system level architecture design in such a way to replicate 'exact' basic anatomical identified operations which embody several key features encountered in the biological system[2]. While there have been numerous research activities on bio-inspired computing and biomimetic computing involving photonics, there has been much fewer work on neuromorphic photonic computing. In fact, the ultimate objective for neuromorphic computing would be to realize what no artificial system has been able to achieve so far--- *to match the immense and flexible learning capabilities at extreme energy efficiencies and scale of the brain*. Previous efforts towards neuromorphic computing have utilized typically CMOS components to imitate neurons and synapses, including dendrites in some cases, to show circuit level demonstrations of training and inference. More recent demonstrations of CMOS based Loihi2 [3] and NorthPole [4] neuromorphic computing with 2.3-22 billion transistor scale neuromorphic computing utilizing digitally encoded electrical spikes communicating with other neurons in four other directions (N-E-W-S). Photonics brings new dimensions to neuromorphic computing in many ways: (1) it offers optical parallelism in wavelength, time, and space domains, (2) matrix-multiplication in optical mesh can be achieved without consumption of energy other than photodetection, (3) information communication can be nearly lossless (~0.1 dB/cm for silicon or silicon nitride waveguides) across extremely broad information bandwidth ( > 10 Thz), (4) there is no parasitic capacitance, inductance, or resistance that limits bandwidth or fidelity of transmission, (5) it is associated with extremely low noise (free of Johnson noise), and (6) it is capable of achieving very fast optical barrier synchronization. For these reasons, photonic neuromorphic computing has prospered in recent years [5]–[8]. On the other hand photonics alone is insufficient to "replicate 'exact' basic anatomical identified operations which embody several key features in biological systems." For one, photonics does not readily offer nonlinearity and time delay at reasonable power or time scales. Secondly, any reasonable optical wavelength scale devices are much larger than biological dendrites or electronic counterparts. A combination of photonic, electronic, and even ionic mechanisms integrated in a neuromorphic computing system, preferably integrated in 3D, can potentially offer neuromorphic computing with potentials to offer basic anatomical identified operations of biological systems. Such efforts in 3D electronic-photonic integrated circuits (3D EPIC) may eventually lead towards ***reverse-engineering the brain*** exploiting best of both worlds of photonic and electronic neuromorphic computing.

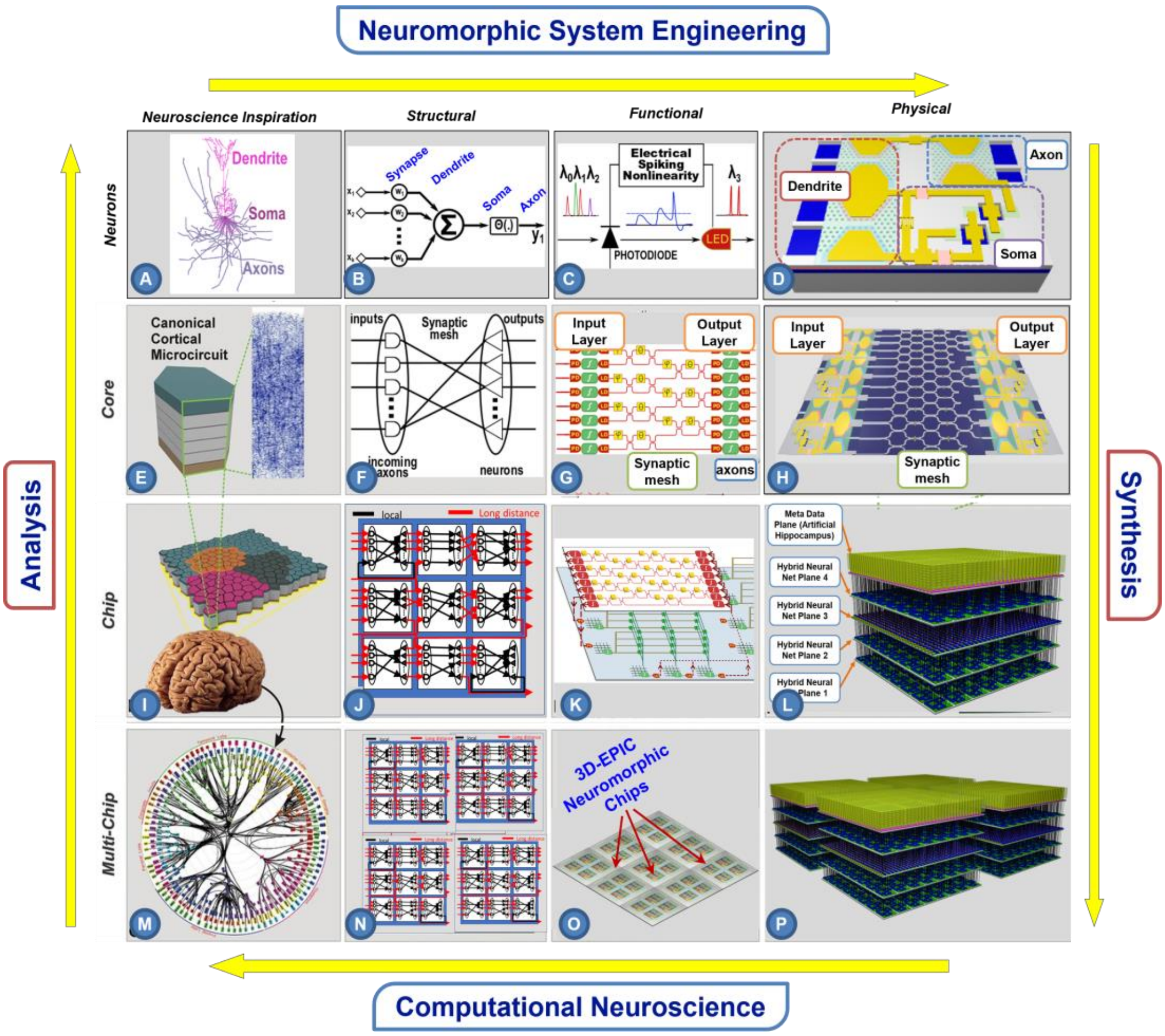

**Figure 1.** A roadmap to develop a conceptual 3D Hierarchical Neuromorphic Nanocomputing architecture extending on the framework by [9]. **(A)** A canonical neuron, **(B)** Neuron's minimal structure, **(C)** Neuron's simplified functional diagram (optoelectronic neuron example), **(D)** A physical schematic for a nanoscale optoelectronic neuron, **(E)** cortical microcircuit, **(F)** Structure of a neurosynaptic core with axons as signal carriers (inputs/output), synapses as directed connection strength, and neurons as nonlinearity. **(G)** Functional view of a photonic synaptic mesh between presynaptic and post synaptic neurons. **(H)** Physical layout of (G). **(I)** A two-dimensional map of cortical columns in a functional network. Multichip scales are both created by interconnecting **(J)** neuron microcircuits reconfigurable optical synaptic interconnects. **(K)** Hybrid optical (red) and electronic (green) neural network forming a hierarchical macrocircuit. **(L)** Schematic of 3D electronic photonic integrated circuit (EPIC) neural network consisting of multiple planes of (K). **(M)** Illustration of long-range connections between cortical regions in the macaque brain [10]. **(N)** Interconnections of many functionally specialized neural macro-circuits (J). **(O)** Multi-3D EPIC chip neural networks emulating functional specializations of interconnected human brain structures. **(P)** Schematic of (O).

## Current and Future Challenges

Some of the fundamental challenges in neuromorphic computing in general can be summarized in the following four 'Gaps' that are observed in typical approaches taken so far.

***Gap 1: Lack of understanding of the principles of learning, plasticity, and dynamics in the context of networked neurons with structured connectivity.*** The brain is a system of networked neurons and synapses with structured connectivity. Detailed physio-chemical mechanisms of learning in such a complex networked system are less understood especially when dynamicity and plasticity of the neuro-physical elements are concerned. In particular, CMOS-based digital circuits with limited dynamicity, plasticity, and networked connectivity failed to achieve neuromorphic computing at the desired energy-efficiency, throughput, scalability, and adaptability.

***Gap 2: Lack of methods to realize neuromorphic dynamics in bio-inspired materials.*** Bio-inspired materials that can fundamentally enable bio-realistic algorithms and models are still not well developed despite the recent progress with metal oxides, polymer composites, photonics and biomolecules. The diverse and rich set of neuromorphic dynamics in the brain are difficult to replicate in the natural and man-made materials. The electro-chemical dynamics of neurotransmitters/ionotropic receivers, morphology changes in the synapses and dendrites, etc are difficult to realize in available materials.

***Gap 3: Lack of methods to realize neuromorphic devices and circuits with the desired dynamicity and interconnectivity.***
The bio-inspired materials with the desired dynamicity need to be built into new artificial electrical, ionic, and/or photonic devices and circuits as required computing elements that can implement the bio-realistic algorithms and models. In particular, the connectivity between the neurons or neuron-like devices should achieve very large interconnectivity (~8000 synaptic connections per neuron in brain). Moreover, bio-realistic circuit designs should be highly nonlinear and rich in dynamics.

***Gap 4: Lack of methods to realize scalable neuromorphic computing system architecture.*** While recently developed Generative Pre-trained Transformer (GPT) [11] show apparent ingenuity based on 175 billion parameters, however, requires pretraining (no online learning) with more than 12 million USD worth of energy on GPU-based systems just for training for GPT-3 [11] and far more for GPT-4. The human brain, in contrast, is capable of remarkably fast learning in a manner that is flexible and enables generalization to new situations and tasks, and it does so with a remarkably low level of energy consumption relative to traditional computational hardware. Scaling to ~100 billion neurons, ~100 trillion synapses, within ~3 litre volume and ~20W power for the learning capability comparable to the brain requires some remarkable system architectures and technologies.

**Advances in Science and Technology to Meet Challenges**

As discussed in the first Section, if we follow the strict definition of neuromorphic computing given in [2] to *replicate 'exact' basic anatomical identified operations which embody several key features encountered in the biological system*, it should pursue **brain-derived neuromorphic computing** beyond brain-inspired computing. Recent advances in photonic memrisitive materials [6] including phase change photonic materials [12] allowed rapid advances and demonstrations of photonic synaptic networks with low (zero or near-zero) static power consumption used as photonic vector-matrix-multipliers or matrix-matrix-multipliers in neural network applications [13]. Spiking photonic and optoelectronic neurons [14], [15] has been developed by utilizing internal laser dynamics or by combining the nonlinear electronic CMOS circuit with optoelectronic photodetectors and modulators (or lasers) to demonstrate in some cases to show sophisticated neuronal heterogeneity [16]. The recent efforts in 3D electronic photonic integrated circuits (3D EPICs) [17] can possibly bring the comparable scalability and interconnectivity within the similar size, weight, and power constraints of the brain. In most cases, the learning mechanisms were based on supervised learning where external element such as FPGAs or CPUs were centrally used to calculate tuning parameters to be applied for individual Mach-Zehnder interferometer (MZI) meshes or micro-resonator-rings (MRR) for backpropagation, gradient descent, etc. Unlike the synapses and the neurons in biological systems where each weight value $W_{ij}$ between the i-th presynaptic neuron and the j-th postsynaptic neuron can be directly modified through the action of Hebbian learning [18], the photonic synaptic meshes with cascaded MZIs and MRRs cannot do the same. Fortunately, distributed learning mechanisms with local learning consistent with contrastive Hebbian Learning was experimentally demonstrated in an MZI mesh network implying that predictive error-learning [19] may be possible in a photonic neural network.

The roadmap for brain-derived neuromorphic computing overcoming the challenges described in the four gaps can pursue realizing the full capability of the human brain, also referred to as (ultimately) 'reverse-engineering the brain'[20]. We believe that this is possible achievable when advanced photonic and electronic technologies are integrated in 3D EPICs following the bio-plausible architecture overcoming the four gaps as part of the roadmap:

- ***Overcoming Gap 1:*** To understand the interplay between neuroscience and neuromorphic computing better, we shall develop a comprehensive simulator and build a novel neuromorphic computing prototype system that incorporates insights from cutting-edge modelling and experiments about synaptic plasticity, network dynamics, and learning in cortical circuits, and fundamental attributes of human learning and memory.
- ***Overcoming Gap 2:*** We shall pursue new photonic and electronic memristive materials that can closely resemble the dynamic mechanisms responsible in the biological neural systems.
- ***Overcoming Gap 3:*** We shall pursue new photonic and electronic devices utilizing such memristive materials as neurons, synapses, and dendrites, and create photonic-electronic circuits that can closely resemble the dynamic mechanisms seen in the biological neural systems.
- ***Overcoming Gap 4:*** We shall pursue 3D photonic-electronic integrated circuits that offer high density and high connectivity with extreme efficiency at scale while supporting hierarchical learning in optical macro-circuits and electronic micro-circuits.

### Concluding Remarks

Brain-inspired photonic computing and photonic neural networks are making rapid advances in demonstrating its utility as matrix-matrix or matrix-vector multiplier accelerators. There are significant challenges in achieving neuromorphic computing beyond brain-inspired computing, as summarized in the four Gaps commonly seen in the approaches taken in the past. The recent technological advances in photonics, electronics, and ionics at nanoscale may offer some possible answers to meet the challenges observed in the four Gaps. Photonics or electronics alone is unlikely to realize the desired neuromorphic computing. Brain-derived neuromorphic computing exploiting photonic, electronic, and perhaps ionic technologies integrated in 3D may overcome the four gaps commonly seen in today's neuromorphic computing. Integration of bio-plausible architecture and learning algorithm on this new hardware platform may lead to flexible and adaptive learning capability with extremely high energy efficiency and throughput, perhaps approaching those of the human brain by overcoming the four Gaps following the new Roadmap.

### Acknowledgements

This material is based upon work supported by the Air Force Office of Scientific Research under award number FA9550-18-1-0186 and award number FA 9550-22-1-0532, and in part by the Office of the Director of National Intelligence, Intelligence Advanced Research Projects Activity under Grant 2021-21090200004. The author is indebted to enlightening discussions with Suman Datta, Shimeng Yu, Jean Anne Incorvia, Alberto Salleo, Volker Sorger, Juejun Hu, Lionel C Kimerling, Kristofer Bouchard, Joy Geng, Rishidev Chaudhuri, Charan Ranganath, Randall C. O'Reilly, Luis El-Srouji, Mehmet On, and Yun-Jhu Lee.

## References

[1] C. A. Mead, *Analog VLSI and Neural Systems*. 1989.

[2] A. Katsiamis and E. Drakakis, "Analogue CMOS Cochlea Systems: A Historic Retrospective," in *Biomimetic Based Applications*, 2011.

[3] G. Orchard, E. P. Frady, D. B. D. Rubin, S. Sanborn, S. B. Shrestha, F. T. Sommer, and M. Davies, "Efficient Neuromorphic Signal Processing with Loihi 2," Nov. 2021.

[4] D. S. Modha *et al.*, "Neural inference at the frontier of energy, space, and time," *Science (1979)*, vol. 382, no. 6668, pp. 329–335, Oct. 2023, doi: 10.1126/science.adh1174.

[5] B. J. Shastri, A. N. Tait, T. de Lima, M. A. Nahmias, H.-T. Peng, and P. R. Prucnal, “Neuromorphic Photonics, Principles of,” in *Encyclopedia of Complexity and Systems Science*, R. A. Meyers, Ed. Berlin, Heidelberg: Springer Berlin Heidelberg, 2018, pp. 1–37.

[6] E. Goi, Q. Zhang, X. Chen, H. Luan, and M. Gu, “Perspective on photonic memristive neuromorphic computing,” *PhotoniX*, vol. 1, no. 1, 2020, doi: 10.1186/s43074-020-0001-6.

[7] A. N. Tait, T. F. de Lima, E. Zhou, A. X. Wu, M. A. Nahmias, B. J. Shastri, and P. R. Prucnal, “Neuromorphic photonic networks using silicon photonic weight banks,” *Sci Rep*, vol. 7, no. 1, p. 7430, 2017, doi: 10.1038/s41598-017-07754-z.

[8] S. J. Ben Yoo, “Brain-Derived 3D NanoPhotonic-NanoElectronic Neuromorphic Computing,” in *2022 IEEE Photonics Conference (IPC)*, Nov. 2022, pp. 1–2, doi: 10.1109/IPC53466.2022.9975516.

[9] P. A. Merolla *et al.*, “A million spiking-neuron integrated circuit with a scalable communication network and interface,” *Science (1979)*, vol. 345, no. 6197, pp. 668–673, Aug. 2014, doi: 10.1126/science.1254642.

[10] D. S. Modha and R. Singh, “Network architecture of the long-distance pathways in the macaque brain,” *Proceedings of the National Academy of Sciences*, vol. 107, no. 30, pp. 13485 LP – 13490, Jul. 2010, doi: 10.1073/pnas.1008054107.

[11] J. Sevilla, L. Heim, A. Ho, T. Besiroglu, M. Hobbhahn, and P. Villalobos, “Compute Trends Across Three Eras of Machine Learning,” in *2022 International Joint Conference on Neural Networks (IJCNN)*, 2022, pp. 1–8, doi: 10.1109/IJCNN55064.2022.9891914.

[12] Z. Fang, R. Chen, J. Zheng, A. I. Khan, K. M. Neilson, S. J. Geiger, D. M. Callahan, M. G. Moebius, A. Saxena, M. E. Chen, C. Rios, J. Hu, E. Pop, and A. Majumdar, “Ultra-low-energy programmable non-volatile silicon photonics based on phase-change materials with graphene heaters,” *Nat Nanotechnol*, vol. 17, no. 8, pp. 842–848, 2022, doi: 10.1038/s41565-022-01153-w.

[13] N. Youngblood, “Coherent Photonic Crossbar Arrays for Large-Scale Matrix-Matrix Multiplication,” *IEEE Journal of Selected Topics in Quantum Electronics*, vol. 29, no. 2: Optical Computing, pp. 1–11, 2023, doi: 10.1109/JSTQE.2022.3171167.

[14] Y.-J. Lee, M. B. On, X. Xiao, R. Proietti, and S. J. Ben Yoo, “Photonic spiking neural networks with event-driven femtojoule optoelectronic neurons based on Izhikevich-inspired model,” *Opt Express*, vol. 30, no. 11, p. 19360, May 2022, doi: 10.1364/OE.449528.

[15] M. Nahmias, H. Peng, T. F. D. Lima, C. Huang, A. Tait, B. Shastri, and P. Prucnal, “A Laser Spiking Neuron in a Photonic Integrated Circuit.,” *arXiv: Applied Physics*, 2020.

[16] Y.-J. Lee, M. B. On, L. El Srouji, L. Zhang, M. Abdelghany, and S. J. Ben Yoo, “Demonstration of Programmable Brain-Inspired Optoelectronic Neuron in Photonic Spiking Neural Network With Neural Heterogeneity,” *Journal of Lightwave Technology*, pp. 1–12, 2024, doi: 10.1109/JLT.2024.3368450.

[17] Y. Zhang, A. Samanta, K. Shang, and S. J. B. Yoo, “Scalable 3D Silicon Photonic Electronic Integrated Circuits and Their Applications,” *IEEE Journal of Selected Topics in Quantum Electronics*, vol. 26, no. 2, pp. 1–10, Mar. 2020, doi: 10.1109/JSTQE.2020.2975656.

[18] M. C. W. Van Rossum, G. Q. Bi, and G. G. Turrigiano, “Stable Hebbian Learning from Spike Timing-Dependent Plasticity,” 2000.

[19] R. C. O’Reilly, “Biologically Plausible Error-Driven Learning Using Local Activation Differences: The Generalized Recirculation Algorithm,” *Neural Comput*, vol. 8, no. 5, pp. 895–938, Jul. 1996, doi: 10.1162/neco.1996.8.5.895.

[20] S. J. Ben Yoo, L. El-Srouji, S. Datta, S. Yu, J. A. Incorvia, A. Salleo, V. Sorger, J. Hu, L. C. Kimerling, K. Bouchard, J. Geng, R. Chaudhuri, C. Ranganath, and R. C. O’Reilly, “Towards Reverse-Engineering the Brain: Brain-Derived Neuromorphic Computing Approach with Photonic, Electronic, and Ionic Dynamicity in 3D integrated circuits.”

# WEIGHTS AND MEMORY

## Non-Volatile Materials for Photonic Computing

**Nathan Youngblood**, Department of Electrical and Computer Engineering, University of Pittsburgh, Pittsburgh, PA 15261 USA
**Carlos A. Ríos Ocampo**, Department of Materials Science and the Institute for Research in Electronics and Applied Physics, University of Maryland, College Park, MD 20742, USA.
**Juejun Hu**, Department of Materials Science and Engineering, Massachusetts Institute of Technology, Cambridge, MA 02139-4307, USA
Email: nathan.youngblood@pitt.edu, riosc@umd.edu, hujuejun@mit.edu

### Status

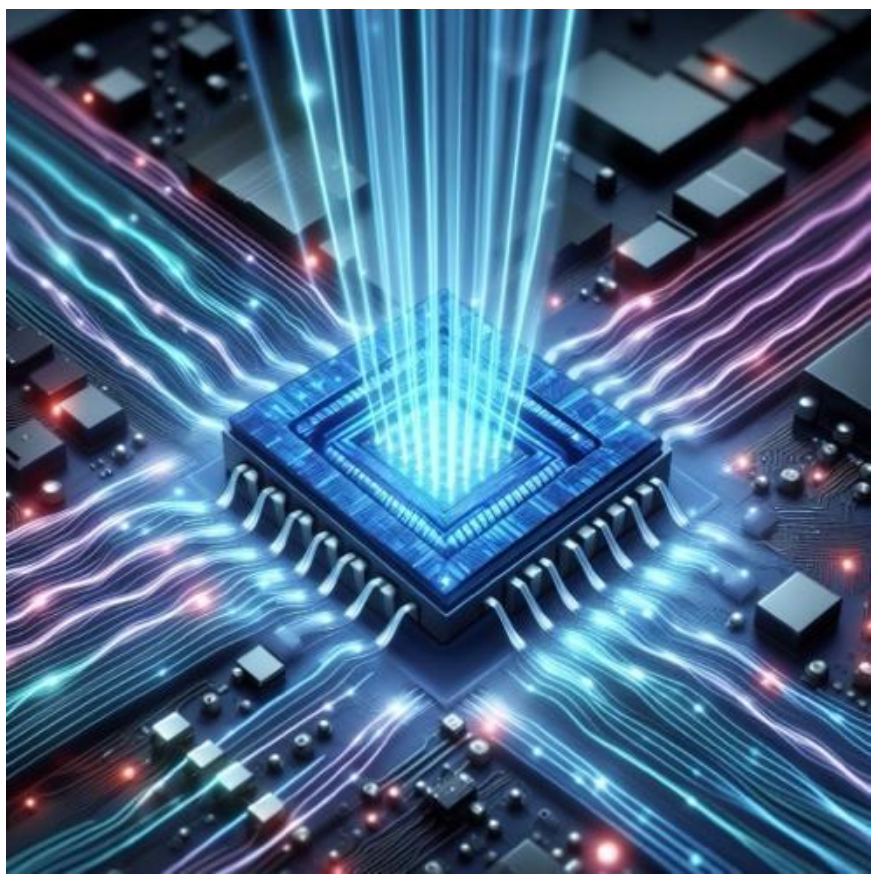

An ever-growing demand for computing power fueled by the rise of artificial intelligence (AI) systems has caused renewed interest in photonic computing. However, a critical challenge for realizing large-scale photonic processors lies in the development of non-volatile devices which are needed for on-chip photonic memory and data storage. Unlike traditional optical components that require continuous power to maintain their state based on volatile mechanisms, such as electro-optic or thermo-optic effects, non-volatile devices offer the ability to retain information even after power is removed. This capability has driven significant research into non-volatile optical materials and devices that can function as memory elements for photonic computing [1], [2], [3], [4].

| Memory Technology | Footprint ($\mu m^2$) | Programming Speed (ns) | Programming Energy (pJ) | Bit Precision | Cycling Endurance | Example Ref. |
|---|---|---|---|---|---|---|
| Phase-change | 5 ~ 25 | 10 ~ 1,000 (am)<br>100 ~ 1M (cry) | 1,000 ~ 10k<br>1,000 ~ 1M | 2 ~ 4 | 250 ~ 20k | Zhang et al. (2019) |
| Ferroelectric | 20k ~ 150k | 1,000 ~ 10M | 30 ~ 3,000 | ~3 | ~300 | Geler-Kremer et al. (2022) |
| MEMS | 5,000 | 0.2 ~ 1,000 | 1 ~ 100 | 1 ~ 2 | 100 ~ 1G | Wen et al. (2023) |
| Memristive | 2 ~ 315 | 0.3 ~ 1,000 | 0.013 ~ 0.4 | 1 ~ 1.5 | ~1,000 | Tossoun et al. (2024) |
| Magneto-optic | 7,850 | 1,000 | ~100k | 3.3 | - | Murai et al. (2020) |
| Charge trapping | 315 ~ 28k | 1M ~ 1G | 10 ~ 1M | 2 ~ 4 | >30 | Song et al. (2016) |

**Table 1: Survey of recent progress in electrically programmable, integrated photonic memory**

In **Table 1**, we summarize the current state of electrically programmable, waveguide-integrated photonic memory technologies grouped by material platform. Here we see a wide range of technologies and mechanisms which can serve as non-volatile photonic memories, such as mechanical, electro-optic, electro-absorptive, phase-change, and ferroelectric. This wide range of effects provides a rich design space in which photonic memories can be optimized to achieve metrics of interest for computing applications. Importantly, we find experimental examples where these electrically programmable memories can offer 7 bits of precision [5], sub-ns programming speeds [6], and sub-pJ programming energies [6], [7], [8]. However, while recent experimental demonstrations have achieved a subset of key metrics needed for such a memory, no single material system or device has successfully “checked all the boxes.” For example, phase-change materials are well known for their stability, multi-level operation, and compact footprint, yet programming endurance has remained an outstanding challenge. While endurance is not an issue for photonic MEMS devices with bistability, achieving fast operation with multiple intermediate levels (with the notable exception of [9]) limits their usefulness as photonic memory elements for computation.

Addressing these limitations are particularly important for photonic computing architectures which are diffraction limited and do not have the spatial scaling advantage of electronics as seen from Moore’s Law. Thus, large-scale photonic computations using weight-stationary methods require agile, efficient, and robust photonic memory arrays. This is critical because these arrays have lower storage density compared to electronic memory and need to be reprogrammed during intermediate computation steps. Previous work has highlighted the fact that without fast, efficient, and scalable memory, photonic computing approaches will actually lead to worse latency and efficiency performance since they are trading the “memory access” bottleneck of digital electronics for a “memory update” bottleneck [10], [11].

**Current and Future Challenges**

To better understand these limitations, it is useful to explore the outstanding technical challenges of each major memory platform which we discuss here in more detail.

**Phase-change materials.** This type of material offers a unique combination of non-volatile response and strong optical modulation—either pure phase (in the C-L bands) with alloys such as $Sb_2Se_3$ [12] and $Sb_2S_3$ [13] or amplitude modulation with $Ge_2Sb_2Te_5$ [14], [15]. This combination leads to ultra-compact devices with form factors under 10 µm leading to a π phase shift or 10 dB extinction ratio in PICs, arguably one of the most compact phase and amplitude modulators in PICs. Lesser-known properties of PCMs, like having a large thermo-optic coefficient [16] and being radiation hard [17], makes them a great candidate for spaceborne applications [18] and for combining volatile and non-volatile response in a single device [19]. Integrating PCMs into PICs for photonic memories and in-memory computing has seen a very productive last decade. The field shifted from hard-to-scale all-optically switched memories [20] to backend-of-the-line (BEOL) integration to foundry-processed chips with electro-thermal switching of PCM memories with <0.1 dB insertion loss [12], [13], [21], [22]. While the CMOS integration has been partially sorted out, the ideal scenario is integrating PCMs into the fabrication process to take advantage of the full layer structure and the multiple waveguide possibilities (currently, an etched window down to the waveguide is required). Regardless, PCM technology is undoubtedly closer to impactful PCM-based optical memory architectures, although, many outstanding challenges remain:

- The voltages required to switch PCMs remain high (5-20 V), limited by the amorphization process which requires melt-quenching [23]. Thus, needing electronics that can make addressing many memories a much harder process involving multiple amplifiers. With demonstrated mature semiconductor and even graphene microheaters already reaching their theoretical limits, the open question is whether PCMs with lower melting temperatures are possible.
- With the integration of electro-thermal devices, PCMs gain scalability and integration but lose the fine multi-level tunability and high endurance achieved with optical switching, thus compromising their reliability and response. Toward this end, few groups are proposing mechanisms to achieve deterministic control of intermediate levels (i.e. controllable amorphous-crystalline spatial distributions) by engineering the heater [12], [24] or adding a series of them [25], [26].

- Much-needed materials science-oriented research has started shining a light into the failure mechanisms of PCM photonic devices, mainly those employing electro-thermal switching, revealing issues such as element segregation, void formation, and inhibition of crystal nucleation and, more importantly, providing mitigation strategies that have informed, among others, the choice of capping material [27], [28] and its thickness [23], [29], [30]. High-endurance and reliable readout remain at the top of current challenges; thus, understanding the nucleation dynamics and the failure mechanisms are priorities in the field.

**Ferroelectrics.** Non-volatile optical control has been demonstrated based on ferroelectric materials in two ways: a direct approach modulating the refractive index of ferroelectric crystals and hence optical phase with voltage biasing [31], [32], and a hybrid approach using a ferroelectric field effect transistor to provide non-volatile control over a separate optical phase shifter [33]. By leveraging the field-driven ferroelectric domain switching effect, the first approach offers the advantage of low insertion loss (0.07 dB [31]) and switching energy (27 pJ [31]). However, the approach requires a high voltage to drive ferroelectric domain switching and a DC bias to readout. Moreover, the small refractive index perturbation afforded by the Pockels effect results in large device footprint. The hybrid scheme is compatible with CMOS driving voltages and readily scalable to large crossbar arrays. To retain the energy benefits of non-volatile operation, the optical phase shifters used in such a hybrid design are limited to field-driven devices, for example those based on metal-oxide-semiconductor capacitors or liquid crystals. This ultimately bounds the performance such as loss and speed.

**MEMS.** A wide variety of latching MEMS structures can be adapted to realize non-volatile photonic functions [34]. Key advantages of MEMS-based non-volatile photonic devices include minimal power consumption, small footprint, extremely insertion loss and crosstalk, as well as scalability to large arrays [35], [36]. High driving voltages remains as a main limitation of MEMS non-volatile photonic devices. Moreover, realizing multilevel operation while retaining nonvolatility presents a nontrivial challenge with only a handful of demonstrations [37], [38]. Finally, MEMS fabrication involves special processing steps such as suspension and vacuum sealing, which are not traditionally available in silicon photonic fabrication processes and their integration into standard foundry manufacturing is only starting to surface [39], [40].

**Electronic Memristors.** Several groups have successfully demonstrated non-volatile optical control using waveguide-integrated electronic memristors [6], [7], [8]. These approaches primarily use nanoscale filamentation to either: 1) increase optical loss in the device through scattering effects or 2) modulate the carrier concentration in the waveguide core or cladding. In the first case, the optical mode must be confined to a geometry of similar scale to the filament, which necessitates the use of plasmonic confinement [8]. While this has been shown to be an effective strategy for highly efficient optical modulation, the insertion losses of such devices are typically quite high. Additionally, using scattering only allows for amplitude modulation without control of phase. More recently, oxide-based memristors integrated into microring resonators have achieved phase modulation by introducing a reversible short circuit between two sides of a hybrid InP/Si PN junction separated by either $HfO_2$ or $Al_2O_3$ [6], [7]. While the effect is indeed non-volatile, these programmable PN junctions require an applied bias during optical readout to observe the state of the memory cell. Multilevel operation beyond 2 bits is also challenging for both filament-based memristor technologies.

**Magneto-Optics.** Typically used for optical isolation [41], waveguide integration of transparent magneto-optical materials, such as Ce-substituted YIG, has shown promise for on-chip routing and modulation of optical signals [42], [43]. When combined with patterned ferromagnetic thin-films and integrated electromagnets, non-volatile programmability can be achieved [43]. While this could enable fast, high endurance photonic memory, this approach requires advanced fabrication due to the need for heterogeneous integration of the magneto-optic cladding. The relatively weak magneto-optic effect of current materials also typically requires devices with larger footprints than carrier-based modulators. Finally, while generating a magnetic field via integrated electromagnets is simple, they require relatively high electrical current to induce optical switching which leads to thermal heating of the photonic circuit.

**Charge Trapping.** While a fundamental mechanism in mass-produced electronic memory technologies, charge trapping has not yet been explored in depth in the context of optical memory. Originally proposed in 2006 by Barrios et al. [44], it was not until a decade later that a floating gate memory was demonstrated in silicon photonics experimentally [45]. Very recently, a semiconductor-insulator-semiconductor capacitor (SISCAP) photonic memory cell was demonstrated using a multi-layer $Al_2O_3/HfO_2$ dielectric stack [46]. While promising for CMOS compatibility and efficient, multilevel storage, this technology requires much longer programming pulses compared to any of the other memory platforms mentioned above. Additionally, the pulse amplitude required to for write and erase operations is in the range of 6V to 20V, which is outside of the typical operating range of CMOS circuitry.

### Advances in Science and Technology to Meet Challenges

Integration into scalable foundry manufacturing remains one major roadblock for adoption of most of these technologies, which involve new materials or processes. Some promising solutions include substitution with materials that have already been validated in CMOS processes, for instance using $HfO_2$, a commonly used high-k dielectric as ferroelectric media [32]. Alternatively, BEOL integration of new non-volatile materials with minimal-to-no changes to standard foundry fabrication processes have recently been demonstrated [21], [26], [47], and new BEOL integration schemes are also being actively investigated [48]. Another intriguing case is the introduction of suspended optomechanical structures through BEOL processing of foundry-manufactured devices [40]. Finally, new process design kits (PDKs) that incorporate unconventional materials or processes are also being developed in foundries, provided that sufficient demands exist to justify these efforts. One such example is the recent demonstration of a silicon photonic MEMS platform, which encompasses MEMS devices fully integrated alongside standard silicon photonics components on the wafer-scale [39]. We foresee that these efforts will ultimately enable seamless integration of non-volatile photonic components into commercial foundry manufacturing and catalyze their widespread adoption. Another important advance needed to address all the challenges is developing high-throughput discovery processes to continue searching for non-volatile materials with more stable levels, larger optical contrast, lower switching energies, and better endurance. The goal is to find an alloy/modulating phenomenon that meets all the desired criteria and check all the boxes in Table 1.

### Concluding Remarks

It is certainly an exciting time for the integration of new materials with large-scale photonic circuits. Advances in material performance coupled with the need for new computing strategies have led to important scientific breakthroughs in photonic memory technology. While many technical challenges which remain—endurance being at the top of the list, significant progress has been made over the last decade. Even as non-volatile photonic memory matures, it is likely that no single technology will address all needs for every computing application but will rather be organized into a memory "hierarchy" like the evolution of electronic memory. Continued success will certainly depend on the continued contributions and collaborations from a wide range of communities (material science, physics, computing architecture, etc.) to help shape the future of this important technology.

## Acknowledgements

*This work has been supported by AFOSR (FA9550-24-1-0064) and the National Science Foundation (2210168/2210169, 2337674, 2225968, and 2329087/2329088) and is supported in part by funds from federal agency and industry partners as specified in the Future of Semiconductors (FuSe) program.*

## References

[1] N. Youngblood, C. A. Ríos Ocampo, W. H. P. Pernice, and H. Bhaskaran, “Integrated optical memristors,” *Nat Photonics*, vol. 17, pp. 561–572, May 2023, doi: 10.1038/s41566-023-01217-w.

[2] J. Mao, L. Zhou, X. Zhu, Y. Zhou, and S. Han, “Photonic Memristor for Future Computing: A Perspective,” *Adv Opt Mater*, vol. 7, no. 22, Nov. 2019, doi: 10.1002/adom.201900766.

[3] T. Alexoudi, G. T. Kanellos, and N. Pleros, “Optical RAM and integrated optical memories: a survey,” *Light Sci Appl*, vol. 9, no. 1, p. 91, May 2020, doi: 10.1038/s41377-020-0325-9.

[4] J. Parra, I. Olivares, A. Brimont, and P. Sanchis, “Toward Nonvolatile Switching in Silicon Photonic Devices,” *Laser Photon Rev*, vol. 15, no. 6, p. 2000501, Jun. 2021, doi: 10.1002/LPOR.202000501.

[5] J. Meng *et al.*, “Electrical programmable multilevel nonvolatile photonic random-access memory,” *Light Sci Appl*, vol. 12, no. 1, p. 189, Aug. 2023, doi: 10.1038/s41377-023-01213-3.

[6] B. Tossoun *et al.*, “High-speed and energy-efficient non-volatile silicon photonic memory based on heterogeneously integrated memresonator,” *Nat Commun*, vol. 15, no. 1, p. 551, Jan. 2024, doi: 10.1038/s41467-024-44773-7.

[7] Z. Fang *et al.*, “Fast and Energy-Efficient Non-Volatile III-V-on-Silicon Photonic Phase Shifter Based on Memristors,” *Adv Opt Mater*, Oct. 2023, doi: 10.1002/adom.202301178.

[8] A. Emboras *et al.*, “Atomic Scale Plasmonic Switch,” *Nano Lett*, vol. 16, no. 1, 2016, doi: 10.1021/acs.nanolett.5b04537.

[9] Y. H. Wen *et al.*, “High-speed photonic crystal modulator with non-volatile memory via structurally-engineered strain concentration in a piezo-MEMS platform,” Oct. 2023.

[10] G. Yang, C. Demirkiran, Z. E. Kizilates, C. A. R. Ocampo, A. K. Coskun, and A. Joshi, “Processing-in-Memory Using Optically-Addressed Phase Change Memory,” in *ACM/IEEE International Symposium on Low Power Electronics and Design*, 2023.

[11] C. Demirkiran *et al.*, “An Electro-Photonic System for Accelerating Deep Neural Networks,” *ACM J Emerg Technol Comput Syst*, vol. 19, no. 4, pp. 1–31, Oct. 2023, doi: 10.1145/3606949.

[12] C. Ríos *et al.*, “Ultra-compact nonvolatile phase-shifter based on electrically reprogrammable transparent phase change materials,” *PhotoniX*, vol. 3, no. 26, May 2022.

[13] R. Chen *et al.*, “Non-volatile electrically programmable integrated photonics with a 5-bit operation,” Jan. 2023, doi: 10.1038/s41467-023-39180-3.

[14] Z. Fang *et al.*, “Ultra-low-energy programmable non-volatile silicon photonics based on phase-change materials with graphene heaters,” *Nat Nanotechnol*, vol. 17, no. 8, pp. 842–848, 2022, doi: 10.1038/s41565-022-01153-w.

[15] C. Ríos *et al.*, “In-memory computing on a photonic platform,” *Sci Adv*, vol. 5, no. 2, p. eaau5759, 2019, doi: 10.1126/sciadv.aau5759.

[16] M. Stegmaier, C. RÍos, H. Bhaskaran, and W. H. P. Pernice, “Thermo-optical Effect in Phase-Change Nanophotonics,” *ACS Photonics*, vol. 3, no. 5, pp. 828–835, 2016, doi: 10.1021/acsphotonics.6b00032.

[17] K. Konstantinou, T. H. Lee, F. C. Mocanu, and S. R. Elliott, “Origin of radiation tolerance in amorphous Ge 2 Sb 2 Te 5 phase-change random-access memory material,” *Proceedings of the National Academy of Sciences*, vol. 115, no. 21, pp. 5353–5358, May 2018, doi: 10.1073/pnas.1800638115.

[18] H. J. Kim *et al.*, “Versatile spaceborne photonics with chalcogenide phase-change materials,” *NPJ Microgravity*, vol. 10, no. 1, 2024, doi: 10.1038/s41526-024-00358-8.

[19] M. Wei *et al.*, “Electrically programmable phase-change photonic memory for optical neural networks with nanoseconds in situ training capability,” *Advanced Photonics*, vol. 5, no. 4, p. 46004, 2023, doi: 10.1117/1.AP.5.4.046004.

[20] C. Rios *et al.*, “Integrated all-photonic non-volatile multi-level memory,” *Nat Photonics*, vol. 9, no. 11, pp. 725–732, 2015, doi: 10.1038/nphoton.2015.182.

[21] M. Wei *et al.*, “Monolithic back-end-of-line integration of phase change materials into foundry-manufactured silicon photonics,” *Nat Commun*, vol. 15, no. 1, Dec. 2024, doi: 10.1038/s41467-024-47206-7.

[22] W. Zhou *et al.*, “In-memory photonic dot-product engine with electrically programmable weight banks,” *Nat Commun*, vol. 14, no. 1, 2023, doi: 10.1038/s41467-023-38473-x.

[23] J. Li *et al.*, “Performance Limits of Phase Change Integrated Photonics,” *IEEE Journal of Selected Topics in Quantum Electronics*, vol. 30, no. 4, 2024, doi: 10.1109/JSTQE.2024.3360526.

[24] C. Zhang *et al.*, “Nonvolatile Multilevel Switching of Silicon Photonic Devices with In2O3/GST Segmented Structures,” *Adv Opt Mater*, vol. 11, no. 8, 2023, doi: 10.1002/adom.202202748.

[25] J. Xia *et al.*, “Seven Bit Nonvolatile Electrically Programmable Photonics Based on Phase-Change Materials for Image Recognition,” *ACS Photonics*, vol. 11, no. 2, 2024, doi: 10.1021/acsphotonics.3c01598.

[26] R. Chen *et al.*, “Deterministic quasi-continuous tuning of phase-change material integrated on a high-volume 300-mm silicon photonics platform,” 2023.

[27] L. Lu, W. Dong, J. K. Behera, L. Chew, and R. E. Simpson, “Inter-diffusion of plasmonic metals and phase change materials,” *J Mater Sci*, vol. 54, no. 4, pp. 2814–2823, Feb. 2019, doi: 10.1007/s10853-018-3066-x.

[28] T. Y. Teo *et al.*, “Capping Layer Effects on $Sb_2S_3$-Based Reconfigurable Photonic Devices,” *ACS Photonics*, vol. 10, no. 9, pp. 3203–3214, Sep. 2023, doi: 10.1021/acsphotonics.3c00600.

[29] C. C. Popescu *et al.*, “An Open-Source Multifunctional Testing Platform for Optical Phase Change Materials,” *Small Science*, vol. 3, no. 12, 2023, doi: 10.1002/smsc.202300098.

[30] L. Martin-Monier *et al.*, “Endurance of chalcogenide optical phase change materials: a review,” *Opt. Mater. Express*, vol. 12, no. 6, pp. 2145–2167, Jun. 2022, doi: 10.1364/OME.456428.

[31] J. Geler-Kremer *et al.*, “A ferroelectric multilevel non-volatile photonic phase shifter,” *Nat Photonics*, May 2022, doi: 10.1038/s41566-022-01003-0.

[32] K. Taki *et al.*, “Non-volatile optical phase shift in ferroelectric hafnium zirconium oxide,” *arXiv:2309.01967*, Sep. 2023, Accessed: May 01, 2024. [Online]. Available: https://arxiv.org/abs/2309.01967v1

[33] R. Tang *et al.*, “Non-Volatile Hybrid Optical Phase Shifter Driven by a Ferroelectric Transistor,” *Laser Photon Rev*, vol. 17, no. 11, p. 2300279, Nov. 2023, doi: 10.1002/LPOR.202300279.

[34] C. Errando-Herranz, A. Y. Takabayashi, P. Edinger, H. Sattari, K. B. Gylfason, and N. Quack, “MEMS for Photonic Integrated Circuits,” *IEEE Journal of Selected Topics in Quantum Electronics*, vol. 26, no. 2, p. 8200916, Mar. 2020, doi: 10.1109/JSTQE.2019.2943384.

[35] X. Zhang, K. Kwon, J. Henriksson, J. Luo, and M. C. Wu, “A large-scale microelectromechanical-systems-based silicon photonics LiDAR,” *Nature*, vol. 603, no. 7900, pp. 253–258, Mar. 2022, doi: 10.1038/s41586-022-04415-8.

[36] M. C. Wu, T. J. Seok, J. Luo, K. Kwon, and J. Henriksson, “Wafer-scale silicon photonic switches beyond die size limit,” *Optica*, vol. 6, no. 4, pp. 490–494, Apr. 2019, doi: 10.1364/OPTICA.6.000490.

[37] A. Unamuno and D. Uttamchandani, “MEMS variable optical attenuator with vernier latching mechanism,” *IEEE Photonics Technology Letters*, vol. 18, no. 1, pp. 88–90, 2006, doi: 10.1109/LPT.2005.860395.

[38] Y. H. Wen *et al.*, “High-speed photonic crystal modulator with non-volatile memory via structurally-engineered strain concentration in a piezo-MEMS platform,” *arXiv:2310.07798*, Oct. 2023, Accessed: May 01, 2024. [Online]. Available: https://arxiv.org/abs/2310.07798v2

[39] N. Quack *et al.*, “Integrated silicon photonic MEMS,” *Microsyst Nanoeng*, vol. 9, p. 27, Dec. 2023, doi: 10.1038/S41378-023-00498-Z.

[40] V. Deenadayalan *et al.*, “Foundry-processed optomechanical photonic integrated circuits,” *OSA Contin*, vol. 4, no. 4, pp. 1215–1222, Apr. 2021, doi: 10.1364/OSAC.419410.

[41] P. Pintus, D. Huang, C. Zhang, Y. Shoji, T. Mizumoto, and J. E. Bowers, “Microring-Based Optical Isolator and Circulator with Integrated Electromagnet for Silicon Photonics,” *Journal of Lightwave Technology*, vol. 35, no. 8, pp. 1429–1437, Apr. 2017, doi: 10.1109/JLT.2016.2644626.

[42] P. Pintus *et al.*, “An integrated magneto-optic modulator for cryogenic applications,” *Nat Electron*, vol. 5, no. 9, pp. 604–610, Sep. 2022, doi: 10.1038/s41928-022-00823-w.

[43] T. Murai, Y. Shoji, N. Nishiyama, and T. Mizumoto, “Nonvolatile magneto-optical switches integrated with a magnet stripe array,” *Opt Express*, vol. 28, no. 21, p. 31675, Oct. 2020, doi: 10.1364/OE.403129.

[44] C. A. Barrios and M. Lipson, “Silicon photonic read-only memory,” *Journal of Lightwave Technology*, vol. 24, no. 7, pp. 2898–2905, Jul. 2006, doi: 10.1109/JLT.2006.875964.

[45] J.-F. Song *et al.*, “Integrated photonics with programmable non-volatile memory,” *Sci Rep*, vol. 6, no. 1, p. 22616, Mar. 2016, doi: 10.1038/srep22616.

[46] S. Cheung *et al.*, “Energy efficient photonic memory based on electrically programmable embedded III-V/Si memristors: switches and filters,” *Communications Engineering*, vol. 3, no. 1, p. 49, Mar. 2024, doi: 10.1038/s44172-024-00197-1.

[47] S. A. Vitale *et al.*, “ Phase Transformation and Switching Behavior of Magnetron Plasma Sputtered Ge 2 Sb 2 Se 4 Te ,” *Adv Photonics Res*, vol. 3, no. 10, 2022, doi: 10.1002/adpr.202200202.

[48] L. Ranno, J. X. B. Sia, K. P. Dao, and J. Hu, “Multi-material heterogeneous integration on a 3-D photonic-CMOS platform,” *Opt Mater Express*, vol. 13, no. 10, pp. 2711–2725, Oct. 2023, doi: 10.1364/OME.497245.

# Energy-efficient volatile and non-volatile memory for neuromorphic photonics

- **Fabio Pavanello**, Univ. Grenoble Alpes, Univ. Savoie Mont Blanc, CNRS, Grenoble INP, CROMA, Grenoble, France [fabio.pavanello@cnrs.fr]
- **Bassem Tossoun**, Hewlett Packard Labs, Hewlett Packard Enterprise, Santa Barbara, CA 93013, USA [bassem.tossoun@hpe.com]
- **Alessio Lugnan**, Nanoscience Laboratory, Department of Physics, University of Trento, Italy [alessio.lugnan.1@unitn.it]
- **Federico Marchesin**, Ghent University – imec, Ghent, Belgium [federico.marchesin@ugent.be]
- **Matěj Hejda**, Hewlett Packard Labs, Hewlett Packard Enterprise, 1831 Diegem, Belgium [matej.hejda@hpe.com]
- **Stanley Cheung**, Department of Electrical and Computer Engineering, North Carolina State University, Raleigh, NC 27606, USA [scheung3@ncsu.edu]
- **Benoit Charbonnier**, Univ. Grenoble Alpes, CEA, LETI, Grenoble, France [benoit.charbonnier@cea.fr]
- **Peter Bienstman**, Ghent University – imec, Ghent, Belgium [peter.bienstman@ugent.be]
- **Thomas Van Vaerenbergh**, Hewlett Packard Labs, Hewlett Packard Enterprise, 1831 Diegem, Belgium [thomas.van-vaerenbergh@hpe.com]

**Status**

Over the past decade, optical and photonic platforms have been receiving significant amount of research interest to enable high-speed, energy efficient application-specific computing architectures. With the growing prospects of unconventional computing paradigms such as non-von Neumann and in-memory computing, the role of memory in optical computing systems becomes more critical.

Among the technologies for optical memories, the integration of thin phase-change material (PCM) films within optical waveguide cross-sections is regarded as one of the most promising [1]. Phase-change transitions between amorphous and crystalline phases can be optically, electrically, or electro-thermally triggered [2]. The achievable contrast for the real part of the bulk refractive index between amorphous and crystalline phases can be much larger for PCMs (0.5 – 5) compared to thermo-optic or free-carrier effects (approx. $10^{-3} – 10^{-2}$), thus leading to more compact devices [3]. Besides, their non-volatile character makes them ideal for developing energy-efficient systems where no power consumption is needed to retain a given state (e.g. a weight configuration after training of optical neural networks (ONNs)), providing advantage over other conventional approaches where e.g., power is continuously dissipated in heaters [4]. Besides, PCM-based integrated optical devices with 6-bit resolution, $10^7$ cycles endurance, and sub-nJ switching energy have been previously demonstrated [5], [6], [7].

Furthermore, compact, non-volatile phase shifters capable of being switched at high-speeds and low-energy not only for zero static power consumption during inference, but also for the potential of on-chip training are essential for several neuromorphic architectures. As an alternative to PCMs, non-volatile photonic memory can be introduced using memristive materials, where external bias leads to resistive switching via reversible formation of rupturing conductive filaments (CFs) [8], [9], [10]. Based on the resistance of the oxide material within the optical waveguide, the carrier density within the waveguide varies causing the plasma dispersion effect and a subsequent change in the effective refractive index of the waveguide. Furthermore, oxide-based memristors have been integrated within silicon photonic microring resonators and Mach-Zehnder Interferometers (MZIs). Previously, memresonators and memristive MZIs have been measured with a high endurance (1,000 cycles), long retention times (~24 hours) over multiple states as well as sub-nanosecond switching times using sub-pJ switching energy. Furthermore, these memristive devices can also operate as MOS-capacitor carrier-accumulation-based modulators, which can be used as volatile phase shifters used for trimming to account for device variability caused by fabrication non-uniformity, or for high-speed modulation of input data and on-chip training of weights [11].

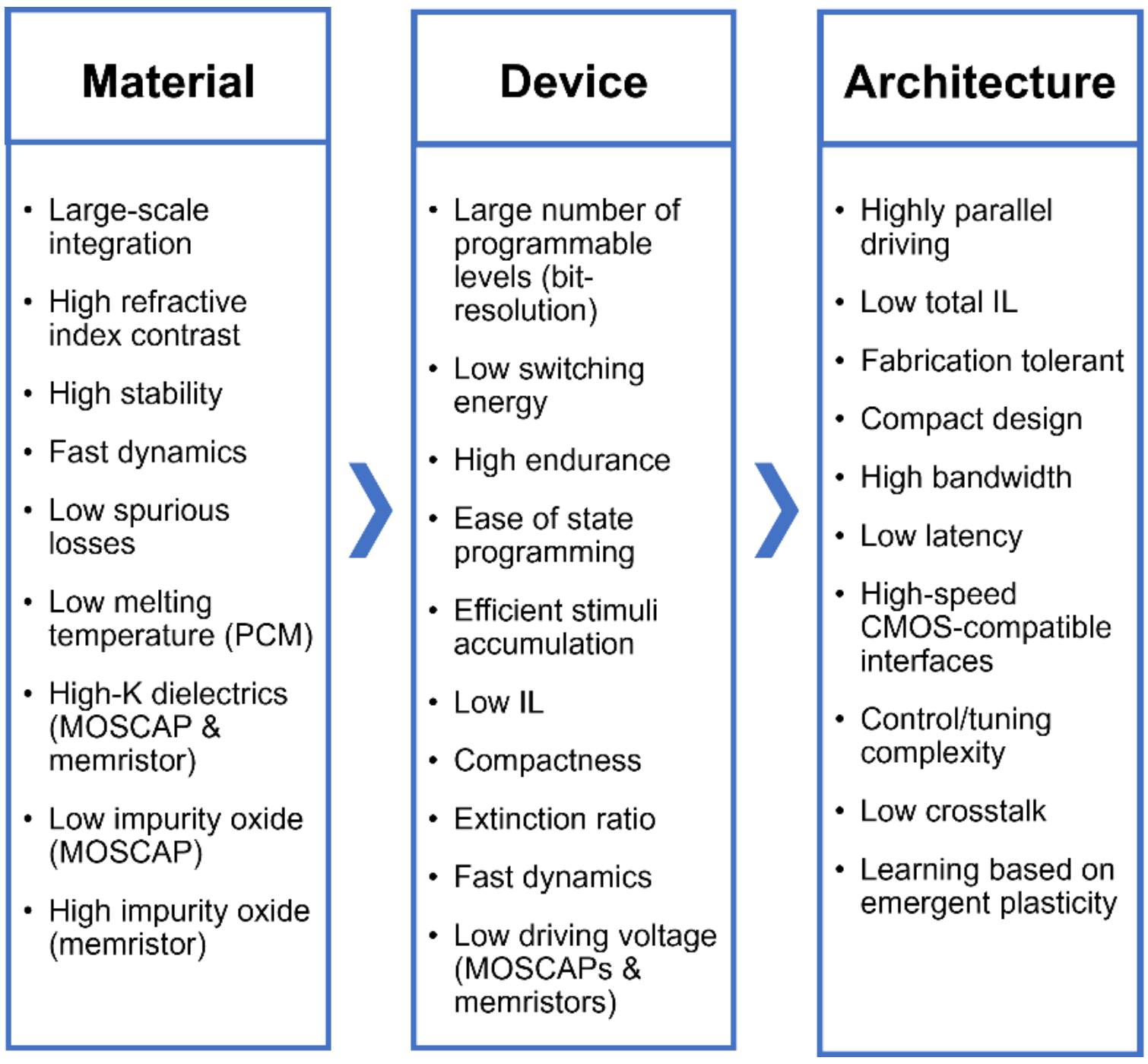

**Figure 1.** Key challenges at a material, device, and architecture levels for large-scale energy-efficient neuromorphic photonic systems

### Current and Future Challenges

Figure 1 summarizes the key challenges at a material, device, and architecture levels that shall be addressed to achieve large-scale neuromorphic photonic systems based on the previously discussed technologies. While various computing architectures based on PCMs have been recently shown experimentally [1], [12], [13], there are still no demonstrations of architectures integrated within complex CMOS-compatible platforms, e.g., implementing high-performance building blocks such as high-speed modulators and detectors in 200- or 300-mm wafers. Such large-scale integration is essential to build complex, highly programmable, and controllable energy-efficient neuromorphic systems [14]. However, no degradation of material stability or device endurance shall be incurred through the integration. For state-of-the-art accuracies across a wide range of tasks, larger bit resolutions per device are highly beneficial, making the improvement of the currently reported 6 bits precision as desirable [6]. Besides, PCM-based devices with very low insertion loss (IL) for coherent approaches are critical to achieve multiple cascaded layers in e.g., MZI-based ONNs [15].

Furthermore, PCM emergent plasticity as a learning mechanism has only recently started to being investigated [1], [13], although it holds a large potential thanks to the fast optically switchable dynamics (sub-ns), while working out of equilibrium [16]. In fact, self-adaptation in PICs is key to a major goal in neuromorphic computing, namely self-learning hardware. Here, the main challenge is to develop scalable learning systems compatible with powerful biologically-plausible training procedures.

Other devices that have gathered a strong interest for reconfigurable neuromorphic systems are III-V/Si MOSCAP- and memristor-based phase shifters. However, there are still numerous challenges for these devices to reach a technology readiness level right for large-scale PICs and high-volume production. Regarding MOSCAP and memristive III-V on Si photonic non-volatile phase shifters, one of the key challenges is in reducing their $V_{\pi}L$ and consequent footprint for the overall system, while keeping a switching voltage below 5V for CMOS driving compatibility.

Reducing the switching voltage is, however, challenging since it may require increasing the semiconductor contact resistance, which simultaneously increases the free-carrier absorption loss inside of the waveguide [11]. Another challenge is in improving the reliability of these devices. At the current stage, the endurance has been measured in 1,000 cycles, before permanent breakdown of the oxide material. Lastly, in the low resistance state, the memresonator can draw up to 10-100 µW of leakage power when data is being read, thus impairing its energy efficiency [8].

**Advances in Science and Technology to Meet Challenges**

Novel PCM compounds such as GeSe, GSST, and GeTe are currently under investigation with more suitable properties for large-scale neuromorphic systems compared to the well-known GST [3], [17]. They achieve lower insertion loss (IL) in both their crystalline and amorphous phases, thus providing an ideal solution for coherent multi-layer photonic architectures [4].

Besides, by combining optically switchable PCMs such as GST within Si microring resonator, it is possible to significantly enhance energy efficiency, speed, and optical contrast of bidirectional PCM weight changes thanks to resonance effects [18]. Moreover, silicon nonlinear effects in microrings can be combined to concurrently achieve nonlinearity, short- and long-term volatile memory, and non-volatile memory from PCMs as shown in Figure 2(a). These devices can be coupled together to build energy efficient and self-adaptive ONNs with low on-chip footprint and high data throughput [13].

One approach to also reduce the power consumed by memristive photonic devices that are set to a low resistance state, is to operate them using voltage pulses. In this case, the device only consumes power during the duration of the voltage pulse. Furthermore, to improve the reliability of memristive phase shifters, one can design an on-chip transistor to limit the current being applied to the memristor. Another approach to better the endurance of these memristive devices is to intentionally dope the oxide material with oxygen vacancies. The resultant doped materials are more electrically conductive because these dopants are electrically charged and mobile under an electric field that may be aided by Joule heating. Additionally, to improve the tuning efficiency of MOSCAP phase tuners, there has been extensive research on using high-k dielectrics such as $HfO_2$ to help reduce the peak-to-peak drive voltage [19].

Advancements in photonics and deep neural networks (DNNs) can be also obtained by decomposing feed-forward DNNs into smaller, modular blocks and reorganizing them to achieve enhanced performance architectures with fewer components such as in Figure 2(b) which reports the schematic of a tensorized ONN (TONN) architecture. A 1024x1024 TONN using 79x less MZIs compared to traditional ONN implementations based on MOSCAP technology was demonstrated capable to achieve above 95% accuracy for handwritten digit classification tasks [20]. TONNs allow for tailored, optimized networks that improve energy efficiency and adaptability, while taking advantage of the reduced IL at a system-level introduced by PCM or memristor-based synapses. This is due to the fewer components employed, thus improving further the overall energy efficiency.

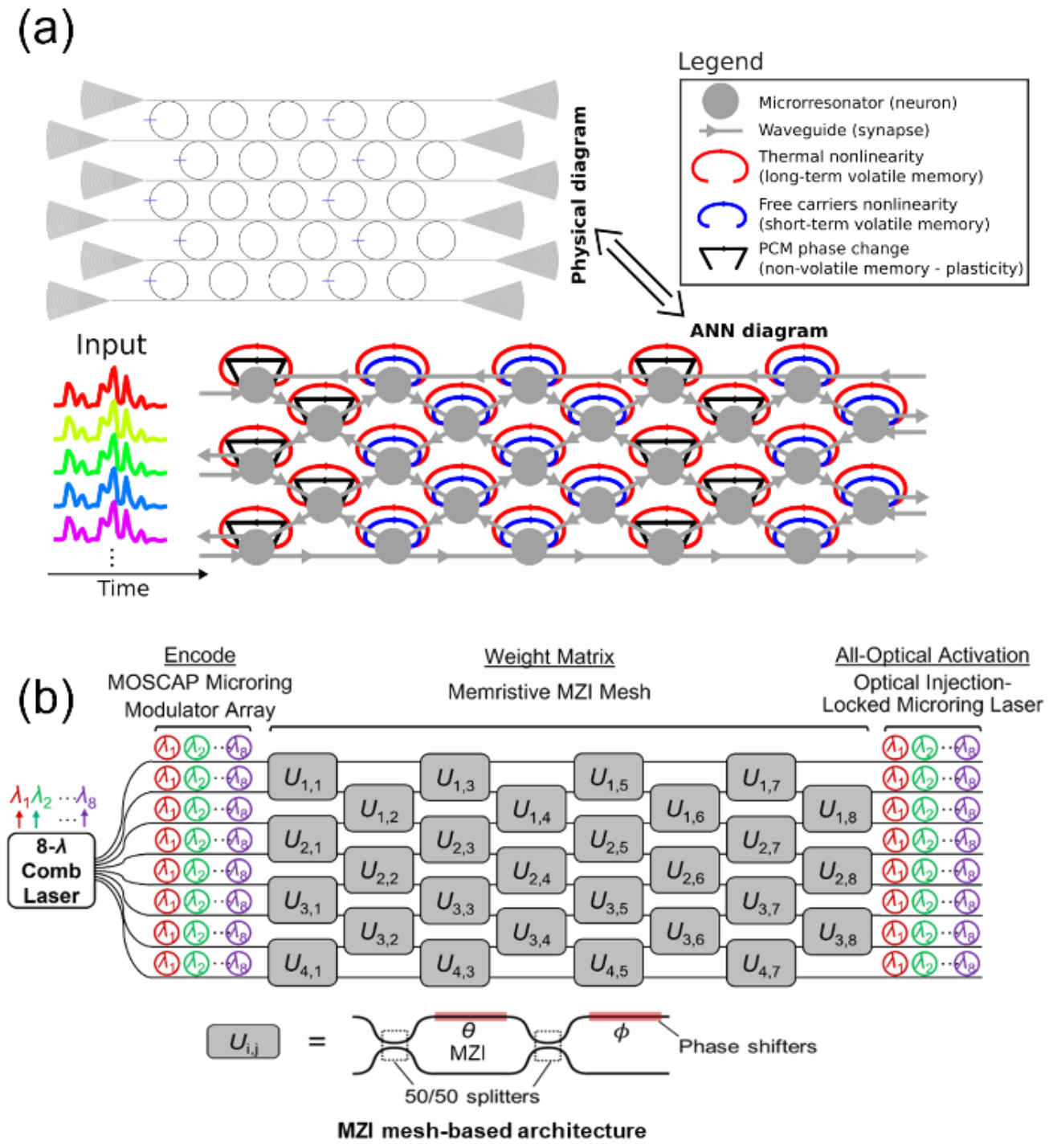

**Figure 2** Examples of architectures leveraging PCMs and MOSCAP technologies. (a) Network of silicon microresonators with PCMs and corresponding neural network diagram; (b) Schematic of the architecture of a MOSCAP-based 1024 × 1024 TONN-MW.

## Concluding Remarks

Next-generation energy-efficient neuromorphic photonic systems will benefit from technologies such as PCM-based devices, memristors, MOSCAPs, and other photonic non-volatile memory devices capable of switching and holding their state with negligible power consumption. However, to achieve superior performance compared to other neuromorphic technologies, several aspects shall be tackled from a material, device, and architecture perspectives. Large-scale integration of these approaches is fundamental to achieve high yield, on-chip access to key functionalities, and volume-scale production.

Major efforts are undergoing in this direction to enhance existing photonic platforms with these technologies, but also to develop energy-efficient devices which can guarantee best performance for neuromorphic computing architectures such as fast dynamics, large refractive index contrasts, and ultra-low insertion losses. Ongoing efforts are also addressing more compact and energy-efficient neuromorphic photonic circuits that will further benefit from enhanced device performance as well as from novel approaches requiring a lower number of devices or relying on training using novel approaches such as emergent PCM plasticity and tensor decomposition algorithms. Specific application-based requirements will dictate which of the key discussed challenges will have priority over the others. However, large-scale and robust integration will be an essential requirement to all the applications aiming at volume production.

## Acknowledgements

*This work has received funding from the European Union's Horizon Europe research and innovation program under grant agreement No. 101070238. Views and opinions expressed are however those of the author(s) only and do not necessarily reflect those of the European Union. Neither the European Union nor the granting authority can be held responsible for them.*

## References

[1] J. Feldmann, N. Youngblood, C. D. Wright, H. Bhaskaran, and W. H. P. Pernice, "All-optical spiking neurosynaptic networks with self-learning capabilities," *Nature*, vol. 569, no. 7755, pp. 208–214, May 2019, doi: 10.1038/s41586-019-1157-8.

[2] L. Martin-Monier, C. C. Popescu, L. Ranno, B. Mills, S. Geiger, D. Callahan, M. Moebius, and J. Hu, "Endurance of chalcogenide optical phase change materials: a review," *Opt. Mater. Express, OME*, vol. 12, no. 6, pp. 2145–2167, Jun. 2022, doi: 10.1364/OME.456428.

[3] P. Prabhathan, K. V. Sreekanth, J. Teng, J. H. Ko, Y. J. Yoo, H.-H. Jeong, Y. Lee, S. Zhang, T. Cao, C.-C. Popescu, B. Mills, T. Gu, Z. Fang, R. Chen, H. Tong, Y. Wang, Q. He, Y. Lu, Z. Liu, H. Yu, A. Mandal, Y. Cui, A. S. Ansari, V. Bhingardive, M. Kang, C. K. Lai, M. Merklein, M. J. Müller, Y. M. Song, Z. Tian, J. Hu, M. Losurdo, A. Majumdar, X. Miao, X. Chen, B. Gholipour, K. A. Richardson, B. J. Eggleton, M. Wuttig, and R. Singh, "Roadmap for phase change materials in photonics and beyond," *iScience*, vol. 26, no. 10, p. 107946, Oct. 2023, doi: 10.1016/j.isci.2023.107946.

[4] Y. Shen, N. C. Harris, S. Skirlo, M. Prabhu, T. Baehr-Jones, M. Hochberg, X. Sun, S. Zhao, H. Larochelle, D. Englund, and M. Soljačić, "Deep learning with coherent nanophotonic circuits," *Nature Photon*, vol. 11, no. 7, pp. 441–446, Jul. 2017, doi: 10.1038/nphoton.2017.93.

[5] C. Wu, H. Yu, S. Lee, R. Peng, I. Takeuchi, and M. Li, "Programmable phase-change metasurfaces on waveguides for multimode photonic convolutional neural network," *Nat Commun*, vol. 12, no. 1, p. 96, Jan. 2021, doi: 10.1038/s41467-020-20365-z.

[6] D. Lawson, S. Blundell, M. Ebert, O. L. Muskens, and I. Zeimpekis, "Optical switching beyond a million cycles of low-loss phase change material $Sb_2Se_3$," *Opt. Mater. Express, OME*, vol. 14, no. 1, pp. 22–38, Jan. 2024, doi: 10.1364/OME.509434.

[7] C. Ríos, M. Stegmaier, P. Hosseini, D. Wang, T. Scherer, C. D. Wright, H. Bhaskaran, and W. H. P. Pernice, "Integrated all-photonic non-volatile multi-level memory," *Nature Photon*, vol. 9, no. 11, Art. no. 11, Nov. 2015, doi: 10.1038/nphoton.2015.182.

[8] B. Tossoun, D. Liang, S. Cheung, Z. Fang, X. Sheng, J. P. Strachan, and R. G. Beausoleil, "High-speed and energy-efficient non-volatile silicon photonic memory based on heterogeneously integrated memresonator," *Nat Commun*, vol. 15, no. 1, p. 551, Jan. 2024, doi: 10.1038/s41467-024-44773-7.

[9] S. Cheung, B. Tossoun, Y. Yuan, Y. Peng, Y. Hu, W. V. Sorin, G. Kurczveil, D. Liang, and R. G. Beausoleil, "Energy efficient photonic memory based on electrically programmable embedded III-V/Si memristors: switches and filters," *Commun Eng*, vol. 3, no. 1, pp. 1–12, Mar. 2024, doi: 10.1038/s44172-024-00197-1.

[10] Z. Fang, B. Tossoun, A. Descos, D. Liang, X. Huang, G. Kurczveil, A. Majumdar, and R. G. Beausoleil, "Fast and Energy-Efficient Non-Volatile III-V-on-Silicon Photonic Phase Shifter Based on Memristors," *Advanced Optical Materials*, vol. 11, no. 24, p. 2301178, 2023, doi: 10.1002/adom.202301178.

[11] S. Srinivasan, D. Liang, and R. G. Beausoleil, "Heterogeneous SISCAP Microring Modulator for High-Speed Optical Communication," in *2020 European Conference on Optical Communications (ECOC)*, Dec. 2020, pp. 1–3. doi: 10.1109/ECOC48923.2020.9333221.

[12] J. Feldmann, N. Youngblood, M. Karpov, H. Gehring, X. Li, M. Stappers, M. Le Gallo, X. Fu, A. Lukashchuk, A. S. Raja, J. Liu, C. D. Wright, A. Sebastian, T. J. Kippenberg, W. H. P. Pernice, and H. Bhaskaran, "Parallel convolutional processing using an integrated photonic tensor core," *Nature*, vol. 589, no. 7840, pp. 52–58, Jan. 2021, doi: 10.1038/s41586-020-03070-1.

[13] A. Lugnan, S. Aggarwal, F. Brückerhoff-Plückelmann, C. D. Wright, W. H. P. Pernice, H. Bhaskaran, P. Bienstman, "Emergent Self-Adaptation in an Integrated Photonic Neural Network for Backpropagation-Free Learning," *Adv. Sci.* 2024, 2404920. doi: 10.1002/advs.202404920.

[14] F. Pavanello, C. Marchand, I. O'Connor, R. Orobtchouk, F. Mandorlo, X. Letartre, S. Cueff, E. I. Vatajelu, G. D. Natale, B. Cluzel, A. Coillet, B. Charbonnier, P. Noe, F. Kavan, M. Zoldak, M. Szaj, P. Bienstman, T. V. Vaerenbergh, U. Ruhrmair, P. Flores, D. Gizopoulos, G. Papadimitriou, V. Karakostas, A. Brando, F. J. Cazorla, R. Canal, P. Closas, A. Gusi-Amigo, P. Crovetti, A. Carpegna, T. M. Carmona, S. D. Carlo, and A. Savino, "NEUROPULS: NEUROmorphic energy-efficient secure accelerators based on Phase change

materials aUgmented siLicon photonicS," *2023 IEEE European Test Symposium (ETS)*, May 2023, doi: 10.1109/ETS56758.2023.10173974.

[15] R. Soref, J. Hendrickson, H. Liang, A. Majumdar, J. Mu, X. Li, and W.-P. Huang, "Electro-optical switching at 1550 nm using a two-state GeSe phase-change layer," *Opt. Express*, vol. 23, no. 2, p. 1536, Jan. 2015, doi: 10.1364/OE.23.001536.

[16] M. Stegmaier, C. Ríos, H. Bhaskaran, C. D. Wright, and W. H. P. Pernice, "Nonvolatile All-Optical 1 × 2 Switch for Chipscale Photonic Networks," *Advanced Optical Materials*, vol. 5, no. 1, p. 1600346, 2017, doi: 10.1002/adom.201600346.

[17] M. Miscuglio, J. Meng, O. Yesiliurt, Y. Zhang, L. J. Prokopeva, A. Mehrabian, J. Hu, A. V. Kildishev, and V. J. Sorger, "Artificial Synapse with Mnemonic Functionality using GSST-based Photonic Integrated Memory," p. 8.

[18] A. Lugnan, S. G.-C. Carrillo, C. D. Wright, and P. Bienstman, "Rigorous dynamic model of a silicon ring resonator with phase change material for a neuromorphic node," *Opt. Express, OE*, vol. 30, no. 14, pp. 25177–25194, Jul. 2022, doi: 10.1364/OE.459364.

[19] X. Huang, D. Liang, C. Zhang, G. Kurczveil, X. Li, J. Zhang, M. Fiorentino, and R. Beausoleil, "Heterogeneous MOS microring resonators," in *2017 IEEE Photonics Conference (IPC)*, Oct. 2017, pp. 121–122. doi: 10.1109/IPCon.2017.8116031.

[20] X. Xiao, M. B. On, T. Van Vaerenbergh, D. Liang, R. G. Beausoleil, and S. J. B. Yoo, "Large-scale and energy-efficient tensorized optical neural networks on III–V-on-silicon MOSCAP platform," *APL Photonics*, vol. 6, no. 12, p. 126107, Dec. 2021, doi: 10.1063/5.0070913.

# Memory in scalable neuromorphic silicon photonic computing systems

**Simon Bilodeau** (sbilodeau@princeton.edu)
**Paul Prucnal** (prucnal@princeton.edu)
Princeton University, NJ, USA

### Status

Conventional general-purpose computing systems are based on the so-called "von Neumann" architecture, where compute is physically separate from memory. The recent proliferation of data-centric applications such as artificial intelligence is straining this model, since data movement is more latency and energy-intensive than the computation itself [1]. As a response, colocation of compute circuits and memory, from near-compute memory to in-memory computing [2], is being increasingly pursued for such specialized applications. More tentatively, there is also a push to investigate computing devices with built-in memory, also known as memristors [3].

The maturation of a commercial silicon photonics fabrication ecosystem has brought photonics into consideration as a scalable platform to implement neuromorphic computation models [4]. The primary motivator are advantages in latency and RF processing bandwidth over electronics [5], with net advantages in energy efficiency still elusive [6]. However, with a few custom-fabricated exceptions [7], demonstrations of neuromorphic photonic systems have so far been von Neumann-like in nature, with on-chip photonic components controlled from off-package drivers, themselves configured from external data converters with off-board memory/control.

Given trends towards memory colocation in neuromorphic-like electronic systems, in this section we evaluate challenges and opportunities for similar developments in neuromorphic photonic systems. Fig. 1 summarizes different requirements for memories in neuromorphic computers [8]. While long-term storage and inference engines benefit from true non-volatility, processors used for training update their weights more regularly, suggesting dynamic (capacitive) memory as an attractive option. In this section, we focus on the hardware, although advances in algorithms will also be required. As an aside, it is interesting to note that the densest assemblies of active photonic components commercially manufactured, that is active matrix liquid crystal and organic light-emitting diode displays with over a million of individually-controlled elements (i.e. large-scale modulator or emitter arrays), intrinsically rely on the presence of local memory in each photonic component or their driver to simplify the drive architecture [9]. Hence, better colocation of memory in neuromorphic photonic systems may be key to their scaleup even without near or in-memory computing.

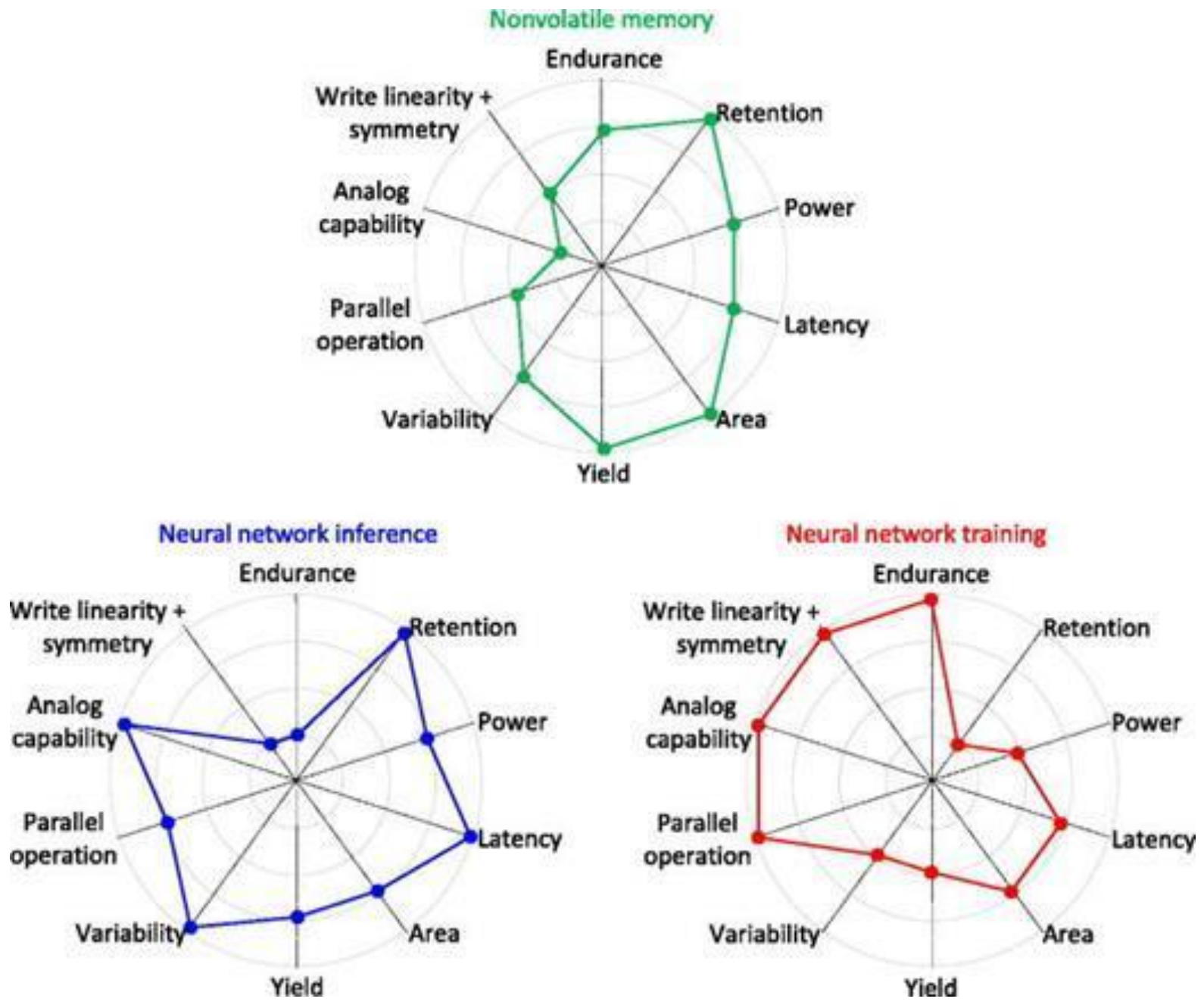

**Figure 1.** Relative importance of different metrics for memories depending on their use in storage (nonvolatile memory), inference, or training. Farther from center is more important. Reprinted from [8], with the permission of AIP Publishing.

## Current and Future Challenges

In the near term, for large photonic systems where yield is critical, mature commercial silicon photonics platforms are the safest option. However, such "standard" silicon photonic technologies are typically limited in functionality to passive waveguide structures, thermo-optic actuation, free carrier-based modulation, and germanium epitaxy for photodetection [10]. Of these building blocks, reverse-biased PN junctions are capacitive, and have been shown to act as dynamic neuromorphic photonic memories, at the cost of optical power-dependent leakage [11]. More robust memories and/or dedicated drivers are not typically available in these silicon photonic platforms, and therefore colocation of memory either requires interfacing with separate (electronic) chips, developing more advanced photonic platforms using established technologies, or foundries qualifying new CMOS-compatible materials with tolerable yield.

A more insidious problem when designing large-scale photonic systems with cointegrated memory is the disparate workflows of photonic and electronic/system designers. The layout-centric approach of photonic designers, necessary to capture the effect of geometry on the physics of light, clashes with the schematic, device detail-agnostic mindset of electronic and higher-level system designers. This is further exacerbated by the lack of standard electronic design automation flow for photonics.

## Advances in Science and Technology to Meet Challenges

To implement photonic neuromorphic computing with collocated memory, existing PICs can benefit from advances in packaging. While not a fundamental hurdle (3D multi-chip integration has been demonstrated in electronics), current public demonstrations of neuromorphic photonic systems have been small-scale, and practical problems will need to be solved in larger-scale demonstrations. Another radically different technological solution would be to leverage the photonic nature of the system to "collocate" conventional physically-separated processing and memory modules through high speed, low latency optical links, in a form of disaggregated memory [12]. More critically, either cost-insensitive applications will need to be demonstrated, or the associated costs will need to be accounted for and continuously improved, to make these multi-die approaches appealing, especially for low-volume R&D.

Beyond assembly of disparate technologies, advances in the commercial photonic integration platforms themselves can enable new high-yield devices and circuits which could better cointegrate memory. Monolithic co-integration of electronics and photonics, as in the GlobalFoundries' 45SPCLO platform [13], in theory enables conventional charge-based memories such as SRAM and DRAM, addressing logic, and other drive circuits to be integrated in the same silicon as the photonic devices. However, large-scale circuits of photonic synapses with memory in this technology have yet to be demonstrated. The presence of transistor design layers also allows implementation of more advanced opto-electronic memory devices such as MOSCAP modulators and flash memory-type modulators, which have been demonstrated in custom processes, but again not in large systems. Actively-developed on-chip III-V integration of silicon photonics, beyond enabling sources and gain, can also enable on-chip SOA all-optical RAM [14].

At longer time horizons, research into new electrical and optical memory materials for inclusion into commercial processes should also keep being explored. Integration of phase-change, ferroelectric, and magneto-optic materials has been explored in waveguides [15]. Improvements in switching energy, cyclability, and switching speed (power, endurance, and latency if borrowing the language of Fig. 1) will continue to be important depending on the application of the memory. Yield and reliability, often ignored in academic publications, will be a key concern for integration in large-scale systems. Ferroelectrics, specifically, have been very recently integrated into a commercial 300 mm silicon photonic process [16]. Back end-of-line-compatible "phase-change" organic memory materials [17] or capacitive polymer or liquid crystal modulators are also promising [18], although they are less mature than the above.

Improvements in the design process will need to continue to enable large memory-collocated systems. More reliable multiphysics simulations (for the new materials above) and the distillation of device behaviour into compact models will enable a hierarchical design flow amenable to the design of large-scale systems that can effectively leverage the local memory. Longer term, as foundry development kits stabilize, models and their variability can be defined from measurements.

**Concluding Remarks**

In conclusion, noting trends in neuromorphic-adjacent electronic processors towards increased colocation of compute and memory, we posit that similar developments should arise in neuromorphic silicon photonic processors. To this end, we identify as challenges the limitations of today's conventional silicon photonic platforms and the immaturity of system-level design. Advances that can remedy this include demonstrations of multi-chip systems, demonstrations of collocated memory in more advanced commercial platforms, development of new foundry-compatible memory materials, and design flow improvements. Such developments will be crucial to fully unlock the potential of large-scale photonic processors.

**Acknowledgements**

This work was supported by the Office of Naval Research (ONR) (25812 G0001 10014800 to P.P.).

## References

[1] O. Mutlu, S. Ghose, J. Gómez-Luna, and R. Ausavarungnirun, "Processing data where it makes sense: Enabling in-memory computation," *Microprocess. Microsyst.*, vol. 67, pp. 28–41, Jun. 2019, doi: 10.1016/j.micpro.2019.01.009.

[2] A. Sebastian, M. Le Gallo, R. Khaddam-Aljameh, and E. Eleftheriou, “Memory devices and applications for in-memory computing,” *Nat. Nanotechnol.*, vol. 15, no. 7, pp. 529–544, Jul. 2020, doi: 10.1038/s41565-020-0655-z.

[3] M.-K. Song, J.-H. Kang, X. Zhang, W. Ji, A. Ascoli, I. Messaris, A. S. Demirkol, B. Dong, S. Aggarwal, W. Wan, S.-M. Hong, S. G. Cardwell, I. Boybat, J. Seo, J.-S. Lee, M. Lanza, H. Yeon, M. Onen, J. Li, B. Yildiz, J. A. del Alamo, S. Kim, S. Choi, G. Milano, C. Ricciardi, L. Alff, Y. Chai, Z. Wang, H. Bhaskaran, M. C. Hersam, D. Strukov, H.-S. P. Wong, I. Valov, B. Gao, H. Wu, R. Tetzlaff, A. Sebastian, W. Lu, L. Chua, J. J. Yang, and J. Kim, “Recent Advances and Future Prospects for Memristive Materials, Devices, and Systems,” *ACS Nano*, vol. 17, no. 13, pp. 11994–12039, Jul. 2023, doi: 10.1021/acsnano.3c03505.

[4] B. J. Shastri, A. N. Tait, T. Ferreira de Lima, W. H. P. Pernice, H. Bhaskaran, C. D. Wright, and P. R. Prucnal, “Photonics for artificial intelligence and neuromorphic computing,” *Nat. Photonics*, vol. 15, no. 2, pp. 102–114, Feb. 2021, doi: 10.1038/s41566-020-00754-y.

[5] W. Zhang, J. C. Lederman, T. Ferreira de Lima, J. Zhang, S. Bilodeau, L. Hudson, A. Tait, B. J. Shastri, and P. R. Prucnal, “A system-on-chip microwave photonic processor solves dynamic RF interference in real time with picosecond latency,” *Light Sci. Appl.*, vol. 13, no. 1, p. 14, Jan. 2024, doi: 10.1038/s41377-023-01362-5.

[6] A. N. Tait, “Quantifying power use in silicon photonic neural networks,” *ArXiv210804819 Phys.*, Aug. 2021, Accessed: Aug. 19, 2021. [Online]. Available: http://arxiv.org/abs/2108.04819

[7] J. Feldmann, N. Youngblood, M. Karpov, H. Gehring, X. Li, M. Stappers, M. Le Gallo, X. Fu, A. Lukashchuk, A. S. Raja, J. Liu, C. D. Wright, A. Sebastian, T. J. Kippenberg, W. H. P. Pernice, and H. Bhaskaran, “Parallel convolutional processing using an integrated photonic tensor core,” *Nature*, vol. 589, no. 7840, Art. no. 7840, Jan. 2021, doi: 10.1038/s41586-020-03070-1.

[8] J. D. Kendall and S. Kumar, “The building blocks of a brain-inspired computer,” *Appl. Phys. Rev.*, vol. 7, no. 1, p. 011305, Jan. 2020, doi: 10.1063/1.5129306.

[9] W. den Boer and W. Den Boer, *Active Matrix Liquid Crystal Displays: Fundamentals and Applications*. Oxford, UNITED STATES: Elsevier Science & Technology, 2005.

[10] S. Y. Siew, B. Li, F. Gao, H. Y. Zheng, W. Zhang, P. Guo, S. W. Xie, A. Song, B. Dong, L. W. Luo, C. Li, X. Luo, and G.-Q. Lo, “Review of Silicon Photonics Technology and Platform Development,” *J. Light. Technol.*, vol. 39, no. 13, pp. 4374–4389, Jul. 2021, doi: 10.1109/JLT.2021.3066203.

[11] S. Lam, A. Khaled, S. Bilodeau, B. A. Marquez, P. R. Prucnal, L. Chrostowski, B. J. Shastri, and S. Shekhar, “Dynamic Electro-Optic Analog Memory for Neuromorphic Photonic Computing.” arXiv, Jan. 29, 2024. doi: 10.48550/arXiv.2401.16515.

[12] M. S. Nezami, T. F. de Lima, M. Mitchell, S. Yu, J. Wang, S. Bilodeau, W. Zhang, M. Al-Qadasi, I. Taghavi, A. Tofini, S. Lin, B. J. Shastri, P. R. Prucnal, L. Chrostowski, and S. Shekhar, “Packaging and Interconnect Considerations in Neuromorphic Photonic Accelerators,” *IEEE J. Sel. Top. Quantum Electron.*, vol. 29, no. 2: Optical Computing, pp. 1–11, Mar. 2023, doi: 10.1109/JSTQE.2022.3200604.

[13] K. Giewont, K. Nummy, F. A. Anderson, J. Ayala, T. Barwicz, Y. Bian, K. K. Dezfulian, D. M. Gill, T. Houghton, S. Hu, B. Peng, M. Rakowski, S. Rauch, J. C. Rosenberg, A. Sahin, I. Stobert, and A. Stricker, “300-mm Monolithic Silicon Photonics Foundry Technology,” *IEEE J. Sel. Top. Quantum Electron.*, vol. 25, no. 5, pp. 1–11, Sep. 2019, doi: 10.1109/JSTQE.2019.2908790.

[14] T. Alexoudi, G. T. Kanellos, and N. Pleros, “Optical RAM and Integrated Optical Memories: A Survey,” *Light Sci Appl*, vol. 9, no. 1, p. 91, May 2020, doi: 10.1038/s41377-020-0325-9.

[15] N. Youngblood, C. A. Ríos Ocampo, W. H. P. Pernice, and H. Bhaskaran, “Integrated optical memristors,” *Nat. Photonics*, vol. 17, no. 7, pp. 561–572, Jul. 2023, doi: 10.1038/s41566-023-01217-w.

[16] K. Alexander, A. Bahgat, A. Benyamini, D. Black, D. Bonneau, S. Burgos, B. Burridge, G. Campbell, G. Catalano, A. Ceballos, C.-M. Chang, C. J. Chung, F. Danesh, T. Dauer, M. Davis, E. Dudley, P. Er-Xuan, J. Fargas, A. Farsi, C. Fenrich, J. Frazer, M. Fukami, Y. Ganesan, G. Gibson, M. Gimeno-Segovia, S. Goeldi, P. Goley, R. Haislmaier, S. Halimi, P. Hansen, S. Hardy, J. Horng, M. House, H. Hu, M. Jadidi, H. Johansson, T. Jones, V. Kamineni, N. Kelez, R. Koustuban, G. Kovall, P. Krogen, N. Kumar, Y. Liang, N. LiCausi, D. Llewellyn, K. Lokovic, M. Lovelady, V. Manfrinato, A. Melnichuk, M. Souza, G. Mendoza, B. Moores, S.

Mukherjee, J. Munns, F.-X. Musalem, F. Najafi, J. L. O'Brien, J. E. Ortmann, S. Pai, B. Park, H.-T. Peng, N. Penthorn, B. Peterson, M. Poush, G. J. Pryde, T. Ramprasad, G. Ray, A. Rodriguez, B. Roxworthy, T. Rudolph, D. J. Saunders, P. Shadbolt, D. Shah, H. Shin, J. Smith, B. Sohn, Y.-I. Sohn, G. Son, C. Sparrow, M. Staffaroni, C. Stavrakas, V. Sukumaran, D. Tamborini, M. G. Thompson, K. Tran, M. Triplet, M. Tung, A. Vert, M. D. Vidrighin, I. Vorobeichik, P. Weigel, M. Wingert, J. Wooding, and X. Zhou, "A manufacturable platform for photonic quantum computing." arXiv, Apr. 26, 2024. doi: 10.48550/arXiv.2404.17570.

[17] S. Bilodeau, E. Doris, J. Wisch, M. Gui, B. Rand, B. Shastri, and P. Prucnal, "All-optical organic photochemical integrated nanophotonic memory: lossless, continuously-tunable, non-volatile." Oct. 14, 2022. doi: 10.21203/rs.3.rs-2106924/v1.

[18] I. Taghavi, M. Moridsadat, A. Tofini, S. Raza, N. A. F. Jaeger, L. Chrostowski, B. J. Shastri, and S. Shekhar, "Polymer modulators in silicon photonics: review and projections," *Nanophotonics*, vol. 11, no. 17, pp. 3855–3871, Sep. 2022, doi: 10.1515/nanoph-2022-0141.

# All-optical neural network training by programmable holographic weights in integrated photonics

**E.A. Vlieg[1], B.J. Offrein[2], F. Horst[3]**
IBM Research Europe – Zürich
[1] VLI@zurich.ibm.com
[2] OFB@zurich.ibm.com
[3] FHO@zurich.ibm.com

## Status

The increasing compute cost of machine learning could be alleviated by accelerating the synaptic signal compute operations in artificial neural networks by an analog crossbar array processor (Figure 1 **a)**). In this concept, input neural signal amplitudes **S** are transferred to output neural signal lines by a crossbar of tunable coupling elements into which the synaptic weight matrix W is programmed to yield the synaptic signal transfer operation W**S**. Considering a vector size N, matrix-vector multiplication processing in the analog domain is promising since the $O(N^2)$ workload can be performed at O(N) energy cost: each crossbar output simply requires a constant neural signal energy budget.

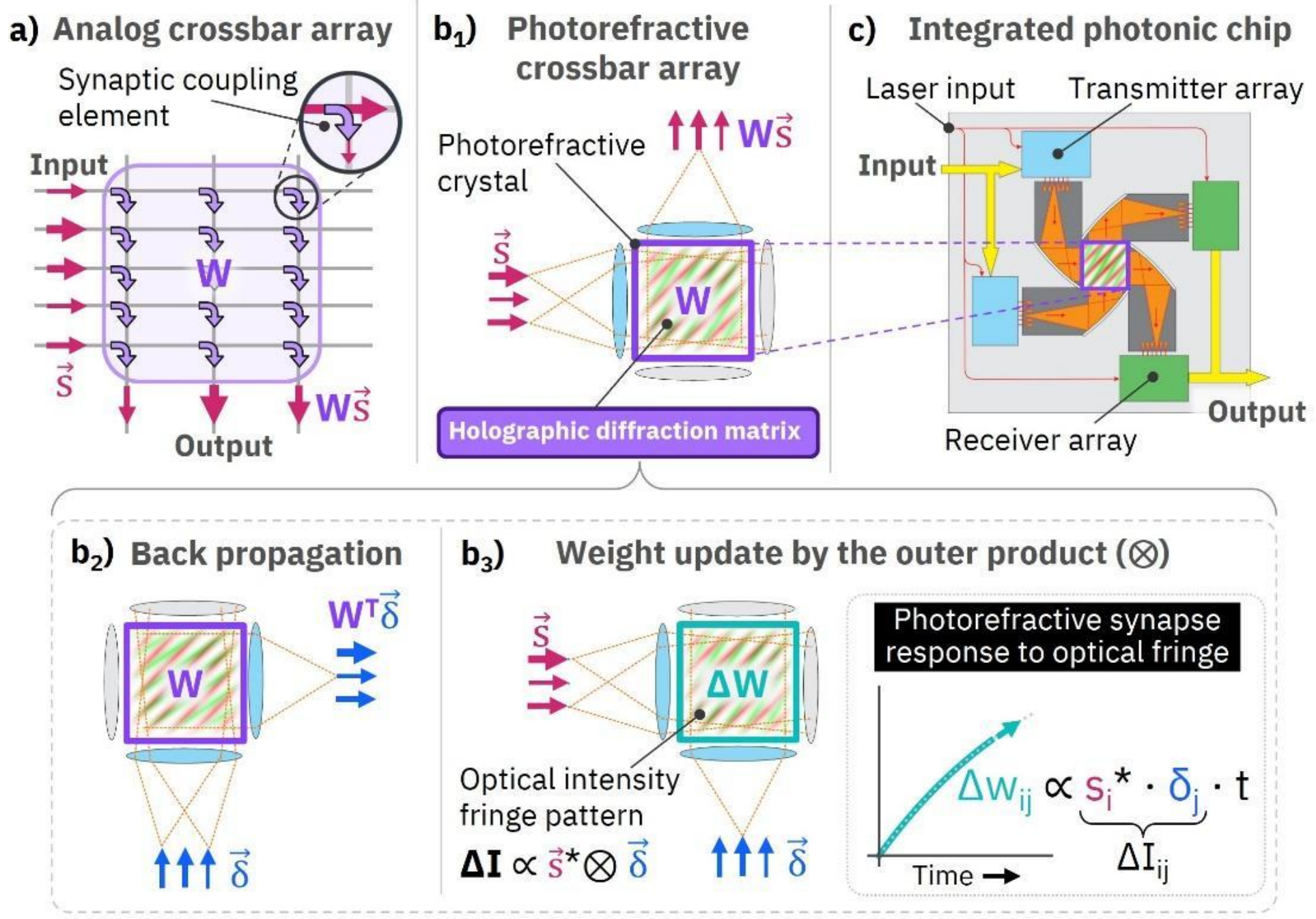

**Figure 1**. a) The analog crossbar array concept. b1) A crossbar array can be realized in photonics by leveraging the photorefractive effect to yield a hologram. The photorefractive crossbar array supports neural network back propagation training by b2) the transposed matrix operation and b3) weight programming by the outer product rule. c) A photorefractive crossbar array in integrated photonics.

The ideal hardware implementation of the analog crossbar array is a topic of scientific investigation, featuring both electronic and photonic approaches. In this work, we pursue a photonic photorefractive crossbar design (Figure 1 **b)**). The synaptic functionality is realized by superimposed Bragg gratings inside a photorefractive crystal that link input and output optical beam channels by diffraction [1,2]. The diffraction strengths can be tuned to yield the desired synaptic weight matrix by applying optical interference fringe patterns to precisely redistribute trapped electric charge by local photoexcitation.

An important merit of the photorefractive crossbar array is that it represents a full-fledged artificial synaptic interconnect layer by supporting backpropagation training in addition to inference (Figure 1 **b)**) [2]. By contrast, most crossbar designs simulate the forward signal propagation only. Additionally, the photorefractive synapses update near-continuously to assist training convergence.

Next, the photorefractive crossbar has special attributes that assist scaling. Firstly, the absence of dedicated hardware per superimposed synapse makes the design more robust against manufacturing variability and yield issues by avoiding weakest link synapses. Indeed, if each synapse is instead realized as an individual device, significant variation may occur [3]. Next, the photorefractive electron redistribution is non-damaging yielding perfect synapse endurance, meaning that the array does not have to be partitioned for an extended lifetime. Lastly, holographic storage boasts a high photonic memory density, thereby minimizing the processor spatial footprint compared to other photonic designs [4].

On the order of $10^7$ photorefractive weights have been experimentally demonstrated and trained using free space optics [5]. However, free space photonics are ultimately limited in terms of stability, footprint, and economics [6]. Accordingly, to push the technology forwards, our scientific goal is to implement the photorefractive crossbar array in integrated photonics, including the main peripheral optical components (Figure 1 **c)**). As a trade off, by going from 3D to 2D holographic memory, the processor area per synapse will now in principle stay constant with array size [7]. We estimate order $10^2$ $\mu m^2$/synapse for the full integrated crossbar array chip, which corresponds to 1-2 orders improvement compared to free-space optics ignoring the laser.

**Current and Future Challenges**

The realization of an integrated photorefractive crossbar array requires two main innovations: the addition of a photorefractive material to an integrated photonic platform and the design of a 2D beam interference network. To this end, prototype photorefractive interaction hardware has been realized in photorefractive semi-insulating GaAs for the demonstration of individual synapses (Figure 2 **a)**) [2].

The key performance indicator of the 2D interaction network is a low transmission loss, as this implies both a good beam collimation quality and low optical crosstalk. In terms of the photorefractive quality of the circuits, the maximum hologram amplitude is key to reach the maximum recoverable optical input power of approximately 10%. This minimizes the optical power consumption while ensuring an evenly illuminated hologram. In this context, our first objective is to demonstrate that integrated photonic circuitry can be realized of adequate photorefractive quality compared to bulk crystal in the face of potential damaging interface and processing effects.

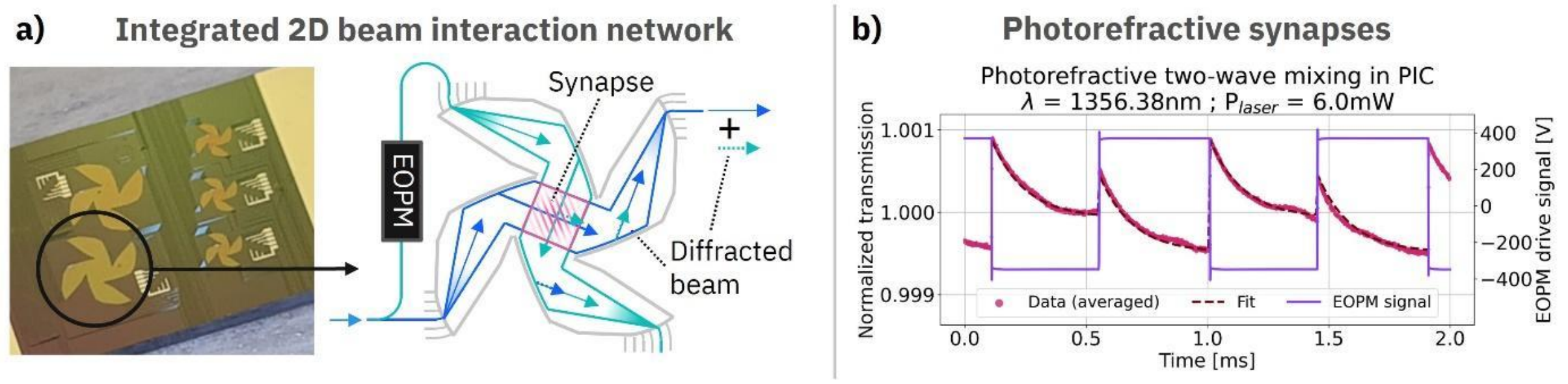

**Figure 2**. a) Integrated photonic hardware made from photorefractive semi-insulating GaAs. A source and destination optical beam enter the 2D interaction network and interfere to form a photorefractive synapse. b) Photorefractive synapse writing.

In the prototype hardware, photorefractive synapses were successfully demonstrated, proving the viability of the integrated photorefractive crossbar design and the desired photorefractive properties of the thin film GaAs (Figure 2 **b)**). Specifically, Figure 2 **b)** shows the periodic formation of a photorefractive synapse each time the source beam is π-phase shifted by an electro-optical phase modulator (EOPM): the interference between the diffracted and propagated beam is temporarily constructive until a new synapse is formed via an exponential saturation process. Although the design has been validated, the diffraction efficiency of order 1e-6 may be improved.

**Advances in Science and Technology to Meet Challenges**

The diffraction efficiency of the crossbar processor can be increased by an enhanced photorefractive material. Assuming an adequate photorefractive charge trap density, the hologram amplitude is mostly governed by the material Pockels coefficient [8]. Then, trap-optimized lithium niobate ($LiNbO_3$) and barium titanate ($BaTiO_3$) are primary material candidates for a performant photorefractive crossbar array processor. The integrated photonic interaction circuitry needs to be redesigned accordingly.

**Concluding Remarks**

An integrated photorefractive crossbar array has the potential to compute the synaptic signal operations in artificial neural networks for both inference and training with millions of parameters and with an insertion loss as low as 10dB. Its basic functionality has been experimentally demonstrated in GaAs prototype hardware. The extension of this work to a fully-fledged crossbar array involves the inclusion of a programmable multi-channel input vector stage. We identify lithium niobate and barium titanate as primary material candidates for a next generation chip with an enhanced diffraction efficiency.

**Acknowledgements**

This project has received funding from the European Union's Horizon 2020 research and innovation program under grant agreement number 828841 (ChipAI), 860360 (POST-DIGITAL), and 101070690 (PHOENIX). We thank the Binnig and Rohrer Nanotechnology Center (BRNC).

## References

[1] D. Psaltis, D. Brady, and K. Wagner, "Adaptive optical networks using photorefractive crystals," Appl. Opt., vol. 27, no. 9, p. 1752, 1988, doi: 10.1364/ao.27.001752.

[2] E. A. Vlieg, L. Talandier, R. Dangel, F. Horst, and B. J. Offrein, "An Integrated Photorefractive Analog Matrix-Vector Multiplier for Machine Learning," Appl. Sci., vol. 12, no. 9, pp. 1–11, 2022, doi: 10.3390/app12094226.

[3] T. Stecconi *et al.*, "Analog Resistive Switching Devices for Training Deep Neural Networks with the Novel Tiki-Taka Algorithm," *Nano Lett.*, vol. 24, no. 3, pp. 866–872, 2024, doi: 10.1021/acs.nanolett.3c03697.

[4] D. Psaltis and G. W. Burr, "Holographic data storage," Computer (Long. Beach. Calif)., vol. 31, no. 2, pp. 52–60, Feb. 1998, doi: 10.1109/2.652917.

[5] Y. Owechko and B. H. Soffer, "Holographic Neurocomputer Utilizing Laser Diode Light Source," SPIE, vol. 2565, pp. 12–19, 1995.

[6] N. Margalit, C. Xiang, S. M. Bowers, A. Bjorlin, R. Blum, and J. E. Bowers, "Perspective on the future of silicon photonics and electronics," *Appl. Phys. Lett.*, vol. 118, no. 22, 2021, doi: 10.1063/5.0050117.

[7] D. Psaltis, D. Brady, X. G. Gu, and S. Lin, "Holography in artificial neural networks," *Nature*, vol. 343, no. 6256, pp. 325–330, 1990, doi: 10.1038/343325a0.

[8] M. Cronin-golomb and M. B. Klein, "Photorefractive Materials and Devices," in *Handbook of Optics*, 2001, pp. 1–42.

# Training optical neural networks

**Jeremie Laydevant[1,2], Peter L. McMahon[1,3]**
1: School of Applied and Engineering Physics, Cornell University, Ithaca, NY 14853, USA
2: USRA Research Institute for Advanced Computer Science, Mountain View, CA 94035, USA
3: Kavli Institute at Cornell for Nanoscale Science, Cornell University, Ithaca, NY 14853, USA

jl3668@cornell.edu, pmcmahon@cornell.edu

## Status

Much of the focus on photonic machine-learning (ML) systems over the past decade has been on the potential speed and energy benefits in inference [1]. Any optical neural network (ONN) [2] needs to be trained, even if it is only intended for inference. However, ultimately, we would like to take advantage of the benefits of optics [3] not only for inference but also for training, to reduce the costs of training large ML models. This motivates the development of methods of training ONNs that are primarily based on computations performed in optical systems rather than in digital-electronic processors. Another motivation is that training methods using the photonic hardware might also allow benefits for *inference*: by not restricting oneself to hardware that can be accurately modelled on a digital-electronic computer and harnessing the natural dynamics of the photonic hardware [4] [5], the hardware might also be made faster or more efficient for inference.

The predominant method to train ONNs is backpropagation, most typically performed in a digital-electronic computer, where the trained parameters are transferred to the photonic hardware for use at inference time. This works provided that the digital-electronic computer faithfully models the behavior of the photonic hardware. The requirements on the accuracy of the digital model for backpropagation can be relaxed if training is performed in a hybrid manner [4] [6],where the forward pass to compute the loss (objective) function is done with the photonic hardware, and the backward pass is done on a digital-electronic computer with automatic differentiation of the digital model [4].

How can we minimize the amount of computation that needs to happen on a digital-electronic computer during training, hence allowing for the possibility of advantages from performing the training primarily with photonic hardware? We will discuss both backpropagation-based and backpropagation-free algorithms for training.

## Current and Future Challenges

A key challenge is to develop training methods that are as close to optimal as possible in all the metrices of “goodness” shown in Figure 1. While ideally there exists a method that is optimal in every metric, it is more likely that some trade-offs will be required.

One approach to training is to perform backpropagation directly in photonic hardware. Backpropagation (BP) is costly to execute in digital-electronic hardware:

1. BP requires storing both the intermediate neural activations and their derivatives.
2. The final error vector is “back-projected” to the intermediate layers by the means of chained transposed matrix-vectors-multiplications.
3. Finally, the actual gradient of the cost function with respect to the intermediate parameters is computed by performing a last vector-vector multiplication.

$$\frac{\partial \mathcal{L}}{\partial \overline{\overline{W}}_{\ell,\ell+1}} = \underbrace{f(\overline{a_\ell})^T . \frac{\partial f(\overline{a_{\ell+1}})}{\partial \overline{a_{\ell+1}}}}_{(1)\ \&\ (3)} . \underbrace{\overline{\overline{W}}_{\ell+1,\ell+2} . \delta_{l+2}}_{(2)}$$

Optical backpropation, proposed in Refs. [7] [8], has recently been demonstrated experimentally [9,10]. In Ref. [9], the authors optically realize steps (2) and (3) in an MZI mesh with a three-step procedure where the parameters updates are experimentally measured using a thermal camera. The scheme in Ref. [10] realizes (1) and (2) optically, but not (3) – one needs to compute the later vector-vector multiplication on a co-processor. A challenge for the further development of optical backpropagation is how to design a method that simultaneously performs (1), (2), and (3) optically.

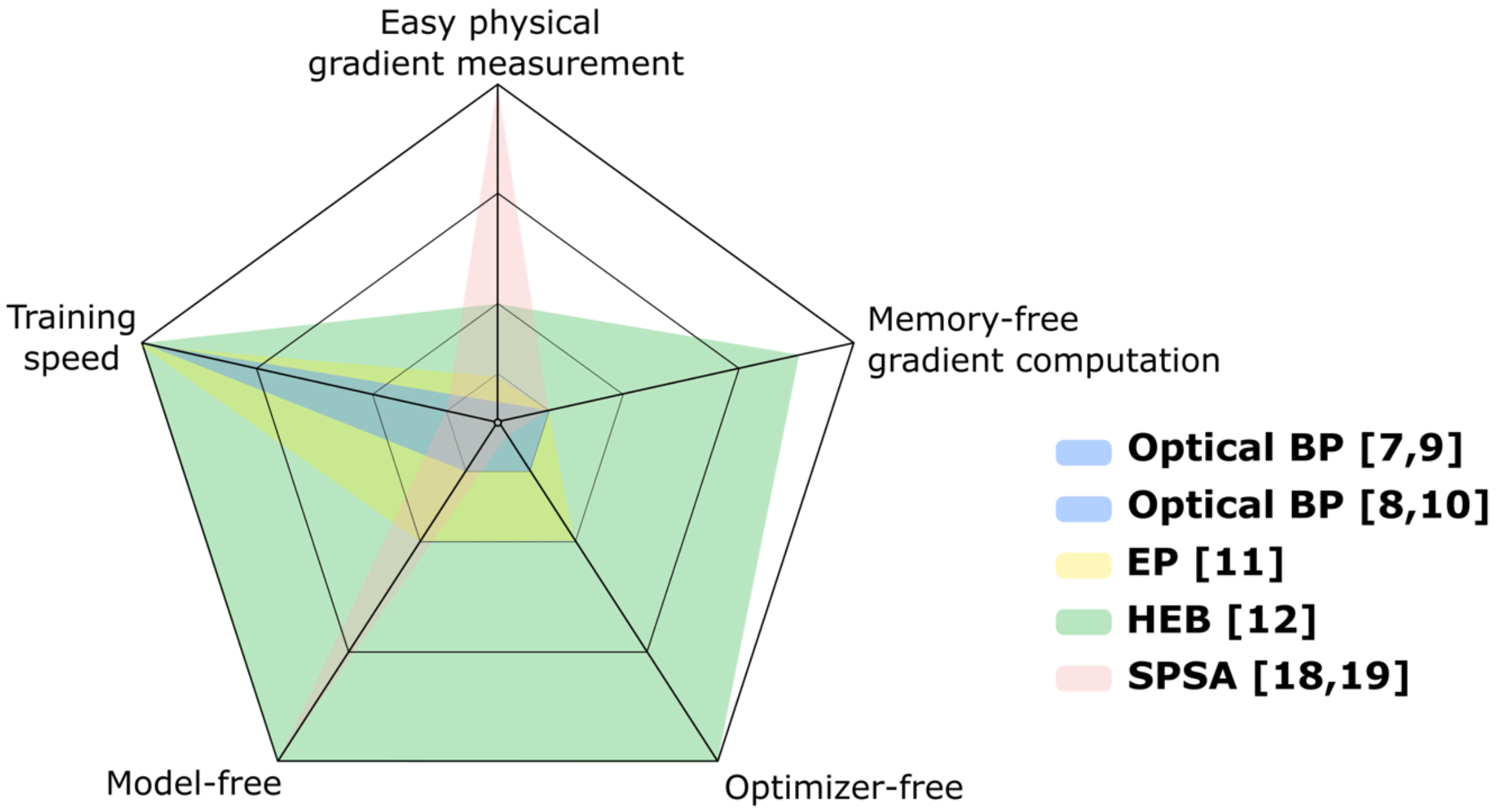

**Figure 1.** Tradeoffs in training algorithms for photonic systems. No known training algorithm is optimal in all metrics. A user must pick method depending on the exact hardware to be trained and the user's computational constraints. On the web, external location is alway better than central location, as such, the ideal training algorithm should fill in the full area while a bad one should barely overlap with th area. Training speed denotes the number of parameters' updates required before the model is trained at low training error. Model-fre denotes how much the training method relaxes the requirements about knowing perfectly the model to train. Optimizer-free denotes th non-requirement of an external optimizer, hence ancillary variables, to perform the update step after the gradients has been computec Memory-free gradient computation refers to the non-requirement of storing intermediate results of the computation to later compute th gradient. Easy physical gradient measurement grades the easiness of measuring a physical gradient based on the number of measurement required to estimate the gradients of all parameters per input data. SPSA: Simultaneous Perturbation Stochastic Optimization.

Equilibrium Propagation [11] (EP) is a training algorithm that is an alternative to backpropagation. It applies to physical systems obeying recurrent non-linear dynamics that converge to an equilibrium or steady state when presented with fixed input data. Training proceeds similarly to BP with two sequential phases: one "free phase" where the system is let free to evolve according to its internal dynamics given some input data, and a second "nudge phase" where the system is slightly nudged toward a target state[2]. Thanks to the recurrent dynamics of the system, the error will naturally but implicitly "backpropagate" in the entire system giving rise to a very simple contrastive learning rule. However, despite this promise of simple and "for free" gradient computation, deriving this learning rule depends on the possibility to derive a Lyapunov – or energy – function ($E(x, s, W)$) [3] that the system minimizes when evolving towards a steady state as it reads: $\frac{\partial L}{\partial W} \propto \frac{\partial E}{\partial W}(x, s^{*,0}\ W) - \frac{\partial E}{\partial W}\left(x, s^{*,\beta}\ W\right)$ where $W$ are the trainable parameters. While well-suited for processing static data such as images, an open challenge is to extend EP to the case of time-dependent inputs. An example of an issue that arises is that the system size may depend on the input sequence length, and so cannot be described by the same energy function for each length. EP also typically suffers from a vanishing gradient with depth. A layer-wise learning rate can be used to compensate for this issue [11] but it is expected that the addition of noise to the already vanishing gradient will complicate the training of very deep architectures.

[2] Ex.: We nudge the output units towards the one-hot encoded target vector by applying a spring-like force to the output units that force the output units to align with the target. All internal nodes will subsequently adjust their activity to realize this alignment.

Hamiltonian Echo Backprop [12] (HEB) applies to Hamiltonian systems that also obey time-reversal symmetry. In contrast to EP, the learning rule for HEB does not depend on knowing the Hamiltonian. However, it requires a time-reversal operation at the output layer. While phase conjugation in optics gives a time-reversal operation and has been experimentally demonstrated, HEB also requires the ability to nudge the output field toward the target value. BP and EP support hierarchical neural-network architectures, but another open challenge for HEB is to find out if and how it can be used to train hierarchical models. Finally, there remains the challenge of designing a practical but energy-efficient experimental implementation, especially since the trainable parameters appear in optical fields rather than naturally persistent system properties.

Another emerging alternative to backpropagation are training methods based on optimizing intermediate objectives – "forward-forward" – rather than a single global objective [13] [14, 1] [15] [16]. These methods don't directly address how to compute gradients at the level of each layer of a neural network but avoid the backwards propagation of information between layers that backpropagation has. A challenge is to merge this class of method, which has shown promising results in simulations [17] [16], with an approach for computing the gradients of each layer without relying on within-layer backpropagation.

Finally, a major open challenge for training algorithms and ONNs is to show how to scale them to state-of-the-art neural-network sizes and task complexities (e.g., ImageNet image classification, or language tasks at the level of GPT) on practical ONN hardware.

**Advances in Science and Technology to Meet Challenges**

Noise in (analog) ONNs is inevitable and might impede end-to-end training using the optical hardware itself, as the error signal could be easily confused with noise. We see here an opportunity to combine local learning methods with the physical training algorithms.

Co-designing the ONN hardware and training algorithm may help to bridge the gap between the various training approaches we have discussed and ONN hardware that can be built at scale. Much of the emphasis on ONN demonstrations has been with ONN hardware that is well-suited for backpropagation-based training using a digital-electronic computer, but this form of ONN hardware may not be best-suited to other training methods. For example, there have been no reports to date of ONNs being trained with EP, which we attribute to the need for new ONN hardware to be developed that realizes recurrent non-linear convergent dynamics in an energy-efficient manner. Similarly, a demonstration of HEB will require substantially different ONN hardware than what is typically used.

[3] Here x,s,W denote respectively the input data, the internal nodes states and the trainable parameters (weights)

Realizing forward-forward training at scale could benefit from improved black-box-optimization methods, so that within-layer backpropagation can be eliminated. Perturbation-based methods can be used [18] [19] but are generally believed to be impractical for models with billions of parameters. However, recent advances showing perturbation-based fine-tuning of models with up to 30 billion parameters [20] could revitalize perturbation-based training methods.

**Concluding Remarks**

We have briefly described several training methods for ONNs, each with its own advantages and limitations. It is an open question what combination of training algorithm and ONN hardware will ultimately enable training of ONNs at large scale with minimal usage of digital-electronic hardware during training; it may well be the case that neither the currently known training methods nor the known hardware architectures and designs are what we will use. However, with the diversity of ideas that have been developed over the past few years, there are many promising avenues to explore.

**Acknowledgements**

J.L. and P.L.M. acknowledge funding from the National Science Foundation (award no. CCF-1918549).

## References

[1] G. Wetzstein, A. Ozcan, S. Gigan, S. Fan, D. Englund,, M. Soljačić, C. Denz, D. A. B. Miller and D. Psaltis , “Inference in artificial intelligence with deep optics and photonics.,” *Nature,* vol. 588, p. 39–47, 2020.

[2] B. J. Shastri, A. N. Tai, T. Ferreira de Lima, W. H. P. Pernice, H. Bhaskaran, C. D. Wright and P. R. Prucnal , “Photonics for artificial intelligence and neuromorphic computing.,” *Nat. Photonics,* vol. 15, p. 102–114, 2021.

[3] P. L. McMahon, “The physics of optical computing.,” *Nat Rev Phys,* vol. 5, p. 717–734, 2023.

[4] L. G. Wright, T. Onodera, M. M. Stein, T. Wang, D. T. Schachte, Z. Hu and P. L. McMahon, “Deep physical neural networks trained with backpropagation.,” *Nature,* vol. 601 , p. 549–555, 2022.

[5] J. Laydevant, L. G. Wright, T. Wang and P. L. McMahon, “The hardware is the software,” *Neuron,* vol. 112, no. 2, pp. 180-183,, 2024.

[6] T. Onodera, M. M. Stein, B. A. Ash, M. Sohoni, M. Bosch, R. Yanagimoto, M. Jankowski, T. P. McKenna, T. Wang, G. Shvets, M. R. Shcherbakov, L. G. Wright and P. L. McMahon, “Scaling on-chip photonic neural processors using arbitrarily programmable wave propagation,” no. arxiv:2402.17750, 2024.

[7] T. W. Hughes, M. Minkov, Y. Shi and S. Fan, “Training of photonic neural networks through in situ backpropagation and gradient measurement.,” *Optica,* vol. 5, no. 7, pp. 864-871, 2018.

[8] X. Guo, T. D. Barrett, Z. M. Wang and A. I. Lvovsky, “Backpropagation through nonlinear units for the all-optical training of neural networks.,” *Photon. Res.,* vol. 9, no. 3, pp. B71-B80, 2021.

[9] S. Pai, Z. Sun, T. W. Hugues, T. Park, B. Bartlett, I. A. D. Williamson, M. Minkov, M. Milanizadeh, N. Abebe, F. Moricheti, A. Melloni, S. Fan, O. Solgaard and D. A. B. Miller, “Experimentally realized in situ backpropagation for deep learning in photonic neural networks.,” *Science,* pp. 398-404, 2023.

[10] J. Spall, X. Guo and A. I. Lvovsky, “Training neural networks with end-to-end optical backpropagation,” no. arxiv:2308.05226, 2023.

[11] B. Scellier and Y. Bengio, “Equilibrium propagation: Bridging the gap between energy-based models and backpropagation.,” *Frontiers in computational neuroscience,* vol. 11, p. 24, 2017.

[12] V. Lopez-Pastor and F. Marquardt, “Self-Learning Machines Based on Hamiltonian Echo Backpropagation,” *Phys. Rev. X,* vol. 13, no. 3, p. 031020, 2023.

[13] G. Hinton, “The forward-forward algorithm: Some preliminary investigations.,” no. arXiv preprint arXiv:2212.13345, 2022.

[14] A. Momeni, B. Rahmani, M. Malléjac, P. Del Hougne and R. Fleury, “Backpropagation-free training of deep physical neural networks.,” *Science,* vol. 382, pp. 1297-1303, 2023.

[15] I. Oguz, J. Ke, Q. Weng, F. Yang, M. Yildirim, N. U. Dinc, J.-L. Hsieh, C. Moser and D. Psaltis, "Forward–forward training of an optical neural network," *Opt. Lett.,* vol. 48, pp. 5249-5252, 2023.

[16] J. Laydevant, A. Lott, D. Venturelli and P. L. McMahon, "The Benefits of Self-Supervised Learning for Training Physical Neural Networks," *Machine Learning with New Compute Paradigms,* 2023.

[17] S. Siddiqui, D. Krueger, Y. LeCun and S. Deny, "Blockwise Self-Supervised Learning at Scale," *Transaction on Machine Learning Research,* 2024.

[18] S. Bandyopadhyay, A. Sludds, S. Krastanov, R. Hamerly, N. Harris, D. Bunandar, M. Streshinsky, M. Hochberg and D. Englund, "Single chip photonic deep neural network with accelerated training," no. arxiv:2208.01623, 2022.

[19] M. Goldmann, A. Skalli and D. Brunner, "Online learning strategies for optical neural networks," in *Proc. SPIE PC12655, Emerging Topics in Artificial Intelligence (ETAI)*, San Diego, 2023.

[20] S. Malladi, T. Gao, E. Nichani, A. Damian, J. D. Lee, D. Chen and S. Arora, "Fine-Tuning Language Models with Just Forward Passes," in *Thirty-seventh Conference on Neural Information Processing Systems*, New Orleans, 2023.

[21] T. Wang, S.-Y. Ma, L. G. Wright, T. Onodera, B. O. Richard and P. L. McMahon, "An optical neural network using less than 1 photon per multiplication.," *Nat Commun,* vol. 13, p. 123, 2022.

[22] C. Roques-Carmes, S. Fan and D. A. B. Miller, "Measuring, processing, and generating partially coherent light with self-configuring optics," Light: Science & Applications 13 (1), 260 (2024)

# Photonic Online Learning

**Sonia Buckley, Adam McCaughan, Bakhrom Oripov**
National Institute of Standards and Technology, 325 Broadway, Boulder, CO 80305
sonia.buckley@nist.gov, adam.mccaughan@nist.gov, bakhrom.oripov@nist.gov

**Status**

Training in machine learning necessarily involves more operations than inference, in addition to requiring higher precision, more memory, and added computational complexity. Many implementations side-step this issue by designing “inference-only” hardware that is trained separately in a one-time simulation. The resulting weights and biases from the training simulation are then transferred to hardware. This is called “offline” or “in-silico” training. While this approach is well-suited to digital systems, in systems with analog components there is often significant degradation in performance accuracy between the simulation and the hardware implementation due to noise, device-to-device variations, and drift. In addition, once trained, a new simulation is required if the application or hardware parameters change over time. A promising alternative approach is “online learning”. We define online learning as any training process that involves making measurements on the physical system itself during training [1]. Online learning techniques have enabled experimental demonstrations of photonic networks that can solve large-scale problems. For example, as early as the 1990s, online learning was used for facial recognition [2], a task that was at the cutting edge for neural networks at that time. More recently, there has been an explosion of experimental demonstrations that include some form of online learning [3], [4], [5], [6], [7], [8], [9], [10]. Another application of online learning is the development of optical accelerators specifically for the training of machine learning models [11]. These accelerators aim to reduce the energy required for training, which takes significantly more energy and time than inference alone.

Traditionally, training in machine learning is almost exclusively achieved with the stochastic gradient descent optimization technique using the backpropagation algorithm. While some online learning has either been proposed or demonstrated using backpropagation or an adaptation [9], [12], [13], [14], [15], [16], a variety of other algorithms have been developed that are easier to implement in hardware or specifically more suited to photonic implementations. Optical hardware is of particular interest for the implementation of biologically plausible algorithms, as the connections between “neurons” are typically physical due to the ease of connectivity in optics compared to that in electronic systems, where these connections are often time-multiplexed. Therefore, the types of appropriate learning algorithms may be very different than those in electronic neuromorphic hardware. Development of algorithms specifically suited to optical hardware is an area of active research.

The brain is capable of fully autonomous online learning, since it is a system that can train itself in the presence of teaching signals without the use of a digital computer. An optical system capable of fully autonomous online learning would be extremely adaptable; it would therefore be useful for applications in dynamic and/or remote environments. However, experimental demonstrations of photonic online learning have thus far involved a computer to perform at least some of the computations in the training process. We refer to this as “computer-in-the-loop” training. A continuum of possibilities exists for training that ranges from offline to fully autonomous online, with computer-in-the-loop techniques in the middle; this is shown in Fig. 1.

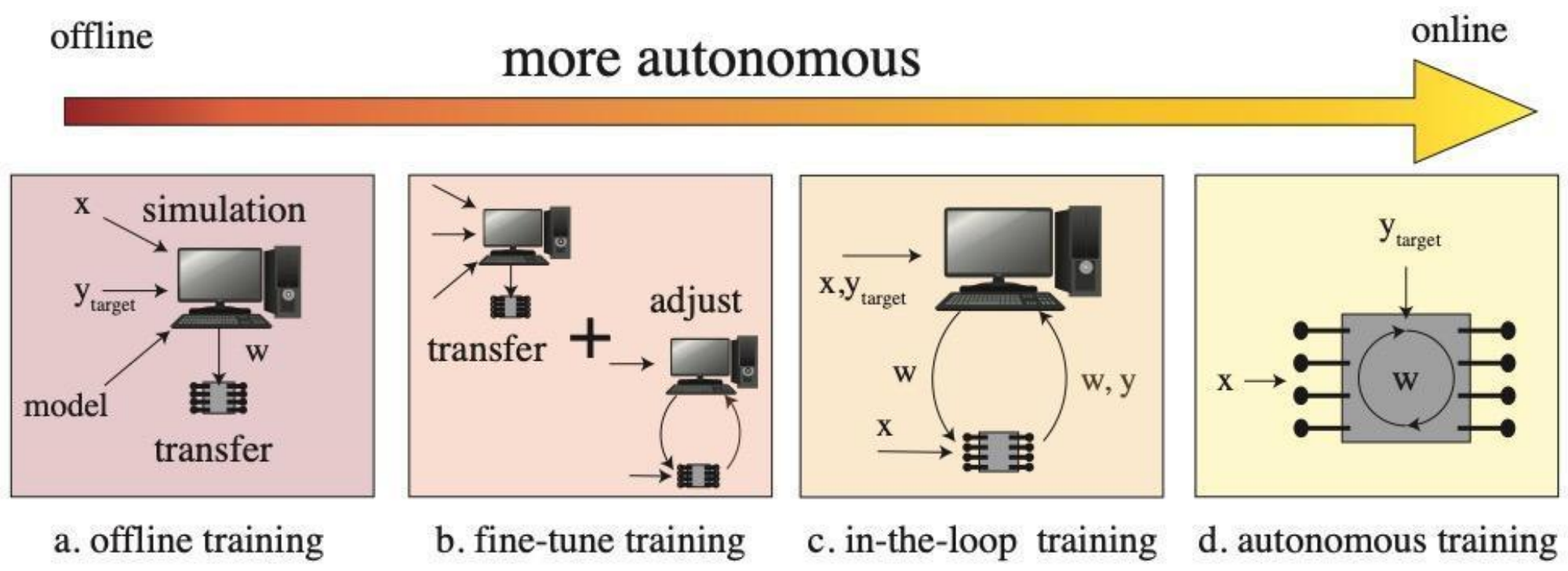

**Figure 1.** (a) Offline training. (b) Fine-tune training involves offline training followed by a small amount of online training. (c) System-in-the-loop combines system measurements with computer processing during training . (d) Fully autonomous online training. Reproduced from [1].

**Current and Future Challenges**

***Input/Output (I/O) –*** Fully-autonomous online learning remains very challenging to obtain with photonic hardware, and it is likely that computer-in-the-loop techniques will continue to be used. This has the benefit of allowing the optics to perform the functions at which they excel (e.g. fast matrix-vector multiplications) while allowing the computer to perform logic functions that are challenging in optics. However, keeping the computer in the loop involves significant I/O between the optical hardware and the computer. Many online training techniques still involve a complex simulation or model, for example surrogate training techniques, and are likely to remain computer-in-the-loop techniques even with improvements in integration.

***Algorithmic challenges –*** Existing machine learning training is incredibly effective, due to years of optimization and experimentation, with enormous academic and industry investment. To be competitive with these implementations is a huge challenge. One way to address this is to "piggyback" on their results and map architectures and training techniques onto the new hardware. Unfortunately, this significantly limits the types of optical implementations that can be employed. New and promising hardware architectures including reservoir computers, highly nonlinear systems with complex network architecture, or spiking implementations [17], [18] are not directly mappable to standard neural network algorithms. These architectures will therefore require new training algorithms that must be developed alongside the hardware, where proof of scaling and utility for modern problems is not guaranteed.

***Fair evaluation -*** Few optical implementations are at a maturity level where they can be applied to state-of-the-art machine learning problems. Finding appropriate benchmark problems on which to test new hardware and choosing the correct metrics for comparison are therefore a major challenge. For example, some common simple benchmark problems, like MNIST, can be solved to high accuracy with a linear network only. Therefore, demonstrating that a network can be trained to solve MNIST does not necessarily prove that it will work on more complex machine learning problems [19]. Other basic benchmarks with similar issues are commonly used.

***Hardware challenges –*** Online learning involves changing the weights during training. When changing the weights is a slow process, for example in on-chip thermal modulators or in spatial light modulator (SLM) arrays, it can significantly increase the time needed for training. The size of photonic components and systems also poses a major challenge to neuromorphic photonic learning in general. The larger size of photonic components must be compensated by the higher speed and ability to frequency-multiplex to compete with electronic implementations to keep systems a manageable size. As with inference-only optical devices, implementing nonlinear functions at low power is also a challenge.

## Advances in Science and Technology to Meet Challenges

***Electro-optical integration –*** Bringing electronic logic further and further on-chip toward tighter electro-optical integration will significantly improve our ability to implement high speed photonic training. Examples could be in "smart-pixel" technologies for free-space optical implementations or on-chip integration of digital logic to perform the "in-the-loop" part of the computation that currently requires I/O through a bottleneck. New foundry fabrication for direct electro-optic integration on chip, and new smart-pixel technology being developed will lead to improvements in these I/O bottlenecks.

***Volatile and non-volatile memory –*** High speed modulators exist for optical modulation, including electro-optic modulators, electro-absorption modulators and carrier injection modulators. Tradeoffs between power consumption, speed and modulation depth can make implementation of a competitive system with a single technology for applying weights. Systems that combine a fast volatile memory for training and a slower non-volatile memory [20], [21] for the final trained hardware or for less frequent updates are being developed to counteract this issue.

***Miniaturization in 3D –*** One of the major advantages of free space optics is that connections can utilize three spatial dimensions. The planar fabrication process used in integrated photonics loses this advantage. On-chip 3D integration technologies are being developed that may allow higher on-chip connectivity [22], [23].

***New algorithms -*** Algorithms that are mathematically equivalent or approximately equivalent to existing machine learning techniques but are much simpler to implement in hardware will allow optical implementations to take advantage of existing machine learning training techniques. Surrogate training techniques and zero-order optimization techniques fall into this category. Fig. 2 and Table 1 illustrate how multiplexed gradient descent [24], a framework for implementing zero-order optimization, could be used to simply implement an autonomous training algorithm, thus eliminating the I/O bottleneck. Table 1 shows the results of simulations determining training times for hardware operating at different rates and shows that this simple online training algorithm could train at a competitive rate for realistic hardware parameters. Related algorithms that combine online learning with compression techniques can further reduce the number of parameters required and also the size of the network and number of photonic components [25].

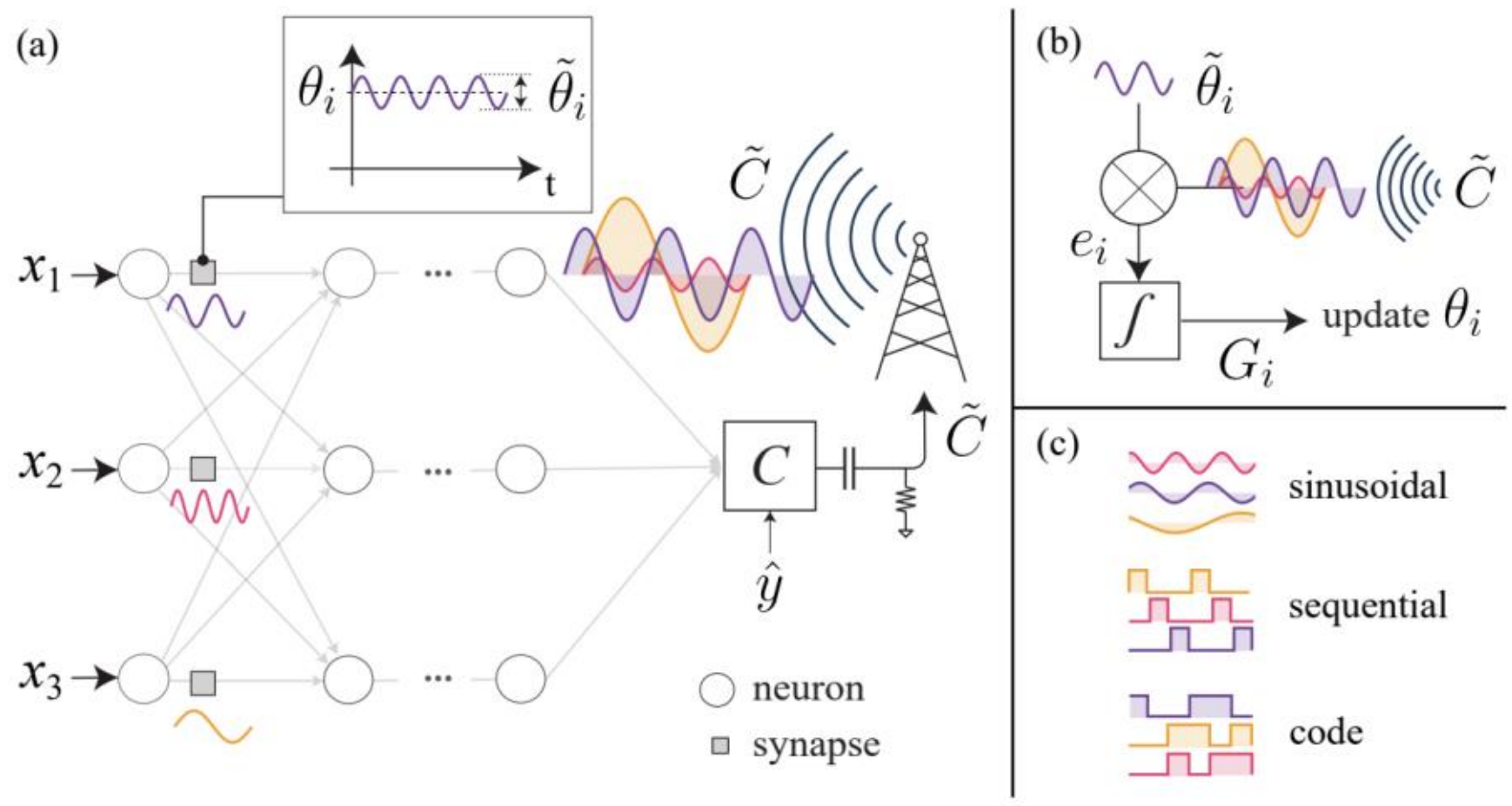

**Fig.2**. A schematic of how multiplexed gradient descent works (reproduced from [24]).

| | Hardware 1 | Hardware 2 | Hardware 3 | Backprop reference |
|---|---|---|---|---|
| **Input sample update time** | 100 ns | 1 ns | 10 ps | n/a |
| **Perturbation time** | 1 ms | 10 ns | 200 ps | n/a |
| **Parameter update time** | 1 ms | 200 µs | 200 ps | n/a |
| **Fashion-MNIST training time** | 33 min | 20 ms | 400 µs | 54 s |
| **CIFAR-10 training time** | 5.6 hours | 200 ms | 4 ms | 480 s |
| **Examples of hardware with relevant time constants** | Computer-in-the-loop, integrated photonics with thermos-optic tuning | Mem-compute devices, analog VLSI | Superconducting devices, athermal resonant silicon photonic modulator | GPU |

**Table 1**. Time to train using MGD technique on different hardware platforms operating at different speeds. See Ref. [24] for details.

## Concluding Remarks

Online learning is a promising technique for training optical hardware that can overcome the issues with device-to-device variability inherent in analog devices. Online learning could also enable optical accelerators for ML training. At present, the speed and utility of most online learning demonstrations have been limited by the need to convert from optical to electronic signals for computation of appropriate weight updates. Additionally, weight changes during training are required of online learning algorithms, and they limit the speed for many optical platforms. Future hardware advances that combine volatile and nonvolatile memories and integrate electronic logic with optical hardware will greatly speed up training of optical hardware. Newer advanced training techniques more suited to optical hardware will lead to novel architectures and devices that can be retrained in the field.

## Acknowledgements

The authors acknowledge helpful discussions with the NIST NCSG community. This research was funded by NIST and the University of Colorado Boulder.

## References

[1] S. M. Buckley, A. N. Tait, A. N. McCaughan, and B. J. Shastri, "Photonic online learning: a perspective," *Nanophotonics*, vol. 12, no. 5, pp. 833–845, Mar. 2023, doi: 10.1515/NANOPH-2022-0553/ASSET/GRAPHIC/J_NANOPH-2022-0553_FIG_003.JPG.

[2] H.-Y. S. Li, Y. Qiao, and D. Psaltis, "Optical network for real-time face recognition," *Applied Optics, Vol. 32, Issue 26, pp. 5026-5035*, vol. 32, no. 26, pp. 5026–5035, Sep. 1993, doi: 10.1364/AO.32.005026.

[3] S. Bandyopadhyay *et al.*, "Single chip photonic deep neural network with accelerated training," Aug. 2022, doi: 10.48550/arxiv.2208.01623.

[4] J. Lim and D. Psaltis, "MaxwellNet: Physics-driven deep neural network training based on Maxwell's equations," *APL Photonics*, vol. 7, no. 1, Jan. 2022, doi: 10.1063/5.0071616/2835095.

[5] C. Moser *et al.*, "Forward–forward training of an optical neural network," *Optics Letters, Vol. 48, Issue 20, pp. 5249-5252*, vol. 48, no. 20, pp. 5249–5252, Oct. 2023, doi: 10.1364/OL.496884.

[6] X. Porte, A. Skalli, N. Haghighi, S. Reitzenstein, J. A. Lott, and D. Brunner, "A complete, parallel and autonomous photonic neural network in a semiconductor multimode laser," *Journal of Physics: Photonics*, vol. 3, no. 2, p. 024017, Apr. 2021, doi: 10.1088/2515-7647/ABF6BD.

[7] L. G. Wright *et al.*, "Deep physical neural networks trained with backpropagation," *Nature 2022 601:7894*, vol. 601, no. 7894, pp. 549–555, Jan. 2022, doi: 10.1038/s41586-021-04223-6.

[8] T. Onodera *et al.*, "Scaling on-chip photonic neural processors using arbitrarily programmable wave propagation," 2024, Accessed: Apr. 01, 2024. [Online]. Available: http://arxiv.org/abs/2402.17750

[9] S. Pai *et al.*, "Experimentally realized in situ backpropagation for deep learning in nanophotonic neural networks," 2022.

[10] M. J. Filipovich *et al.*, "Monolithic Silicon Photonic Architecture for Training Deep Neural Networks with Direct Feedback Alignment," Nov. 2021, doi: 10.48550/arxiv.2111.06862.

[11] J. Launay *et al.*, "Hardware Beyond Backpropagation: a Photonic Co-Processor for Direct Feedback Alignment.", arXiv:2012.06373 (2020) doi: 10.48550/arXiv.2012.06373.

[12] L. G. Wright *et al.*, “Deep physical neural networks trained with backpropagation,” *Nature 2022 601:7894*, vol. 601, no. 7894, pp. 549–555, Jan. 2022, doi: 10.1038/s41586-021-04223-6.

[13] T. Zhou *et al.*, “In situ optical backpropagation training of diffractive optical neural networks,” *Photonics Research, Vol. 8, Issue 6, pp. 940-953*, vol. 8, no. 6, pp. 940–953, Jun. 2020, doi: 10.1364/PRJ.389553.

[14] M. Hermans, M. Burm, T. Van Vaerenbergh, J. Dambre, and P. Bienstman, “Trainable hardware for dynamical computing using error backpropagation through physical media,” *Nature Communications 2015 6:1*, vol. 6, no. 1, pp. 1–8, Mar. 2015, doi: 10.1038/ncomms7729.

[15] X. Liu, Y. Gao, Z. Huang, and Z. Gu, “Training optronic convolutional neural networks on an optical system through backpropagation algorithms,” *Optics Express, Vol. 30, Issue 11, pp. 19416-19440*, vol. 30, no. 11, pp. 19416–19440, May 2022, doi: 10.1364/OE.456003.

[16] M. Minkov, S. Fan, T. W. Hughes, and Y. Shi, “Training of photonic neural networks through in situ backpropagation and gradient measurement,” *Optica, Vol. 5, Issue 7, pp. 864-871*, vol. 5, no. 7, pp. 864–871, Jul. 2018, doi: 10.1364/OPTICA.5.000864.

[17] J. Robertson, P. Kirkland, G. Di Caterina, and A. Hurtado, “VCSEL-based photonic spiking neural networks for ultrafast detection and tracking,” *Neuromorphic Computing and Engineering* , vol. 4, no. 1, p. 014010, Mar. 2024, doi: 10.1088/2634-4386/AD2D5C.

[18] S. Khan *et al.*, “Superconducting optoelectronic single-photon synapses,” *Nature Electronics 2022 5:10*, vol. 5, no. 10, pp. 650–659, Oct. 2022, doi: 10.1038/s41928-022-00840-9.

[19] Tim Hargreaves, “‘Is it Time to Ditch the MNIST Dataset?,’” TTested. https://www.ttested.com/ditch-mnist.

[20] Z. Fang, J. Zheng, A. Saxena, J. Whitehead, Y. Chen, and A. Majumdar, “Non-Volatile Reconfigurable Integrated Photonics Enabled by Broadband Low-Loss Phase Change Material,” *Adv Opt Mater*, vol. 9, no. 9, p. 2002049, May 2021, doi: 10.1002/ADOM.202002049.

[21] S. Lam *et al.*, “Dynamic Electro-Optic Analog Memory for Neuromorphic Photonic Computing,” Jan. 2024, Accessed: Apr. 01, 2024. [Online]. Available: https://arxiv.org/abs/2401.16515v1

[22] J. Chiles, S. M. Buckley, S. W. Nam, R. P. Mirin, and J. M. Shainline, “Design, fabrication, and metrology of 10 × 100 multi-planar integrated photonic routing manifolds for neural networks,” *APL Photonics*, vol. 3, no. 10, 2018, doi: 10.1063/1.5039641.

[23] J. Moughames *et al.*, “Three-dimensional waveguide interconnects for scalable integration of photonic neural networks,” *Optica*, vol. 7, no. 6, p. 640, Jun. 2020, doi: 10.1364/optica.388205.

[24] A. N. McCaughan, B. G. Oripov, N. Ganesh, S. W. Nam, A. Dienstfrey, and S. M. Buckley, “Multiplexed gradient descent: Fast online training of modern datasets on hardware neural networks without backpropagation,” *APL Machine Learning*, vol. 1, no. 2, p. 26118, Jun. 2023, doi: 10.1063/5.0157645/2900191.

[25] Y. Zhao *et al.*, “Real-Time FJ/MAC PDE Solvers via Tensorized, Back-Propagation-Free Optical PINN Training,” Dec. 2023, Accessed: Apr. 29, 2024. [Online]. Available: https://arxiv.org/abs/2401.00413v2

# How to train an optical neural network

**James Spall[1], Xianxin Guo[1] and A. I. Lvovsky[2]**
[1] Lumai Ltd., Wood Centre for Innovation, Oxford, UK
[2]Clarendon Laboratory, University of Oxford, Parks Road, Oxford OX1 3PU, UK
Alex.Lvovsky@physics.ox.ac.uk

## Status

A key challenge in developing optical or, more generally, analog neural networks (NNs) is training [1]. Most of the existing analog NN implementations are trained in silico: a simulation ("digital twin") of the analog NN is constructed in a digital computer which is then trained using the error backpropagation algorithm [2]. The weight matrices calculated in this manner are then transferred into the analogue twin in the form of relevant control parameters.

This approach has significant shortcomings. The first one is its speed. According to some estimates [3], training occupies 10 to 15% in the lifetime of a neural network. This may appear to be an insignificant value; however, the inference time in an analogue NN is expected to be reduced by several orders of magnitude compared to the digital counterpart, hence the share of training time, if still done digitally, can be expected to reach above 99%. The second shortcoming is accuracy. No simulation is perfect, so an analogue NN equipped with the weight matrices computed in silico is likely to exhibit significant errors.

A partial solution to the latter issue consists in the so-called physics-aware or hybrid training [4, 5]. To understand this method, we recall that, in the backpropagation algorithm, the gradient o of the loss function with respect to each weight matrix $\widehat{W}^{(i)}$ is the outer product

$$\frac{\partial L}{\partial \widehat{W}^{(i)}} = \delta^{(i)} \otimes a^{(i-1)}, \qquad (1)$$

of two vectors: the neuron activation vector $a^{(i-1)}$ and the error vector calculated via the backpropagation algorithm, i.e. in the reverse order of layers:

$$\delta^{(i-1)} = \left(\widehat{W}^{(i)^T} \delta^{(i)}\right) g'\left(z^{(i-1)}\right), \qquad (2)$$

where $g(\cdot)$ is the activation function and $z^{(i)} = \widehat{W}^{(i)} a^{(i-1)}$ is the pre-activation. The last layer error vector $\delta^{(L)}$ is obtained from the disparity between the NN output vector and the label vector; the specific expression depends on the loss function choice.

The idea of the physics-aware training is that the activation vector is computed via the analog network (which is readily available) while the backpropagation leading to the error vector is still done in silico. The presence of "physics in the loop" helps reduce the effect of discrepancy between the physical analog network and its digital twin. This is particularly effective in the case of static noise, i.e. discrepancies that do not vary in time.

## Current and Future Challenges

Hybrid training does not address the primary issue: the need to create a digital twin for the analogue NN and the resulting the poor speed and efficiency of training it. This challenge can be addressed by developing the so-called *in-situ* backpropagation. The idea is to compute the reverse sequence of error vectors by physically allowing the analogue signal to propagate backwards through the NN. Comparing Eq. (2) with the forward-propagation chain

$$a^{(i)} = g\left(\widehat{W}^{(i)} a^{(i-1)}\right), \qquad (3)$$

we observe a symmetry in the linear part of the equation: while the forward propagating vectors are multiplied by the weight matrix, the backpropagating vectors are multiplied by its transpose. In an optical NN (ONN), the latter multiplication can be realized by sending the optical field representing the error vector backwards through the same linear optical arrangement as is used to implement the forward propagation. It is perhaps ironic that the term “backpropagation”, initially coined to denote a pure mathematical operation [2], acquires its literal meaning in the context of ONNs.

A further challenge however arises when one compares the features of Eqs. (2) and (3) related to the activation function. While forward propagation requires application of a nonlinear function to each element of the pre-activation vector $z^{(i)}$, backward propagation involves a linear operation: multiplication of the error vector by $g'\left(z^{(i-1)}\right)$ . If a nonlinear optical effect is used in an ONN to implement the activation, in situ backpropagation requires this effect to feature (i) nonlinear response for the forward input; (ii) linear response for backward input; (iii) multiplication of the backward input by the derivative of the activation function. Multiple attempts to address this challenge resulted either in schemes being too complicated for practical realization [6], or only functional for specific network structures, not compatible with conventional modern architectures [7, 8].

**Advances in Science and Technology to Meet Challenges**

Our group has developed a solution to this long-standing problem [9, 10]. We found that the nonlinear optical phenomena of saturable absorption or saturable gain can play the role of the activation function satisfying the above requirements. Specifically, the conditions (i) and (ii) are addressed by making the forward propagating field sufficiently strong to invoke the nonlinearity, but attenuating the backward field significantly below that level. An important further insight of our work is that specific nonlinear phenomena listed above also satisfy the requirement (iii) with the accuracy that is sufficient for training.

We tested our scheme through extensive simulation on multiple machine intelligence problems and in a variety of realistic settings (which included optical noise) and found the quality of training to be at a level of state-of-the-art digital backpropagation [9]. Subsequently we completed an experimental test [10], constructing a two-layer ONN with three input, five hidden and two output nodes and computing the training signal by backward propagation of the optical field therein. The ONN was implemented in free space with coherent data encoding. Matrix-vector multiplication [11] followed the Stanford scheme and the activation function utilized saturable absorption in atomic rubidium vapour. The ONN was trained to classify a point in the 2D plane into two classes according to the values of the (*x, y*) coordinates supplied as input.

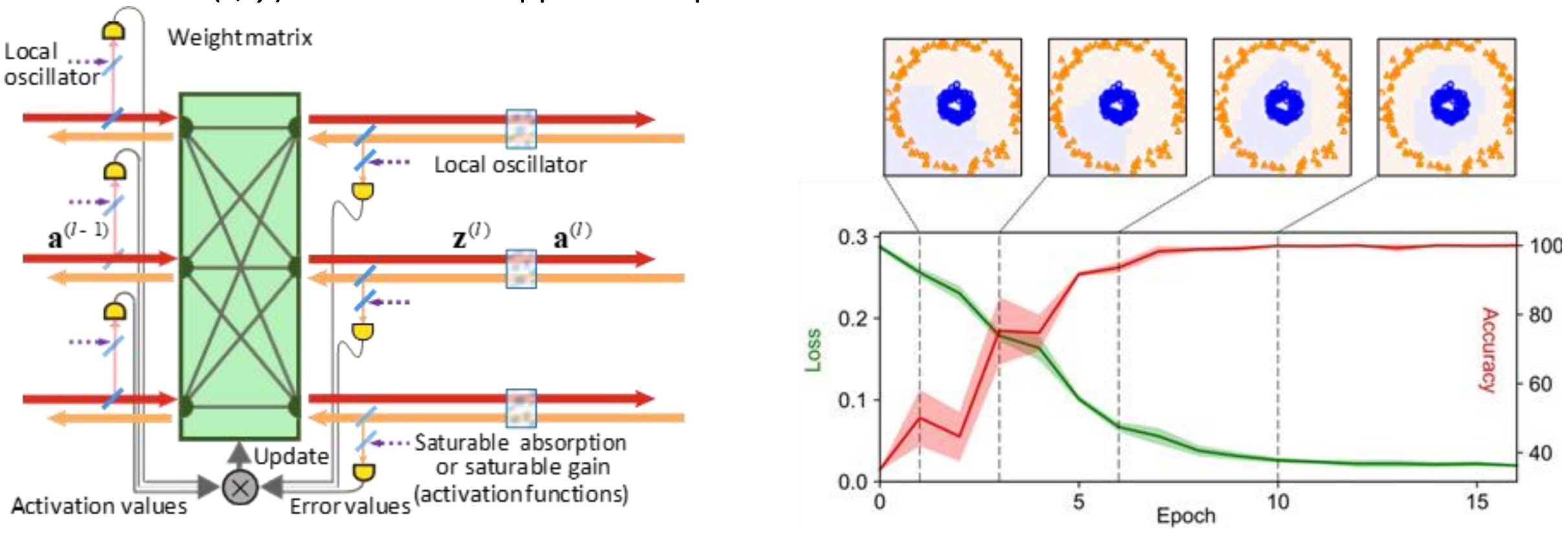

*Fig. 1*

We started with injecting the input vector, which contained a training set element, into the ONN and letting it propagate forward through the layers, obtaining the activation vector for each layer (at the last layer, the activation function is not applied). We then calculated the error vector $\delta^{(L)} = z^{(L)} - t$ of the last layer, where $\boldsymbol{t}$ is the output target value vector (label) for the given input. The error signals were reinjected back into the ONN and travelled backwards along the same path as the feedforward signal, optically computing the error vector $\delta^{(l)}$, which was used to digitally calculate the weight matrix training gradient (1) for each layer.

We stress the difference between our work and the proposal of Hughes et al. [12], subsequently carried out experimentally [13]. That work, reported in a separate article of this Roadmap, does not implement backpropagation through the nonlinear units. It focuses on linear sub-layers implemented as interferometer meshes on an integrated photonic chip, and uses backpropagation to train the control parameters of individual interferometers within the mesh. Hence the two studies pursue different goals and can be seen as complementary to each other.

Although these results show conceptual possibility of in situ training via backpropagation, they also highlight the challenges associated with its scaling [1]. Backpropagating optical field must follow in the exact footsteps of the forward field, which is hard to achieve in practice. The activation function is typically not known exactly, hence it is difficult to precisely satisfy the three aforementioned requirements. The dynamic range of the backpropagating field is limited due to physical constraints, which exacerbates the problem of exploding and vanishing gradients. As the error backpropagates through the entire deep computational graph, its inaccuracy accumulates.

**Concluding Remarks**

These concerns raise the question whether backpropagation is at all a suitable method for training optical and analogue NNs. This question is reminiscent of that raised by a team led by Hinton in their influational paper [14], which argues that backpropagation is unlikely to be the main process driving the training in the brain cortex for a set of reasons that largely overlap with the aforementioned challenges arising in analogue NNs. They argue that the learning process taking place in the brain is more likely to “do away with the explicit propagation of error derivatives and instead compute them locally through differences in propagated activities” of neurons. The term “locally” here implies that the weight matrix gradient in each layer is calculated based on the neuron activation values within that layer only.

The success of analogue NN technology therefore largely depends on our ability to develop such local training mechanisms. Existing ideas include target propagation [15], equilibrium propagation [16], direct feedback alignment [17], the forward- forward algorithm [18], coupled learning [19], prospective configuration [20] and many others. While many of these algorithms showed good performance in simulation, and some were demonstrated in experiment [17, 21], none have yet performed on par with backpropagation, even in simulation. Reaching that performance level is an outstanding problem and important milestone in the roadmap towards practical neuromorphic photonics.

**Acknowledgements**
This work is supported by Innovate UK Smart Grant 10043476.

## References

[1] S. M. Buckley, A. N. Tait, A. N. McCaughan, and B. J. Shastri, “Photonic online learning: a perspective,” *Nanophotonics*, vol. 12, no. 5, pp. 833–845, 2023.

[2] D. E. Rumelhart, G. E. Hinton, and R. J. Williams, “Learning representations by back-propagating errors,” *nature*, vol. 323, no. 6088, pp. 533–536, 1986.

[3] D. Patterson, J. Gonzalez, Q. Le, C. Liang, L.-M. Munguia, D. Rothchild, D. So, M. Texier, and J. Dean, “Carbon emissions and large neural network training,” *arXiv preprint arXiv:2104.10350*, 2021.

[4] L. G. Wright, T. Onodera, M. M. Stein, T. Wang, D. T. Schachter, Z. Hu, and P. L. McMahon,

"Deep physical neural networks trained with backpropagation," *Nature*, vol. 601, no. 7894, pp. 549–555, 2022.
[5] J. Spall, X. Guo, and A. I. Lvovsky, "Hybrid training of optical neural networks," *Optica*, vol. 9, no. 7, pp. 803–811, 2022.
[6] K. Wagner and D. Psaltis, "Multilayer optical learning networks," *Appl. Opt.*, vol. 26, no. 23, pp. 5061–5076, Dec 1987.
[7] J. E. Steck, S. R. Skinner, A. A. Cruz-Cabrara, and E. C. Behrman, "A lagrangian formulation for optical backpropagation training in kerr-type optical networks," vol. 7, pp. 771–778, 1995.
[8] A. A. Cruz-Cabrera, M. Yang, G. Cui, E. C. Behrman, J. E. Steck, and S. R. Skinner, "Reinforcement and backpropagation training for an optical neural network using self-lensing effects," *IEEE Trans. Neural Netw. Learn. Syst.*, vol. 11, no. 6, pp. 1450–1457, 2000.
[9] X. Guo, T. D. Barrett, Z. M. Wang, and A. Lvovsky, "Backpropagation through nonlinear units for the all-optical training of neural networks," *Photonics Research*, vol. 9, no. 3, pp. B71–B80, 2021.
[10]James Spall, Xianxin Guo, A. I. Lvovsky, "Training neural networks with end-to-end optical backpropagation", arXiv:2308.05226J. Spall, X. Guo, T. D. Barrett, and A. I. Lvovsky, "Fully reconfigurable coherent optical vector–matrix multiplication," *Opt. Lett.*, vol. 45, no. 20, pp. 5752–5755, Oct 2020.
[11] T. W. Hughes, M. Minkov, Y. Shi, and S. Fan, "Training of photonic neural networks through in situ backpropagation and gradient measurement," *Optica*, vol. 5, no. 7, pp. 864–871, Jul 2018.
[12] S. Pai, Z. Sun, T. W. Hughes, T. Park, B. Bartlett, I. A. Williamson, M. Minkov, M. Milanizadeh, N. Abebe, F. Morichetti *et al.*, "Experimentally realized in situ backpropagation for deep learning in photonic neural networks," *Science*, vol. 380, no. 6643, pp. 398–404, 2023.
[13] T. P. Lillicrap, A. Santoro, L. Marris, C. J. Akerman, and G. Hinton, "Backpropagation and the brain," *Nature Reviews Neuroscience*, vol. 21, no. 6, pp. 335–346, 2020.
[14] Y. Le Cun, "Learning process in an asymmetric threshold network," in *Disordered systems and biological organization*. Springer, 1986, pp. 233–240.
[15] B. Scellier and Y. Bengio, "Equilibrium propagation: Bridging the gap between energy-based models and backpropagation," *Frontiers in computational neuroscience*, vol. 11, p. 24, 2017.
[16] M. Nakajima, K. Inoue, K. Tanaka, Y. Kuniyoshi, T. Hashimoto, and K. Nakajima, "Physical deep learning with biologically inspired training method: gradient-free approach for physical hardware," *Nature communications*, vol. 13, no. 1, p. 7847, 2022.
[17] G. Hinton, "The forward-forward algorithm: Some preliminary investigations," *arXiv preprint arXiv:2212.13345*, 2022.
[18] M. Stern, D. Hexner, J. W. Rocks, and A. J. Liu, "Supervised learning in physical networks: From machine learning to learning machines," *Phys. Rev. X*, vol. 11, p. 021045, May 2021. [Online]. Available: https://link.aps.org/doi/10.1103/PhysRevX.11.021045
[19] Y. Song, B. Millidge, T. Salvatori, T. Lukasiewicz, Z. Xu, and R. Bogacz, "Inferring neural activity before plasticity as a foundation for learning beyond backpropagation," *Nature Neuroscience*, pp. 1–11, 2024.
[20] A. Momeni, B. Rahmani, M. Mall´ejac, P. Del Hougne, and R. Fleury, "Backpropagation-free training of deep physical neural networks," *Science*, vol. 382, no. 6676, pp. 1297–1303, 2023.

# *In situ* training of on-chip neural networks

**Charles Roques-Carmes and Shanhui Fan**
Edward L. Ginzton Laboratory, Stanford University, Stanford, California 94305, USA
chrc@stanford.edu, shanhui@stanford.edu

**Status**

Integrated photonics has emerged as a promising platform for inference in machine learning. Photonics benefits from massive parallelism in the spectral and spatial domain, low-power computation and processing, as well as mature optical interconnect technology [1]–[3]. Several candidates for on-chip photonic computing have been demonstrated in the past decade, including arrays of Mach-Zehnder interferometers (MZI) [4]–[6], wavelength-division multiplexing [7], optical frequency combs [8], and diffractive networks [9], with applications across image classifications [10] and vowel recognition [4]. All these architectures implement a cascade of computational layers, where a linear photonic operation is performed in the spatial or spectral domain, followed by a nonlinear optical, electronic, or optoelectronic activation function. While photonics presents some key benefits in inference tasks, the training of photonic chips to learn a specific task has remained challenging. Several approaches emerged in the past few years, such as *in situ* training backpropagation [11], physics-aware training [12], and physical dithering of network parameters [6], [13]. Efficient training of photonic chips may not only benefit photonic machine learning, but also applications in trainable modal decompositions of light fields [14], matrix optimization [15], and heuristic combinatorial optimizers [5], [16].

*In situ* backpropagation is a method that relies on a connection between time reversal and adjoint fields (related to network weight sensitivity), as illustrated in Figure 1, and can be in principle implemented in various integrated photonic architectures. The key idea, as proposed in Ref. [11], is implemented on each layer of the neural network consisting of a waveguide array made of MZIs, and consists of three steps: In the first step, inference is realized by sending an input signal $X$ through the array (see inset of Figure 1b). One records the output of the array. We denote the field inside the mesh as $E$. In the second step, an error signal $\delta_n$ is determined from the output of the array recorded in the first step and the prescribed cost function, and sent through the network backwards. One again records the output of the network, denoted as $X_{TR}$ (time reversal adjoint fields). We denote the field inside the mesh as $E_{adj}$. In the third step, one sends an input of $X + X_{TR}^*$ along the forward direction of the array. The gradient information is then contained in the intensity of the fields at the arm of each of the waveguide segments, since the gradient is proportional to $Re\left(E_{adj}^T E\right)$, as can be derived by the adjoint variable method [11]. The gradient of a given cost function with respect to all parameters of the networks (e.g., the phase of the MZI) can be determined by a single iteration of these three steps.

*In situ* backpropagation was experimentally realized in Ref. [17], following the procedure described in the previous paragraph, and with an experimental setup shown in Figure 2. The experimental realization of *in situ* backpropagation requires the continuous monitoring of field intensities throughout the network at all steps of the training. This can be achieved by using grating taps placed at the waveguide arms of the MZI's that scatters part of the light out of the chip. The scattered light is then detected by an infrared camera [17] or integrated detectors. The intensity distribution of the scattered light provides the information about the gradient of the cost function.

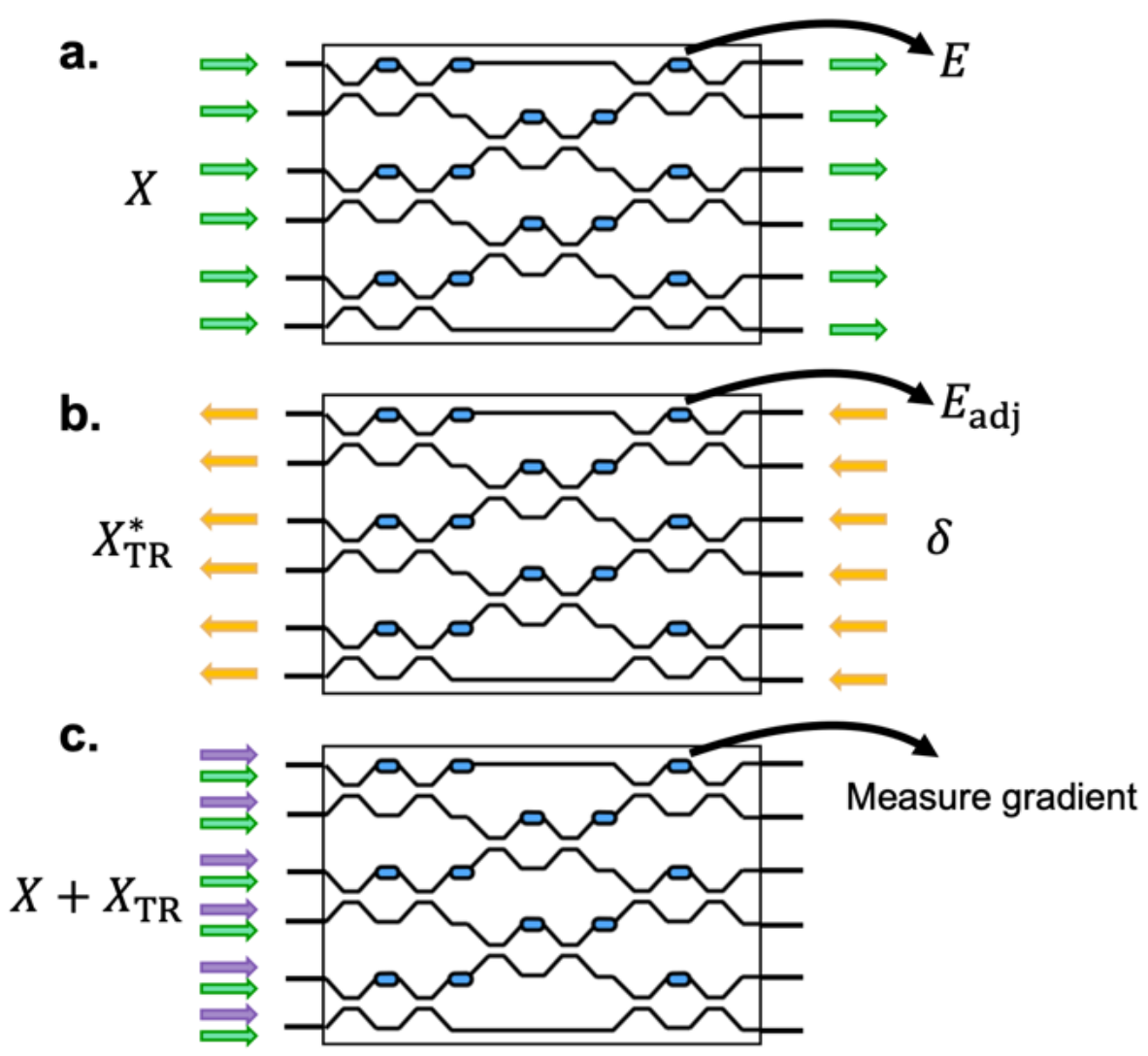

**Figure 1.** Analogy between time reversal and adjoint field measurement in photonic neural networks. a. Forward signal propagation for inference. The field in the waveguide network corresponds to the forward propagation field E. b. Backward signal propagation for gradient measurement and training. The field measured in the waveguide network corresponds to the adjoint field Eajd. c. Sum propagation step, where the sum of the incident and time-reversal adjoint fields is propagated through the network. The local field intensity can be measured to determine the gradient.

## Current and Future Challenges

*In situ* backpropagation presents some fundamental advantages compared to other gradient calculation methods, such as finite difference approximation schemes [6]. These advantages include higher efficiency and better scalability, as gradients are obtained directly through parallel intensity measurements. Additionally, *in situ* backpropagation ensures greater accuracy since the gradients are derived from the physical response of the system instead of numerical approximations. However, to ensure that this advantage persists at a large scale, one should ensure that the *in situ* backpropagation method can handle large, complex networks without significant overhead. This will entail the development of advanced photonic components capable of maintaining high fidelity and low loss across many layers and nodes. In addition, generalizing *in situ* backpropagation to optical nonlinear activation functions remains challenging [17], although this might not be required to demonstrate practical advantage compared to digital architectures.

For any practical applications, it is worthwhile to ask how to scale such integrated photonic neural networks to thousands or tens of thousands of inputs. This question holds for scaling up both inference and training tasks since realizing the true potential of optical computing may require ~$10^4$ inputs [1]. For such large scales, reconfigurable grating taps may be needed to mitigate the influence of additional losses due to power monitoring. Addressing these challenges without introducing additional noise and sacrificing computational accuracy is critical for *in situ* backpropagation, since training is in general more sensitive than inference to computational accuracy.

While integrated photonics can in principle achieve such large scales, the true potential of optical computing relies in ultralow latency computation. Concretely, this means that one should aim to achieve inference (and training) in integrated photonic platforms for reasonable matrix sizes (say > 100 x 100) with latency on the order of a few nanoseconds to compete with digital architectures (with a fundamental bound eventually imposed by the speed of light in integrated waveguides). For reference, a recent work achieved sub-nanosecond latency with vector size *N* = 6 in a 3-layer deep neural network including nonlinear activation functions [6]. Practically, this also means that one should mitigate delays due to (1) data movement, conversion, and storage between units; (2) MZI reconfiguration time; (3) data conversion from electronic to photonic domain. This type of low latency architecture most likely requires the development of co-integrated electronic and photonic integrated circuits, as described in the next section. Since *in situ* backpropagation requires continuous tuning of the MZI weights (at each training step), scaling also entails the development of fast and low-loss MZI [18].

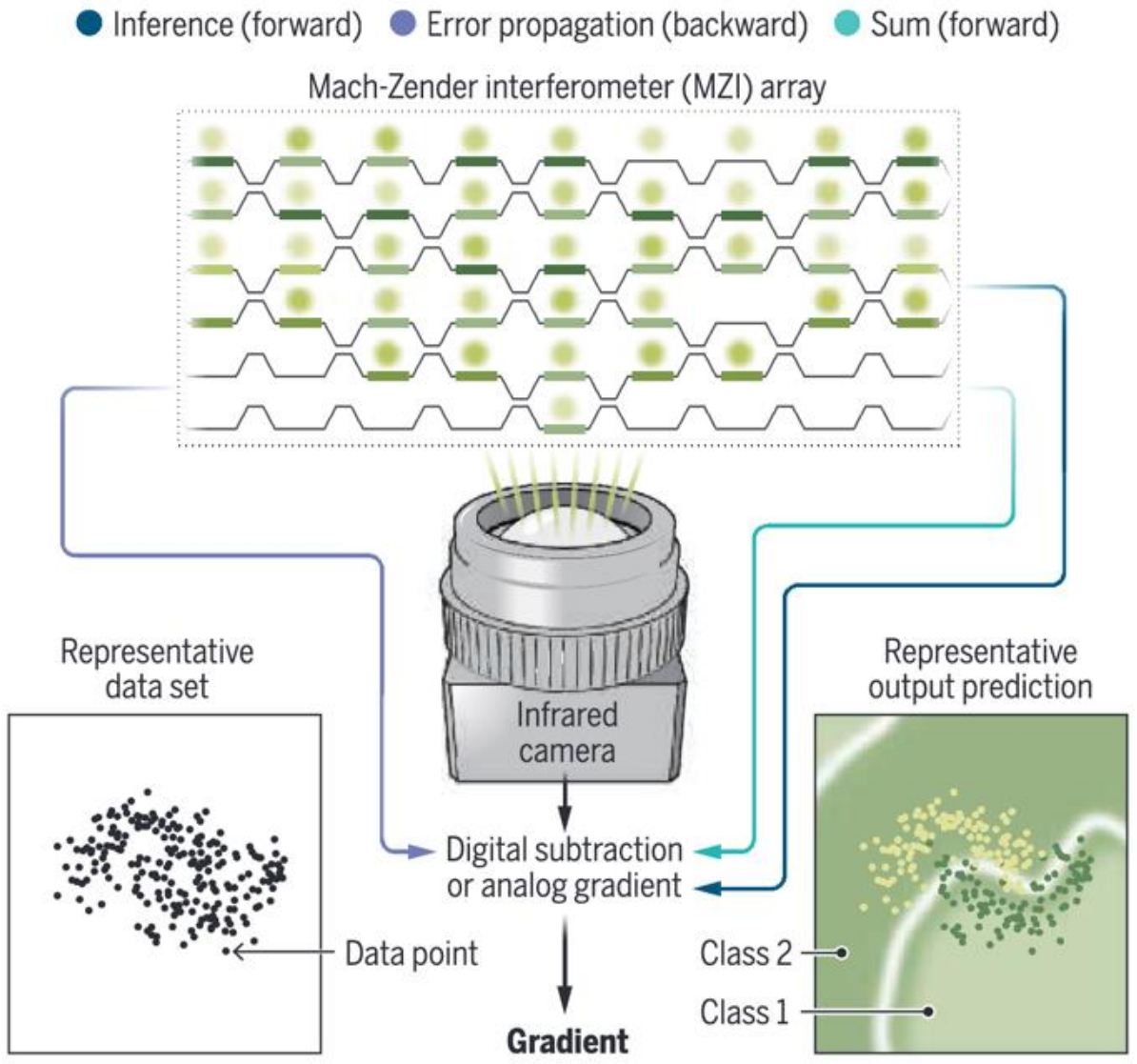

**Figure 2.** Learning gradients *in situ* with photonic chips. The IR camera over the chip imaged all grating tap monitors necessary for backpropagation. A representative data set is shown as well as the representative output prediction from the photonic chip after *in situ* training. Reproduced with permission from Ref. [3]

## Advances in Science and Technology to Meet Challenges

One of the key features of *in situ* backpropagation is that it relies on a direct connection between a physical operation (time reversal) and the desired mathematical operation (calculating gradients, as is usually performed for inverse design based on the adjoint method [11]). A broader question is whether there exist other such analogies that could be applied to integrated photonic computation. Potential avenues include variational principles of wave optics (e.g., light propagation solves an optimization problem per Fermat's principle [1]) or using self-configuring optics [14].

One could also look for inspiration in the field of biologically inspired machine learning, where alternative training methods based on the dynamics of physical systems have been proposed [19]. Another avenue that could facilitate the realization of scalable inference and training in integrated photonic circuits is the development of application-specific algorithms for photonic chips – for instance, that exploit the wave nature of photon propagation on a chip [20].

To achieve real scaling advantages with integrated photonic computing, a few advances in hardware are required. While they rely on existing technology, this will necessitate a significant effort (where both academic groups and industry partners are to play a paramount role). One such challenge is the co-integration of photonic and electronic chips, since electronics are still required for data storage, control, and nonlinear activation functions (in optoelectronic platforms). This must be achieved without sacrificing computational accuracy or latency. An anticipated milestone of the field would be to demonstrate a large-scale integrated photonic processor (even with limited reconfigurability) that could beat a state-of-the-art GPU (in terms of latency) and in real-world conditions.

## Concluding Remarks

Integrated photonics presents a few key features, such as bandwidth and spatial parallelism, that makes it a unique platform for the next generation of machine learning hardware, both for inference and training. While several architectures have been proposed to realize low latency inference with integrated photonic circuits, the realm of proposals for trainable photonic architectures is scarcer. Physics-inspired methods such as *in situ* backpropagation provide a promising route towards *in situ* training of integrated photonic circuits, as was recently experimentally demonstrated.

However, the real potential of integrated photonic computing (for inference or training) will require the development of very-large-scale integration and volume manufacturing technologies that allow the co-integration of photonic and electronic chips and could eventually overturn digital computing platforms (on some specific tasks). The first anticipated milestones will be some comparative advantage over digital architectures on inference. Then, using physics-inspired training methods, such as *in situ* backpropagation, one could also expect advantages in terms of training.

We also believe that novel physics-inspired training algorithms may facilitate the scaling up of *in situ* training. Such methods may rely on variational principles encountered in wave physics, self-configuring photonic networks, mapping to dynamics of physical systems, or a combination thereof.

**Acknowledgements**
*The authors acknowledge the support of a MURI project from the U. S. Air Force Office of Scientific Research (Grant No. FA9550-21-1-0312). C. R.-C. is supported by a Stanford Science Fellowship.*

## References

[1] P. L. McMahon, “The physics of optical computing,” *Nat. Rev. Phys. 2023 512*, vol. 5, no. 12, pp. 717–734, Oct. 2023, doi: 10.1038/s42254-023-00645-5.
[2] G. Wetzstein *et al.*, “Inference in artificial intelligence with deep optics and photonics,” *Nat. 2020 5887836*, vol. 588, no. 7836, pp. 39–47, Dec. 2020, doi: 10.1038/s41586-020-2973-6.
[3] C. Roques-Carmes, “Learning photons go backward,” *Science (80-. ).*, vol. 380, no. 6643, pp. 341–342, Apr. 2023, doi: 10.1126/SCIENCE.ADH0724/ASSET/8CC3FB1B-697C-4A7E-8E9A-9C92ED86C741/ASSETS/GRAPHIC/SCIENCE.ADH0724-F1.SVG.
[4] Y. Shen *et al.*, “Deep Learning with Coherent Nanophotonic Circuits,” *Nat. Photonics*, Oct. 2017, doi: 10.1038/nphoton.2017.93.
[5] M. Prabhu *et al.*, “Accelerating recurrent Ising machines in photonic integrated circuits,” *Optica*, vol. 7, no. 5, 2020, doi: 10.1364/optica.386613.
[6] S. Bandyopadhyay *et al.*, “Single chip photonic deep neural network with accelerated training,” Aug. 2022, Accessed: May 21, 2024. [Online]. Available: https://arxiv.org/abs/2208.01623v1.
[7] M. A. Nahmias, T. F. De Lima, A. N. Tait, H. T. Peng, B. J. Shastri, and P. R. Prucnal, “Photonic Multiply-Accumulate Operations for Neural Networks,” *IEEE J. Sel. Top. Quantum Electron.*, vol. 26, no. 1, Jan. 2020, doi: 10.1109/JSTQE.2019.2941485.
[8] J. Feldmann *et al.*, “Parallel convolutional processing using an integrated photonic tensor core,” *Nat. 2020 5897840*, vol. 589, no. 7840, pp. 52–58, Jan. 2021, doi: 10.1038/s41586-020-03070-1.
[9] T. Fu *et al.*, “Photonic machine learning with on-chip diffractive optics,” *Nat. Commun. 2023 141*, vol. 14, no. 1, pp. 1–10, Jan. 2023, doi: 10.1038/s41467-022-35772-7.
[10] F. Ashtiani, A. J. Geers, and F. Aflatouni, “An on-chip photonic deep neural network for image classification,” *Nat. 2022 6067914*, vol. 606, no. 7914, pp. 501–506, Jun. 2022, doi: 10.1038/s41586-022-04714-0.
[11] T. W. Hughes, M. Minkov, Y. Shi, and S. Fan, “Training of photonic neural networks through in situ backpropagation and gradient measurement,” *Optica*, vol. 5, no. 7, p. 864, Jul. 2018, doi: 10.1364/optica.5.000864.
[12] L. G. Wright *et al.*, “Deep physical neural networks trained with backpropagation,” *Nat. 2022 6017894*, vol. 601, no. 7894, pp. 549–555, Jan. 2022, doi: 10.1038/s41586-021-04223-6.
[13] M. Milanizadeh *et al.*, “Coherent self-control of free-space optical beams with integrated silicon photonic meshes,” *Photonics Res. Vol. 9, Issue 11, pp. 2196-2204*, vol. 9, no. 11, pp. 2196–2204, Nov. 2021, doi: 10.1364/PRJ.428680.
[14] C. Roques-Carmes, S. Fan, and D. A. B. Miller, “Measuring, processing, and generating partially coherent light with self-configuring optics,” *Light: Science & Applications 13 (1), 260 (2024)*

[15] S. Pai, B. Bartlett, O. Solgaard, and D. A. B. Miller, "Matrix Optimization on Universal Unitary Photonic Devices," *Phys. Rev. Appl.*, vol. 11, no. 6, p. 064044, Jun. 2019, doi: 10.1103/PHYSREVAPPLIED.11.064044/FIGURES/14/MEDIUM.

[16] C. Roques-Carmes *et al.*, "Heuristic recurrent algorithms for photonic Ising machines," *Nat. Commun.*, vol. 11, no. 1, p. 249, 2020, doi: 10.1038/s41467-019-14096-z.

[17] S. Pai *et al.*, "Experimentally realized in situ backpropagation for deep learning in photonic neural networks," *Science (80-. ).*, vol. 380, no. 6643, pp. 398–404, Apr. 2023, doi: 10.1126/SCIENCE.ADE8450/SUPPL_FILE/SCIENCE.ADE8450_MOVIES_S1_AND_S2.ZIP.

[18] C. Haffner *et al.*, "All-plasmonic Mach–Zehnder modulator enabling optical high-speed communication at the microscale," *Nat. Photonics 2015 98*, vol. 9, no. 8, pp. 525–528, Jul. 2015, doi: 10.1038/nphoton.2015.127.

[19] B. Scellier and Y. Bengio, "Equilibrium propagation: Bridging the gap between energy-based models and backpropagation," *Front. Comput. Neurosci.*, vol. 11, p. 246298, May 2017, doi: 10.3389/FNCOM.2017.00024/BIBTEX.

[20] T. W. Hughes, I. A. D. Williamson, M. Minkov, and S. Fan, "Wave physics as an analog recurrent neural network," *Sci. Adv.*, vol. 5, no. 12, Dec. 2019, doi: 10.1126/SCIADV.AAY6946/SUPPL_FILE/AAY6946_SM.PDF.

# Noise in Analog Photonic Neural Networks

**Egor Manuylovich, Morteza Kamalian-Kopae, Pedro Freire, Sergei K. Turitsyn**
AiPT, Aston University, UK
[e.manuylovich@aston.ac.uk, kamalian_morteza@yahoo.com, p.freiredecarvalhosourza@aston.ac.uk, s.k.turitsyn@aston.ac.uk ]

**Status**

The modern expansion of AI across diverse fields goes along with the increasing complexity of the algorithms and hardware solutions. Increasing computational complexity to achieve superior performance makes hardware implementation more challenging, and it directly affects both power consumption and the accumulation of signal processing latency, which is a critical issue in many applications. Power consumption can be potentially reduced using analog neural networks, the performance of which, however, is limited by noise aggregation. Optical computing, more specifically, photonic neural networks (PNNs) implemented on silicon integration platforms, stand out as a promising candidate to endow neural network (NN) hardware, offering the potential for low power consumption, ultra-fast computations exploiting advantages of photonics, i.e., energy efficiency, massive parallelization, THz bandwidth, and low-latency [1].

Noise limitations have been shown to saturate the computing accuracy of PNNs, posing a significant challenge in achieving high tolerance to noise [2]. While noise can be exploited to train NNs [3], there is a need to address the impact of noise on the effective bit resolution during inference, which is a key challenge in designing noise-tolerant PNNs with an optimal trade-off between signal-to-noise ratio (SNR) and power consumption [4, 5, 6]. Developing noise-resilient deep learning PNN layouts has been demonstrated, offering high-speed operation and improved performance [7]. In Ref. [8], a new noise injection approach was introduced leveraging the Bayesian optimization method, enhancing the robustness of analog NNs without any hardware modifications or sacrifice of accuracy. Furthermore, harnessing intrinsic noise sources in the photonic system might offer potential benefits, such as improving diversity in generated images, albeit with a trade-off in fidelity [6]. A new type of NN has been proposed that incorporates stochastic resonance as an inherent part of its architecture [9], demonstrating the potential for increased robustness against the impact of noise.

Other approaches to mitigate noise and increase the classification accuracy in analog networks include: leveraging network-inherent assets to suppress uncorrelated noise, the use of ghost neurons to address correlated noise across populations of neurons, and the pooling of neuron populations as an efficient approach to suppress uncorrelated noise [10]. Further advances in practical implementations of noise mitigation strategies in analog photonic NNs can lead to improved performance, higher resilience, and expanded applications [6]. These challenges highlight the complexity of developing photonic NNs that can effectively tolerate high levels of noise and the need for innovative solutions to address these limitations.

**Current and Future Challenges**

The key research issues in this area revolve around understanding, quantifying and mitigating the impact of noise on the accuracy and performance of analog photonic NNs, as well as developing techniques to enhance the robustness of these networks to analog hardware errors. One of the important challenges is developing noise-resilient networks by increasing the noise tolerance of the trained models, thereby maintaining inference accuracy under noisy analog computation [11]. This also accounts for the noise arising from discrepancies between the physically implemented neural network and its digital twin, which is commonly used during training [12]. Noise-aware training along with the introduction of novel dynamic nodes serve as potential solutions to the issue of noise in analog computers. Another challenge is understanding the fundamental aspects of noise in analog circuits and developing the appropriate theoretical framework describing noise propagation in photonic neuromorphic systems.
[13].

While the first challenge primarily addresses the issue of unwanted noisy input data, the second focuses on noise originating from the analog computer itself. These two sources of noise and uncertainty differently affect system performance. Nevertheless, designing noise-tolerant APNNs requires a comprehensive understanding of both the randomness of the input data and the uncertainty introduced by the hardware design and operation.

In general, the area shares multiple challenges with other optical computing fields, due to the relative immaturity of technology for implementing nonlinear activation functions optically and the difficulty in controlling analog weights [14]. Similarly, neuromorphic photonics encounters problems in implementing all-optical nonlinear transfer functions without high optical power and in achieving efficient nonlinear processing and data storage. Photonic implementations of nonlinear functions could be one of the potential solutions to the noise mitigation problem as some nonlinear functions such as tanh and sigmoid show potential for binarizing the input signal, which could mitigate the impact of noise similar to the regeneration of digital electronic signals.

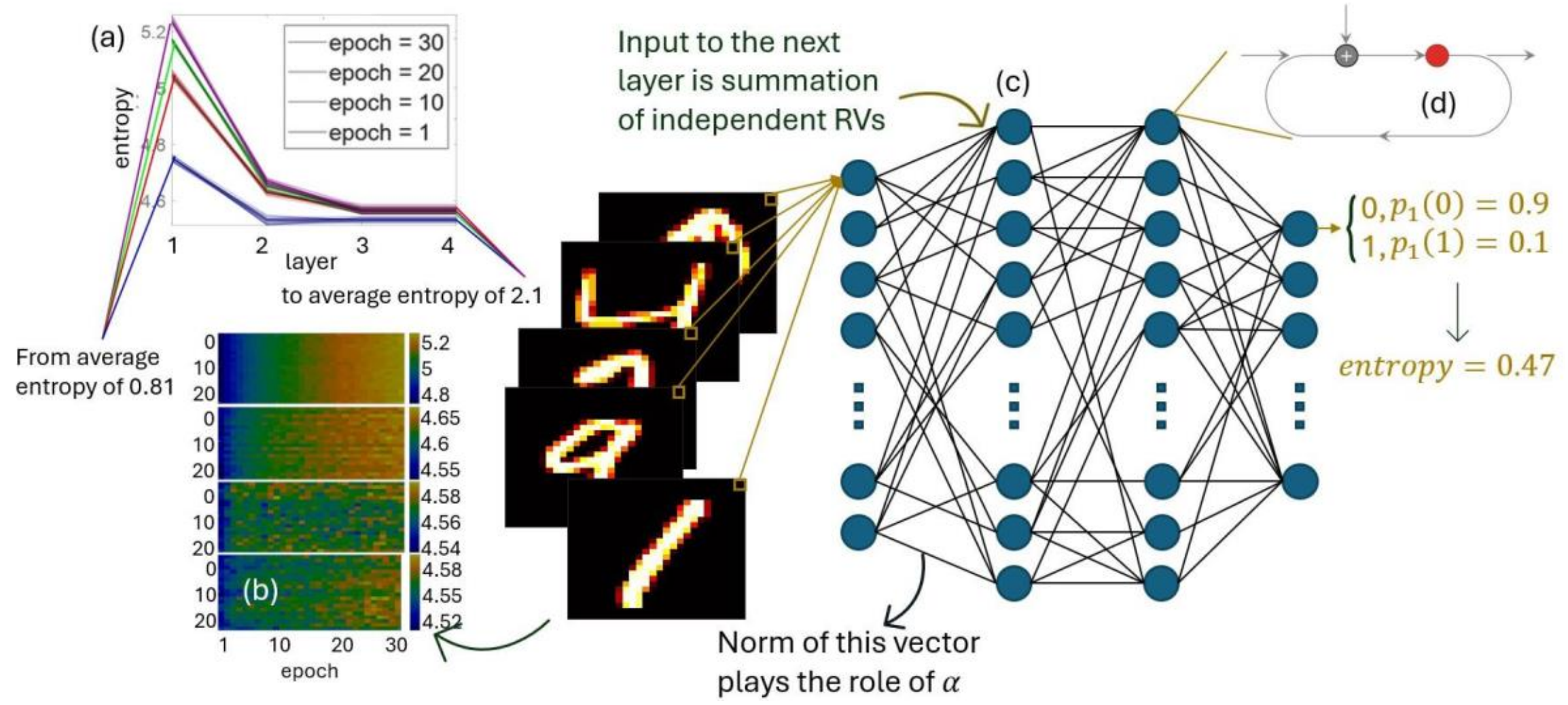

**Figure 1**: a) evolution of the average entropy (across nodes) for all layers and for various epochs, b) entropy of all nodes across epochs, c) the ANN structure with 4 layers of 20 tanh nodes, and d) the model for each nonlinear node including the additive noise. The task is to classify the hand-written digits of the MNIST dataset. Numerical differential entropy is used, and the schematic shown for the nonlinear mode is a general implementation of a dynamic activation function with memory, nonlinearity and intrinsic additive noise.

## Advances in Science and Technology to Meet Challenges

An important step in advancing our understanding of the impact of noise in analog ANNs and devising strategies to mitigate it is quantifying the information flow in the networks. Considering input features and noise as random elements in analog PNNs, entropy can be used to quantify noise impact. The signal's entropy flows in two directions: from the input and feature-generating layers to the output layer, and along the training process, following gradient descent (Fig. 1). At any epoch, entropy trajectory can be derived as a function of network weights and the nonlinear activation function. This derivation has two fixed boundary conditions: input data uncertainty and output layer uncertainty, varying by problem. Our observations suggest training converges towards weight characteristics that maximize entropy.

A prevalent method for estimating the impact of noise on predictive uncertainty involves differentiating between uncertainties attributable to the model (epistemic or model uncertainty) and those due to the data (aleatoric or data uncertainty) [15]. The former can be reduced by refining the model implemented in the NN, whereas the latter is inherently irreducible. We believe that using bi-stable optical elements capable of producing regenerative functions might benefit future designs of analog PNNs. Of particular interest could be the application of the so-called Stochastic Resonance (SRs) [16, 9], extensively studied across various physical systems, including climate modeling, electronic circuits, neural models, and chemical reactions [16]. SR is typically observed in nonlinear systems where noise significantly influences one of the characteristic time scales [17]. In such systems, the input noise level can be adjusted to optimize the signal-to-noise ratio (SNR), enhancing system performance under specific noisy conditions. The dynamics of SR are commonly modeled using a bi-stable system, which receives two types of inputs: a coherent signal and random noise [18, 19]:
(ξ(t) )̇ =-(dU_0 (ξ))/dξ+s(t)+σN(t) (1)

Here s(t) represents the input signal that will be transformed by the dynamical system into the output signal ξ(t), σ is the Gaussian noise amplitude. Figure 2 shows the shape of the bi-stable SR potential U0(ξ)−ξs(t) depending on the input signal of the SR node and the corresponding evolution of the internal state and the transfer function of the node.

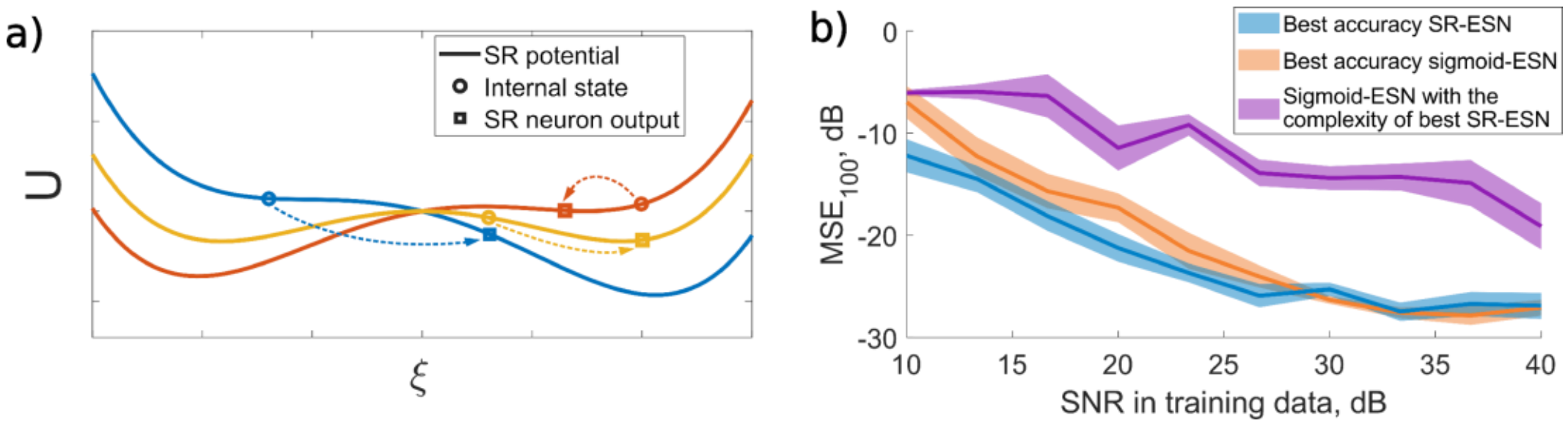

**Figure 2**: a) The landscape of SR system $U(\xi,t) = U_0(\xi) - \xi s(t)$ and evolution of internal state $\xi$ depending on input signal s(t); and b) benefit of using SR activation function when training ESNs on noisy data.

In Ref. [9], the advantage of using SRs at nonlinear nodes within an echo state network (ESN) was demonstrated. When subjected to training with noisy data, the proposed methodology outperforms conventional sigmoid-based ESN models in both accuracy and computational efficiency, highlighting the potential of integrating SR into nonlinear node architectures.

## Concluding Remarks

The progress in analog photonic neural networks, despite their considerable advantages over conventional counterparts, faces a critical challenge: noise mitigation. Whether originating from data contamination or inherent to the computing hardware, noise must be effectively managed during the training phase of any artificial neural network and later during the inference phase. This necessitates a comprehensive understanding of noise's effects on various network characteristics. Such insights can then inform the development of noise-mitigating nodes, such as, e.g., stochastic resonance, where noise power influences the mapping of inputs to stable outputs with enhanced SNR. In addition to investigating the SNR, further exploration will focus on entropy as a more comprehensive metric for quantifying the robustness and performance of analog photonic neural networks. This holistic approach acknowledges that robustness extends beyond SNR alone, encompassing the network's capacity to manage uncertainty and complexity effectively. Further research is imperative to explore the photonic realization of the regenerative dynamic nodes in the PNNs and to tailor noise models to accommodate the intricacies of these innovative designs.

## Acknowledgements
This is a contribution of NIST, an agency of the U.S. government, not subject to copyright.

## References

[1] A. Tsakyridis, M. Moralis-Pegios, G. Giamougiannis, M. Kirtas, N. Passalis, A. Tefas, and N. Pleros, "Photonic neural networks and optics-informed deep learning fundamentals," *APL Photonics*, vol. 9, no. 1, 2024.

[2] C. Huang, V. J. Sorger, M. Miscuglio, M. Al-Qadasi, A. Mukherjee, L. Lampe, M. Nichols, A. N. Tait, T. Ferreira de Lima, B. A. Marquez, *et al.*, "Prospects and applications of photonic neural networks," *Advances in Physics: X*, vol. 7, no. 1, p. 1981155, 2022.

[3] M. Zhou, T. Liu, Y. Li, D. Lin, E. Zhou, and T. Zhao, "Toward understanding the importance of noise in training neural networks," in *International Conference on Machine Learning*, pp. 7594– 7602, PMLR, 2019.

[4] K. Liao, T. Dai, Q. Yan, X. Hu, and Q. Gong, "Integrated photonic neural networks: Opportunities and challenges," *ACS Photonics*, 2023.

[5] T. F. de Lima, A. N. Tait, H. Saeidi, M. A. Nahmias, H.-T. Peng, S. Abbaslou, B. J. Shastri, and P. R. Prucnal, "Noise analysis of photonic modulator neurons," *IEEE Journal of Selected Topics in Quantum Electronics*, vol. 26, no. 1, pp. 1–9, 2019.

[6] C. Wu, X. Yang, H. Yu, R. Peng, I. Takeuchi, Y. Chen, and M. Li, "Harnessing optoelectronic noises in a photonic generative network," *Science Advances*, vol. 8, no. 3, p. eabm2956, 2022.

[7] G. Mourgias-Alexandris, M. Moralis-Pegios, A. Tsakyridis, S. Simos, G. Dabos, A. Totovic, N. Passalis, M. Kirtas, T. Rutirawut, F. Gardes, *et al.*, "Noise-resilient and high-speed deep learning with coherent silicon photonics," *Nature Communications*, vol. 13, no. 1, p. 5572, 2022.

[8] N. Ye, L. Cao, L. Yang, Z. Zhang, Z. Fang, Q. Gu, and G.-Z. Yang, "Improving the robustness of analog deep neural networks through a bayes-optimized noise injection approach," *Communications Engineering*, vol. 2, no. 1, p. 25, 2023.

[9] E. Manuylovich, D. A. Ron, M. Kamalian-Kopae, and S. Turitsyn, "Stochastic resonance neurons in artificial neural networks," *arXiv preprint arXiv:2205.10122*, 2022.

[10] N. Semenova and D. Brunner, "Noise-mitigation strategies in physical feedforward neural networks," *Chaos: An Interdisciplinary Journal of Nonlinear Science*, vol. 32, no. 6, 2022.

[11] C. Zhou, P. Kadambi, M. Mattina, and P. N. Whatmough, "Noisy machines: Understanding noisy neural networks and enhancing robustness to analog hardware errors using distillation," *arXiv preprint arXiv:2001.04974*, 2020.

[12] L. G. Wright, T. Onodera, M. M. Stein, T. Wang, D. T. Schachter, Z. Hu, and P. L. McMahon, "Deep physical neural networks trained with backpropagation," *Nature*, vol. 601, no. 7894, pp. 549–555, 2022.

[13] N. Semenova, J. Moughames, X. Porte, M. Kadic, L. Larger, and D. Brunner, "Scalability and noise in (photonic) hardware neural networks," in *AI and Optical Data Sciences II*, vol. 11703, p. 117030I, SPIE, 2021.

[14] F. Morichetti, "Grand challenges in neuromorphic photonics and photonic computing," 2024.

[15] J. Gawlikowski, C. R. N. Tassi, M. Ali, J. Lee, M. Humt, J. Feng, A. Kruspe, R. Triebel, P. Jung, R. Roscher, *et al.*, "A survey of uncertainty in deep neural networks," *Artificial Intelligence Review*, vol. 56, no. Suppl 1, pp. 1513–1589, 2023.

[16] N. B. Harikrishnan and N. Nagaraj, "When noise meets chaos: Stochastic resonance in neurochaos learning," *Neural Networks*, vol. 143, pp. 425–435, 2021.

[17] V. S. Anishchenko, A. B. Neiman, F. Moss, and L. Shimansky-Geier, "Stochastic resonance: noise-enhanced order," *Physics-Uspekhi*, vol. 42, no. 1, p. 7, 1999.

[18] H. N. Balakrishnan, A. Kathpalia, S. Saha, and N. Nagaraj, "Chaosnet: A chaos based artificial neural network architecture for classification," *Chaos: An Interdisciplinary Journal of Nonlinear Science*, vol. 29, no. 11, p. 113125, 2019.

[19] G. P. Harmer, B. R. Davis, and D. Abbott, "A review of stochastic resonance: Circuits and measurement," *IEEE Transactions on Instrumentation and Measurement*, vol. 51, no. 2, pp. 299– 309, 2002.

# Noise-aware Training of Photonic Generative Adversarial Networks

**Changming Wu[1] and Mo Li[1,2]**

cmwu@uw.edu
moli96@uw.edu

[1]*Department of Electrical and Computer Engineering, University of Washington, Seattle, WA 98195, USA*
[2]*Department of Physics, University of Washington, Seattle, WA 98195, USA*

## Status

Optical computing promises significant advantages in power efficiency, parallelism, and computational speed [1]. Numerous photonic neural network platforms have been developed, utilizing either free-space optics with diffractive optical elements[2], [3] or integrated silicon photonics with coherent[4] or incoherent light[5], [6]. However, the most notable achievements in photonic neural networks to date have predominantly involved discriminative models[7], [8], [9], which classify high-dimensional, rich sensory inputs. Typically, these photonic neural network systems operate offline: training occurs on a digital computer, and the learned parameters are subsequently transferred to the photonic processors for inference tasks.

Compared to electronic neural networks that employ state-of-the-art digital processors, the analog nature of optical computing with photonic processors could potentially limit their applications[10]. Accumulated computational random errors can significantly degrade performance. Furthermore, these errors during data processing diminish the tolerable error margin, particularly when photonic processors exhibit high precision. Although the capacity for training within current photonic neural networks is substantial, their inference accuracy remains inferior to that of conventional digital processors unless post-digital signal processing is incorporated, even for simple tasks. Consequently, enhancing error tolerance and system reliability under model perturbations is critical for advancing photonic neural network technology.

## Current and Future Challenges

Inferencing on photonic computing hardware is currently constrained to the neural network's forward path, with physical noises affecting inferencing accuracy while leaving the backward path and weight updates unaffected. An on-chip, in-situ training algorithm that adapts to system imperfections could mitigate these issues. However, training directly on photonic processors is a complex task [11], [12]. While training an arbitrary physical system as a neural network has been explored, even minor system mismatches can severely impair training outcomes [13]. Under such circumstances, finding neural network models that can fully leverage these random errors could broaden the application scope of photonic neural networks.

A generative adversarial network (GAN) is a good example showing how photonic processors can effectively utilize random hardware imperfections and physical noises in practice. A GAN comprises two competing sub-networks: a generator and a discriminator (see Fig. 1). These models engage in a zero-sum game where the discriminator tries to differentiate between "fake" instances produced by the generator and "real" instances from the training dataset. Conversely, the generator strives to deceive the discriminator by creating new instances that mimic the real ones. This competitive dynamic enhances both networks' capabilities until they reach a state of equilibrium. Distinct from previously discussed discriminative models, the uniqueness of GANs stems from their use of stochastic random noise as input. This approach means that inherent random errors in the computing kernel augment the noise from the input vector. Additionally, the network leverages this noise to generate diverse patterns designed to trick the discriminator.

**Advances in Science and Technology to Meet Challenges**

Wu et al. explored photonics-based generative adversarial networks (GANs), introducing a weight compensatory training method that effectively leverages noise to produce diverse handwritten image patterns [14]. The GAN's generator incorporates a photonic processor consisting of an array of programmable phase-change mode converters (PMMC), functioning as a tensor core [7]. This phase-change-based photonic GAN follows an offline training configuration, where the inference task is executed only after all pre-trained parameters have been transferred to the physical devices, which is a process that inherently introduces programming errors. When input signals advance through each layer, they are processed by the photonic tensor core and converted into the electrical domain after photodetection. Subsequent postprocessing reconvenes the data into the optical domain for transmission to the next layer, further introducing implementation errors. Harnessing the errors, Wu et al. employed the weight compensatory training method, as illustrated in Fig. 1a, which anticipates the error distribution to follow a Gaussian pattern with a specified standard deviation of noise integrated during the forward training path.

Experimental results, highlighted in Fig. 1c, demonstrate the efficacy of this noise-aware training and noisy inference approach over traditional noiseless methods in terms of image quality and diversity. A comparison of the handwritten number "7" shows that images produced by the noise-aware trained GAN exhibit distinct handwritten features with clear outlines, whereas those from the noiseless approach display noisy backgrounds. These findings emphasize that the noise-influenced weight-compensatory GAN, built on photonic hardware, delivers superior inference performance compared to its noiseless counterparts. Additionally, it is noted that the inference accuracy of discriminative networks deteriorates with noisier hardware [15], marking a stark contrast in performance dynamics compared to GAN. Despite unavoidable optoelectronic disturbances and faults, this improved performance highlights the potential of photonic neural networks in generative models.

Surprisingly, the image quality achieved using non-ideal hardware even surpasses that from ideal (error-free software baseline) setups, demonstrating the potential of conducting more complex inferences on photonic GANs using non-ideal analog photonic computing hardware, such as PMMC devices. To quantitatively evaluate GAN performance, the Frechet Inception Distance (FID) was calculated, which evaluates both the fidelity and diversity of generated images by comparing their feature distribution to that of images from the training dataset. A lower FID score indicates better GAN performance. As shown in Fig.2, the FIDs of patterns generated by an ideal, error-free processor (trained using a noise-free approach) are higher compared with those from a practical processor with errors (trained using a noise-aware approach), illustrating that the noise-aware trained GAN consistently outperforms the noise-free approach across all tested error levels. Notably, even at a practical error level of 5% (shown in Fig. 2), the FID for the GAN trained using the noise-aware approach on a practical photonic processor remains below the FID from the software baseline, which assumes the operation of a noise-free approach trained GAN on an ideal, errorless photonic processor.

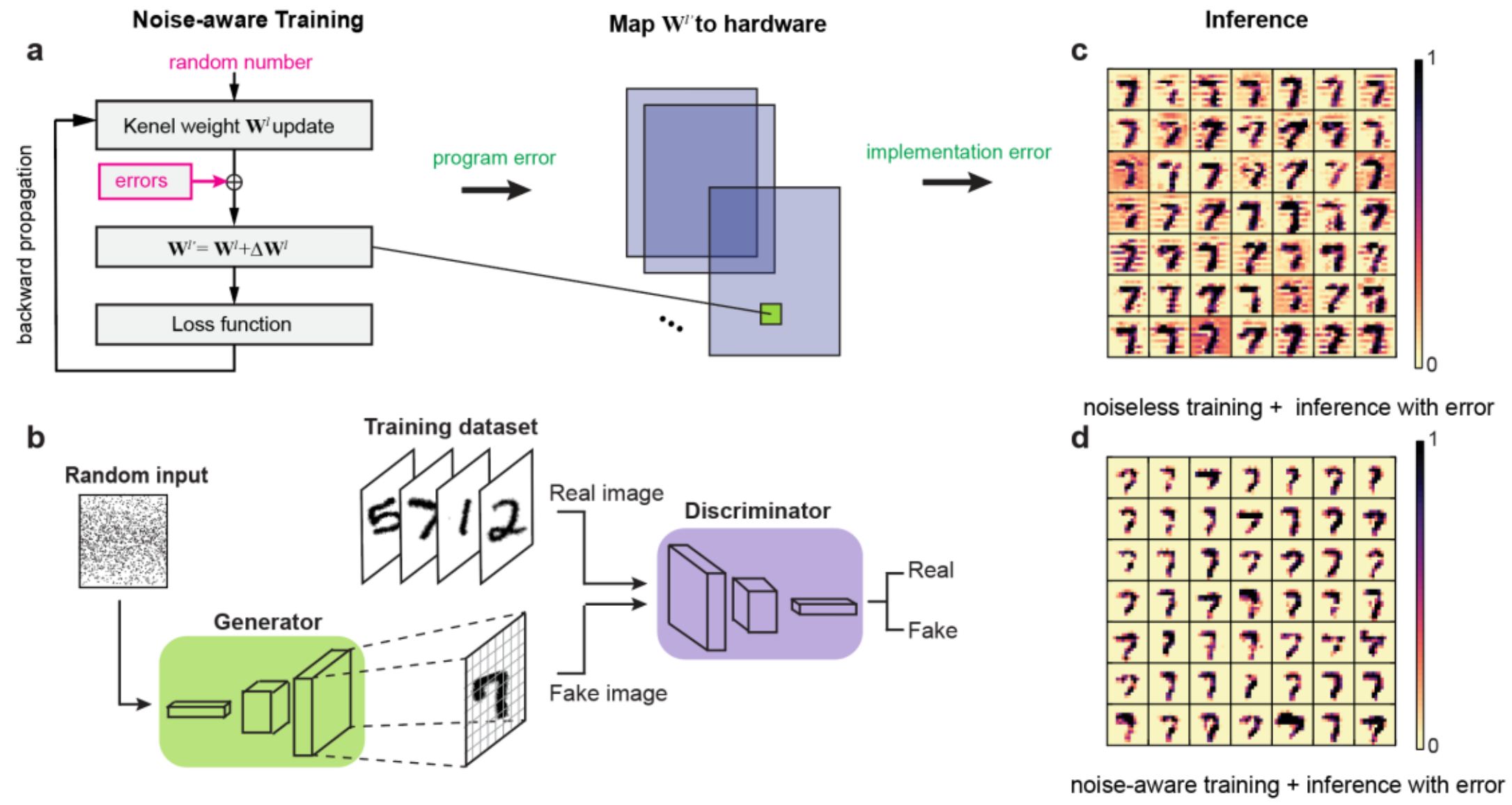

**Figure 1. a.** The offline noise-aware training and inference processes flow of the generator. The process of mapping the trained weight to the hardware during implementation inevitably introduces errors. **b.** A GAN architecture is composed of two sub-network models, a generator, and a discriminator. **c** and **d.** Handwritten images generated by the GAN (c) trained using noiseless approach and (d) trained using noise-aware approach, while perform inference using practical photonic processors with errors. Reproduced with permission. Copyright 2022, AAAS.

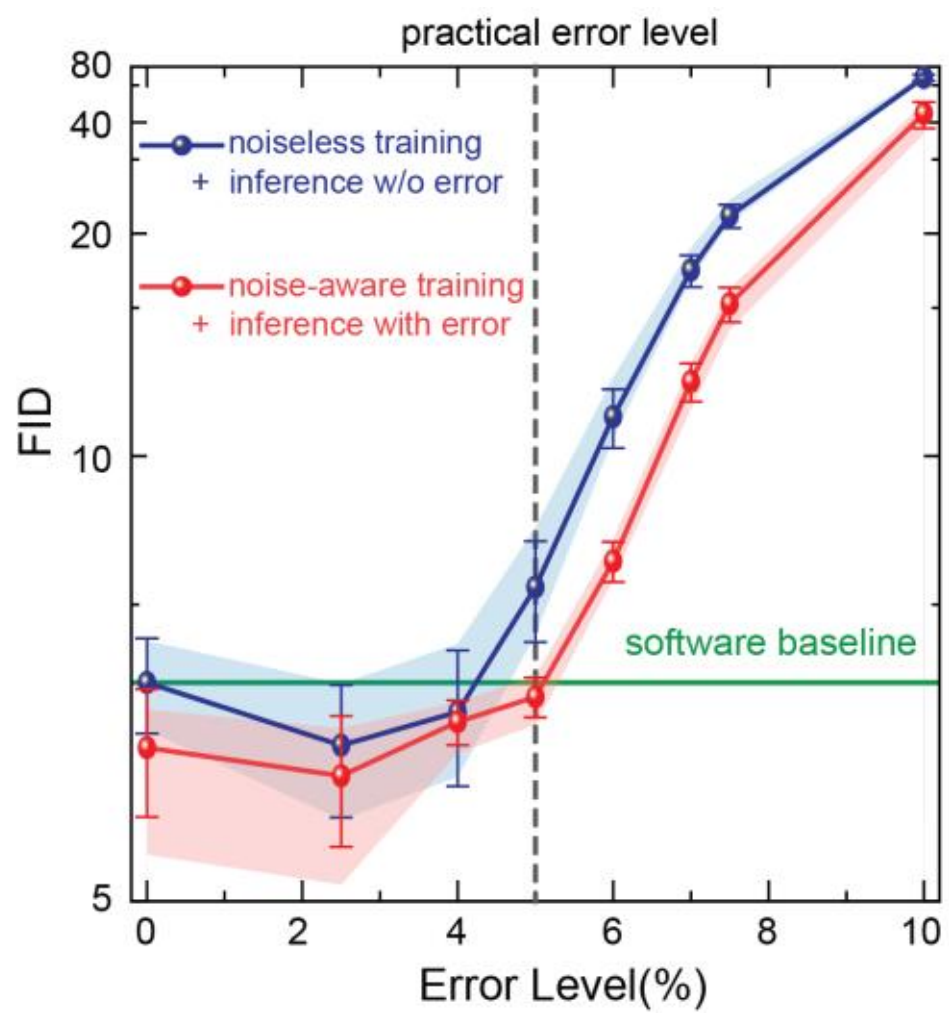

**Figure 2.** The FID of the generated images by the ideal, error-free processor (trained using a noise-free approach) and the practical processor with errors (trained using a noise-aware approach), respectively, under various error level ranging from 0% to 10%. Reproduced with permission. Copyright 2022, AAAS.

## Concluding Remarks

In conclusion, here we highlight a photonic GAN as a prime example of how error tolerance can be enhanced under model perturbations, thereby advancing the utility of photonic processors for computing tasks. Unlike traditional discriminative networks, which are adversely affected by hardware errors, a properly trained GAN can surpass the performance of ideal processors. The results not only demonstrate the capability of non-ideal analog photonic computing hardware to handle more complex inferences in photonic GANs but also highlight a significant advantage: such hardware can surpass the performance of traditional digital computing processors and even that of theoretically perfect processors. This underscores the unique benefits and opportunities of employing photonic computing processors in practical applications, where they can leverage inherent operational imperfections to achieve superior outcomes.

This finding extends the application of photonic neural networks to generative models, demonstrating that errors can not only be mitigated but can also be strategically harnessed. The noise-aware training approaches introduced here are versatile, making them applicable to a wide range of optoelectronic neuromorphic computing platforms and schemes. Moreover, the enhanced noise resilience of these models suggests their potential for scalability in large-scale photonic neural networks, where electronics and photonics are tightly integrated.

**Acknowledgments**
The authors acknowledge the funding support provided by ONR MURI (award no. N00014-17-1-2661) and the NSF (award no. CCF-2105972).

**References**

[1] P. L. McMahon, "The physics of optical computing," *Nature Reviews Physics*, vol. 5, no. 12, pp. 717–734, 2023.
[2] X. Lin *et al.*, "All-optical machine learning using diffractive deep neural networks," *Science (1979)*, vol. 361, no. 6406, pp. 1004–1008, 2018.
[3] T. Zhou *et al.*, "Large-scale neuromorphic optoelectronic computing with a reconfigurable diffractive processing unit," *Nat Photonics*, vol. 15, no. 5, pp. 367–373, 2021.
[4] Y. Shen *et al.*, "Deep learning with coherent nanophotonic circuits," *Nat Photonics*, vol. 11, no. 7, pp. 441–446, 2017.
[5] A. N. Tait *et al.*, "Neuromorphic photonic networks using silicon photonic weight banks," *Sci Rep*, vol. 7, no. 1, p. 7430, 2017.
[6] C. Huang *et al.*, "A silicon photonic–electronic neural network for fibre nonlinearity compensation," *Nat Electron*, vol. 4, no. 11, pp. 837–844, 2021.
[7] C. Wu, H. Yu, S. Lee, R. Peng, I. Takeuchi, and M. Li, "Programmable phase-change metasurfaces on waveguides for multimode photonic convolutional neural network," *Nat Commun*, vol. 12, no. 1, p. 96, 2021.
[8] H. Zhang *et al.*, "An optical neural chip for implementing complex-valued neural network," *Nat Commun*, vol. 12, no. 1, p. 457, 2021.
[9] F. Ashtiani, A. J. Geers, and F. Aflatouni, "An on-chip photonic deep neural network for image classification," *Nature*, vol. 606, no. 7914, pp. 501–506, 2022.
[10] X. Yang, C. Wu, M. Li, and Y. Chen, "Tolerating Noise Effects in Processing-in-Memory Systems for Neural Networks: A Hardware–Software Codesign Perspective," *Advanced Intelligent Systems*, vol. 4, no. 8, p. 2200029, 2022.
[11] T. Zhou *et al.*, "In situ optical backpropagation training of diffractive optical neural networks," *Photonics Res*, vol. 8, no. 6, pp. 940–953, 2020.
[12] S. Pai *et al.*, "Experimentally realized in situ backpropagation for deep learning in photonic neural networks," *Science (1979)*, vol. 380, no. 6643, pp. 398–404, 2023.
[13] L. G. Wright *et al.*, "Deep physical neural networks trained with backpropagation," *Nature*, vol. 601, no. 7894, pp. 549–555, 2022.
[14] C. Wu *et al.*, "Harnessing optoelectronic noises in a photonic generative network," *Sci Adv*, vol. 8, no. 3, p. eabm2956, 2022.
[15] V. Joshi *et al.*, "Accurate deep neural network inference using computational phase-change memory," *Nat Commun*, vol. 11, no. 1, p. 2473, 2020.

# APPLICATIONS

## Image sensing with optical neural networks

**Mandar M. Sohoni[1], Logan G. Wright[2], Tianyu Wang[3], Peter L. McMahon[2,4]**
[1] School of Applied and Engineering Physics, Cornell University, Ithaca, NY 14853, USA
[2] Department of Applied Physics, Yale University, New Haven, CT 06511, USA
[3] Department of Electrical and Computer Engineering, Boston University, Boston, MA 02215, USA
[4] Kavli Institute at Cornell for Nanoscale Science, Cornell University, Ithaca, NY 14853, USA

E-mails: mms477@cornell.edu logan.wright@yale.edu wangty@bu.edu pmcmahon@cornell.edu

### Status

Optical images are one of the most established tools used to encode information about properties of physical objects and systems. In image sensing, an inference is made about an entity, or entities, within an optical image. Conventionally, this is done by running an algorithm on a digitized version of the optical image. Artificial neural networks (ANNs) have rapidly become the gold standard to extract information from these optical images [1, 2] and have hence become the backbone of computer vision. A typical computer vision pipeline involves an imaging system (such as a lens and a camera sensor) and an image-processing algorithm (such as an ANN). Generally, the imaging system tries to capture all of the optical information, even if irrelevant for the image's end use, which can lead to inefficiencies in imaging resources such as the number of pixels, or photons to be collected. For example, consider the scenario in Figure 1a - an image of the speed limit sign contains thousands of pixels, but less than ten pixels' worth of information is needed to encode the actual speed limit, a data recording inefficiency greater than two orders of magnitude! The development of alternate image-sensing paradigms which attempt to re-envision the imaging system as an encoder, i.e., a pre-processor that extracts relevant information, have seen some success in circumventing these inefficiencies. Examples of such techniques include active perception [3], compressed sensing [4, 5], end-to-end optimization [6], and in-sensor computing [7, 8, 9].

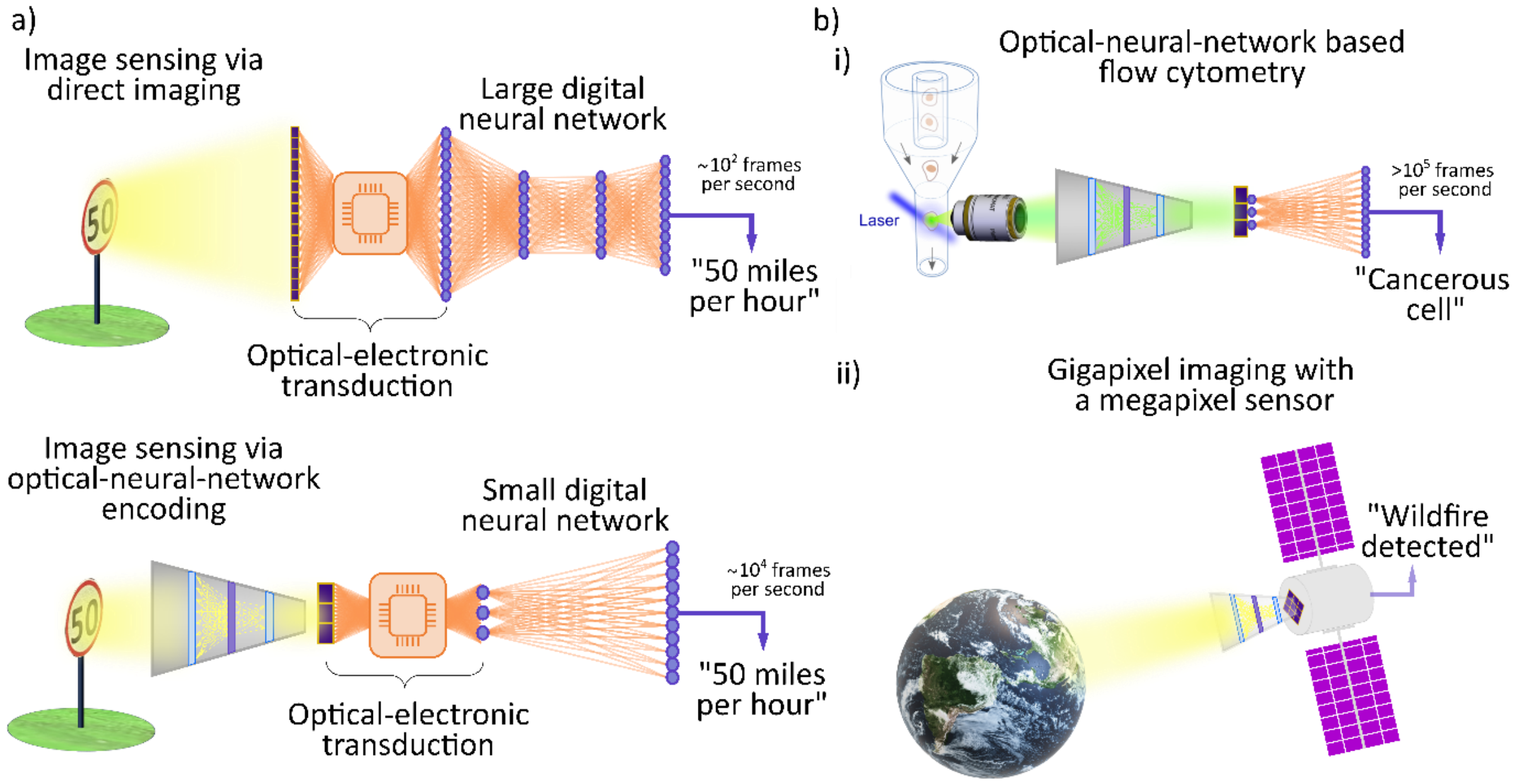

**Figure 1. Image sensing via direct imaging versus optical encoding**. **a.** In conventional image sensing, an image is collected by a camera and then processed, often using a neural network to extract a small piece of relevant information such as the text of a sign. There is an alternative approach: rather than reproduce the full image of a scene on a sensor array which contains irrelevant information, an ONN encoder can instead pre-process the image, compressing and extracting only the image information necessary for its end use, allowing a much smaller (fewer pixel) sensor array. By using a smaller number

of pixels, and so reducing the bandwidth bottleneck between optics and electronics, the overall frame rate of the image-sensing system can be higher. Figure adapted from Ref. [10]. **b.** Potential applications of ONN-based image sensors. Panel (i) shows a potential application of ONN-based flow cytometry. An ONN preprocesses images of cells and the compressed information is used to decide whether the cell should be retained or discarded. The compression of cell-image information will potentially allow very low-latency inferences. Panel (ii) shows another potential application of ONN image sensors in satellite imagery. The ONN image sensor scans over a large field-of-view, potentially up to a resolution of a few gigapixels, and compresses the optical information to bandwidths current cameras can handle.

Through the lens of deep learning, most, if not all of these optical encoders can be thought of as single- or multi-layer optical neural networks (ONNs). Thus, we argue that a compelling application of ONNs is developing ONN image sensors [10]. Instead of accelerating computation on data originating in the digital domain, ONN image sensors perform sophisticated optical processing on data originating in the optical domain which allows them to provide advantages in sensing performance (see Figure 1b). A few examples of possible advantages are: (1) very low-latency inferences enabled by optimized feature extraction [11], (2) optical-domain computation that is impractical to perform after digitization (such as hyperspectral feature extraction [12, 13]), and (3) optimized downsampling of high-resolution images [14], potentially up to a few gigapixels. Broadly speaking, most sensors are starting to generate more data than can efficiently be stored and processed [15]. ONNs, and more generally physical neural networks, seem aptly positioned to address this issue. Given the myriad of ONN platforms, there appear to be ample opportunities to explore novel ONN sensors that optimally extract information from light's spatial, spectral, and temporal degrees of freedom.

## Current and Future Challenges

### *A suitable nonlinear activation function*

A neural network's nonlinear activation functions allow it to have depth, which consequently enables hierarchical feature extraction - an important prerequisite for the conditional compression of image data. One of the most prominent challenges in the field of ONNs, and ONN image sensors by extension, has been developing a scalable, low-energy, and low-latency optical (or opto-electronic) nonlinear activation function. While there have been many promising proof-of-concept demonstrations of nonlinear activation functions to enable multi-layer ONNs for machine-learning acceleration purposes [16, 17], field-ready ONN image sensors will require a nonlinear activation function that has the following unique properties:

1. The ability to operate with broadband, incoherent light. Although ONN image sensors for applications involving coherent illumination (e.g. LIDAR) can be imagined [18], in most image sensing applications objects are illuminated by broadband, incoherent light.
2. The ability to provide gain. Even the most well-designed linear optical layers suffer from some loss, and gain is required in order to achieve deep networks while maintaining a certain amount of precision.
3. The ability to support highly multimodal operation since one of the major benefits of ONN image sensors is their ability to distill information from a large number of spatial modes.
4. The ability to operate at speeds greater than a few megahertz.

With an ideal nonlinearity, we believe that ONN image sensors should be able to achieve a continuous sampling rate that is either impossible or extremely difficult for any conventional machine vision system to achieve.

### *Developing and training large scale ONNs*

Generally, ANNs designed for sophisticated image processing always downsample high-resolution images to reduce computational complexity. To paraphrase R. Hesse et al., this strategy comes with the drawback of sacrificing granularity in the produced feature maps, thus constraining the networks' capacities to accurately comprehend intricate details in dense prediction tasks [19]. As mentioned earlier, one of the most alluring aspects of optical hardware is its ability to handle high-dimensional data with relative ease. One can imagine end-to-end optimized ONN image sensors that compress gigapixel images to the required input sizes for state-of-the-art ANNs, thereby greatly increasing their efficiency and performance.

The question then arises: How does one design and train an ONN image sensor that can effectively compress gigapixel images, or more generally, optical information contained in a large number of spatial, spectral, and temporal modes? Most ONNs are either trained *in silico* [20], with finite-difference-like methods [17], or with hybrid optical-forward-pass-digital-backward-pass algorithms [21]. While these have been quite successful, they rely on having a faithful digital model of the optical encoder (with the exception of finite-difference-like methods) which becomes increasingly difficult as the model size increases. The development of novel, digital-model-free training algorithms for large and/or deep ONNs is a hurdle that needs to be overcome before the potential benefits of ONN image sensors can truly be explored.

***Optimizing the spatial footprint of ONN image sensors***

Arguably, most ONN image sensors will be deployed on the edge, for example in unmanned devices such as satellites or drones. In cases like these, ONN image sensors need to be as compact as possible. One approach that minimizes the spatial footprint of ONN image sensors is the use of metasurfaces [22, 23, 24, 25] and diffractive elements [26] instead of conventional optical matrix-vector multipliers. However, even with these sorts of improvements, there is a trade-off between compactness and the complexity of mathematical operations (such as connectivity) that optical systems can implement [27, 28]. This this is not very deleterious since the first few layers of image-processing neural networks typically extract local features of images, but a major challenge in the field remains the design of space-efficient ONN image sensors that retain the benefits of optical pre-processing.

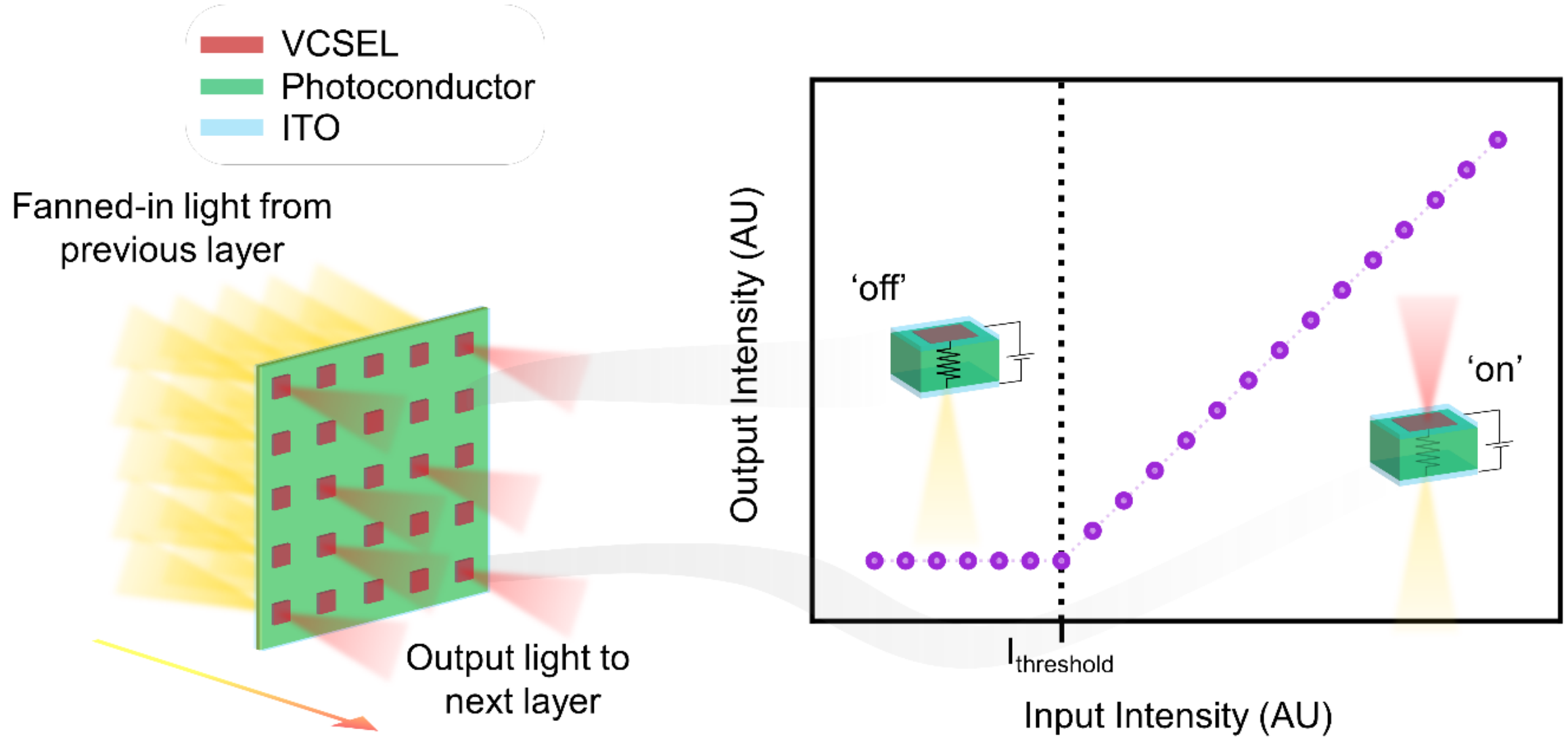

**Figure 2. High-level schematic of an example device realizing a high-speed optical-to-optical non-linear activation function.** Consider an activation function implemented through a simple circuit consisting of a photoconductor, an array of vertical-cavity surface-emitting lasers (VCSELs), and two layers of Indium Tin Ox-ide (ITO). For an individual neuron, when the intensity from the previous optical layer is below a certain level, the current supplied to the VCSEL is below the lasing threshold and it does not emit any light. However, when the intensity is large enough, the VCSEL begins lasing. This results in a nonlinear relation between the input and output light intensity, in this case a rectified linear unit. Similar device concepts involving light-emitting diodes and transistors instead of VCSELs, and photodiodes instead of a photoconductor may also provide plausible solutions.

## Advances in Science and Technology to Meet Challenges

There have been multiple proof-of-concept demonstrations of performing energy-efficient and fast linear computations in devices that have moderately-sized input dimensions [29]. However, from the perspective of a commercially viable product, they still leave a lot to be desired. Questions such as "How does one perform high-fidelity, large-scale linear optical operations in the smallest volume possible?" and "How does the lack of real-valued weights affect ONN image sensors when dealing with intensity-encoded, incoherent-light inputs?" need to be clearly answered before ONN image sensors can even hope to become market ready.

As for the development of a suitable nonlinear activation function that will enable deep ONN image sensors, we argue that a promising research direction is the development of optical-to-optical nonlinear activation (OONA) functions that involve local analog electronic processing [30, 31]. Such devices consist of three main components: (1) high-speed photodetectors, (2) local analog-electronic circuits to provide gain and, if required, nonlinearity, and (3) emitters such as light-emitting diodes (LEDs) or vertical-cavity surface-emitting lasers (VCSELs). As an example of such a device, consider an activation function implemented through a simple circuit consisting of a photoconductor and an array of VCSELs, as shown in Figure 2. When the intensity from the previous optical layer is below a certain level, the current supplied to the VCSEL is below the lasing threshold and it does not emit any light. However, when the intensity is large enough, the VCSEL begins lasing. This results in a nonlinear relation between the input and output light power, in this case a rectified linear unit. Opto-electronic devices similar to those mentioned above are already being developed commercially [32], albeit for the purpose of chip-to-chip communication. With little change, it is likely they could be repurposed to design OONAs.

**Concluding Remarks**

Data processing and compression in the optical domain is important in optical sensing for the following reasons: (1) there is an information-transfer bottleneck between the optical and electronic domains, and (2) many applications of imaging ultimately have the goal to perform an inference about entities in the optical image, thus eliminating the need to record and store the entirety of the optical data. ONNs provide a natural way to create optical encoders: an ONN at the start of an image-sensing pipeline allows sophisticated data processing to be performed in the optical domain, which can alleviate resolution and speed limits arising from the optical-electronic bottleneck.
Looking to the future, ONN image sensors could find applications in scenarios where desired information requires extremely high resolution, such as astronomy or satellite imaging, in high-speed control scenarios such as plasma stabilization or missile defence applications, and in situations where power and digital compute are scarce such as unmanned aerial vehicles, satellites, and other devices on the edge.

**References**

[1] G. E. Hinton and R. R. Salakhutdinov, "Reducing the dimensionality of data with neural networks," Science, vol. 313, p. 504–507, 2006.
[2] A. Krizhevsky, I. Sutskever and G. E. Hinton, "ImageNet Classification with Deep Convolutional Neural Networks," in Advances in Neural Information Processing Systems, 2012.
[3] R. Bajcsy, "Active perception," Proceedings of the IEEE, vol. 76, p. 966–1005, 1988.
[4] M. F. Duarte, M. A. Davenport, D. Takhar, J. N. Laska, T. Sun, K. F. Kelly and R. G. Baraniuk, "Single-pixel imaging via compressive sampling," IEEE signal processing magazine, vol. 25, p. 83–91, 2008.
[5] N. Antipa, G. Kuo, R. Heckel, B. Mildenhall, E. Bostan, R. Ng and L. Waller, "DiffuserCam: lensless single-exposure 3D imaging," Optica, vol. 5, p. 1–9, 2018.
[6] G. Wetzstein, A. Ozcan, S. Gigan, S. Fan, D. Englund, M. Soljačić, C. Denz, D. A. B. Miller and D. Psaltis, "Inference in artificial intelligence with deep optics and photonics," Nature, vol. 588, p. 39–47, 2020.
[7] P. J. Burt, "Smart sensing within a pyramid vision machine," Proceedings of the IEEE, vol. 76, p. 1006–1015, 1988.
[8] L. Mennel, J. Symonowicz, S. Wachter, D. K. Polyushkin, A. J. Molina-Mendoza and T. Mueller, "Ultrafast machine vision with 2D material neural network image sensors," Nature, vol. 579, p. 62–66, 2020.
[9] F. Zhou and Y. Chai, "Near-sensor and in-sensor computing," Nature Electronics, vol. 3, p. 664–671, 2020.
[10] T. Wang, M. M. Sohoni, L. G. Wright, M. M. Stein, S.-Y. Ma, T. Onodera, M. G. Anderson and P. L. McMahon, "Image sensing with multilayer nonlinear optical neural networks," Nature Photonics, vol. 17, p. 408–415, 2023.

[11] F. Ashtiani, A. J. Geers and F. Aflatouni, "An on-chip photonic deep neural network for image classification," Nature, vol. 606, p. 501–506, 2022.
[12] K. Monakhova, K. Yanny, N. Aggarwal and L. Waller, "Spectral DiffuserCam: lensless snapshot hyperspectral imaging with a spectral filter array," Optica, vol. 7, p. 1298–1307, 2020.
[13] S.-H. Baek, H. Ikoma, D. S. Jeon, Y. Li, W. Heidrich, G. Wetzstein and M. H. Kim, "Single-shot hyperspectral-depth imaging with learned diffractive optics," in Proceedings of the IEEE/CVF International Conference on Computer Vision, 2021.
[14] V. Sitzmann, S. Diamond, Y. Peng, X. Dun, S. Boyd, W. Heidrich, F. Heide and G. Wetzstein, "End-to-end optimization of optics and image processing for achromatic extended depth of field and super-resolution imaging," ACM Transactions on Graphics (TOG), vol. 37, p. 1–13, 2018.
[15] SIA/SRC Decadal Plan for Semiconductors (SRC 2021).
[16] K. Wagner and D. Psaltis, "Multilayer optical learning networks," Applied Optics, vol. 26, p. 5061–5076, 1987.
[17] S. Bandyopadhyay, A. Sludds, S. Krastanov, R. Hamerly, N. Harris, D. Bunandar, M. Streshinsky, M. Hochberg and D. Englund, "Single chip photonic deep neural network with accelerated training," arXiv preprint arXiv:2208.01623, 2022.
[18] S. A. Skirlo, "Photonics for technology: circuits, chip-scale LIDAR, and optical neural networks," 2017.
[19] R. Hesse, S. Schaub-Meyer and S. Roth, "Content-Adaptive Downsampling in Convolutional Neural Networks," in Proceedings of the IEEE/CVF Conference on Computer Vision and Pattern Recognition, 2023.
[20] T. Zhou, X. Lin, J. Wu, Y. Chen, H. Xie, Y. Li, J. Fan, H. Wu, L. Fang and Q. Dai, "Large-scale neuromorphic optoelectronic computing with a reconfigurable diffractive processing unit," Nature Photonics, vol. 15, p. 367–373, 2021.
[21] J. Spall, X. Guo and A. I. Lvovsky, "Hybrid training of optical neural networks," Optica, vol. 9, p. 803–811, 2022.
[22] Q. Guo, Z. Shi, Y.-W. Huang, E. Alexander, C.-W. Qiu, F. Capasso and T. Zickler, "Compact single-shot metalens depth sensors inspired by eyes of jumping spiders," Proceedings of the National Academy of Sciences, vol. 116, p. 22959–22965, 2019.
[23] G. Côté, F. Mannan, S. Thibault, J.-F. Lalonde and F. Heide, "The differentiable lens: Compound lens search over glass surfaces and materials for object detection," in Proceedings of the IEEE/CVF Conference on Computer Vision and Pattern Recognition, 2023.
[24] K. Wei, X. Li, J. Froech, P. Chakravarthula, J. Whitehead, E. Tseng, A. Majumdar and F. Heide, "Spatially varying nanophotonic neural networks," arXiv preprint arXiv:2308.03407, 2023.
[25] L. Huang, Q. A. A. Tanguy, J. E. Fröch, S. Mukherjee, K. F. Böhringer and A. Majumdar, "Photonic advantage of optical encoders," Nanophotonics, vol. 13, p. 1191–1196, 2024.
[26] J. Li, D. Mengu, N. T. Yardimci, Y. Luo, X. Li, M. Veli, Y. Rivenson, M. Jarrahi and A. Ozcan, "Spectrally encoded single-pixel machine vision using diffractive networks," Science Advances, vol. 7, p. eabd7690, 2021.
[27] D. A. B. Miller, "Why optics needs thickness," Science, vol. 379, p. 41–45, 2023.
[28] K. Shastri and F. Monticone, "Nonlocal flat optics," Nature Photonics, vol. 17, p. 36–47, 2023.
[29] B. Pezeshki, R. Kalman, A. Tselikov and C. Danesh, "High speed light microLEDs for visible wavelength communication," in Light-Emitting Devices, Materials, and Applications XXV, 2021.
[30] P. L. McMahon, "The physics of optical computing," Nature Reviews Physics, vol. 5, p. 717–734, 2023.
[31] A. J. Cortese, C. L. Smart, T. Wang, M. F. Reynolds, S. L. Norris, Y. Ji, S. Lee, A. Mok, C. Wu, F. Xia and others, "Microscopic sensors using optical wireless integrated circuits," Proceedings of the National Academy of Sciences, vol. 117, p. 9173–9179, 2020.
[32] Z. Chen, A. Sludds, R. Davis III, I. Christen, L. Bernstein, L. Ateshian, T. Heuser, N. Heermeier, J. A. Lott, S. Reitzenstein and others, "Deep learning with coherent VCSEL neural networks," Nature Photonics, vol. 17, p. 723–730, 2023.

# Photonic neuromorphic sensory processing

**Satoshi Sunada**

Institute of Science and Engineering, Kanazawa University, Kakuma-machi, Kanazawa, Ishikawa, 920-1192, Japan
sunada@se.kanazawa-u.ac.jp

## Status

The rapid progress in Internet of Things (IoT) technology has led to a substantial proliferation of sensory nodes, which generate vast amounts of data. Transmitting data from sensory nodes to cloud computing presents significant challenges in time-delay-sensitive applications. Consequently, the demand for in-sensor or near-sensor processing has significantly increased. Biological sensory systems acquire vast amounts of data, facilitate efficient processing, and employ high-level intelligent functions, even in dynamic, noisy, and complex environments. Notably, studies have shown that the retina of the human eye performs neural computations to extract essential features of visual data [1,2]. This near-sensor processing significantly reduces the data volume before transmission to the brain, enhancing robustness, energy efficiency, and processing speed.

Photonic neuromorphic computing integrates photonic and neuromorphic processing principles at the forefront of technological innovation. Through combination with recent optical sensing techniques, this approach can harness the speed, bandwidth, and energy efficiency of photonic systems to process sensory information in a manner analogous to biological brains. A key advantage of such systems is their capability for direct optical manipulation, which enables the efficient processing of optical sensing information and substantially reduces energy consumption and latency during transmission. This capability is valuable in fields requiring ultra-low latency processing, such as optical communications [3,4] and real-time visual information processing [5–7]. Specifically, this technology holds promise for real-time processing of large amounts of image data in applications, such as flow cytometry or autonomous driving. The compression of real-world visual information [5] and low-latency on-chip image recognition [7] have been demonstrated based on nonlinear optical neural networks. Additionally, this technology is compatible with computational imaging techniques and can be extended to ultrahigh-speed imaging with sub-nanosecond temporal resolution by using a straightforward photonic neuromorphic system combined with wavelength-division multiplexing techniques [8].

Photonic reservoir computing, a distinct approach within neuromorphic computing, also holds the potential for low-latency processing of sensory information. This computational paradigm exploits the inherent properties of materials, devices, or systems for information processing and enables the use of a reservoir as a sensing unit as well as a processing unit. Several studies have demonstrated the seamless integration of optical sensing with information processing based on this concept [9,10].

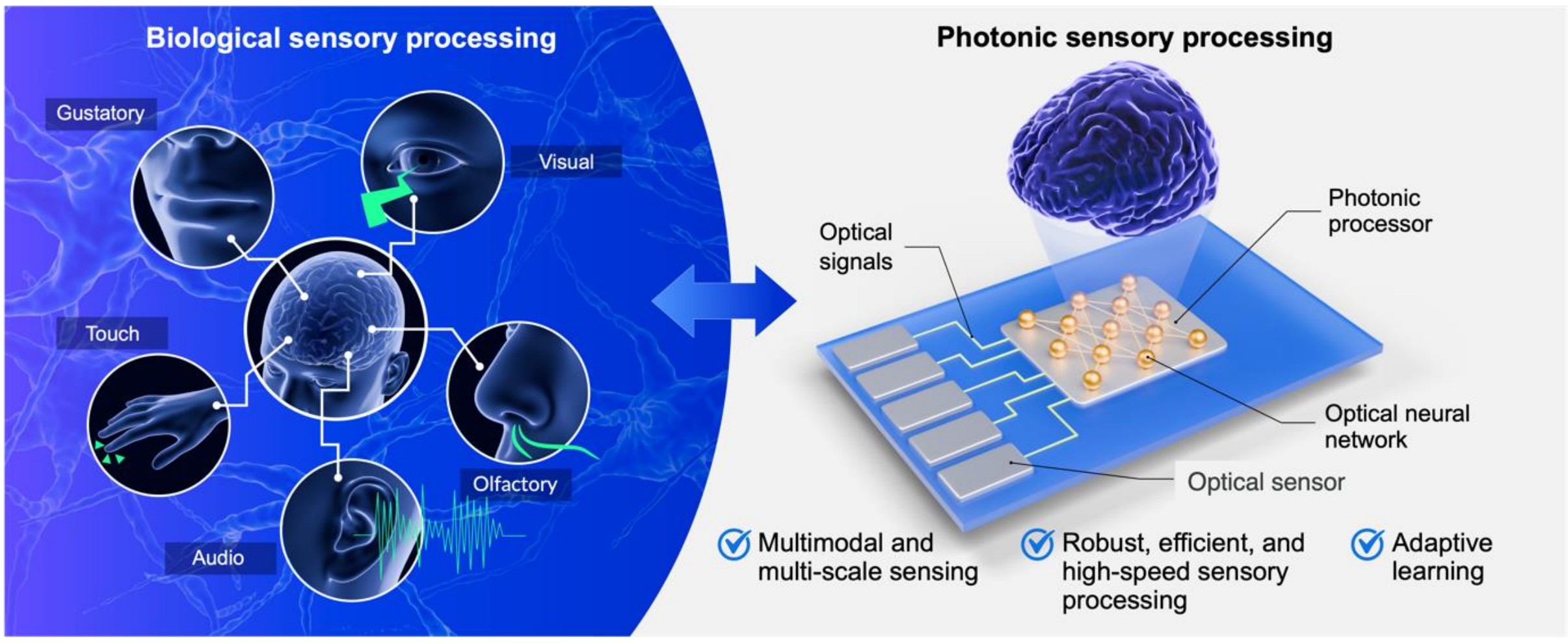

**Figure 1.** Conceptual schematics of biological sensory system and photonic sensory processing system.

## Current and Future Challenges

Biological brains have evolved to efficiently process information from multiple sensory organs, which suggests that neuromorphic systems should seamlessly integrate sensing and processing functions. However, research in neuromorphic photonics has predominantly focused on computational aspects and has not specialized in processing optical sensing data. Optical sensing technologies generally offer high-precision and high-speed acquisition of various physical quantities. Developing a neuromorphic system integrating optical sensing and processing (i.e., photonic in-sensor intelligent processing) could maximize the benefits of existing optical sensing technologies while minimizing reliance on digital electronics, thereby making it advantageous for edge devices with limited energy resources. The conceptual schematic is shown in Fig. 1. A significant challenge is the development of a neuromorphic system architecture, which is capable of directly processing optical sensing information that is typically encoded across multiple domains, such as optical phase, intensity, polarization, or wavelength. Photonic domain transformation, such as wavelength-to-space conversion or time-to-space conversion [8], could be useful for seamlessly integrating sensing and processing units.

Another challenge lies in comprehensively understanding and implementing the sensing and processing mechanisms of biological sensory systems. These challenges include:

- Robust sensory processing: Given the integration of the sensory system with complex, noisy, and dynamic environments, robustness is paramount. This depends on sensor sensitivity and resistance to noise. Developing new materials, device structures, and information-processing techniques to enhance sensor responsiveness to stimuli is crucial.
- Wide dynamic range and multi-scale sensing: Obtaining sensory signals across various scales with a limited dynamic range is challenging. While dynamic gain control can effectively expand the dynamic range, it is important to explore bioinspired approaches.
- Multimodal information acquisition and sensory processing: Biological systems acquire and encode various multimodal information types from each sensory organ as neuronal activity, thereby enabling reliable processing even in complex situations. Developing sensing principles for effectively acquiring different sensory stimuli using a single sensor (without sensor fusion) is appealing. Currently, technologies are emerging to convert diverse physical information, such as auditory [11] and tactile inputs [12,13], into visual information and allow for common processing. Multimodal optical sensory processing is of interest.
- Adaptive and autonomous learning: While the sensory system is expected to offer adaptive intelligence through online learning capabilities, the learning stage can consume significant energy and time. Developing methods for energy-efficient and autonomous learning is crucial yet challenging.

## Advances in Science and Technology to Meet Challenges

Further advancements in this field would be propelled by embracing biological mimicry, interdisciplinary insights, and progress in machine learning, non-equilibrium and nonlinear physics, and neuroscience.

A notable hypothesis suggests that biological systems gain crucial functional advantages by operating near the edge of instability, specifically at the critical point of phase transitions between order and disorder [14]. Criticality has been argued to offer an optimal balance between robustness to perturbations and adaptability to changing conditions, high sensitivity to stimuli, and optimal computational power [14]. For instance, the inner ears of vertebrates demonstrate extraordinary sensitivity and exquisite frequency selectivity, and they can detect acoustic stimuli across multiple scales such that hair cells operate near the Hopf bifurcation [15]. Dynamic systems at the edge of chaos are deemed suitable for complex computations and maximizing memory capacity. A power-law behavior, suggesting a self-organized critical phenomenon, has recently been observed in optical dynamical systems, such as lasers with delayed feedback [16]. Additionally, photonic spiking neural networks show potential for emerging critical behaviors. Criticality can also manifest in non-Hermitian photonic systems operating at an exceptional point, such as asymmetric microcavities and parity-time symmetric systems, and this enhances sensing sensitivity [17], whereas non-Hermitian systems are linear and lack scale-free behavior. Photonic systems exhibiting criticality hold promise for multi-scale sensory processing.

Robust and multimodal sensory processing can be achieved using a high-dimensional distributed representation of sensory information, which forms the basis of hyperdimensional computing [18] and reservoir computing. This distributed representation enables energy-efficient and noise-robust computations with low-precision and basic arithmetic operations. The spatial degrees of freedom of light enable an optical hyperdimensional distributed representation of multimodal information in a parallel and energy-efficient manner. For example, the speckle phenomenon resulting from scatterers can be used for an optically distributed representation containing information on various physical quantities, such as wavelength, shape, temperature, and pressure [12,13]. In other words, it can effectively encode multimodal external stimuli. Leveraging high optical dimensionality, optical multiplexing, and novel computing paradigms such as hyperdimensional computing holds promise for sensing processing.

Nonlinear operations are crucial for processing complex, high-dimensional data. For example, large-area vertical cavity surface emitting lasers (VCSELs) [19], which are capable of operating at low thresholds and high speeds, show promise for efficiently performing nonlinear processing of high-dimensional information.

## Concluding Remarks

Photonic neuromorphic sensory processing can efficiently process large volumes of data in noisy, dynamically changing, and complex environments. Its development heralds a new era of computing technology by melding biological inspiration with the unique advantages of photonics. However, significant accuracy, energy efficiency, and higher-level intelligence improvements are urgently required. Improving and integrating optoelectronic components and optical memory, alongside exploring system architectures that optimize light utilization, are paramount.

## Acknowledgments

This study received partial support from JSPS KAKENHI (Grant No. JP23H03467, JP22K18792), Grant-in-Aid for Transformative Research Areas A (JP22H05198), and JST, CREST (Grant No. JPMJCR24R2). The author expresses gratitude to T. Niiyama for fruitful discussions on criticality.

## References

[1] F. Liao, F. Zhou, and Y. Chai, "Neuromorphic vision sensors: Principle, progress and perspectives," *Journal of Semiconductors*, vol. 42, no. 1, 013105, 2021.

[2] T. Gollisch and M. Meister, "Eye smarter than scientists believed: neural computations in circuits of the retina." *Neuron*, vol. 65, no. 2, pp. 150–164, 2010.

[3] S. Sackesyn, C. Ma, J. Dambre, and P. Bienstman, "Experimental realization of integrated photonic reservoir computing for nonlinear fiber distortion compensation," *Opt. Express*, vol. 29, no. 20, pp. 30991–30997, 2021.

[4] C. Huang, S. Fujisawa, T. F. de Lima, A. N. Tait, E. C. Blow, Y. Tian, S. Bilodeau, A. Jha, F. Yaman, H.-T. Peng, H. G. Batshon, B. J. Shastri, Y. Inada, T. Wang, and P. R. Prucnal, "A silicon photonic–electronic neural network for fibre nonlinearity compensation," *Nature Electronics*, vol. 4, no. 11, pp. 837–844, 2021.

[5] T. Wang, M. M. Sohoni, L. G. Wright, M. M. Stein, S.-Y. Ma, T. Onodera, M. G. Anderson, and P. L. McMahon, "Image sensing with multilayer nonlinear optical neural networks," *Nature Photonics*, vol. 17, no. 5, pp. 408–415, 2023.

[6] X. Lin, Y. Rivenson, N. T. Yardimci, M. Veli, Y. Luo, M. Jarrahi, and A. Ozcan, "All-optical machine learning using diffractive deep neural networks," *Science*, vol. 361, no. 6406, pp. 1004–1008, 2018.

[7] F. Ashtiani, A. J. Geers, and F. Aflatouni, "An on-chip photonic deep neural network for image classification," *Nature*, vol. 606, pp. 501–506, 2022.

[8] T. Yamaguchi, K. Arai, T. Niiyama, A. Uchida, and S. Sunada, "Time-domain photonic image processor based on speckle projection and reservoir computing," *Communications Physics*, vol. 6, no. 1, 250, 2023.

[9] S. Sunada and A. Uchida, "Photonic reservoir computing based on nonlinear wave dynamics at microscale," *Scientific Reports*, vol. 9, no. 1, 19078, 2019.

[10] N. Fang, S. Wang, and C. Wang, "Distributed optical fiber vibration sensing implemented with delayed feedback reservoir computing," Optics & Laser Technology, vol. 162, 109244, 2023.

[11] M. Sheinin, D. Chan, M. O'Toole, and S. G. Narasimhan, "Dual-shutter optical vibration sensing," in *2022 IEEE/CVF Conference on Computer Vision and Pattern Recognition (CVPR)*, June 2022, pp. 16303–16312.

[12] S. Shimadera, K. Kitagawa, K. Sagehashi, Y. Miyajima, T. Niiyama, and S. Sunada, "Speckle-based high-resolution multimodal soft sensing," *Scientific Reports*, vol. 12, no. 1, 13096, 2022.

[13] K. Kitagawa, K. Tsuji, K. Sagehashi, T. Niiyama, and S. Sunada, "Optical hyperdimensional soft sensing: speckle-based touch interface and tactile sensor," *Opt. Express*, vol. 32, no. 3, pp. 3209–3220, 2024.

[14] M. A. Muñoz, "Colloquium: Criticality and dynamical scaling in living systems," *Rev. Mod. Phys.,* vol. 90, 031001, 2018.

[15] A. J. Hudspeth, "Integrating the active process of hair cells with cochlear function," *Nature Reviews Neuroscience*, vol. 15, no. 9, pp. 600–614, 2014.

[16] T. Niiyama and S. Sunada, "Power-law fluctuations near critical point in semiconductor lasers with delayed feedback," *Phys. Rev. Res.*, vol. 4, 043205, 2022.

[17] J. Wiersig, "Review of exceptional point-based sensors," *Photon. Res.*, vol. 8, no. 9, pp. 1457–1467, 2020.

[18] P. Kanerva, "Hyperdimensional computing: An introduction to computing in distributed representation with high-dimensional random vectors," *Cognitive Computation*, vol. 1, no. 2, pp. 139–159, 2009.

[19] A. Skalli, J. Robertson, D. Owen-Newns, M. Hejda, X. Porte, S. Reitzenstein, A. Hurtado, and D. Brunner, "Photonic neuromorphic computing using vertical cavity semiconductor lasers," *Opt. Mater. Express*, vol. 12, no. 6, pp. 2395–2414, 2022.

# Optoelectronic Visual Computing

**Suyeon Choi[1], Gordon Wetzstein[1]**
[1] Department of Electrical Engineering, Stanford University
350 Jane Stanford Way, Stanford, CA, 94305, USA

E-mails: suyeon@stanford.edu, gordon.wetzstein@stanford.edu

## Status

The field of optoelectronic visual computing has been actively developed by both academia and industry overcoming the limitations of many traditional imaging and display applications, including microscopy, astrophotography, and photography for consumer devices. Early works date back to the 1980s, which enabled measuring X-ray radiation emitted from astronomical objects using coded apertures, and the extended depth-of-field imaging through wavefront coding. Later on, this idea has been pursued by the computational photography community over the last 20 years or so, contributing to consumer electronics, with features including high dynamic range, super-resolution, and light fields, among others.

Similarly, the joint design of optics, electronics, and algorithms led to the renaissance of holography and optical computing. Holography, which is widely believed as the ultimate display technology, was introduced in the late 1940s. The development of the laser enabled the first optical holograms, while digital computers and spatial light modulators (SLMs) have facilitated holographic video through computer-generated holography (CGH) algorithms. These algorithms encode 3D scene information using the principle of wave optics and have been developed for decades to adapt various graphics formats [2]. Holographic display offers paths to solve some of the biggest remaining challenges for wearable computing systems, such as focus cues, dynamic steering capabilities, vision correction, device form factors, as well as image resolution and brightness. Optical computing operates massively in parallel at the speed of light, presenting significant potential to resolve the computational requirements of modern artificial intelligence (AI) applications. In this article, we provide an overview of recent examples of such optoelectronic visual computing systems.

## Current and future challenges

While the fields of optics design, low-level image processing algorithm development, and high-level computer vision network design have co-evolved over decades, the traditional approaches to imaging optics design still largely rely on heuristic methods, and proxy metrics like the point spread function in a compartmentalized fashion. However, these do not lead to the optimal solutions for domain-specific tasks, which are indeed the dominant use cases for imaging. Moreover, while neural networkbased image-processing approaches excel across a wide range of computer vision applications, they often require substantial computational resources. These demands limit their effectiveness and applicability in real-world scenarios where computational resources are constrained, such as in edge devices.

Still, there are many challenges presenting major roadblocks to unlocking the full potential of holographic near-eye displays for wearable computing. One of these challenges is the limited experimental image quality that has never met computer graphics standards. This is largely due to the model mismatch between the simulation model used for CGH and the physical model in the setup. Moreover, the computational complexity of CGH algorithms is often very high due to the fundamental principle, where one pixel at a plane affects a region at another plane, unlike in conventional geometric optics where pixels are associated with rays. This poses a fundamental trade-off between algorithm runtime and achieved image quality, which has prevented the generation of high-quality 3D holograms at fast speeds. Other limitations of current holographic display technologies include the limited degrees of freedom that hinder accurate depiction of 3D scenes over a large étendue, the lack of optical architecture suitable for eyeglasses form factors for all-day usage, and accompanying engineering challenges. Advances in science and technology to meet challenges

Recently, advancements in artificial intelligence have greatly expanded the potential of optoelectronic visual computing systems. For imaging and computing systems, this advancement enables the true end-to-end optimization of optical elements, image signal processors (ISPs), and computer vision networks for learning application-specific cameras. For display systems, it provides opportunities to co-design the hardware with display-specific rendering, modeling algorithms.

Sitzmann et al. were the first to propose the end-to-end optimization of optics and algorithms for camera design ([17], Fig. 1A). This methodology has improved performances for various imaging applications, including depth estimation, single-shot high-dynamic-range imaging, and hyperspectral imaging [1, 4, 7, 9, 13, 20], by jointly optimizing a diffractive optical element and algorithms. These joint optimizations of nanophotonic or diffractive optical elements significantly advance the modern optical and photonic computing systems. In these systems, either single or multiple stacked diffractive layers, or nanophotonic circuits, are optimized for all-optical or hybrid optical-electronic computing with modern machine learning techniques ([3, 11], Fig. 1B).

The advanced hybrid bonding capabilities enable multi-layer stacked CMOS image sensors, having an increasing amount of processing power directly on-chip. The emerging on-sensor compute was employed to jointly optimize low-level in-pixel processing and irradiance encoding, along with higher-level algorithms running off-chip, for applications in high-dynamic-range and high-speed compressive video sensing (Fig. 1C). These methodologies can be used to reduce the bandwidth between the processor and sensors, by deploying the first layer(s) of a convolutional neural network (CNN) on chip. [12, 14, 18, 19].

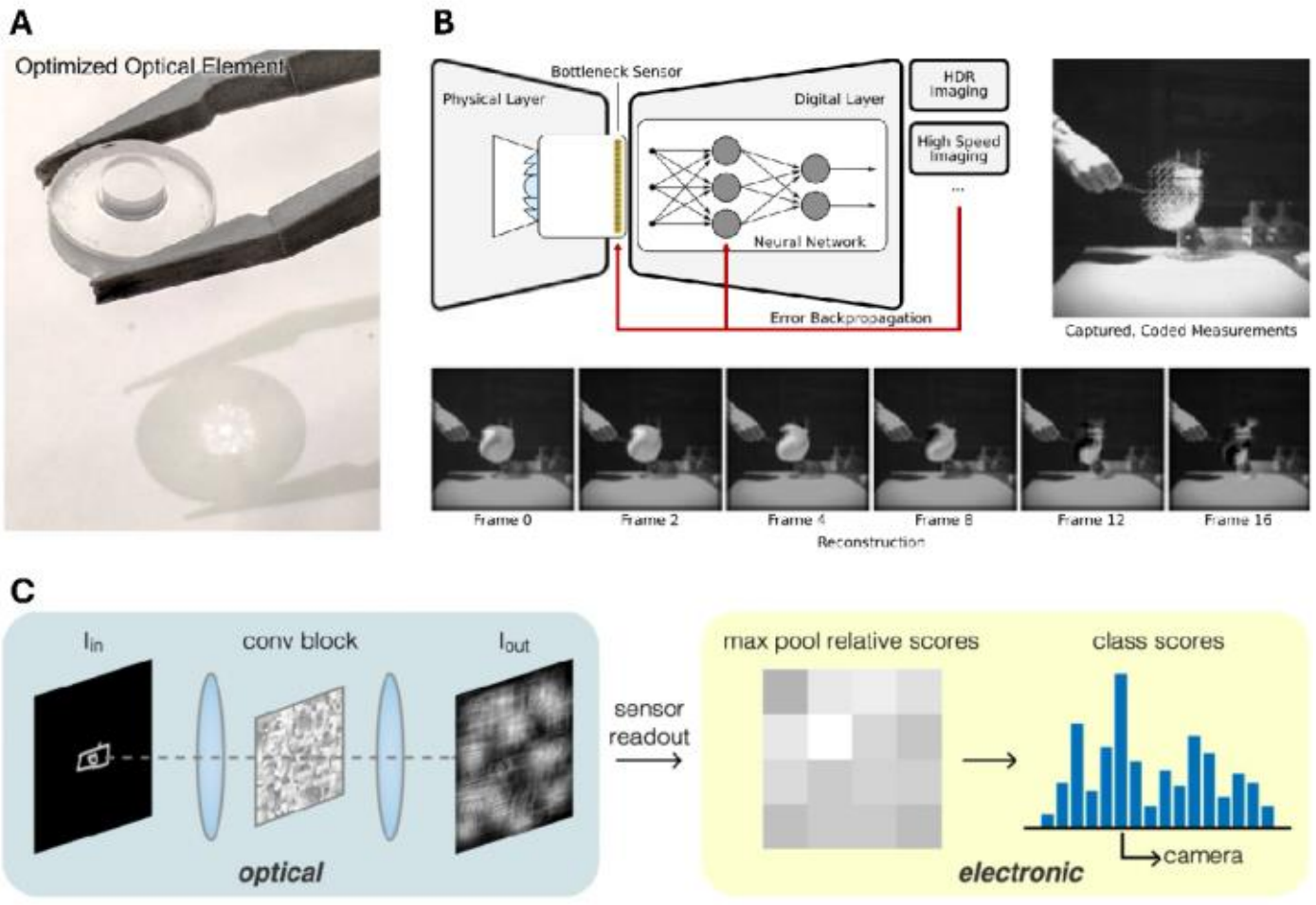

**Figure 1. A.** Image of a refractive optical element designed using the end-to-end optimization. **B.** Illustrations of the end-to-end optimization framework for senses (*top left*). A single coded exposure (*top right*), and several frames of the high-speed video reconstructed from this image showing an exploding balloon (*bottom*). **C.** Schematic of a hybrid optical-electronic computing. Figures adapted from [3, 12, 17]

The recent advancement of holographic displays using AI has focused on addressing the aforementioned limitations, offering a breakthrough for near-eye displays. For example, Kim et al. proposed an ultra-thin holographic display architecture using waveguide that is co-designed with CGH algorithms that optimize the higher-order noise [8, 10] (Fig. 2 A). Peng et al. introduced the camera-in-the-loop optimization that achieved the state-of-the-art image quality. The image quality is further improved by training differentiable wave propagation models, by massively reducing the gap between simulated models and physical optics [5, 6] (Fig. 2 D, E). Once trained, the learned model provides insights into physical optics aspects, such as phase distortion and gradient of the system, or the shape of optical elements in the middle of the setup (e.g., iris) (Fig. 2 B-C). Moreover, neural networks can be trained to enable real-time CGH algorithms [15, 16].

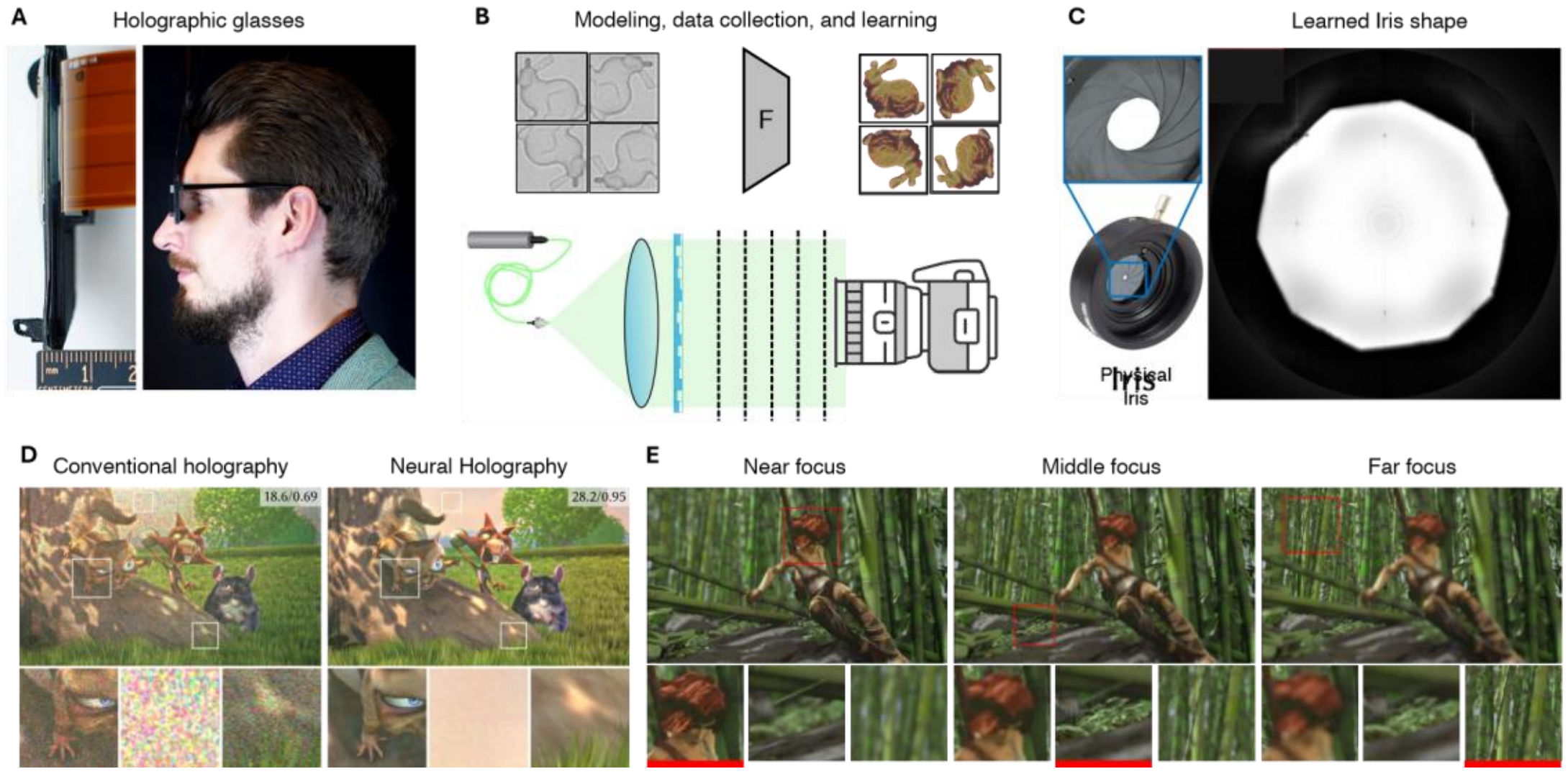

**Figure 2. A.** Images of the holographic glasses, which can provide full-color 3D holographic images with 2.5 mm thick optics. **B.** Illustrations for modeling, data collection, and training for holographic near-eye displays. **C.** A photo of a physical iris in the setup and the iris shape as learned from the model. **D.** Experimentally captured results using conventional CGH and the CGH using the learned model from the Neural Holography framework. **E.** Experimentally captured results using 3D CGH and the learned model. Figures reproduced from [5, 10]

## Concluding remarks

The evolving paradigm of end-to-end optimization of physical optics, in-pixel sensor processing capabilities, and tailored algorithms for image reconstruction and inference, represents a groundbreaking shift from the conventional, compartmentalized approach to component design. A number of applications have rapidly progressed by adapting this methodology, notably in developing cutting-edge cameras, optical computing systems, programmable CMOS image sensors, and computational displays. Future work in this area includes exploring new application domains, such as 3D printing, and microscopy, and designing more efficient optimization algorithms for the large-scale problems of nanophotonic imaging and computing, among many other directions.

## Acknowledgements

The authors would like to thank Yifan “Evan” Peng, Manu Gopakumar, Jonghyun Kim, and Brian Chao for fruitful discussions.

## References

[1] Baek, S.-H., Ikoma, H., Jeon, D. S., Li, Y., Heidrich, W., Wetzstein, G., and Kim, M. H. Single-shot hyperspectral-depth imaging with learned diffractive optics. In Proceedings of the IEEE/CVF International Conference on Computer Vision (2021), pp. 2651–2660.

[2] Blinder, D., Birnbaum, T., Ito, T., and Shimobaba, T. The state-of-the-art in computer generated holography for 3d display. Light: Advanced Manufacturing 3, 3 (2022), 572–600.

[3] Chang, J., Sitzmann, V., Dun, X., Heidrich, W., and Wetzstein, G. Hybrid optical-electronic convolutional neural networks with optimized diffractive optics for image classification. Scientific reports 8, 1 (2018), 1–10.

[4] Chang, J., and Wetzstein, G. Deep optics for monocular depth estimation and 3d object detection. In Proceedings of the IEEE/CVF International Conference on Computer Vision (2019), pp. 10193–10202.

[5] Choi, S., Gopakumar, M., Peng, Y., Kim, J., O'Toole, M., and Wetzstein, G. Time-multiplexed neural holography: a flexible framework for holographic near-eye displays with fast heavily-quantized spatial light modulators. In ACM SIGGRAPH 2022 Conference Proceedings (2022), pp. 1–9.

[6] Choi, S., Gopakumar, M., Peng, Y., Kim, J., and Wetzstein, G. Neural 3d holography: Learning accurate wave propagation models for 3d holographic virtual and augmented reality displays. ACM Transactions on Graphics (TOG) 40, 6 (2021), 1–12.

[7] Dun, X., Ikoma, H., Wetzstein, G., Wang, Z., Cheng, X., and Peng, Y. Learned rotationally symmetric diffractive achromat for full-spectrum computational imaging. Optica 7, 8 (2020), 913–922.

[8] Gopakumar, M., Kim, J., Choi, S., Peng, Y., and Wetzstein, G. Unfiltered holography: optimizing high diffraction orders without optical filtering for compact holographic displays. Optics letters 46, 23 (2021), 5822–5825.

[9] Ikoma, H., Nguyen, C. M., Metzler, C. A., Peng, Y., and Wetzstein, G. Depth from defocus with learned optics for imaging and occlusion-aware depth estimation. In 2021 IEEE International Conference on Computational Photography (ICCP) (2021), IEEE, pp. 1–12.

[10] Kim, J., Gopakumar, M., Choi, S., Peng, Y., Lopes, W., and Wetzstein, G. Holographic glasses for virtual reality. In ACM SIGGRAPH 2022 Conference Proceedings (2022), pp. 1–9.

[11] Lin, X., Rivenson, Y., Yardimci, N. T., Veli, M., Luo, Y., Jarrahi, M., and Ozcan, A. All-optical machine learning using diffractive deep neural networks. Science 361, 6406 (2018), 1004–1008.

[12] Martel, J. N., Mueller, L. K., Carey, S. J., Dudek, P., and Wetzstein, G. Neural sensors: Learning pixel exposures for hdr imaging and video compressive sensing with programmable sensors. IEEE transactions on pattern analysis and machine intelligence 42, 7 (2020), 1642–1653.

[13] Metzler, C. A., Ikoma, H., Peng, Y., and Wetzstein, G. Deep optics for single-shot high-dynamic-range imaging. In Proceedings of the IEEE/CVF Conference on Computer Vision and Pattern Recognition (2020), pp. 1375–1385.

[14] Nguyen, C. M., Martel, J. N., and Wetzstein, G. Learning spatially varying pixel exposures for motion deblurring. In 2022 IEEE International Conference on Computational Photography (ICCP) (2022), IEEE, pp. 1–11.

[15] Peng, Y., Choi, S., Padmanaban, N., and Wetzstein, G. Neural holography with camera-in-the-loop training. ACM Transactions on Graphics (TOG) 39, 6(2020), 1–14.

[16] Shi, L., Li, B., Kim, C., Kellnhofer, P., and Matusik, W. Towards real-time photorealistic 3d holography with deep neural networks. Nature 591, 7849 (2021), 234–239.

[17] Sitzmann, V., Diamond, S., Peng, Y., Dun, X., Boyd, S., Heidrich, W., Heide, F., and Wetzstein, G. End-to-end optimization of optics and image processing for achromatic extended depth of field and super-resolution imaging. ACM Transactions on Graphics (TOG) 37, 4 (2018), 1–13.

[18] So, H. M., Bose, L., Dudek, P., and Wetzstein, G. Pixelrnn: In-pixel recurrent neural networks for end-to-end-optimized perception with neural sensors. In CVPR (2024).

[19] So, H. M., Martel, J. N., Wetzstein, G., and Dudek, P. Mantissacam: Learning snapshot high-dynamic-range imaging with perceptually-based in-pixel irradiance encoding. In 2022 IEEE International Conference on Computational Photography (ICCP) (2022), IEEE, pp. 1–12.

[20] Wetzstein, G., Ozcan, A., Gigan, S., Fan, S., Englund, D., Soljačić, M., Denz, C., Miller, D. A., and Psaltis, D. Inference in artificial intelligence with deep optics and photonics. Nature 588, 7836 (2020), 39–47.

# Optical Intelligent Signal Processing for Optical Communications

**Chaoran Huang**
Department of Electronic Engineering, the Chinese University of Hong Kong, Hong Kong, SAR China
[crhuang@ee.cuhk.edu.hk]

**Status**

Today's Internet traffic is growing 30% annually. In the next decade, optical network capacity needs to increase 10 times to sustain the operation of the Internet. Current optical communication systems have already strained the capacity of optical fibers, inevitably leading to increased channel imperfection and signal distortions. Signal distortions are addressed by DSP implemented in high-speed custom application-specific integrated circuits (ASICs) which is subject to the CMOS process node improvement cycle. Today, DSP chips for optical communications have approached the limits of semiconductor technologies in terms of power dissipation and density. The most advanced CMOS nodes are required to handle today's high-speed fiber communication systems. For example, 5nm CMOS PAM4 DSPs are used in 400G and 800G data center applications. For next-generation 1.6T transceivers, the 3nm CMOS node will be required. Despite using state-of-the-art CMOS nodes, DSP chips still must avoid using powerful but computationally expensive algorithms to maintain their power dissipation below the maximum thermal dissipation capacity. For example, digital back-propagation, a powerful algorithm for fiber nonlinearity compensation, cannot be implemented on the DSP chips because its complexity and associated power consumption are too high. The limited signal processing capability has resulted in the so-called nonlinear Shannon limit [1]. In the coming ten years, DSP needs to deal with 10x more data traffic; correspondingly, their energy per bit must be reduced by 10. Unfortunately, as semiconductor technologies are evolving on probably the last bit of Moore's law, DSP will find it increasingly challenging to support the continued growth of internet traffic in the future.

The bottleneck in DSP has motivated research in signal processing using photonics physics and in the photonic domain. Traditional optical signal processing technologies based on dispersion engineering and nonlinear effects, such as wavelength conversion, pulse shaping, multiplexing/demultiplexing, and information broadcasting, directly act on optical signals in the all-optical domain, avoiding the bandwidth and rate limitations of DSP chips. However, these techniques generally rely on bulk optical components such as dispersion-compensating fibers, highly nonlinear fibers, periodically poled lithium niobate (PPLN) waveguides, and semiconductor optical amplifiers (SOAs). These components are bulky, lack reconfigurability, have single functionality, are difficult to integrate on a large scale, and power consuming.

In recent years, on-chip optical signal processing technology has garnered increasing attention. In particular, the emerging field of neuromorphic photonics [2], which combines integrated reconfigurable photonics with intelligent computing frameworks such as reservoir computing and deep learning, has the potential to perform multifunctional, reconfigurable information processing for different communication systems. This approach promises faster and more energy-efficient intelligent signal processing than its digital counterparts. In the meanwhile, integrated photonic fabrication and packaging technologies are advancing rapidly. High-density integrated optoelectronic devices and interconnects make scalable information processing and computing possible.

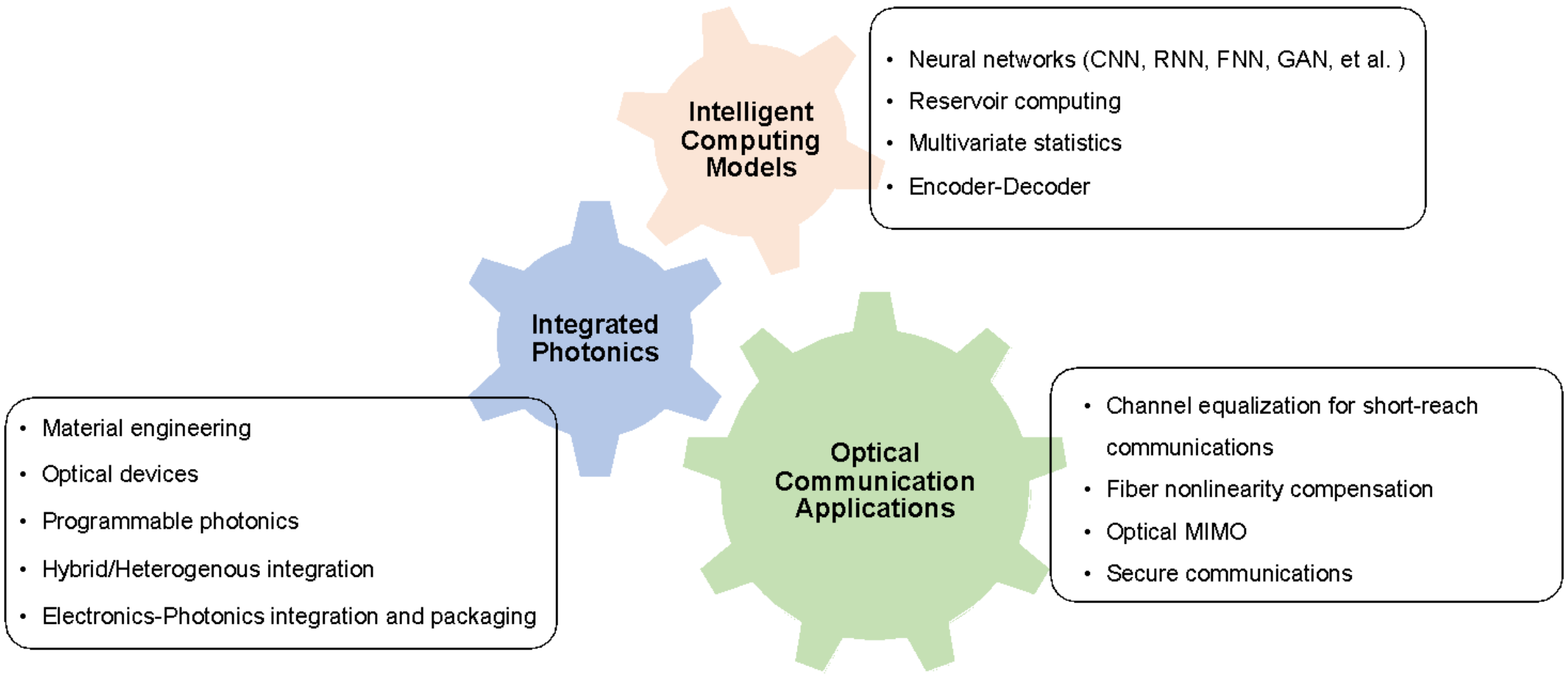

**Figure 1.** Overview of optical intelligent processing for optical communications.

## Current and Future Challenges

There has been a surge of research interest and demonstration of photonic intelligent signal processing systems in recent years. A common approach is to implement artificial neural networks on photonic systems to address transmission distortions. Various types of photonic neural networks have been designed to cater to different communication systems and signal impairments, including reservoir computing [3], [4], [5], [6], [7], feedforward neural networks [8], recurrent neural networks [9], convolutional neural networks [10] et al. These systems have shown their capability in handling various optical communication schemes, such as intensity modulated direct detection (IMDD) [3] and coherent optical communications [8][11], and Kramers–Kronig detection [12]. They effectively address transmission impairments like chromatic dispersion, optical fiber nonlinearities in single-channel and multiwavelength channels, and optical MIMO for spatial division multiplexed systems [13][13] [14]. Additionally, efforts are being made to build secure optical communication systems by leveraging the ability of optics to perform optical encoding and decoding with low latency and low power consumption [16]. Photonic intelligent signal processing systems have also shown their reconfigurability and flexibility to different modulation formats, data rates, and transmission distances by deploying programmable photonic devices and circuits, in contrast to traditional optical signal processing systems.

Despite these potential advantages and successful demonstrations, the development of optical signal processing technologies lags behind current optical communication systems. 1. Many intelligent signal processing systems can only address static impairments, such as chromatic dispersion and fiber nonlinearities, but lack the capability to handle time-varying impairments, such as polarization mode dispersion. Energy efficient and high-speed reconfigurable devices (i.e., phase shifters) are needed. 2. Most systems have only been effective in short-reach communication systems, not long-haul communications. This limitation arises because long and low-loss optical delay lines and many optical nonlinear nodes are required to address accumulated distortions in long-haul communications, which are still challenging to realize on optical chips. 3. Most optical intelligent signal processing technologies primarily focus on a single lightwave dimension and a single type of signal impairment. However, modern optical transmission systems enhance communication capacity by integrating multiple physical dimensions, such as wavelength, polarization, and spatial modes. This inevitably leads to increased and more complex channel imperfections and signal distortions that need to be addressed by optical intelligent processors.

An important factor to consider in photonic processors is the excess loss and noise. Excess loss and noise are unavoidable in photonic processors and cause signal-to-noise ratio degradation. On the other hand, achieving better equalization performance typically requires more optical components, which leads to larger insertion loss. Therefore, these trade-offs must be factored into the overall link design. Low-loss optical components and low-noise on-chip amplifiers are urgently needed to mitigate such trade-offs. In addition, ambient noise and thermal fluctuations can severely affect the computing accuracy of the photonic processor, leading to increased bit error rates. These issues must be carefully controlled with external circuits, and the addition to the overall power consumption need to be considered.

**Advances in Science and Technology to Meet Challenges**

To address these challenges and transform early system demonstrations into practical and fully performant photonic intelligent processors, photonic systems must incorporate novel technologies. One significant challenge is increasing the number of optical components on a single chip to extend the functionality and transmission distance that the photonic processor can handle. Reducing excess losses and noise remains a key prerequisite for realizing reliable large-scale photonic processors that can compete with electronics. Essential components, such as on-chip optical amplifiers, are highly desirable to compensate for insertion loss within the circuits or at the circuit edge. Additionally, the efficiency of devices, especially nonlinear devices necessary for realizing nonlinear nodes in photonic neural networks, needs to be optimized. The insertion loss of silicon photonic components, such as phase shifters, modulators, couplers, and optical delay lines, must be minimized to reduce overall system losses. No single photonic platform provides all the desirable components and features. Silicon photonic integration offers an unprecedented platform for producing large-scale and low-cost photonic systems, but it does not include gain elements and the insertion loss is only moderate. Hybrid and heterogeneous photonic integration represent advanced approaches to building photonic systems by combining different material platforms and components on a single chip[17], [18]. This integration can include components from various material systems, such as silicon, indium phosphide, silicon nitride, and lithium niobate, to leverage the unique advantages of each material. Heterogeneous photonic-electronic integration goes a step further by seamlessly integrating diverse photonic and electronic components onto a single substrate, enabling complex functionalities and scalable control [19]. These integration techniques are important for developing sophisticated, scalable, and efficient photonic processors that can address the limitations of traditional silicon photonics and compete with electronic systems.

In addition to advancing photonic technologies, innovations can also be made in the synthetic design of photonic systems to perform more powerful intelligent algorithms and tailor these algorithms for better implementation on photonic platforms. For example, designing deep optical reservoir computing systems can significantly extend the transmission distance beyond what a single reservoir can handle. Combining optical MIMO with suitable algorithms such as blind source separation can simplify the design of backend digital processors and improve the overall system's energy efficiency [15] [20]. Another direction to enhance functionality and transmission distance is to develop electronic-photonic processors. The concept aims to harness the strengths of both electronics and photonics while mitigating their respective limitations. For instance, photonics can handle optical nonlinearity and inter-channel (wavelength, spatial modes) impairments that require large bandwidth and computing resources, while electronics can manage dynamic impairments that require frequent updates. By optimizing both the photonic and electronic hardware, as well as the software, hybrid processors can achieve more efficient and effective signal processing, pushing the boundaries of what is possible with current technology.

**Concluding Remarks**

Neuromorphic photonics, combining integrated photonics with intelligent computing frameworks, holds promise for fast, energy-efficient intelligent signal processing. However, challenges such as excess loss, noise, and the ability to handle long-distance transmission systems and time-varying impairments remain. To meet these challenges, advancements in photonic technologies are crucial. Increasing the number of optical components on a single chip, reducing excess losses and noise, and optimizing the efficiency of devices are key areas of focus. Hybrid and heterogeneous photonic integration can provide a platform for scalable and efficient photonic processors. Innovations in algorithm-hardware co-design can further enhance photonic systems' functionality and transmission distance. Ultimately, the integration of electronics and photonics in hybrid processors offers a promising path to overcome current limitations and advance signal processing capabilities beyond current boundaries.

**Acknowledgements**
This work was supported RGC ECS 24203724, NSFC 62405258, ITF ITS/237/22, RGC YCRF C1002-22Y, NSFC/RGC Joint Research Scheme N_CUHK444/22 and Optica Foundation."

## References

[1] A. D. Ellis, J. Zhao, and D. Cotter, "Approaching the Non-Linear Shannon Limit," *Journal of Lightwave Technology*, vol. 28, no. 4, pp. 423–433, Feb. 2010, doi: 10.1109/JLT.2009.2030693.

[2] B. J. Shastri *et al.*, "Photonics for artificial intelligence and neuromorphic computing," *Nature Photonics*, vol. 15, no. 2, pp. 102–114, 2021.

[3] S. Sackesyn *et al.*, "Experimental realization of integrated photonic reservoir computing for nonlinear fiber distortion compensation," *Opt. Express, OE*, vol. 29, no. 20, pp. 30991–30997, Sep. 2021, doi: 10.1364/OE.435013.

[4] K. Vandoorne *et al.*, "Experimental demonstration of reservoir computing on a silicon photonics chip," *Nature communications*, vol. 5, no. 1, pp. 1–6, 2014.

[5] D. Brunner, M. C. Soriano, C. R. Mirasso, and I. Fischer, "Parallel photonic information processing at gigabyte per second data rates using transient states," *Nature communications*, vol. 4, no. 1, pp. 1–7, 2013.

[6] A. Bogris, C. Mesaritakis, S. Deligiannidis, and P. Li, "Fabry-Perot Lasers as Enablers for Parallel Reservoir Computing," *IEEE Journal of Selected Topics in Quantum Electronics*, vol. 27, no. 2, pp. 1–7, Mar. 2021, doi: 10.1109/JSTQE.2020.3011879.

[7] F. Da Ros, S. M. Ranzini, H. Bülow, and D. Zibar, "Reservoir-computing based equalization with optical pre-processing for short-reach optical transmission," *IEEE Journal of Selected Topics in Quantum Electronics*, vol. 26, no. 5, pp. 1–12, 2020.

[8] C. Huang *et al.*, "Silicon photonic-electrical neural networks for fiber nonlinearity compensation," *Nature Electronics*, pp. 1–3, 2021.

[9] B. Wang, T. F. de Lima, B. J. Shastri, P. R. Prucnal, and C. Huang, "Multi-Wavelength Photonic Neuromorphic Computing for Intra and Inter-Channel Distortion Compensations in WDM Optical Communication Systems," *IEEE Journal of Selected Topics in Quantum Electronics*, vol. 29, no. 2: Optical Computing, pp. 1–12, Mar. 2023, doi: 10.1109/JSTQE.2022.3213172.

[10] J. Ye *et al.*, "Multiplexed orbital angular momentum beams demultiplexing using hybrid optical-electronic convolutional neural network," *Commun Phys*, vol. 7, no. 1, pp. 1–7, Mar. 2024, doi: 10.1038/s42005-024-01571-3.

[11] M. Sorokina, S. Sergeyev, and S. Turitsyn, "Fiber echo state network analogue for high-bandwidth dual-quadrature signal processing," *Optics express*, vol. 27, no. 3, pp. 2387–2395, 2019.

[12] S. Masaad, E. Gooskens, S. Sackesyn, J. Dambre, and P. Bienstman, "Photonic reservoir computing for nonlinear equalization of 64-QAM signals with a Kramers–Kronig receiver," *Nanophotonics*, vol. 12, no. 5, pp. 925–935, Mar. 2023, doi: 10.1515/nanoph-2022-0426.

[13] A. Annoni *et al.*, "Unscrambling light—automatically undoing strong mixing between modes," *Light Sci Appl*, vol. 6, no. 12, pp. e17110–e17110, Dec. 2017, doi: 10.1038/lsa.2017.110.

[14] K. Choutagunta, I. Roberts, D. A. B. Miller, and J. M. Kahn, "Adapting Mach–Zehnder Mesh Equalizers in Direct-Detection Mode-Division-Multiplexed Links," *Journal of Lightwave Technology*, vol. 38, no. 4, pp. 723–735, Feb. 2020, doi: 10.1109/JLT.2019.2952060.

[15] C. Huang *et al.*, "High-Capacity Space-Division Multiplexing Communications With Silicon Photonic Blind Source Separation," *Journal of Lightwave Technology*, vol. 40, no. 6, pp. 1617–1632, Mar. 2022, doi: 10.1109/JLT.2022.3152027.

[16] Y. Chen *et al.*, "Photonic unsupervised learning variational autoencoder for high-throughput and low-latency image transmission," *Science Advances*, vol. 9, no. 7, p. eadf8437, Feb. 2023, doi: 10.1126/sciadv.adf8437.

[17] C. Xiang *et al.*, "Laser soliton microcombs heterogeneously integrated on silicon," *Science*, vol. 373, no. 6550, pp. 99–103, Jul. 2021, doi: 10.1126/science.abh2076.

[18] T. Komljenovic *et al.*, "Heterogeneous silicon photonic integrated circuits," *Journal of Lightwave Technology*, vol. 34, no. 1, pp. 20–35, 2016.

[19] K. Giewont *et al.*, "300-mm Monolithic Silicon Photonics Foundry Technology," *IEEE Journal of Selected Topics in Quantum Electronics*, vol. 25, no. 5, pp. 1–11, Sep. 2019, doi: 10.1109/JSTQE.2019.2908790.

[20] Zhang, W., Tait, A., Huang, C. et al. "Broadband physical layer cognitive radio with an integrated photonic processor for blind source separation". *Nat Commun* 14, 1107 (2023)

# Microwave Photonics Enables Future Cognitive Radio

**Weipeng Zhang**, Princeton University (Princeton, NJ 08544, USA)
**Joshua C. Lederman**, Princeton University (Princeton, NJ 08544, USA)
**Bhavin Shastri**, Queen's University (Kinston, Ontario K7L 3N6, Canada)
**Paul Prucnal**, Princeton University (Princeton, NJ 08544, USA)
weipengz@princeton.edu, jlederman@princeton.edu, shastri@ieee.com, prucnal@princeton.edu

## Status

Upcoming generations of radio-frequency (RF) protocols aim to fulfill rising wireless communication demands by expanding to higher carrier frequencies and increasing spatial multiplexing. These new features require new RF processing methods, and a promising technology for implementing next-generation RF processing can be found in microwave photonics (MWP). MWP combines RF, optical, and optoelectronic devices to create a powerful cognitive radio platform with high bandwidth, low latency, and high energy efficiency [1]. MWP researchers are actively working to enhance device performance [2, 3], to incorporate novel photonic-enabled functionalities [4-10], and to investigate application domains including beamforming [11-12], interference cancellation [13-15], and general-purpose microwave processing [16]. That said, MWP faces packaging and control challenges that must be addressed, as well as processing limitations that make it infeasible to construct a pure-photonic cognitive radio system. Instead, MWP systems for low-latency, broadband processing can be combined with digital electronics for post-processing, high-level analysis, and control to create unified systems with the advantages of both technologies. Such systems can leverage feedback loops between fast electronics such as field-programmable gate arrays (FPGAs) and tunable photonic devices to enable real-time adaptivity and reconfigurable processing [14, 16].

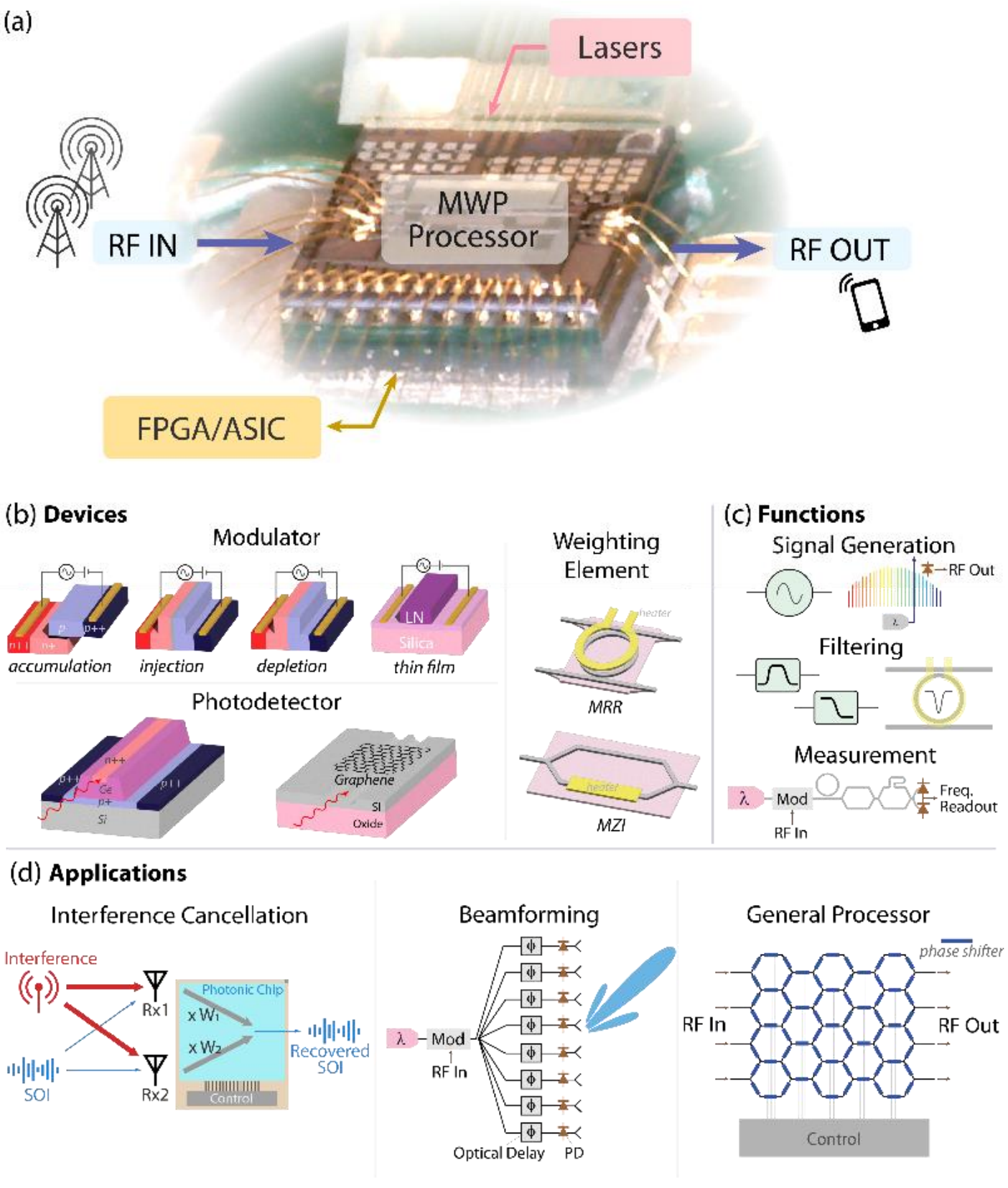

**Figure 1.** (a) An example photonic cognitive radio processor. Latest research progress in MWP: (b) improved devices, (c) newly enabled functions, (d) demonstrated applications. FPGA, field-programmable gate array. ASIC, application-specific integrated circuit. MRR, microring resonator. MZI, Mach-Zehnder interferometer. SOI, Signal of interest.

## Current and Future Challenges

Designing effective MWP systems is made challenging by several factors. Photonic devices, and micro-ring resonators in particular, are sensitive to fabrication variation, resulting in “identical” chips that operate differently. Lasers on silicon photonic chips have traditionally been unavailable, and off-chip lasers result in additional packaging complexity and coupling loss. Each tunable photonic device in a system requires at least one independent analog electronic control signal, adding further packaging complexity and digital-to-analog converter (DAC) cost. On top of this, care must be taken to minimize any thermal, shot, and relative-intensity noise introduced by the photonic components. Implementing photonic tunable time-delay, essential to RF beamforming and cancellation, can be particularly difficult due to the long waveguides and series of switches that are traditionally required.

Applying MWP systems presents additional challenges. Micro-rings, generally essential to wavelength-division multiplexed MWP implementations, are sensitive to temperature variation, reducing processing accuracy and precision unless compensatory strategies are implemented. Not all types of processing are possible with MWP, requiring the introduction of digital electronic processing and performant coordination of the photonics with the electronics. Finally, MWP systems have typically had a fixed architecture, without the capability to reconfigure depending on the target application.

## Advances in Science and Technology to Meet Challenges

The on-chip integration of other technologies—laser sources and complementary metal-oxide-semiconductor (CMOS) digital circuitry—with photonics represents a major avenue for addressing packaging challenges associated with MWP. Multiple methods of on-chip laser integration have recently been demonstrated [17], and though there are costs in footprint, the advantages in packaging and on-chip optical power are significant. On-chip laser comb generation could allow wavelength-division multiplexed processing with a single laser source [18]. CMOS integration would simplify coordinated processing between FPGAs and photonic systems and allow the integration of DACs to provide analog control capacity. Integrated RF amplifiers would also aid optical modulation.

Other solutions include post-fabrication trimming to address fabrication variation and non-volatile device tuning using phase-change or photochromic materials to simplify control. Active feedback stabilization can address the temperature sensitivity of photonic devices. A balanced photonic cancellation architecture can be applied to reduce relative intensity noise, the dominant noise contributor at high optical powers [19]. Tunable phase control can be implemented with low chip area using a series of resonators, or alternatively greater levels of time delay can be achieved with the same amount of chip area with multi-layer chips and taper-shaped transitions [20].

## Concluding Remarks

Recent developments in the design of integrated photonic devices, the integration of photonic and electronic devices, and the coordination and control of electronic/photonic systems have enabled microwave photonics to approach the maturity necessary to address new demands on wireless communication systems, potentially ushering in a transformation of the wireless landscape.

## Acknowledgments

This study is supported by the National Science Foundation (NSF) (ECCS-2128616 to P.R.P.) and the Office of Naval Research (ONR) (N00014-18-1-2527 P.R.P).

## References

[1] D. Marpaung, J. Yao, and J. Capmany, “Integrated microwave photonics,” Nature Photonics, vol. 13, no. 2, pp. 80–90, 2019.

[2] A. Rahim et al., “Taking silicon photonics modulators to a higher performance level: state-of-the-art and a review of new technologies,” Advanced Photonics, vol. 3, no. 2, pp. 024003–024003, 2021.
[3] H. Feng et al., “Integrated lithium niobate microwave photonic processing engine,” Nature, pp. 1–8, 2024.
[4] Y. Yu and X. Sun, “Surface acoustic microwave photonic filters on etchless lithium niobate integrated platform,” in CLEO: Science and Innovations, 2023, pp. SW3L-3.
[5] S. Gertler et al., “Narrowband microwave-photonic notch filters using Brillouin-based signal transduction in silicon,” Nature Communications, vol. 13, no. 1, p. 1947, 2022.
[6] O. Daulay et al., “Ultrahigh dynamic range and low noise figure programmable integrated microwave photonic filter,” Nature Communications, vol. 13, no. 1, p. 7798, 2022.
[7] I.-T. Chen, B. Li, S. Lee, S. Chakravarthi, K.-M. Fu, and M. Li, “Optomechanical ring resonator for efficient microwave-optical frequency conversion,” Nature Communications, vol. 14, no. 1, p. 7594, 2023.
[8] S. Sun et al., “Integrated optical frequency division for microwave and mmWave generation,” Nature, pp. 1–6, 2024.
[9] I. Kudelin et al., “Photonic chip-based low-noise microwave oscillator,” Nature, pp. 1–6, 2024.
[10] Y. Tao et al., “Fully on-chip microwave photonic instantaneous frequency measurement system,” Laser & Photonics Reviews, vol. 16, no. 11, p. 2200158, 2022.
[11] C. Zhu et al., “Silicon integrated microwave photonic beamformer,” Optica, vol. 7, no. 9, pp. 1162–1170, 2020.
[12] P. Martinez-Carrasco, T. H. Ho, D. Wessel, and J. Capmany, “Ultrabroadband high-resolution silicon RF-photonic beamformer,” Nature Communications, vol. 15, no. 1, p. 1433, 2024/
[13] W. Zhang et al., “Broadband physical layer cognitive radio with an integrated photonic processor for blind source separation,” Nature Communications, vol. 14, no. 1, p. 1107, 2023.
[14] W. Zhang et al., “A system-on-chip microwave photonic processor solves dynamic RF interference in real time with picosecond latency,” Light: Science & Applications, vol. 13, no. 1, p. 14, 2024.
[15] J. C. Lederman et al., “Real-time photonic blind interference cancellation,” Nature Communications, vol. 14, no. 1, p. 8197, 2023.
[16] D. Pérez-López et al., “General-purpose programmable photonic processor for advanced radiofrequency applications,” Nature Communications, vol. 15, no. 1, p. 1563, 2024.
[17] C. Xiang et al., “3D integration enables ultralow-noise isolator-free lasers in silicon photonics,” Nature, vol. 620, no. 7972, pp. 78–85, 2023.
[18] P. Del’Haye, A. Schliesser, O. Arcizet, T. Wilken, R. Holzwarth, and T. J. Kippenberg, “Optical frequency comb generation from a monolithic microresonator,” Nature, vol. 450, no. 7173, pp. 1214–1217, 2007.
[19] E. C. Blow et al., “Radio-Frequency Linear Analysis and Optimization of Silicon Photonic Neural Networks,” Advanced Photonics Research, p. 2300306, 2024.
[20] W. D. Sacher et al., “Monolithically integrated multilayer silicon nitride-on-silicon waveguide platforms for 3-D photonic circuits and devices,” Proceedings of the IEEE, vol. 106, no. 12, pp. 2232–2245, 2018.

# Photonic Ising machines: Towards acceleration of industrial optimization problems

**Fabian Böhm, Thomas Van Vaerenbergh**
[1]Hewlett Packard Labs, Hewlett Packard Enterprise, 71034 Böblingen, Germany
[2]Hewlett Packard Labs, Hewlett Packard Enterprise, 1831 Diegem, Belgium

[ fabian.bohm@hpe.com , thomas.van-vaerenbergh@hpe.com , ]

## Status

Combinatorial optimization problems are ubiquitous and highly relevant in various industrial applications, such as logistics, drug design and machine learning. Notably, many of these problems are resource and time-intensive to solve on digital computers, which makes more efficient computing concepts highly desirable. Here, optical Ising machines have emerged as a physical computing concept that promises faster computation at reduced energy consumption [1]. Ising machines exploit the fact that many of these problems can be mapped to the energy function of Ising models, such that (sub)optimal solutions correspond to low-energy configurations of the Ising Hamiltonian (See Fig.1). This simple model with binary states and linear coupling can then be realized in optical hardware, that will rapidly minimize the Ising energy and thereby find solutions to hard optimization problems. While optical Ising machines have been around since the 1980s [2], they have experienced a surge in popularity in the last 10 years. This has been spurred by the discovery that minimization of the Ising energy can be achieved with ultrafast optical processes, such as mode competition in coupled lasers [3]. Since then, a plethora of Ising machines have been proposed and demonstrated. Today's optical Ising machines can be broadly categorized by whether they are purely analog or hybrid analog-digital system. In purely analog systems, state generation, coupling and energy minimization are all realized with analog signals in a time-continuous feedback system. Examples of such systems are optically coupled parametric oscillators [4]. In hybrid systems, part of these functionalities is implemented with digital components, hence requiring analog-digital conversion in a time-discrete feedback system. Examples of this are Ising machines with FPGA-based feedback systems [5]. Both analog and hybrid approaches have been realized as integrated [6,7] and as discrete systems [4,5]. While the former allows for compact and mass-fabricable devices, the latter promises scalability to large problem sizes. Examples of integrated systems are coupled ring resonators that implement Ising machines on a single photonic chip [6]. In discrete systems, the same can for example be achieved in ring fiber cavities [4,5]. At this point, optical Ising machines have demonstrated speed ups over conventional optimization methods, such as simulated annealing, by a factor of x100-x1000 for several hard benchmark tasks [1,5]. Still, it remains a challenge to develop Ising machines into computing systems that can convincingly outperform digital computers in industrial use-cases.

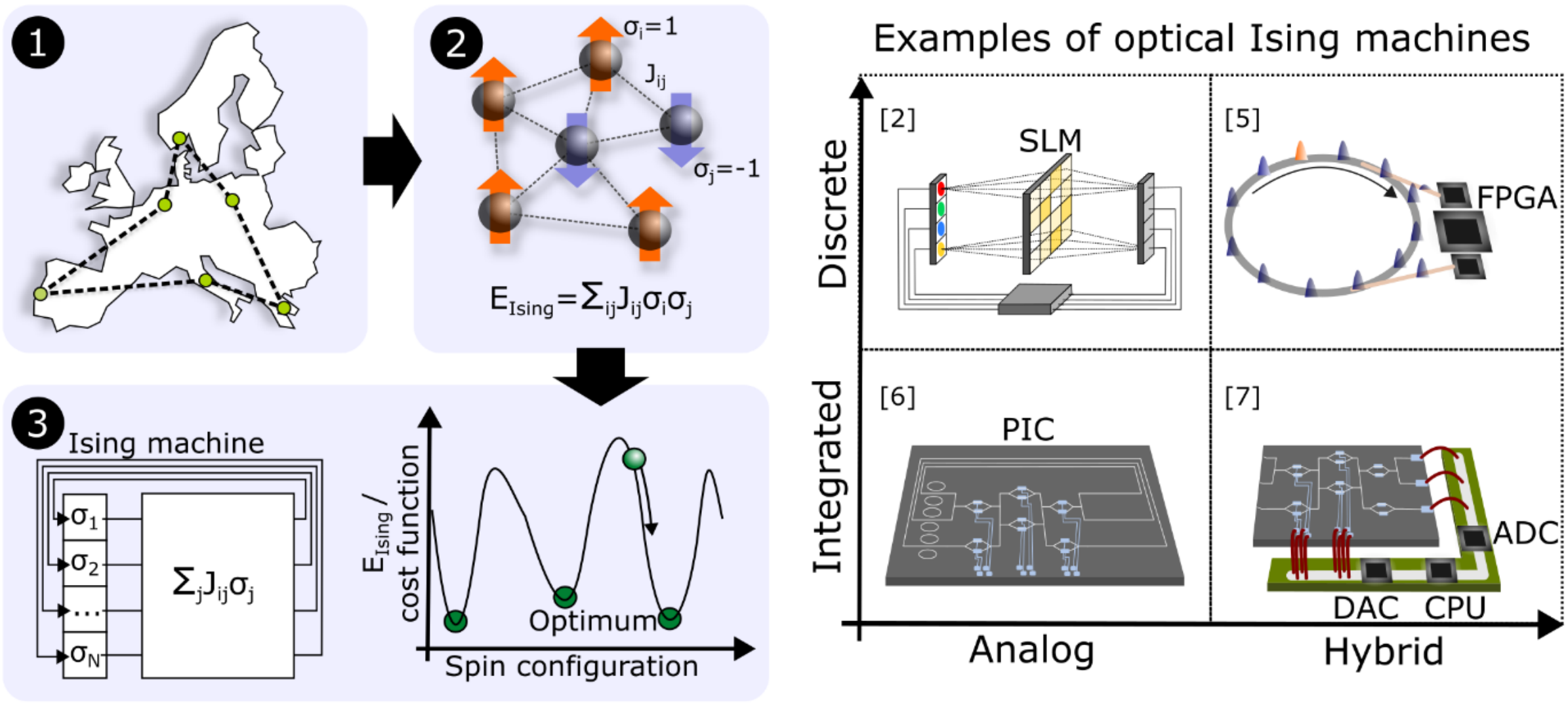

**Figure 1.** (a) Operating principle of Ising machines. An optimization problem (1), such as route planning, is mapped to the energy function of an Ising spin network (2). This Ising model is then implemented in an Ising machine, which will minimize the Ising energy and thereby find the optimal solution (3). (b) Examples of Ising machines for integrated analog [6], integrated hybrid [7], discrete hybrid [5] and discrete analog [2] optical Ising machines.

## Current and Future Challenges

To merit the partial replacement or enhancement of conventional optimization solvers with Ising machines, they must demonstrate superiority over digital computers in several performance metrics, such as the time to solve a problem (time-to-solution) and energy consumption, while being equally affordable and scalable to large problem sizes. Notably, while Ising machines have shown performance advantages for select academic problems, there are still considerable challenges when applying Ising machines to more general industry-relevant use cases.

### Scalability and generalizability to industrial use cases

Contrary to typical academic benchmark instances, industry-relevant use cases can contain millions of problem variables [8]. The implementation and coupling of that many spin variables in an optical system is still difficult to achieve, particularly for integrated Ising machines. Another challenge is that industrial use cases may often have intrinsic structure that can be easily exploited by tailored conventional optimization algorithms for faster computation. It is still an open question whether Ising machines can maintain this advantage when the problem is mapped from its native form to the Ising model.

### Bit resolution and spin amplitude inhomogeneity

Industrial use cases can require large dynamic ranges in the spin coupling weights $J_{ij}$ of the Ising Hamiltonian [9], e.g., when they represent different distance scales (meters vs kilometres) in route planning problems. For Ising machines, where the spin coupling is implemented in the optical domain, the achievable bit resolution of matrix-vector-multipliers cannot yet support these different scales [9,10]. Moreover, in analog Ising machines, the continuous spin variables need to be driven to a binary configuration to correctly embed the Ising model. Deviations from this binarization are inducing errors in the embedding, which considerably deteriorates performance [11,12].

### Embedding overhead of higher-order terms and hard constraints

As most Ising machines minimize the quadratic Ising Hamiltonian, they are not suited to natively embed higher-order variable interactions and hard constraints, which are common in industrial use cases. Embedding such problems currently requires the introduction of additional penalty terms and auxiliary variables in the Ising Hamiltonian. This overhead often increases the number of spin variables, the required resolution, and the complexity of the problem, thereby deteriorating performance [13,14,15].

**Miniaturization and integration in conventional computing systems**

As dedicated accelerators, Ising machines require a conventional computing environment for general purpose computing and storage. To incorporate Ising machines into existing HPC or edge-computing systems, they must become affordable and achieve a compact form factor, e.g., as plug-in cards.

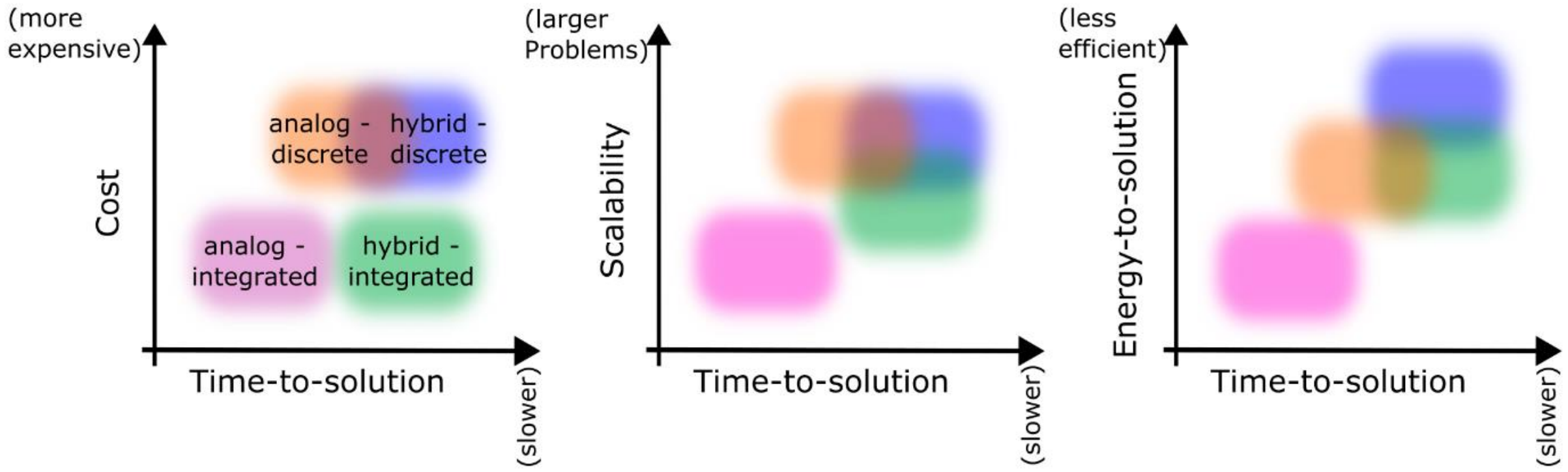

**Figure 2.** Sketch of the potential relative performance of different optical Ising machine concepts with regards to time-to-solution, energy-to-solution, cost, and scalability.

## Advances in Science and Technology to Meet Challenges

Considering the potential of Ising machines to outperform digital computers, analog systems stand out in many of the aforementioned performance metrics (see Fig.2). Whereas hybrid systems can be limited by their digital components (e.g., clocked operation and energy intensive analog-digital conversion), analog system can leverage parallelism and convergence at high analog bandwidths for considerably faster computation [6]. Moreover, integrated Ising machines enable low-energy and low-cost devices that can be mass-manufactured and are incorporable into existing digital computing systems. However, as introduced in the previous section, the performance of analog Ising machines is limited by their scalability, bit resolution, embedding overhead and amplitude inhomogeneity. To address the issue of scalability, considerably efforts are undertaken in designing tiled architectures [16,17]. Here, large-scale optimization methods are split into smaller sub-problems, which are then solved with several small-scale Ising machines, that are interconnected as chiplets. With advances in large-scale photonic integrated circuits, tiled architectures could thus enable scalable integrated Ising machines with millions of variables in a small footprint. With regards to bit resolution, there is growing interest in embedding methods that can lower the required bit resolution by introducing additional variables. For MIMO problems, it has been shown that the resolution and the performance of Ising machine can be enhanced with a linearly increasing overhead [9]. This method can be similarly applied to other problems, hence providing a way to apply Ising machines to error-sensitive applications. To tackle the embedding overhead of the Ising model, there is growing interest in designing Ising machines that natively support higher-order interactions in their energy function. In electronic Ising machines, the inclusion of such higher order terms has already shown to lead to considerable performance improvements [13]. Similar efforts exist to achieve this with optical Ising machines, e.g., by exploiting nonlinear optical processes or optical interference [18,19]. To negate the negative effect of amplitude inhomogeneity, feedback systems have been proposed that enforce binarization through an error correction signal [11]. However, such feedback systems have primarily been demonstrated in simulations or with digital processing systems and experimental realizations with purely analog hardware will be an important challenge in the future. Additionally, there is interest in reducing inhomogeneity through suitable choices of the nonlinear optical system, that facilitates binarization of the spin amplitudes [12]. Moreover, programmable optical nonlinearities can create the opportunity to design transfer functions to yield better performance.

## Concluding Remarks

With the current rapid rise in demand for more computing power, the development of alternative computing architectures is becoming increasingly relevant to ensure a sustainable future computing infrastructure. By tackling some of the most resource intensive applications in optimization, Ising machines promise faster and more efficient computation for important industrial use cases. The past decade of research has already brought with it a plethora of ways for building practical Ising machines, which have partly shown favourable performance compared to digital computers in academic benchmark problems. Crucially, the next decade will likely be devoted to moving past academic benchmarks and towards proving the usefulness of Ising machines in real-world use cases. Next to the various engineering challenges, this will also involve a more critical assessment of benchmarking against conventional optimization methods [20]. As for many unconventional computing methods, benchmarking tools that can facilitate a fair base of comparison are still scarce. However, understanding the best way to evaluate the performance, selecting appropriate benchmark tasks, and performing fair comparisons against state-of-the-art optimization methods across all platforms will be a crucial step to ultimately merit the use of Ising machines.

## Acknowledgements

This work is partially supported by the Defense Advanced Research Projects Agency (DARPA) under the Air Force Research Laboratory (AFRL) Agreement No. FA8650-23-3-7313. The views, opinions, and/or findings expressed are those of the authors and should not be interpreted as representing the official views or policies of the Department of Defense or the U.S. Government.

## References

[1] N. Mohseni, P. L. McMahon and T. Byrnes, “Ising machines as hardware solvers of combinatorial optimization problems,” Nat. Rev. Phys., vol. 4, pp. 363-379, 2022.

[2] N. Farhat, D. Psaltis, A. Prata and E. Paek, “Optical Implementation of the Hopfield model,” Appl. Opt., vol. 24, pp. 1469-1475, 1985.

[3] S. Utsunomiya, K. Takata and Y. Yamamoto, “Mapping of Ising models onto injection-locked laser systems,” Opt. Express, vol. 19, pp. 18091-18108, 2011.

[4] T. Inagaki, K. Inaba, R. Hamerly, K. Inoue, Y. Yamamoto and H. Takesue, “Large-scale Ising spin network based on degenerate optical parametric oscillators,” Nat. Photon., vol. 10, pp. 415-419, 2016.

[5] T. Honjo et al., “100,000-spin coherent Ising machine,” Sci. Adv., vol. 7, eabh0952, 2021.

[6] N. Tezak et al., “Integrated Coherent Ising Machines Based on Self-Phase Modulation in Microring Resonators,” IEEE J. Sel. Top. Quantum Electron., vol. 26, 5900115, 2020.

[7] M. Prabhu et al., “Accelerating recurrent Ising machines in photonic integrated circuits,” Optica, vol. 7, pp. 551-558, 2020.

[8] M. Schlenkirch and S.N. Parragh, “Solving large scale industrial production scheduling problems with complex constraints: an overview of the state-of-the-art,” Procedia Comput. Sci., vol. 217, pp. 1028-1037, 2023.

[9] A. K. Singh and K. Jamieson, “Multi Digit Mapping for Low Precision Ising Solvers,” 2024, arXiv:2404.05631.

[10] S. Garg, J. Lou, A. Jain, Z. Guo, B. J. Shastri and M. Nahmias, “Dynamics Precision Analog Computing for Neural Networks,” IEEE J. Sel. Top. Quantum Electron., vol. 29, no. 2, 6100412, 2023.

[11] T. Leleu, Y. Yamamoto, P. L. McMahon and K. Aihara, “Destabilization of Local Minima in Analog Spin Systems by Correction of Amplitude Heterogeneity,” Phys. Rev. Lett., vol. 122, 040607, 2019.

[12] F. Böhm, T. Van Vaerenbergh, G. Verschaffelt and G. Van der Sande, “Order-of-magnitude” differences in computational performance of analog Ising machines induced by the choice of nonlinearity,” Commun. Phys., vol. 4, 149, 2021.

[13] M. Hizzani et al., “Memristor-based hardware and algorithms for higher-order Hopfield optimization solver outperforming quadratic Ising machines,” 2023, arXiv:2311.01171.

[14] D. Dobrynin et al., “Disconnectivity graphs for visualizing combinatorial optimization problems: challenges of embedding to Ising machines,” 2024, arXiv:2403.01320.

[15] E. Valiante, M. Hernandez, A. Barzegar and H. G. Katzgraber, „Computational overhead of locality reduction in binary optimization problems,” Comput. Phys. Commun., vol. 269, 108102, 2021.

[16] A. Sharma, R. Afoakwa, Z. Ignjatovic and M. Huang, “Increasing Ising machine capacity with multi-chip architectures,” ISCA ’22: Proceedings of the 49th Annual International Symposium on Computer Architecture, pp. 508-521, 2022.

[17] G. H. Hutchinson, E. Sifferman, T. Bhattacharya and D. B. Strukov, “FPIA: Field-Programmable Ising Arrays with In-Memory Computing,” 2024, arXiv2401.16202.

[18] S. Kumar, H. Zhang and Y. Huang, “Large-scale Ising emulation with four body interaction and all-to-all connections,” Commun. Phys., vol. 3, 108, 2020.

[19] M. K. Bashar and N. Shukla, “Designing Ising machines with higher order spin interactions and their application in solving combinatorial optimization,” Sci. Rep., vol. 13, 9558, 2023.

[20] D. E. B. Neira et al., „Benchmarking the Operation of Quantum Heuristics and Ising machines: Scoring Parameter Setting Strategies on Optimization Applications,” 2024, arXiv:2402.10255.